\documentclass[10pt,a4paper,twoside]{report}

\usepackage[square]{natbib}          

\usepackage{notoccite}

\usepackage[utf8]{inputenc}   

\usepackage[english]{babel} 

\usepackage{nomencl}
\makenomenclature
\RequirePackage{ifthen} 
\ifthenelse{\equal{\languagename}{english}}%
    { 
    \renewcommand{\nomgroup}[1]{%
      \ifthenelse{\equal{#1}{R}}{%
        \item[\textbf{Roman symbols}]}{%
        \ifthenelse{\equal{#1}{G}}{%
          \item[\textbf{Greek symbols}]}{%
          \ifthenelse{\equal{#1}{S}}{%
            \item[\textbf{Subscripts}]}{%
            \ifthenelse{\equal{#1}{T}}{%
              \item[\textbf{Superscripts}]}{}}}}}%
    }{%
    \ifthenelse{\equal{\languagename}{french}}%
    { 
    \renewcommand{\nomgroup}[1]{%
      \ifthenelse{\equal{#1}{R}}{%
        \item[\textbf{Symbole romains}]}{%
        \ifthenelse{\equal{#1}{G}}{%
          \item[\textbf{Symboles grecs}]}{%
          \ifthenelse{\equal{#1}{S}}{%
            \item[\textbf{Indices}]}{
            \ifthenelse{\equal{#1}{T}}{%
              \item[\textbf{Exposants}]}{}}}}}
    }}%

\usepackage[acronym,nonumberlist,nopostdot,nogroupskip]{glossaries}
\setglossarystyle{long}
\makeglossaries

\newcommand{\acknowledgments}{@undefined} 
\addto\captionsenglish{\renewcommand{\acknowledgments}{Acknowledgments}}
\addto\captionsenglish{\renewcommand{\acronymname}{Glossary}} 

\newcommand{\coverThesis}{@undefined} 
\newcommand{\coverSupervisors}{@undefined} 
\addto\captionsenglish{\renewcommand{\coverThesis}{A thesis presented for the degree of Doctor of Philosophy in}}
\addto\captionsenglish{\renewcommand{\coverSupervisors}{Supervisor}}
\addto\captionsenglish{}
\addto\captionsenglish{}
\addto\captionsenglish{}
\addto\captionsenglish{}
\addto\captionsfrench{\renewcommand{\coverThesis}{Th\`ese pour l'obtention du Maîtrise des Sciences en}}
\addto\captionsfrench{\renewcommand{\coverSupervisors}{Directeur(s) de th\`ese}}
\addto\captionsfrench{}
\addto\captionsfrench{}
\addto\captionsfrench{}
\addto\captionsfrench{}

\def\FontMn{
  \usefont{T1}{phv}{m}{n}\fontsize{14pt}{14pt}\selectfont}

\usepackage{geometry}	
\usepackage{setspace}

\usepackage{indentfirst}

\usepackage{graphicx}

\usepackage{amsmath}  
\usepackage{amsthm}   
\usepackage{amsfonts} 
\usepackage{amssymb} 
\usepackage{tabularx} 
\usepackage{caption}  
\usepackage[caption=false,listofformat=subsimple,labelformat=simple]{subfig}

\usepackage{ar}

\usepackage{dcolumn}
\newcolumntype{d}{D{.}{.}{-1}} 
\newcolumntype{e}{D{E}{E}{-1}} 

\usepackage{rotating}

\usepackage[pdftex]{hyperref} 
\hypersetup{colorlinks=true,       
           linkcolor=black,    
            anchorcolor=black,
            citecolor=violet,  
            filecolor=black,  
            menucolor=black,  
            urlcolor=cyan,    
	          bookmarks=true,         
	          bookmarksopen=false,    
	          bookmarksnumbered=true, 
	          pdftitle={Thesis},
            pdfauthor={H. T. Moges},
            pdfsubject={Thesis Title},
            pdfkeywords={Thesis Keywords},
            pdfstartview=FitV,
            pdfdisplaydoctitle=true}

\usepackage[figure,table]{hypcap}

\usepackage{multirow}

\usepackage{booktabs}

\usepackage{pdfpages}

\usepackage{algorithm}
\usepackage{algpseudocode}

\usepackage{enumitem}

\newacronym{ode}{ODE}{ordinary differential equation}
\newacronym{dof}{DoF}{degrees of freedom}
\newacronym{eom}{EoM}{equations of motion}
\newacronym{pss}{PSS}{Poincar{\'e} surface of section}
\newacronym{mle}{mLE}{maximum Lyapunov exponent}
\newacronym{le}{LE}{Lyapunov exponent}
\newacronym{nd}{ND}{$N$-dimensional}
\newacronym{sali}{SALI}{smaller alignment index}
\newacronym{gali}{GALI}{generalized alignment index}
\newacronym{cr}{CR}{color and rotation}
\newacronym{rk}{RK}{Runge-Kutta}
\newacronym{ics}{IC}{initial condition}
\newacronym{po}{PO}{periodic orbit}
\newacronym{mfl}{MFL}{magnetic field line}
\newacronym{mf}{MF}{magnetic field}
\newacronym{lar}{LAR}{large aspect ratio}
\newacronym{com}{CoM}{constant of motion}
\newacronym{ds}{DS}{dynamical system}
\newacronym{svd}{SVD}{singular value decomposition}
\newacronym{gc}{GC}{guiding center}
\newacronym{gcm}{GCM}{guiding center motion}
\newacronym{gm}{GM}{galaxy model}
\newacronym{sm}{SM}{standard map}
\newacronym{am}{AM}{accelerator mode}
\newacronym{chpc}{CHPC}{center for high-performance computing}
\newacronym{mhd}{MHD}{magnetohydrodynamics}

\glsaddall

\begin{document}

\pagestyle{plain}

\pagenumbering{roman}


\thispagestyle {empty}


\begin{center}
%
\vspace{2.5cm}
\includegraphics[height=50mm]{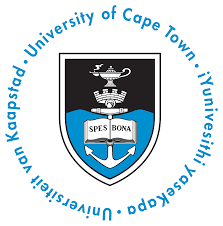}

\vspace{2.0cm}
\rule{\linewidth}{0.1mm} \\[0.4cm]
{\FontMn Hamiltonian Chaos: From Galactic Dynamics to Plasma Physics} \\ 
\rule{\linewidth}{0.6mm} \\[1cm]
\vspace{2.cm}
{\FontMn Henok Tenaw MOGES} \\ 
\vspace{2.6cm}
{ \coverThesis} \\
{the Department of Mathematics and Applied Mathematics} \\ 
\vspace{1.0cm}
  \coverSupervisors: Prof. Haris SKOKOS \\ 
%
\vspace{1.9cm}
{December 2024} \\ 
\end{center}
\clearpage



\subsection*{Declaration}

I, \underline{H.~T.~Moges}, hereby declare that the work on which this thesis is based is my original work (except where acknowledgments indicate otherwise) and that neither the whole work nor any part of it has been, is being, or is to be submitted for another degree in this or any other university. I authorize the University to reproduce for the purpose of research either the whole or any portion of the contents in any manner whatsoever.

	\vspace{2cm}


\clearpage


\section*{\acknowledgments}

\addcontentsline{toc}{section}{\acknowledgments}

I would like to begin by expressing my gratitude to the Department of Mathematics and Applied Mathematics at the University of Cape Town (UCT) for the support throughout my PhD journey. This work was funded by Woldia University and the Ethiopian Ministry of Education, as well as the UCT Science Faculty PhD Fellowship Award (mid-2020 to mid-2023). I am also grateful for the partial support I received from the UCT international students scholarship (2021-2024) as well as the Erasmus+ program, which financed my two research visits to Athens, Greece. A special thanks to the Center for High-Performance Computing (CHPC) for providing all the computational resources that made this research possible.

My PhD journey began with a brief email and a Skype call with Prof.~Haris Skokos, during which he candidly remarked, ``If you want to join my research group, you must be aware that I am demanding''. I agreed, and true to his word, he provided unwavering support and insightful feedback, guiding me throughout my research in the most enriching way possible. Prof.~Skokos, thank you for teaching me everything I know about chaotic dynamics and scientific research and for making this journey an incredible experience. I would also like to thank my colleagues in the `Nonlinear Dynamics and Chaos' group, both past and present, including Dr. Bertin Many Manda, Dr. Malcolm Hillebrand, Dr. Bob Senyange, Dr. Arnold Ngapasare, Jean-Jacq Du Plessis, San{\'e} Erasmus, Dylan Theron, and Cassandra Barbis, for proofreading parts of this work. A big thank you to the former members for the invaluable discussions and the many unforgettable moments we shared at the start of my research journey.

I am also grateful to those who have worked with me on various projects, including Yannis Antonenas, Dr. Giorgos Anastassiou, and Dr. Ovidiu Racoveanu. Special thanks to A/Prof.~Thanos Manos for his patience and support during our collaborations. I would like to thank A/Prof.~Yannis Kominis and his research group, particularly for their useful discussions and for making me feel welcome during my six-month research visits at the National Technical University of Athens, Greece. A special thank you goes to Prof.~Panos Patsis and Dr. Matthaios Katsanikas for your guidance and kindness during our collaboration, as well as for being great hosts during my second visit to Greece at the Research Center for Astronomy at the Academy of Athens. A second round of thanks to Malcolm for all the useful discussions and your help with the derivation of the Ferrers bar derivatives.

Lastly, I want to express my heartfelt thanks to my family, friends, and colleagues for their continuous support. Special thanks goes to Prof.~Abebe Zegeye and Rachel Browne, whose encouragement has been invaluable. Rachel, thank you for being a great friend and for all the fun chats and meals we have shared. To Julia Albu (RIP) and THE FAMILY, especially Katherine and Joost as well as Zambi, thank you for welcoming me into your home and always including me in all the delicious dinners. Joost, I will never forget your kindness and your unforgettable braais. Finally, I want to thank my parents and siblings, especially my partner-in-crime, Eyerusalem Kebede (J), and our Amani. J, even though you were not there at the start of my PhD journey (and probably did not sign up for the chaos), you have embraced it with understanding and unwavering support. Thank you for always being there and for not running away yet! 
\clearpage



\section*{Abstract}

\addcontentsline{toc}{section}{Abstract}

The primary focus of this thesis is the numerical investigation of the influence of chaos in Hamiltonian models describing the behavior of charged particle orbits in plasma, the motion of stars in barred galaxies, and the diffusion of trajectories in multidimensional maps. First, we systematically explore the interplay between magnetic and kinetic chaos in toroidal fusion plasmas, where non-axisymmetric perturbations disrupt smooth magnetic flux surfaces, generating complex particle trajectories. Using the Generalized Alignment Index (GALI) method of chaos detection, we efficiently quantify chaos, compare the behavior of magnetic field lines and particle orbits, visualize the radial distribution of chaotic regions, and offer the GALI method as a valuable tool for studying the dynamics of plasma physics models. Next, we study the evolution of phase space structures in a three-dimensional (\(3D\)) barred galactic potential, following successive \(2D\) and \(3D\) pitchfork and period-doubling bifurcations of periodic orbits. By employing the so-called 'color and rotation' technique to visualize the four-dimensional Poincar\'e surface of sections of the system, we reveal distinct structural patterns. We further investigate the long-term diffusion transport and chaos properties of single and coupled standard maps, focusing on parameters that induce anomalous diffusion through the presence of accelerator modes exhibiting ballistic transport. Using different ensembles of initial conditions in chaotic regions influenced by these modes, we examine asymptotic diffusion rates and their corresponding time scales, identifying conditions that suppress anomalous transport and lead to long-term convergence to normal diffusion across various coupled map arrangements. Lastly, we perform the first comprehensive investigation into the behavior of the GALI indices for various attractors in continuous and discrete-time dissipative systems, extending the application of the method to non-Hamiltonian systems. A key aspect of our work involves analyzing and comparing the performance of the GALI method with the computation of Lyapunov Exponents for non-Hamiltonian dissipative systems exhibiting hyperchaotic motion.
\clearpage


\section*{Dissemination of the results of this PhD research}

The results of this thesis have been disseminated through publications in international peer-reviewed journals:

\begin{itemize}
    \item  \textbf{Moges H.T.}, Manos T., Skokos Ch.: ``Anomalous diffusion in single and coupled standard maps with extensive chaotic phase spaces,'' Physica D, 431, 133120, 2021 

    \item \textbf{Moges H.T.}, Antonenas Y., Anastassiou G., Skokos Ch., Kominis Y.: ``Kinetic versus Magnetic Chaos in Toroidal Plasmas: A systematic quantitative comparison,'' Physics of Plasmas, 31, 012302, 2024    

	\item  \textbf{Moges H.T.}, Katsanikas M., Hillebrand M., Patsis P. A., Skokos Ch.: ``The evolution of phase space along a series of supercritical bifurcations in a 3D galactic bar potential,'' International Journal of Bifurcation and Chaos, 34, 2430013, 2024

    \item \textbf{Moges H.T.}, Racoveanu O., Manos T., Skokos Ch.: ``On the behavior of the Generalized Alignment Index Method (GALI) for non-Hamiltonian dissipative systems,'' International Journal of Bifurcation and Chaos, 2025, (accepted for publication - in press, preprint arXiv:2503.01784)
\end{itemize} 

Conference presentations showcasing parts of the results of this thesis are:
\begin{itemize}
    \item The 12th International Conference on Mathematics and Physics (NoLineal 20-21), Madrid, Spain, June 30-July 2, 2021. Poster presentation: ``On the behavior of the Generalized Alignment Index (GALI) method for regular motion in multidimensional Hamiltonian systems''. 
    
    \item The 3rd International Conference on Integrable Systems and Nonlinear Dynamics and  the School ``Integrable and Nonlinear Days'', Yaroslavl, Russia, October 04-08, 2021. Oral Presentation: ``On the behavior of the Generalized Alignment Index (GALI) method for regular motion in multidimensional Hamiltonian systems''.
    
    \item The International Conference on Nonlinear Science and Complexity, Thessaloniki, Greece, September 26-29, 2022. Oral Presentation: ``Anomalous diffusion in single and coupled standard maps with extensive chaotic phase spaces''

    \item The 10th International Congress on Industrial and Applied Mathematics (ICIAM), Waseda University, Tokyo, Japan, August 20-25, 2023. Oral Presentation: ``Anomalous diffusion in standard maps with extensive chaotic phase spaces''

     \item  The Applied Nonlinear Dynamical Systems and Chaos (ANDSC) Conference, hosted by the Royal Academy of Sciences, Madrid, Spain, July 1-3, 2024. Oral Presentation: ``Quantification and Comparison of Magnetic and Kinetic Chaos in Toroidal Plasmas''
     
     \item The 5th International Conference on Integrable Systems and Nonlinear Dynamics (ISND-2024), Yaroslavl, October 7-11, 2024. Oral Presentation: ``Quantification and Comparison of Magnetic and Kinetic Chaos in Toroidal Plasmas'' 
\end{itemize}

\clearpage


%
\tableofcontents
\hypersetup{
    linkcolor=brown, 
}
\clearpage 



%

%
\phantomsection
\addcontentsline{toc}{section}{\acronymname}
\printglossary[type=\acronymtype,title=\acronymname, toctitle=\acronymname]
\clearpage

%
\setcounter{page}{1}
\pagenumbering{arabic}



\chapter{General introduction}\label{chapter:introduction}
Dynamical systems (DSs) theory has been widely applied to solve complex problems in diverse fields, such as plasma physics \citep{WhiteBook}, galactic dynamics \citep{contopoulos2002order}, fluid dynamics \citep{white2003fluid}, and even machine learning (e.g., artificial neural networks) \citep{brunton2022data}. Thus, due to its broad applicability, DS theory has become integral to modern scientific research. An important approach in this domain is the Hamiltonian formalism, which has proven particularly effective in studying chaotic behavior in many physical systems (e.g.~see \citep{bountis2012complexHAM}). Several numerical methods have been developed to investigate the dynamics and detect chaos in these systems. For systems with low number of degrees of freedom (DoF), techniques such as the Poincar{\'e} surface of section (PSS) (e.g.~see \citep{lichtenberg2013regular}) and the color and rotation (CR) method \citep{patsis1994using} enable us to visualize the dynamics. In order to identify and quantify chaos, tools such as the computation of the maximum Lyapunov exponent (mLE) \citep{Benettin1980a}, the smaller alignment index (SALI) \citep{skokos2001alignment}, and the generalized alignment index (GALI) \citep{skokos2016smaller} have been successfully applied. These chaos detection techniques can be efficiently computed through methods like the tangent map \citep{Skokos2010variational}.

The growing interest in magnetically confined fusion plasmas has significantly impacted the development of Hamiltonian techniques for analyzing the topology of magnetic field lines (MFLs) and charged particle motions in various magnetic field (MF) configurations \citep{AbdullaevBook}. In fusion devices, charged particles predominantly follow MFLs, making the topology of these lines an important component for plasma confinement \citep{WhiteBook}. Under the influence of magnetic perturbations, the topology of MFLs can become chaotic, leading to stability problems related to the destruction of magnetic surfaces \citep{abdullaev2008description}. Particle trajectories may also exhibit chaotic behavior in the presence of magnetic perturbations or electromagnetic fields \citep{kominis2010kinetic}. Interestingly, chaotic MFs do not necessarily introduce chaotic particle motion \citep{ram2010dynamics}. A primary aspect of the investigation performed in this thesis was the quantification and understanding of the relationship between the complexity of MFL topology and the dynamics of charged particles, as well as their implications for the transport properties of fusion plasmas.

For \(2\) DoF Hamiltonian systems, the PSS effectively reduces the four-dimensional (\(4D\)) phase space to a \(2D\) one. In higher-dimensional systems, direct phase space visualization becomes infeasible as the phase space dimensionality is at least \(4D\). The CR method addresses this limitation by projecting the \(4D\) phase space onto a \(3D\) subspace while using color to represent the fourth dimension. Applying this technique, we examined in our work the phase space structures around stable and unstable periodic orbits (POs) and their influence on the physical properties of astronomical models \citep{katsanikas2011structure1,patsis2014phasea}. In particular, we investigated the evolution of the phase space structures in a \(3D\) bar galactic potential, following successive \(2D\) and \(3D\) pitchforks, as well as period-doubling bifurcations of POs. 

In addition, the study of diffusion and transport in conservative Hamiltonian systems and symplectic maps remains a significant area of research \citep{Chirikov1979,ManRob2014PRE,kroetz2016hidden}. In this work, we focused on the dynamics of coupled standard map systems, which provide a flexible framework for exploring chaotic and regular dynamics by varying system parameters, such as nonlinear kick and coupling strength \citep{altmann2008anomalous}. In particular, we studied and understood the long-term diffusion properties (normal versus anomalous) and chaotic behavior of such systems. Finally, we presented the first detailed analysis of the performance of the GALI method in non-Hamiltonian dissipative systems, including systems exhibiting hyperchaotic motion.

Although each system we considered required particular methods to capture its underlying dynamics, the overarching goal of this work was the same: to significantly contribute to our understanding of DSs' behavior through the utilization of efficient numerical tools such as the GALI chaos detection technique. By employing systematic computational approaches, we provided new insights into how chaotic dynamics influence various DSs. 

The remainder of this thesis is organized as follows:

\begin{itemize}
\item In Chapter \ref{chapter:two}, we provide a general overview of Hamiltonian mechanics and chaotic dynamics. It reviews numerical integration methods and introduces chaos detection techniques, including the computation of Lyapunov exponents (LEs) and the GALI method, with a particular focus on the behavior of LEs for dissipative DSs. Special emphasis is given on the properties and numerical computation of the GALI method. 

\item In Chapter \ref{chapter:three}, we systematically compare the chaoticity of MFLs and particle orbits in nonaxisymmetric magnetic perturbations of toroidal MF configurations. It introduces the application of the GALI method to plasma physics models (to the best of our knowledge) for the first time and quantitatively analyzes the link between magnetic and kinetic chaos, emphasizing implications for optimizing plasma confinement.

\item In Chapter \ref{chapter:four}, we investigate the evolution of phase space structures in a \(3D\) bar galactic potential for successive \(2D\) and \(3D\) pitchfork and period-doubling bifurcations of POs. Using the CR visualization technique, we examine the structural patterns appearing in the \(4D\) PSSs of the system. 

\item In Chapter \ref{chapter:five}, we explore the long-term diffusion properties of ensembles of orbits in coupled standard maps. By analyzing diffusion rates and related time scales, we highlight the relationship between phase space dynamics, coupling strength between maps, and chaos strength by implementing indicators such as the mLE and the GALI method.

\item In Chapter \ref{chapter:six},  we examine the behavior of the GALI method in continuous and discrete-time non-Hamiltonian dissipative systems. Using models such as the Lorenz system and the generalized H{\'e}non map, we provide the first comprehensive analysis of the behavior of the GALI indices in hyperchaotic systems, comparing their performance with LEs to classify various dynamical regimes.

\item In Chapter \ref{chapter::summary}, we summarize the main findings of our work and outline future research directions in the study of chaotic dynamics. 
\end{itemize}

\clearpage


\chapter{Dynamical systems and numerical methods} \label{chapter:two}
\section{Introduction} \label{sectionCh2:introduction}  
In this chapter, we explore an overview of dynamical systems (DSs), with a particular emphasis on dissipative systems, and present the definition of chaos and Hamiltonian systems. We also introduce the numerical integration methods, chaos detection techniques and computational processes that were employed throughout this thesis. We focus specifically on the chaos detection techniques, namely the Lyapunov exponents (LEs) and the generalized alignment index (GALI) method, which we use throughout this work. 
\subsection{Overview of dynamical systems} \label{section:DS theory}
A DS is a mathematical model that describes the evolution of a physical process over time. It is characterized by various phenomena that occur in a phase space, which represents all possible configurations of the system. For instance, in a simple pendulum, the phase space is two-dimensional (\(2D\)), with axes representing the pendulum's angular position and angular velocity. Each state in the phase space corresponds to a detailed snapshot of the system's total configuration at a given moment. The evolution of a DS is governed by deterministic laws that guide the transition from one state to the next in the phase space.

There are two types of DSs:

\begin{enumerate}
\item \textbf{Discrete time DSs:} These systems are typically represented by sets of difference equations or iterated maps, and they are mathematically given by:
\begin{equation} \label{eq:Diff}
    \mathbf{x}_{n+1} = \mathbf{f}(\mathbf{x}_n),
\end{equation}
where \(\mathbf{x}_n\) is the state vector \(x\) at a discrete time \(t = n\), with $n$ being a positive integer, and \(\mathbf{f}(\mathbf{x}_n)\) denotes a set of real valued functions computed at \(\mathbf{x}_n\). $\mathbf{x}_{n+1}$ represents the next state vector. 

\item \textbf{Continuous time DSs:} These systems are represented by sets of ordinary differential equations (ODEs) of the form:
\begin{equation} \label{eq:ODE}
\frac{d\mathbf{x}}{dt} = \mathbf{f}(\mathbf{x}(t)),
\end{equation}
where \(\mathbf{f}: \mathbb{R}^m \longrightarrow \mathbb{R}^m\) is a vector field, \(\mathbf{x}(t) \in \mathbb{R}^m\) represents a vector of state variables evolving over continuous time \(t\), and \(\dfrac{d\mathbf{x}}{dt}\) denotes the time derivative. 
\end{enumerate}

\subsubsection{Dissipative systems} \label{section:Dissipative Systems}

DSs can also be categorized as either conservative or dissipative, based on how the phase space volume evolves over time. In conservative systems, the phase space volume remains constant, as described by Liouville's theorem \citep{liouville1838note}. This means that while the shape of a volume element in the phase space may change, its total volume is preserved. In comparison, dissipative systems exhibit a contraction of phase space volume over time. This contraction often results from the system's energy loss or the convergence of trajectories onto a lower-dimensional surface in the phase space, a so-called \textit{attractor}. A volume expansion, on the other hand, would typically lead to unbound motion in the phase space. 

According to \citep{lanford2005strange}, a subset \(S\) of the phase space is defined as an attractor if: 
\begin{enumerate} [label=\textnormal{(\Roman*)}]
\item  \(S\) is invariant under the evolution of the DS. 
\item There exists an open neighborhood (basin of attraction\footnote{The basin of attraction for a subset \(S\) consists of states in phase space that approach \(S\) as time approaches infinity. If multiple attractors exist, almost all initial states (except a set of measure zero) belong to the basin of one of them.}) around \(S\) that contracts toward \(S\) over time. 
\item \(S\) cannot be split into nonoverlapping invariant subsets.
\end{enumerate}

In one-dimensional (\(1D\)) dissipative systems, the only attractors are stable fixed points, where the system remains unchanged over time. A fixed point \( \mathbf{x}^* \) in a DS \eqref{eq:ODE} is a point where \( \frac{d\mathbf{x}}{dt} = 0 \) (e.g.~see \citep[Section 9.2]{strogatz2018nonlinear}). For instance, in a simple atmospheric model, a stable fixed point might represent a persistent weather pattern, like a stable high-pressure system. In \(2D\) systems, stable limit cycles can also exist in addition to stable fixed points. These limit cycles represent periodic behavior, where the system's state oscillates and converges to a fixed amplitude, regardless of the initial state (e.g.~see \citep[Chapter 7]{strogatz2018nonlinear}).  In atmospheric systems, a stable limit cycle might represent the regular seasonal cycle of temperature and rainfall. 

In dissipative systems with three or more dimensions, we can have even more complex attractors known as strange attractors. These attractors exhibit sensitive dependence on initial conditions (ICs), meaning that very small differences in the ICs can result in drastically different long-term behavior. Strange attractors have a very complicated geometric structure (e.g.~see \citep[Sect. 9.3]{strogatz2018nonlinear}). For example, in a limit cycle attractor, a small volume in phase space undergoes a relatively small change in shape. Meaning, it stretches in one direction and contracts in another, resulting in a structure resembling a "very thin strand" of almost a constant length. In contrast, chaotic systems exhibit a continuous process of stretching and folding. An initial small volume in phase space is stretched in one direction and folded back onto itself, resembling a thinner ``ribbon" being stretched and twisted \citep{cencini2010chaos}. This process repeats, resulting in a complex structure. The H{\'e}non map \citep{henon1976two} is a classic example of this stretching and folding behavior. 

In order to determine whether a system is dissipative, conservative, or growing in phase-volume, one needs to evaluate the gradient of the vector field $\mathbf{f}$ in the DS \citep{Goldstein2002}. The classification is based on the sign of the divergence, $\nabla \cdot \mathbf{f} $: if 
\begin{itemize}
\item $\nabla \cdot \mathbf{f}  < 0$, the system is dissipative
\item $\nabla \cdot \mathbf{f}  = 0$, the system is conservative
\item $\nabla \cdot \mathbf{f}  > 0$, the system is growing in phase-volume 
\end{itemize}
Alternatively, we can compute the determinant of the Jacobian matrix associated with the ODE \eqref{eq:ODE} (or the discrete DS \eqref{eq:Diff}).  A negative Jacobian determinant indicates that the system is dissipative. While both the divergence and the Jacobian determinant allow us to classify the system's dynamics, the divergence provides a more direct measure of phase space volume changes over time. 

\subsection{Chaos theory} \label{section:Chaos theory}

Chaos theory generally explores irregular or unpredictable behavior observed in systems governed by deterministic laws. This phenomenon, often known as ``deterministic chaos", emphasizes the interplay between randomness and determinism. Although randomness can be observed in phenomena like the motion of gas molecules or charged particles, deterministic laws offer the fundamental framework for understanding systems that display predictable patterns. Common examples of such systems include planetary orbits and the motion of a pendulum, which can be accurately described by deterministic laws, such as those in Newtonian mechanics \citep{gregersen2010britannica}. In broad terms, both planetary orbits and pendulum motion are governed by gravitational forces and thus follow deterministic principles. Given the ICs, we can predict  their future behavior with certainty. However, these two systems differ significantly in terms of complexity and dynamic behavior. Planetary orbits involve complex gravitational interactions with intricate, long-term dynamics, while a simple pendulum's motion is relatively straightforward and governed by only a few basic parameters.

According to Devaney's \citep{devaney1989introduction}, the definition of chaos includes three key features: 

\begin{enumerate}
\item \textit{Sensitivity on ICs:} A function \( \mathbf{f}: V \to V \) has sensitive dependence on ICs if there exists a positive value \(\delta\) such that, for any point \(\mathbf{x}\) in the domain \(V\) and any neighborhood \(\Delta\) of \(\mathbf{x}\), there are points \(z\) in \(\Delta\) and a non-negative integer \(n\) where the distance between the \(n\)-th iterations of the function \(\mathbf{f}\) applied to \(\mathbf{x}\) and \(\mathbf{z}\) exceeds \(\delta\) (i.e., \( |\mathbf{f}^n(\mathbf{x}) - f^n(\mathbf{z})| > \delta \)).

This feature implies that points initially close to \(\mathbf{x}\) will eventually diverge by at least \(\delta\) after repeated applications of \(\mathbf{f}\). While not all nearby points will diverge, there will always be at least one point in any neighborhood around \(\mathbf{x}\) that does. In essence, the sensitivity on ICs describes how a small variation in the initial state can lead to significant changes in a system's long-term behavior. 

\item \textit{Topological transitivity:} A function \( \mathbf{f}: V \to V \) is topologically transitive if, for any two open sets \( U \) and \( W \) in \( V \), there exists a positive integer \( n \) such that the \( n \)-th iteration of \( \mathbf{f}\ \) applied to \( U \) intersects \( W \), i.e., \( f^n(U) \cap W \neq \emptyset \). 

This property implies that points in \( U \) will eventually enter \( W \) after enough iterations, which means the DS cannot be divided into two distinct, invariant open sets. Consequently, the system's evolution covers the entire phase space over time, keeping the chaotic system from splitting into separate parts that do not interact.

\item \textit{Existence of a dense set of periodic orbits (POs)}: A function \( \mathbf{f}: V \to V \) has a dense set of POs, if for any point \( \mathbf{x} \) in \( V \) and any neighborhood \(\Delta\) of \(\mathbf{x}\), there exists a periodic point \(\mathbf{y}\) in \(\Delta\) such that \( |\mathbf{f}^n(\mathbf{y}) = \mathbf{y}| \) for some integer \(n\).

This feature implies that POs are distributed throughout the phase space, and they can be found arbitrarily close to any point. These POs, often unstable, contribute to the complexity and unpredictable dynamics that characterize chaotic systems. 
\end{enumerate}

Simply put, a chaotic system has three fundamental features: unpredictability caused by sensitive dependence on ICs, indecomposability resulting from topological transitivity, and a form of regularity because chaotic systems have POs that are dense in the system. In practice, the phase space trajectories of a DS in chaotic regions of the DS diverge exponentially fast from nearby orbits over time, highlighting the first feature of the definition of Devaney. This sensitive dependence on ICs is often regarded as the main characteristic of chaos in a DS, and it is crucial for identifying and quantifying chaotic behaviors in a DS. Although the other two definitions are relevant from a mathematical perspective, we will focus on sensitivity on ICs as the defining feature of chaos throughout this document.

\section{Hamiltonian systems} \label{section:General Hamiltonian}
The Hamiltonian formalism is an important tool for studying DSs, including those that display chaotic behaviors. It provides a solid mathematical framework that has helped in advancing our understanding of chaos and its effects in various physical problems \citep{Goldstein2002}. The Hamiltonian formulation provides a fundamentally unique perspective by describing the system's dynamics through a set of ODEs. In this formalism, if we have \( N \) number of generalized coordinates \( \mathbf{q} = (q_1, \ldots, q_N) \) defining a DS's position in an \( N \)-dimensional (\(ND\)) configuration space, and \( N \) generalized momenta \( \mathbf{p} = (p_1, \ldots, p_N) \) which are conjugate to the coordinates \( \mathbf{q}\), then the \textit{time-dependent} Hamiltonian function is a scalar value that represents the total energy of the DS, and it is given by, the real valued function:

\begin{equation}\label{eq:Gen TD Ham}
H(\mathbf{q}, \mathbf{p}, t).  
\end{equation}

The Hamiltonian \eqref{eq:Gen TD Ham} is referred to as \textit{autonomous} when there is no explicit time dependency, i.e., it only depends on \((\mathbf{q}, \mathbf{p})\) not \(t\). Furthermore, the Hamiltonian function $H$ (representing the total energy $E$) is an integral of motion and acts as a conserved quantity along the trajectories of the DS:

\begin{equation}\label{eq:Gen Aut Ham}
E = H(\mathbf{q}, \mathbf{p}). 
\end{equation}

The evolution of the DS is governed by \( 2N \) equations of motion (EoM) that describe \( N \) orbits \( \mathbf{x}(t)  =  (q_j(t),  p_j(t), t)\), where \(j=1,\cdots,N \), in a phase space defined by the \( 2N \) independent variables. The EoM of \eqref{eq:Gen TD Ham} governing the system's dynamics are given by:
\begin{equation}\label{eq:Gen Ham EoMs}
\begin{aligned}
\frac{d\mathbf{q}}{dt} &= \frac{\partial H}{\partial \mathbf{p}}, \\
\frac{d\mathbf{p}}{dt} &= - \frac{\partial H}{\partial \mathbf{q}}, \\
\frac{d\mathbf{H}}{dt} &= \frac{\partial H}{\partial \mathbf{t}}.
\end{aligned}
\end{equation}

To compute chaos indicators such as the LEs and the GALI, which will be defined in Sect.~\ref{section:Chaos Indicators}, we follow the time evolution of small initial deviation vectors, \( \mathbf{v}(t_0) = \delta \mathbf{x}(t_0) \), from a reference point using the so-called \textit{variational equations} and the tangent map method \citep{contopoulos1978number}.

The tangent map describes how perturbations in the state of a DS evolve over time, and it is  derived from the variational equations. These equations represent the linearized dynamics of small perturbations around an IC, capturing how deviations evolve according to the system's EoM. By calculating the second derivative of the orbit's tangent dynamics \( \mathbf{x}(t) \), we obtain the tangent vectors. These tangent vectors are then used to evolve the initial deviation vectors \( \mathbf{v}(t_0) \) forward in time (see e.g., \citep{Skokos2010variational}).

In particular, the variational equations for a Hamiltonian system with \(N\) DoF can be written as:
\begin{equation}\label{eq:Gen Ham VoEs}
\dot{\mathbf{v}}(t) = 
\begin{bmatrix}
\delta q_j(t) \\
\delta p_j(t)
\end{bmatrix}
=
\left[ \mathbf{J}_{2N} \cdot \mathbf{D}^2_H(\mathbf{x}(t)) \right] \cdot \mathbf{v}(t_0), \quad j=1,2,\ldots,N,
\end{equation}
where $\mathbf{J}_{2N} =
\begin{bmatrix}
    \mathbf{0}_N & \mathbf{I}_N \\
-\mathbf{I}_N & \mathbf{0}_N
\end{bmatrix}
$ is the Jacobian matrix with \(\mathbf{I}_N\) and \(\mathbf{0}_N\) representing the \(N \times N\) identity and zero matrices, respectively. The matrix \(\mathbf{D}^2_H(\mathbf{x}(t))\) is the \(2N \times 2N\) Hessian matrix evaluated at the position \(\mathbf{x}(t)\) of the orbit in the system's phase space. Its entries are given by $\mathbf{D}^2_H(\mathbf{x}(t))_{k,j} = \frac{\partial^2 H}{\partial x_k \partial x_j} \bigg|_{\mathbf{x}(t)}$, for all \(k, j = 1, 2, \ldots, 2N\).

The coefficient matrix in Eq.~ \eqref{eq:Gen Ham VoEs} depends on how the orbit \(\mathbf{x}(t)\) evolves over time, independent of the specific ICs and perturbations introduced by \(\mathbf{v}(t)\). This is because the matrix elements are derived from the second derivatives of the Hamiltonian, which describe the dynamics of the system itself. Therefore,  the variational equations \eqref{eq:Gen Ham VoEs} form a set of linear equations with respect to \(\mathbf{v}(t)\). We must solve these equations along with the system's Hamilton EoM \eqref{eq:Gen Ham EoMs} to make sure a consistent evolution of both the orbit's trajectory and the perturbations. 
\section{Numerical integration and computational process} \label{section:Numerical Methods}
\subsection{General integration schemes} \label{section:NumeicalScheme}
The general purpose Runge-Kutta (RK) methods are among the most widely used numerical integration techniques for solving DEs. Runge and Heun improved Euler's simple integration method to produce a more advanced numerical integration technique. Later, Kutta developed a more general approach, which we now refer to as the RK methods \citep{Hairer1993,hairer2006geometric}.

The \(p\)-stage explicit RK method for solving the ODE \eqref{eq:ODE} is defined as follows \citep{hairer2006geometric}:
\begin{equation} \label{eq:RK}
\begin{aligned}
    \mathbf{x}_{n+1} &= \mathbf{x}_n + h \sum_{j=1}^{p} b_j k_j, \\
    k_1 &= \mathbf{f}(t_n, \mathbf{x}_n), \\
    k_2 &= \mathbf{f}(t_n + c_2 h, \mathbf{x}_n + h(a_{21} k_1)), \\
    k_3 &= \mathbf{f}(t_n + c_3 h, \mathbf{x}_n + h(a_{31} k_1 + a_{32} k_2)), \\
    &\vdots \\
    k_p &= \mathbf{f}(t_n + c_p h, \mathbf{x}_n + h(a_{p1} k_1 + a_{p2} k_2 + \cdots + a_{p,p-1} k_{p-1})).
    \end{aligned}
\end{equation}
where \( a_{21}, a_{31}, a_{32}, \ldots, a_{p1}, a_{p2}, \ldots, a_{p,p-1}, b_1, \ldots, b_p, c_2, \ldots, c_p \) are real number coefficients satisfying \( c_i = \sum_{j=1}^{i-1} a_{ij} \) for \(i = 2, 3, \dots, p\).   

There are various RK schemes with different orders to solve the DS \eqref{eq:ODE},  including the commonly used fourth order. The order of a scheme indicates the power of the step size \(h\) in its leading error term. Higher-order RK methods, such as the eighth order DOP853 \citep{hairer2006geometric}, can further provide more accurate solutions. For Hamiltonian systems like \eqref{eq:Gen Aut Ham}, symplectic integrators \citep{Skokos2010variational, senyange2018computational} are particularly well-suited, as they preserve the symplectic structure of  the phase space. This preservation is important for long-term numerical stability and accurately representing the system's overall behavior.  For a detailed analysis of the performance of various symplectic and non-symplectic integrators, readers can refer to \citep{danieli2018computational} and references therein.

Throughout this document, unless explicitly stated, we utilize the simplicity and robustness of the well-known and commonly used fourth-order RK (RK4) scheme to numerically solve the EoM of the DEs, along with their respective variational equations.

\subsection{Computational process}\label{section:Compuration Process}
The computations for this thesis were carried out in the center for high-performance computing (CHPC) of South Africa's Lengau cluster, \url{https://chpc.ac.za}. This high-performance system features powerful multicore CPU nodes (with up to 48 cores), allowing for efficient simulations without requiring internode communication. All simulations were primarily written in Fortran 90, utilizing Intel Fortran and GCC gfortran compilers at optimization level 3 (-O3). Data analysis was mainly performed in Fortran 90 and MATLAB R2023b, with additional support from Mathematica,  Python, and Gnuplot. Linux shell tools such as bash were also employed to manage files, organize data, automate tasks, and significantly enhance workflow efficiency.

Computational optimization played a crucial role in our research. We used multithreading tools like GNU Parallel and OpenMP, combined with code profiling in order to significantly increase the speed of our numerical simulations. For example, GNU Parallel \citep{tange2018gnu} allowed us to run multiple independent simulations simultaneously on different CPU cores by efficiently using the resources allocated by the cluster. This approach drastically cut down computational time, enabling a deeper investigation of the system's long-term dynamics.

\section{Chaos indicators} \label{section:Chaos Indicators}
In order to understand the chaotic behavior of various DSs and to investigate the effect of different physical factors on this behavior, we need reliable methods to identify chaos and quantify its strength. In general, there are two main categories of chaos detection techniques. The first category includes methods based on the analysis of the studied orbit itself, such as frequency map analysis (FMA) \citep{laskar1999introduction,laskar2003frequency} and the 0-1 test \citep{gottwald2009implementation,gottwald2016}. The second category includes methods that involve the study of the evolution of initial nearby orbits or deviation vectors, such as the computation of the maximum Lyapunov exponent (mLE) \citep{Benettin1980a,Benettin1980b,skokos2010lyapunov}, the fast Lyapunov indicator (FLI) and its variants \citep{froeschle1997fasta,froeschle1997fastb, Barrio2016}, the mean exponential growth of nearby orbits (MEGNO) \citep{Cincotta2000,Cincotta2016}, the relative Lyapunov indicator (RLI) \citep{sandor2004relative}, the smaller alignment index (SALI) \citep{skokos2001alignment,skokos2004detecting}, and its generalization, the GALI method \citep{skokos2007geometrical,manos2012probing,skokos2016smaller}. For more detailed information on various chaos indicators and a comparison of their performance, the reader is referred to \citep{skokos2016chaos,bazzani2023performance} and references therein. 

In our study, we mainly apply two chaos indicators, the mLE and the GALI, which have been shown (e.g.~see \citep{skokos2010lyapunov,skokos2016chaos}) to be very successful even for systems with many degrees of freedom (DoF), where the phase space visualization is not of much help. Now, for simplicity,  let us look at the definition and properties of these indicators in the context of a continuous DS of the form \eqref{eq:ODE}. It is worth noting that the following definitions also apply to the discrete-time DS \eqref{eq:Diff}. 

\subsection{The Lyapunov exponents}\label{section:LEs}
Aleksandr Lyapunov, in his 1892 doctoral dissertation, introduced a method to analyze the stability of non-stationary solutions in DS by linearizing their EoM, a technique that eventually led to the development of what are now  known as LEs \citep{lyapunov1992general}. These exponents  have become the most widely used chaos indicators, and they have been applied to successfully distinguish between regular and chaotic motion of DSs. LEs provide a qualitative estimate of the exponential rate at which nearby trajectories diverge or converge in the system's phase space. In other words, they give us information for effectively quantifying the system's sensitivity to ICs.

The mLE, denoted as \(\Lambda_1\), is the most frequently studied LE. Mathematically, it is defined by the following expression (see \citep{skokos2010lyapunov} and references therein):

\begin{equation}\label{eq:mLEs}
    \Lambda_1 = \lim_{t \to \infty} \sigma_1(t).
\end{equation}
The mLE is typically approximated by the \textbf{finite-time maximum Lyapunov exponent (ftmLE)}, \(\sigma_1\) as a function of time $t$:
\begin{equation}  \label{eq:ftmLE}
    \sigma_1(t) = \frac{1}{t} \ln \frac{\|\mathbf{v}(t)\|}{\|\mathbf{v}(0)\|},
\end{equation}
where \(\mathbf{v}(0)\) and \(\mathbf{v}(t)\) represent the deviation vectors at times \(t=0\) and \(t>0\), respectively, and \(||.||\) denotes the Euclidean norm. 

In practice, estimating the mLE \eqref{eq:mLEs} can be challenging due to the slow convergence of \(\sigma_1(t)\) to its asymptotic value, especially for weakly chaotic orbits, which display a slow separation rate between nearby trajectories. In order to accurately identify their chaotic nature, we must integrate the system for an extended computational time. However, as integration time increases, the size of the Euclidean norm of the deviation vectors \(|| \mathbf{v}(t) ||\) may in general increase exponentially, leading to  numerical instability \citep{Benettin1980a}. This instability complicates obtaining reliable mLE estimates for any DSs.  To address these issues, we can use the linearity and compositional properties of the evolution operator, \(d_x\Pi^t\) \citep{Froeschle1973}, by periodically rescaling the deviation vectors, which helps to avoid numerical overflow while preserving their exponential growth rate. This approach enables accurate mLE estimation, even for systems with weak chaos, and it is implemented as follows:

\begin{equation}
\Lambda_1 = \lim_{j \to \infty} \frac{1}{j\tau} \sum_{k=1}^{j} \ln(\gamma_k),
\text{ where } \gamma_k = \frac{\|\mathbf{v}(k\tau)\|}{\|\mathbf{v}(0)\|}, 
\end{equation} 
with \(\tau\) being a small-time interval of the integration, and \(t=j\tau\). Renormalizing the deviation vector after each interval \(\tau\) helps prevent numerical overflow. For an autonomous Hamiltonian system \eqref{eq:Gen Aut Ham}, a positive \(\Lambda_1\) indicates chaotic behavior, while a zero value (\(\Lambda_1 = 0\)) corresponds to regular orbits. In such systems, the ftmLE for regular orbits generally follows a power decay law, \(\sigma_1 (t) \approx t^{-1}\) \citep{skokos2010lyapunov}.

While the mLE indicates whether a dynamical orbit is regular (\(\Lambda_1 =0\)) or chaotic (\(\Lambda_1 >0\)), the entire LEs spectrum (\(\Lambda_1, \Lambda_2, \cdots, \Lambda_{2N}\) for an \(ND\) DS) provides additional information into the system's dynamics and statistical properties. These include determining the dimension of strange attractors and classifying different types of trajectories in dissipative systems.

\subsubsection*{Characterizing orbits in dissipative systems:} Consider a dissipative system with an \(ND\) phase space. The spectrum of LEs plays an instrumental role in classifying various types of dynamical trajectories in this system \citep{cencini2010chaos}. Different orbits in a dissipative system can be characterized based on their corresponding LEs as follows:
\begin{enumerate}
	\item \textit{Stable fixed point}: All LEs are negative, denoted as \((-, -, \cdots ,-)\).
	\item \textit{Stable limit cycle}: One LE (the mLE) is zero, while the rest are all negative, represented as \((0, -,\cdots, -)\).
	\item \textit{\(k\)-dimensional stable torus}: The first \(k\) LEs asymptotically approach zero, while the remaining LEs are negative, represented as \((k(0,\cdots, 0), -,\cdots,-)\). This behavior shows quasi-periodic motion, where the system follows a \(k\)-dimensional toroidal structure in its phase space. While the motion repeats, the pattern in each cycle is unique. 
	\item \textit{Chaotic attractor}: The mLE is positive, with one zero LE and the rest negative, represented as \((+, 0, -,\cdots,-)\). The presence of a zero (the second largest) LE indicates a strange attractor. 
	\item \textit{Hyperchaotic attractor}: There are at least two positive LEs. Hyperchaos, introduced by R\"{o}ssler \citep{rossler1979equation}, describes a chaotic system with at least two positive LEs, which require the phase space dimension of the ODEs \eqref{eq:ODE} to be at least \(4D\). Unlike systems with a single positive LE, such as chaotic attractors, hyperchaotic systems exhibit more complex dynamics that expand in multiple directions. This increased complexity enhances the system's effectiveness in various chaos based applications, such as 
    image processing and secure communication \citep{grassi1999system,brogliato2007dissipative,sprott2010elegant}. We will discuss hyperchaotic attractors and in general hyperchaotic systems further in Chap. \ref{chapter:six}.
\end{enumerate}

For \(2ND\) Hamiltonian systems, the LEs \(\Lambda_k\) of any given orbit in a Hamiltonian phase space comes in pairs as (\(\Lambda_k\), \(-\Lambda_{2N-k+1}\)), for \(1 \le k \le N\). By default, at least one pair of LEs is zero, i.e., \(\Lambda_N = \Lambda_{N+1} = 0\), which is associated with the conserved quantity (i.e., the Hamiltonian) and reflects the time invariance of the system. The order of the LEs spectrum is as follows:
\begin{equation} \label{eq:LEs order}
 \Lambda_1 \geq \Lambda_2 \geq \ldots \geq \Lambda_N \geq -\Lambda_N \geq \ldots \geq -\Lambda_2 \geq -\Lambda_1.
\end{equation}
The sum of all LEs, \(\sum_{k=1}^{2N} \Lambda_k\),  represents the rate of volume change in the phase space of the DS. For dissipative systems, this sum is negative, indicating exponential volume shrinkage. On the other hand, for autonomous Hamiltonian systems and symplectic maps, this sum is zero, implying the preservation of volume. 

In the case of \(2ND\) Hamiltonian systems, the first \(k\) LEs, \(1<k \le 2N\), are typically computed using the so-called standard method as described in \citep{Benettin1980a}. This method involves the time evolution of \(k\) initial, linearly independent, and orthonormal deviation vectors. To maintain numerical stability, the Gram-Schmidt orthonormalization procedure is periodically applied. However, frequent orthonormalization can slow down numerical simulations and is often unnecessary. Instead, a more efficient approach is to replace the evolved deviation vectors with a new set of orthonormal vectors while simultaneously numerically integrating the EoM. 

The mLE has been used as a chaos indicator for many years (see for e.g., \citep{skokos2010lyapunov,pikovsky2016lyapunov}). However, it can be computationally expensive, especially for multidimensional systems and those with weakly chaotic or sticky orbits. This slow convergence arises due to the fact that the mLE depends on the entire evolution of the deviation vectors, which requires accurately following the system's dynamics over an extended period.  

In this work, we will implement the GALI method, introduced by \cite{skokos2007geometrical}, which identifies chaos by determining the alignment of more than one initial normalized deviation vector. This method has proven to be efficient and has been successfully applied to investigate chaos in various DSs, see for e.g., \citep{bountis2009application,manos2013interplay,Moges2020,ghanbari2021detecting,huang2022application}.

\subsection{The generalized alignment index method} \label{section:GALI}
Considering  a \(2ND\) phase space of a Hamiltonian system of \(N\) DoF or a \(2ND\) symplectic map, the GALI of order \(k\) (GALI\(_k\)) measures the volume of a generalized parallelogram defined by the evolution of \(k\) unit deviation vectors, denoted as \(\hat{\mathbf{v}}_1, \hat{\mathbf{v}}_2, \cdots, \hat{\mathbf{v}}_k\) where \(2 \le k \le 2N\), at any given time and  \(\hat{\mathbf{v}}_j = \dfrac{\mathbf{v}_j}{\lVert \mathbf{v}_j \Vert}, \, j =1, 2, \cdots, 2N\). More specifically, the GALI\(_k\) is defined as the norm of the wedge (or exterior) product of these \(k\) unit vectors \citep{skokos2007geometrical}
\begin{equation}  \label{eq:GALI}
    \mbox{GALI}_k(t) = \lVert \hat{\mathbf{v}}_1(t) \wedge \hat{\mathbf{v}}_2(t) \wedge \hat{\mathbf{v}}_3(t) \wedge \dots \wedge \hat{\mathbf{v}}_k(t) \Vert.
\end{equation}

If the number of deviation vectors \(k\) exceeds the dimension \(2N\) of the system's phase space, the vectors are by default linearly dependent, resulting in a zero volume, i.e., GALI\(_k = 0\), for \(k > 2N\). 

In practice, we determine the GALI\(_k\) \eqref{eq:GALI} as the product of the singular values, \(s_j\), \(j=1, \dots, k\), of the \(2N \times k\) matrix whose columns are the entries of the \(k\) normalized deviation vectors \citep{skokos2008detecting} 
\begin{equation} \label{Def:SVD}
    \mbox{GALI}_k =\prod_{j=1}^{k} s_j.
 \end{equation}  

 In order to numerically compute the GALI\(_k\) of a particular orbit, one should follow the time evolution of the orbit itself using Hamilton's EoM [or the map's equations] and simultaneously evolve the $k$ initially unit (typically random) deviation vectors using the corresponding variational equations [tangent map]. 

The most commonly used and simplified form of the GALI index is GALI\(_2\), which is equivalent to the SALI method. The SALI is defined by the area of the parallelogram formed by two deviation vectors \(\mathbf{v}_1(t)\) and \(\mathbf{v}_2(t)\) \citep{skokos2001alignment,skokos2016smaller}, and it is computed by:

\begin{equation}    \label{eq:SALI}
    \text{SALI}(t) = \min \left\{ \| \hat{\mathbf{v}}_1(t) + \hat{\mathbf{v}}_2(t) \|, \| \hat{\mathbf{v}}_1(t) - \hat{\mathbf{v}}_2(t) \| \right\}.
 \end{equation}
If the two unit deviation vectors (\(\hat{\mathbf{v}}_1(t)\) and \(\hat{\mathbf{v}}_2(t)\)) are aligned in the same direction, then the SALI \(= 0\) (GALI\(_2=0\)). It can be easily seen that the index attains its maximum value of SALI\(_2=\sqrt{2}\) (i.e., GALI\(_2 = 1\)), when the two vectors are perpendicular. 

The behavior of the GALI for \(N\) DoF Hamiltonian systems can be summarized as follows:

For regular orbits, the \(k \le N\) linearly independent initial deviation vectors will eventually fall in the \(ND\) tangent space of the torus \citep{skokos2007geometrical}. However, they are unlikely to become linearly dependent, which results in an asymptotic GALI\(_k\) value that remains practically constant, as the resulting volume does not vanish. On the other hand, when starting with \(k > N\) initially linearly independent deviation vectors, the asymptotic GALI\(_k\) value will be zero. This occurs because some deviation vectors will become linearly dependent as they all will fall in the \(ND\) tangent space of the torus. This behavior leads to a power law decay of GALI\(_k\). Hence, the general behavior of GALI\(_k\) for regular orbits lying on an \(ND\) torus is \citep{skokos2007geometrical}

\begin{equation}\label{prop:GALI_regular}
\text{GALI}_k(t) \propto \begin{cases} \text{constant} & \text{if } 2 \leq k \leq N, \\ \dfrac{1}{t^{2(k-N)}} & \text{if } N < k \leq 2N. \end{cases}
\end{equation}

In the case of chaotic orbits, all deviation vectors will eventually align in the direction determined by the mLE, leading to an exponential decay of GALI\(_k\) towards zero. The rate of GALI\(_k\)'s decay depends on the values of the \(k\) largest LEs of the orbit, as shown in \citep{skokos2007geometrical,manos2012probing}.

\begin{equation}\label{eq:GALI_chaos}
 \text{GALI}_k(t) \propto \displaystyle e^{-[(\Lambda_1 - \Lambda_2) + (\Lambda_1 - \Lambda_3) + \ldots + (\Lambda_1 - \Lambda_k)]t}.
\end{equation}
In particular, we obtain from \eqref{eq:GALI_chaos} that GALI\(_2 \propto \displaystyle e^{-[(\Lambda_1 - \Lambda_2)] t}\). 

A different behavior of the GALI\(_2\) (or equivalently the SALI) can be observed in the case of the $2D$ area preserving maps. For such systems, the two initially linearly independent deviation vectors (\(\mathbf{v}_1(t)\) and \(\mathbf{v}_2(t)\)) will align at the tangent space of a $1D$ torus, which has the same dimensionality as the torus itself, i.e., this space is \(1D\). As a result, both vectors will eventually become linearly dependent. This behavior leads the GALI\(_2\) to asymptotically tend to zero, similar to the behavior observed in chaotic orbits. However, the rate at which GALI\(_2\) goes to zero for regular orbits of $2D$ maps is not exponential, as in the case of chaotic orbits, but rather follows the relation \citep{manos2007studying, skokos2016smaller}
\begin{equation}    \label{Prop:GALI_2 for SM}
    \mbox{GALI}_2(n) \propto \frac{1}{n^2}.
\end{equation}
where \(n\) is the number of iterations on the \(2D\) map. The numerical algorithms for the computation of the SALI and the GALI methods using the singular value decomposition (SVD) procedure can be found in \citep{skokos2016smaller}.

One of the key questions that can arise is whether the GALI method can effectively characterize dynamical trajectories in dissipative systems, similar to how LEs are used in Sect.~\ref{section:LEs}. We will address this in Chap.~\ref{chapter:six} by comparing the GALI and LE spectrum performance in classifying and quantifying various attractors in both discrete and continuous dissipative models. 
\clearpage


\chapter{Magnetic and kinetic chaos in toroidal plasmas}
\label{chapter:three}

\section{Introduction} \label{section:introductionCh3}
Fusion is a fundamental process present across the universe. When we look at the stars at night or the Sun during the day, we are basically witnessing fusion in action. In the Sun, hydrogen nuclei fuse to form helium, releasing energy through nuclear fusion. While this process serves as a simple example, other types of fusion reactions also occur. For instance, large stellar bodies often generate energy via the carbon-nitrogen-oxygen (CNO) cycle. Here on Earth, researchers are exploring deuterium-tritium (D-T) fusion as a promising source of clean energy \citep{FreidbergBook}.

Fusion energy research aims to develop power plants to generate clean energy by controlling and confining fusion reactions. This broad field of science and engineering aims to create an environmentally-friendly and reliable power source \citep{FreidbergBook}. The most energy efficient fusion reaction occurs between the isotopes of hydrogen-deuterium and tritium as the fourth state of matter, \textit{plasma}. Plasma is an ionized gas of charged particles (electrons and protons) in a quasi-neutral state. It requires extremely high temperatures to sustain control over its fusion productivity, which is the rate of fusion reactions. If the plasma cools rapidly or loses too many charged particles (due to situations like leaks or insufficient confinement), the fusion process can become less efficient and unsustainable for practical power generation. 

Plasma confinement is crucial in order to maintain fusion reactions. There are two major approaches for plasma confinement: inertial confinement, which involves heating and compressing the plasma, and magnetic confinement, which uses the properties of charged particles moving in a magnetic field (MF). The magnetic confinement approach, as applied in toroidal plasmas, has been the most successful method in fusion experiments to date \citep{WhiteBook}.

\begin{figure}[!htb]
  \centering
  \includegraphics[width=0.8\textwidth]{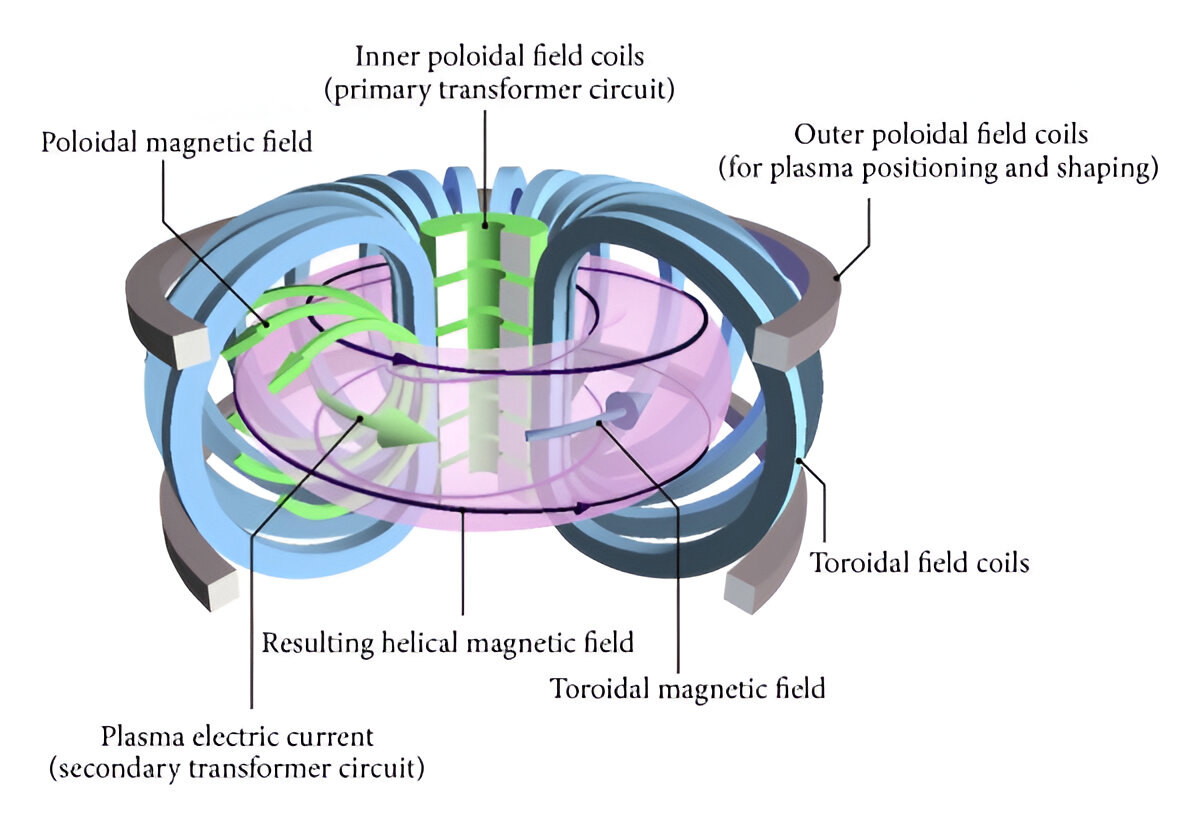}
  \caption{Basic structure of a Tokamak [image source: \href{www.euro-fusion.org}{euro-fusion}].}
  \label{fig:Tokamak}
\end{figure}

To effectively describe a toroidal magnetic configuration, we commonly use a coordinate system aligned with the MF \cite[Sect.~1.2]{WhiteBook}. This approach simplifies the field representation and results in a general theoretical framework for various toroidal fusion devices, such as tokamak and stellarators. The tokamak, a Russian acronym that loosely translates as ``donut-shaped vessel surrounded by magnets", is currently the most advanced and practical device for the production of energy from fusion. Figure \ref{fig:Tokamak} illustrates the basic structures of a tokamak, highlighting the plasma current (the region inside the torus shown in pink), the magnetic coils responsible for generating the field, and the twisted MF created by the plasma particle interaction.

In a tokamak, we can create a poloidal MF using toroidal coils arranged in a circular configuration. These MF coils confine plasma particles and may allow the plasma to reach the conditions necessary for fusion. One set of coils (the toroidal coils shown in light blue in Fig.~\ref{fig:Tokamak}) generates a strong MF that runs around the long way of the torus (the ``toroidal" field indicated by light blue arrows). This arrangement generates a strong MF in the toroidal direction of the torus. 

However, a purely toroidal MF leads to charged particle drifting due to gradients in the MF strength (which may be related to factors such as coil imperfections or plasma instabilities) and centrifugal forces, which can cause plasma to escape the confinement area. In order to prevent the plasma escape, a second MF in the poloidal direction is introduced. A central solenoid (the green coils shown at the center of Fig.~\ref{fig:Tokamak}), which functions as a magnet carrying electric current, creates this ``poloidal'' field that runs around the short way of the torus (indicated by green arrows in Fig.~\ref{fig:Tokamak}). This poloidal field, together with the toroidal field, generates helical field lines (black curves with arrows in Fig.~\ref{fig:Tokamak}) that effectively cancel out drifts and improve particle confinement.

The combination of both twisted toroidal and poloidal MFs effectively confines the plasma particles. Furthermore, an additional set of coils (the two outer poloidal field coils in gray in Fig.~\ref{fig:Tokamak}) can be added to balance the outward plasma pressure. This outer MF shapes and positions the plasma in the confinement area. 

The interplay of toroidal, poloidal, and outer MFs produces strong plasma confinement \citep{FreidbergBook}. However, we also need to maintain the plasma confined in a stable configuration in some form of equilibrium. Magnetohydrodynamics (MHD) is the study of these equilibria and dynamics of plasmas in MFs by accounting for the balancing of forces acting on the plasma while making sure its stability and confinement in a specific region \citep{FreidbergBook}. 


\subsection{Representation of magnetic field lines}
We follow the framework outlined in \citep{WhiteBook} for describing the general toroidal coordinate system and the toroidal and poloidal surfaces of all toroidal devices, such as Tokamak. Using a coordinate system that is well aligned with the MF gives significant advantages over standard coordinate systems. Aligning the coordinates simplifies the equations governing charged particles, which can improve the efficiency of theoretical analysis and understanding of plasma dynamics. This alignment also minimizes the complexity of the system by increasing the efficiency and accuracy of our numerical simulations. 

While MFs and plasma equilibrium must eventually be described in laboratory coordinate systems, such as the typical Cartesian or cylindrical coordinates, the general toroidal coordinates are especially introduced to correspond to the toroidal geometry of the MF surfaces. These coordinates are functions of Euclidean coordinates, and they are particularly useful to represent the geometry of the MFs. By introducing general coordinates \((\psi, \theta, \zeta)\), where \(\psi\) defines nested toroidal surfaces, \(\theta\) represents a generalized poloidal angle, and \(\zeta\) is the toroidal direction, we can create a right-handed coordinate system (see Fig.~\ref{fig3:White,Fig1.2}).

\begin{figure}[!htb]
    \centering
    \includegraphics[width=0.6\textwidth]{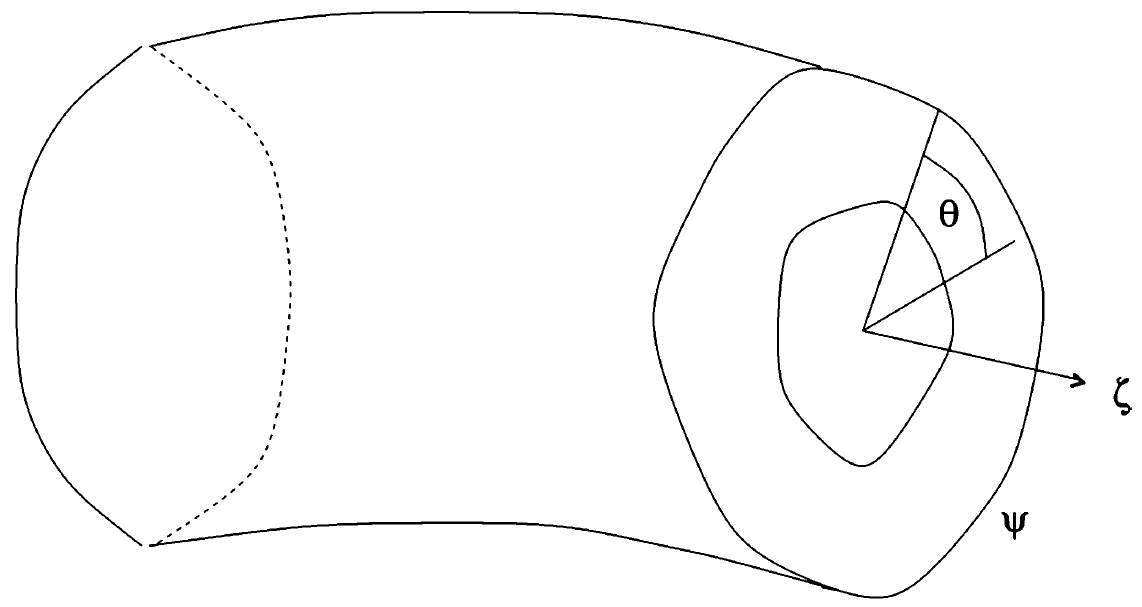}
    \caption{The general toroidal coordinate system used to describe the MF surfaces in a toroidal fusion device \cite[Fig.~1.2]{WhiteBook}.}
    \label{fig3:White,Fig1.2}
  \end{figure}

A magnetic surface is a \(2D\) region in space where magnetic field lines (MFLs) are entirely confined. The defining property of a magnetic surface is that the MFLs remain tangent to the surface at every point, preventing them from crossing it \cite[Sect.~1.6]{WhiteBook}. This feature is essential for understanding how plasma and charged particles behave in magnetic confinement systems.

According to Kolmogorov's theorem \citep[Appendix 8]{arnol2013mathematical}, a small perturbation in the strength of a symmetric system (typically exhibiting rotational symmetry around the major axis \(\zeta\) in Fig.~\ref{fig3:White,Fig1.2}) maintains well-defined magnetic surfaces, with the exception in some localized regions where magnetic islands and chaotic behavior may exist (e.g.~see \citep{dumas2014kam}). 

To investigate the chaotic dynamics of turbulent plasmas in fusion devices, it is important to analyze how non-axisymmetric perturbations influence MFLs and flux surfaces. These perturbations can disrupt the concentric closed flux surfaces of magnetic equilibria, which result in chaotic behavior \citep{WhiteBook,abdullaev2008description}. 

The flux surfaces most affected by the perturbations are determined by a resonance condition involving the radial profile of the safety factor \(q\) and the poloidal \(m\) and toroidal \(n\) mode numbers. This perturbation is commonly represented by a spatial dependence of the form \(a_{mn}\exp[i(m\theta-n\zeta)]\).

When resonance conditions are met, chains of magnetic stability islands form at the resonant flux surfaces. For small perturbation strengths, chaos is localized near the separatrices of these islands. However, as the perturbation strength increases, chaotic behavior can extend to larger regions of the plasma. In these chaotic regions, the dynamics of MFLs are influenced by the complex structure of the homoclinic and heteroclinic lobes of the unstable (saddle) periodic field lines \citep{evans2005experimental}.

Figure \ref{fig3:White,Fig1.3} illustrates a poloidal surface (constant \( \theta \)) and a toroidal surface (constant \( \zeta \)), which together form a topologically toroidal volume. The magnetic axis acts like the central boundary for these surfaces. For reference, the total toroidal and poloidal magnetic flux through each surface, denoted as \( \psi \) and \( \psi_p \), respectively, is normalized to zero at the magnetic axis. This normalization guarantees that the surfaces are defined by their relative magnetic flux values rather than their absolute magnitudes. The periodicity of the angular variables \(\theta \) and \(\zeta \) (with period \(2\pi \)) maintains the topology of the toroidal volume.

\begin{figure}[!htb]
    \centering
    \includegraphics[width=0.35\textwidth]{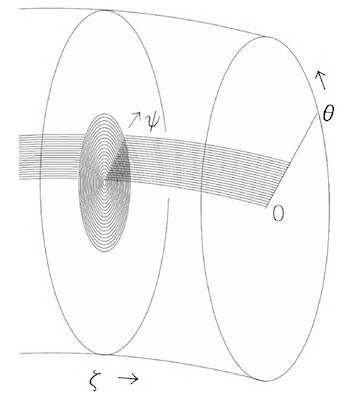}
    \caption{Toroidal surface (constant \( \zeta \)) and poloidal surface (constant \( \theta \)) defining the magnetic flux \(\psi\) and \(\psi_p\), respectively. Both  \(\psi\) and \(\psi_p\) are zero at the magnetic axis (gray region)  \cite[Fig.~1.3]{WhiteBook}.}
    \label{fig3:White,Fig1.3}
  \end{figure}
Tokamak uses toroidal configurations to contain high energy plasma particles and retain energy. The MFs created by the surrounding coils are crucial for confining the plasma. However, the chaotic behavior of high energy particles, which experience large drifts relative to the MFLs, can greatly affect the confinement of the device, and this chaos goes beyond what can be explained by the MFLs alone \citep{heidbrink2020mechanisms}.

To quantify and analyze this chaotic behavior, we employ the general alignment index (GALI) method in Sect.~\ref{section:GALI}. Chapter \ref{chapter:three} marks the main contribution of this thesis, introducing the first application of the GALI method in the realm of plasma physics models. Furthermore, this chapter provides the first quantitative comparison between kinetic and magnetic chaos in toroidal plasma, particularly in tokamak. By effectively measuring chaos, the GALI method offers a value that characterizes the chaotic nature of each orbit in the space defined by three constants of motion (CoM): energy, magnetic moment, and toroidal momentum.  Our fresh approach could significantly contribute to how we understand plasma particle dynamics and to optimize confinement in magnetic confinement devices in general. 

The content of this chapter is based around the findings presented in \cite{moges2024kinetic}.
\section{Hamiltonian descriptions} \label{section:HamiltonianCh3}
\subsection{Magnetic field lines}
The Lorentz force governs the interaction between the MFs and plasma particles, resulting in different dynamics for the MFs and the particles' motion. In order to characterize MFLs, we consider a DS described by an ODE:
\begin{equation} \label{eq3:ODE}
\dfrac{ds}{dx} = \mathbf{B}(x),
\end{equation}
which represents the relationship between the arc length \(s\) along a curve and the MF vector \(\mathbf{B}(x)\) at a point \(x\) on that curve. The Eq.~\eqref{eq3:ODE} indicates that the tangent vector to the MFL at any given point, \(\frac{ds}{dx}\), is equal to the MF vector at the same point \(\mathbf{B}(x)\), which means the MFL follows the direction of the MF. Although the ODE \eqref{eq3:ODE} does not explicitly include a time variable, \(\mathbf{B}(x)\) can still vary with time, and thus the solution \(s(x)\) can change over time.

There are different approaches to represent a MF, and according to White's book \citep{WhiteBook}, we choose the vector field representation for two key reasons. First, this approach allows for a clear understanding of the Hamiltonian nature of MFLs. Second, it can be useful in studying both the particle trajectories and the MHD equilibria. In this representation, and according to Gauss's law for the MF, the MF \( \mathbf{B} \) is defined as the curl of a vector potential \(\mathbf{A}\) 
\begin{equation} \label{eq3:MF}
\mathbf{B} = \nabla \times \mathbf{A}.  
\end{equation}

By applying Poincar\'e's Lemma \citep{henon1982numerical}, we can represent a vector field with zero curl as the gradient of a scalar potential function. This representation inherently satisfies Gauss's law for magnetism, making sure that \( \nabla \cdot \mathbf{B} = 0 \), as the divergence of a curl is always zero. Using this, we can express the vector potential \( \mathbf{A} \) in terms of the toroidal coordinates \( (\psi, \theta, \zeta) \) as follows:

\begin{equation} \label{eq3:curl vec pot 1}
\mathbf{A} = \mathbf{A}_{\psi} \nabla \psi + \mathbf{A}_{\theta} \nabla \theta + \mathbf{A}_{\zeta} \nabla \zeta,
\end{equation}
where the components \(\mathbf{A}_{\psi}, \mathbf{A}_{\theta},\) and  \(\mathbf{A}_{\zeta} \) describe how the vector potential varies along the respective directions in the radial coordinate \(\psi\), the poloidal angle \(\theta\), and the toroidal angle \(\zeta\). Combining these three components gives us the overall vector potential \(\mathbf{A}\). 

Let us define a scalar function \(G\) with the condition \(\frac{\partial G}{\partial \psi} =\mathbf{A}_\psi\). The gradient of \(G\) is then given by 
\begin{equation} \label{eq3:curl scalalr G}
    \nabla G = A_{\psi} \nabla \psi + \frac{\partial G}{\partial \theta} \nabla \theta + \frac{\partial G}{\partial \zeta} \nabla \zeta.
\end{equation}
By subtracting \eqref{eq3:curl scalalr G} from \eqref{eq3:curl vec pot 1}, we obtain 
\begin{equation} \label{eq3:curl vec pot 2}
  \mathbf{A} = \nabla G + \left( \mathbf{A}_{\theta} - \frac{\partial G}{\partial \theta} \right) \nabla \theta + \left( \mathbf{A}_{\zeta} - \frac{\partial G}{\partial \zeta} \right) \nabla \zeta. 
\end{equation}
We can define the toroidal flux \( \psi \) and poloidal flux \(  \psi_p \) using the function \( G \) in \eqref{eq3:curl scalalr G} as follows
\begin{equation}\label{eq: canonical repre. for flux}
\begin{aligned}
\psi &=  \mathbf{A}_{\theta} - \frac{\partial G}{\partial \theta},  \\
\psi_p &= - \left( \mathbf{A}_{\zeta} - \frac{\partial G}{\partial \zeta} \right).
\end{aligned}
\end{equation}
Substituting \eqref{eq: canonical repre. for flux} into \eqref{eq3:curl vec pot 2} yields a simplified expression for the vector potential \(\mathbf{A}\)
\begin{equation} \label{eq3:vec pot A}
\mathbf{A} \equiv \nabla G + \psi \nabla \theta - \psi_p \nabla \zeta.
\end{equation}
Now substituting \eqref{eq3:vec pot A} on the MF expression in \eqref{eq3:MF}, we obtain 
\begin{equation}
\mathbf{B} = \nabla \times \left( \nabla G + \psi \nabla \theta - \psi_p \nabla \zeta \right).
\end{equation}
Applying the vector calculus identity for the curl of a sum, we have
\begin{equation} \label{eq3:MF 2}
\mathbf{B} = \nabla \times \left( \nabla G \right) + \nabla \times \left( \psi \nabla \theta \right) - \nabla \times \left( \psi_p \nabla \zeta \right).
\end{equation}
From the identity that the curl of a gradient is zero (\( \nabla \times \nabla G  = 0 \)), \eqref{eq3:MF 2} simplifies to
\begin{equation} \label{eq3:MF 3}
\mathbf{B} = \nabla \times \left( \psi \nabla \theta \right) - \nabla \times \left( \psi_p \nabla \zeta \right).
\end{equation}
Applying the vector calculus identity to both right side terms of \eqref{eq3:MF 3}, i.e., \(\nabla \times \left( \psi \nabla \theta \right) = \nabla \psi \times \nabla \theta,\) and \(\nabla \times \left( \psi_p \nabla \zeta \right) = \nabla \psi_p \times \nabla \zeta\), the MF can now be expressed in terms of the Boozer coordinates  (\(\psi, \theta, \zeta\)) \citep{boozer1981plasma} 
\begin{equation}\label{eq:gen mag field} 
\mathbf{B} = \nabla \psi \times \nabla \theta - \nabla \psi_p \times \nabla \zeta.
\end{equation}
where \(\theta\) and \(\zeta\) are the poloidal and toroidal angles, and the terms \(\psi\) and \(\psi_p\) represent the toroidal and poloidal magnetic fluxes, respectively. Eq.~\eqref{eq:gen mag field} is referred to as the Clebsch representation of a MF (see \citep{d2012flux}).
In order to find the Hamiltonian expression of MFLs, we need to derive \( \dfrac{d\theta}{d\zeta} \) and \( \dfrac{d\psi}{d\zeta} \) from \eqref{eq:gen mag field}. Let us first start with:
\begin{equation}
  \begin{aligned}
    \mathbf{B} \cdot \nabla \psi &= (\nabla \psi \times \nabla \theta - \nabla \psi_p \times \nabla \zeta) \cdot \nabla \psi \\
    &= -(\nabla \psi_p \times \nabla \zeta) \cdot \nabla \psi, \\[10pt]
    \mathbf{B} \cdot \nabla \theta &= \left( \nabla \psi \times \nabla \theta - \nabla \psi_p \times \nabla \zeta \right) \cdot \nabla \theta \\[10pt]
    &= -\left( \nabla \psi_p \times \nabla \zeta \right) \cdot \nabla \theta, \\[10pt]
    \mathbf{B} \cdot \nabla \zeta &= (\nabla \psi \times \nabla \theta - \nabla \psi_p \times \nabla \zeta) \cdot \nabla \zeta \\
    &= (\nabla \psi \times \nabla \theta) \cdot \nabla \zeta,
  \end{aligned}
  \end{equation}  
where the terms \((\nabla \psi \times \nabla \theta) \cdot \nabla \psi\), \((\nabla \psi \times \nabla \theta) \cdot \nabla \theta\) and \((\nabla \psi_p \times \nabla \zeta) \cdot \nabla \zeta\) are zero since the cross products are perpendicular to \(\nabla \psi\), \(\nabla \theta\) and \(\nabla \zeta\), respectively.

Now, we can write 
\begin{equation}\label{eq3:MFL Ham 1}
\begin{aligned}
  \frac{d \psi}{d \zeta} &= \frac{\mathbf{B} \cdot \nabla \psi}{\mathbf{B} \cdot \nabla \zeta} 
  = -\frac{(\nabla \psi_p \times \nabla \zeta) \cdot \nabla \psi}{(\nabla \psi \times \nabla \theta) \cdot \nabla \zeta}, \\
  \frac{d \theta}{d \zeta} &= \frac{\mathbf{B} \cdot \nabla \theta}{\mathbf{B} \cdot \nabla \zeta} 
  = -\frac{(\nabla \psi_p \times \nabla \zeta) \cdot \nabla \theta}{(\nabla \psi \times \nabla \theta) \cdot \nabla \zeta}.
\end{aligned}
\end{equation}
In Boozer coordinates, the gradient of a scalar function \(\psi_p\) can be expressed as 
\begin{equation}
\nabla \psi_p = \left( \frac{\partial \psi_p}{\partial \psi} \right) \nabla \psi 
+ \left( \frac{\partial \psi_p}{\partial \theta} \right) \nabla \theta 
+ \left( \frac{\partial \psi_p}{\partial \zeta} \right) \nabla \zeta. 
\end{equation}
Substituting this expression into \eqref{eq3:MFL Ham 1}, we obtain:
\begin{equation}\label{eq3:MFL Ham 2}
  \begin{aligned}
    \frac{d\psi}{d\zeta} &= -\frac{\left[ \left( \frac{\partial \psi_p}{\partial \psi} \nabla \psi + \frac{\partial \psi_p}{\partial \theta} \nabla \theta \right) \times \nabla \zeta \right] \cdot \nabla \psi}{\left( \nabla \psi \times \nabla \theta \right) \cdot \nabla \zeta} \\
    &= -\frac{\left( \frac{\partial \psi_p}{\partial \theta} \left( \nabla \theta \times \nabla \zeta \right) \cdot \nabla \psi\right)}{\left( \nabla \psi \times \nabla \theta \right) \cdot \nabla \zeta}, \\[10pt]
    \frac{d\theta}{d\zeta} &= -\frac{\left[ \left( \frac{\partial \psi_p}{\partial \psi} \nabla \psi + \frac{\partial \psi_p}{\partial \theta} \nabla \theta \right) \times \nabla \zeta \right] \cdot \nabla \theta}{\left( \nabla \psi \times \nabla \theta \right) \cdot \nabla \zeta} \\
    &= \frac{\left( \frac{\partial \psi_p}{\partial \psi} \left( \nabla \psi \times \nabla \zeta \right) \cdot \nabla \theta \right)}{\left( \nabla \psi \times \nabla \theta \right) \cdot \nabla \zeta}.
  \end{aligned}
    \end{equation}  
Since the coordinate system \((\psi, \theta, \zeta)\) is orthogonal, i.e., the gradient vectors \(\nabla \psi\), \(\nabla \theta\), and \(\nabla \zeta\) are mutually perpendicular. Hence, we can simplify the dot products in both the numerator and the denominator of \eqref{eq3:MFL Ham 2}, leading to the Hamiltonian EoM for the MFLs \citep{WhiteBook}:
\begin{equation}\label{eq:unper Ham MFL}
\begin{aligned}
  \frac{d\psi}{d\zeta} &= -\frac{\partial \psi_p }{\partial \theta}, \\
  \frac{d\theta}{d\zeta} &= \frac{\partial \psi_p}{\partial \psi},
\end{aligned}
\end{equation}
where the poloidal flux \( \psi_p(\psi, \theta, \zeta)\) represents the Hamiltonian function of the MF, with \(\zeta\) (the toroidal angle) acting as the independent variable equivalent to time. Meanwhile, \( \theta\) and \(\psi\) represent the canonical coordinate and its conjugate momentum, respectively. Although the system involves three independent variables (\(\psi, \theta, \zeta\)), the evolution of \(\theta\) and \(\psi\) are described in terms of the independent variable \(\zeta \). The Hamiltonian EoM \eqref{eq:unper Ham MFL} thus represents a time-dependent, one and a half dimension continuous DS. 

The system \eqref{eq:unper Ham MFL} is generally non-integrable due to the potential explicit dependence of  \( \psi_p \) on \( \zeta \). If \( \psi_p \) is independent of \( \zeta \), then the system becomes integrable and describes an axisymmetric MF configuration.

The topology of MFLs is strongly influenced by the explicit dependence of the Hamiltonian \( \psi_p \) on the poloidal \(\theta\) and toroidal \(\zeta\) angles \citep{AbdullaevBook}. This dependence occurs from perturbation factors (such as plasma current and plasma shape), which alter the Hamiltonian. These potential changes can disrupt the magnetic surfaces, leading to chaotic behavior. In our study, we refer to this chaotic behavior, which happens from the complex topology of the MFLs, as \textit{magnetic chaos}. The perturbed Hamiltonian \( \psi_p \) can be expressed as:

\begin{equation}\label{eq:per MFL Ham}
\psi_p = \int \dfrac{d\psi}{q(\psi)} - \sum_{m,n} a_{mn}(\psi) \sin(m\theta - n\zeta),
\end{equation}
where the safety factor \(q(\psi) = \dfrac{d\zeta}{d\theta}\) is the ratio of toroidal to poloidal turns for the MFL. The summation term represents the non-axisymmetric perturbations to the MF, where the coefficients  \( a_{mn}(\psi) \) determine the amplitude of these perturbations, while \(m\) and \(n\) denote the poloidal and toroidal mode numbers, respectively. For instance, \((m, n)=(2, 1)\) indicates a perturbation with twice the poloidal wavelength and a single toroidal wavelength. This implies that the perturbation changes significantly in the poloidal direction compared to the toroidal direction.

Substituting the Hamiltonian \(\psi_p\) from \eqref{eq:per MFL Ham} into the MFL EoM from \eqref{eq:unper Ham MFL}, we obtain the EoM for the perturbed MFLs, which is given by 
\begin{equation}\label{eq:per MFL EoM}
\begin{aligned}
\frac{d\psi}{d\zeta} &= \sum_{m,n} m a_{mn}(\psi) \cos(m\theta - n\zeta), \\
\frac{d\theta}{d\zeta} &= \dfrac{1}{q(\psi)}. 
\end{aligned}
\end{equation}

The integrable part of the perturbed EoM \eqref{eq:per MFL EoM} is defined by the safety factor profile. The safety factor term \(q\) characterizes the helicity of the MF on a specific magnetic surface \(\psi\). In simple terms, it counts how many times a MFL revolves around the short direction (poloidally) compared to the long direction (toroidally) in the torus [see Fig.~\ref{fig3:Fig1}] and it is generally given by the expression \citep[Sect. 2.9]{WhiteBook}:

\begin{equation}\label{eq:gen q factor}
q(\psi) = q_{\text{ma}} \Bigg [1 +  \left( \left( \dfrac{q_{\text{w}}}{q_{\text{ma}}} \right)^{N}  - 1  \right) \left(\frac{\psi}{\psi_{\text{w}}}\right)^{N}  \Bigg ]^{1/N},
\end{equation}
where \(q_{\text{ma}}\) and \(q_{\text{w}}\) are the safety factor values at the magnetic axis (where \(\psi = 0\)) and the wall (i.e., the plasma edge where \(\psi = \psi_\text{w}\)), respectively. The quantity \(\psi_\text{w}\) is the corresponding toroidal flux at the wall, and \(N\) is the component that controls the radial shape of the \(q\) profile. A larger value of \(N\) produces a more peaked safety factor profile around the magnetic axis.

The safety factor profile \(q(\psi)\) determines the local helicity of the unperturbed field lines as they revolve around a constant \(\psi\) magnetic surface, which is a \(2D\) plane where the poloidal magnetic flux remains constant. Resonance conditions between the local helicity and the helicity of specific modes \((m/n)\) can disrupt these magnetic surfaces, leading to the formation of magnetic island chains and the introduction of localized chaotic behavior. Here, \(m\) and \(n\) represent the poloidal and toroidal mode numbers, respectively, which define the spatial structure of the perturbation.

Chirikov's criterion \citep{littlejohn1983variational,Chirikov1979} describes how the overlap of neighboring magnetic islands can result in the formation of extended chaotic regions. These chaotic regions significantly affect plasma transport and the performance of confinement devices. Although extended MFL chaos is often associated with increased particle transport and particle loss, it is important to note that this relationship is not always direct. For instance, previous studies have shown that $10 keV$ electrons in a pre-disruptive ITER scenario \footnote{A pre-disruptive ITER scenario refers to the conditions in the International Thermonuclear Experimental Reactor (\href{https://www.iter.org/}{ITER}) before a disruption event.} can exhibit chaotic transport events in the absence of extended MF chaos \citep{Spizzo2019}.

\subsection{Guiding center motion}
In order to understand the evolution of particle confinement and to quantify particle diffusion in fusion devices, we need to follow the long-term behavior of charged particle orbits. To simplify this task, we average the fast gyro-motion\footnote{Gyro-motion averaging is a technique in plasma physics that simplifies the equations of motion for charged particles by averaging over their rapid circular motion around MFLs, allowing focus on their slower drift dynamics (e.g.~see \citep{FreidbergBook}).} of particles around MFLs, leading to the derivation of guiding center motion (GCM) equations \citep{WhiteBook}. The GCM approximation describes simplified particle dynamics by representing a particle's motion as a quick circular gyration around a MFL, together with slower parallel motion along the field line. This approximation also accounts for additional drifts, such as gravitational drift, that can cause the guiding center (GC) to deviate from the MFL. The GC approximation is valid when the electromagnetic field is weakly inhomogeneous; in particular, when the gyro radius (\(\rho\)) is significantly smaller than the characteristic length scales of the MF. This condition can be expressed as:

\begin{equation}
\rho \approx 10^2 \dfrac{mE}{\mathbf{B}Z} cm,
\end{equation}
where \(E\) is the particle's energy in electron volts, \(Z\) is its charge, and \(\mathbf{B}\) is the MF strength in Gauss.

Here, we will provide a brief overview of the GC averaging transformation, which leads to the derivation of GC EoM in a Hamiltonian formalism. In order to derive these EoM, we begin from the Littlejohn Lagrangian \citep{littlejohn1983variational}, which describes the motion of a charged particle in a MFL and is given as:
\begin{equation}
L = (\mathbf{A}+\rho_{||}\mathbf{B}) \cdot \mathbf{u}+\mu\dot{\xi}-H,
\end{equation}
where \(\mathbf{A}\) and \(\mathbf{B}\) represent the vector potential and MF, respectively, \(\mathbf{u}\) is the GC velocity, \(\mu\) is  the magnetic moment, \(\xi\) is the gyro-phase, \(\rho_{||}\) is the velocity component parallel to the MF \(\mathbf{B}\), and \(H\) is the Hamiltonian function. The corresponding Hamiltonian \(H\) is a function of \( \mathbf{B} \), \( \mu \), and \( \rho_{\parallel} \) and  expressed as: 

\begin{equation}\label{eq:GC H}
    H = \frac{\rho_{||}^2}{2} \mathbf{B}^2 + \mu \mathbf{B}.
\end{equation}

The dynamics of a charged particle can be simplified by using a single parameter, the magnetic moment \(\mu\) (which is a constant value when there are no collisions or time-dependent electric fields). By applying normalizing time with the inverse gyro-frequency \( \omega_0^{-1} \) and distance with the major radius \( R \), we can directly compare the drift velocities of various particle types. The GCM approximation holds when the gyro-radius \( \rho \) (defined as the ratio of the particle's perpendicular velocity \( u_{\perp} \) to the MF strength \( \mathbf{B} \)) is much smaller than the characteristic length scale of the MF, i.e., \(\rho = u_{\perp} \mathbf{B} \ll 1\). Under this condition, the magnetic moment \( \mu \) is conserved, and the cross-field drift scales with \( \rho^2 \) \citep[Sect. 3.1]{WhiteBook}. 

A small resonant perturbation to an equilibrium MF can significantly disrupt MF surfaces, resulting in the formation of magnetic islands and chaotic regions. To simplify the study of particle motion in such complex systems, the GC formalism can be extended to accommodate perturbations with localized spatial behavior, thus avoiding the complexities associated with arbitrary perturbations. In axisymmetric tokamak, the MF primarily depends on the toroidal angle \(\zeta\).  However, even in tokamak, small perturbations, such as those emerging from discrete toroidal field coils, can introduce complex MF structure. These perturbations can lead to magnetic mirroring effects, where particles may become trapped between regions of strong MFs. 

In addition, tearing and shear Alfv{\'e}n perturbations\footnote{Tearing and shear Alfv{\'e}n perturbations are plasma instabilities that can greatly influence the behavior of magnetically confined plasmas, particularly in tokamak (e.g.~see \citep{White2015,FreidbergBook}).}, which primarily affect the component of the MF orthogonal to the equilibrium field, can be characterized by an arbitrary function of spatial coordinates (\( \theta, \psi , \zeta \)) and time \( t \) \citep{White2013a, White2013b, WhiteBook}. Such perturbations are expressed as:
\begin{equation}
    \delta \mathbf{B} = \nabla\times\alpha\mathbf{B},
\end{equation}
where  \( \delta \mathbf{B}\) represents the perturbed MF and \( \alpha = \sum_{m,n} \alpha_{m,n}(\psi) e^{i(m\theta-n\zeta)} \) is a scalar function of \( \theta \), \( \psi \), \( \zeta \) and \( t \) that represents the amplitude of the perturbations. 

The Lagrangian for the GCM of charged particles in MFL using the standard Boozer coordinates \(\psi, \theta, \zeta\) \citep{boozer1981plasma} takes the following form:
\begin{equation} \label{eq:Lan MFL}
L = \left[ \psi + (\rho_{\parallel} + \alpha) I \right] \dot{\theta} + \left[ (\rho_{\parallel} + \alpha) g - \psi_p \right] \dot{\zeta} + \mu \dot{\xi} - H.
\end{equation}

The Lagrangian \eqref{eq:Lan MFL} is in simplified form that assumes orthogonal coordinate systems that involve the poloidal angle \(\theta\), the toroidal angle \( \zeta\), and the gyro-phase angle \(\xi\) \citep{WhiteBook, Bierwage2022}. The corresponding conjugate momenta are defined as: 
\begin{equation} \label{eq: general momentum}
    \begin{aligned}
      p_{\theta} &= \frac{\partial L}{\partial \theta} = \psi + (\rho_{\parallel} + \alpha) I, \\
      p_{\zeta} &= \frac{\partial L}{\partial \zeta} = (\rho_{\parallel} + \alpha) g - \psi_p, \\
      p_{\xi} &= \frac{\partial L}{\partial \xi} = \mu.
    \end{aligned}
\end{equation}

Hence, the Hamiltonian governing the GCM system's dynamics \eqref{eq:GC H} can be expressed in terms of the canonical variables (\(\theta, p_\theta\)), (\(\zeta, p_\zeta\)), (\(\xi, \mu\)) as follows:
\begin{equation} \label{eq:per GC H}
H = \frac{\left[ p_{\zeta} + \psi_p(\psi) - \alpha(\psi, \theta, \zeta) \right]^2}{2}B^2(\psi, \theta) + \mu B(\psi, \theta),
\end{equation}
where \(\psi\) and \(\psi_p\) depend on the canonical momenta \(p_{\theta}\) and \(p_{\zeta}\), respectively. The perturbed Hamiltonian \eqref{eq:per GC H} for GCM can become highly nonlinear when influenced by a perturbation \(\alpha\). This nonlinearity can lead to chaotic behavior, which we refer to as \textit{kinetic chaos}.

In order to simplify the analysis and accurately describe the system's dynamics, we employ the large aspect ratio (LAR) approximation, which is a common approach in plasma physics  \citep[Sect. 3.6]{WhiteBook}. This approximation assumes that the torus major radius (\( R \)) is relatively larger than the cross-sectional radius (\( r \)),  \( \frac{R}{r} > 1 \). For instance, in Fig.~\ref{fig3:White,Fig1.2}, \(R\) denotes the distance from the center of the torus to its outer edge, while \(r\) represents the distance from the center of the plasma cross-section to its edge. The LAR approximation simplifies the computation of both the MF and the GC equations \citep{FreidbergBook, WhiteBook}. Instead of the standard cylindrical approximation, we apply the actual toroidal geometry, characterized to leading order terms by \(g \simeq 1 \), a negligible plasma current \( I \simeq 0 \), and a MF \( \mathbf{B} = 1 - \sqrt{2 \psi} \cos \theta \), where the field is normalized to its on-axis value. In this case, the canonical momentum \( p_{\theta} \) is equivalent to the toroidal flux, i.e., \( p_{\theta} = \psi \). 

The corresponding EoMs of the Hamiltonian \eqref{eq:per GC H} are given by:
\begin{equation}\label{eq:per GC EoM}
  \begin{aligned}
    \frac{d\zeta}{dt} &= \rho_{||} \mathbf{B}^2, \\ 
    \frac{d\theta}{dt} &= \frac{1}{q(p_{\theta})} \left( \rho_{||} \mathbf{B}^2 + (\mu + \rho_{||}^2 \mathbf{B}) \frac{\partial \mathbf{B}}{\partial \psi_{p}} \right), \\
    \frac{dp_{\zeta}}{dt} &= \rho_{||} \mathbf{B}^2 \frac{\partial \alpha}{\partial \zeta}, \\
    \frac{dp_{\theta}}{dt} &= - (\mu + \rho_{||}^2 \mathbf{B}) \frac{\partial \alpha}{\partial \theta} + \rho_{||} \mathbf{B}^2 \frac{\partial \alpha}{\partial \theta}.
  \end{aligned}
  \end{equation}

It was shown that in the case of low energy particles, where both magnetic moment (\( \mu \)) and the parallel velocity (\( \rho_{\parallel} \)) approach zero, the particle's motion along the MFL (parallel motion) and its perpendicular drifts become negligible. As a result, the particle's motion closely follows the MFLs. This means that the EoM of the GC Hamiltonian \eqref{eq:per GC EoM}, which describes the particle's motion, becomes equivalent to the EoM of the Hamiltonian of the MFLs \eqref{eq:per MFL EoM} (e.g.~see \citep{ram2010dynamics,samanta2017energization,antonenas2021analytical}).

In the absence of perturbations [\(\alpha=0\) in \eqref{eq:per GC H}], the GC Hamiltonian exhibits axisymmetry, leading to integrability with three independent CoM \((E, p_{\zeta}, \mu)\) which characterize each unperturbed orbit. All orbits are classified as either trapped or passing and confined or lost in the \(3D\) CoM phase space. 

Trapped particles are confined in a fixed area of the magnetic confinement devices due to the MF configurations, i.e., they cannot escape from this specific region. On the other hand, passing particles are not restricted to a fixed area, and they can travel across the entire plasma volume. Confined particles, however, are not restricted to a particular region, but they remain in the plasma volume for extended periods, which allow them to interact more freely with the plasma. Nevertheless, particles may be lost from the plasma due to collisions or other mechanisms.

In the LAR approximation, the loss boundary is defined by the following equation:
\begin{equation}\label{eq:walls}
    E = \frac{\left(p_{\zeta} + \psi_{p}(\psi_\text{w})\right)^2}{2}\left(1\mp\sqrt{2\psi_\text{w}}\right)^2 + \mu\left(1\mp\sqrt{2\psi_\text{w}}\right),  
\end{equation}
where the first term represents the kinetic energy corresponding to the particle's toroidal motion, while the second term represents the potential energy due to the particle's interaction with the MF. The positive signs in \eqref{eq:walls} correspond to co-passing particles that are moving in the same direction as the MFLs, while negative signs refer to counter-passing particles that are moving opposite to the MFLs.

Furthermore, orbits passing through the magnetic axis ($\psi=0$ in Fig.~\ref{fig3:White,Fig1.3}), satisfy the following energy condition:
\begin{equation}\label{eq:magnetic axis}
    E = \frac{p_{\zeta}^2}{2} + \mu,   
\end{equation}
and the trapped and passing particles' boundary is defined by 
\begin{equation}\label{eq:trapped passing boundary}
    E = \mu\left(1\mp\sqrt{2\psi(p_{\zeta})}\right).
\end{equation}  
                
The three equations defining the loss boundary \eqref{eq:walls}, magnetic axis \eqref{eq:magnetic axis}, and trapped-passing boundary \eqref{eq:trapped passing boundary} generate parabolic curves in the \((E, p_\zeta)\) space when considering a fixed \(\mu\) value in the \(3D\) \((E, p_\zeta, \mu)\) CoM space. Although trapped particles are confined to a specific region, they can still be lost if their energy (\(E\)) or magnetic moment (\(\mu\)) changes due to collisions or some other process. Passing particles, on the other hand, remain confined or are lost depending on their energy levels and the particular configuration of the MF. 

In order to visually characterize chaos in both MFLs \eqref{eq:per MFL EoM} and particle orbits \eqref{eq:per GC EoM}, we apply a Poincar{\'e} surface of section (PSS) following, for example, \citep{lichtenberg2013regular}. Low energy particles in magnetic confinement devices typically follow MFLs, exhibiting similar phase space structures. However, for higher energy particles, there is no clear relation between kinetic and magnetic chaos. While the PSS provides insights into the GC orbits at a specific energy level, investigating the comparison between kinetic and magnetic chaos requires an efficient chaos detection technique, especially in order to globally characterize and quantify the chaotic nature of many particles at different energy levels in the \((E, p_\zeta)\) space. 

Our study aims to systematically compare the chaotic behavior of MFLs and particle orbits caused by non-axisymmetric magnetic perturbations. To efficiently achieve this goal, we employ the generalized alignment index (GALI) method \eqref{eq:GALI}.  The GALI index enables the characterization of particle orbits according to their chaotic behavior in the system's phase space. It also provides insights into how specific perturbations affect each particle with different kinetic properties. It is important to point out that the maximum Lyapunov exponent (mLE) \eqref{eq:mLEs} has been effectively used in plasma physics for years as a chaos indicator (e.g.~see \citep{falessi2015lagrangian,veranda2017magnetohydrodynamics}). In order to validate GALI's efficacy in detecting chaos in both the GC \eqref{eq:per GC EoM} and MFL \eqref{eq:per MFL EoM} EoMs, we will compare and contrast its performance with the well established and commonly used mLE. This comparison will guarantee the reliability of GALI before proceeding to our primary objective, which is to compare and quantify the chaotic behavior of MFLs and particle orbits.

\section{Numerical results} \label{section:ResultsCh3}
\subsection{Chaos detection and quantification} \label{section:Ch3R1}
In order to showcase the chaos indicators that we use in our study (mLE and GALI) to detect chaos, we begin by considering a simplified form of the GC Hamiltonian model \eqref{eq:GC H}. In the simple model, we assume a radially uniform \(\psi\)-independent additive perturbation term in the LAR GC Hamiltonian, setting \(\mathbf{B}=1-\sqrt{2p_\theta} \cos \theta\) and \(\psi_{p}(p_{\theta}) = p_{\theta}\), while also considering a constant safety factor \(q(\psi)=1\). As a result, the GC Hamiltonian \eqref{eq:GC H} simplifies to: 
\begin{equation}\label{eq:GCM-Ham-q=1}
H = \frac{1}{2} \left( p_{\zeta} + p_{\theta} \right)^2 \left( 1 - \sqrt{2 p_\theta} \cos \theta \right)^2 + \mu \left( 1 - \sqrt{2 p_\theta} \cos \theta \right) + \epsilon \left[ \sin(m_1 \theta - n_1 \zeta) + \sin(m_2 \theta - n_2 \zeta) \right],
\end{equation} 
where the magnetic moment \(\mu\) is subject to two perturbation modes with poloidal mode numbers \(m_1, m_2\), and toroidal mode numbers \(n_1, n_2\), along with perturbation amplitude \(\epsilon\). The corresponding EoM can be derived as follows:
\begin{equation} \label{eq:GCM-EqnM-q=1}
  \begin{aligned}
    \frac{d\zeta}{dt} &= (1 - \mathbf{B})^2 (p_\zeta + p_\theta), \\ 
    \frac{d\theta}{dt} &= (1 - \mathbf{B})^2 (p_\zeta + p_\theta) 
                      - \frac{\mu \cos \theta}{\sqrt{2 p_\theta}} 
                      - \frac{\cos \theta \, (1 - \mathbf{B}) (p_\zeta + p_\theta)^2}{\sqrt{2 p_\theta}}, \\ 
    \frac{dp_\zeta}{dt} &= - (p_\zeta + p_\theta)^2 (1 - \mathbf{B}) \sqrt{2 p_\theta} \sin \theta \\ 
    &\quad - \mu \sqrt{2 p_\theta} \sin \theta 
    + \epsilon \left[ m_1 \cos (n_1 \zeta - m_1 \theta) 
    + m_2 \cos (n_2 \zeta - m_2 \theta) \right], \\ 
    \frac{dp_\theta}{dt} &= - \epsilon \left[ n_1 \cos (n_1 \zeta - m_1 \theta) 
    + n_2 \cos (n_2 \zeta - m_2 \theta) \right].
  \end{aligned}	
\end{equation}

To demonstrate the efficiency of the GALI method \eqref{eq:GALI} and its relation with the mLE \eqref{eq:mLEs} we consider various types of orbits of the simplified $2D$ GC Hamiltonian \eqref{eq:GCM-Ham-q=1}. We present the computations of these indices for specific parameters of the system. In particular, we consider particles with normalized energy \(E = 8.131\times10^{-6}\) and magnetic moment \(\mu = 8.1423\times10^{-6}\) subject to two perturbation modes with mode numbers \(m_1 = 1\), \(n_1 = 5\), \(m_2 = 1\), \(n_2 = 3\) and perturbation amplitude of \(\epsilon = 0.135\times10^{-8}\). While these parameters are chosen for computational efficiency, they can be related to physical quantities in a real tokamak system. For example, the normalized energy and magnetic moment perturbation can be connected to corresponding physical energy and MF perturbation values by considering a reference energy scale and MF strength.

For a given IC, we begin by numerically integrating the system's EoM \eqref{eq:GCM-Ham-q=1} to follow the time evolution of the orbit itself. The orbit intersects with the PSS, defined by the conditions \( \theta = 0 \) and \( p_{\theta} > 0 \), at an energy level \( E = 8.131 \times 10^{-6} \) multiple times. This process generates a set of points in the \(2D\) \((\zeta, p_{\zeta})\) space. 

\begin{figure}[!htb]
    \centering
    \includegraphics[width=1\textwidth]{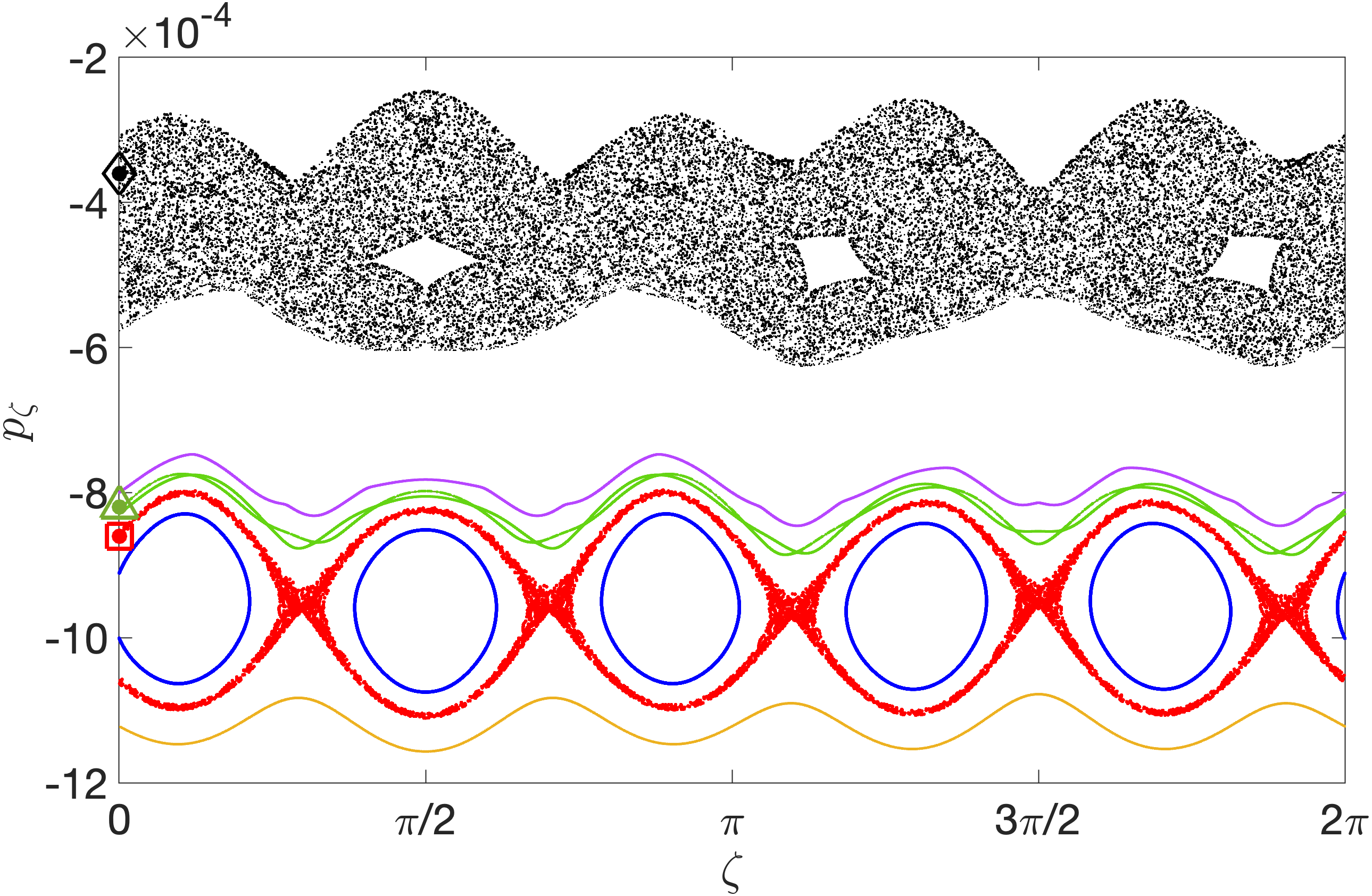}
    \caption{The PSS \((\theta = 0; p_\theta > 0)\) of the GC Hamiltonian \eqref{eq:GCM-Ham-q=1} for normalized energy \(E = 8.131 \times 10^{-6}\) and magnetic moment \(\mu = 8.1423\times10^{-6}\), subject to perturbations with mode numbers  \((m_1, n_1) = (1, 5)\), \((m_2, n_2) = (1, 3)\), and perturbation amplitude \(\epsilon = 0.135\times10^{-8}\). Regular orbits correspond to blue, orange, and purple points, while chaotic orbits are shown by green, red, and black points. The ICs for the chaotic orbits are \(p_{\zeta_0} = -0.85 \times 10^{-3}\) (red square point), \(p_{\zeta_0} = -0.81 \times 10^{-3}\) (green triangle point), and \(p_{\zeta_0} = -0.36 \times 10^{-3}\) (black diamond point) with \(\zeta = 0\).}
    \label{fig3:Fig1}
  \end{figure}

Figure~\ref{fig3:Fig1} displays the system's PSS, where we can easily identify both regular and chaotic GC orbits. In the phase space portrait, we see regular orbits (blue small islands and curves in orange and purple) and chaotic ones (black scattered points, alongside red and green points at the borders of stability islands). The PSS is generated by integrating the EoM of the Hamiltonian \eqref{eq:GCM-Ham-q=1} up to \(t=10^9\) time units in order to clearly reveal the true nature of the studied orbits.

The considered chaotic orbits are characterized by the following ICs: \(p_{\zeta_0} = -0.85 \times 10^{-3}\) (red square point), \(p_{\zeta_0} = -0.81 \times 10^{-3}\) (green triangle point), and \(p_{\zeta_0} = -0.36 \times 10^{-3}\) (black diamond point). All three chaotic orbits we consider exist around chains of stable resonant islands. The location of these resonances is influenced by the safety factor, which in this case is \(q=1\), and it depends on the poloidal and toroidal mode numbers, in particular \(m=(1, 1)\) and \(n=(5, 3)\), of the perturbation. This relationship determines the number of specific resonances and their positions based on the value of the momentum \(p_\zeta\). As a result, in Fig.~\ref{fig3:Fig1}, we observe two stable island resonance chains with multiplicity one (\(m_1 = m_2 = 1\)): five resonances (\(n_1=5\)) at lower \(p_\zeta\) values (which are surrounded by the red chaotic regions), and three smaller resonances (\(n_2=3\)) at the higher \(p_\zeta\) values (which are surrounded by the black scattered points).

In order to compute chaos indicators like the mLE \eqref{eq:mLEs} and the GALI \eqref{eq:GALI}, the system's so-called variational equations \eqref{eq:Gen Ham VoEs} need to also be integrated along with the EoM \eqref{eq:GCM-EqnM-q=1}. From \eqref{eq:Gen Ham VoEs} these equations will have the following form:
\[
\dot{\mathbf{v}} = [\delta \zeta , \delta \theta, \delta p_\zeta, \delta p_\theta]^T  = \left[ \mathbf{J}_{4} \cdot \mathbf{D}^2_H \right] \cdot \mathbf{v_0}
\]
where 
\[
D^2H = 
\begin{bmatrix}
\frac{\partial^2 H}{\partial \zeta \partial \zeta} & \frac{\partial^2 H}{\partial \zeta \partial \theta} & \frac{\partial^2 H}{\partial \zeta \partial p_\zeta} & \frac{\partial^2 H}{\partial \zeta \partial p_\theta} \\
\frac{\partial^2 H}{\partial \theta \partial \zeta} & \frac{\partial^2 H}{\partial \theta \partial \theta} & \frac{\partial^2 H}{\partial \theta \partial p_\zeta} & \frac{\partial^2 H}{\partial \theta \partial p_\theta} \\
\frac{\partial^2 H}{\partial p_\zeta \partial \zeta} & \frac{\partial^2 H}{\partial p_\zeta \partial \theta} & \frac{\partial^2 H}{\partial p_\zeta \partial p_\zeta} & \frac{\partial^2 H}{\partial p_\zeta \partial p_\theta} \\
\frac{\partial^2 H}{\partial p_\theta \partial \zeta} & \frac{\partial^2 H}{\partial p_\theta \partial \theta} & \frac{\partial^2 H}{\partial p_\theta \partial p_\zeta} & \frac{\partial^2 H}{\partial p_\theta \partial p_\theta}
\end{bmatrix}.
\]
The variational equations are given by the following expressions:
\begin{equation}\label{eq:GCM-VarEqn-q=1} 
  \begin{aligned} 
    \frac{d \delta \zeta}{dt} &= \big[2(p_\zeta + p_\theta) \mathbf{B} \sqrt{2 p_\theta} \sin\theta \big] \delta \theta 
  + \mathbf{B}^2 \delta p_\zeta \\ 
  &\quad + \big[ \mathbf{B}^2 - \sqrt{2} (p_\zeta + p_\theta) \mathbf{B} \frac{\cos\theta}{\sqrt{p_\theta}} \big] \delta p_\theta, \\ 
    \frac{d \delta \theta}{dt} &= \bigg[ -\cos\theta \sin\theta (p_\zeta + p_\theta)^2 
  + \frac{\sin\theta \mathbf{B} (p_\zeta + p_\theta)^2}{\sqrt{2 p_\theta}} 
  + \frac{\mu \sin\theta}{\sqrt{2 p_\theta}} \bigg] \delta \theta \\ 
  &\quad + \bigg[\mathbf{B}^2 - \frac{\sqrt{2}\cos\theta \mathbf{B} (p_\zeta + p_\theta)}{\sqrt{p_\theta}} \bigg] \delta p_\theta \\ 
  &\quad + \bigg[ \mathbf{B}^2 
  + \frac{\mu \cos\theta}{2 p_\theta^{1.5}} 
  - \frac{\sqrt{2}\cos\theta \mathbf{B} (p_\zeta + p_\theta)}{\sqrt{p_\theta}} 
  + \frac{\cos\theta \mathbf{B} (p_\zeta + p_\theta)^2}{2\sqrt{2}p_\theta^{1.5}} 
  + \frac{\cos^2\theta (p_\zeta + p_\theta)^2}{2 p_\theta} \bigg] \delta p_\theta, \\ 
    \frac{d \delta p_\zeta}{dt} &= \epsilon \bigg[ n_1 \sin (n_1 \zeta - m_1 \theta) (n_1 \delta \zeta - m_1 \delta \theta) 
  + n_2 \sin (n_2 \zeta - m_2 \theta) (n_2 \delta \zeta - m_2 \delta \theta) \bigg], \\ 
    \frac{d \delta p_\theta}{dt} &= \big[ 2 p_\theta \cos\theta (p_\zeta + p_\theta)^2 
  - \sqrt{2 p_\theta} (\mu + (p_\zeta + p_\theta)^2) \big] \delta \theta \\ 
  &\quad + 2 p_\theta \sin (2\theta) (p_\zeta + p_\theta) \delta p_\zeta 
  + (p_\zeta + 3 p_\theta) \delta p_\theta \\ 
  &\quad - \sin\theta (\mu \delta p_\theta) 
  + \frac{(p_\zeta + p_\theta)}{\sqrt{2 p_\theta}} (4 p_\theta \delta p_\zeta) 
  + (p_\zeta + 5 p_\theta) \delta p_\theta \\ 
  &\quad + \epsilon \bigg[ m_1 \sin (n_1 \zeta - m_1 \theta) (-n_1 \delta \zeta + m_1 \delta \theta) 
  - m_2 \sin (n_2 \zeta - m_2 \theta) (-n_2 \delta \zeta + m_2 \delta \theta) \bigg]. 
  \end{aligned}		
\end{equation}

In our analysis of the simplified GC model, we perform all numerical simulations by simultaneously integrating both the EoM \eqref{eq:GCM-EqnM-q=1} and the variational equations \eqref{eq:GCM-VarEqn-q=1} using the RK4 scheme \eqref{eq:RK}. The integration time step is chosen carefully to ensure that the relative energy error remains below \(10^{-8}\) throughout the entire simulation period.

\begin{figure}[!htb]
    \centering
    \subfloat[ftmLE ($t$)\label{fig3:Fig2a}]{\includegraphics[width=0.51\textwidth]{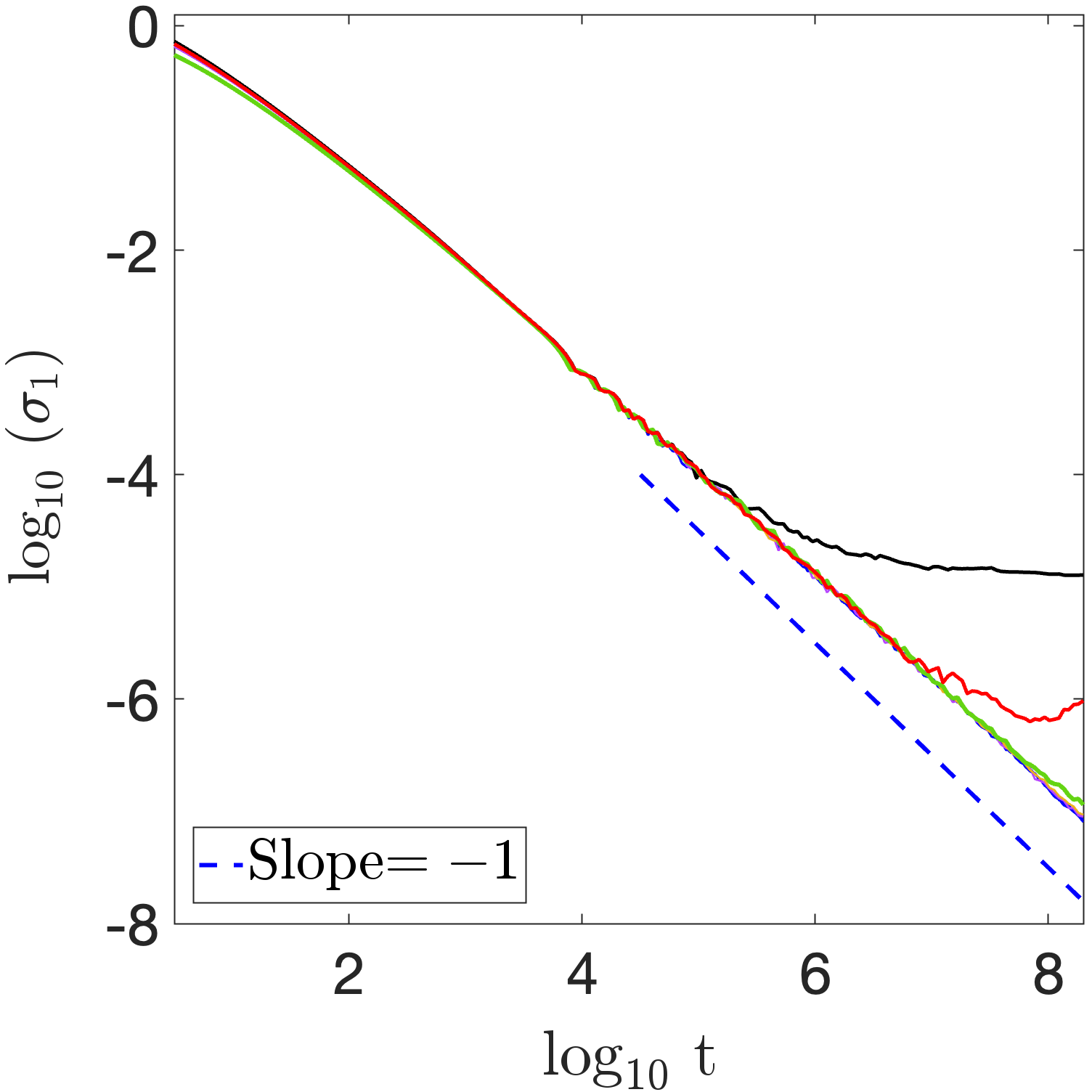}}\hfill 
    \subfloat[GALI\(_2 (t)\)\label{fig3:Fig2b}] {\includegraphics[width=0.49\linewidth]{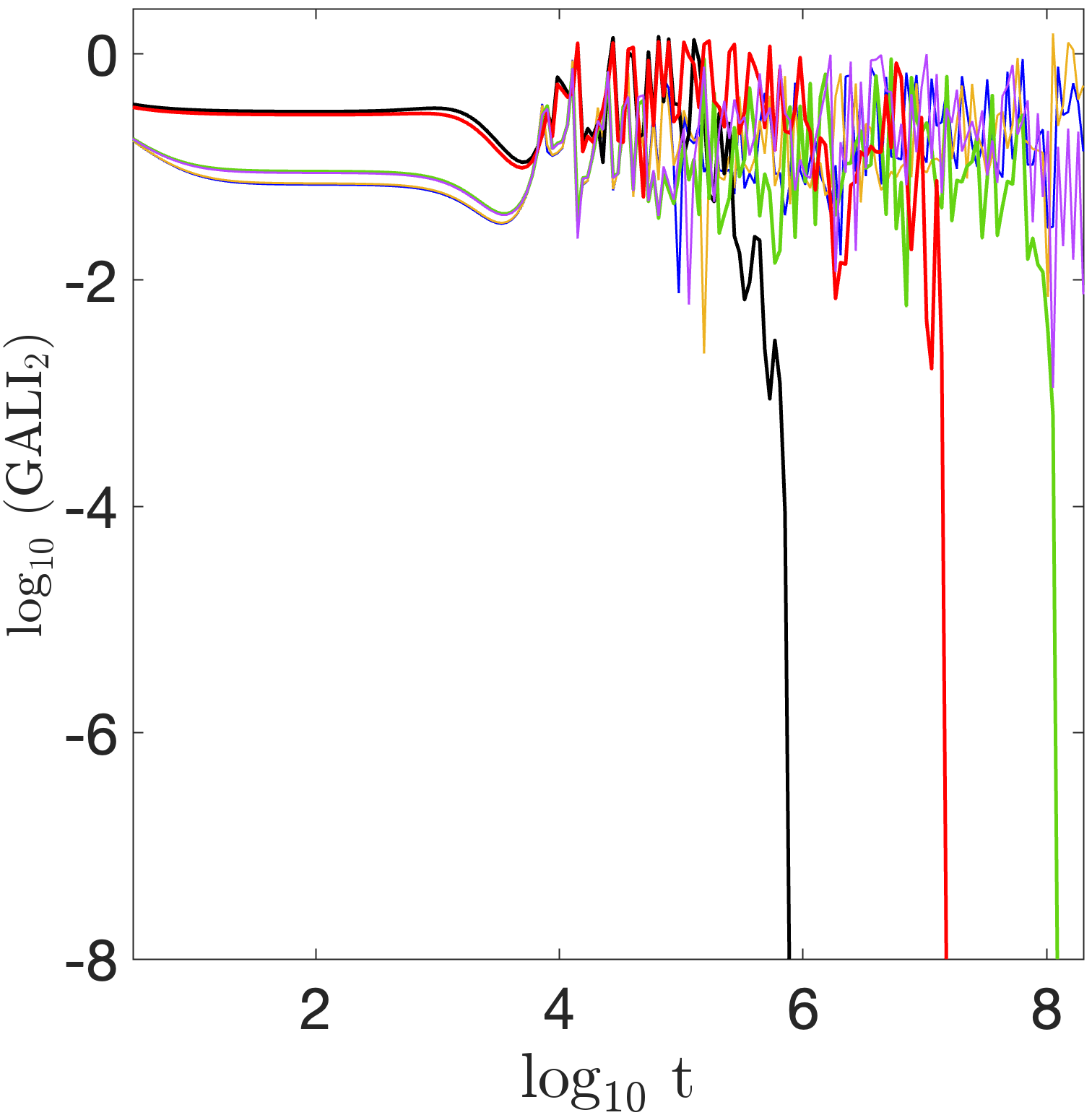}}     
    \caption{The time evolution of (a) the ftmLE, \( \sigma_1 \) \eqref{eq:ftmLE}, and (b) the GALI\(_2\) \eqref{eq:GALI} for the six GC orbits shown in Fig.~\ref{fig3:Fig1} using the same color scheme. The dashed line in (a) represents a function proportional to \(t^{-1}\).}
    \label{fig3:Fig2}
  \end{figure}

Figures \ref{fig3:Fig2a} and (b), respectively, present the time evolution of the ftmLE, \( \sigma_1 \), and the GALI\(_2\) for the six orbits shown in Fig.~\ref{fig3:Fig1}, using the same color scheme. It is noteworthy that we have chosen to focus only on the GALI\(_2\) method (practically equivalent to the SALI \citep{skokos2001alignment}) as our chaos indicator, which is the lowest order GALI indicator. While higher order GALI indices could potentially be used to detect chaos in multidimensional DSs, the GALI\(_2\) is quite sufficient for our study as our primary goal is to distinguish between chaotic and regular orbits. 

For regular orbits (represented by blue, orange, and purple curves), \( \sigma_1 \) tends to zero following a power law decay proportional to the function \( t^{-1} \) [indicated by a dotted line in Fig.~\ref{fig3:Fig2a}], while for the chaotic orbits (black, red, and green curves) it eventually saturates to a positive number. In Fig.~\ref{fig3:Fig2a}, we observe that this saturation of \( \sigma_1 \) for chaotic orbits occurs at different times, depending on the strength of chaos. The black curve starts saturating at \( t \approx 10^6 \), indicating a higher degree of chaos, whereas the red curve saturates later, at \( t \approx 4 \times 10^7 \), which indicates a lower strength of chaos (i.e., weakly chaotic behavior). It is worth mentioning that the third chaotic orbit (green curve) does not display any clear signs of saturation within the considered simulation time of \( 2 \times 10^8 \). 

On the other hand, as seen in Fig.~\ref{fig3:Fig2b}, the GALI\(_2\) remains practically constant for the regular orbits (represented by blue, orange, and purple curves), while it decays exponentially fast to zero for the chaotic orbits.  GALI\(_2\) clearly distinguishes the strength of chaos for the three chaotic orbits (black, red, and green curves) under considerably shorter computational time compared to the ftmLE. This efficiency is because chaos quickly causes GALI\(_2\) values to deviate significantly (by large orders of magnitude) from those of regular orbits, which provides a distinct and immediate indication of chaotic dynamics. 

To illustrate how the GALI\(_2\) differentiate chaos more efficiently compared to the mLE, let us consider the first two [black and red curves in Fig.~\ref{fig3:Fig2b}] cases of chaotic orbits. For the orbit shown in black, the GALI\(_2\) decays to zero [\(\log_{10} (\text{GALI}_2) \approx -8.697\)] at \(\log_{10} (t) \approx 5.89\). For the red orbit, GALI\(_2\) decays to \(\log_{10} (\text{GALI}_2) \approx-8.562\) at \(\log_{10} (t) \approx  7.18\). In both cases, the ftmLE saturates to its positive asymptotic value:  \(\log_{10} (\sigma_1) \approx -4.862\) at \(\log_{10} (t) \approx 7.60\) for the black orbit, and \(\log_{10} (\sigma_1) \approx -6.044\) at \(\log_{10} (t) \approx 8.26\) for the red orbit [Fig.~\ref{fig3:Fig2a}]. This observation reveals that the GALI\(_2\) detects chaos at least two orders of magnitude faster for the black orbit and one order of magnitude faster for the red orbit in comparison with the ftmLE. Note that, in Fig.~\ref{fig3:Fig2a}, the ftmLE for the blue orbit deviates from the \(t^{-1}\) power law behavior at \(\log_{10} (t) \approx 6\) and finally saturates to its limiting value only after another order of magnitude, at \(\log_{10} (t) \approx 7.60\). 

It is important to highlight that the green curve in Fig.~\ref{fig3:Fig2a}, which corresponds to a weakly chaotic orbit, could be mistakenly classified as belonging to a regular orbit. This is due to the fact that the ftmLE fails to show a clear sign of divergence from the $t^{-1}$ power law decay (associated with regular orbits) up to the final integration time we considered, \( \log_{10}(t) \approx 8.301 \). On the other hand, GALI\(_2\) abruptly decreases to practically zero in Fig.~\ref{fig3:Fig2b}  [\( \log_{10}(\text{GALI}_2) \approx -8.819 \) at \( \log_{10}(t) \approx 8.09 \)], conclusively identifying the green orbit as chaotic. This example clearly demonstrates that the GALI\(_2\) method is more efficient for detecting weakly chaotic orbits in our system than the estimation of the mLE.
                    
In order to obtain a global understanding of the GC Hamiltonian \eqref{eq:GCM-Ham-q=1}, it is impractical to analyze individual orbits to determine their regular and chaotic nature, as we have done in Fig.~\ref{fig3:Fig2}. A broader and more holistic approach is necessary. The objectives of this chapter are to provide a more comprehensive analysis of chaos in the GC Hamiltonian \eqref{eq:GC H}. The GALI\(_2\) method has an important advantage over the computation of the mLE, as it provides numeral values in a specific range \([0, 1]\). In particular, we can identify chaotic orbits when GALI\(_2 = 0\) [e.g., for the orbits corresponding to green, black, and red curves in Fig.~\ref{fig3:Fig2b}] and regular orbits when GALI\(_2\) is significantly larger than zero [e.g., for orbits associated with blue, orange, and purple curves in Fig.~\ref{fig3:Fig2b}]. These well-defined differences in the GALI\(_2\) values allow for a more global analysis of the orbital behavior over sets of many ICs. Let us explore this feature property by taking a grid of ICs within the \((z, p_{\zeta})\) space of the simplified GC Hamiltonian \eqref{eq:GCM-Ham-q=1}. 

For each considered IC of the simple GC Hamiltonian \eqref{eq:GCM-Ham-q=1} we compute multiple trajectory intersections with the PSS, defined on the \(\theta=0 \) plane, in the direction where the plasma current flow has \(p_{\theta} > 0\). We note that \(p_{\theta}\) is always determined by the Hamiltonian equation \eqref{eq:GCM-Ham-q=1} and the \((z, p_{\zeta})\) values of each IC. This process generates a $2D$ set of points in the \((z, p_{\zeta})\) space. 

\begin{figure}[!htb]
    \centering
    \subfloat[ftmLE\label{fig3:Fig3a}]{\includegraphics[width=0.49\textwidth]{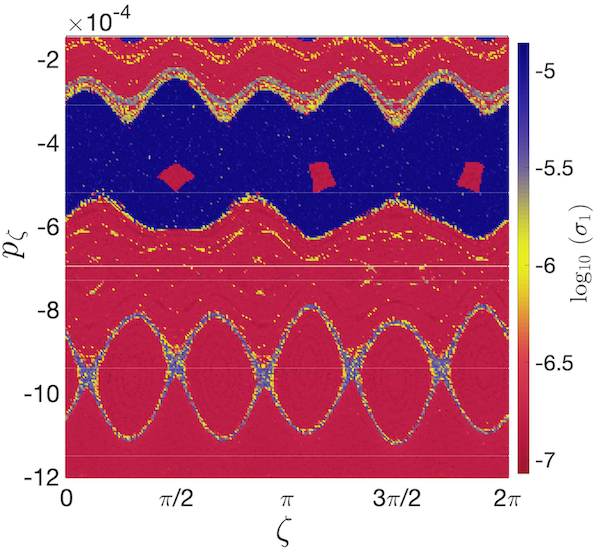}}\hfill
    \subfloat[GALI\(_2\)\label{fig3:Fig3b}] {\includegraphics[width=0.49\linewidth]{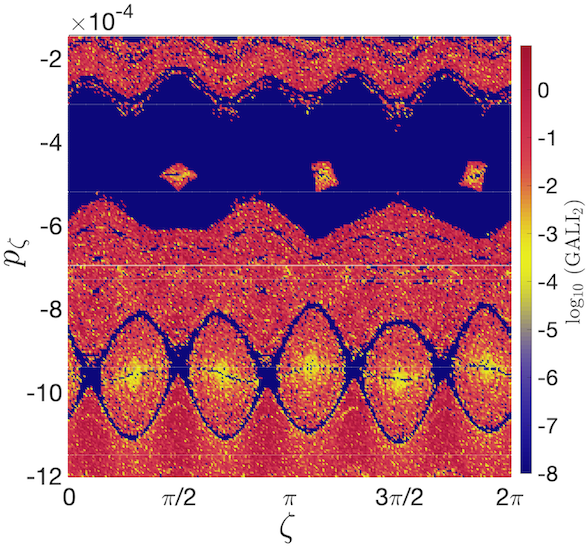}}\hfill
    \subfloat[GALI\(_2 \le 10^{-8}\)\label{fig3:Fig3c}]{\includegraphics[width=0.5\textwidth]{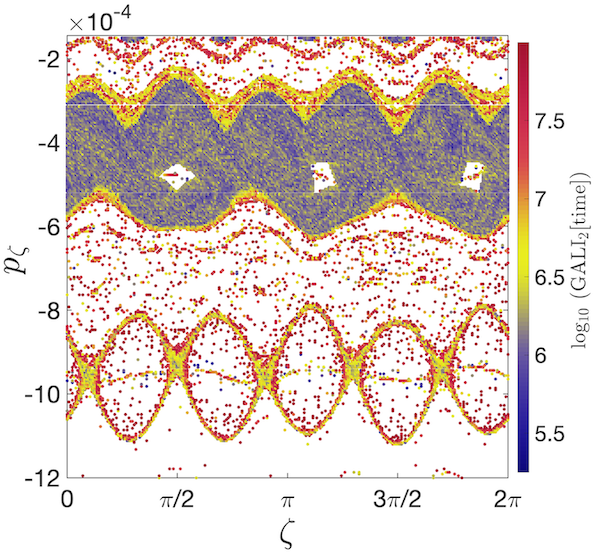}}
    \caption{The PSS \((\theta = 0; p_\theta > 0)\) of the GC Hamiltonian \eqref{eq:GCM-Ham-q=1} where points are colored according to the values of (a) \(\sigma_1\) and (b) GALI\(_2\) after \(t=10^8\) integration time units. (c) The time required for GALI\(_2\) to detect chaos, i.e., for GALI\(_2 \le 10^{-8}\). We consider ICs uniformly distributed in \((\zeta, p_\zeta)\) space with \( \zeta \in [0, 2\pi]\) and \(p_\zeta \in [-1.2 \times 10^{-3}, -1.5 \times 10^{-4}]\), while the energy  \(E\), magnetic moment \(\mu\) and perturbation parameter values are the same as in Fig.~\ref{fig3:Fig1}.}
    \label{fig3:Fig3}
  \end{figure}

In Fig.~\ref{fig3:Fig3a} and (b), we depict the PSS of the GC Hamiltonian \eqref{eq:GCM-Ham-q=1}, where the ICs are colored according to their ftmLE and GALI\(_2\) values after \(t=10^8\) integration time units, respectively. These figures are generated for a dense set of uniformly distributed ICs, i.e., \(62,751\) points, in the  \((\zeta, p_\zeta)\) space defined by \( \zeta \in [0, 2\pi]\) and \(p_\zeta \in [-1.2 \times 10^{-3}, -1.5 \times 10^{-4}]\). Each IC is assigned a logarithmic color scale on the right side of each panel. In Fig.~\ref{fig3:Fig3a}, chaotic orbits, which are characterized by positive ftmLE values (around \(10^{-5}\)), are shown in dark blue. Similarly, in Fig.~\ref{fig3:Fig3b}, chaotic orbits are depicted in dark blue, but here they are identified by small GALI\(_2\) values (approximately below \(10^{-8}\)). On the other hand, red color regions in the PSS portraits correspond to regular orbits. These orbits exhibit much smaller ftmLE values (around \(10^{-7}\)) in Fig.~\ref{fig3:Fig3a}, and significantly positive GALI\(_2\) values (approximately \(1\)) in Fig.~\ref{fig3:Fig3b}. Intermediate colors (light gray-yellowish regions) represent weakly chaotic orbits that require a longer integration time to fully reveal their chaotic nature. 

The GALI\(_2\) plot in Fig.~\ref{fig3:Fig3b} provides a more detailed representation of the GC phase space chaoticity, identifying narrow chaotic regions around separatrices of higher order resonant island chains. This demonstrates a significant advantage of the GALI\(_2\) index for accurately identifying chaotic orbits compared to the ftmLEs. Lastly, Fig.~\ref{fig3:Fig3c} displays the time required for GALI\(_2\) to converge to (practically) zero values [in reality we consider the threshold value \(\log_{10} (\text{GALI}_2) = -8\)], indicating the detection of chaos. White regions represent areas where GALI\(_2\) has not reached this threshold within the considered simulation time, highlighting areas of lower strength of chaos or regular orbits.  

It is important to note that orbits categorized as regular based on the ftmLE and the GALI\(_2\) computations up to time $t=10^{8}$ [red regions in Figs.~\ref{fig3:Fig3a} and (b)] may eventually exhibit chaotic behavior if we integrate the system \eqref{eq:GCM-Ham-q=1} for longer times. Fig.~\ref{fig3:Fig3} clearly illustrates the limitation of ftmLE [Fig.~\ref{fig3:Fig3a}] in capturing fine chaotic structures compared to GALI\(_2\) [Fig.~\ref{fig3:Fig3b}]. The smoother appearance of the ftmLE color map is due to its inability to accurately detect weakly chaotic regions under the considered periods. On the other hand, GALI\(_2\) has the advantage of quickly identifying these orbits, such as those in the dark blue areas around the five main resonances in Fig.~\ref{fig3:Fig3b}, which are not clearly identified by the ftmLE in Fig.~\ref{fig3:Fig3a}. These regions correspond to the separatrices of small island chains associated with higher order resonances between the unperturbed motion and the perturbing modes.

Furthermore, we observe the computational CPU time required to create both the ftmLE and GALI\(_2\) color plots is comparable, despite the fact that GALI\(_2\) requires the evolution of an orbit and two initially linearly independent deviation vectors, while ftmLE requires only an orbit and a single deviation vector. More specifically, producing the ftmLE and GALI\(_2\) color plots in Figs.~\ref{fig3:Fig3a} and (b), respectively, was computationally quite high. It required approximately \(1.29 \times 10^6\) minutes (\(\approx 895.06\) days) to create the GALI\(_2\) color plot, while the ftmLE color plot took around \(1.23 \times 10^6\) minutes (\(\approx 853.19\) days). These computations were carried out up to the final integration time of \(t = 10^8\) on the Lengau cluster of the CHPC of South Africa (see Sect.~\ref{section:Compuration Process}).  

\begin{figure}[!htb]
    \centering
    \subfloat[PSS\label{fig3:Fig4a}]{\includegraphics[width=0.52\textwidth]{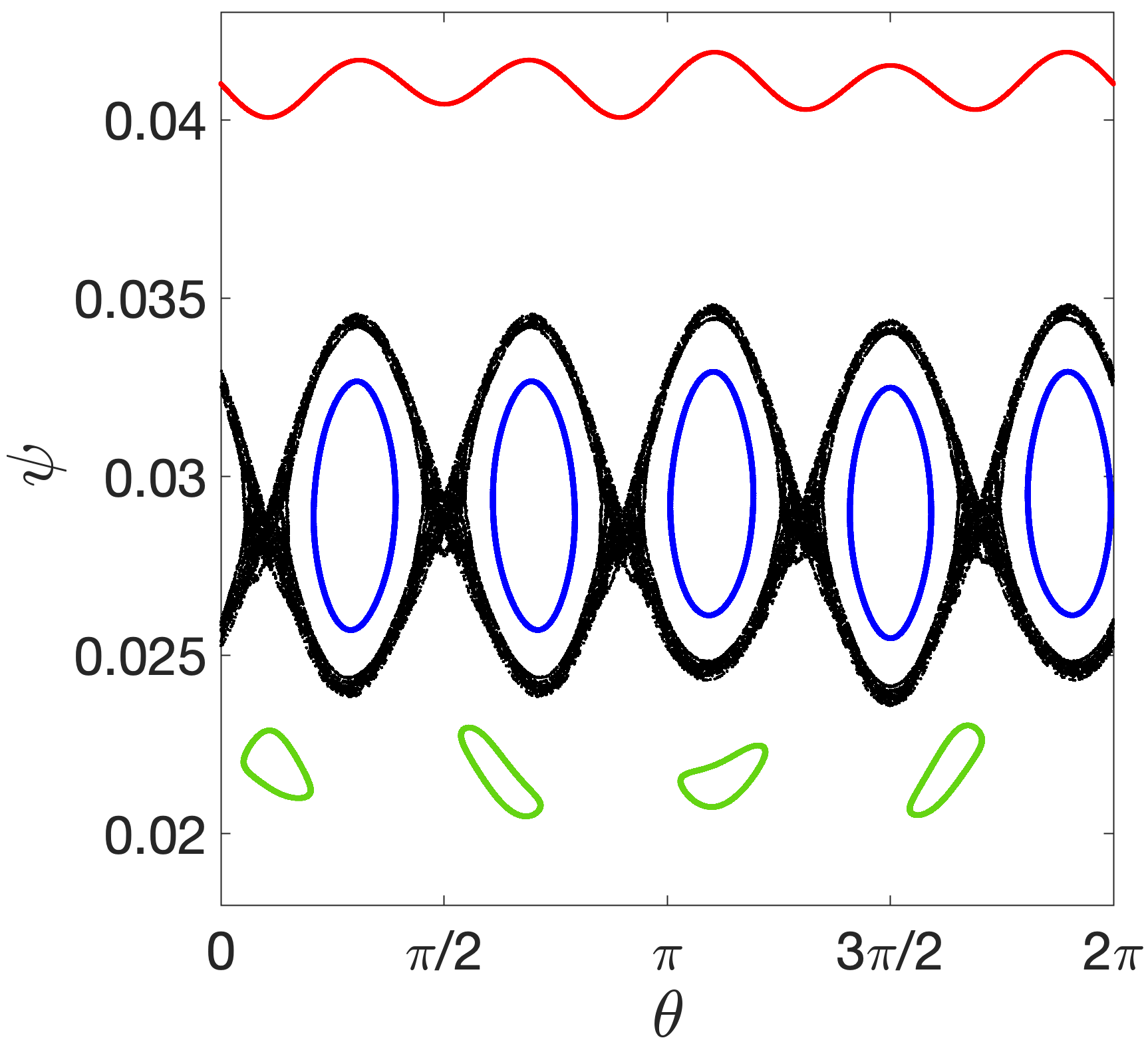}}\hfill
    \subfloat[ftmLE\(n\)\label{fig3:Fig4b}] {\includegraphics[width=0.48\linewidth]{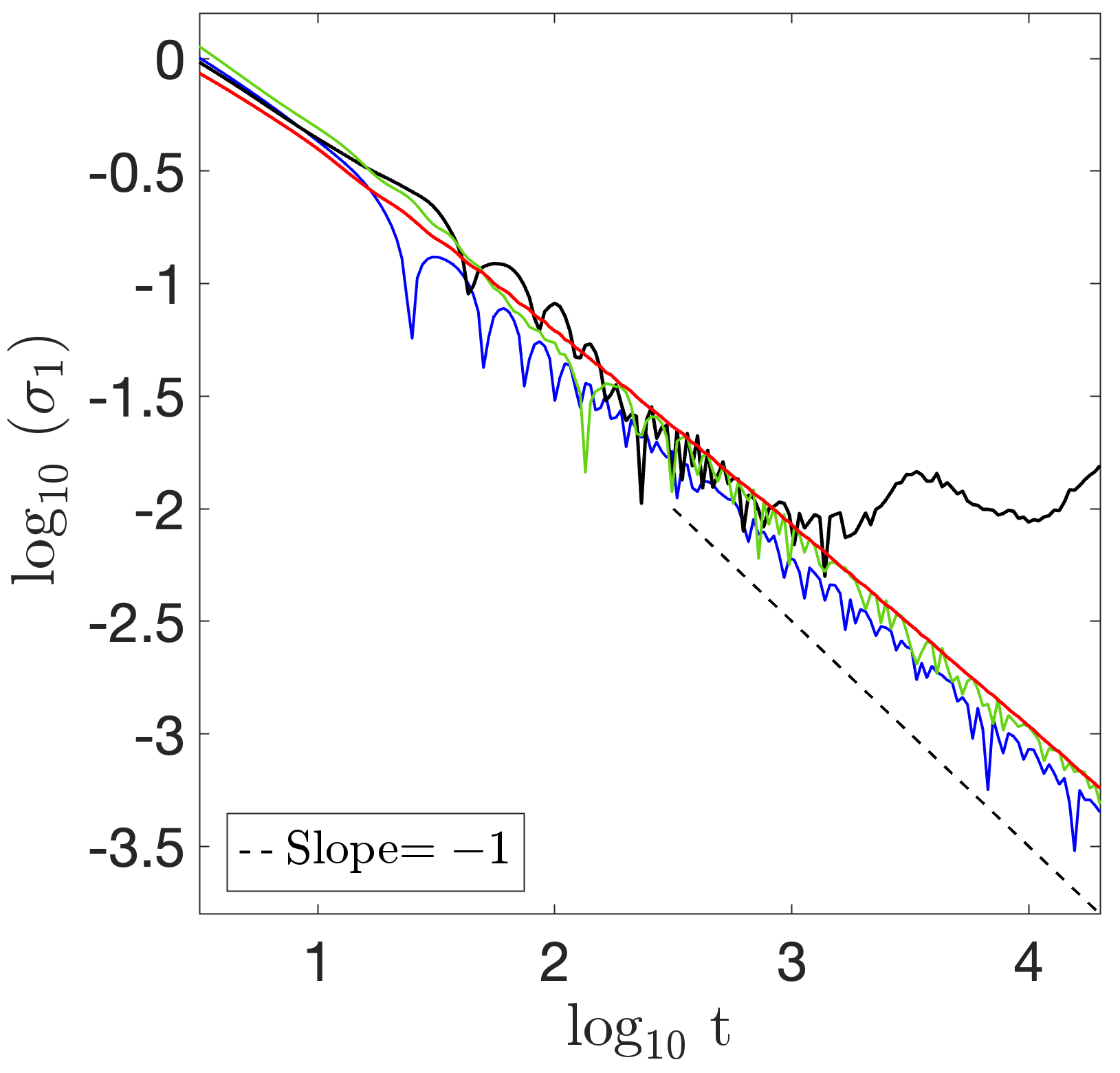}}\hfill
    \subfloat[GALI\(_2(n)\)\label{fig3:Fig4c}]{\includegraphics[width=0.5\textwidth]{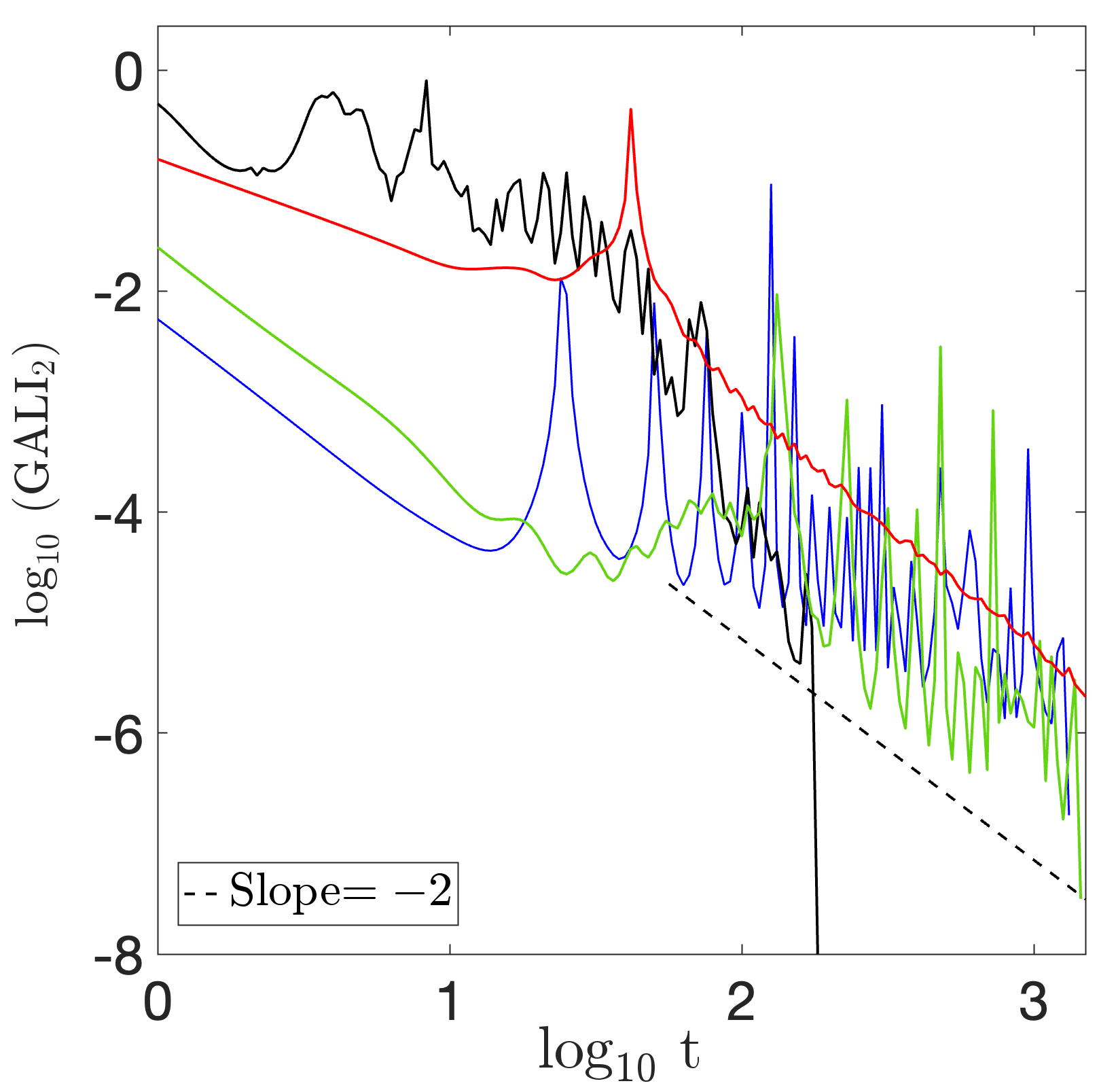}}
    \caption{(a) The PSS ($\zeta = 0; \mod 2\pi$) orbits of the MFL Hamiltonian \eqref{eq:per MFL Ham}. In particular, we consider regular orbits, which are plotted by green, blue, and red points, as well as one chaotic orbit denoted by black points. The time evolution of the (b) ftmLE  \(\sigma_1\) \eqref{eq:ftmLE}, and (c) the GALI\(_2\) \eqref{eq:GALI} for these orbits. The dashed lines in panels (b) and (c) represent functions proportional to the power laws \(n^{-1}\) and \(n^{-2}\), respectively.}
    \label{fig3:Fig4}
  \end{figure}

Having examined the efficiency of GALI\(_2\) for the GCM system, we now shift our focus to the magnetic Hamiltonian system \eqref{eq:per MFL Ham} to perform a similar analysis. In Fig.~\ref{fig3:Fig4a}, we present the PSS of the system, showcasing some representative regular and chaotic orbits. In particular, we consider three different ICs (\(\theta, \psi\)) of regular orbits: \((0.03,0.03)\) shown by green point \((0.12,0.0212)\) displayed by blue point and \((0,0.04)\) presented by red point. In addition, we consider a chaotic orbit with IC \((\theta, \psi) = (0.032, 0.033)\), depicted by black points, which is located at the border of the five main resonance islands of stability. We previously demonstrated how GALI\(_2\) effectively identifies chaotic orbits with varying strengths of chaoticity in Fig.~\ref{fig3:Fig2}. Here, we considered three different regular orbits for the magnetic system \eqref{eq:per MFL Ham} to illustrate the power law decay of GALI\(_2\) index for these orbits. 

The time evolution of the ftmLE \(\sigma_1\) \eqref{eq:ftmLE} and the GALI\(_2\) \eqref{eq:GALI} of these orbits is presented in Figs.~\ref{fig3:Fig4b} and (c), respectively. For regular orbits, the \(\sigma_1\) tends to zero following a power law \(n^{-1}\), while for chaotic orbits, it saturates at a constant positive value. On the other hand, the GALI\(_2\) eventually decreases to zero exponentially fast for the chaotic orbit [black curve in  Fig.~\ref{fig3:Fig4c}], while it follows a power law \(n^{-2}\) for the regular orbits [green, blue, and red curves in  Fig.~\ref{fig3:Fig4c}]. This behavior differs from that observed for the GC Hamiltonian in Fig.~\ref{fig3:Fig4c}, where GALI\(_2\) for regular orbits eventually saturates to a constant non-zero value.

 The observed GALI\(_2\) behavior for regular orbits in Fig.~\ref{fig3:Fig4c} is in agreement with the theoretical prediction for the index behavior in the case of the $2D$ area preserving maps in \eqref{Prop:GALI_2 for SM}. In fact, the MFLs Hamiltonian \eqref{eq:per MFL Ham} can be transformed to a $2D$ area preserving map, like the Chirikov standard map \citep{Chirikov1979}. In the case of a regular orbit on a $2D$ map, the two initially linearly independent deviation vectors eventually fall on the tangent space of the $1D$ invariant curve on which the motion takes place. As a result, these two deviations will eventually align, leading to a power law decay \(n^{-2}\) \eqref{Prop:GALI_2 for SM}, where \(n\) represents the number of iterations of the map, causing the GALI\(_2\) to decay accordingly \citep{manos2007studying}. For a chaotic orbit [represented by black points located near the boundary of the five main resonance islands of stability in Fig.~\ref{fig3:Fig4a}], the two deviation vectors align with the direction defined by the mLE. Thus, the GALI\(_2\) decays to zero exponentially fast, following the form described in \citep{skokos2004detecting}, which is similar to the behavior observed for chaotic orbits in the GC Hamiltonian [Fig.~\ref{fig3:Fig2b}].

 Similar to our observation for the GC Hamiltonian in Figs.~\ref{fig3:Fig2} and \ref{fig3:Fig3}, the distinct behavior of the GALI\(_2\) index allows for a fast and clear distinction between chaotic and regular orbits in the phase space of the EoM of the MFLs \eqref{eq:per MFL EoM}. From Fig.~\ref{fig3:Fig4c}, we see that even at \(n \approx 2000\) time units, GALI\(_2\) values for various regular orbits (represented by the blue, green, and red curves) remain well above \(10^{-8}\) threshold. On the other hand, for a chaotic orbit, GALI\(_2\) decays to zero exponentially fast, falling below the threshold within approximately \(n \approx 180\) time units (black curve). 
 
 In general, orbits with GALI\(_2\) values smaller than \(10^{-8}\) can be characterized as chaotic within relatively short integration times, whereas the ftmLE requires significantly longer integration times to reveal the true chaotic nature of these orbits. For example, in Fig.~\ref{fig3:Fig4b}, the ftmLE for the black curve has not yet fully saturated to its positive limit value at \(t \approx 2 \times 10^4\), while GALI\(_2\) goes to zero [\(\log_{10} (\text{GALI}_2) \approx -8.724\)] at \(t \approx 180\) for the same orbit [black curve in Fig.~\ref{fig3:Fig4c}]. This efficiency of GALI\(_2\) highlights its advantage in quickly identifying chaotic behavior in the MF Hamiltonian \eqref{eq:per MFL Ham}, significantly reducing the required computation time for distinguishing between orbit types.

\subsection{Magnetic versus kinetic chaos in toroidal plasmas}
We will now systematically compare the features of magnetic and kinetic chaos across a wide range of particle orbit energies. Magnetic flux surfaces undergo disruption at resonant locations  \(\psi_{\text{res}}\) where the safety factor, \(q(\psi_{\text{res}})\), satisfies the resonance condition \(q(\psi_{\text{res}}) = \dfrac{m}{n}\), with \(m\) and \(n\) being the perturbation mode numbers. Non-axisymmetric perturbations also change the GC phase space, creating resonant islands where the kinetic \(q\) factor, $q_{\text{kin}}$, satisfies the resonance condition \(q(\psi_{\text{kin}}) = \dfrac{m}{n}\). The kinetic factor, \(q_{\text{kin}}\), is defined as the ratio of the bounce toroidal rotational frequency to the bounce frequency \citep{shinohara2018estimation,White2015}. It is determined by the particle's kinetic properties over the three CoM expressed as \(q_{\text{kin}} = \dfrac{m}{n}\).

Intuitively, since low energy particles closely follow MFLs, we expect their kinetic factor \(q_{\text{kin}}\) to closely approach the safety factor, \(q\), of the MFLs \eqref{eq:per MFL Ham}. As a result, these particles exhibit chaotic behavior similar to that observed in the MFLs, with the chaotic behavior mostly pronounced around resonant stability island separatrices and overlapping resonant regions.

By carefully choosing the safety factor profile \(q(\psi)\) \eqref{eq:gen q factor}, we can keep the topology of the MF while maintaining the formation of magnetic islands and other disruptions. In our study, we use a typical \(q\)-factor profile defined by \eqref{eq:gen q factor} (see \citep[Page 56]{WhiteBook}), with parameters \(N = 2\), \(q_{\text{ma}} = 1.1\), \(q_\text{w} = 4.0\), and \(\psi_\text{w} = 0.05\). We also consider two perturbation modes with mode numbers \((m, n) = (3, 2)\) and \((5, 2)\). For the sake of simplicity, we assume a radially uniform perturbation profile by setting the coefficients \(\alpha_{m, n}\) independent of \(\psi\) and equal to \(\epsilon = 7.5 \times 10^{-5}\) for both modes. In addition, we consider hydrogen particles to be in a MF with \(\mu \mathbf{B_0} = 1\text{T}\).

Our goal is to investigate the influence of perturbations on particle orbits and compare the resulting chaotic behavior to that of the MFLs over a range of particle energies. As particle energy increases, the kinetic impact increases, leading to distinct resonance patterns and chaotic behaviors in the kinetic phase space, which may be different from those observed in the magnetic phase space. To investigate the impact of particle energy on chaotic behavior, we consider three different energy levels. In particular, we consider a low energy case for \(\mu \mathbf{B_0} = 2.6\) \([eV]\), an intermediate energy arrangement for \(\mu \mathbf{B_0} = 2.0\) \([KeV]\), and a high energy case with \(\mu \mathbf{B_0} = 33.1\) \([KeV]\). In Table.~\ref{tab:particle_cases}, we summarize the values of all model parameters for each case. At the lowest energy level, we can directly compare between kinetic and magnetic phase space using the LAR approximation, allowing for a clear understanding of how the two spaces can relate under a small perturbation. 

\begin{figure}[!htb]
  \centering
  \includegraphics[width=1\textwidth]{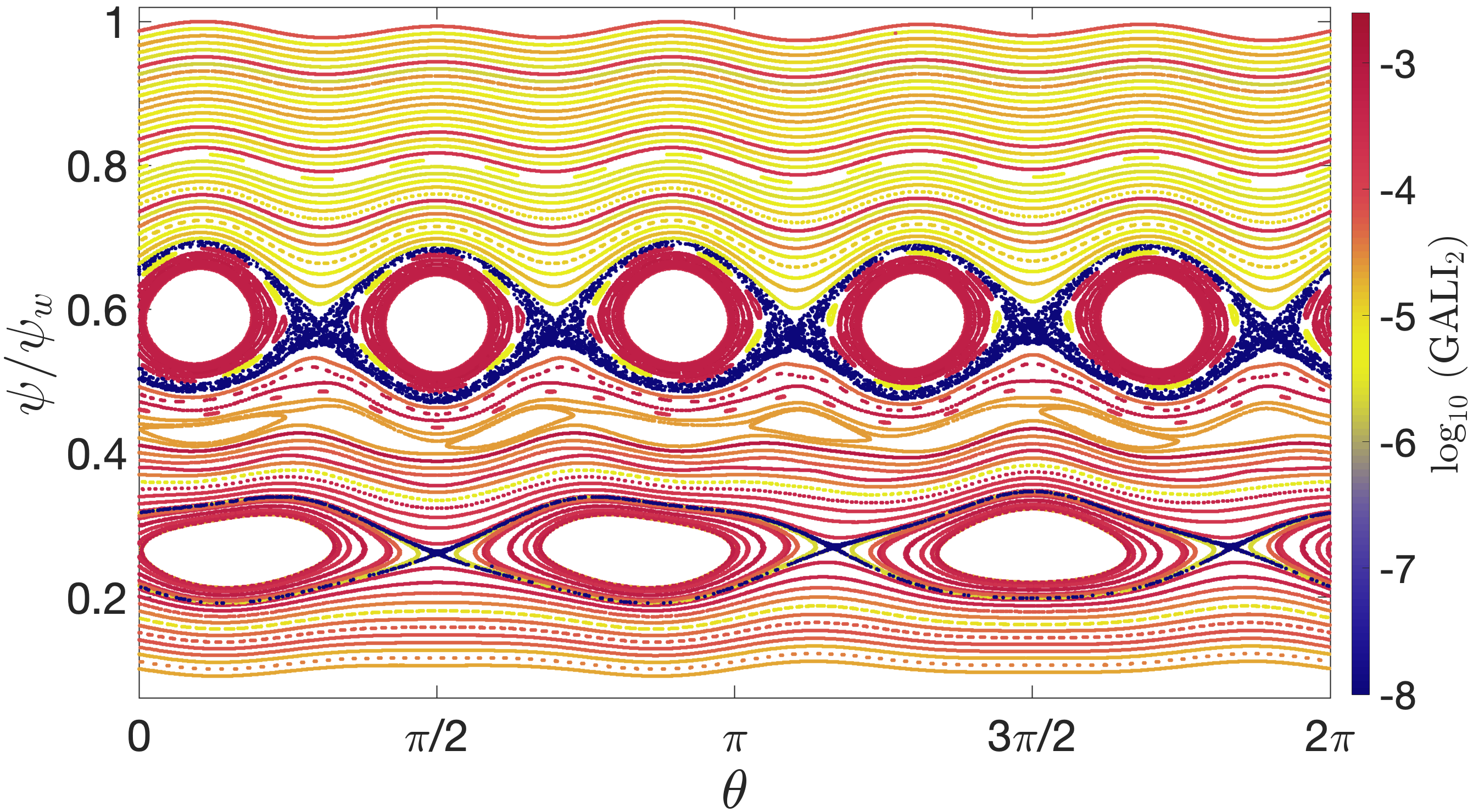}
  \caption{The PSS ($\zeta = 0; \mod 2\pi$) of the magnetic Hamiltonian  \eqref{eq:per MFL Ham} where the ICs are colored according to their final GALI\(_2\) values after \(10^8\) integration time units. We consider several ICs in the \((\theta, \psi)\) space \(\theta \in [0, 2\pi]\) and \( \psi \in [4.5 \times 10^{-3}, 5.03 \times 10^{-2}]\) with two perturbation modes \((m,n) = (3, 2)\) and \((5, 2)\), each having equal amplitudes ($\alpha_{3,2} = \alpha_{5,2} = \epsilon = 7.5 \times 10^5$). The toroidal flux \( (\psi) \) shown on the y-axis is in normalized form \( \left(\psi/\psi_{\text{w}}\right) \).}
  \label{fig3:Fig5}
\end{figure}

In order to make the comparison between the kinetic and magnetic phase space, we begin by creating the PSS of the magnetic case with Hamiltonian \eqref{eq:per MFL Ham} at \(\zeta=0\) for multiple ICs in the \((\theta, \psi/\psi_\text{w})\) plane, as shown in Fig.~\ref{fig3:Fig5}. It is worth noting that for a direct comparison with the corresponding kinetic phase space, we are normalizing the canonical momentum (toroidal flux, \(\psi\)) of the MFLs. This normalization allows for a direct comparison of the PSS of the EoM of the MFLs \eqref{eq:per MFL EoM} with the PSS of the EoM of the GC Hamiltonian \eqref{eq:Gen Ham EoMs}, based on the earlier setup in \eqref{eq:per GC H}, where we have set \(p_\theta = \psi\). We color each IC according to its final GALI\(_2\) value at time $t=10^8$, which highlights various resonant islands, island chains, and chaotic regions in the magnetic phase space. Dark blue regions denote chaotic motion, yellowish regions correspond to weakly chaotic ICs that may require longer computation time to fully exhibit their chaotic nature, and red regions indicate regions of regular motion. The magnetic PSS plot in Fig.~\ref{fig3:Fig5} will serve as the main reference for systematically comparing magnetic chaos with the corresponding kinetic chaos for the various energy levels we consider.

\begin{figure}[!htb]
  \centering
  \subfloat[ \((\zeta, p_\zeta)\) PSS\label{fig3:Fig6a}]{\includegraphics[width=0.49\textwidth]{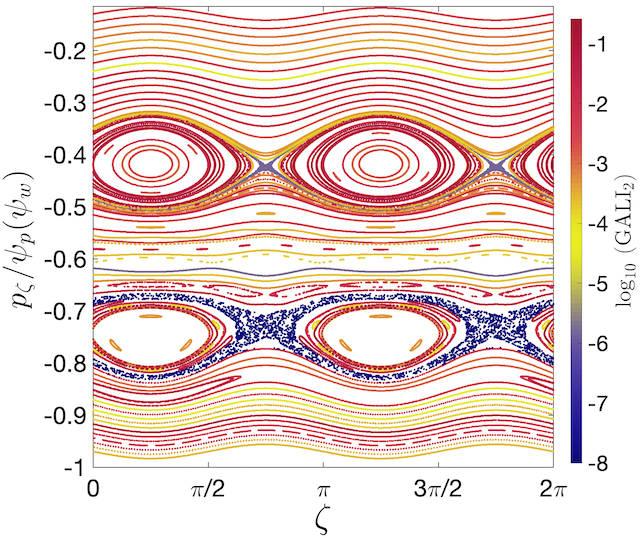}}\hfill
  \subfloat[\((\theta, p_\zeta)\) PSS\label{fig3:Fig6b}] {\includegraphics[width=0.49\linewidth]{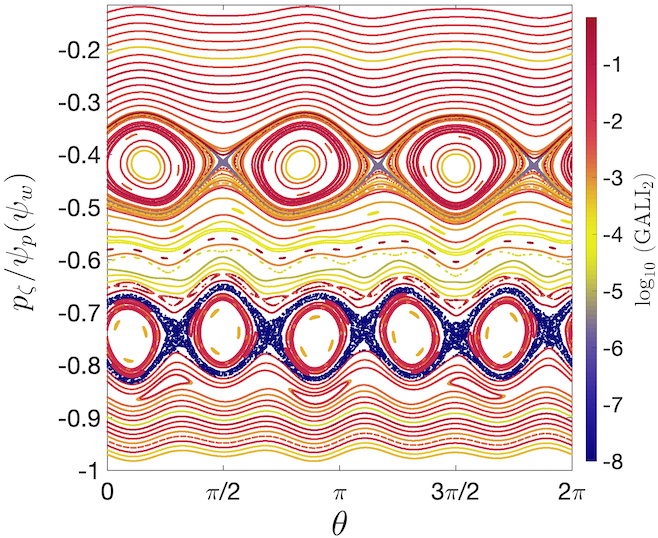}}\hfill
  \subfloat[\((\theta, p_\theta)\) PSS\label{fig3:Fig6c}]{\includegraphics[width=0.5\textwidth]{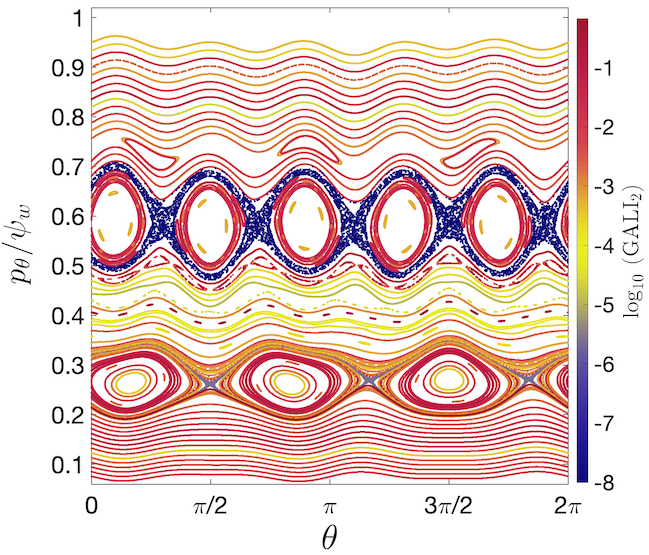}}
\caption{The low energy level case of Table.~\ref{tab:particle_cases}. The PSS (a) \((\zeta, p_\zeta)\) (where $\theta = 0;  p_\theta > 0$), (b) \((\theta, p_\zeta)\) (where $\zeta = 0;  p_\theta > 0$), and (c) \((\theta, p_\theta)\) where ($\zeta = 0;  p_\zeta > 0$) of the Hamiltonian \eqref{eq:GC H} in which the ICs are colored according to their final GALI\(_2\) values at \(10^8\) time units. We consider a grid of ICs of the phase space, with \(\zeta\), and  \(\theta\) defined in\(\mod \, 2\pi\), \(p_\zeta \in [-2.41 \times 10^{-2}, 2 \times 10^{-4}]\) in panels (a) and (b), and \(p_\theta \in [2 \times 10^{-4}, 4.45 \times 10^{-2}]\) in (c). The y-axis in each panel shows canonical momenta \(p_\zeta\) and \(p_\theta\) in respective normalized form. See the text in the description for details.}
  \label{fig3:Fig6}
\end{figure}

Figure \ref{fig3:Fig6} illustrates the behavior of low energy particles with energy \(E=3.4 eV\) and a MF intensity \(\mu \mathbf{B_0} = 2.6 eV\). Fig.~\ref{fig3:Fig6a} and (b) display PSS in \((\zeta, p_\zeta)\) and \((\theta, p_\zeta)\) coordinates for the low energy case outlined in Table~\ref{tab:particle_cases}, respectively. The GC orbits are colored according to their final GALI\(_2\) values at \(t = 10^8\). Note that we normalize the canonical toroidal and poloidal momenta, i.e.~ \(p_\zeta\) by \(p_\zeta/\psi_{p}(\psi_\text{w})\) and \(p_\theta\) by \(p_\theta/\psi_\text{w}\), respectively, for the kinetic cases, similar to what we have done for the MFLs in Fig.~\ref{fig3:Fig5}. This normalization allows a clear comparison and contrast of phase space structures across the different cases we consider, regardless of the specific numerical values of the momenta. Our primary interest is on the qualitative aspects, such as the phase space structures, including radial orbital positions, and the presence of stability islands, rather than on the exact values of the toroidal and poloidal momenta. Notably, the number of stability islands in each island chain corresponds to their respective mode numbers.

In addition, Fig.~\ref{fig3:Fig6c} shows the PSS \((\theta, p_\zeta/\psi_{p}(\psi_\text{w}))\) of the kinetic Hamiltonian EoM \eqref{eq:per GC EoM} for direct comparison with its corresponding PSS \((\theta, \psi/\psi_\text{w})\) of the magnetic Hamiltonian EoM \eqref{eq:per MFL EoM} presented in Fig.~\ref{fig3:Fig5}. This comparison highlights a strong correlation between kinetic and magnetic chaotic regions, in particular around the separatrices of island chains. Both the kinetic and magnetic domains exhibit similar chaotic behavior, typically located at the same radial ($\psi$) positions. The degree of chaotic behavior in both phase spaces is indicated by the GALI\(_2\) values in the respective color bars.

\begin{figure}[!htb]
  \centering
  \subfloat[ \((\zeta, p_\zeta)\) PSS\label{fig3:Fig7a}]{\includegraphics[width=0.49\textwidth]{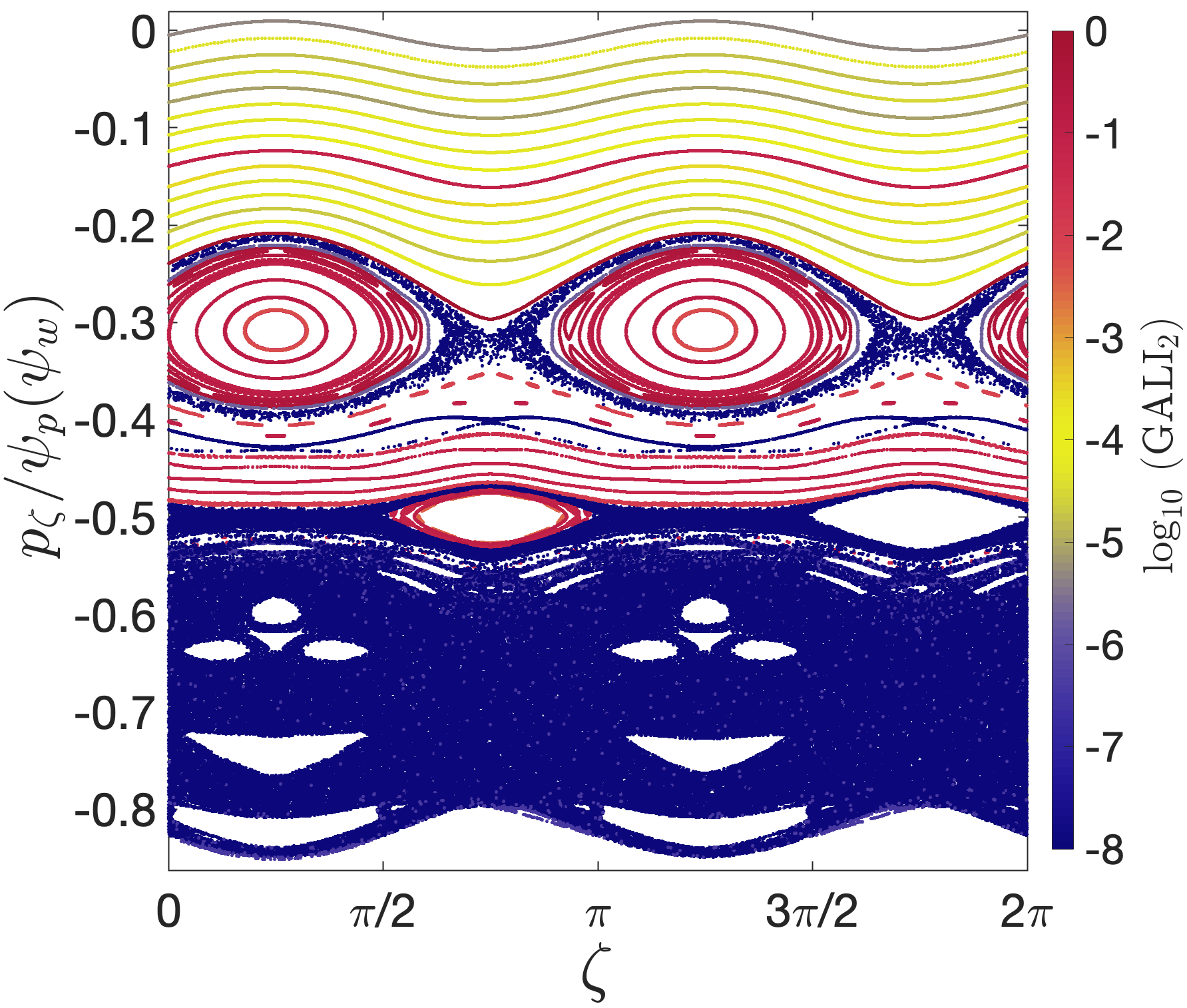}}\hfill
  \subfloat[\((\theta, p_\zeta)\) PSS\label{fig3:Fig7b}] {\includegraphics[width=0.49\linewidth]{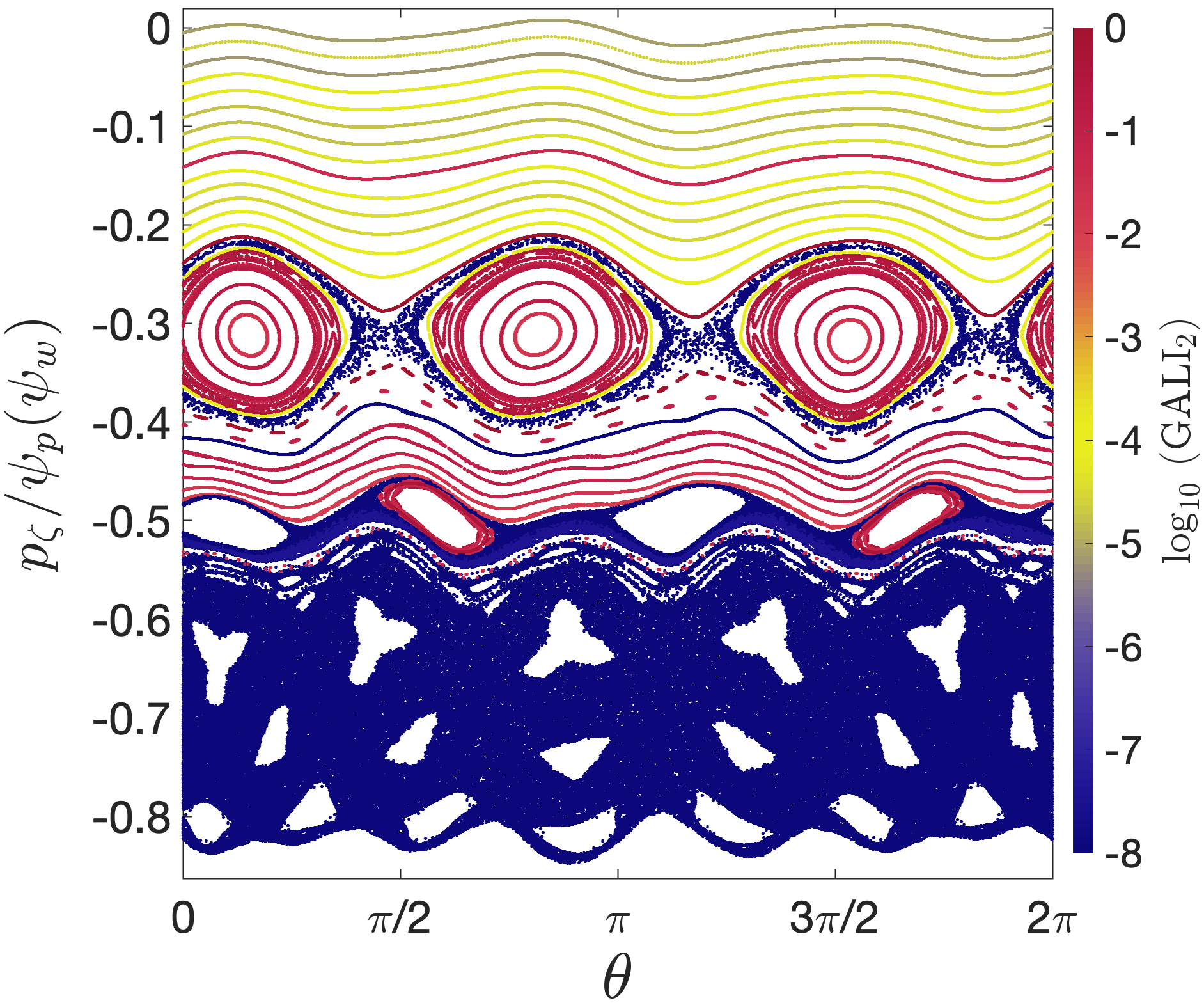}}\hfill
  \subfloat[\((\theta, p_\theta)\) PSS\label{fig3:Fig7c}]{\includegraphics[width=0.5\textwidth]{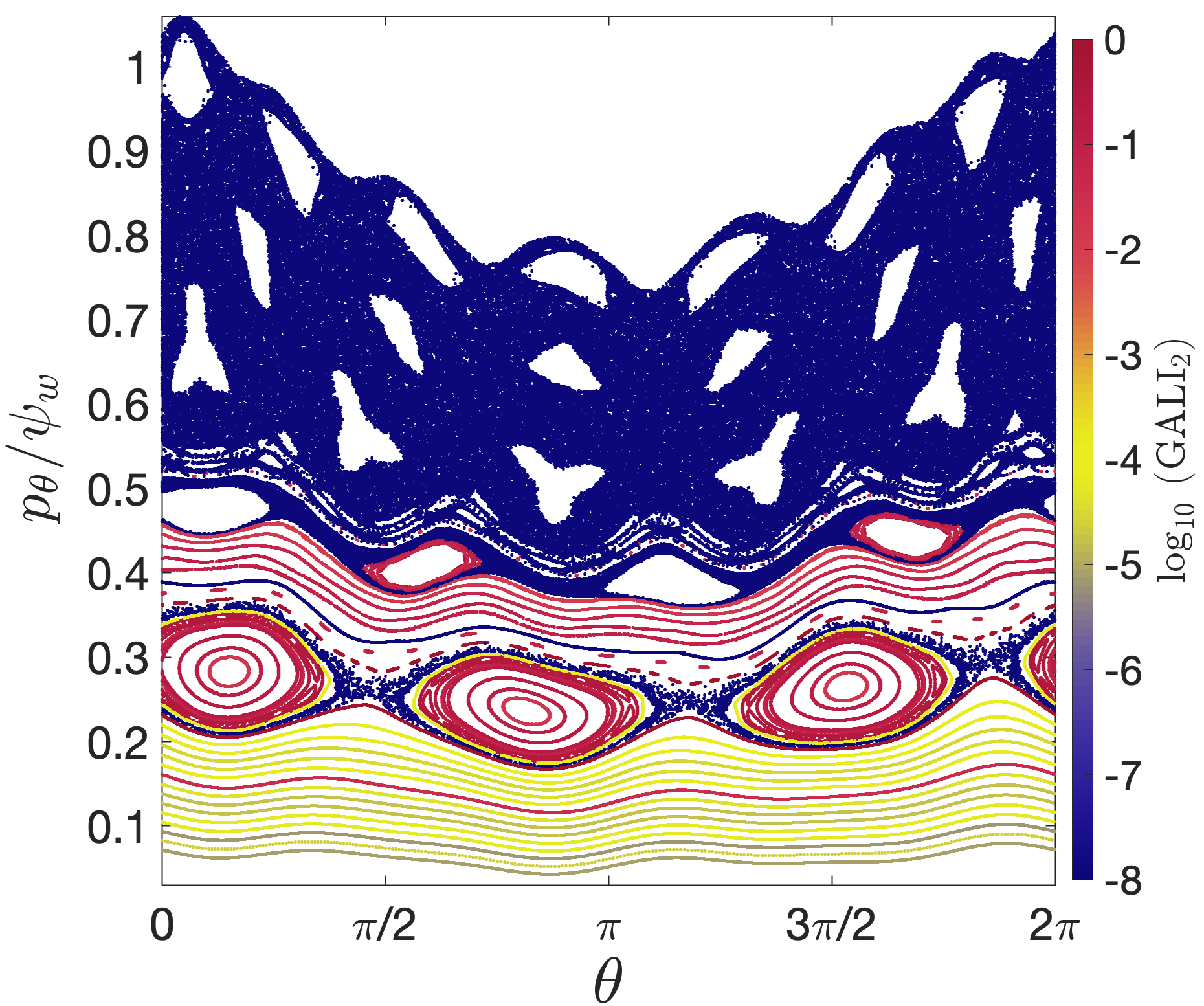}}
  \caption{Similar to Fig.~\ref{fig3:Fig6} but for the thermal energy level case of Table.~\ref{tab:particle_cases}. The ICs are selected from the intervals: \(p_\zeta \in [-2 \times 10^{-2}, 6 \times 10^{-4}]\) and \(p_\theta \in [1.8 \times 10^{-3}, 5.31 \times 10^{-2}]\), with \(\zeta\), and \(\theta\) defined in\(\mod \, 2\pi\).}
  \label{fig3:Fig7}
\end{figure}

In Fig.~\ref{fig3:Fig7}, we examine the behavior of thermal particles with an energy of \(E = 2.9 keV\) and a MF intensity of \(\mu \mathbf{B_0} = 2.0 keV\) similar to Fig.~\ref{fig3:Fig6}. Figs.~\ref{fig3:Fig7a} and (b) present the PSS of the thermal energy case of Table~\ref{tab:particle_cases} with Hamiltonian \eqref{eq:GC H}. These results reveal resonant islands at different \(p_\zeta\) values, with a considerably larger chaotic region observed for smaller \(p_\zeta\) values [around the location of the three resonances in the \((\theta, p_\zeta)\) plane] compared to the lower energy level in Figs.~\ref{fig3:Fig6a} and (b). 

The kinetic chaos associated with larger \(p_\zeta\) values is located at a wider range of radial positions where the main five resonant island chains are located [Fig.~\ref{fig3:Fig7c}]. These chaotic regions are more extended than the one observed in Fig.~\ref{fig3:Fig5}. We note that the chaotic regions extend to the wall \((p_\theta / \psi_\text{w} = 1)\), indicating potential particle loss due to adding small perturbations. In addition, at the thermal energy level, we can observe the introduction of new multiple island chains within the existing chaotic regions, particularly in the areas where \(p_\zeta/\psi_{p}(\psi_\text{w}) < 0.5\) [Figs.~\ref{fig3:Fig7a} and (b)], and \(p_\theta / \psi_\text{w} > 0.5\) [Fig.~\ref{fig3:Fig7c}].

\begin{figure}[!htb]
  \centering
  \subfloat[ \((\zeta, p_\zeta)\) PSS\label{fig3:Fig8a}]{\includegraphics[width=0.49\textwidth]{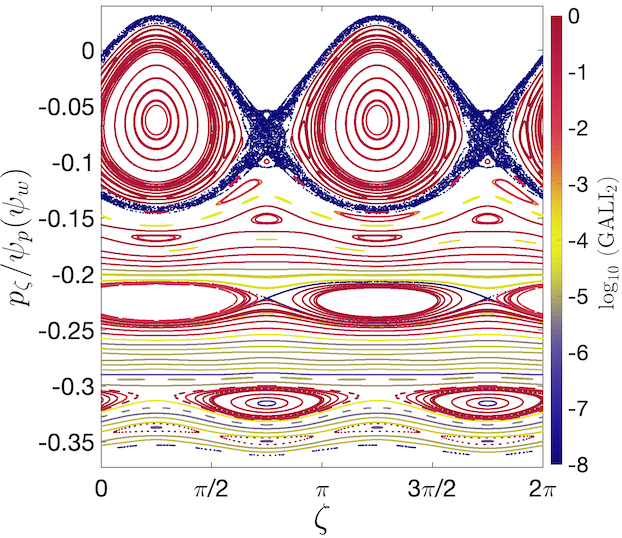}}\hfill
  \subfloat[\((\theta, p_\zeta)\) PSS\label{fig3:Fig8b}] {\includegraphics[width=0.49\linewidth]{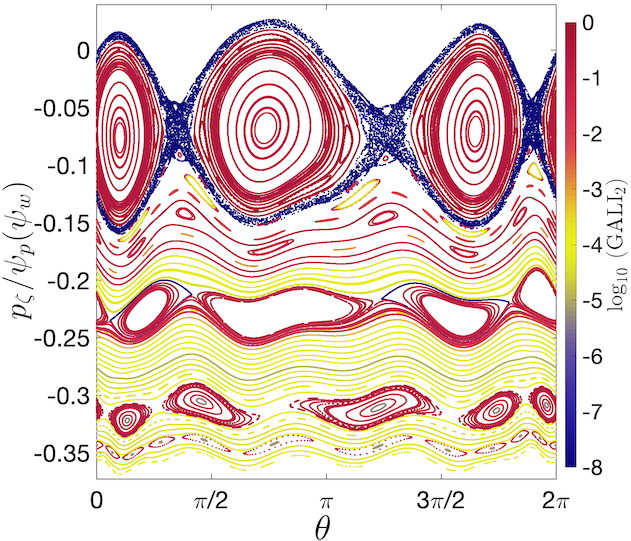}}\hfill
  \subfloat[\((\theta, p_\theta)\) PSS\label{fig3:Fig8c}]{\includegraphics[width=0.5\textwidth]{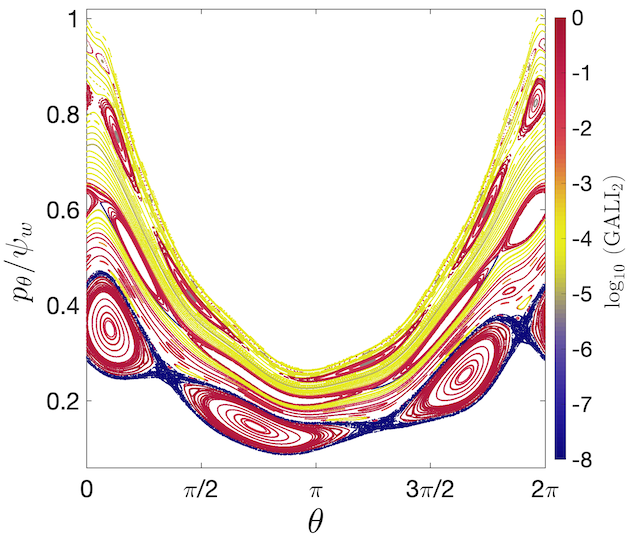}}
  \caption{Similar to Fig.~\ref{fig3:Fig6} but for the high energy level case of Table.~\ref{tab:particle_cases}. The ICs selected from the intervals: \(p_\zeta \in [-9 \times 10^{-3}, 1 \times 10^{-3}]\) and \(p_\theta \in [4.1 \times 10^{-3}, 4.78 \times 10^{-2}]\), with \(\zeta\), and \(\theta\) defined in\(\mod \, 2\pi\).}
  \label{fig3:Fig8}
\end{figure}

The behavior of the system \eqref{eq:GC H} for the case of higher energy particles, specifically \(E = 39.1\) $keV$ and \(\mu \mathbf{B_0} = 33.1\) $keV$, is presented in Fig.~\ref{fig3:Fig8}. In the kinetic PSS Figs.~\ref{fig3:Fig8a} and (b), we observe that the resonance chains occur at significantly different \(p_\zeta\) values compared to those seen for the lower energy particles in Fig.~\ref{fig3:Fig6} \citep{antonenas2021analytical,anastassiou2024role}. Notably, the extent of kinetic chaos is considerably reduced at this energy level relative to the intermediate case with thermal particles shown in Fig.~\ref{fig3:Fig7}. The chaotic zones are primarily confined to areas located near the separatrices of the main five island chains [see Fig.~\ref{fig3:Fig8c}].

\begin{table}[h!]
  \renewcommand{\arraystretch}{1.5} 
  \centering
\caption{The details for the three different energy level cases we consider: energy values \(E\), magnetic moment values \(\mu \mathbf{B_0}\), and the associated figures illustrating the results for each case. In all cases, we use the perturbation modes $(m, n) = (3, 2)$ and $(5, 2)$ and equal amplitudes $a_{3,2} = a_{5,2} = e = 7.5 \times 10^{-5}$.}
\vspace{0.2cm} 
\label{tab:particle_cases}
\begin{tabular}{|c|c|c|c|}
\hline
\textbf{Case (particles)}           & \textbf{Energy ($E$)} & \textbf{$\mu \mathbf{B_0}$} & \textbf{Figures} \\
\hline
Low energy     & 3.4 eV               & 2.6 eV              & Fig.~\ref{fig3:Fig5}, Figs.~\ref{fig3:Fig9a} \& \ref{fig3:Fig10a} \\
\hline
Thermal energy       & 2.9 eV               & 2.0 keV             & Fig.~\ref{fig3:Fig6}, Figs.~\ref{fig3:Fig9b} \& \ref{fig3:Fig10b} \\
\hline
High energy     & 39.1 keV             & 33.1 keV            & Fig.~\ref{fig3:Fig7}, Figs.~\ref{fig3:Fig9c} \& \ref{fig3:Fig10c} \\
\hline
\end{tabular}
\end{table}

The results illustrated in Figs.~\ref{fig3:Fig6} to \ref{fig3:Fig8} demonstrate that similar magnetic perturbations can generate different effects on particles with different kinetic properties. While an individual PSS color plot at specific energy levels can give valuable insights for a characteristic case, a comprehensive understanding of perturbation effects necessitates a systematically global phase space analysis. Such an approach allows us to not only identify the dynamics of each GC orbit but also to classify it as trapped, confined, or lost depending on how the particles interact with the MFs and plasma conditions. 

Trapped particles remain restricted to specific regions in the magnetic confinement devices due to the MF configuration. In contrast, confined particles are not limited to a particular region but still remain in the plasma volume for extended periods. This long term confinement allows the particles to significantly influence plasma behavior and contribute to fusion processes. Lost particles, on the other hand, are those that escape magnetic confinement and leave the plasma volume. This loss may reduce the efficiency of the confinement device and potentially impact plasma performance. In general, by assigning a chaos measure to each orbit, we can further quantify the chaotic behavior of GC orbits over the entire particle energy range in the phase space defined by the CoM.

\begin{figure}[!htbp]
  \centering
  \subfloat[Low energy level\label{fig3:Fig9a}]{\includegraphics[width=0.48\textwidth]{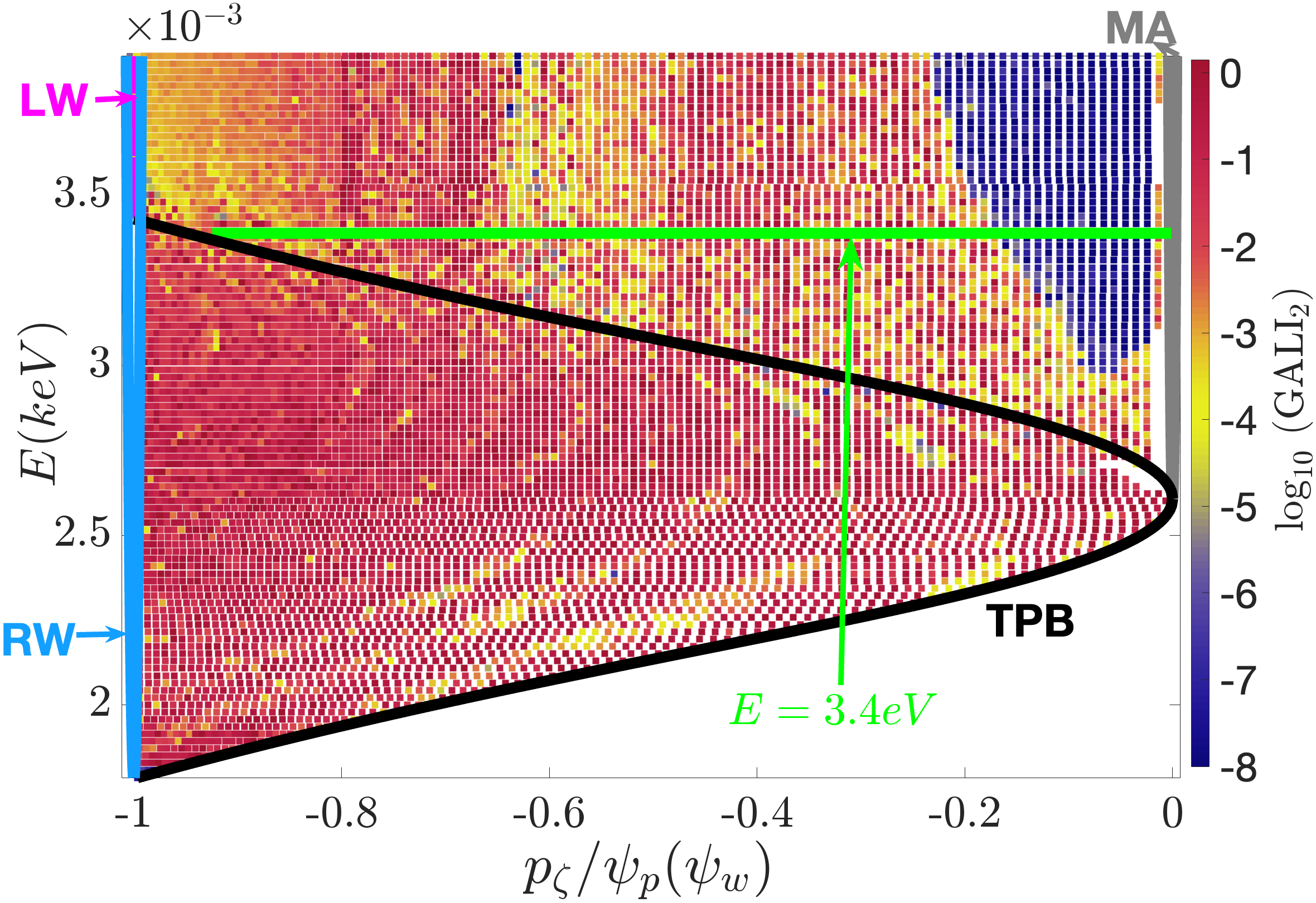}}\hfill
  \subfloat[Thermal energy level\label{fig3:Fig9b}] {\includegraphics[width=0.52\linewidth]{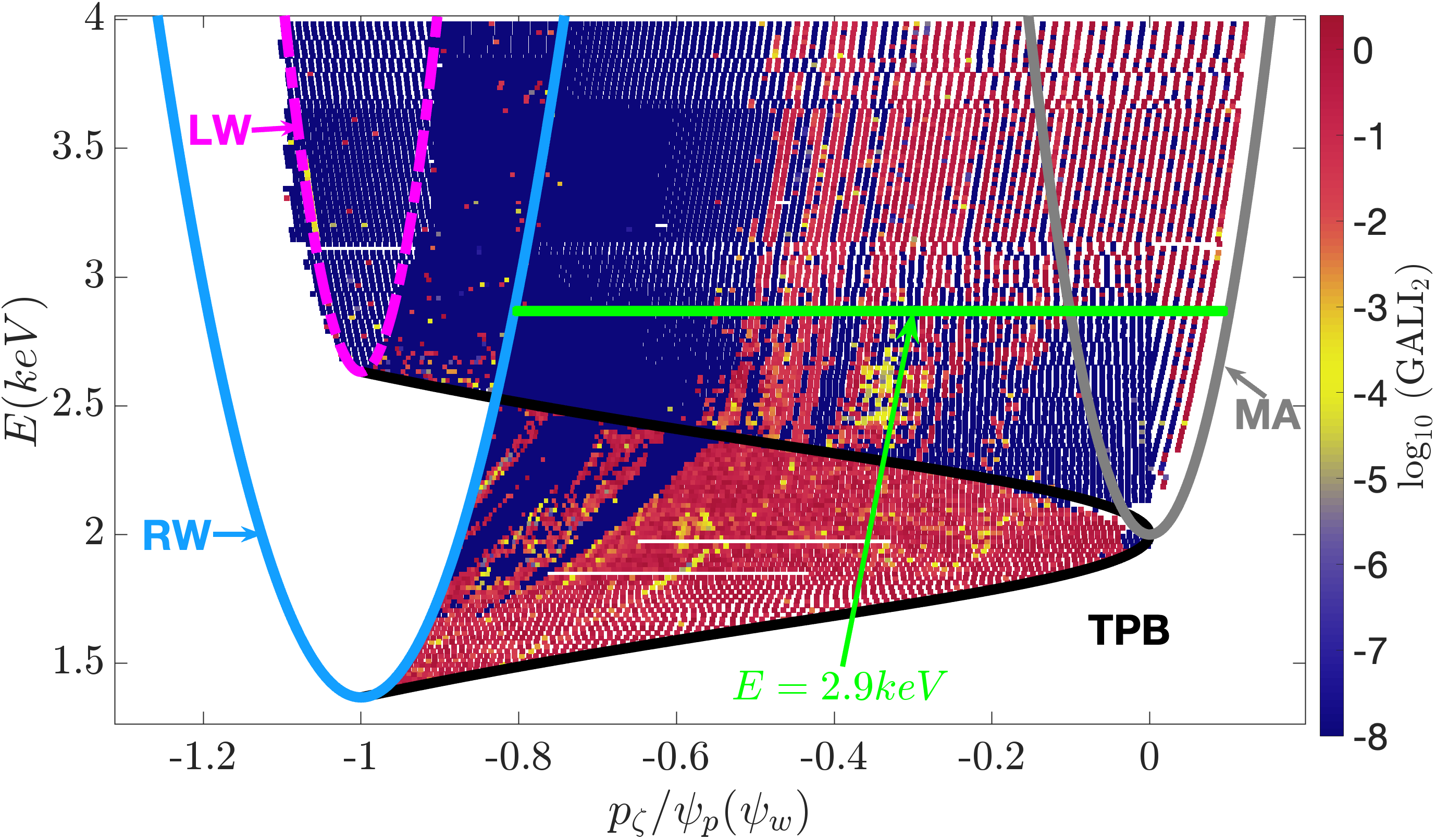}}\hfill
  \subfloat[High energy level\label{fig3:Fig9c}]{\includegraphics[width=0.5\textwidth]{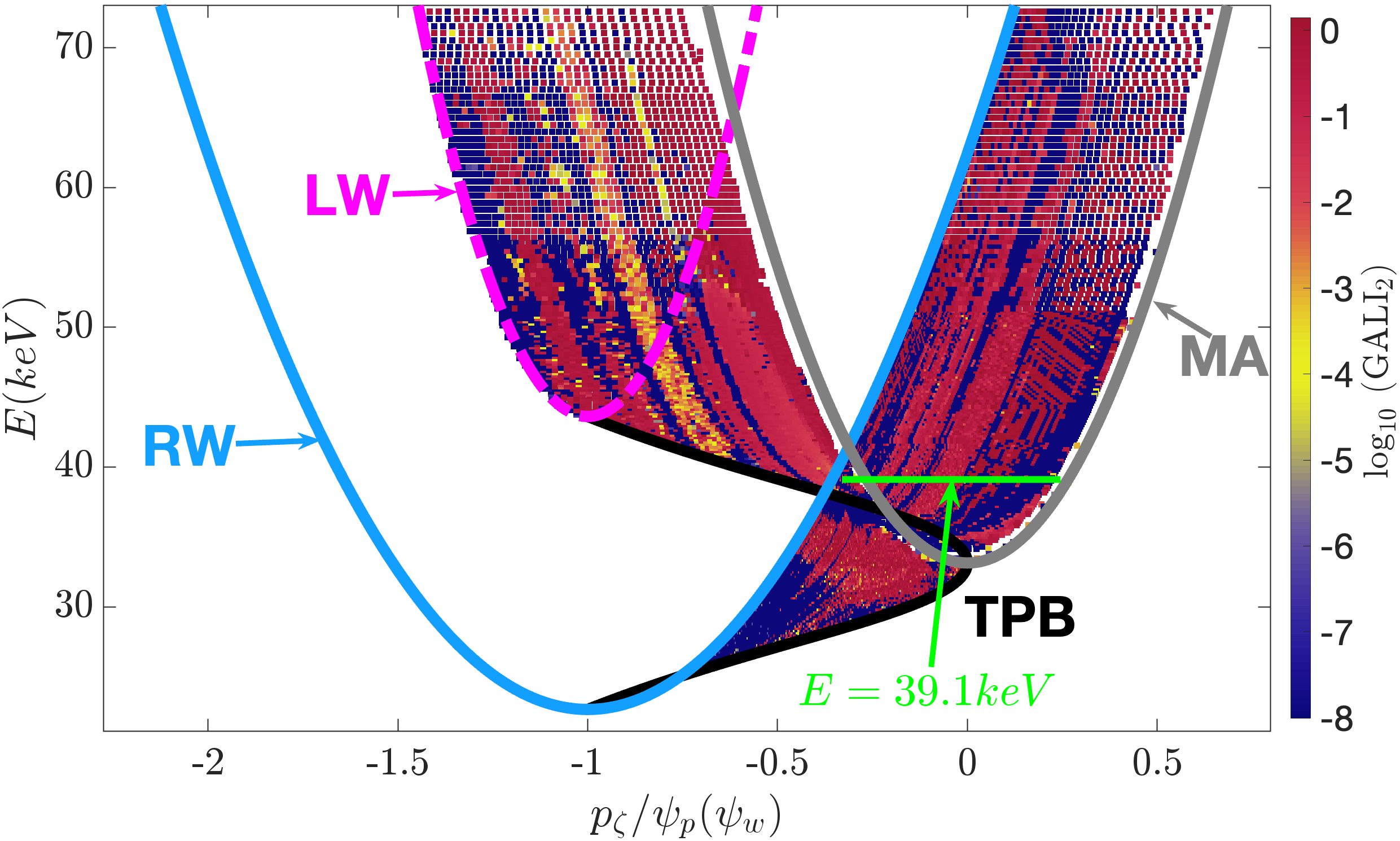}}
  \caption{The CoM space \((E, p_\zeta, \mu=\text{constant})\) for a set of GC orbits where each point represents an orbit colored according to their respective GALI\(_2\) value at $10^{8}$ integration time units. The energy values considered in the kinetic PSS (Figs.~\ref{fig3:Fig6}, ~\ref{fig3:Fig7} and ~\ref{fig3:Fig8}) are represented by a horizontal green line in panels (a), (b), and (c), respectively. Key features of the system in each panel are shown by different parabolas: the right wall (RW - solid light blue), the left wall (LW - dashed light pink), the magnetic axis (MA - solid gray), and the trapped/passing boundary (TPB -  solid black). Note that there are some missing points (forbidden areas of motion) around the border of the MA due to negative values of \(p_\theta\) in the expression \(\mathbf{B}=1-\sqrt{2p_\theta} \cos \theta\) in the Hamiltonian \eqref{eq:per GC H}.} 
  \label{fig3:Fig9}
\end{figure}

For each GC orbit of the Hamiltonian \eqref{eq:GC H} in the CoM space, the orbit's chaotic nature is quantified using the associated GALI\(_2\) value at \(t = 10^8\). Figs.~\ref{fig3:Fig9a}, (b) and (c) illustrate this analysis for low, thermal, and higher energy particles, respectively. The primary objective of this representation is to identify the chaotic nature of these orbits and categorize them as trapped or passing, as well as confined or lost particles. 

An orbit is classified as trapped or passing based on its location relative to the black parabola [in Fig.~\ref{fig3:Fig9}], which is plotted using the definition in \eqref{eq:trapped passing boundary}. In the CoM space depicted in Fig.~\ref{fig3:Fig9}, all points can be considered as confined, except those in the white spaces, which represent lost particles. These lost particles are located between the right wall (light blue solid parabolas) and the left wall (pink dotted parabolas) of the MF, as defined in \eqref{eq:walls}. In addition, we aim to identify the positions of each orbit relative to the magnetic axis \eqref{eq:magnetic axis} (gray solid parabolas) and the magnetic wall. 

By classifying each orbit in this manner, we gain valuable insights into the behavior of individual particles, which are essential for optimizing the efficiency of energy fusion devices. Overall, our results in Figs.~\ref{fig3:Fig9a}, (b), and (c) illustrate that similar magnetic perturbations (with the same modes) result in significantly different levels of kinetic chaos, depending on the particles' kinetic properties.

The \(3D\) CoM space (\(E, p_\zeta, \mu=\text{constant}\)) is illustrated in Fig.~\ref{fig3:Fig9} for a set of ICs on a grid $(p_{\zeta}, E)$ space, where $p_{\zeta} \in [-0.0255, 0]$, $E \in [6.85 \times 10^{-9}, 1.5 \times 10^{-8}]$ with $ \mu = 10^{-8}$ [Fig.~\ref{fig3:Fig9a}]; $p_{\zeta} \in [-0.025, 0]$, $E \in [5.25 \times 10^{-6}, 1.2 \times 10^{-5}]$ with $\mu = 7.67 \times 10^{-6}$ [Fig.~\ref{fig3:Fig9b}]; as well as $ p_{\zeta} \in [-0.0255, 0]$, $E \in [9.6 \times 10^{-5}, 1.6 \times 10^{-4}]$ with $\mu = 1.27 \times 10^{-4}$ [Fig.~\ref{fig3:Fig9c}]. Each point in the CoM space represents a GC orbit, colored according to its respective GALI\(_2\) value. Dark blue indicates chaotic motion, while red indicates regular motions. 

From the low energy case [Fig.~\ref{fig3:Fig9a}] to the highest energy level [Fig.~\ref{fig3:Fig9c}], we can observe how specific perturbations may influence each orbit's chaotic behavior, momentum transport, and confinement in the system. In particular, small chaotic regions appear at the boundaries of distinct resonant island chains, indicating confined chaos. On the other hand, more extensive chaotic regions suggest the presence of resonance overlap, which can lead to a wide range of particles exhibiting chaotic motion. The location of chaotic regions relative to the wall and the trapped/passing boundary provides insights into particle losses and the complex dynamics associated with particle trapping and passing boundaries. 

For the low energy level, small chaotic regions are present near the magnetic axis in the passing boundary [blue regions in the top right corner of Fig.~\ref{fig3:Fig9a}]. In this case, orbits tend to remain confined as expected. For the thermal energy case, chaotic regions are more extensive, in which they appear both near the passing boundary and the trapped boundary [blue regions inside and outside the black parabolas in Fig.~\ref{fig3:Fig9b}]. At the highest energy level, chaotic regions are present on both trapped and passing boundaries [blue regions and stripes in Fig.~\ref{fig3:Fig9c}]. While the chaotic regions are more widespread for thermal energy particles, orbits remain relatively confined. However, at the highest energy level (and to a lesser extent for the thermal energy level), we observe an increase in particle loss, as indicated by the white spaces between the magnetic walls and the axis in Figs.~\ref{fig3:Fig9b} and (c). 

We further made a direct comparison between the information presented in Figs.~\ref{fig3:Fig9a}, (b) and (c) and the corresponding PSS plots in Figs.~\ref{fig3:Fig6}, ~\ref{fig3:Fig7} and ~\ref{fig3:Fig8}, respectively, by drawing a horizontal green line in Fig.~\ref{fig3:Fig9}. The horizontal green line at \(E = 39.1 keV\) [Fig.~\ref{fig3:Fig9a}] corresponds to the low-energy case in Fig.~\ref{fig3:Fig6}. Similarly, the horizontal green line at \(E = 2.9 keV\) [Fig.~\ref{fig3:Fig9b}] represents the thermal energy level case in Fig.~\ref{fig3:Fig7}, and the horizontal green line at \(E=3.4 eV\) [Fig.~\ref{fig3:Fig9c}] corresponds to the high-energy level case in Fig.~\ref{fig3:Fig8}.

For example, by comparing the horizontal green line in Fig.~\ref{fig3:Fig9b} (representing the thermal energy level case) to the PSS in Figs.~\ref{fig3:Fig7a} and (b) for the same energy, we can clearly identify extended chaos at the beginning, approximately in the range \((-1 <  p_\zeta/\psi_{p}(\psi_\text{w}) < -0.55)\), which is reflected in the PSS approximately in the range \(( p_\zeta/\psi_{p}(\psi_\text{w}) < -0.55)\). Following this, narrow chaotic stripes occur, such as around \( p_\zeta/\psi_{p}(\psi_\text{w}) \approx -0.5\) and \(-0.3\), which correspond to the chaotic zones around the five and three resonances observed in the PSS (Fig.~\ref{fig3:Fig7}), respectively. This demonstrates a strong correlation in quantifying chaos between the two figures. As a result, Fig.~\ref{fig3:Fig7} can be regarded as a representative PSS for a specific thermal energy level (to be exact at \(E = 2.9 keV\)) in Fig.~\ref{fig3:Fig9b}, where the global chaotic dynamics of various orbits for multiple energy levels are depicted. It is worth noting that the efficient quantification of chaos using GALI\(_2\) allows for the assignment of a chaos measure to each orbit, corresponding to points on a dense grid. 

Determining the exact radial position and extent of kinetic chaos in magnetic confinement devices has practical challenges since unperturbed particle orbits are not confined to a single flux surface, especially for higher energies due to GC drifts. Nevertheless, we can introduce a reference flux surface, \(\psi_0\), for each unperturbed orbit, which can provide an estimation of the approximate radial position where the kinetic chaos may occur. This reference flux surface is defined as follows:
\begin{equation} \label{eq:ref flux surf 1}
 \psi_p(\psi_0) = \langle \psi_p(\psi) \rangle = \Biggl \langle \dfrac{\pm g(\psi)}{\mathbf{B}(\psi,\theta)}  \sqrt{2\Big( E - \mu \mathbf{B}(\psi,\theta) \Big)} \Biggl \rangle  - p_\zeta. 
\end{equation}
In \eqref{eq:ref flux surf 1}  \(\langle \cdot \rangle\) represents orbit averaging over the magnetic surface. For trapped particles, the reference flux surface \(\psi_0\) can be approximated by the \(\psi\) value at the points of radial reversal, which are the locations along the particle's trajectory where the particle changes direction of motion perpendicular to the MFLs. In the lowest order LAR configuration, \(\psi_0\) can be defined for all particle orbit types using the CoM \((E, p_\zeta, \mu)\). 

For trapped particles, the range of radial motion is confined by the MF configuration, their energy, and their magnetic moment, satisfying the condition \(\mathbf{B}(\psi) \geq \frac{E}{\mu}\) \citep{pinches1998hagis}. This condition leads to the conclusion that the average of an imaginary function over a real interval [the square root term in \eqref{eq:ref flux surf 1}] is zero, which implies that \(\psi_p(\psi_0) = - p_\zeta\). In simple terms, trapped particles are restricted to a particular region of the MF, and their motion is predominantly influenced by the momentum \(p_\zeta\). 

For passing particles (either co- or counter-passing), we can use the three constants \(E, p_\zeta\) and \(\mu\) to simplify the expression in \eqref{eq:ref flux surf 1} to \(\sqrt{2(E - \mu)} - p_\zeta\). This simplification happens because the MF strength \(\mathbf{B}(\psi) \) can be approximated as constant for passing particles, where the particles closely follow the MFLs. Furthermore, the variation of the poloidal flux \(g(\psi)\) is negligible since the orbits are mainly determined by the toroidal components of the MF. Thus, we can express this relation as follows:
\begin{equation} \label{eq:ref flux surf 2}
  \psi_p(\psi_0) =
\begin{cases} 
  - p_\zeta, & \text{for trapped particles} \\  
  \pm \sqrt{2(E - \mu)} - p_\zeta, & \text{for passing particles.} 
\end{cases}
\end{equation}

The reference flux surface \(\psi_0\) \eqref{eq:ref flux surf 2}, determined using the CoM \(E\), \(\mu\), and \(p_\zeta\), can replace \(p_z\) for uniquely identifying each orbit when used alongside \(E\) and \(\mu\). Combining \(E\), \(\psi_0\), and \(\mu\) will provide valuable information regarding the radial position and the extent of kinetic resonance islands of stability and chaotic regions \citep{pinches1998hagis,antonenas2021analytical}. 

\begin{figure}[!htbp]
  \centering
  \subfloat[Low energy level\label{fig3:Fig10a}]{\includegraphics[width=0.51\textwidth]{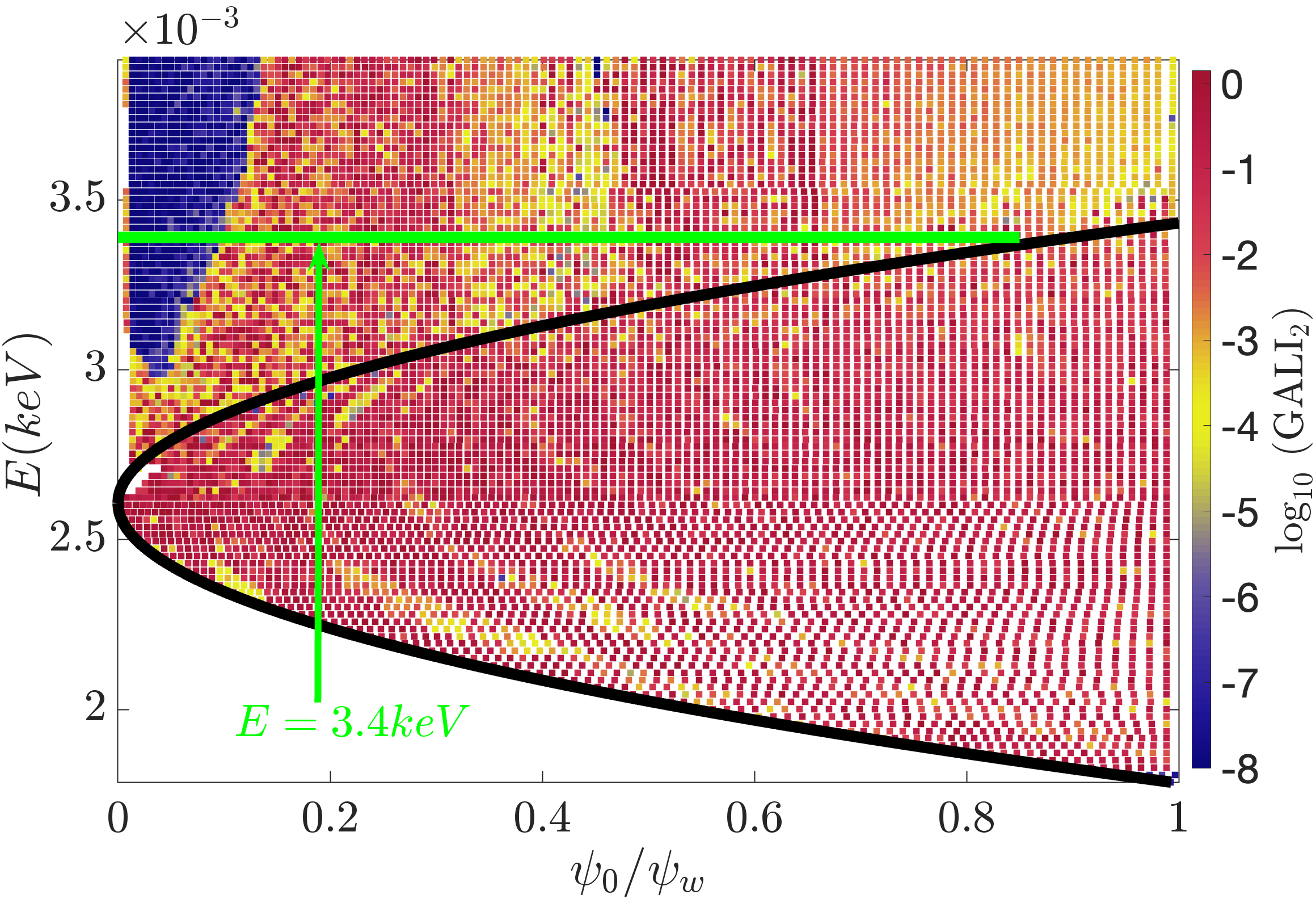}}\hfill
  \subfloat[Thermal energy level\label{fig3:Fig10b}] {\includegraphics[width=0.49\linewidth]{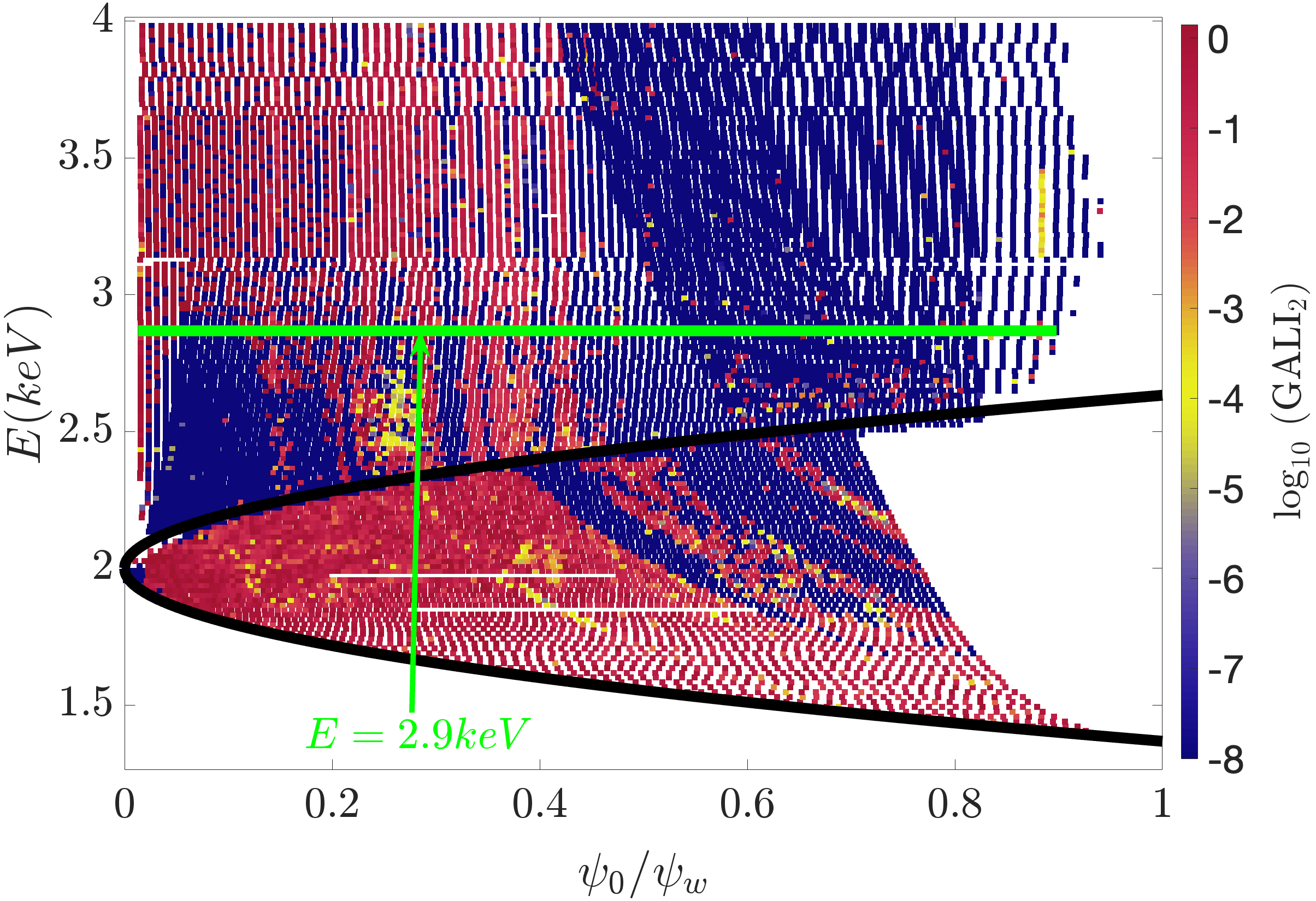}}\hfill
  \subfloat[High energy level\label{fig3:Fig10c}]{\includegraphics[width=0.5\textwidth]{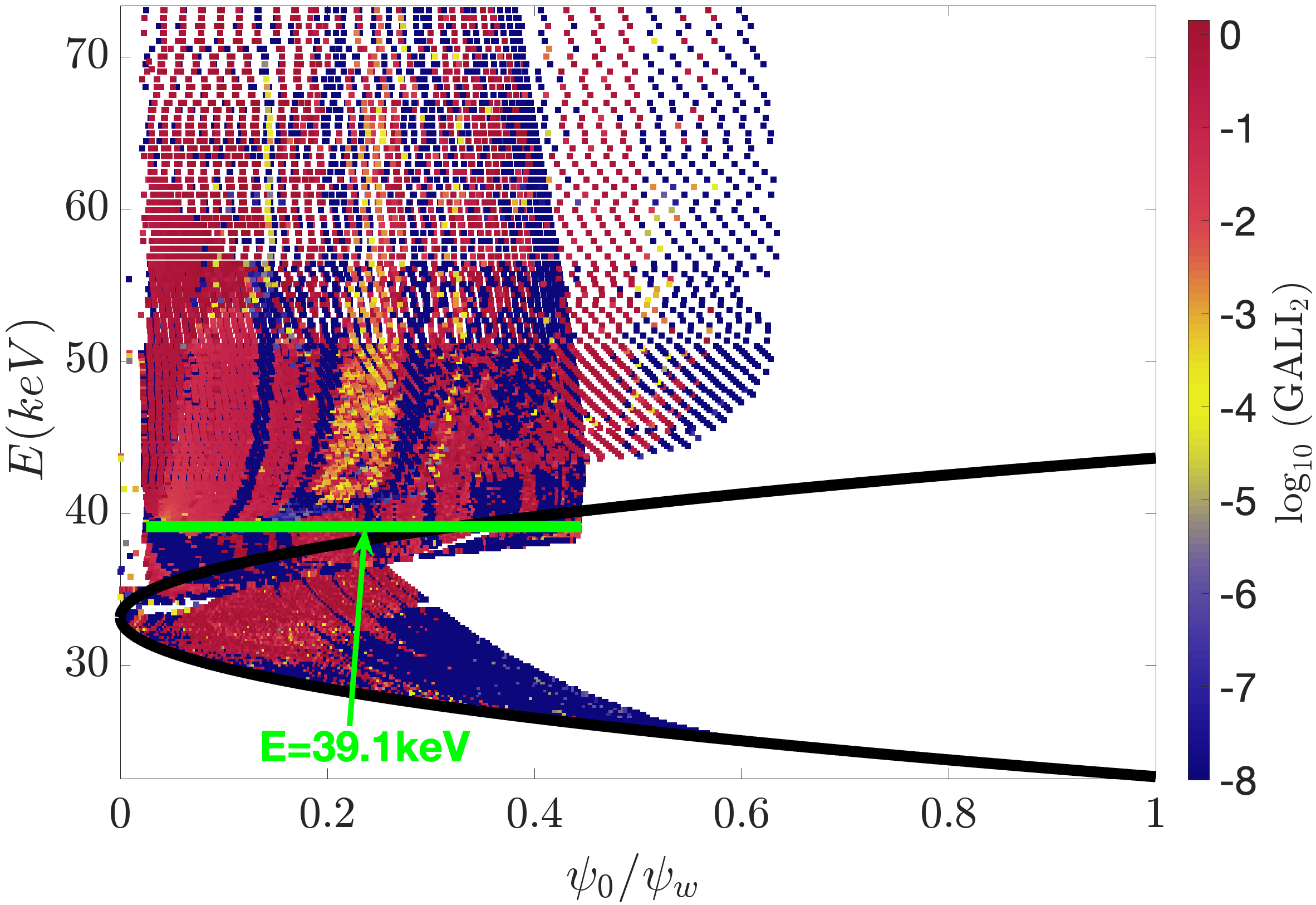}}
  \caption{Similar to Fig.~\ref{fig3:Fig9} but presents the CoM space \((E, \psi_0, \mu=\text{constant})\). The horizontal green line in panels (a), (b) and (c) represents the energy values considered in the kinetic PSS shown in Figs.~\ref{fig3:Fig6}, ~\ref{fig3:Fig7}, and ~\ref{fig3:Fig8}, respectively.} 
  \label{fig3:Fig10}
\end{figure}

Figures \ref{fig3:Fig10a}, (b) and (c) present the \((E, \psi_0)\) space instead of the \((E, p_\zeta)\) shown in Figs.~\ref{fig3:Fig10a}, (b) and (c) for the same arrangements of ICs, where each panel represents the lower, thermal, and higher energy levels case of Table \ref{tab:particle_cases}, respectively. The radial position where the kinetic chaos occurs can be determined in Fig.~\ref{fig3:Fig10}, using the reference flux surface \(\psi_0\) instead of \(p_\zeta\). Due to the definition of \(\psi_0\) in \eqref{eq: canonical repre. for flux}, the figures here are flipped from left to right, but the overall structure of the CoM space remains consistent with that observed in Fig.~\ref{fig3:Fig9}. It is worth noting that this is not the only alternative representation of the CoM. Readers can refer to, for e.g., \citep{Bierwage2022,benjamin2023distribution} for additional representations of CoM space.

Let us again consider the thermal case as an example. Comparing Fig.~\ref{fig3:Fig10b} (green horizontal line) and the PSS in Fig.~\ref{fig3:Fig7c} at the same energy level, we observe that chaos extends towards the end of the phase space, approximately in the range \(0.55 <  \psi_0/\psi_\text{w} < 1\), as indicated in the PSS approximately in the range \( p_\zeta/\psi_{p}(\psi_\text{w}) < -0.55\). On the other hand, narrow, chaotic stripes are observed at the beginning of the phase space. This comparison demonstrates a close resemblance between the ICs colored according to the GALI\(_2\) value [regions represented by the horizontal line in Fig.~\ref{fig3:Fig10b}] and the PSS created using these ICs in Fig.~\ref{fig3:Fig7c} for the same energy level in quantifying chaos. Therefore, Fig.~\ref{fig3:Fig10} can also serve as a valuable tool for globally characterizing chaos in the GC Hamiltonian \eqref{eq:GC H} across various energy levels. 

\section{Summary and conclusions} \label{section:SummaryCh3}
In this chapter, we introduced the GALI\(_2\) as a new tool for quantifying and studying chaos in plasma physics. We demonstrated its effectiveness for detecting and quantifying both magnetic and kinetic chaos in a toroidal plasma model, comparing it to the mLE. Our study showed that these indices are effective in distinguishing between chaotic and regular orbits [Figs.~\ref{fig3:Fig2b} and \ref{fig3:Fig4b}], as well as identifying weakly chaotic orbits [for e.g., the green curve in Fig.~\ref{fig3:Fig2b} and yellowish regions in Fig.~\ref{fig3:Fig3b}], which may require longer integration times to fully reveal their true chaotic nature. These results illustrate the indices' ability to accurately characterize the dynamical behavior of different orbits in these systems.

In toroidal plasma, the locations of kinetic and magnetic resonant islands along with the conditions for kinetic and magnetic chaos show significant differences. The chaotic transport of high-energy particles can negatively impact confinement limits and the overall performance of fusion devices (see \citep{anastassiou2024role} and references therein). Therefore, understanding the relationship between magnetic and kinetic resonances and their chaotic behaviors is crucial. In order to systematically investigate this relationship, we need to (i) identify and quantify chaos and (ii) represent particle orbits in a compact kinetic parameter space.

To compare kinetic and magnetic chaos, we construct PSS for representative cases of the MF Hamiltonian \eqref{eq:per MFL Ham} (Fig.~\ref{fig3:Fig5}) and for three distinct energy levels: low (Fig.~\ref{fig3:Fig6}), thermal (Fig.~\ref{fig3:Fig7}), and high (Fig.~\ref{fig3:Fig8}) energy particles of the GCM Hamiltonian \eqref{eq:per GC H}, representing plasma particles. For low energy particles, the kinetic [Fig.~\ref{fig3:Fig6c}] and magnetic (Fig.~\ref{fig3:Fig5}) PSS show strong resemblance. In particular, chaos appears near the separatrices of the stable island chains, which indicates that chaos is localized at the same radial (\(\psi\)) position. Furthermore, the strength of chaos, as quantified by small GALI\(_2\) values [GALI\(_2 < 10^{-8}\)], is similar in both cases.  As a result, the chaotic nature of particle orbits and their transport characteristics are directly related to the configuration of the MFLs for low energy particles. On the other hand, for the thermal case, the kinetic chaos associated with the stability of the resonant island chain extends over a wider range of radial positions [Fig.~\ref{fig3:Fig7c}] compared to the magnetic chaos. These stability resonant island chains are located at significantly different radial positions, and the kinetic chaos is reduced due to larger drifts across the MFLs for energetic particles [Fig.~\ref{fig3:Fig7c}].

This comparison led to the need for a systematically detailed analysis in the \(3D\) kinetic space of the CoM. Each point in this space (\(E, p_\zeta, \mu = \text{constant}\)) represents an unperturbed GC orbit of the Hamiltonian \eqref{eq:per GC H}, which allows for the classification of orbits as trapped or passing and confined or lost according to the definition in \eqref{eq:trapped passing boundary} and \eqref{eq:walls}, respectively (Fig.~\ref{fig3:Fig9}). When perturbations are introduced, a specific GALI\(_2\) value can be assigned to each point, providing detailed information about the kinetic characteristics of the orbits that become chaotic due to perturbations, as well as their position relative to the torus wall. These diagrams (Fig.~\ref{fig3:Fig9}) provide a detailed phase space resolution of kinetic chaos and contribute to a deeper understanding of the role that specific perturbation modes play in particle, energy, and momentum transport.  

Furthermore, we have also introduced an alternative representation of the diagrams in Fig.~\ref{fig3:Fig9} by introducing a reference flux surface (\(\psi_0\)) to replace the toroidal momentum, \(p_z\). This reference flux surface provides an estimate of the approximate radial position where kinetic chaos may occur for each unperturbed orbit. This is illustrated in the CoM space \((E, \psi_0, \mu=\text{constant})\) depicted in Fig.~\ref{fig3:Fig10}.

\clearpage





\chapter{Evolution of phase space structures at bifurcations of periodic orbits in a \(3D\) galactic bar potential}
\label{chapter:four}
\section{Introduction}
\label{section:introductionCh4}
\begin{figure}[!htb]
  \centering
  \includegraphics[width=0.6\textwidth]{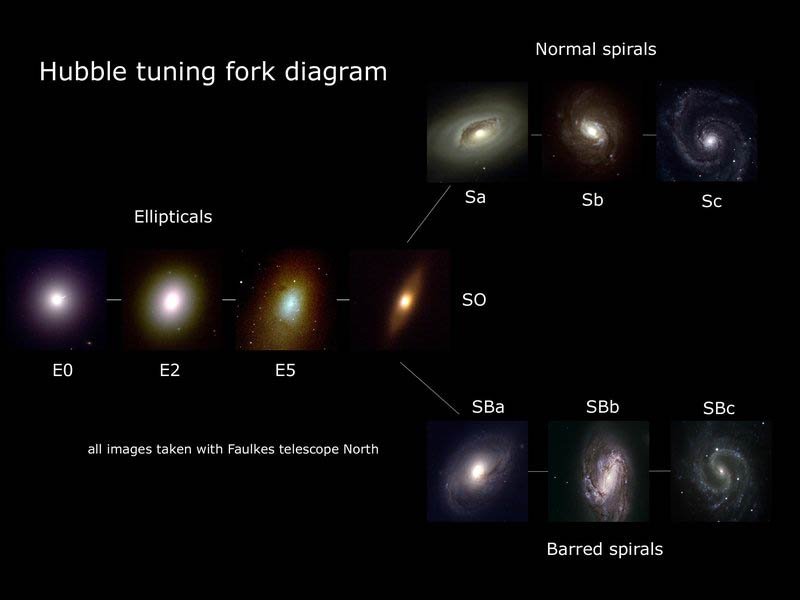}
  \caption{A basic representation of the Hubble tuning fork diagram [image source: \href{https://noirlab.edu/public/images/noao-ngc1365/}{Las Cumbres Observatory (LCO)}].}
  \label{fig:Hubble TFD}
\end{figure}

The renowned Edwin Hubble classified galaxies into four morphological types: elliptical, spiral, barred spiral, and irregular (see, for example, \citep{hubble1982realm}, and \citep[Section 1.1.3]{binney2011galactic}). He further divided spiral and barred spiral galaxies based on the relative size of their central bulge and the prominence of their spiral arms. Galaxies with a large central bulge and broad spiral arms are typically classified as type `a', while those featuring a small central bulge and well-defined, tightly twisted spiral arms are generally classified as type `c'. A basic schematic representation of these classifications is illustrated in Fig.~\ref{fig:Hubble TFD}.

Elliptical galaxies are smooth, oval-shaped structures with a bright center. Hubble assigned the letter E followed by a number to indicate their roundness, which ranges from almost circular (E0) to quite elliptical (E6) (see Fig.~\ref{fig:Hubble TFD}). On the other hand, disc galaxies share a round center similar to elliptical galaxies but also have a thin, flat disc of stars. Most disc galaxies have either spiral arms (Sa, Sb, Sc in Fig.~\ref{fig:Hubble TFD}) or bars near the center (barred spirals denoted as SBa, SBb, SBc in Fig.~\ref{fig:Hubble TFD}). Barred spiral galaxies are a subtype of spiral galaxies that are characterized by their central bar-shaped structure. Some disc galaxies that do not have spiral arms are referred to as lenticular galaxies (for instance, S0 in Fig.~\ref{fig:Hubble TFD}). Finally, there are irregular galaxies that do not fit into any of these shapes. In addition to Hubble's classification of galaxies, there are more recent alternatives, such as de Vaucouleurs' method \citep{de1959classification}, which examines more detailed aspects of a galaxy's shape. The study presented in this chapter focuses specifically on galaxies with bars at their centers.

Understanding the motion of objects in galaxies, such as stars, gas, and dark matter, is vital for understanding the galaxies' fundamental dynamics. These motions, known as orbits, can exhibit periodic, quasiperiodic, or chaotic behavior. Many galaxies often feature central bars, like NGC1365, one of the most prominent barred spiral galaxies visible in the sky (for more details, see for example \href{https://noirlab.edu/public/images/noao-ngc1365/}{NOIRLab}). Models of these barred galaxies have been invaluable for studying their dynamics and evolution. Researchers have extensively explored the periodic orbits (POs) that exist in these models. Ferrers' bar galaxy models (GMs), in particular, have proven to be effective for investigating the key properties of real bar galaxies, including star and gas distribution, bar strength, and system stability. The detection and stability of POs in these models have been thoroughly studied in the literature (e.g.~see \citep{skokos2002orbitala, skokos2002orbitalb,patsis2019orbital, manos2022orbit}).

One of the most important families of orbits in bar GMs is the so-called x1 family (e.g.~see \citep{contopoulos1989orbits}), typically characterized by an elliptical-like morphology in non-axisymmetric galactic systems. The major axis of these orbits aligns with that of the bar. Stable members of this family are often regarded as the backbone of bars. These POs, both in two- (\(2D\)) and three-dimensions (\(3D\)), play an important role in shaping the system's phase space. In \(3D\) bar potentials, additional families of POs emerge that are related to the x1 family \citep{skokos2002orbitala}.

As the system's energy increases, x1 POs may develop cusps or even double loops at their edges. The x1 family undergoes a series of bifurcations, which in \(3D\) systems, lead to the formation of the so-called x1-tree (see \citep{skokos2002orbitala,skokos2002orbitalb} for more details). A pitchfork bifurcation of a stable PO family results in the creation of a pair of new stable POs with the same period, while the original family becomes unstable, while a period-doubling bifurcation destabilizes the parent family, and it gives rise to a new stable PO with double the period of the original family (e.g.~see \citep[Sect.~7.1b]{lichtenberg2013regular}).

The primary objective of the work presented in this chapter is to systematically study the evolution of phase space structures in a \(3D\) bar galactic potential, both before and after successive \(2D\) and \(3D\) pitchfork and period-doubling bifurcations of families in the x1-tree. The phase space structure before and after a \(3D\) pitchfork bifurcation of individual members of the x1 family has been previously studied by \citep{katsanikas2011structure1}. Building upon and extending that research, this chapter explores how the phase space of a \(3D\) system evolves through successive pitchfork bifurcations while also incorporating period-doubling bifurcations into the analysis. Specifically, we focus on the evolution of the phase space structure in the following cases: (A) a \(2D\) pitchfork bifurcation of the x1 family leading to the creation of a new \(2D\) family, (B) this new \(2D\) family undergoes a bifurcation, transforming into a \(3D\) family through a similar pitchfork process, and (C) we analyze the changes in the phase space as the \(3D\) family undergoes two successive \(3D\) period-doubling bifurcations.

In our study, we use the color and rotation technique developed by \citep{patsis1994using} to visualize the four-dimensional (\(4D\)) Poincar{\'e} surface of section (PSS) and gain insights into the dynamics of the different types of orbits in our \(3D\) galactic type Hamiltonian system. This technique has also been successfully applied in other studies, including \citep{katsanikas2011structure1,katsanikas2022phase,zachilas2013structure,patsis2014phasea,patsis2014phaseb,agaoglou2021visualizing}.

Furthermore, although our primary objective in this chapter is to investigate the evolution of the phase space structure of different families of orbits, we will also examine how the generalized alignment index (GALI) \eqref{eq:GALI} behaves in comparison to the maximum Lyapunov exponent (mLE) \eqref{eq:mLEs}. By considering specific cases of regular and chaotic orbits, we demonstrate that our findings from Chap. \ref{chapter:three} (e.g.~see Figs.~\ref{fig3:Fig2} and \ref{fig3:Fig4}) regarding the effectiveness of the GALI of order \(2\) (GALI\(_2\)) index for detecting and quantifying chaos in Hamiltonian systems in plasma model are also applicable to galactic Hamiltonian systems. This performance of the index confirms that our conclusion that the GALI\(_2\) method is both accurate and fast in quantifying chaos is valid regardless of the specific Hamiltonian system. 

The content of this chapter is based around the findings presented in \cite{moges2024evolution}.

\section{Hamiltonian bar galaxy model}\label{section:HamiltonianCh4}
In order to study the orbital dynamics of a \(3D\) galactic model, we use the Hamiltonian formalism, a mathematical framework for describing the evolution of a galactic dynamical system (DS). In particular, we consider the autonomous Hamiltonian system,

\begin{equation} \label{eq:BG H} 
H = \frac{1}{2} \left( p_x^2 + p_y^2 + p_z^2 \right) + V(x, y, z) - \Omega_b \left( x p_y - y p_x \right), 
\end{equation}
which describes the dynamics of a test particle (i.e., a star) in a \(3D\) rotating bar GM. In \eqref{eq:BG H}, the bar rotates around the \(z\)-axis, with its major axis being aligned along the \(y\)-axis and the minor axis along the \(x\)-axis (see, for example, \citep{skokos2002orbitala} and reference therein). The canonical momenta \(p_x\), \(p_y\), and \(p_z\) correspond to the Cartesian coordinates \(x\), \(y\), and \(z\), respectively. The total potential \(V(x, y, z)\) includes both axisymmetric and non-axisymmetric components. \(\Omega_b\) represents the bar's rotation speed, and the Jacobi constant, denoted as \(E_j\), is the total energy in the rotating reference frame of the Hamiltonian \(H\).

The corresponding EoM of the Hamiltonian \eqref{eq:BG H} can be derived as follows:
\begin{equation}\label{eq:BG EoM} 
  \begin{aligned}
      \frac{dx}{dt} &= p_x + \Omega_b y, \\
      \frac{dy}{dt} &= p_y - \Omega_b x, \\
      \frac{dz}{dt} &= p_z, \\
      \frac{dp_x}{dt} &= -\frac{\partial V}{\partial x} + \Omega_b p_y, \\
      \frac{dp_y}{dt} &= -\frac{\partial V}{\partial y} - \Omega_b p_x, \\
      \frac{dp_z}{dt} &= -\frac{\partial V}{\partial z}.
  \end{aligned}
  \end{equation}  
where the explicit form of the first-order derivatives of the various components of the potential \(V(x, y, z)\) is given in the Appendix \ref{chapter:appendixA}. 

The potentials of the galactic model \eqref{eq:BG H} we consider in our study consist of three primary components: the bulge potential \(V_S\), the disc potential \(V_D\), and the bar component of the galaxy \(V_B\). These components collectively define the total galactic potential (\(V = V_S + V_D + V_B\)), and they are defined as follows:

\begin{enumerate} [label=\textnormal{(\Roman*)}]
\item \textbf{Bulge Potential}: Galaxies may feature a rounded component known as the bulge or spheroid. These galactic spheroids are smaller than the disk part of the galaxy and have distinct features, such as a roughly spherical shape of stars with different compositions.  The bulge potential we consider is based on a Plummer sphere \citep{plummer1911problem} which is given by the equation

\begin{equation}\label{eq:Plummer}
V_S(x, y, z) = -\frac{GM_S}{\sqrt{x^2 + y^2 + z^2 + \epsilon_s^2}},
\end{equation}
where \(M_S\) represents the total mass of the system, and \(G\) is the gravitational constant. The linear scale of the system, which generates the potential \(V_S\), is determined by the Plummer scale length of the bulge, \(\epsilon_s\).

\item \textbf{Disk Potential}: The Miyamoto-Nagai potential \citep{miyamoto1975three}, is commonly used to model galactic disk components, and it is expressed as
\begin{equation} \label{eq:MiyamotoNagai}
    V_D(x, y, z) = -\frac{GM_D}{\sqrt{x^2 + y^2 + \left(A + \sqrt{B^2 + z^2}\right)^2}},
\end{equation}
where \(M_D\) is the total mass of the disk, and \(A\) and \(B\) are the disc's horizontal and vertical scale lengths, respectively. This potential has been used by Pfenniger \citep{pfenniger19843d} for numerical studies of periodic orbit (PO) structures and stability in more realistic \(3D\) bar GMs.

Note that when $A=0$, $V_D$ reduces to the Plummer's spherical potential $V_S$, with scale length  $B=\epsilon_s$. As a result, depending on the choice of the two parameters $A$ and $B$,  $V_D$ can represent the potential of anything from a very thin disk to a spherical component. 

\item \textbf{Bar Potential}: In our study, we represent the bar component using the Ferrers model \citep{ferrers1877potentials}, which is a widely studied model (e.g.~see \citep{pfenniger19843d,skokos2002orbitalb,bountis2012complex}), that describes the bar as a finite, uniform ellipsoid with a specific density distribution. The mathematical expression for the bar potential is given by 

\begin{equation} \label{eq:Vbar}
V_B = -\pi G a b c \frac{\rho_c}{3} \int_{\lambda}^{\infty} \frac{du}{\Delta(u)} (1 - m^2(u))^{3},
\end{equation}
where \(\rho_c = \frac{105}{32\pi} \frac{GM_B}{abc}\) is the central density, \(M_B\) is the total mass of the bar, while \( m^2(u) = \frac{y^2}{a^2 + u} + \frac{x^2}{b^2 + u} + \frac{z^2}{c^2 + u}\), and \(\Delta^2(u) = (a^2 + u)(b^2 + u)(c^2 + u)\). \(\lambda\) is the unique positive solution to \(m^2(\lambda) = 1\) for regions outside the bar (\(m(u) \geq 1\)) and \(\lambda = 0\) for regions inside the bar (\(m(u) < 1\)), while \(a > b > c\) are the semi-axes of the bar \citep{pfenniger19843d}. The corresponding mass density of the bar potential is given as follows:
\[
\rho =
\begin{cases}
\rho_c (1 - m^2)^2 & \text{for } m \leq 1, \\
0 & \text{for } m > 1,
\end{cases}
\]
with \(m^2 = \frac{y^2}{a^2} + \frac{x^2}{b^2} + \frac{z^2}{c^2}\). This model, characterized by a central density peak that gradually declines to zero at a finite radius, closely aligns with the observational density profile of barred galaxies and the outcomes of \(N\)-body simulations. See Sect.~\ref{sec:Ferrers Bar Pot} for more details. 
\end{enumerate}

Choosing the appropriate parameter values is important to make sure that the GM's \eqref{eq:BG H} rotation curve aligns with the real-world observations. While the exact values may vary, they must be carefully selected to accurately represent the observed rotational patterns of galaxies. This model incorporates a superposition of a Ferrers bar (a specific type of potential distribution - see Sect.~\ref{sec:Ferrers Bar Pot}), the bulge potential, and a Miyamoto-Nagai disc. The units and parameter values used in this work are similar to those employed in the pioneering study by Pfenniger \citep{pfenniger19843d}, and the specific configuration considered corresponds to the `model D' in \citep{skokos2002orbitalb}. The total mass is set to $1$, with \( G(M_S + M_D + M_B) = 1 \). The length unit is \(1\) kpc, the time unit is \( 2 \times 10^{6} \) yr, and the mass unit is \( 2 \times 10^{11}\) \(M_\odot \) (solar masses). Using these units, the chosen parameter values are: bar axial ratios \( A = 6 \), \( B = 1.5 \), \( C = 0.6 \); disc parameters \( A = 3 \), \( B = 1 \), \( \epsilon_s = 0.4 \); bulge mass \( M_B = 0.2 \); disk mass \( M_D = 0.72 \); and bar pattern speed \( \Omega_b = 0.054\), which results in a corotation at \( R_c = 6 \).

\section{Stability of periodic orbits and the method of color and rotation}
\label{section:CR Method}
There are numerous studies that have explored various types of periodic orbits (POs) near the center of a barred galaxy (e.g.~see \citep{contopoulos1980orbits,athanassoula1983orbits,skokos2002orbitala,katsanikas2011structure1,patsis2014phasea,manos2022orbit}). These orbits form the fundamental framework that shapes the bar's structure. The exploration began with \citep{athanassoula1983orbits}, which identified the main x\(1\) family of POs in order to explain the morphology and the characteristics of barred galaxies. The morphology of \(3D\) bar galaxies, including an extensive list of both \(2D\) and \(3D\) families of POs, can be found in \citep{skokos2002orbitala}. 

\subsection{Stability types of periodic orbits in Hamiltonian systems}
A PO occurs when a system's trajectory revisits the same state after a fixed period. The multiplicity of a PO corresponds to the number of times the orbit's trajectory intersects a PSS in one period. For example, in our Cartesian coordinates \((x, y, z, p_x, p_y, p_z)\), if the PSS is defined by \(y = 0\) and \(p_y > 0\), an orbit with a multiplicity of two means that the set of trajectories will cross the successive intersection point on the plane \(y = 0\) twice before completing one full period. Higher multiplicity orbits have also been observed, and a detailed catalog of these families is presented in \citep{patsis2019orbital}.
 
 In order to study the phase space structure of the Hamiltonian \eqref{eq:BG H} near a PO, we first locate the initial conditions (ICs) for this PO using an iterative method that analyzes successive intersections (e.g.~see \citep{pfenniger1993computational}). We initially consider small deviations from the IC and then integrate the perturbed orbit to the next upward intersection point (i.e., \(y = 0; p_y > 0\) ). This approach gives us the \(4D\) \((x, z, p_x, p_z)\) PSS. Through this process, we create a \(4D\) Poincar\'e map, which relates the initial and final points. A PO is identified when the difference between the initial and final coordinates must be less than \(10^{-10}\) in absolute value.

The relationship between the final deviations of the neighboring orbit from the periodic one and the initially introduced deviations is expressed in vector form as:
 \begin{equation}\label{eq:monod}
 \xi = M(T) \cdot \xi_0,
\end{equation}
where \( \xi_0 \) and \( \xi \) represent the initial and final deviations, respectively, and \( M(T) \) denotes the so-called monodromy matrix for the Hamiltonian, which is a \( 4 \times 4 \) matrix in the case of the Hamiltonian \eqref{eq:BG H}. This matrix \(M(T)\), also known as the fundamental solution matrix of the variational equations, is determined at a time equal to one period \(T\) of the orbit (see \citep{skokos2001stability} and references therein for more details). The characteristic polynomial matrix \(M\) can be expressed as
\begin{equation}\label{eq:monod xcs}
\lambda^4 + \alpha \lambda^3 + \beta \lambda^2 + \alpha \lambda + 1 = 0
\end{equation}
where \(\lambda\) represents the eigenvalues of the matrix \(M\) and the coefficients \(\alpha\) and \(\beta\) are parameters that depend on the system.
 
\begin{figure}[!htb]
  \centering
  \includegraphics[width=0.6\textwidth]{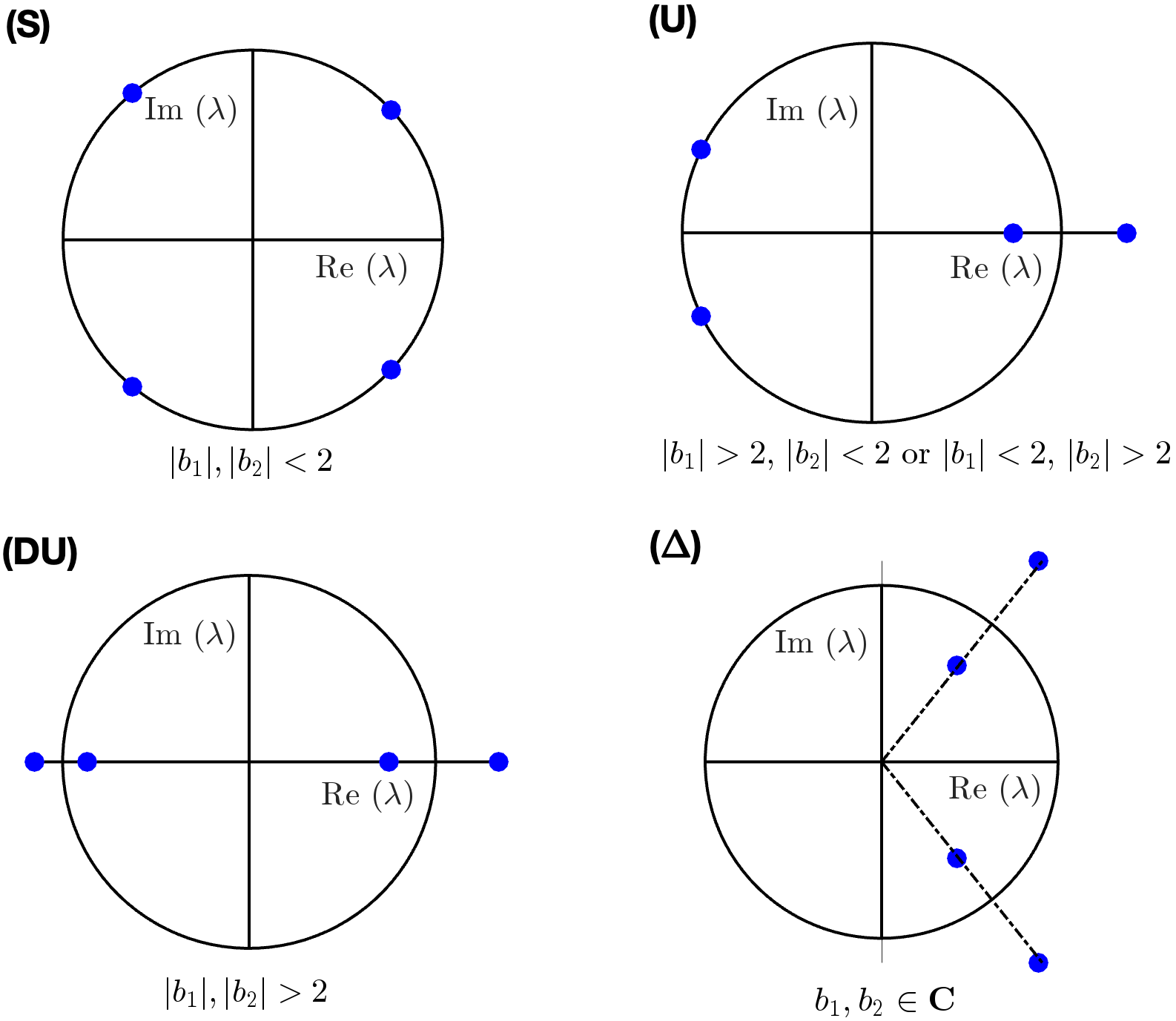}
  \caption{A visual representation of the eigenvalue configurations on the complex plane relative to the unit circle for the stable (S), simple unstable (U), double unstable (DU), and complex unstable (\(\Delta\)) cases. Note that the specific location of the eigenvalues on the complex plane may vary depending on the values of the POs' stability indices  \(b_1\) and \(b_2\) in \eqref{eq:monod xsoln}. For instance, if \(b_1 < -2\) while \(|b_2| < 2\) in the simple unstable (U) case, the eigenvalues will rather be located on the left (negative) side of the real axis.}
  \label{fig4:Fig_Eig}
\end{figure}

The solutions of \eqref{eq:monod xcs} \( \lambda_j\), \(j = 1, 2, 3, 4\), must satisfy the relations $\lambda_1 \lambda_2 = 1$, and $\lambda_3 \lambda_4 = 1$, which indicate that the eigenvalues of \(M\) come in reciprocal pairs. These relationships occur naturally from the symplectic structure inherent to the Hamiltonian system. We can further express each pair of eigenvalues as follows:
\begin{equation}\label{eq:monod xsoln}
  \lambda_j, \frac{1}{\lambda_j} = \frac{1}{2} \left[ -b_j \pm \sqrt{b_j^2 - 4} \right]
\end{equation}
where \(b_j\), \(j=1, 2\) are called the PO's stability indices and are defined as \(b_j = \frac{1}{2} \left( \alpha \pm \frac{1}{2} \right)\), and \(\Delta\) is the discriminant given by \(\Delta = \alpha^2 - 4 (\beta - 2)\). The stability types of POs can be classified based on the values of the stability indices \(b_1\) and \(b_2\) as well as the discriminant \(\Delta\), which are determined from the eigenvalues of $M(T)$ in Eq.~\eqref{eq:monod xsoln}. More specifically, following \citep{contopoulos1985simple}, we classify a PO as follows, with all cases shown in Fig.~\ref{fig4:Fig_Eig}: 

\begin{itemize}
    \item \textbf{Stable (S):} if the discriminant is positive (\(\Delta > 0\)) and both stability indices satisfy \(|b_1| < 2\) and \(|b_2| < 2\). Under these conditions, all four eigenvalues \( (\lambda_j, j = 1, 2, 3, 4) \) lie on the unit circle in the complex plane.
    \item \textbf{Simple Unstable (U):} if \(\Delta > 0\) and \(|b_1| > 2\) while \(|b_2| < 2\) or \(|b_1| < 2\) while \(|b_2| > 2\). In this case, two eigenvalues are on the real axis and the other two are on the unit circle in the complex plane. For example, when \( b_1 > 2 \), there are two real positive eigenvalues: e.g., \( \lambda_1 > 1 \), and its complex conjugate, \( \frac{1}{\lambda_2} < 1\). On the other hand, for \( b_1 < -2 \), \( \lambda_1 < -1\) and \( -1 < \frac{1}{\lambda} < 0 \).
    \item \textbf{Double Unstable (DU):} if \(\Delta > 0\) and \(|b_1| > 2\) and \(|b_2| > 2\). In this case, all four eigenvalues lie on the real axis of the complex plane.  
    \item \textbf{Complex Unstable (\(\Delta\)):} if \(\Delta < 0\), which means all four eigenvalues are complex numbers and off the unit circle in the complex plane. These eigenvalues correspond to two complex conjugate stability indices, with two of the eigenvalues located inside the unit circle and the other two outside. 
\end{itemize}

The distinction between different types of instability was initially introduced by Broucke \citep{broucke1969stability} and Hadjidemetriou \citep{hadjidemetriou1975continuation}, where it has been used to study the stability of POs in Hamiltonian systems with three degrees of freedom (DoF). For a comprehensive overview of the stability analysis and classification of POs in \(3D\) Hamiltonian systems, readers can refer to \citep{pfenniger19843d} and \citep{contopoulos1985simple}. In addition, for a general understanding of different instabilities in Hamiltonian systems with \(N\) DoF, the reader may refer to  \citep{skokos2001stability}. 

In this chapter, we will systematically investigate the evolution of the phase space structure of the \(3D\) GM \eqref{eq:BG H} through a series of PO's bifurcations as the parent family transitions from stability to simple instability (S $\rightarrow$ U), leading to the emergence of a new family of POs. In our cases, the parent \(2D\) or \(3D\) family becomes simple unstable as the system's energy increases. This results in the creation of a stable family of either \(2D\) or \(3D\) POs through this bifurcation. 

Before analyzing particular bifurcation, let us first look at the behavior of the mLE and the GALI for the orbits of the \(3D\) rotating bar GM \eqref{eq:BG H}.

\subsection{Chaos detection and quantification} \label{section:Ch4R1}
\begin{figure}[!htb]
    \centering
    \includegraphics[width=0.6\textwidth]{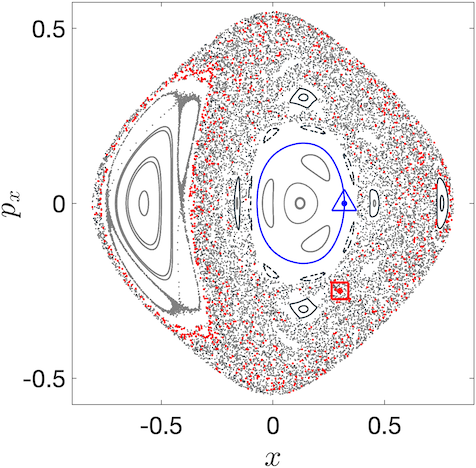}
    \caption{The PSS \((y = 0; p_y > 0)\) for the simplified \(2D\) system of \eqref{eq:BG H} [setting \((z,p_z)=(0,0)\)] at energy \(E_j=0.41\). The evolution of several orbits is shown in gray. A regular orbit with IC \((x, p_x) = (0.32, 0)\) (blue triangular point) is indicated by a smooth blue closed curve, while a chaotic orbit with IC \((x, p_x) = (0.3, -0.25)\) (red square point) is depicted as red scattered points.}
    \label{fig4:Fig_A0}
  \end{figure}

To continue demonstrating the efficiency of the GALI\(_2\) method \eqref{eq:GALI} and to compare it with the mLE \eqref{eq:mLEs}, let us briefly examine the $2D$ counterpart of our bar GM by simplifying the \(3D\) system \eqref{eq:BG H} to a \(2D\) one. We accomplish this reduction by considering motion confined to the galactic plane \(z=0\), and setting both the coordinate \(z\) and the corresponding momentum \(p_z\) to zero. We first create a \(2D\) PSS for \(y = 0; p_y > 0\) to visually identify representative regular and chaotic orbits of the system. In Figure \ref{fig4:Fig_A0}, we illustrate a PSS of the \(2D\) bar GM, showing a regular orbit with IC \((x, p_x) = (0.32, 0)\)  (blue triangular point), which leads to the creation of a smooth closed blue curve. In addition, we depict a chaotic orbit with IC \((x, p_x) = (0.3, -0.25)\)  (red square point), resulting in red scattered points. The PSS is generated by integrating several orbits of the system for \(10^5\) time units using the Runge-Kutta four-order (RK4) scheme \eqref{eq:RK}. 
In order to compute our chaos indicators, the mLE and the GALI\(_2\) index, we need to integrate the EoM \eqref{eq:BG EoM} along with the variational equations \eqref{eq:Gen Ham VoEs} of the system. The corresponding variational equations of the bar GM \eqref{eq:BG H} are given by 
\begin{equation} \label{eq:BG VarEq} 
\begin{aligned}
  \frac{d\delta x}{dt} &= \delta p_x + \Omega_b \delta y, \\
  \frac{d\delta y}{dt} &= \delta p_y - \Omega_b \delta x, \\
  \frac{d\delta z}{dt} &= \delta p_z, \\
  \frac{d\delta p_x}{dt} &= -\frac{\partial^2 V}{\partial x^2} \delta x - \frac{\partial^2 V}{\partial x \partial y} \delta y - \frac{\partial^2 V}{\partial x \partial z} \delta z + \Omega_b \delta p_y, \\
  \frac{d\delta p_y}{dt} &= -\frac{\partial^2 V}{\partial y \partial x} \delta x - \frac{\partial^2 V}{\partial y^2} \delta y - \frac{\partial^2 V}{\partial y \partial z} \delta z - \Omega_b \delta p_x, \\
  \frac{d\delta p_z}{dt} &= -\frac{\partial^2 V}{\partial z \partial x} \delta x - \frac{\partial^2 V}{\partial z \partial y} \delta y - \frac{\partial^2 V}{\partial z^2} \delta z.
\end{aligned}
\end{equation}
where the second-order derivatives of the potentials \(V\) are provided in Appendix \ref{chapter:appendixA}

Figures \ref{fig4:Fig_A1a} and (b) present the time evolution of the finite-time maximum Lyapunov exponent (ftmLE), \( \sigma_1 \) \eqref{eq:ftmLE}, and the GALI\(_2\) \eqref{eq:GALI} for the highlighted ICs given in Fig.~\ref{fig4:Fig_A0} of the \(2D\) system, respectively. For the regular orbit (blue curves in Fig.~\ref{fig4:Fig_A0}), \( \sigma_1 \) tends to zero following a power law proportional to the function \( t^{-1} \) (indicated by a blue dashed line in Fig.~\ref{fig4:Fig_A1a}), while for the chaotic orbit (red scattered points in Fig.~\ref{fig4:Fig_A0}), \( \sigma_1 \) eventually saturates to a positive value (red curve in Fig.~\ref{fig4:Fig_A1a}). On the other hand, the GALI\(_2\) remains constant for the regular orbit (blue curve in Fig.~\ref{fig4:Fig_A1b}) but it decays to zero exponentially fast for the chaotic orbit (red curve in Fig.~\ref{fig4:Fig_A1b}). For the chaotic orbit (red curves in Fig.~\ref{fig4:Fig_A1}), the GALI\(_2\) decays to zero, obtaining \(\log_{10} (\text{GALI}_2) \approx -8.13\) at \(\log_{10} (t) \approx 2.64\). In contrast, the ftmLE starts to saturate to its positive limiting value of \(\log_{10} (\sigma_1) \approx -1.62\) at \(\log_{10} (t) \approx 2.4\) (indicated by the dotted black vertical lines in Fig.~\ref{fig4:Fig_A1a}) where the ftmLE deviates from the \(-1\) slope. At this stage, from the ftmLE we cannot confirm chaos definitively as more time is needed to be completely certain that the ftmLE saturates to its positive value. On the other hand, the GALI\(_2\) decays exponentially very fast to \(10^{-8}\) (i.e., at  \(\log_{10} (t) \approx 2.4\), the dotted black vertical lines in Fig.~\ref{fig4:Fig_A1b}). Thus, similar to Fig.~\ref{fig3:Fig2} in Chap. \ref{chapter:three} for the GC model, the GALI\(_2\) index accurately characterizes the chaotic orbit in a shorter computational time compared to the ftmLE. The underlying reason is that as soon as the evolution of the ftmLE begins to deviate from the \(-1\) slope [dotted black vertical line in Fig.~\ref{fig4:Fig_A1a}], the GALI\(_2\) immediately drops to zero [dotted black vertical line in Fig.~\ref{fig4:Fig_A1b}]. The vertical dotted black line indicates the time at which the deviation vector deviates from the IC.

\begin{figure}[!htb]
  \centering
  \subfloat[ftmLE ($t$)\label{fig4:Fig_A1a}]{\includegraphics[width=0.49\textwidth]{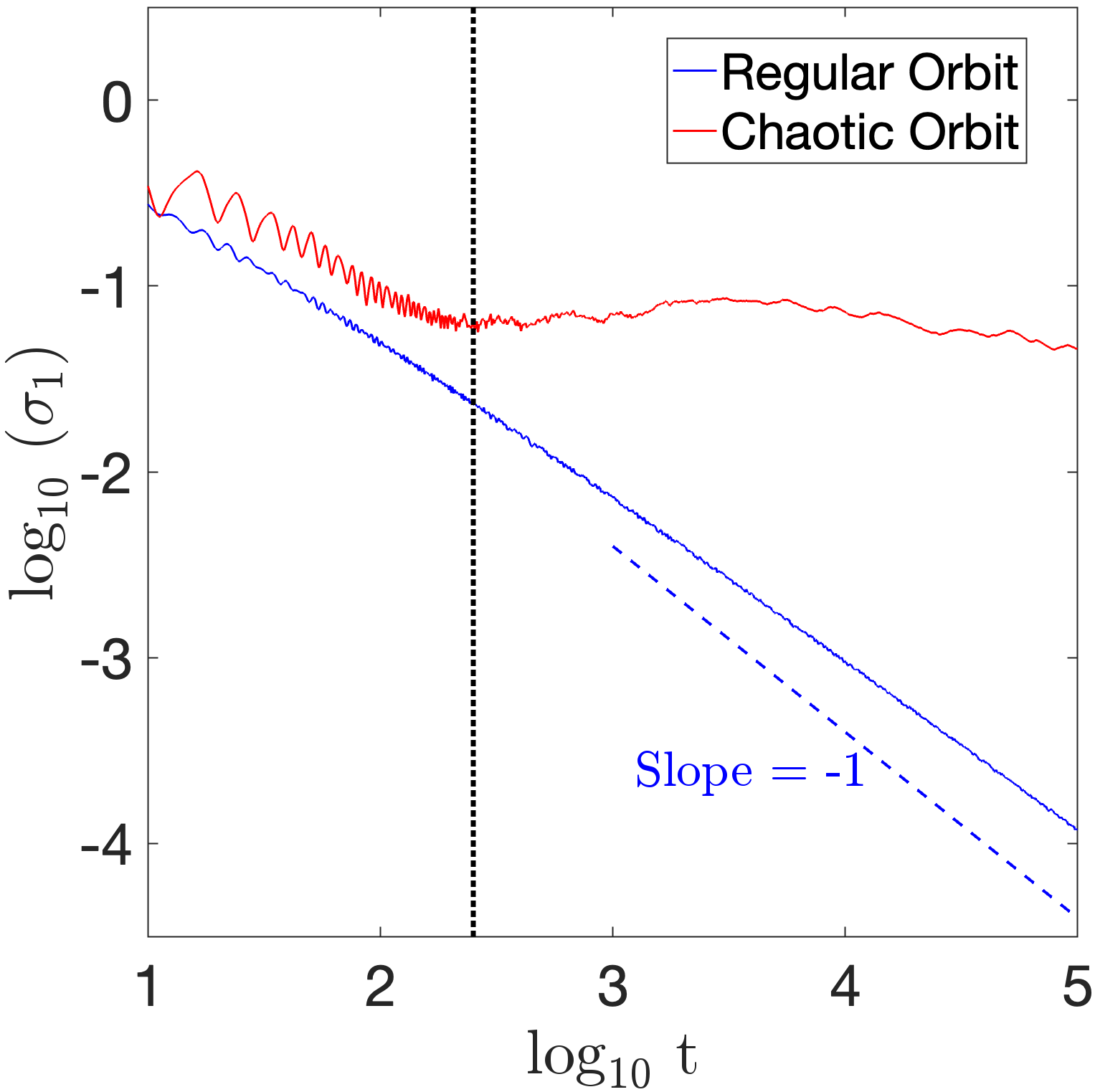}}\hfill 
  \subfloat[GALI\(_2 (t)\)\label{fig4:Fig_A1b}] {\includegraphics[width=0.49\linewidth]{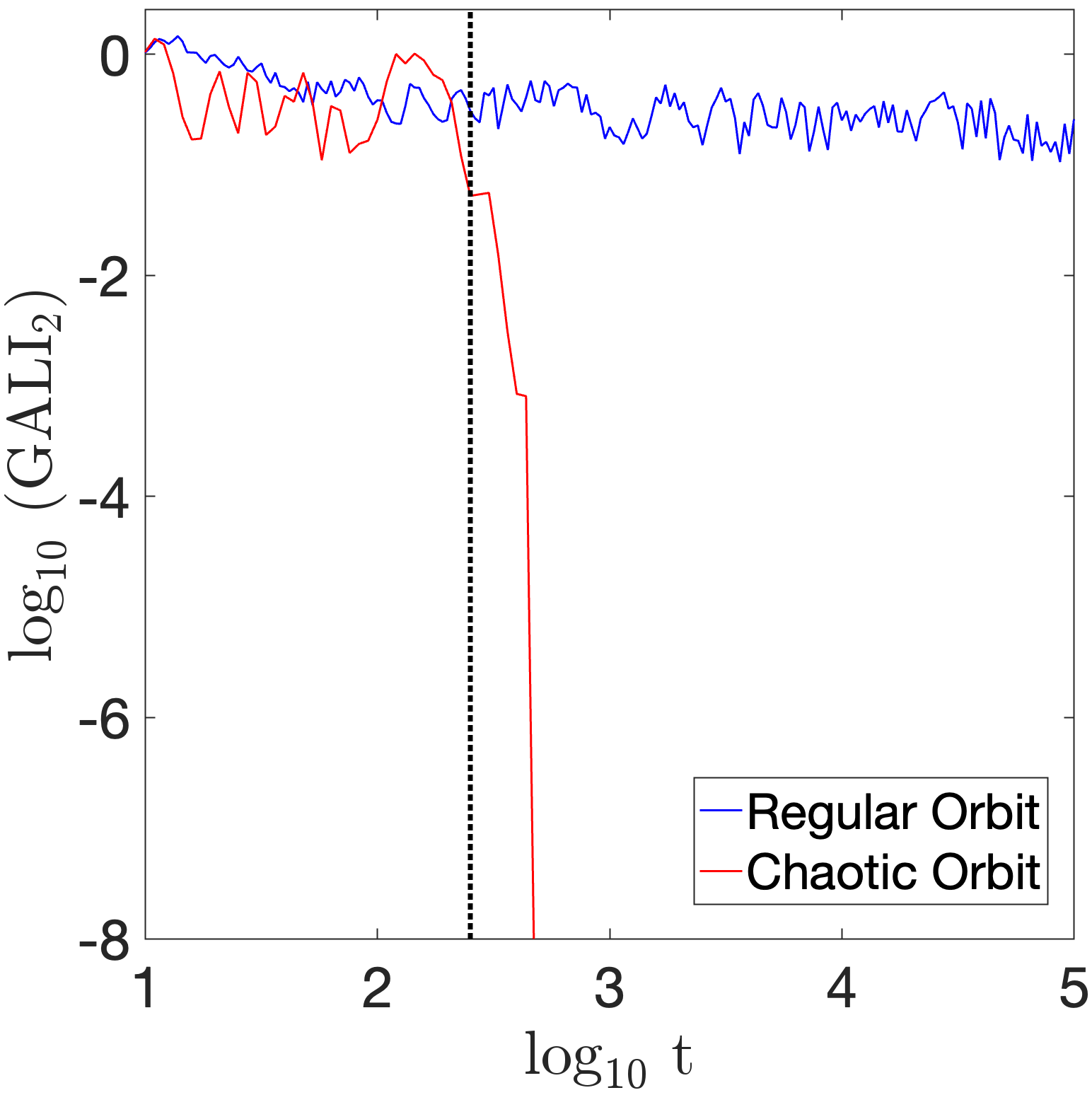}}     
  \caption{The time evolution of (a) the ftmLE \( \sigma_1 \) \eqref{eq:ftmLE}, and (b) the GALI\(_2\) \eqref{eq:GALI} for the regular (blue curves) and chaotic (red curves) orbits shown in Fig.~\ref{fig4:Fig_A0}. The blue dashed line in (a) corresponds to a function proportional to \(t^{-1}\), while the dotted black vertical lines in both panels indicate $\log_{10} (t)=2.4$ time units.}
  \label{fig4:Fig_A1}
\end{figure}

The results of Fig.~\ref{fig4:Fig_A1} strongly support our previous observation in Chap. \ref{chapter:three}, that compared to the commonly employed mLE approach, the GALI\(_2\) index is a very efficient method for distinguishing between regular and chaotic orbits in Hamiltonian systems. However, in this chapter, our main goal is to investigate the evolution of phase space structures in the \(3D\) Hamiltonian system \eqref{eq:BG H} as the POs undergo a series of \(2D\) and \(3D\) pitchfork and period-doubling bifurcations. Thus, in order to gain a deeper understanding of these complex phase space structures, we will use a visualization technique which we now will introduce.

\subsection{The method of color and rotation}
While the construction of the standard \(2D\) PSS as well as phase space projections are effective techniques to visualize the dynamics of \(2D\) Hamiltonian systems [such as the GC system \eqref{eq:GC H} in Chapter~\ref{chapter:three}], and \(2D\) maps [such as the standard map in Chap.~\ref{chapter:five}], their application becomes impractical for higher-dimensional systems. This is especially true for \(3D\) Hamiltonian models or \(4D\) maps, as these lead to the creation of spaces with dimension greater than three. For such systems, we can use a technique proposed by Patsis and Zachilas \citep{patsis1994using} the so-called \textit{method of color and rotation} (CR). As its name suggests, this method involves color and rotation on a \(4D\) PSS. The CR method has been proven to be effective in investigating the behaviors of various phase space structures in \(3D\) Hamiltonian systems, including the \(3D\) rotating GM system \eqref{eq:BG H} (see, for example, \citep{contopoulos2002order,katsanikas2011structure1,katsanikas2013instabilities, patsis2014phasea}). Furthermore, the method has been successfully applied to study the phase space dynamics of $4D$ symplectic maps \citep{zachilas2013structure}. 

The CR method focuses on one or a few orbits by using a \(3D\) projection of the consequents. We then apply a color scale to effectively show the value of the fourth dimension, which is the one coordinate not included in the \(3D\) subspace representation. This visualization technique offers more details into the system's behavior, with the color variations revealing patterns or structures that may not be immediately visible in a standard \(3D\) projection, especially when rotating the \(3D\) colored structures in order to observe them from different angles. Moreover, the CR method enables us to explore the complex dynamics of the \(4D\) phase space of a system and understand how all dimensions influence the system's behavior. We implement the CR method through the following steps:

\begin{enumerate} [label=\textnormal{(\Roman*)}]
    \item \textbf{Create a PSS:} We start with an IC of our \(3D\) Hamiltonian \eqref{eq:BG H} and calculate several intersections of the related trajectory with the \( y=0 \) plane (with \( p_y > 0\) always determined from the Hamiltonian) to get a dense set of points in the PSS. This results in a \(4D\) set of points in the \((x, p_x, z, p_z)\) space.
    \item \textbf{Choose a \(3D\) subspace:} After obtaining the PSS, we choose a \(3D\) subspace such as \((x, p_x, z)\) or \((p_x, z, p_z)\), and plot the three-tuple of points. The axes can be scaled based on the minimum and maximum values of each coordinate for clarity.
    \item \textbf{Color code the fourth dimension:} Each three-tuple of points is colored based on the value of its fourth coordinate. To better understand the \(3D\) data space, we simply implement a rather straightforward approach of rotating the \(3D\) objects using plotting software. In particular, we use MATLAB's latest version (R2024a) \(3D\) plot toolbox. Rotating the subspace enhances our understanding of the generated phase space structures. By doing so, we manage to analyze the created structures from different angles and reveal features that may not be apparent from a single point of view.
    \item \textbf{Normalize the color scale:} Finally, we normalize the resulting interval of the fourth dimension values (which is used to color the points) into the range $[0, 1]$ for simplicity. This normalization allows us to focus on the created color patterns rather than on the exact values of the fourth coordinate. It also ensures that the color scale in the produced color plots remains consistent and easy to interpret. This approach has been previously used in various studies (e.g.~see \citep{katsanikas2011structure1})
\end{enumerate}

Assuming we choose the \((x, z, p_z)\) subspace for presentation, then the color of the triplet point is determined by its corresponding \(p_x\) value. This allows us to analyze how the points in the \((x, z, p_z)\) subspace are distributed along the \(p_x\) direction. If the \(p_x\) value remains constant, the final \(3D\) projection will show a uniform color, i.e., in practice only one color will appear in the color scale.

Implementing the CR technique on a \(4D\) PSS, the presence of a torus with a smooth color variation on its surface indicates a regular orbit near a stable PO. On the other hand, the chaotic nature of an orbit is manifested on a \(4D\) PSS through the irregular behavior of points in the \(3D\) projection and/or by the mixing of colors representing the fourth dimension along the created \(3D\) structure \citep{patsis1994using,katsanikas2011structure1,katsanikas2013instabilities}.

Unless an orbit quickly escapes to infinity for the given IC of the \(3D\) GM \eqref{eq:BG H} before reaching the final simulation time we consider (typically \(t=10^6\) time units), the \(4D\) PSS for the consequents is colored based on the respective values of the momentum \(p_x\) (or sometimes the coordinate \(x\) for better visualization). While presenting the \(3D\) PSS according to other possible projections may provide alternative perspectives, this choice does not change the overall outcome of our analysis. In our study, we choose to present a specific \(3D\) projection of the \(4D\) PSS that offers more information on the arrangement of the various structures and provides a higher resolution.

We numerically evolve the system's EoM \eqref{eq:BG EoM} for ICs \((x_0, p_{x0}, z_0, p_{z0})\) on the PSS \(y = 0\), with the corresponding upward momentum \(p_{y0}\) being determined by the given \(E_j\) value. The integration is done by using the RK4 scheme \eqref{eq:RK}. We select an appropriate integration time step for each considered galactic orbit in our investigation to ensure that the relative energy error remains smaller than \(10^{-10}\) throughout the orbit's evolution.

\section{Numerical results} \label{section:ResultsCh4}
We thoroughly examine the phase space structure evolution before and after a series of \(2D\) and \(3D\) bifurcations of POs in the \(3D\) Hamiltonian \eqref{eq:BG H}. In particular, we systematically investigate four cases: a \(2D\) pitchfork bifurcation, a \(3D\) pitchfork bifurcation, and two \(3D\) period doubling bifurcations. In our study, a pitchfork bifurcation is a type of bifurcation that occurs in the Hamiltonian \eqref{eq:BG H} when a stable PO of one family of orbits give birth to two symmetrically stable POs of a new family of orbits. This bifurcation is also referred to as ``direct” or “supercritical”  (see, for example, \citep[Sect.~2.4.3]{contopoulos2002order}, and \citep[Sect.~3.4]{strogatz2018nonlinear}).

\begin{figure}[!htbp]
    \centering
    \subfloat[The \(3D\) \((E_j , x, z)\) space for \( -0.45 < E_j < 0.23\)\label{fig4:Fig1a}]{\includegraphics[width=1\textwidth]{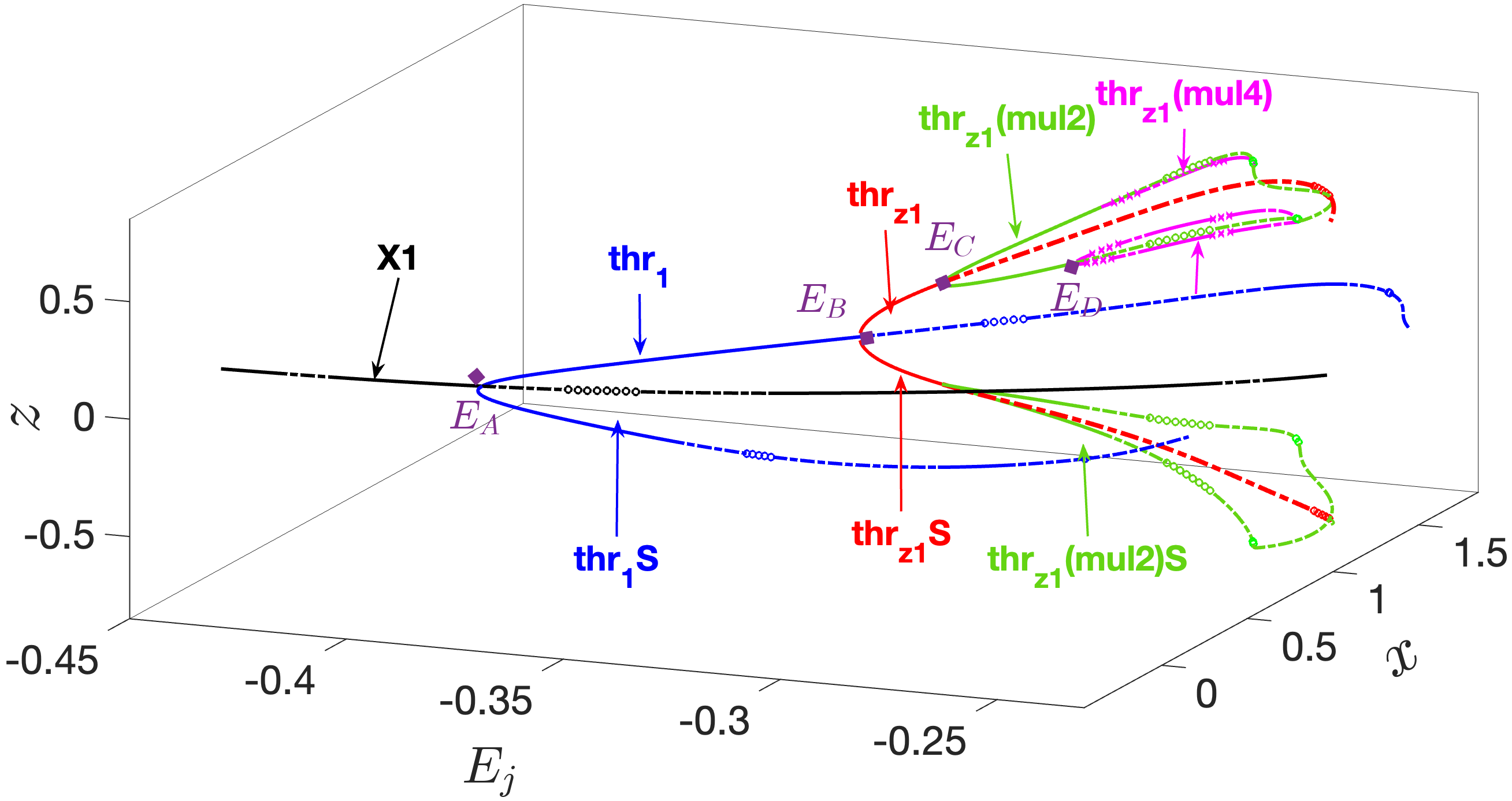}} \\
    \subfloat[The \(2D\) \((E_j , x)\) space for \( -0.45 < E_j < 0.3\)\label{fig4:Fig1b}] {\includegraphics[width=0.49\linewidth]{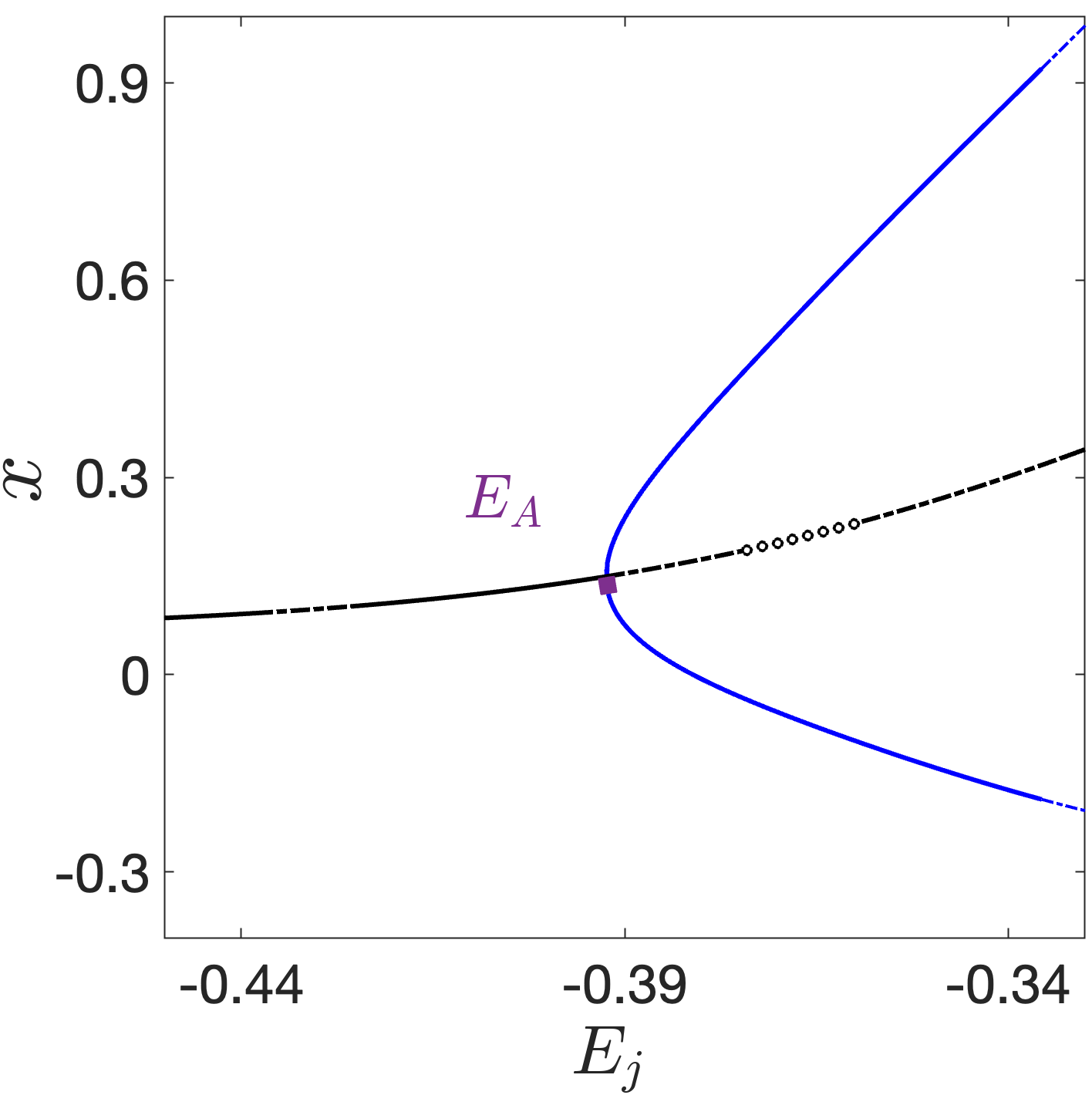}}
    \subfloat[The \(2D\) \((E_j , z)\) space for \( -0.35 < E_j < 0.23\)\label{fig4:Fig1c}] {\includegraphics[width=0.49\linewidth]{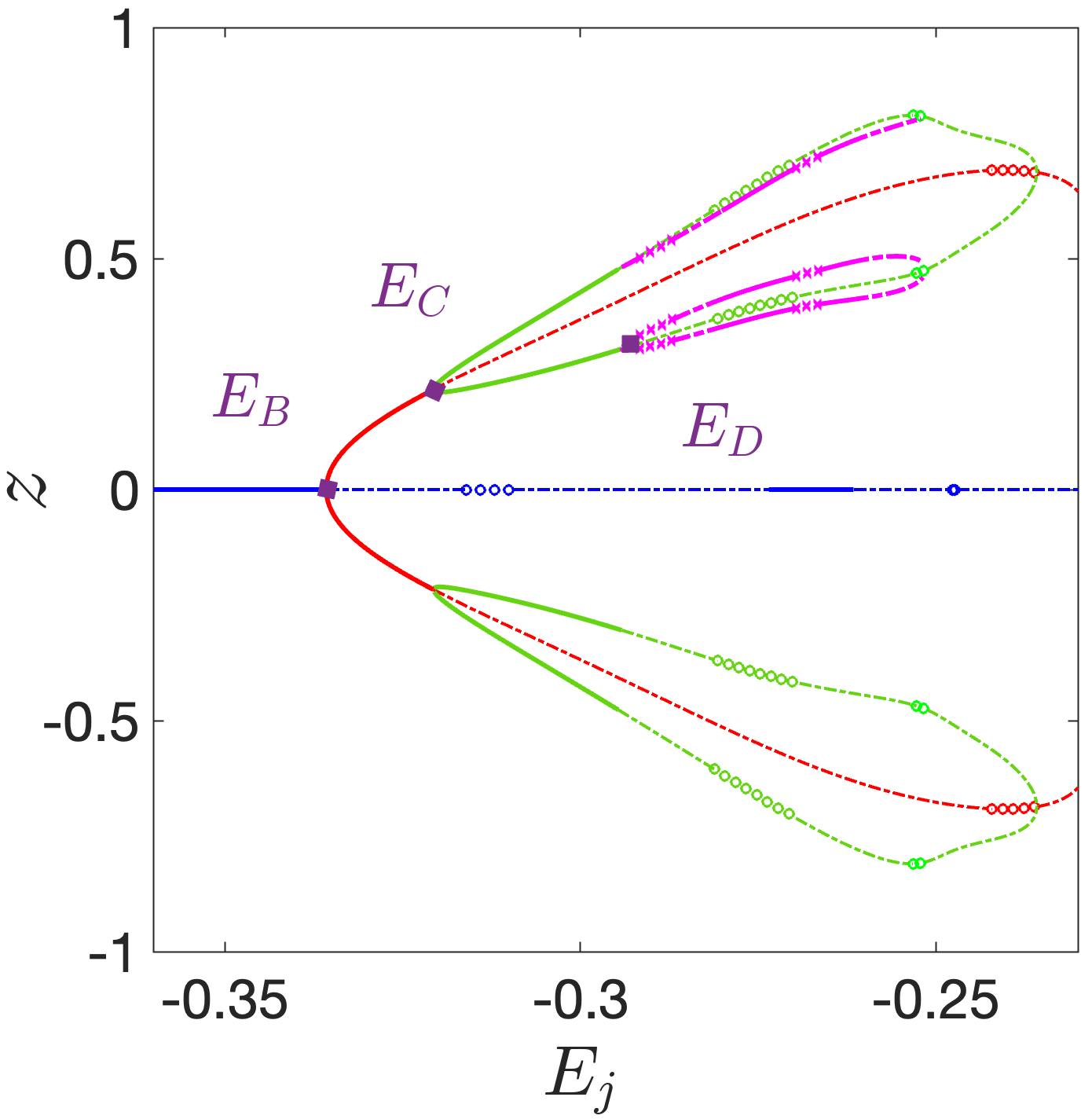}}
    \caption{The characteristic diagrams of various families of POs for the Hamiltonian \eqref{eq:BG H} as the energy level, \(E_j\), varies. Each family is represented by different colored curves: thr\(_1\) (blue), thr\(_{z1}\) (red), thr\(_{z1}\)(mul2) (green), and thr\(_{z1}\)(mul4) (magenta), along with their symmetric counterparts denoted by adding an ``S" at the end of the family's name. The main \(2D\) family x\(1\) is represented by black curves. Different curve styles indicate the stability types of the POs: solid curves indicate stable POs, while dashed line curves, open circles, and ``x" symbol curves denote simple unstable, double unstable, and complex unstable POs, respectively. The  energy \(E_j\) values where bifurcations occur are: $E_A=-0.3924$,  $E_B=-0.3356$,  $E_C=-0.3203$ and $E_D=-0.2943$.  Note that the thr\(_{z1}\)(mul4) family has four branches, with the upper two branches (higher positive \(z\) values) overlapping in (a) and (c).} 
    \label{fig4:Fig1}
  \end{figure}

Figure \ref{fig4:Fig1} illustrates the so-called \textit{characteristic diagram} (see, for example, \citep{skokos2002orbitala}) of the families of POs we consider in our study. This diagram shows the ICs of the POs as a function of the Jacobi constant (\(E_j\)), which is a measure of the total energy of the system \eqref{eq:BG H}. The characteristic diagram illustrates the relationship and the evolution of the families of POs under consideration as \(E_j\) increases. For each IC, four coordinates define its position in the \(4D\) PSS of the Hamiltonian \eqref{eq:BG H}. Thus, we consider the \(3D\) \((E_j, x, z)\) subspace [Fig.~\ref{fig4:Fig1a}] to provide a more detailed characteristic diagram. Additionally, the corresponding \(2D\) projections \((E_j, x)\) [Fig.~\ref{fig4:Fig1b}] and \((E_j, z)\) [Fig.~\ref{fig4:Fig1c}] are shown for the considered \(2D\) and \(3D\) families, respectively. We take \(E_j\) values in the interval $-0.45 < E_j < 0.23$, where all four families of POs, which we study in our work, are fully observed. 

Considering the main planar family x\(1\) (black curves), we observe a bifurcation occurring at energy level \(E_A=-0.3924\). This bifurcation leads to the birth of two new stable planar families, i.e., the orbits are on the \((x, y)\) plane, denoted as thr\(_1\) and its symmetric counterpart with respect to the \(y-\)axis, thr\(_1\)S (blue curves in Fig.~\ref{fig4:Fig1}). Then, the stable planar family, thr\(_1\), loses its vertical stability and undergoes a bifurcation at energy level \(E_B=-0.3356\). This bifurcation results in the creation of two new, initially stable, \(3D\) families: thr\(_{z1}\) and its symmetric counterpart with respect to the system's equatorial plane, denoted by thr\(_{z1}\)S [red curves Figs.~\ref{fig4:Fig1a} and Fig.~\ref{fig4:Fig1c}]. This bifurcation represents the \(3D\) pitchfork bifurcation. Then, the \(3D\) family thr\(_{z1}\) undergoes a period-doubling bifurcation \citep[Sect.~2.11.2]{contopoulos2002order} at \(E_C=-0.3203\), which results in a family of multiplicity two, denoted by thr\(_{z1}\)(mul2), along with its symmetric family thr\(_{z1}\)(mul2)S (green curves in Figs.~\ref{fig4:Fig1a} and Fig.~\ref{fig4:Fig1c}). Finally, at energy \(E_D=-0.2943\), the thr\(_{z1}\)(mul2) stable  \(3D\) family undergoes another period-doubling bifurcation. As a result, the family thr\(_{z1}\)(mul2) becomes simple unstable, creating a new \(3D\) stable family we call thr\(_{z1}\)(mul4) [along with its symmetric counterpart thr\(_{z1}\)(mul4)S] of multiplicity four. These newly formed families [i.e., thr\(_{z1}\)(mul4) and thr\(_{z1}\)(mul4)S] are shown by magenta curves in Figs.~\ref{fig4:Fig1a} and (c). Note that due to our choice of projection to present the characteristic diagrams in Figs.~\ref{fig4:Fig1a} and (c), some curves, in particular the upper two branches of the magenta curve representing thr\(_{z1}\)(mul4) family of POs, actually consist of two overlapping curves. The different curve styles in Fig.~\ref{fig4:Fig1} indicate the stability type of each PO family; namely, solid curves represent stable POs, dashed line curves indicate simple unstable POs, open circles denote double unstable POs, while curves containing the ``x'' symbol correspond to complex unstable POs. We will delve into the evolution of each family in detail in the follow-up subsections.

\begin{figure}[!htb]
    \centering
    \includegraphics[width=1\textwidth]{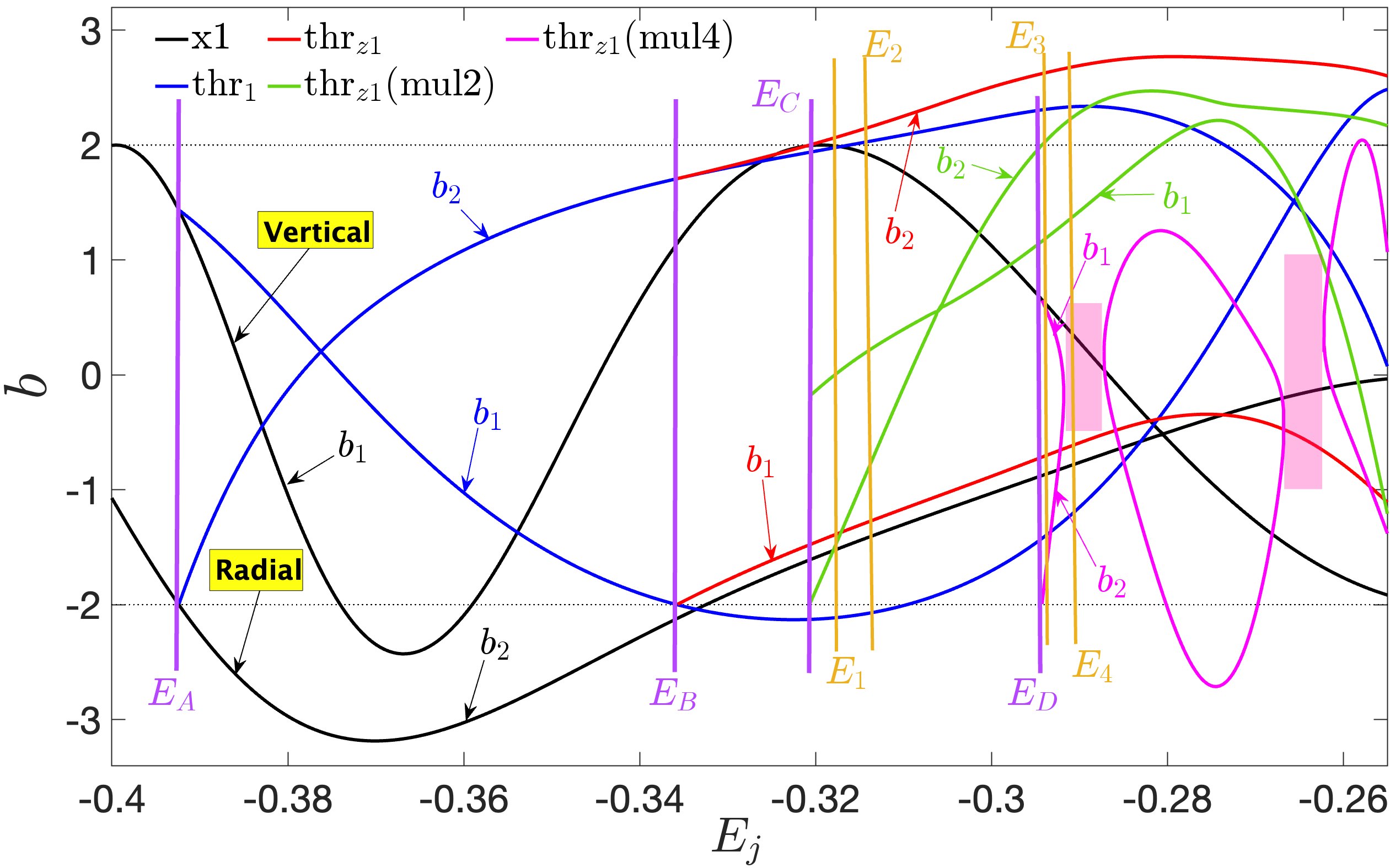}
    \caption{The evolution of the vertical (\(b_1\)) and radial (\(b_2\)) stability indices as a function of the Jacobi constant (\(E_j\)) for the families of POs considered in our study. The stability indices of the various families are depicted by different colored curves: x\(1\) (black), thr\(_{1}\) (blue), thr\(_{z1}\) (red), thr\(_{z1}\)(mul2) (green), and thr\(_{z1}\)(mul4) (magenta). The four vertical purple lines indicate the bifurcation points ($E_A=-0.3924$,  $E_B=-0.3356$,  $E_C=-0.3203$ and $E_D=-0.2943$), while the four vertical orange lines represent specific energy levels \(E_1 = -0.3183\), \(E_2 = -0.3157\), \(E_3 = -0.2941\), and \(E_4 = -0.2907\), which will be further discussed in Sect.~\ref{sec:Period-doubling}.}
    \label{fig4:Fig2}
  \end{figure}

Figure \ref{fig4:Fig2} depicts the so-called ``stability diagram" (e.g.~see \citep{skokos2002orbitala,patsis2014phasea}), which shows how the stability of different families of POs changes with respect to the energy level, \(E_j\). This diagram illustrates the evolution of the two stability indices, \(b_1\) and \(b_2\) \eqref{eq:monod xsoln} (which are related to vertical and radial perturbations of the POs, respectively) as a function of \(E_j\). Again, we represent each family and the associated symmetric counterpart by a single color curve.

We can analyze the stability transitions observed in Fig.~\ref{fig4:Fig2} through the behavior of the \(b_1\) and \(b_2\) indices as follows: Initially, when the index \(b_2\) of the main family x\(1\) (represented by the black curves) crosses the \(b = -2\) line at energy \(E_A=-0.3924\), the x\(1\) PO loses its radial stability, resulting in the creation of the two new stable families thr\(_{1}\) and thr\(_{1}\)S (blue curves). At the second bifurcation at \(E_B=-0.3356\), when the vertical stability index \(b_1\) crosses the \(b = -2\) line, the stable family thr\(_{1}\) transitions to simple instability. At the same time, we observe the birth of the new stable \(3D\) families thr\(_{z1}\) and thr\(_{z1}\)S (red curves in Fig.~\ref{fig4:Fig2}). At the third observed bifurcation at \(E_C=-0.3203\), the radial stability index  \(b_2\) of the thr\(_{z1}\) family crosses the  \(b = -2\) line, and the family becomes simple unstable. This transition leads to the first \(3D\) family formed through a period-doubling bifurcation we study; namely, the thr\(_{z1}\)(mul2) family (green curves). Lastly, the fourth bifurcation takes place at energy \(E_D=-0.2943\), where the  \(b_2\) index of the thr\(_{z1}\)(mul2) family crosses the  \(b = -2\) line, leading to the creation of the stable multiplicity four family thr\(_{z1}\)(mul4) and its symmetric counterpart thr\(_{z1}\)(mul4)S (magenta curves in Fig.~\ref{fig4:Fig2}).

The stability indices for each family (thr\(_{1}\), thr\(_{z1}\), etc.) are the same for their respective symmetric counterparts (thr\(_{1}\)S, thr\(_{z1}\)S, etc.). Furthermore, in Fig.~\ref{fig4:Fig2}, the stability curves of the families introduced in the system by period-doubling [thr\(_{z1}\)(mul2) and thr\(_{z1}\)(mul4)] do not originate from the intersection of the parent family's stability index with the \(b = 2\) line. This intersection occurs only if we compute the stability indices of the parent family as having double multiplicity. To avoid overloading the figure with many curves, we chose to not present the stability diagrams of these families (the reader can refer to \citep[Appendix A]{skokos2002orbitalb} and \citep{patsis2019orbital} for more details). 

In Fig.~\ref{fig4:Fig2}, the four vertical purple lines, denoted by \(E_A\), \(E_B\), \(E_C\), and \(E_D\), respectively, indicate the energy levels where critical bifurcations occur. These bifurcations lead to the birth of new families:  thr\(_{1}\) from x\(1\) ($E_A=-0.3924$), thr\(_{z1}\) from thr\(_{1}\) ($E_B=-0.3356$), thr\(_{z1}\)(mul2) from thr\(_{z1}\) through period-doubling ($E_C=-0.3203$), and thr\(_{z1}\)(mul4) from thr\(_{z1}\)(mul2) through another period-doubling bifurcation ($E_D=-0.2943$). The magenta-shaded areas in Fig.~\ref{fig4:Fig2} depict energy intervals where the thr\(_{z1}\)(mul4) family becomes complex unstable (the discriminant \(\Delta < 0\) in Eq.~\eqref{eq:monod xsoln}, and their \(b_1\) and \(b_2\) are undefined). Additionally, the four vertical orange lines indicate the energy levels, which correspond to the specific cases discussed in Sect.~\ref{sec:Period-doubling} (see Table \ref{tab:stability} for more details).

\begin{figure}[!htb]
    \centering
    \includegraphics[width=1\textwidth]{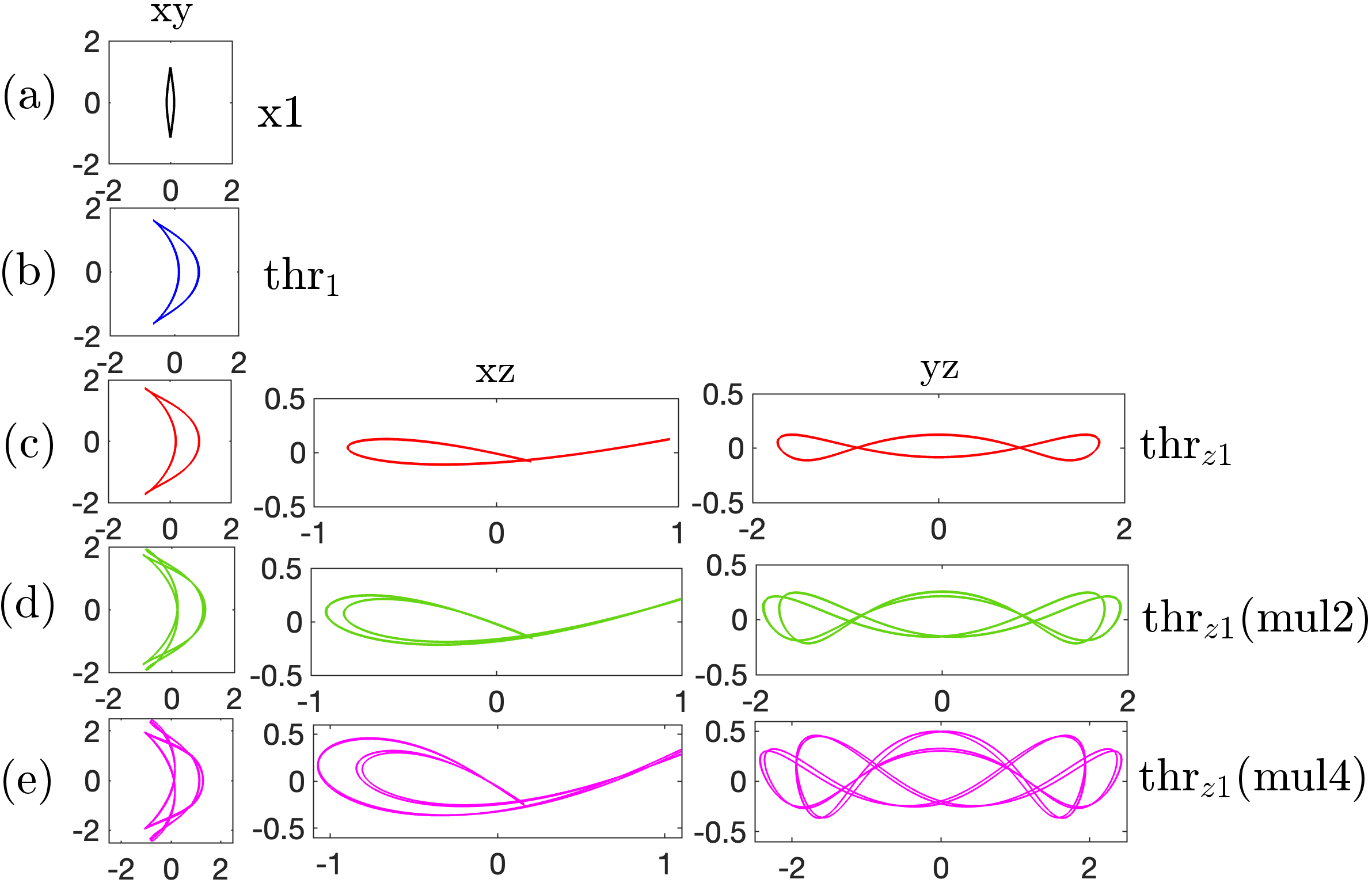}
    \caption{The morphology of representative stable POs from the five families of POs we consider. We plot three multiplicity one POs: (a) the x\(1\) PO with \(E_j = -0.41\), (b) the thr\(_1\) PO with \(E_j = -0.35\), and (c) the thr\(_{z1}\) PO with \(E_j = -0.3306\). A multiplicity two thr\(_{z1}\)(mul2) PO for  \(E_j = -0.3183\), and a multiplicity four thr\(_{z1}\)(mul4) PO for \(E_j = -0.2831\) are, respectively, shown in (d) and (e). From left to right, the families are shown in different \(2D\) projections, namely \((x, y)\), \((x, z)\), and \((y, z)\).}
    \label{fig4:Fig3}
  \end{figure}

The morphology of representative stable members of the families of POs is shown in Fig.~\ref{fig4:Fig3}. The orbits are displayed in different \(2D\) projections: the \((x, y)\), \((x, z)\), and \((y, z)\) planes from left to right. We present a stable PO of each family in a separate panel from top to bottom: x\(1\) [black curve, Fig.~\ref{fig4:Fig3}(a)], thr\(_1\) [blue curve, Fig.~\ref{fig4:Fig3}(b)], thr\(_{z1}\) [red curves, Fig.~\ref{fig4:Fig3}(c)], thr\(_{z1}\)(mul2) [green curves, Fig.~\ref{fig4:Fig3}(d)], and thr\(_{z1}\)(mul4) [magenta curves, Fig.~\ref{fig4:Fig3}(e)]. The POs of the x\(1\), thr\(_1\), and thr\(_{z1}\) families are of multiplicity one, which means these orbits cross the \(y = 0\) axis (with \(p_y > 0\)) only once per orbital period. On the other hand, the PO of the thr\(_{z1}\)(mul2) and the thr\(_{z1}\)(mul4) shown by the green and magenta curves in Fig.~\ref{fig4:Fig3}, respectively, have two and four such crossings per period, which means that their multiplicity is two and four, respectively. 

Next, we will use the information provided by the characteristic diagram in Fig.~\ref{fig4:Fig1}, the stability diagram of Fig.~\ref{fig4:Fig2}, and the morphology of representative stable POs depicted in Fig.~\ref{fig4:Fig3} as a base for our further investigations. These figures serve as a reference in our analysis of the dynamical evolution of phase space structures in the \(4D\) PSS across a series of successive bifurcations. 

\subsection{Pitchfork bifurcations} \label{SubSec:pitchforkCh4}
A pitchfork bifurcation occurs when a family of stable POs becomes unstable, at the same time generating a pair of new stable POs with the same multiplicity as the parent family. In this section, we study in detail the evolution of the phase space structure before and after a \(2D\) and \(3D\) pitchfork bifurcation, which takes place in the Hamiltonian \eqref{eq:BG H}.

\subsubsection{A 2D pitchfork bifurcation}\label{sec:2D pitchfork bif.}
From the results depicted in Fig.~\ref{fig4:Fig2}, we observe that the main planar x\(1\) family (black curves) remains stable for small \(E_j\) values, with both of its stability indices falling in the range of $(-2, 2)$. As illustrated in Fig.~\ref{fig4:Fig3}(a), the morphology of the x\(1\) PO exhibits an elliptical-like shape aligned along the major axis. As the system's energy \(E_j\) increases, the x\(1\) family undergoes a transition to simple instability at a critical point \(E_A = -0.3924\). This instability leads to the creation of two initially stable \(2D\) families of POs: the thr\(_1\) [a representative stable PO is shown in Fig.~\ref{fig4:Fig3}(b)] and its symmetric counterpart thr\(_1\)S. The thr\(_1\) family (along with the thr\(_1\)S) remains stable for energies \(E_A < E_j < E_B = -0.3356\), as indicated by its stability indices (blue curves in Fig.~\ref{fig4:Fig2}), which lie between \(-2\) and \(2\). On the other hand, the x\(1\) family is simple unstable throughout the interval [\(E_A, E_B\)], except for a small energy window (\(-0.374<E_j<-0.359\)) where x\(1\) exhibits double instability.

To illustrate the \(3D\) projection of the DS's \(4D\) phase structure for a quasiperiodic orbit close to the x\(1\) stable PO, we consider a representative case of the DS \eqref{eq:BG H} \(4D\) PSS with \(E_j < E_A\), particularly for \(E_j = -0.41\). Furthermore, this example allows us to effectively demonstrate how the CR method we defined in Sect.~\ref{section:CR Method} can be used to visualize the phase space dynamics and interpret the underlying properties of the system's behavior.

\begin{figure}[!htb]
  \centering
  \subfloat[Perturbation of the x\(1\)(S) PO\label{fig4:Fig4a}] {\includegraphics[width=1\linewidth]{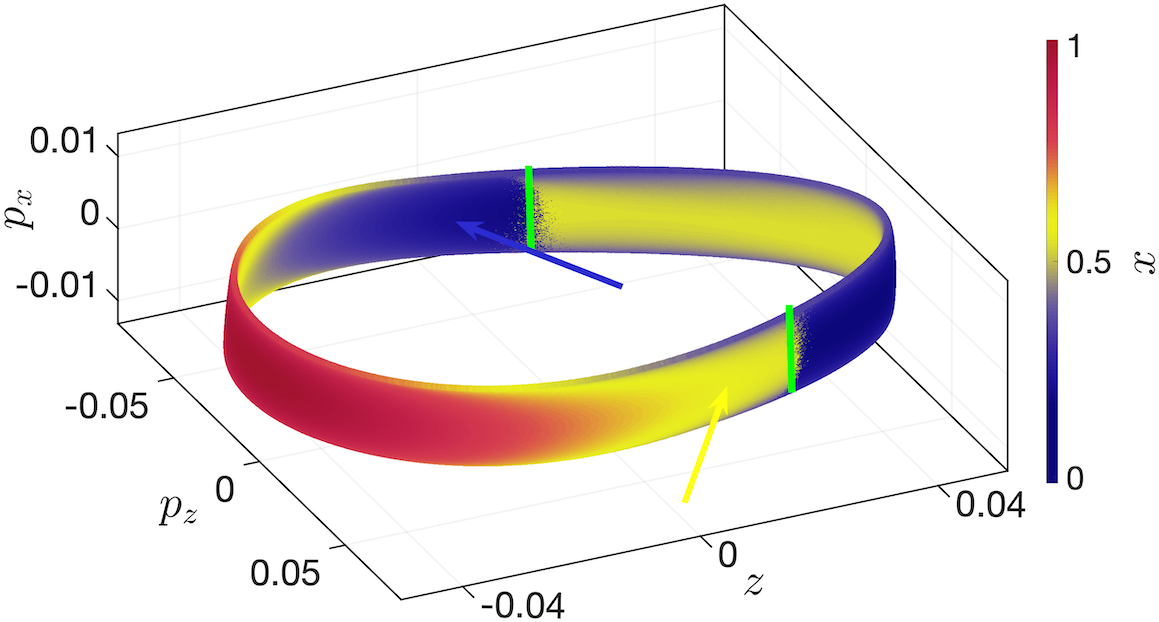}}\\
  \subfloat[Similar to (a) but from a different viewing angle\label{fig4:Fig4b}] {\includegraphics[width=0.55\linewidth]{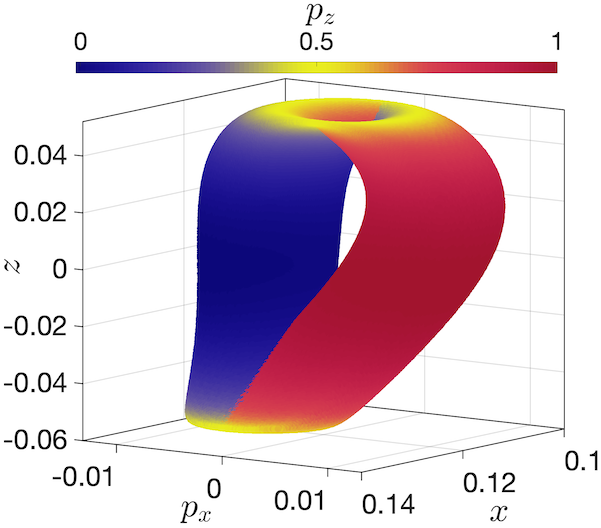}}
  \subfloat[GALI\(_2 (t)\)\label{fig4:Fig4c}] {\includegraphics[width=0.44\linewidth]{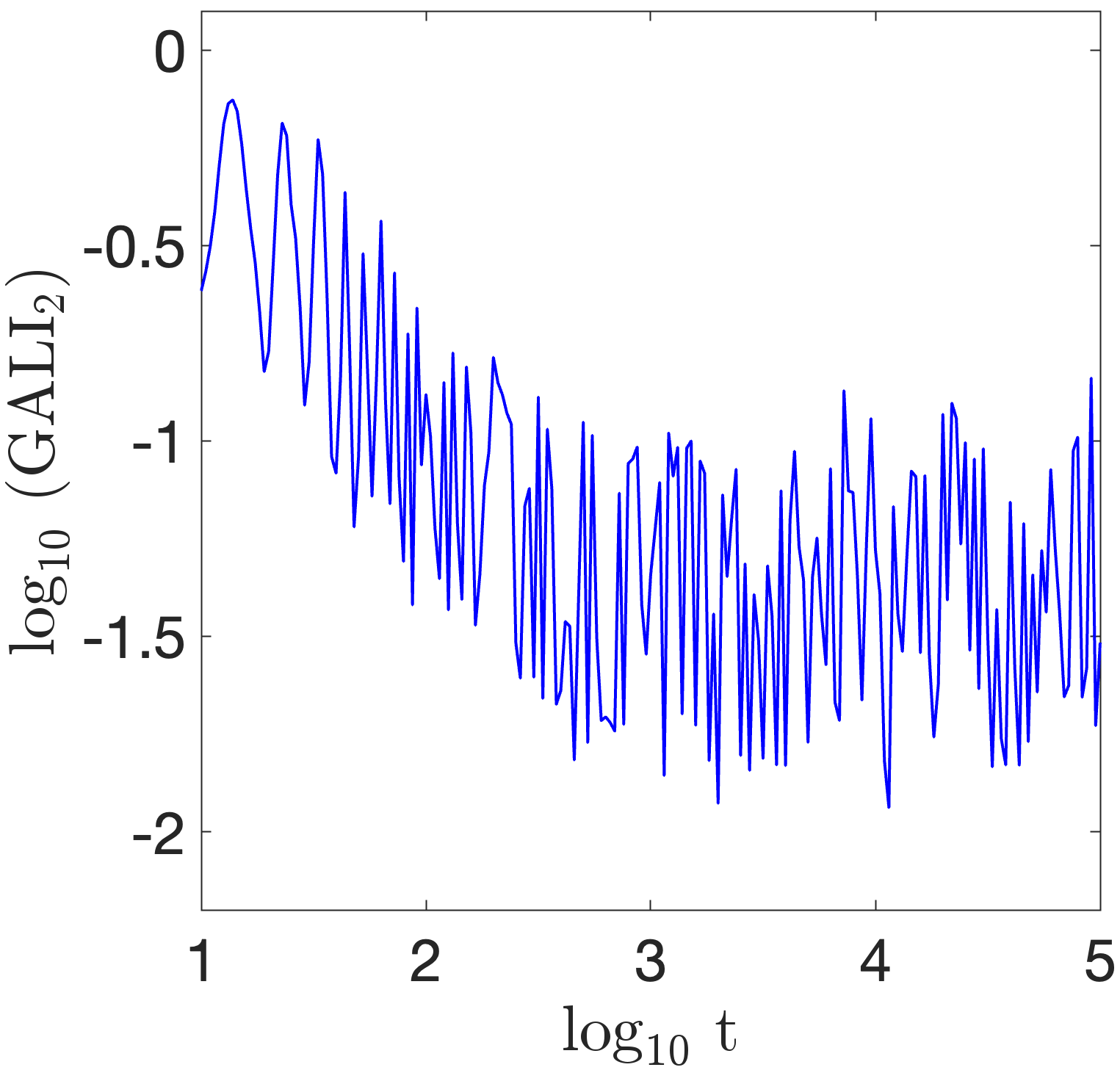}}
  \caption{The \(3D\) projection \((p_x, z, p_z)\) of the system's \eqref{eq:BG H} \(4D\) PSS for a quasiperiodic torus around the stable x\(1\) PO at energy \(E_j = -0.41\) colored according to the values of the fourth coordinate \(x\). The color scale, associated with the \(x\) values, is normalized on the range $[0, 1]$ (see Sect.~\ref{section:CR Method}). The IC for this orbit is obtained by a small perturbation \((\Delta z = 5\times10^{-2})\) of the PO's IC along the $z-$axis, and an appropriate adjustment of the \(p_y > 0\) value. Points on the torus maintain their color as they transition between the interior and exterior surfaces of the torus at regions indicated by the vertical green lines. The yellow and dark blue arrows, described in the text, represent the structure of the torus and  indicate where these surfaces intersect. (b) Similar to panel (a) but for the \(3D\) \((x, p_x, z)\) projection. (c) The time evolution of the GALI$_{2}$ for the same orbit, demonstrating the orbit's regular nature.}
  \label{fig4:Fig4}
\end{figure}

Figure \ref{fig4:Fig4} depicts the \(3D\) colored projection of the system's \(4D\) PSS of a quasiperiodic orbit close to the x\(1\) PO at \(E_j = -0.41\). At this energy, the x\(1\) PO is stable, and hence it is surrounded by invariant tori in the \(4D\) PSS. In particular, the torus depicted in Fig.~\ref{fig4:Fig4} is obtained by a small perturbation of the x\(1\) PO's IC along the \(z\)-axis, specifically by \(\Delta z = 5\times10^{-2}\). In order to make sure that both the PO itself and the perturbation in the \(z\) direction have exactly the same energy, we also adjust the momentum along the \(y\)-axis of the perturbed orbit by considering \(p_y > 0\). The \(3D\) subspace \((p_x, z, p_z)\) projection of the \(4D\) PSS in Fig.~\ref{fig4:Fig4} reveals the presence of an invariant torus around the stable x\(1\) PO. The smooth color variation on the torus itself suggests that orbits obtained from small \(z\) perturbations of the stable x\(1\) PO remain close to this orbit. 

By analyzing the structure of the torus in Fig.~\ref{fig4:Fig4}, we can make the following two key observations: 
\begin{enumerate}
\item The smoothness of the color variation on the torus suggests that the  \((p_x, z, p_z)\) subspace maintains a similar configuration in its fourth dimension (\(x\)). This smooth color variation on the projected torus indicates that the orbit exhibits the characteristic of a regular orbit. The regular nature of the orbit is further confirmed by the computed time evolution of its GALI\(_2\) index \eqref{eq:GALI} in Fig.~\ref{fig4:Fig4c}. As we can see from this figure, the GALI\(_2\) oscillates around a constant positive value, a behavior typically observed for regular orbits of conservative Hamiltonian systems [e.g.~see the blue curve in Fig.~\ref{fig4:Fig_A1b}]. The important characteristic here is the smooth variation of the color on the surface of the form, rather than the specific color pattern itself, which suggests a similar smooth arrangement in the not plotted fourth dimension of the \(3D\) subspace (in this case, \(x\)) of the system's \(4D\) PSS. This feature is evident in Fig.~\ref{fig4:Fig4b} where we plot the \(3D\) projection of the same orbit on the \((x, p_x, z)\) subspace. In a nutshell, taking any possible \(3D\) projection of the system's \eqref{eq:BG H} \(4D\) PSS for an orbit close to the stable x\(1\) PO results in a torus-like structure with smooth color variation in the fourth dimension. 

\item Consider moving anticlockwise on the torus in Fig.~\ref{fig4:Fig4a}, starting from the point indicated by the yellow arrow on the torus's exterior surface, or clockwise starting from the point indicated by the dark blue arrow on the interior surface. By doing so, we see that the color remains consistent as we transition between the interior and exterior surfaces of the torus. In fact, the torus's external and internal surfaces intersect at specific regions, which are represented by vertical green lines in Fig.~\ref{fig4:Fig4a}. These smooth color transitions observed on the \(3D\) projected torus have already been discussed in detail previously (e.g.~see \citep{katsanikas2011structure1}). 
\end{enumerate}

\begin{figure}[!htb]
  \centering
  \subfloat[\label{fig4:Fig5a}] {\includegraphics[width=0.45\linewidth]{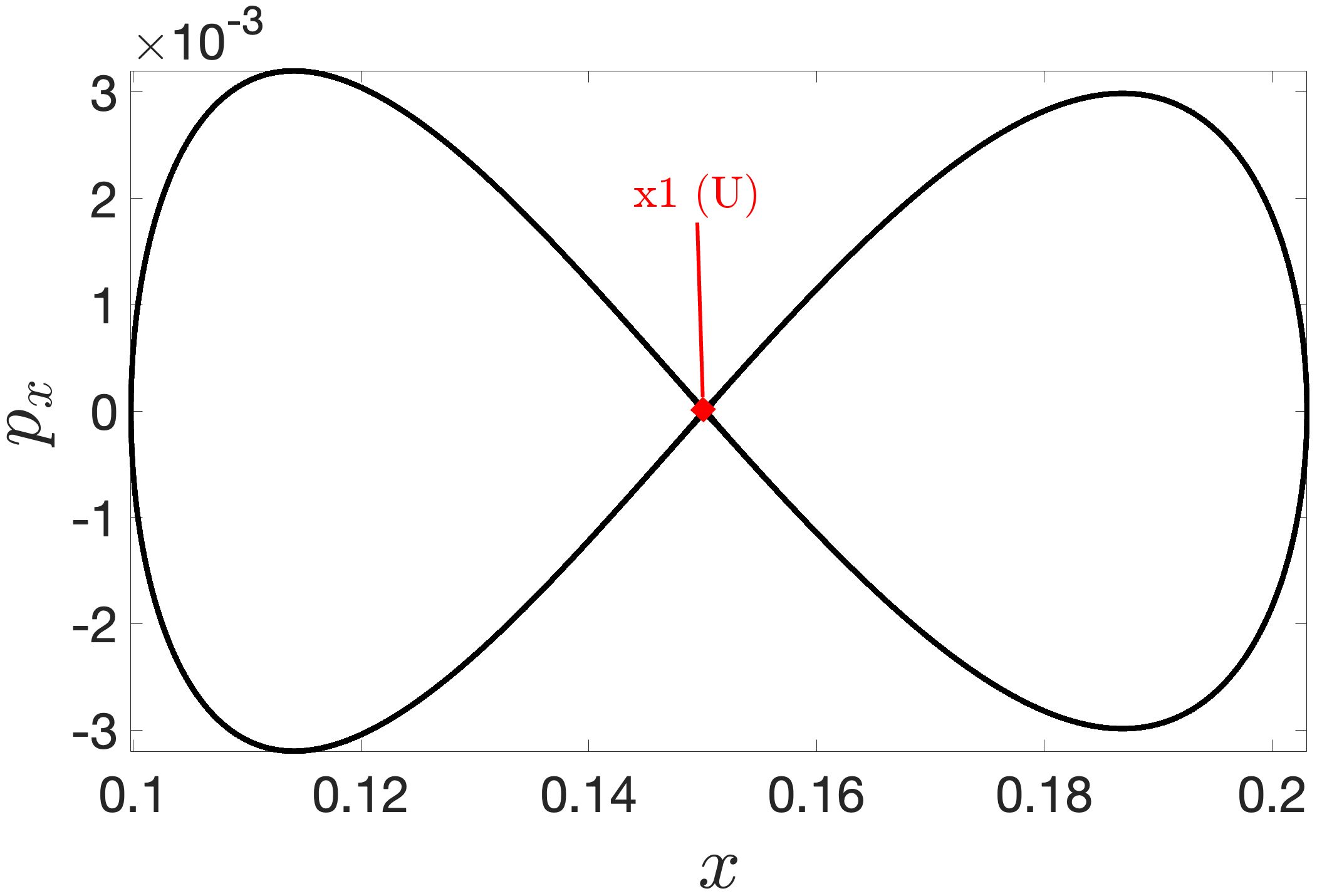}}
  \subfloat[\label{fig4:Fig5b}] {\includegraphics[width=0.55\linewidth]{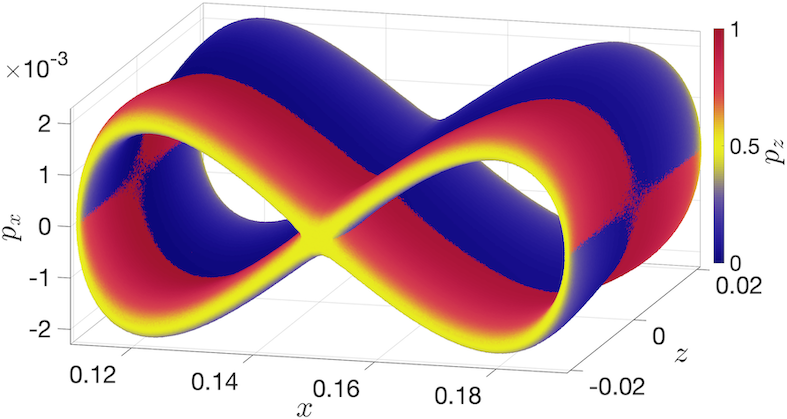}}
  \caption{The (a)  \(2D\) \((x , p_x)\) PSS and (b) \(3D\) colored projection \((p_x, z, p_z)\) of the system's \eqref{eq:BG H} \(4D\) PSS for orbits created through perturbations of the unstable x\(1\) PO along the (a) \(x\) by \(\Delta x = 10^{-3}\) and (b) \(z\) coordinates by \(\Delta z = 2 \times 10^{-2}\) for \(E_j = -0.3919\) (this energy value is between the critical energies \(E_A \) and \(E_B\)). The red arrow in (a) indicates the IC of the unstable x\(1\) PO, which is marked by a red point.}
  \label{fig4:Fig5}
\end{figure}

Slightly increasing \(E_j\) beyond \(E_A\) results in the transition of the x\(1\) PO family to simple instability, while the newly bifurcated planar POs of the thr\(_1\) family (and its symmetric counterpart thr\(_1\)S) are stable. The instability in the x\(1\) family is radial, as indicated by the \(b_2\) index (blue curve in Fig.~\ref{fig4:Fig2}) crossing below the critical stability threshold \(b = -2\). Thus, applying radial perturbations to the unstable x\(1\) POs by changing their \(x\) coordinates, we get planar chaotic orbits creating a ``Figure-8" structure on the system's \(4D\) PSS. Fig.~\ref{fig4:Fig5a} illustrates the planar  \((x, p_x)\) projection of the system's \eqref{eq:BG H} \(4D\) PSS (i.e., \(z = p_z = 0\)) for the orbit created by perturbing the IC of the simple unstable x\(1\) PO at \(E_j = -0.3919\) by \(\Delta x = 10^{-3}\). The consequents of this orbit form a Figure-8 structure in the \((x, p_x)\) plane. The unstable x\(1\) PO (indicated by the red arrow) is located at the center of the Figure-8 structure. This configuration is a typical pattern observed in \(2D\) PSSs for radial perturbations of the unstable x\(1\) POs (e.g.~see \citep[Fig.~2.45]{contopoulos2002order}). It is worth mentioning that for the \(3D\) GM \eqref{eq:BG H}, we observe that the \(2D\)  projection of the system's \(4D\) PSS for the simple unstable x\(1\) POs retains the Figure-8 morphology depicted in Fig.~\ref{fig4:Fig5a}, even when we apply small radial changes (such as \(\Delta x = 10^{-5}\) and  \(\Delta x = 10^{-8}\)). 

On the other hand, perturbing the stable planar x\(1\) POs along the \(z\) coordinate introduces additional dimensions to the space visited by the orbit. Fig.~\ref{fig4:Fig5b} illustrates the \((p_x, z, p_z)\) projection of the system's \(4D\) PSS for a particular perturbation of \(\Delta z = 2 \times 10^{-2}\) at \(E_j = -0.3919\) of the unstable x\(1\) POs. In general, when we perturb the unstable x\(1\) PO by \(\Delta z \le 2 \times 10^{-2}\), the \((x, z, p_z)\) consequents do not diffuse in the \(z\)-direction. Instead, these consequents form a ribbon-like structure in the \(4D\) PSS. This is indicated by the smooth color variation of the fourth coordinate (\(x\)) of the \(3D\) projection in Fig.~\ref{fig4:Fig5b}. Similar to the torus of the quasiperiodic orbit in Fig.~\ref{fig4:Fig4}, we observe characteristic color transitions from the interior to the exterior sides of this structure. Furthermore, one can observe that the \((x, p_x)\) projection in Fig.~\ref{fig4:Fig5b} still maintains the Figure-8 structure that we observed for the same unstable \(x1\) PO perturbation by \(\delta x = 10^{-3}\) in Fig.~\ref{fig4:Fig5a}. To the best of our knowledge, this is the first reported analysis of the Figure-8 structure associated with vertical perturbations of a radially unstable (i.e., \( b_2  \notin [-2, 2] \)) and vertically stable [i.e., \( b_1  \in (-2, 2) \)] PO of the \(3D\) Hamiltonian system \eqref{eq:BG H}. While our study in this chapter primarily focuses on the \(3D\) phase space structures around POs that extend beyond the x\(1\) family, the results provided in Fig.~\ref{fig4:Fig5} may serve as a starting point for future studies in this area. For practical considerations about studying  \(3D\) orbits near simple unstable POs that maintain vertical stability, the reader is referred to Appendix A of \citep{moges2024evolution}. 

In Fig.~\ref{fig4:Fig52a}, we present the evolution of the computed GALI\(_2\) for the orbits corresponding to Figs.~\ref{fig4:Fig5a} and (b). For the orbit created by perturbing the IC of the simple unstable x\(1\) PO at \(E_j = -0.3919\) by \(\Delta x = 10^{-3}\), the GALI\(_2\) (red curve) eventually decays to zero exponentially fast but at a slower rate (around \( t \approx 5\times 10^5 \)) compared to, for example, \(2D\) system of \eqref{eq:BG H} at \(E_j=0.41\) [red curve in Fig.~\ref{fig4:Fig_A1b}], indicating weakly chaotic behavior. On the other hand, for the same unstable \(x1\) PO perturbation by \(\Delta z \le 2 \times 10^{-2}\) (blue curve), the GALI\(_2\) oscillates around a constant positive value, characteristic of regular motion.

\begin{figure}[!htb]
  \centering
  \includegraphics[width=0.8\textwidth]{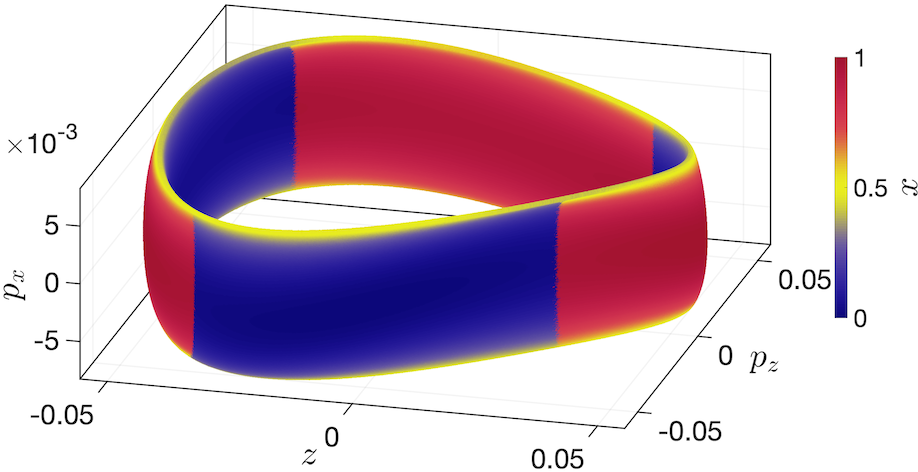}
  \caption{The \(3D\) colored projection \((p_x, z, p_z)\) of the equation of the Hamiltonian \eqref{eq:BG H} system's \(4D\) PSS, which is created by perturbing the stable thr\(_{1}\) PO by  \(\Delta z = 5\times 10^{-2}\) at \(E_j = -0.3919\) as in Fig.~\ref{fig4:Fig5}.}
  \label{fig4:Fig6}
\end{figure}

\begin{figure}[!htb]
  \centering
  \subfloat[\label{fig4:Fig52a}] {\includegraphics[width=0.49\linewidth]{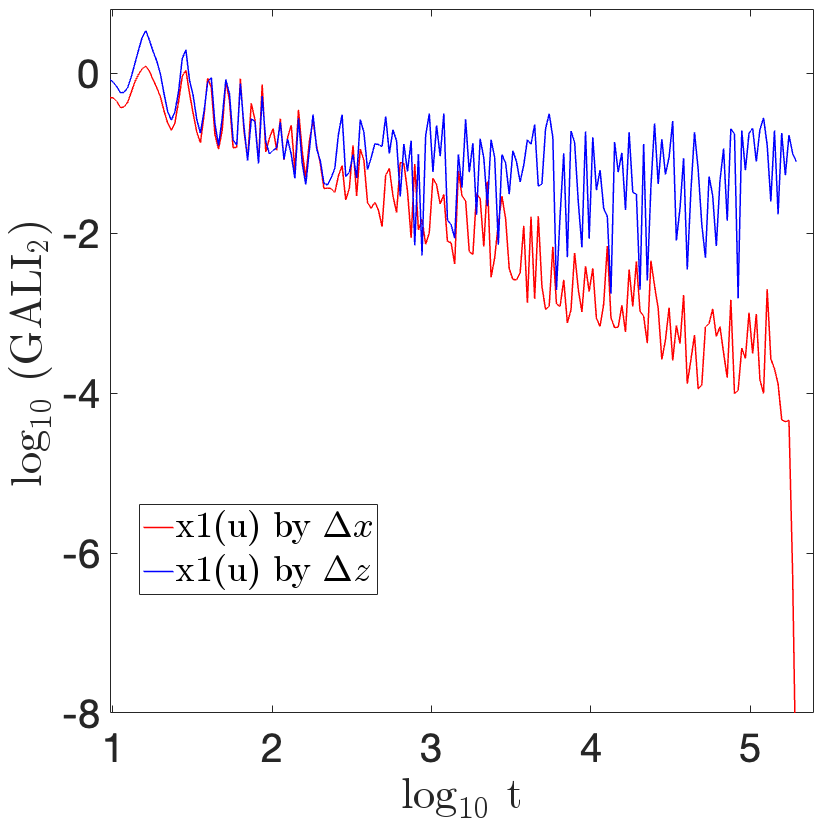}}
  \subfloat[\label{fig4:Fig52b}] {\includegraphics[width=0.49\linewidth]{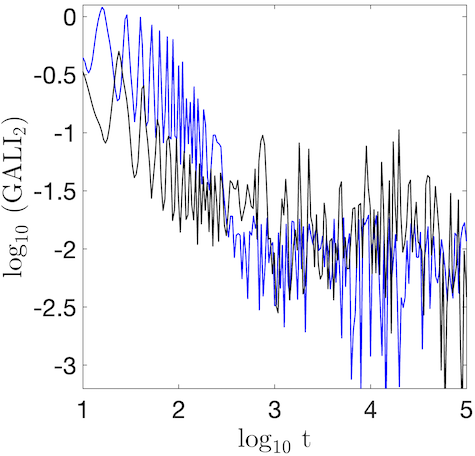}}
  \caption{(a) The time evolution of the GALI\(_2\) for the orbits corresponding to Fig.~\ref{fig4:Fig5}. Specifically, the red and blue curves represent the orbits shown in Fig.~\ref{fig4:Fig5a} and Fig.~\ref{fig4:Fig5b}, respectively. (b) The time evolution of  the GALI\(_2\) for the orbits corresponding to the thr\(_1\) (blue curve) [the perturbed stable thr\(_{1}\) PO in Fig.~\ref{fig4:Fig6}] and its symmetric counterpart thr\(_1\)S (black curve) POs by  \(\Delta z = 5\times 10^{-2}\).}
  \label{fig4:Fig52}
\end{figure}

Furthermore, perturbing the newly bifurcated stable planar thr\(_1\) PO in the \(z\)-direction creates an invariant torus around the PO (Fig.~\ref{fig4:Fig6}). This torus is again characterized by a smooth color variation, similar to what was shown in Fig.~\ref{fig4:Fig4}. We note that according to the evolution of their GALI\(_2\) depicted in Fig.~\ref{fig4:Fig52b}, the motion is regular, where the blue and black curves correspond to the stable thr\(_1\) and thr\(_1\)S POs, respectively, with the same perturbation \(\Delta z = 5\times 10^{-2}\). In both cases, the GALI\(_2\) remains practically constant. As expected, the dynamical behavior for orbits near the stable thr\(_1\) and its symmetric counterpart thr\(_1\)S is virtually the same (i.e., the blue and black curves in Fig.~\ref{fig4:Fig52b} converge to practically the same value). 

In Fig.~\ref{fig4:Fig6}, we can also observe distinct color transitions (in this case, between dark blue and red regions) moving from the exterior to the interior surface of the torus and vice versa, which occur along regions where the darker blue and red areas intersect. Similar tori to those around the stable thr\(_1\) POs presented in Fig.~\ref{fig4:Fig6} can also be formed around the stable thr\(_1\)S POs. Both of these tori are located within the left and right lobes of the Figure-8 structure illustrated in Fig.~\ref{fig4:Fig5b}. We will explore these behaviors in more detail in the following subsections. 

\subsubsection{A 3D pitchfork bifurcation} \label{sec:3D pitchfork bif.}
From the results of Fig.~\ref{fig4:Fig2}, we observe that as the energy level \(E_j\) of the system \eqref{eq:BG H} further increases, the stability type of the newly formed thr\(_1\)/thr\(_1\)S families, which emerge from the main x\(1\) bifurcation, also changes. At the critical energy level \(E_B = -0.3356\), the initially stable planar thr\(_1\) family transitions to simple instability (blue curves in Fig.~\ref{fig4:Fig1}). This transition creates a pitchfork bifurcation, resulting in two new \(3D\) families of POs: the thr\(_{z1}\) family and its symmetric counterpart, thr\(_{z1}\)S (red curves in Fig.~\ref{fig4:Fig1}). For now, let us just focus our attention on thr\(_{z1}\), as its symmetric counterpart family thr\(_{z1}\)S exhibits a similar dynamical behavior. The morphology of a representative stable PO belonging to the thr\(_{z1}\) family at \(E_j = -0.3306\) is shown in Fig.~\ref{fig4:Fig3} (red curves). The orbit's $(x, y)$ projection shows a crescent-like shaped structure, which is similar to the structure observed for the representative stable \(2D\) thr\(_{1}\) PO in Fig.~\ref{fig4:Fig3}(b). The thr\(_{z1}\) family becomes simple unstable at \(E_C = -0.3203\) and retains this instability over a range of energy values (see red curves in Fig.~\ref{fig4:Fig2}). From Fig.~\ref{fig4:Fig2} we observe that the vertical stability index \(b\) of the thr\(_1\) family (blue curves) crosses the \(b = -2\) line at \(E_j = E_B\), indicating that thr\(_1\) becomes vertically unstable beyond \(E_B\). On the other hand, the radial stability remains within the stable range (\(-2 < b_2 < 2\)) on the energy interval \((E_B, E_C)\).

\begin{figure}[!htb]
  \centering
  \subfloat[Perturbations of the thr\(_1\) (U), thr\(_{z1}\) (S) and thr\(_{z1}\)S (S) POs\label{fig4:Fig7a}]{\includegraphics[width=1\textwidth]{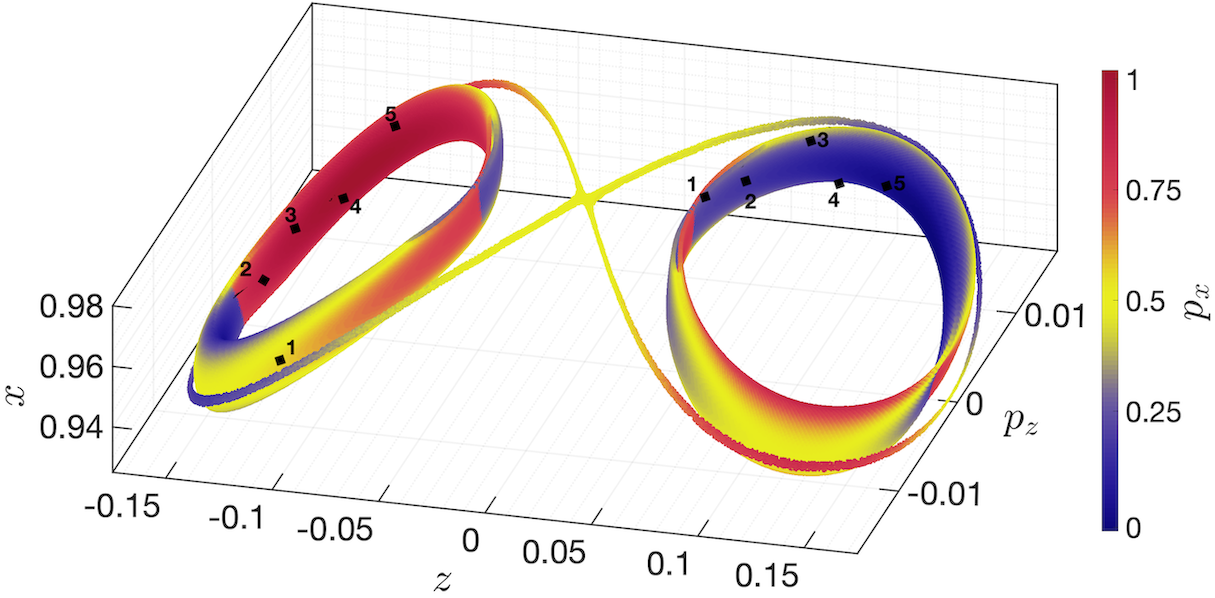}} \\
  \subfloat[Right torus in (a)\label{fig4:Fig7b}] {\includegraphics[width=0.47\linewidth]{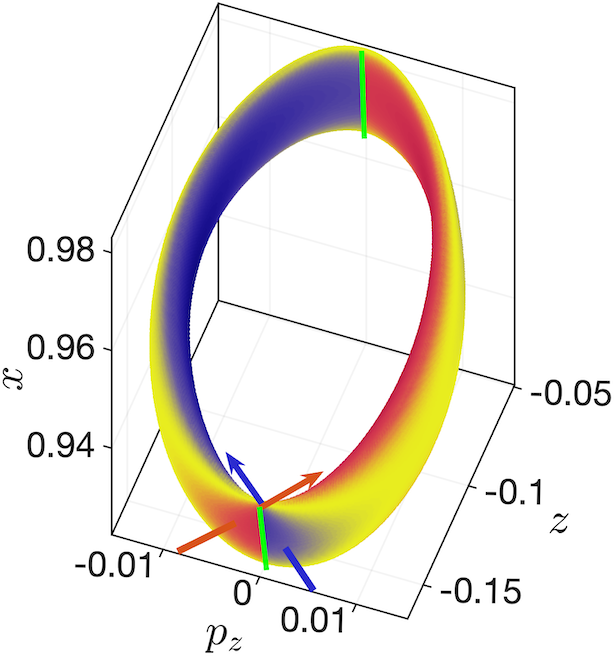}}
  \subfloat[Similar to (b) but from a different viewing angle\label{fig4:Fig7c}] {\includegraphics[width=0.53\linewidth]{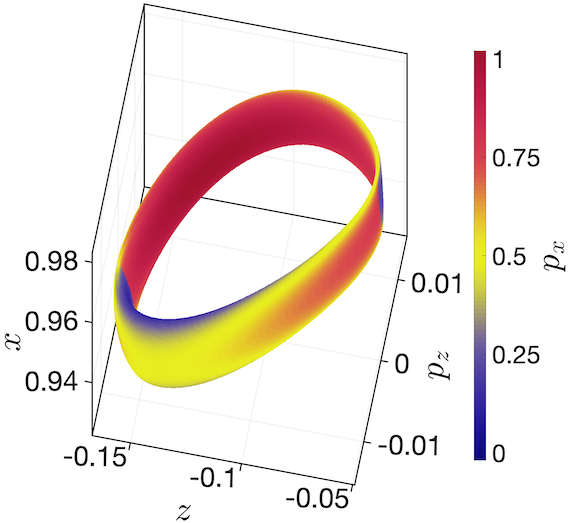}}
  \caption{(a) The \(3D\) colored \((x, z, p_z)\) projection of the system's \(4D\) PSS at \(E_j = -0.3307\) for three orbits, where color represents the \(p_x\) value. A small \(z\)-direction perturbation \((\Delta z = 10^{-5})\) applied to the unstable \(2D\) thr\(_1\) PO results in the Figure-8 structure, while perturbation of the stable \(3D\) POs thr\(_{z1}\) and thr\(_{z1}\)S with \(\Delta z = 5 \times 10^{-2}\) generates two invariant tori, with the right torus corresponding to the thr\(_{z1}\) orbit and the left to the thr\(_{z1}\)S. (b) The right torus in (a) projected in the \(3D\) \((p_z, z, x)\) subspace. (c) Similar to panel (b) but for the \(3D\) \((z, p_z, x)\) projection. The black diamond points in (a) represent the first five consequents of orbits starting from a specific point (marked as ``1") on each torus. In panel (b), points on each torus maintain their color as they transition between the interior and exterior surfaces of the torus at regions indicated by the green lines. The blue and red arrows, described in the text, represent the structure of the torus and indicate where these surfaces intersect. A similar observation can be made for panel (c).}
  \label{fig4:Fig7}
\end{figure}

To understand the phase space structures for energies just above \(E_B\) (where the thr\(_1\) family is unstable and the thr\(_{z1}\) and thr\(_{z1}\)S ones are stable), we present in Fig.~\ref{fig4:Fig7a} the \(3D\) \((x, z, p_z)\) projection of the system's \(4D\) PSS for orbits obtained by small perturbations along the \(z\) axis of the unstable thr\(_1\) PO, as well as the stable thr\(_{z1}\), and thr\(_{z1}\)S POs at \(E_j = -0.3307\). Each \((x, z, p_z)\) consequent is colored based on its \(p_x\) value.

\begin{figure}[!htb]
  \centering
  \includegraphics[width=0.6\textwidth]{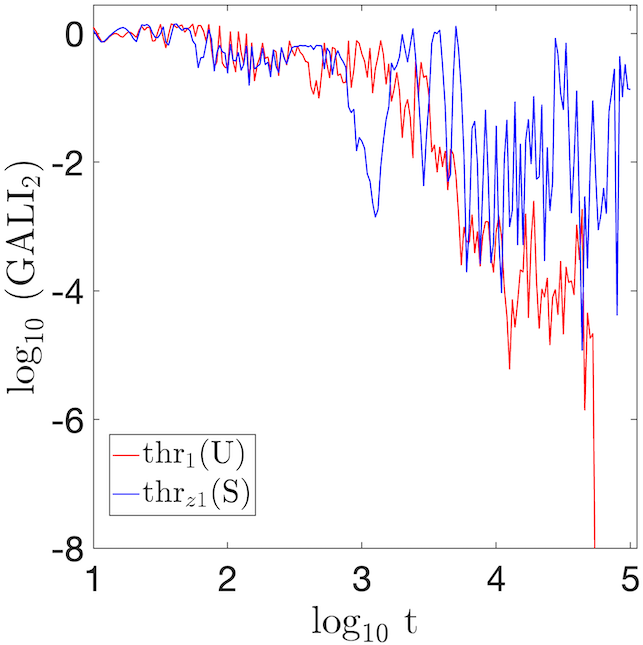}
  \caption{The time evolution of the GALI\(_2\) for the orbits corresponding to Fig.~\ref{fig4:Fig7}. In particular the red and blue curves represent the perturbations of the thr\(_1\) (U) and  thr\(_{z1}\) (S) POs, respectively.}
  \label{fig4:Fig72}
\end{figure}

From the findings of Fig.~\ref{fig4:Fig7}, we observe that applying a small \(z\) perturbation (\(\Delta z = 10^{-5}\)) to the ICs of the unstable \(2D\) thr\(_1\) PO results in the creation of a thin Figure-8 structure in the (\(x, z, p_z\)) projection shown in Fig.~\ref{fig4:Fig7a}. Due to the instability of the thr\(_1\) PO at this specific energy value, even a very small perturbation in the \(z\)-direction causes the orbit to diverge from the PO's location in the phase space, where the IC of this orbit is located at the intersection of the two halves of the Figure-8 structure. Nonetheless, there is a smooth color variation across this Figure-8 structure [Fig.~\ref{fig4:Fig7a}], which suggests the presence of similar Figure-8 structures with smooth color variations in all possible \(3D\) projections of the system's \(4D\) PSS. Such structures have also been observed in previous works (e.g.~see \citep{patsis1994using,katsanikas2013instabilities}). The time evolution of the GALI\(_2\) for this orbit is depicted in Fig.~\ref{fig4:Fig72} (red curve), with the characteristic exponential decrease confirming its chaotic nature.   

Perturbing the stable thr\(_{z1}\) and thr\(_{z1}\) POs in the \(z\)-direction by \(\Delta z = 5 \times 10^{-2}\) leads to quasiperiodic motion observed on the two invariant tori inside the Figure-8 lobes [Fig.~\ref{fig4:Fig7a}]. The right torus corresponds to the perturbations of the stable thr\(_{z1}\) PO, while the left torus relates to an orbit in the vicinity of the stable thr\(_{z1}\)S orbit. Since both newly formed families (i.e., thr\(_{z1}\) and thr\(_{z1}\)S) are of multiplicity one (similar to their parent thr\(_1\) family), the two tori remain independent. This means that the orbits on the left torus do not visit the right torus (and vice versa), as there is a distinct separation between the two sets of quasiperiodic orbits. The first five consequents of each quasiperiodic orbit are denoted by black diamond symbols numbered `1' to `5'  in Fig.~\ref{fig4:Fig7a}. The configuration of these tori resembles what was observed for the quasiperiodic orbits around the stable POs of \(x1\) (Fig.~\ref{fig4:Fig4}) and thr\(_1\) (Fig.~\ref{fig4:Fig6}) families. These orbits are regular, as shown by the evolution of the GALI\(_2\) for the perturbations of the stable thr\(_{z1}\) PO case (represented by the blue curve in Fig.~\ref{fig4:Fig72}) which remain practically constant. While the perturbations of the three POs orbits (\(x1\), thr\(_1\) and thr\(_1\)S) of Fig.~\ref{fig4:Fig7} appear different in different \(3D\) projections, they maintain the Figure-8 structure for the x\(1\) perturbed orbit and exhibit toroidal structures for the perturbation of the thr\(_1\) and thr\(_1\)S.

We note that the color transition from the exterior to the interior surface of a torus, observed for perturbations of the planar POs (as seen in Fig.~\ref{fig4:Fig4}), is also observed for perturbations of the stable \(3D\) POs. In Fig.~\ref{fig4:Fig7b}, we present the \((p_z, z, x)\) projection of the system's \(4D\) PSS for a quasiperiodic orbit close to the thr\(_{z1}\) PO colored according to the corresponding \(p_x\) values [this is actually the right torus in Fig.~\ref{fig4:Fig7a}]. We observe color transitions when moving from the exterior to the interior surface of the torus [following the directions indicated by the red and blue arrows in Fig.~\ref{fig4:Fig7b}], and vice versa. These transitions occur in the regions where the blue and red areas intersect. These regions are indicated by vertical green lines in Fig.~\ref{fig4:Fig7b}. This color transition is also clearly visible in Fig.~\ref{fig4:Fig7c}, where the structure of Fig.~\ref{fig4:Fig7b} is seen from a different angle. 

\begin{figure}[!htb]
  \centering
  \includegraphics[width=0.8\textwidth]{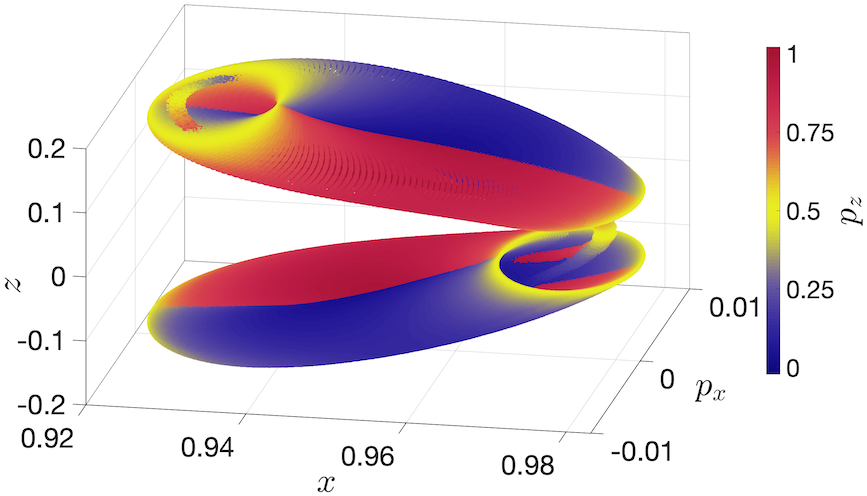}
  \caption{Similar to Fig.~\ref{fig4:Fig7a}, but for the \((x, p_x, z)\) \(3D\) projection when points are colored according to their \( p_z\) value.}
  \label{fig4:Fig8}
\end{figure}

\begin{figure}[!htb]
  \centering
  \subfloat[Perturbation of the thr\(_1\) PO (U) integrated up to \(t = 10^4\)\label{fig4:Fig9a}]{\includegraphics[width=0.8\textwidth]{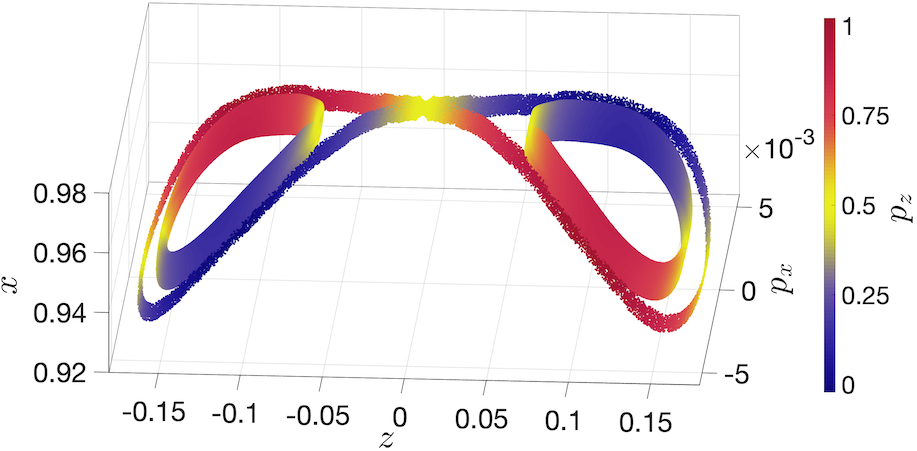}}\\ 
  \subfloat[Perturbation of the thr\(_1\) PO (U) integrated up to \(t = 10^5\)\label{fig4:Fig9b}] {\includegraphics[width=0.8\linewidth]{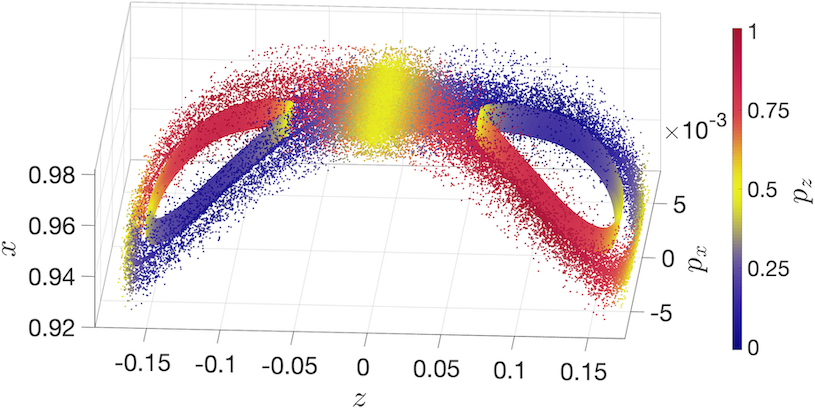}}
  \caption{Similar to Fig.~\ref{fig4:Fig7a} but for perturbations of the unstable thr\(_1\) PO by \(\Delta z = 10^{-5}\), and the stable thr\(_{z1}\) and thr\(_{z1}\)S POs by \(\Delta p_x = 2.5 \times 10^{-3}\) and \(\Delta p_z = 1.2 \times 10^{-2}\), respectively. The perturbation of the unstable thr\(_1\) PO is integrated up to (a) \(t = 10^4\) and (b) \(t = 10^5\) time units, while the tori within the Figure-8 lobes are generated by integrating the perturbations of the stable thr\(_{z1}\) and thr\(_{z1}\)S POs up to  \(t = 10^6\) time units.}
  \label{fig4:Fig9}
\end{figure}

Let us now briefly examine what happens when we view the Figure-8 structure of Fig.~\ref{fig4:Fig7a} in a different \(3D\) projection. In Fig.~\ref{fig4:Fig8}, we present the same orbits as in Fig.~\ref{fig4:Fig7a}, but in the \((x, p_x, z)\) \(3D\) projection, where the \(p_z\) coordinate determines the color of each consequent. This \(3D\) projection of the system's \(4D\) PSS reveals that the tori of the quasiperiodic orbits surrounding the newly formed \(3D\) families, thr\(_{z1}\) and thr\(_{z1}\)S, form an eight-shaped ribbon-like structure, with the unstable thr\(_1\) orbit located at the intersection of its lobes. Thus, we see that the specific arrangement of the tori and Figure-8 structures heavily depends on how we select the \(3D\) projection of the system's \(4D\) PSS and the type of perturbations applied to the POs. In Figs.~\ref{fig4:Fig7a} and \ref{fig4:Fig8}, we applied the following perturbations: \(\Delta z = 10^{-5}\) to the unstable thr\(_1\) PO, which resulted in the creation of Figure-8 structures, and \(\Delta z = 5 \times 10^{-2}\) for the stable thr\(_{z1}\) and thr\(_{z1}\)S POs. Now, let us explore different perturbations to observe how the phase space structure changes. Fig.~\ref{fig4:Fig9} depicts the \(3D\) projection of the system's \(4D\) PSS under different perturbations, namely \(\Delta z = 10^{-5}\) for the unstable thr\(_1\) PO and \(\Delta p_x = 2.5 \times 10^{-3}\) and \(\Delta p_z = 1.2 \times 10^{-2}\) for its stable \(3D\) bifurcations, thr\(_{z1}\) and thr\(_{z1}\)S, respectively. The consequents of these orbits form the 8-shaped structure and the two tori when the orbits are integrated up to \(t \approx 10^4\) time units [Fig.~\ref{fig4:Fig9}]. For larger integration times, the Figure-8 structure begins to diffuse in the \(4D\) PSS, occupying a larger phase space volume [see Fig.~\ref{fig4:Fig9b}, where the orbits were integrated up to \(t = 10^5\) time units]. 

The Figure-8 structures are clearly visible when we apply any \(\Delta z\) perturbations to the planar simple unstable thr\(_1\) PO family [Figs.~\ref{fig4:Fig7a}, \ref{fig4:Fig8} and \ref{fig4:Fig9}]. On the other hand, when small radial perturbations (such as \(\Delta x\) or \(\Delta p_x\)) are applied to the ICs of this PO, we observe that the resulting orbits remain planar. These planar orbits then form closed curves around the simple unstable PO in the \((x, p_x)\) projection of the system's \(4D\) PSS, resembling what has been observed for the \(x_1\) family in Fig.~\ref{fig4:Fig5a}  (see also \citep{patsis2014phasea} for more details).

\begin{figure}[!htb]
  \centering
  \includegraphics[width=1\textwidth]{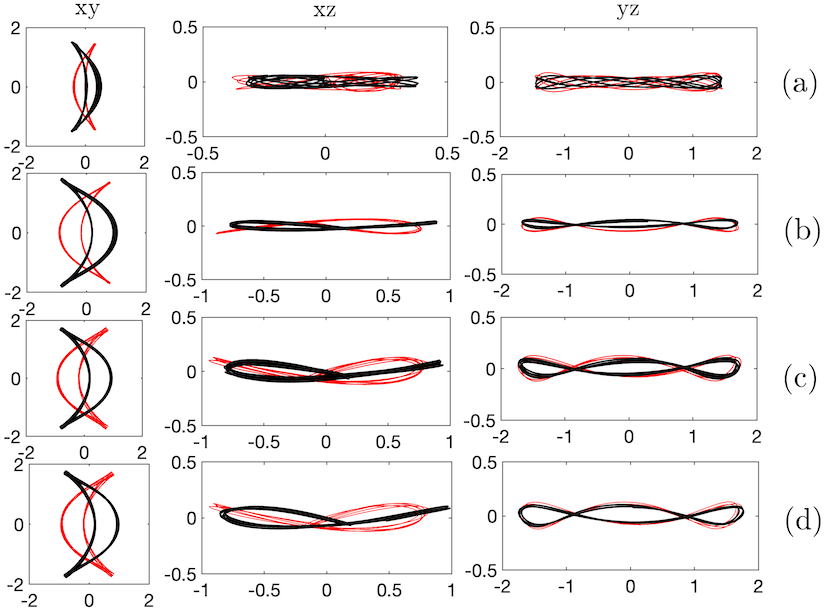}
  \caption{The morphology of representative quasiperiodic orbits around the \(2D\) thr\(_1\) and \(3D\) thr\(_{z1}\) POs obtained by a \(\Delta z = 5 \times 10^{-2}\) perturbation of the PO's ICs (black curves). The left, middle, and right columns show projections in the \((x, y)\), \((x, z)\), and \((y, z)\) planes. The orbits are created by perturbing the thr\(_1\) [thr\(_{z1}\)] PO at two different energy levels: (a) \(E_j = -0.38\) and (b) \(E_j = -0.338\) [(c) \(E_j = -0.3346\) and (d) \(E_j = -0.33\)]. Red curves represent the morphologies corresponding to the symmetric counterparts of each orbit.}
  \label{fig4:Fig10}
\end{figure}

Let us now shift our focus to the system's \eqref{eq:BG H} configuration space \((x, y, z)\) and examine the morphologies given by quasiperiodic orbits around the thr\(_1\) and thr\(_{z1}\) POs near the bifurcation point \(E_B=-0.3356\) (see Figs.~\ref{fig4:Fig1} and \ref{fig4:Fig2}). Fig.~\ref{fig4:Fig10} illustrates the \((x, y)\), \((x, z)\), and \((y, z)\) projections for perturbations of both the thr\(_{z1}\) and thr\(_{z1}\)S POs (black curves) and their corresponding symmetric counterparts thr\(_1\)S and thr\(_{z1}\)S (red curves). For simplicity, we will limit our analysis to the thr\(_1\) and thr\(_{z1}\) orbits, as there are no apparent morphological differences between these orbits and their symmetric counterparts. 

When \(E_j < E_B\), the \(2D\) thr\(_1\) family is stable while the \(3D\) thr\(_{z1}\) family is not yet to be created. Figs~\ref{fig4:Fig10}(a) and (b) show the \(2D\) \((x, y)\), \((x, z)\), and \((y, z)\) projections of the orbit obtained by perturbing the planar thr\(_1\) PO along the \(z\)-axis by \(\Delta z = 5 \times 10^{-2}\) for \(E_j = -0.38\) and \(E_j = -0.338\), respectively. On the other hand, when \(E_j > E_B\), the \(3D\) thr\(_{z1}\) and thr\(_{z1}\)S families are stable, unlike the \(2D\) families thr\(_{z1}\) and thr\(_{z1}\)S, which become unstable (see Fig.~\ref{fig4:Fig1}). We have already shown all possible \(2D\) projections of the \(3D\) thr\(_{z1}\) PO for \(E_j = -0.3306\) in Fig.~\ref{fig4:Fig3}(c). Now, in Figs~\ref{fig4:Fig10}(c) and (d), we present quasiperiodic orbits that result from a \(\Delta z = 5 \times 10^{-2}\) perturbation of the thr\(_{z1}\) POs (black curves) for \(E_j = -0.3346\) and \(E_j = -0.33\), respectively.

Comparing the \(3D\) quasiperiodic orbits near the stable \(2D\) thr\(_1\) PO \(E_j \lesssim  E_B\) in Figs~\ref{fig4:Fig10}(a) and (b) with the \(3D\) thr\(_{z1}\) PO (\(E_j \gtrsim E_B\)) [Fig.~\ref{fig4:Fig3}(c)] and its corresponding perturbations for \(E_j = -0.3346\) [Fig.~\ref{fig4:Fig10}(c)] and slightly larger energy value, \(E_j = -0.33\) [Fig.~\ref{fig4:Fig10}(d)], we observe an interesting feature regarding the system's morphologies near the \(3D\) bifurcation point. The \((x, y)\) projections (left columns in Fig.~\ref{fig4:Fig10}) of all these orbits exhibit similar morphologies resembling a crescent-like shaped structure when both orbits from the symmetric families are considered. However, the configuration changes when examining the other two projections on the \((x, z)\) and \((y, z)\) planes (middle and right columns in Fig.~\ref{fig4:Fig10}). The perturbed thr\(_1\) orbit for energies farther from the \(3D\) bifurcation point (\( E_j \ll E_B\)) [Fig.~\ref{fig4:Fig10}(a)] differs significantly from both the \(3D\) thr\(_{z1}\) PO [Fig.~\ref{fig4:Fig3}(c)] and its nearby quasiperiodic orbits [Figs.~\ref{fig4:Fig10}(c) and (d)]. This contrast emphasizes the distinct behaviors of the system's morphologies as the energy crosses the bifurcation value \(E_B\), particularly in the \((x, z)\) and \((y, z)\) projections. Furthermore, a noteworthy observation is that for energies approaching, but still below, the bifurcation point (\(E_j < E_B\)), the \((x, z)\) and \((y, z)\) projections of the \(3D\) perturbed \(2D\) stable thr\(_1\) PO [Fig.~\ref{fig4:Fig10}(b)] closely resemble those of the thr\(_{z1}\) PO [Fig.~\ref{fig4:Fig3}(c)] and its perturbed orbits [Figs~\ref{fig4:Fig10}(c) and (d)]. This similarity is present despite the absence of the thr\(_{z1}\) family for energies below \(E_B\). This observation emphasizes a significant finding: the configuration space morphology of \(3D\) perturbations applied to the \(2D\) stable thr\(_1\) PO just before the bifurcation (\(E_j < E_B\)) foreshadows the shape of the \(3D\) thr\(_{z1}\) PO that will emerge at the bifurcation point (\(E_j = E_B\)). In essence, the system's \(2D\) quasiperiodic orbits gradually evolve towards the morphology of the \(3D\) bifurcating family. A similar behavior, but for different orbits, has been previously reported in \citep{patsis2014phasea}.

\subsection{3D period-doubling bifurcations} \label{sec:Period-doubling}

In this section, we consider \(3D\) period-doubling bifurcations leading to multiplicity-two and multiplicity-four orbits. We examine the evolution of the phase space structures before and after these bifurcations, as well as the bifurcations of the parent families.

\subsubsection*{A. A period-doubling bifurcation leading to a multiplicity two orbit}
As we increase the energy \(E_j\) of the system, the \(3D\) thr\(_{z1}\) family of POs transitions from stable to simple unstable at the critical point \( E_C = -0.3203 \). This change occurs when its radial (\( b_2 \)) stability index goes above the stability threshold \(b = 2\) (red curve in Fig.~\ref{fig4:Fig2}). As a result, a period-doubling bifurcation occurs, which leads to the creation of the \(3D\) thr\(_{z1}\)(mul2) family and its symmetric counterpart thr\(_{z1}\)(mul2)S. A representative stable PO of the thr\(_{z1}\)(mul2) family is shown in Fig.~\ref{fig4:Fig3}(d). This orbit exhibits a form similar to one of the parent family thr\(_{z1}\) of multiplicity one, but with two distinct `loops' instead of just one. The initially stable thr\(_{z1}\)(mul2) family becomes simple unstable at \( E_j = -0.2943\) and double unstable for \( -0.2807 < E_j < -0.2692 \). 

\subsubsection*{B. A period-doubling bifurcation leading to a multiplicity four orbit}
The thr\(_{z1}\)(mul2) family of POs (green curves in Fig.~\ref{fig4:Fig1}), transitions from stability to instability at energy \(E_D = -0.2943\). This change of stability results in the creation of the multiplicity four \(3D\) family, thr\(_{z1}\)(mul4) along with its symmetric counterpart thr\(_{z1}\)(mul4)S, depicted by the magenta curves in Fig.~\ref{fig4:Fig1}. The initially stable thr\(_{z1}\)(mul4) family becomes complex unstable in two small energy intervals: \( -0.2917 < E_j < -0.2872 \) and \( -0.2667 < E_j < -0.2622 \) (which are represented by the shaded magenta regions in Fig.~\ref{fig4:Fig2}). In between these complex instability intervals, the multiplicity four family is simple unstable on the interval  \( -0.28 < E_j < -0.2698\) and remains stable otherwise. The morphology of a representative stable PO member of the thr\(_{z1}\)(mul4) family for \( E_j = -0.2831 \), is shown in Fig.~\ref{fig4:Fig3}(e).

\subsubsection*{C. Investigation of the phase space structure}
In order to understand how the multiplicity of two and four POs created by the two period-doubling bifurcations influence the phase space structure of system \eqref{eq:BG H}, we systematically examine four different cases with energies larger than the third critical point, i.e. \( E_j > E_C \), where  \( j = 1, 2, 3, 4\). These four energy levels are denoted by orange vertical lines in Fig.~\ref{fig4:Fig2}. The first two energies are between the \( E_C \) and \( E_D \) values, where the thr\(_{z1}\)(mul4) PO has not yet been created, while the latter two values represent energies greater than \( E_D \). 

The four cases are carefully selected to explore all possible stability combinations of the four major bifurcated families of POs we have observed thus far: the \(2D\) thr\(_{1}\) as well as the \(3D\) thr\(_{z1}\), the thr\(_{z1}\)(mul2), and the thr\(_{z1}\)(mul4) family. Table \ref{tab:stability} presents the \( E_j \) values, the stability characteristics of the POs of each family, and the figures where the associated phase space structures are shown. For simplicity, we denote stable, simple unstable, double unstable, and complex unstable POs as ``S," ``U," ``DU," and ``$\Delta$", respectively. A dash ``-" indicates that the specific family does not exist at the particular energy value, as discussed in Figs \ref{fig4:Fig1} and \ref{fig4:Fig2}. 

\begin{table}[ht]
  \centering
  \caption{Details for the four specific cases with energies \( E_j \), \( j = 1, 2, 3, 4 \), which are discussed in Sect.~\ref{sec:Period-doubling}, along with the stability types of the POs belonging to the thr\(_{1}\), thr\(_{z1}\), thr\(_{z1}\)(mul2), and thr\(_{z1}\)(mul4) families at those energies with S, U, DU, and \(\Delta\), respectively, denoting stable, simple unstable, double unstable, and complex unstable POs. Note that ``-'' denotes that the family of POs does not exist for the specific energy value. The table also includes the figure numbers of the \(3D\) projections of the system's \(4D\) PSS presented for each energy.}

  \label{tab:stability}
  \begin{tabular}{|c|c|c|c|c|c|c|}
  \hline
  \textbf{Case} & \textbf{Energy} & \textbf{thr\(_1\)} & \textbf{thr\(_{z1}\)} & \textbf{thr\(_{z1}\)(mul2)} & \textbf{thr\(_{z1}\)(mul4)} & Figure\\
  \hline
  1 & -0.3183 & U & U & S & - & Fig.~\ref{fig4:Fig11}\\
  2 & -0.3157 & DU & U & S & - & Fig.~\ref{fig4:Fig12}\\
  3 & -0.2941 & U & U & U & S & Fig.~\ref{fig4:Fig14}\\
  4 & -0.2907 & U & U & U & $\Delta$ & Fig.~\ref{fig4:Fig15}\\
  \hline
  \end{tabular}
  \end{table}
    
  \subsubsection*{\textbf{Case 1: \boldmath\(  E_1 = -0.3183, \quad \text{thr}_1 \text{ (U)}, \quad \text{thr}_{z1} \text{ (U)}, \quad \text{thr}_{z1}\mbox{(mul2)} \text{ (S)}, \quad \text{thr}_{z1}\mbox{(mul4)} \text{ (-)} \)}}

\begin{figure}[!htbp]
  \centering
  \subfloat[Perturbations of the thr\(_{1} \) (U), the thr\(_{z1} \) (S) and the thr\(_{z1} \)S (S) POs\label{fig4:Fig11a}]{\includegraphics[width=0.8\textwidth]{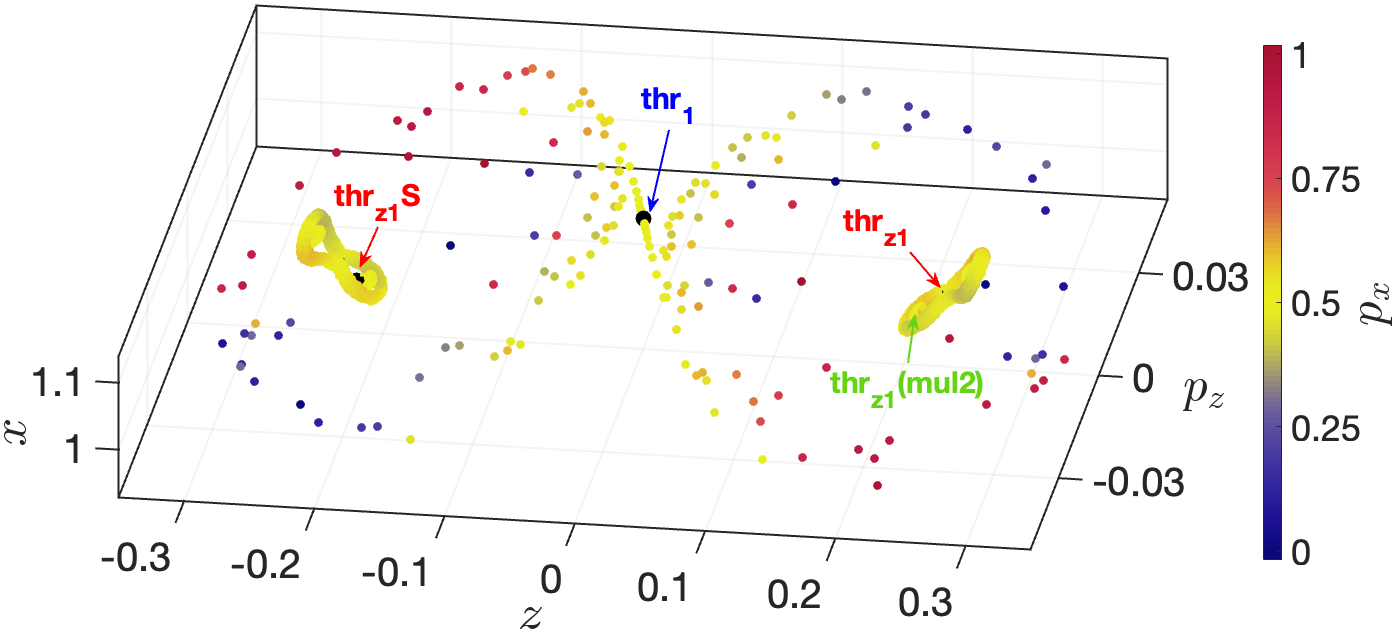}} \\
  \subfloat[Zoom in of the right Figure-8 structure in (a)\label{fig4:Fig11b}] {\includegraphics[width=0.8\linewidth]{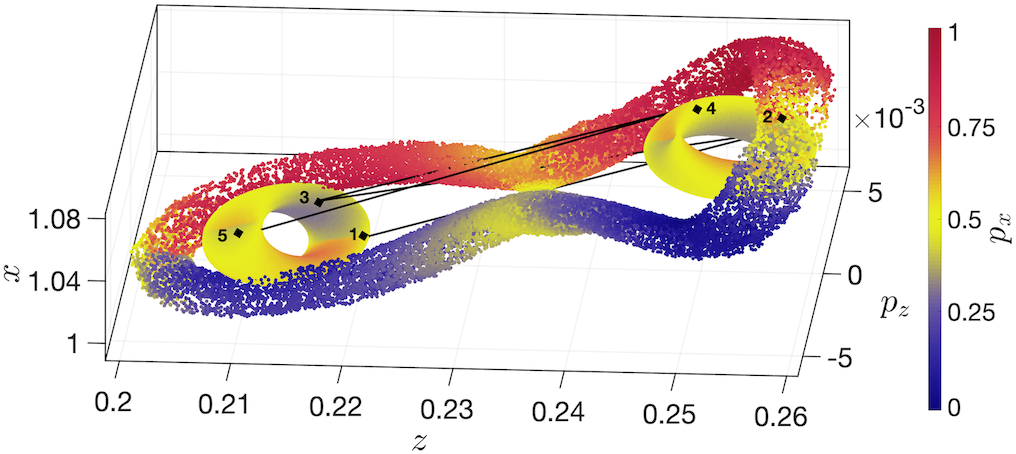}}
\caption{The \(3D\) colored \((x, z, p_z)\) projection of the \(4D\) PSS system \eqref{eq:BG H} at \(E_1 = -0.3183\), where points are colored according to their \( p_z \) values. (a) The large, loosely defined Figure-8 structure is generated by perturbing the simple unstable thr\(_{1} \) PO by \( \Delta z = 10^{-5} \), while the smaller Figure-8 structures on the right and left are produced by perturbing the simple unstable thr\(_{z1} \) and thr\(_{z1} \)S POs by \( \Delta z = 2.4 \times 10^{-6} \), respectively. (b) A zoomed-in view of the right small Figure-8 structure in (a), surrounding two tori generated by a \( \Delta z = 2.3 \times 10^{-2} \) perturbation of the stable thr\(_{z1} \)(mul2) PO. The first five consequents of the quasiperiodic orbit that form these tori are represented by a black diamond symbol connected by lines.}
  \label{fig4:Fig11}
\end{figure}

At the energy \( E_1 = -0.3183 \) (the first orange vertical line in Fig.~\ref{fig4:Fig2}, just after the third bifurcation point \( E_C \)), we observe that both the thr\(_1\) (blue curves in Fig.~\ref{fig4:Fig2}) and thr\(_{z1}\) (red curves in Fig.~\ref{fig4:Fig2}) families of POs are simple unstable, while the thr\(_{z1}\)(mul2) family (green curves in Fig.~\ref{fig4:Fig2}) is stable. Note that at this energy, the thr\(_{z1}\)(mul4) family has not been yet created. Fig.~\ref{fig4:Fig11a} depicts the \(3D\) projection of the system's \eqref{eq:BG H} \(4D\) PSS for case 1. A \( \Delta z = 10^{-5} \) perturbation of the simple unstable thr\(_1\) PO results in a scattered set of points vaguely forming a Figure-8 structure. This structure is less well-defined compared to the one seen in Fig.~\ref{fig4:Fig7a} because not only the thr\(_1\) PO is unstable, but the thr\(_{z1}\) PO is also no longer stable. As a result, there are no surrounding tori where clear Figure-8 structures can form for the perturbed unstable thr\(_1\) PO at \(E_1\). The presence of invariant tori around stable POs of the newly bifurcated families plays an important role in maintaining well-defined typical Figure-8 structures when a parent family transitions from stability to simple instability. The consequents of orbits created from perturbing unstable POs are influenced by the unstable manifolds (e.g.~see \citep[Fig.~12]{katsanikas2013instabilities}). This leads to the formation of the standard Figure-8 structures [such as in Fig.~\ref{fig4:Fig7a}] around tori associated with quasiperiodic orbits near stable POs of the bifurcated families.

It is important to highlight that the scattered points of the perturbed unstable thr\(_1\) PO in Fig.~\ref{fig4:Fig11a} show a smooth color variation along the less well-defined 8-shaped structure. This indicates a ``sticky behavior", which is a transient phase in the orbit's evolution that could significantly influence the long-term dynamics of disk galaxies. The thr\(_1\) orbit perturbed by \( \Delta z = 10^{-5} \) at \( E_j = E_1 \) behaves differently from the \( \Delta z \) perturbed thr\(_1\) PO at \( E_j = -0.3307 \) [Fig.~\ref{fig4:Fig7a}]. This shows how the dynamics near a PO (in this case, simple unstable thr\(_1\)) can change as we move away from where its bifurcation point occurred. Thus, the structure of the phase space around unstable POs is closely related to their immediate surroundings. 

On the other hand, the orbit created by a \(\Delta x\) perturbation of the planar thr\(_1\) PO at \( E_j = E_1 \) remains planar. The behavior of these perturbed orbits on the (\(x,  p_x \)) projection is influenced by the strength of the perturbation, leading to planar quasiperiodic motion or rapid diffusion in the system's phase space. Our analysis shows that the consequents of such orbits cover the entire phase space when \( \Delta x = 10^{-1} \), whereas a smaller radial perturbation (such as \( \Delta x = 10^{-2} \)) results in quasiperiodic motion. The structure of the 2D \((x, p_x)\) projection generated for this quasiperiodic orbit at \(E_1\) is similar to the one we observed in Fig.~\ref{fig4:Fig5a}. 

Now, let us consider the \(3D\) thr\(_{z1}\) PO. Introducing a small vertical perturbation of \( \Delta z = 2.4 \times 10^{-6} \) to the simple unstable thr\(_{z1}\) and thr\(_{z1}\)S POs results in an interesting outcome. The perturbed unstable orbits now display a well-defined Figure-8 structure with smooth color variation. The simple unstable thr\(_{z1}\) (thr\(_{z1}\)) PO is located at the intersection of the two loops of the right (left) well-defined Figure-8 structure in Fig.~\ref{fig4:Fig11a}. These Figure-8 structures develop around invariant tori associated with quasiperiodic orbits near the stable thr\(_{z1}\)(mul2) and thr\(_{z1}\)(mul2)S POs. The tori depicted in Fig.~\ref{fig4:Fig11b} are the outcome of a \( \Delta z = 2.3 \times 10^{-2} \) perturbation applied to the thr\(_{z1}\)(mul2) PO. These tori coexist with the Figure-8 structure generated by the \( \Delta z \) perturbation of the simple unstable thr\(_1\) PO [essentially, the right small Figure-8 structure in Fig.~\ref{fig4:Fig11a}]. A similar configuration is observed in Fig.~\ref{fig4:Fig7a}, but with a basic difference: the tori in Fig.~\ref{fig4:Fig11b} originate from a single (multiplicity two) orbit, whereas those in Fig.~\ref{fig4:Fig7a} belong to two distinct (multiplicity one) orbits. 

The connection between the two tori in Fig.~\ref{fig4:Fig11b} becomes evident when following the consequents starting from the IC marked by ``1" (black diamond symbol on the left torus). Then, the second consequent appears on the right torus (black diamond symbol marked by ``2''), and this pattern continues. In other words, the first five consequents (line-connected black diamond symbols in Fig.~\ref{fig4:Fig11b}) alternate between the left and right torus, which indicates that the tori are generated by a quasiperiodic orbit around a stable PO of multiplicity two. However, the consequents in the left and right torus in Fig.~\ref{fig4:Fig7a} remain confined to their respective tori.

\subsubsection*{\textbf{Case 2: \boldmath\( E_2 = -0.3157, \quad \text{thr}_1 \text{ (DU)}, \quad \text{thr}_{z1} \text{ (U)}, \quad \text{thr}_{z1}\mbox{ (mul2)} \text{ (U)}, \quad \text{thr}_{z1}\mbox{(mul4)} \text{ (-)} \)}}
We now consider the case \( E_j = E_2 \) corresponding to the second orange vertical line in Fig.~\ref{fig4:Fig2}, which occurs just after the transition of the thr\(_1\) family from simple to double instability. At this energy level, the thr\(_1\) planar family (blue curves in Fig.~\ref{fig4:Fig2}) is double unstable, while the \(3D\) families, thr\(_{z1}\) and thr\(_{z1}\)(mul2) (red and green curves in Fig.~\ref{fig4:Fig2}, respectively), retain the same stability types observed in Case 1 for \( E_j = E_1\), i.e. thr\(_{z1}\) is simple unstable and thr\(_{z1}\)(mul2) is stable. We note that, as in Case 1, the thr\(_{z1}\)(mul4) does not exist for \( E_j = E_2 \). 

\begin{figure}[!htbp]
  \centering
  \subfloat[Perturbations of the thr\(_{1} \) (DU), the thr\(_{z1} \) (S) and the thr\(_{z1} \)S (S) POs\label{fig4:Fig12a}]{\includegraphics[width=0.8\textwidth]{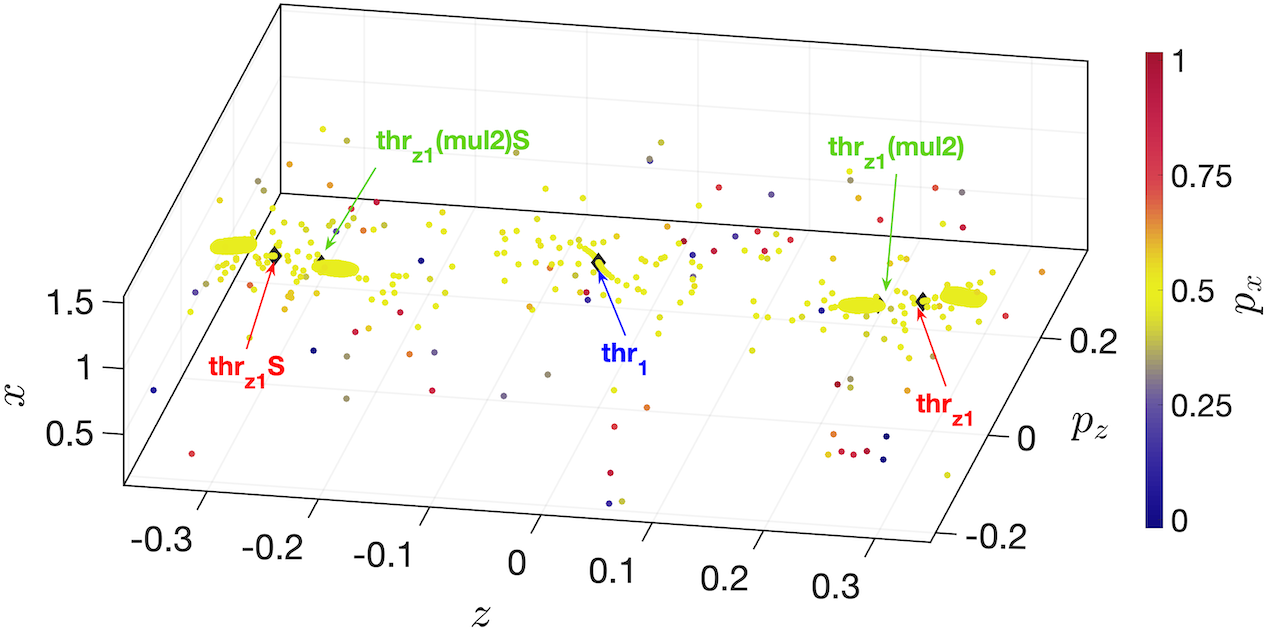}} \\
  \subfloat[Zoom in of the right Figure-8 structure in (a)\label{fig4:Fig12b}] {\includegraphics[width=0.8\linewidth]{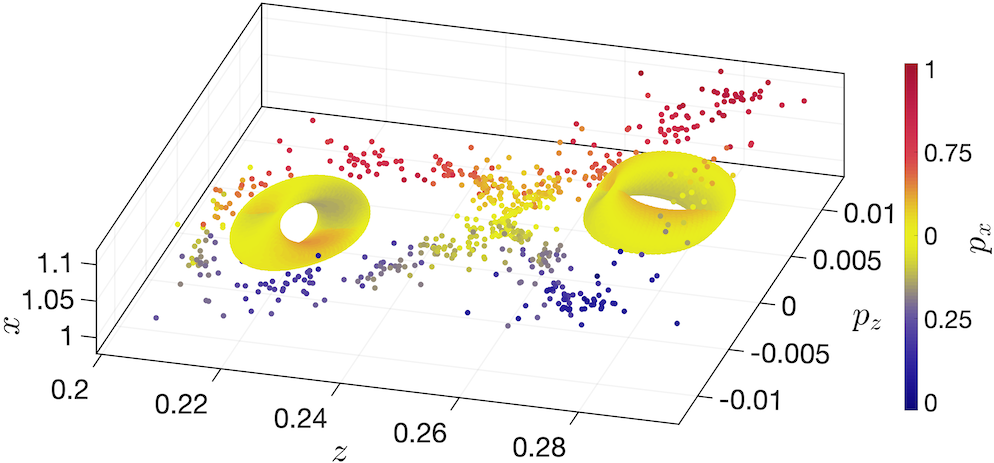}}
\caption{The \(3D\) colored \((x, z, p_z)\) projection of the \(4D\) PSS system \eqref{eq:BG H} at \(E_2 = -0.3157\), where points are colored according to their \( p_z \) values. (a) The cloud of scattered points are generated by perturbing the double unstable thr\(_{1} \) PO (whose perturbation is indicated by a blue arrow) with \( \Delta z = 10^{-5} \). In contrast, the two loosely formed Figure-8 structures are produced by perturbing the simple unstable thr\(_{z1} \) and thr\(_{z1} \)S POs (indicated by red arrows) by \( \Delta z = 2.9 \times 10^{-6} \). Additionally, the four tori surrounded by the two loosely formed Figure-8 structures result from the perturbations of the stable thr\(_{z1} \)(mul2) and thr\(_{z1} \)(mul2)S POs (their location is shown by green arrows) by \( \Delta z = 1.4 \times 10^{-2} \) and \( \Delta p_z = 5 \times 10^{-3} \), for \( z > 0 \) and for \( z < 0 \), respectively. (b) A zoom-in view of the right small Figure-8 structure in (a) surrounding two tori.}
  \label{fig4:Fig12}
\end{figure}

\begin{figure}[!htb]
  \centering
  \includegraphics[width=0.52\textwidth]{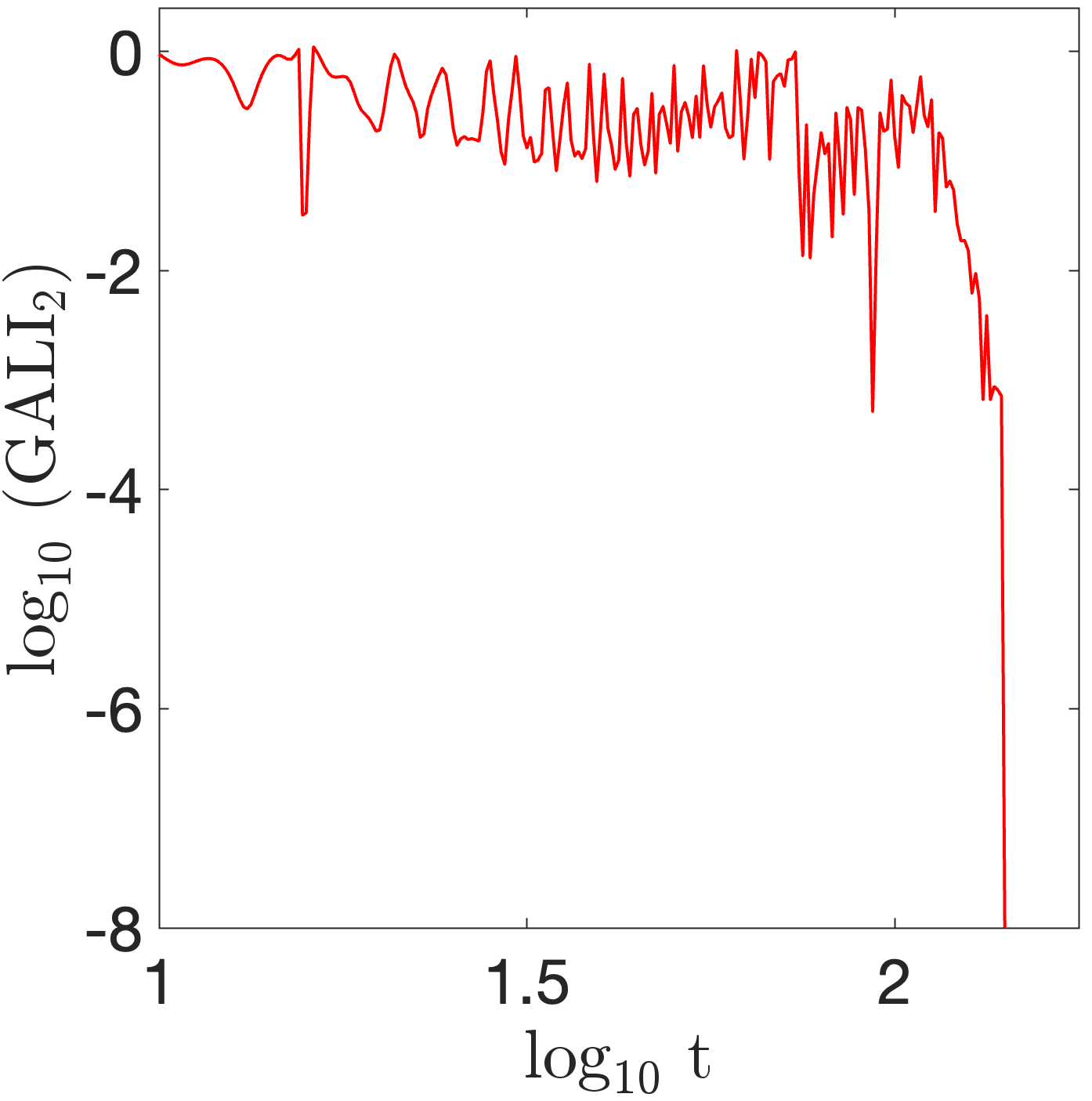}
  \caption{The time evolution of the GALI\(_2\) for perturbations of the thr\(_{1} \) (DU) shown in Fig.~\ref{fig4:Fig12a}. As expected, the GALI\(_2\) rapidly decreases to zero, characteristic of strong chaotic behavior.}
  \label{fig4:Fig12G}
\end{figure}

Introducing a small (\( \Delta z = 10^{-5} \)) perturbation to the IC of the double unstable thr\(_1\) PO results in a pronounced chaotic behavior, as demonstrated by the abrupt decrease of the orbit's GALI\(_2\) to zero in Fig.~\ref{fig4:Fig12G}. Fig.~\ref{fig4:Fig12a} illustrates the corresponding \(3D\) (\(x, z, p_z\)) projection of the system's \(4D\) PSS for this perturbed orbit, whose consequents create a scattered cloud of points with mixed color points. The fact that the orbit rapidly diffuse in the phase space significantly reduces the likelihood of forming a well-defined Figure-8 structure around the simple unstable thr\(_1\) PO that we observed in Fig.~\ref{fig4:Fig11a}.

The phase space structure around the simple unstable thr\(_{z1}\) PO at \( E_j = E_2 \) differs significantly from that at \( E_j = E_1 \). In this case, the application of a small perturbation of \( \Delta z = 2.9 \times 10^{-6} \) to the thr\(_{z1}\) PO and its symmetric counterpart thr\(_{z1}\)S result in the formation of a Figure-8 structure characterized by smooth color variation [the small Figure-8 structures at the right and left parts of Figs.~\ref{fig4:Fig12a} and (b), where the region around the right Figure-8 of \ref{fig4:Fig12a} is shown]. However, this Figure-8 structure is transient. As time grows, the orbit gradually deviates to a larger phase space region. These Figure-8 formations still coexist with smooth invariant tori created by the quasiperiodic orbits resulting from a \( \Delta z = 1.4 \times 10^{-2} \) and \( \Delta p_z = 5 \times 10^{-3} \) perturbation of the stable thr\(_{z1}\)(mul2) and thr\(_{z1}\)(mul2)S POs as shown in Fig.~\ref{fig4:Fig12a}. The consequents of both quasiperiodic orbits lead to the formation of invariant tori surrounding the left and right lobes of the Figure-8 structures that are generated by the unstable thr\(_{z1}\) and thr\(_{z1}\)S perturbed orbits [a zoomed-in view of the right set of tori is depicted in Figs.~\ref{fig4:Fig11b}]. By comparing Figs.~\ref{fig4:Fig11a} and \ref{fig4:Fig12b}, we can deduce that the Figure-8 structures generated by a \(\Delta z\) perturbation of the simple unstable thr\(_{z1}\) and thr\(_{z1}\) POs become less well-defined  as we increase the system's energy.

\begin{figure}[!htb]
  \centering
  \includegraphics[width=0.55\textwidth]{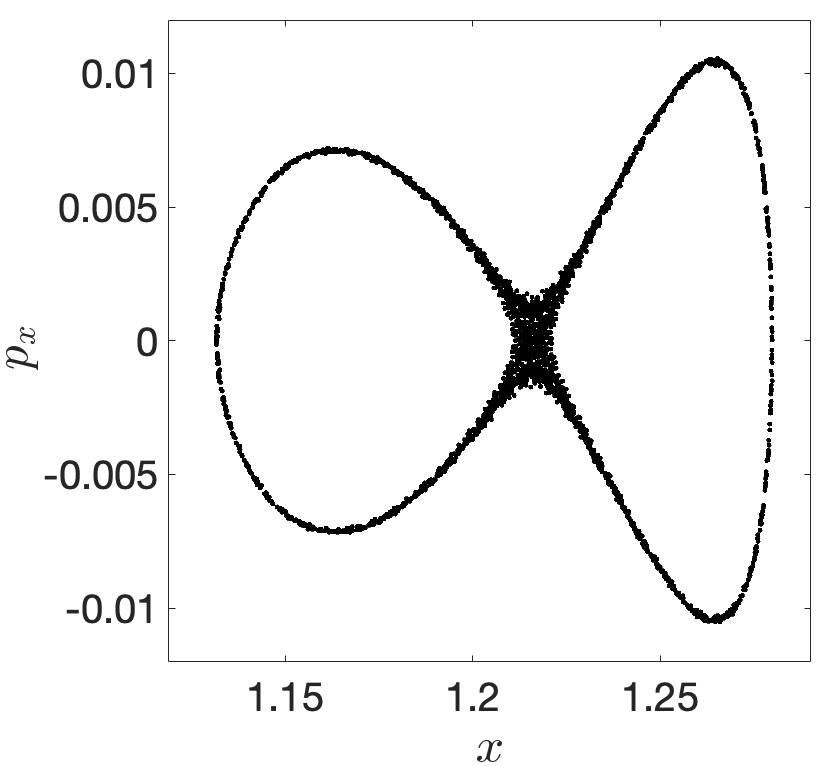}
  \caption{The \(2D\) \((x, p_x)\) projection of an orbit obtained by perturbing the double unstable thr \(_1\) PO by \(\Delta x = 10^{-5}\) at \(E_2 = -0.3157\).}
  \label{fig4:Fig13}
\end{figure}

A detailed examination of the dynamics of the system's \(4D\) PSS at \( E_j = E_2 \) reveals that radial perturbations of the planar thr\(_1\) family exhibit double instability. The \(3D\) projection of the \(4D\) phase space structure shows the chaotic nature near this double instability, often appearing as clouds of points with mixed colors \citep{katsanikas2013instabilities}. However, the unstable thr\(_1\) PO for \( \Delta x \) perturbations, such as \( \Delta x = 10^{-5} \), results in a planar orbit. The consequents of this orbit create a distinct Figure-8 structure on the (\(x, p_x\)) projection shown in Fig.~\ref{fig4:Fig13}, rather than a cloud of points depicted by the \(\Delta z\) perturbations in Fig.~\ref{fig4:Fig12a}. 

We observe that the two lobes of the Figure-8 structure generated by the radial perturbations are asymmetrical (Fig.~\ref{fig4:Fig13}). It is worth noting that small changes in the \(x\) direction for the simple unstable thr\(_{z1}\) POs rapidly diffuse in the entire phase space after initially forming an 8-shaped structure. This is what we expect given the radial instability of the thr\(_{z1}\) PO. 

\subsubsection*{\textbf{Case 3: \boldmath\( E_3 = -0.2941, \quad \text{thr}_1 \text{ (U)}, \quad \text{thr}_{z1} \text{ (U)}, \quad \text{thr}_{z1}\mbox{(mul2)} \text{ (U)}, \quad \text{thr}_{z1}\mbox{(mul4)} \text{ (S)} \)}}
At energy \( E_j =  E_3\) (indicated by the third orange vertical line in Figure~\ref{fig4:Fig2}), the thr\(_1\) PO is again simple unstable, along with the thr\(_{z1}\) family. At this energy, the thr\(_{z1}\) (mul2) and thr\(_{z1}\)(mul2) families also become simple unstable, having their radial stability index \( b_2 > 2 \) (green curves in Figure~\ref{fig4:Fig2}). The transition of the thr\(_{z1}\) (mul2) family from stability to instability leads to the creation of the stable thr\(_{z1}\)(mul4) and its symmetric counterpart, the thr\(_{z1}\)(mul4)S family of POs of multiplicity four (magenta curves in Figure~\ref{fig4:Fig2}).

\begin{figure}[!htbp]
  \centering
  \subfloat[Perturbations of the thr\(_{z1}\)(mul2) (U) and  the thr\(_{z1}\)(mul4) (S) POs\label{fig4:Fig14a}]{\includegraphics[width=1\textwidth]{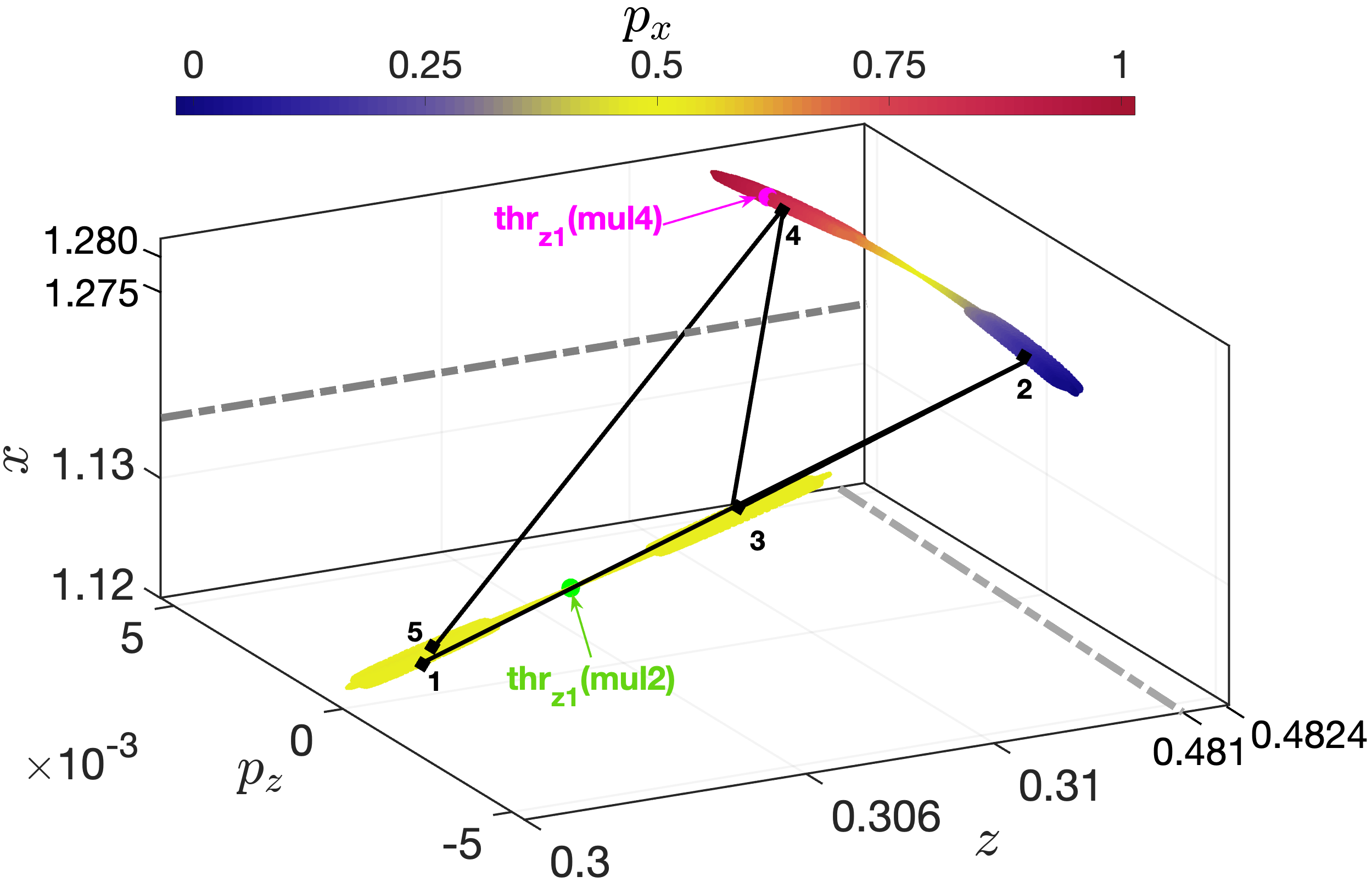}} \\
  \subfloat[Zoom-in of the bottom Figure-8 structure in (a)\label{fig4:Fig14b}] {\includegraphics[width=0.5\linewidth]{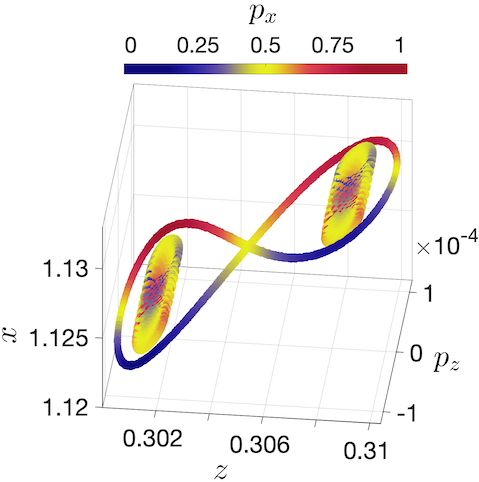}}
  \subfloat[Zoom-in of the top Figure-8 structure in (a)\label{fig4:Fig14c}] {\includegraphics[width=0.5\linewidth]{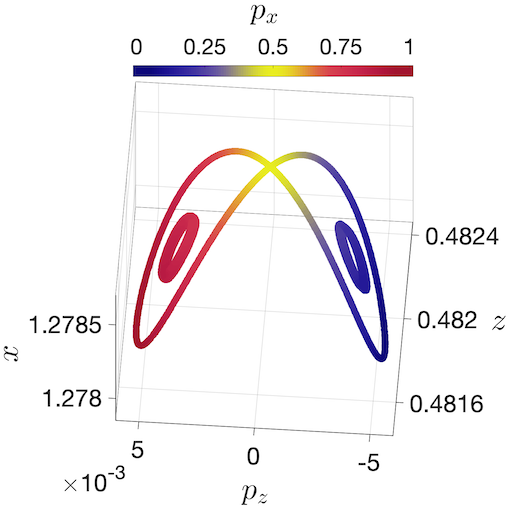}}
    \caption{The \(3D\) colored \((x, z, p_z)\) projection of the \(4D\) PSS system at \(E_3 = -0.2941\), where points are colored according to their \(p_x\) values. (a) The two thin and elongated Figure-8 structures are generated by perturbing the multiplicity two simple unstable thr\(_{z1}\)(mul2) PO by \(\Delta z = 3 \times 10^{-6}\), while the four tori surrounded by these Figure-8 formations result from perturbing the multiplicity four stable thr\(_{z1}\)(mul4) PO by \(\Delta x = 10^{-5}\), \(\Delta z = 6 \times 10^{-5}\), and \(\Delta p_z = 8 \times 10^{-5}\). The gray dashed lines indicate cutoff regions of the \(x\) and \(z\) axes, and the numbered black diamond symbols with connected lines represent the first five consequents of the orbit creating the invariant tori. A zoom-in view of the (b) bottom and (c) top Figure-8 structure of panel (a).} 
  \label{fig4:Fig14}
\end{figure}

When perturbing both multiplicity one, simple unstable thr\(_1\) and thr\(_{z1}\) POs along the \( z \)-axis at \(E_3=-0.2941\), we observe chaotic motion, which is reflected in the creation of scattered clouds of points with mixed color variations in any \(3D\) phase space projection of the system's 4D PSS. In contrast, applying a small \( \Delta z = 3 \times 10^{-6} \) perturbation to the simple unstable thr\(_{z1}\)(mul2) and thr\(_{z1}\)(mul2)S POs of multiplicity two results in the formation of two distinct Figure-8 structures. All these formations are shown in Fig.~\ref{fig4:Fig14a}, where we plot the \(3D\) colored \((x, z, p_z)\) projection of the system's \(4D\) PSS at \(E_3 = -0.2941\), where points are colored according to their \(p_x\) value. These Figure-8 structures are observed near the ICs of the thr\(_{z1}\)(mul2) PO. Since this orbit is of multiplicity two, two Figure-8 structures are formed. Furthermore, these Figure-8 formations in Fig.~\ref{fig4:Fig14a} are relatively thin and located close to the invariant tori surrounding the stable thr\(_{z1}\)(mul4) PO.

The thr\(_{z1}\)(mul4) PO has a multiplicity of four, and consequently small perturbations to its ICs lead to the creation of four separate tori in the phase space. These tori are characterized by smooth color variations on the \(3D\) projections of the \(4D\) PSS. Figs.~\ref{fig4:Fig14b} and (c) depict examples of such tori, which are generated by applying perturbations of \( \Delta x = 10^{-5} \), \( \Delta z = 6 \times 10^{-5} \), and \( \Delta p_z = 8 \times 10^{-5} \) to the stable IC of the thr\(_{z1}\)(mul4) PO. The invariant tori are surrounded by the Figure-8 structures formed by the perturbed unstable thr\(_{z1}\)(mul2) at \(E_3\). Note that we can obtain another four invariant tori around the stable symmetric counterpart PO of the family thr\(_{z1}\)(mul4)S, which will be surrounded by the Figure-8 formations around the unstable thr\(_{z1}\)(mul2)S PO. To ensure that both Figure-8 structures related to the unstable thr\(_{z1}\)(mul2) PO are visible in a single panel, we have appropriately adjusted the $x$ and $z$ axis scales of Fig.~\ref{fig4:Fig14a}. More specifically, we narrowed down both axes by excluding specific axes intervals between the two structures: \(1.135 < x < 1.27\) and \(0.315 < z < 0.48\) (the cutoff regions are indicated by gray dashed lines in Fig.~\ref{fig4:Fig14a}).

The numbered black diamond symbols with connected lines in Fig.~\ref{fig4:Fig14a} illustrate the first five consequents on the tori surrounding the stable thr\(_{z1}\)(mul4) PO. The jumping of these points between each torus clearly demonstrates that the four tori are formed by one single orbit. The zoomed-in views in Figs.~\ref{fig4:Fig14b} and (c) provide a closer look at each Figure-8 structure and the surrounded tori. Fig.~\ref{fig4:Fig14b} focuses on the tori containing points ``1", ``3", and ``5" of Fig.~\ref{fig4:Fig14a}, while Fig.~\ref{fig4:Fig14c} highlights the phase space region around the remaining two tori, which correspond to points,``2" and ``4" in Fig.~\ref{fig4:Fig14a}. 

These findings align with what we have seen so far: the formation of a Figure-8 structure typically occurs near the (parent) simple unstable PO, as long as it is surrounding invariant tori, which are characterized by smooth color variations around the newly bifurcated stable PO. Similar behaviors were also observed in Fig.~\ref{fig4:Fig7a} for the case of a pitchfork bifurcation, before the period-doubling bifurcation occurring at \(E_j = -3307\) and in Figs.~\ref{fig4:Fig12} and \ref{fig4:Fig14} after a period-doubling bifurcation.

Let us now consider radial perturbations at \(E_j = E_3\). Even very small \(\Delta x\) perturbations (of the order of \(10^{-8}\)) applied to the planar simple unstable thr\(_1\) PO result in chaotic behaviors. Similarly, minor \(\Delta x\) perturbations of the \(3D\) simple unstable thr\(_{z1}\) PO lead again to chaotic behavior, which is reflected in the creation of clouds of scattered points with mixed colors in the system's phase space. It is important to note that despite its radial instability, the thr\(_{z1}\)(mul2) family showed similar behaviors when radial perturbations at \(E_3\) were applied to the ones observed for vertical perturbation. Both types of perturbations consistently resulted in the formation of the standard Figure-8 structures with smooth color variations.

\subsubsection*{\textbf{Case 4: \boldmath\( E_4 = -0.2907, \quad \text{thr}_1 \text{ (U)}, \quad \text{thr}_{z1} \text{ (U)}, \quad \text{thr}_{z1}\mbox{(mul2)} \text{ (U)}, \quad \text{thr}_{z1}\mbox{(mul4)} (\Delta) \)}}

\begin{figure}[!htb]
  \centering
  \includegraphics[width=0.85\textwidth]{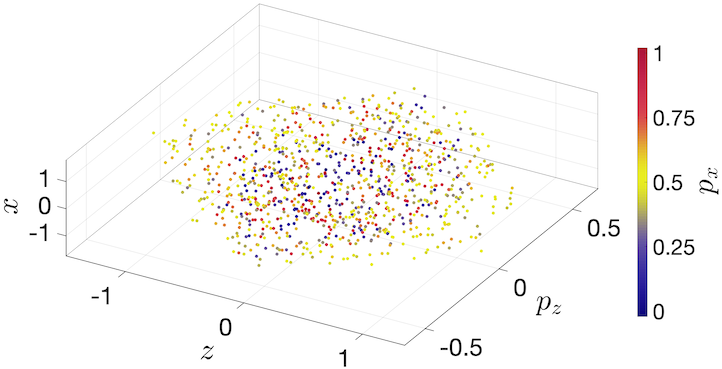}
  \caption{The \(3D\) colored \((p_x, z, p_z)\) projection of the system's \(4D\) PSS at energy \(E_4 = -0.2907\). The scattered clouds of points are generated by perturbing the complex unstable thr\(_{z1}\)(mul4) PO by \(\Delta z = 10^{-5}\). 
  }
  \label{fig4:Fig15}
\end{figure}

At the highest energy level we consider in our study, namely \(E_4 = -0.2907\) (indicated by the fourth orange vertical line in Fig.~\ref{fig4:Fig2}), all studied POs are unstable. The thr\(_{1}\), thr\(_{z1}\), and thr\(_{z1}\)(mul2) families are simple unstable, while the multiplicity four thr\(_{z1}\)(mul4) PO is complex unstable (note that the orange curve denoting \(E_4\) falls within the shaded magenta area in Fig.~\ref{fig4:Fig2}). Unlike the previous three cases, none of the POs we study are surrounded by tori where regular motion takes place. Even small perturbations of any of the unstable POs lead to chaotic behaviors, which are manifested through the creation of clouds of scattered points with mixed color variations in any \(3D\) projection of the system's  \(4D\) PSS. These clouds eventually diffuse over the entire phase space. Fig.~\ref{fig4:Fig15} illustrates this behavior, showing the \(3D\) projection (\(x, z, p_z\)) of the \(4D\) PSS for the orbit that results from a small \(\Delta z = 10^{-5}\) perturbation of the complex unstable thr\(_{z1}\)(mul4) PO. 

\section{Summary and conclusions} \label{section:SummaryCh4}
In this chapter, we investigated the evolution of the phase space structure in a \(3D\) bar galaxy Hamiltonian system \eqref{eq:BG H} before and after \(2D\) and \(3D\) pitchforks as well as two \(3D\) period-doubling bifurcations. Our focus was on bifurcations where a stable parent family of POs becomes unstable, simultaneously giving birth to a new family of stable POs. In particular, the first \(2D\) pitchfork bifurcation occurs when the main x\(1\) family of POs transitions from stability to simple instability at energy \(E_j = E_A\) (Fig.~\ref{fig4:Fig1}). This bifurcation creates two planar, initially stable families: the thr\(_1\) family and its symmetric counterpart along the \(y\)-direction, the thr\(_{1}\)S family. 

The thr\(_1\) family becomes simple unstable at \(E_j = E_B\) (Fig.~\ref{fig4:Fig1}), resulting in the creation of two new \(3D\) stable PO families: the thr\(_{z1}\) and its symmetric counterpart perpendicular to the galactic plane, thr\(_{z1}\)S. Subsequently, the thr\(_{z1}\) changes its stability type from stable to simple unstable at \(E_j = E_C\) (Fig.~\ref{fig4:Fig1}) through a period-doubling bifurcation, giving birth to the stable thr\(_{z1}\)(mul2) family and its symmetric counterpart thr\(_{z1}\)(mul2)S. The final bifurcation we studied occurs when the thr\(_{z1}\)(mul2) family becomes simple unstable at \(E_j = E_D\) (Fig.~\ref{fig4:Fig1}), resulting in the appearance of the multiplicity four stable thr\(_{z1}\)(mul4) PO and its symmetric counterpart thr\(_{z1}\)(mul4)S. 

The color and rotation method introduced by \citep{patsis1994using} was employed in our study to visualize the system's \(4D\) PSS. The perturbations of all stable PO we examined result in the creation of invariant tori with smooth color variations in any \(3D\) projection of the system's \eqref{eq:BG H} \(4D\) PSS. Furthermore, the color pattern is continuous across the entire torus surface, showing a clear transition from either the interior to the exterior surfaces or vice versa [Figs.~\ref{fig4:Fig4} and \ref{fig4:Fig7b}]. These findings aligned with previous studies (e.g.~see \citep{katsanikas2011structure1, zachilas2013structure}), confirming that this behavior is typical around the stable POs of the \(3D\) Hamiltonian system \eqref{eq:BG H}.

We further analyzed the evolution of phase space structures around POs as they transition from stability to simple instability. Pitchfork bifurcations typically result in two new stable families of PO, each associated with distinct sets of tori in the system's \(4D\) PSS [Fig.~\ref{fig4:Fig7a}]. On the other hand, period-doubling bifurcations lead to the formation of interconnected tori around stable POs of double multiplicity relative to the parent family [Fig.~\ref{fig4:Fig11b}]. 

In general, perturbations around simple unstable POs often lead to the formation of Figure-8 structures in the phase space. These structures typically surrounded tori that emerge from perturbations of the bifurcated stable POs at the same energy value. This pattern was also observed for \(3D\) pitchfork bifurcations of all multiplicities we considered: multiplicity one (Fig.~\ref{fig4:Fig7}), multiplicity two (Fig.~\ref{fig4:Fig11}) and multiplicity four (Fig.~\ref{fig4:Fig14}). However, for planar simple unstable POs that were vertically unstable but radially stable, radial perturbations often result in quasiperiodic planar orbits [Fig.~\ref{fig4:Fig5a}]. The appearance and coherence of Figure-8 structures are closely related to the stability of the tori they surround. As energy levels increase further from the bifurcation points, these Figure-8 structures begin to break apart and spread out (Fig.~\ref{fig4:Fig12}), highlighting their dependence on the existence of invariant tori.

A particularly interesting finding was observed in the \(3D\) pitchfork bifurcation we studied in Sect.~\ref{sec:3D pitchfork bif.}. As the system's energy approaches the bifurcation point (\(E = E_B\) in Fig.~\ref{fig4:Fig2}), the perturbed \(2D\) thr\(_1\) orbit begins to exhibit morphologies similar to those observed for the bifurcated thr\(_{z1}\) family, even before this \(3D\) family of POs is created (Fig.~\ref{fig4:Fig10}). This observation suggests that the system, in some sense, anticipates or foresees the dynamical behavior of the upcoming bifurcating family. This behavior, which was also previously reported by \citep{patsis2014phasea}, may require further study to better understand its underlying causes and effects in \(2D\) Hamiltonian systems.

In addition, we compared the behavior of the GALI method \eqref{eq:GALI} to the mLE \eqref{eq:mLEs} by analyzing specific examples of regular and chaotic orbits. We demonstrated that the GALI\(_2\) index effectively detects chaos in the \(3D\) bar galaxy Hamiltonian system \eqref{eq:BG H} (Figure~\ref{fig4:Fig_A1}) and used it to reveal the chaotic nature of several of the orbits we studied. This finding extends our earlier assertion (from Chap. \ref{chapter:three}) that the GALI\(_2\) index is both accurate and efficient for quantifying chaos in various types of Hamiltonian systems. 
\clearpage

\chapter{Long-term diffusion transport and chaos properties of coupled standard maps} \label{chapter:five}

\section{Introduction} \label{section:introductionCh5}
For decades, researchers have been deeply interested in studying diffusion and transport phenomena in conservative Hamiltonian systems and area-preserving symplectic maps (e.g.~see \citep{Chirikov1979,rechester1980calculation,meiss1983correlations,karney1983long,kroetz2016hidden}). These studies are fundamental to understanding processes in nonequilibrium statistical mechanics, where the macroscopic properties of matter emerge from the chaotic motion of microscopic particles, such as atoms or molecules (e.g.~see \citep{klages2007microscopic}). A key objective in this area of research is to understand how the global behavior of trajectories in phase space relates to significant physical phenomena, including chemical reaction rates, fluid mixing efficiency, and confinement of particles in systems like accelerators or fusion plasma devices (see \citep{Meiss2015} for a comprehensive review).

The standard map (SM), also known as the Chirikov map or kicked rotor system \citep{Chirikov1979}, serves as a basic model for understanding diffusion and transport phenomena. Its dynamics depends on the value of the kick-strength parameter value \(K\), which governs the system's behaviors over a range of regimes. The SM describes the evolution of two variables: the angular position and angular momentum. For certain values of \(K\), the angular momentum displays chaotic behavior, often resembling a Gaussian random walk. This behavior, indicative of normal diffusion (see \citep{altmann2008anomalous}), is characterized by a linear increase in squared angular momentum over time. However, under specific parameter conditions, the system exhibits anomalous diffusion, where the transport of particles deviates from the typical Euclidean norm. Such deviations are often associated with the presence of accelerator modes (AMs) \citep{Chirikov1979}, which are stable periodic orbits (POs) surrounded by islands of stability in phase space.

The presence of AMs introduces additional complexity to the dynamics of the SM. Orbits originating in the AM islands display ballistic transport, which is characterized by a continuous increase in angular momentum over time and a diffusion rate significantly higher than that seen in normal diffusion. Meanwhile, trajectories near AM islands often exhibit superdiffusion, influenced by stickiness effects near cantori structures (e.g.~see \citep{dvorak1998stickiness}). On the other hand, trajectories confined to regular stability islands undergo subdiffusion, where transport is significantly suppressed (e.g.~see \citep{zaslavsky2000hierarchical,venegeroles2007leading,venegeroles2008calculation}). A comprehensive study of the general diffusion processes, stability maps, and transport rates for various SM parameters is conducted in \citep{ManRob2014PRE}. In this work, the authors analyzed how different period AMs influence the interplay between chaos and regular motion in the system's phase space.

Researchers have expanded SM systems to coupled and higher-dimensional maps. For instance, studies on coupled SMs have revealed trapping regimes and modified diffusion time scales resulting from interactions between regular and chaotic regions \citep{altmann2008anomalous}. These systems exhibit complex behaviors akin to those found in physical models such as disordered Klein-Gordon chains \citep{antonopoulos2016coupled}. Similarly, diffusion transport in systems more directly related to physical applications, including nanosystems and electron motion in molecular graphene, has been extensively investigated (e.g.~see \citep{sato2019anomalous}). These studies highlight the important role of microscopic chaos in driving macroscopic transport across diverse physical and theoretical contexts.

In this chapter, we explore the diffusion transport properties of ensembles of ICs of SMs in the presence of AMs, with a focus on their long-term (asymptotic) behavior. Building on previous studies, particularly by \citep{ManRob2014PRE}, we conduct a thorough and systematic global diffusion analysis of these ICs, aiming to characterize diffusion behavior, stability maps, and phase space dynamics in the presence of chaotic orbits and AMs of varying periods. This basic study emphasizes diffusion rates and the time scales required for ensembles of ICs to reach their asymptotic diffusion rates, taking into account the relative contributions of chaotic regions and areas surrounding the stable AMs.

Our primary goal is to investigate the globally averaged diffusion rates over the entire phase space, rather than focusing on the diffusion properties of individual ICs. In particular, we consider cases where the kick-strength parameter values lead to the phase space of the SM being predominantly dominated by chaotic orbits, with only small stable islands around the AMs. This approach provides valuable information about the independent dynamics of each map in the coupled SM systems, highlighting the differences in diffusion rates and the time scales needed for various ensembles of ICs to converge to their respective asymptotic diffusion rates. These time scales are shown to depend on the relative sizes of chaotic regions and regions surrounding stable AMs.

Next, we extend our analysis to coupled SM systems, considering both equal and varying kick-strength parameters for the coupled maps. In these coupled setups, we investigate: (I) the impact of coupling strength on global diffusion rates and (II) the influence of different ensembles of ICs, particularly those with varying proportions of chaotic regions around the stable AMs, on global diffusion rates and the time scales required to reach their long-term values.

Finally, we relate these diffusion timescales to the underlying dynamics of the system, which we quantify by measuring the extent of chaotic and regular orbits. Specifically, we employ the maximum Lyapunov exponent (mLE) \eqref{eq:mLEs} and the generalized alignment index (GALI) method \eqref{eq:GALI_chaos} chaos detection techniques. Both methods have been successfully used to distinguish between regular and chaotic behaviors of coupled SMs  (see Sect.~\ref{section:Chaos Indicators} for more detailed discussions). 

The content of this chapter is based around the findings presented in \cite{moges2022anomalous}.

\section{The standard map model} \label{section:modelCh5}
\begin{figure}[!htb]
    \centering
    \includegraphics[width=0.5\textwidth]{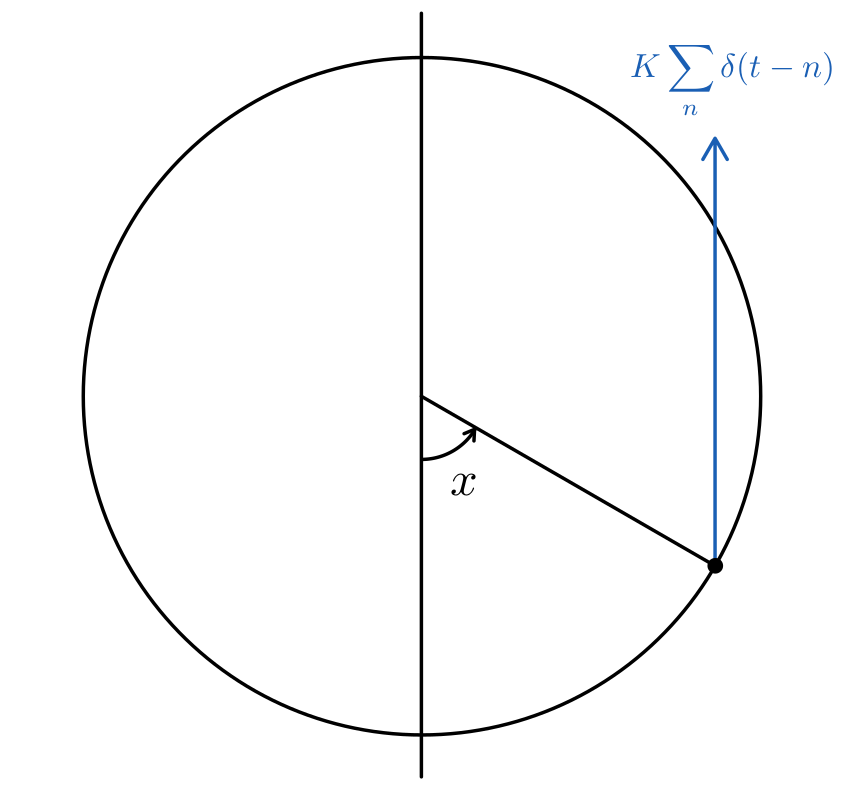}
    \caption{A simple schematic representation of kicked rotor.}
    \label{fig5:FigA0}
  \end{figure}

We aim to investigate the long-term diffusion transport and chaotic characteristics of both SM and coupled SMs. The SMs describe the dynamics of the well-known, extensively studied kicked rotor system. Consider a simple example of a single kicked rotor, shown in Figure \ref{fig5:FigA0}. The figure represents a rotating object, such as an electron orbiting around an atom in a frictionless environment, where no energy is lost. The angular position \(x\) of the object (the black dot) changes continuously over time, with \(x\) indicating the object's position relative to the reference point. Now, imagine that this rotating object is periodically ``kicked" by a uniform field, such as a gravitational field, at discrete time intervals (i.e., at integer time steps). Each time the field acts, it applies a force on the particle with a strength denoted by $K$. In Figure \ref{fig5:FigA0}, the expression \(K \sum_{n} \delta(t - n)\) represents this force or perturbation, which occurs at specific time intervals \(t = n\).  

The setup of the system in Fig.~\ref{fig5:FigA0} can be described by the following Hamiltonian \citep{delande2013kicked}:
  \begin{equation} \label{eq:Hsm}
    H(x, p, t) = \frac{p^2}{2} + K \cos x \sum_{n=-\infty}^{\infty} \delta(t - n)
\end{equation}
where \(x\) represents the angular position, \(p\) is the momentum, and \(K\) is the kick strength. The term \(\delta (t - n)\) is the Dirac delta function, which mathematically describes the discrete application of kicks at time intervals \(t=n\), where \(n\) is an integer. 

The corresponding EoM of \eqref{eq:Hsm} are given by:
\begin{equation} \label{eqm:sm}
    \begin{aligned}
\frac{dx}{dt} &= \frac{\partial H}{\partial p} = p, \\
\frac{dp}{dt} &= -\frac{\partial H}{\partial x} = K \sin x \sum_{n=-\infty}^{\infty} \delta(t - n).
\end{aligned}
\end{equation}
These equations indicate that between two consecutive kicks, the rotator moves freely, with its momentum \( p \) remaining constant and the angular position \( x \) increasing linearly with time. When a kick happens, the momentum changes instantly by an amount  \(K \sin x \), where \(x \) is the angular position at the time of the kick. 

The dynamics of the kicked rotator are described by the so-called \textbf{Chirikov-Taylor} SM \citep{Chirikov1979}, which we refer to as the SM in this chapter. These dynamics are governed by the following set of difference equations: 

\begin{equation} \label{eq:ssm}
    \begin{aligned}
x_{n+1} - x_n &= p_{n+1} \\
p_{n+1} - p_n &= K \sin(x_n)
\end{aligned}
\end{equation}
where \( x_{n+1} \) and \( p_{n+1} \) are the canonical coordinates at time \( t \). Here \(x\) [Fig.~\ref{fig5:FigA0}] corresponds to the angular position of the particle, and \(p\) denotes the radial velocity or the momentum after the \( n \)-th kick, and \(n = 0, 1, 2, \ldots\) denotes the discrete time steps, or number of iterations of the map. The force function is often described using sine or cosine functions. In our case, we set the force as \( K \sin(x_n) \). This formation expresses the time derivative of the position as equal to the momentum and describes the force as the time derivative of the momentum. Note that in the SM model \eqref{eq:ssm} we consider, the mass is set to \(m=1\) to simplify the EoM without affecting the fundamental dynamics of the system. In our study, due to the system's periodicity, adding or subtracting multiples of \(2\pi\) to the angular coordinate \(x\) does not affect the physical state of the kick rotor, so we impose modulo \(2\pi\) for \(x\) in \eqref{eq:ssm}. This further simplifies our analysis by ensuring equal positions on the circle in Fig.~\ref{fig5:FigA0}. On the contrary, the momentum \(p\) in \eqref{eq:ssm} can take any value, allowing unbounded motion in the momentum space, which is essential for observing chaotic dynamics and diffusion properties. 

We typically interpret of this map as describing the evolution of the rotation angle \(x_n\) and the angular momentum \(p_n\) of a periodically `kicked' pendulum rotating in a free field. The kick by a nonlinear force occurs at each time unit. The constant \(K\) measures the intensity of the nonlinear kick where the variables \(x_n\), \(p_n\), and the parameter \(K\) are all dimensionless. A set of \( (x_n, p_n) \) for a given \( K \) and several distinct ICs form what we define as a phase space plot, which we use in order to provide basic insight into the dynamics of the map \eqref{eq:ssm}. The relatively simple \(2D\) discrete model \eqref{eq:ssm} mimics complex dynamical behaviors of an autonomous Hamiltonian system with two DoF, such as chaos and bifurcation (e.g.~see \citep{lichtenberg2013regular}). Despite the obvious simplicity, the SM exhibits a wide variety of both regular and chaotic motions depending on the nonlinear kick strength, $K$.

\begin{figure}[!htbp]
     \centering
     \subfloat[\( K = 0.5\)\label{fig5:FigA1a}]{\includegraphics[width=0.33\textwidth]{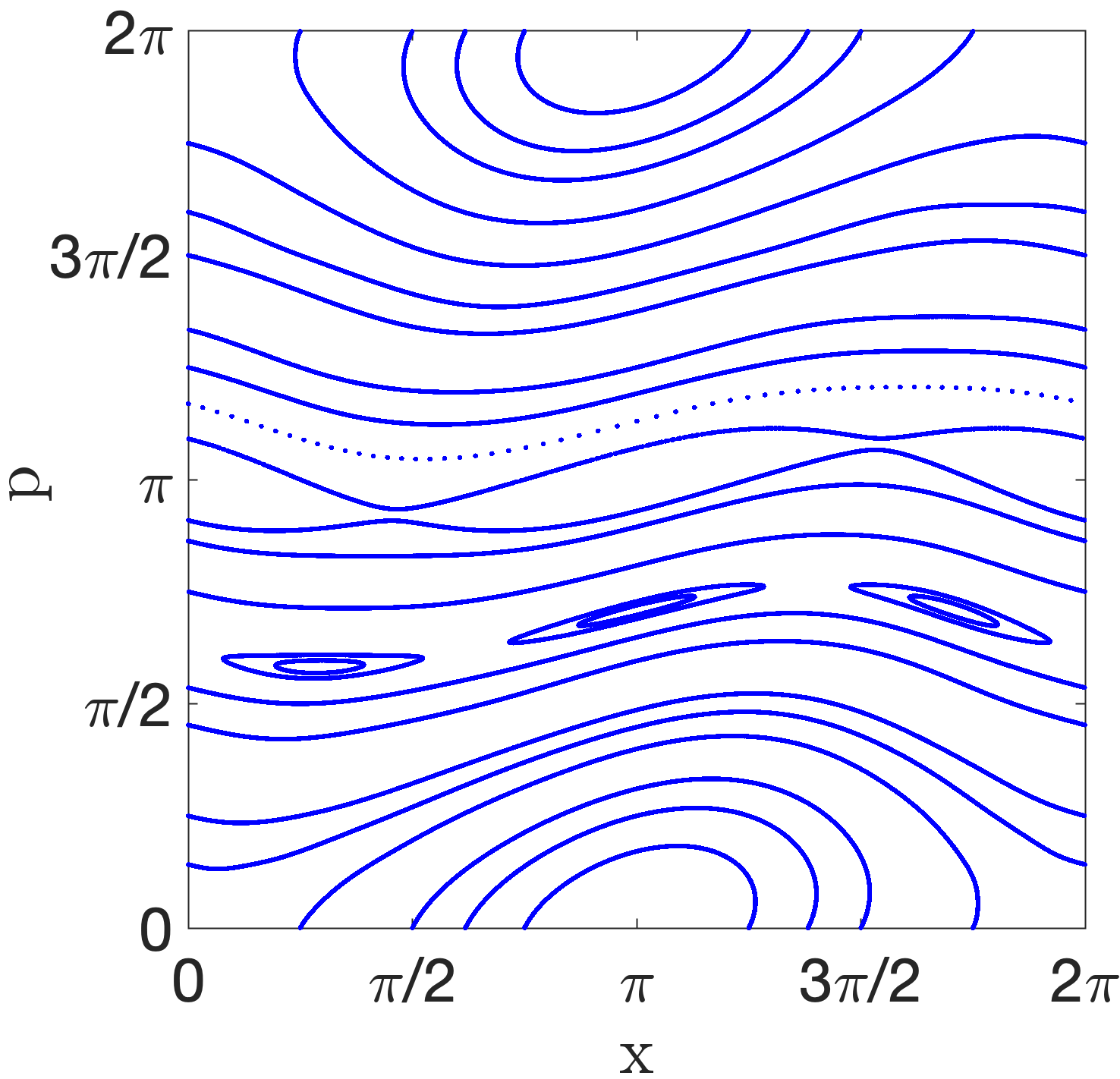}} 
     \subfloat[\( K = 1.5\)\label{fig5:FigA1b}]{\includegraphics[width=0.33\textwidth]{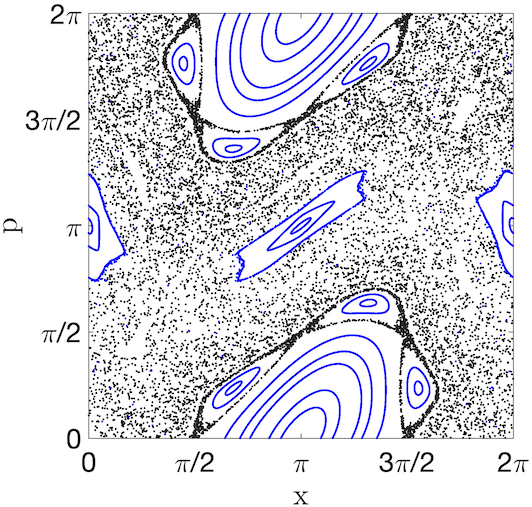}} 
     \subfloat[\( K = 6.5\)\label{fig5:FigA1c}]{\includegraphics[width=0.33\textwidth]{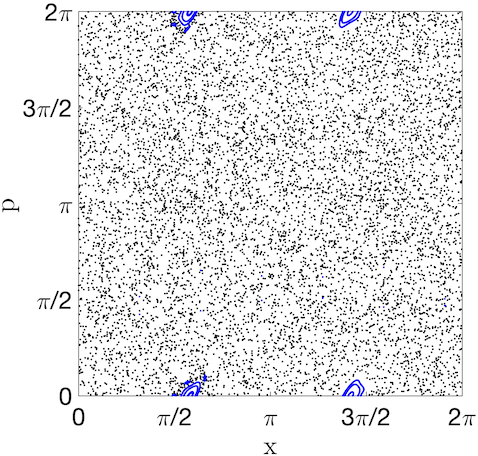}} 
       \caption{Representative phase space portraits of the SM \eqref{eq:ssm} for (a) \(K=0.5\), (b) \(K=1.5\) and (c) \(K=6.5\). The phase space portraits are produced using \(n = 10^5\) iterations of several ICs, where blue and black points represent regular and chaotic orbits, respectively.}
       \label{fig5:FigA1}
 \end{figure}

In the absence of external forces (\( K = 0 \)), the SM simplifies to a linear map, resulting in either quasiperiodic motion or the appearance of POs covering only portions of the \(2D\) phase plane. However, with the introduction of a non-zero nonlinearity kick \( K > 0 \), the dynamics of the map exhibit more complex behavior. These POs can be categorized into stable and unstable ones. Stable POs are surrounded by quasiperiodic orbits, while unstable POs coexist with chaotic regions. Figures \ref{fig5:FigA1a}, (b), and (c) present the phase space portraits of the SM \eqref{eq:ssm} for increasing nonlinearity kick strengths \(K=0.5\), \(K=1.5\) and \(K=6.5\), respectively, created by $10^5$ iterations of various ICs. In Fig.~\ref{fig5:FigA1a}, for a small parameter value \(K=0.5\), we observe that the phase space is practically filled with stability islands (blue points). As we increase the parameter value to \(K=1.5\) [Fig.~\ref{fig5:FigA1b}], a chaotic sea represented by the scattered points (black points) begins to occupy portions of the phase space. For higher \(K=6.5\) [Fig.~\ref{fig5:FigA1c}], we see that chaotic regions expand further, dominating a significant portion of the system's phase space. It is worth noting that with an additional increase in \(K\) value, the entire phase space of the SM \eqref{eq:ssm} becomes covered by scattered points, indicating complete chaotic behavior.

\section{Diffusion measures and properties} \label{section:HamiltonianCh5}
We can quantify the rate at which the system \eqref{eq:ssm} orbits spread out in phase space, or diffuse, mathematically. A common measure for this diffusion is the mean squared displacement (variance) of the angular momentum, \(\langle (\Delta p)^2 \rangle\). This quantity is expressed as a function of the number of iterations, \(n\), through the following power law relationship \citep{Chirikov1979,Meiss2015}:

\begin{equation} \label{eq:pvar}
\langle (\Delta p)^2 \rangle = D_\mu(K) \cdot n^\mu,
\end{equation}
where \(\langle (\Delta p)^2 \rangle = p_n - p_0\), with \(p_0\) being the initial angular momentum of the orbit, and \(\langle  (\Delta p)^2 \) is the average \((\Delta p)^2  \rangle\) over an ensemble of ICs at each iteration \(n\). 

The diffusion exponent \(\mu\), which we numerically estimate, lies in the interval \([0,2]\), reflecting the range of physically meaningful diffusion behaviors, from no diffusion (\(\mu = 0\)) to ballistic motion (\(\mu = 2\)). In general, \(\mu\) determines the type of diffusion: no diffusion when \(\mu = 0\), normal diffusion occurs when \(\mu = 1\), subdiffusion when \(0 < \mu < 1\), and superdiffusion when \(1 < \mu \leq 2\). The extreme case of \(\mu = 2\) represents ballistic transport, which is strongly associated with the presence of accelerator modes (AMs) \citep{Chirikov1979} in the system. AMs are stable structures that differ from the usual stability islands in that they are boosted by a constant amount in momentum with each iteration. In order to observe these AMs, we apply the \(\bmod \, 2\pi\) only on the \(x\) coordinate of the SM \eqref{eq:ssm}, while allowing the momentum \(p\) direction to take unrestricted positive values. 

\begin{figure}[!htbp]
    \centering
    \subfloat[Phase space portrait for \( K = 6.5\)\label{fig5:FigA2a}]{\includegraphics[width=0.49\textwidth]{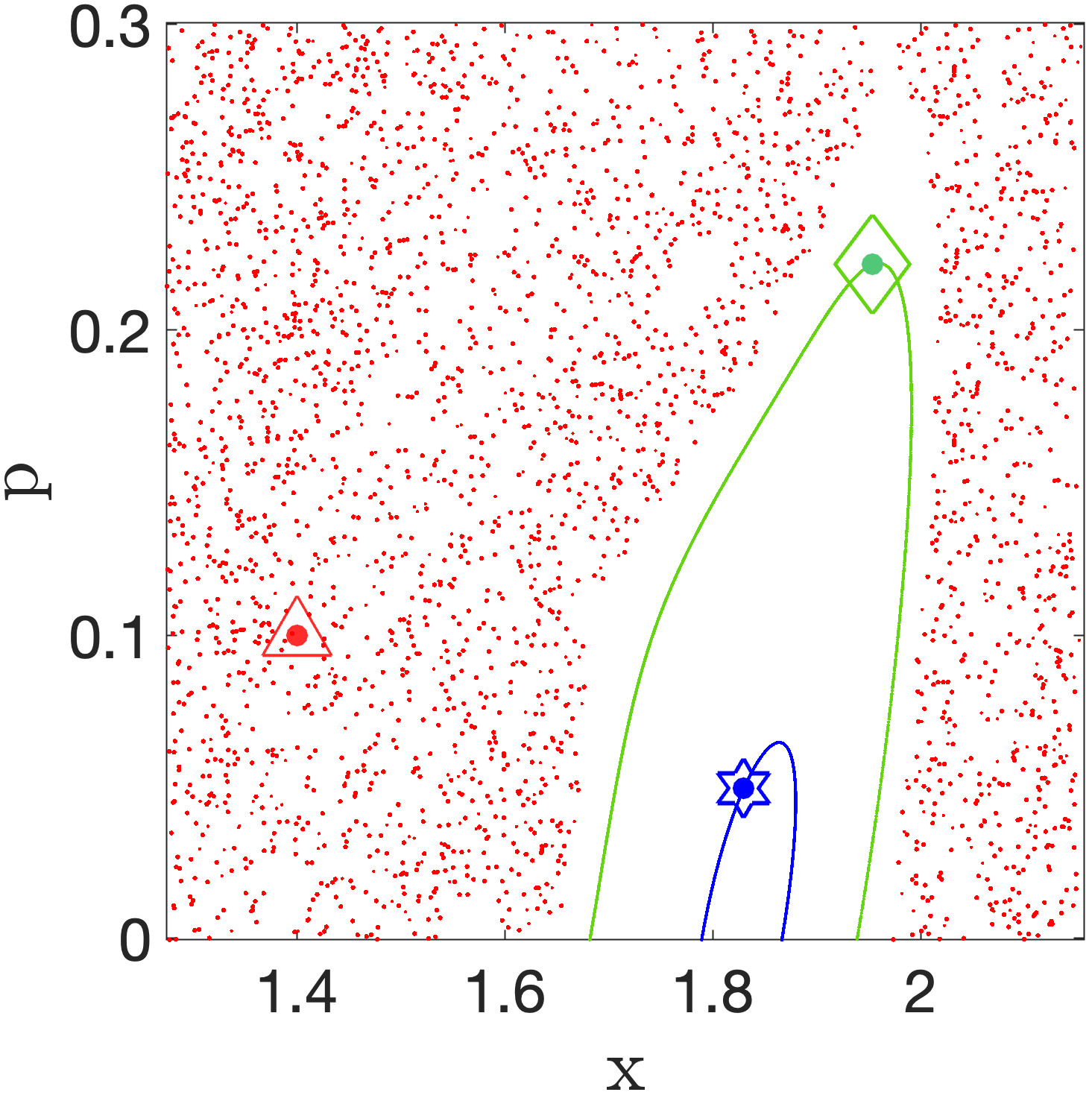}} 
    \subfloat[$\langle (\Delta p)^2 (n)\rangle$\label{fig5:FigA2b}]{\includegraphics[width=0.49\textwidth]{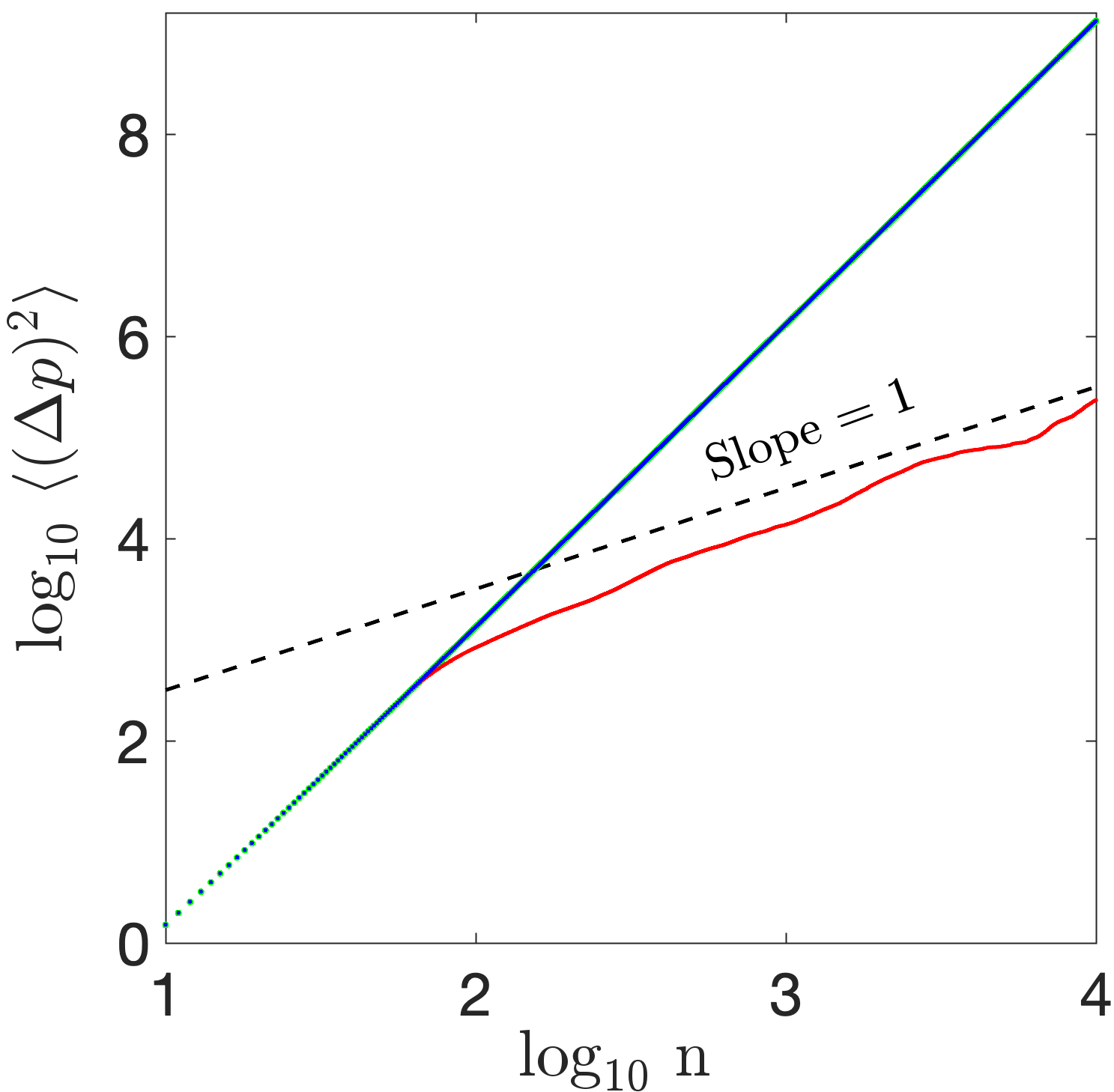}} 
    \label{fig5:A2}
    \caption{(a) The phase space portrait of the SM \eqref{eq:ssm} for \(K=6.5\), showing a chaotic orbit with IC  $(x_0, p_0) = (1.4, 0.1)$ (red triangle) and two obits near the stable period \(p=1\) AM of the system with ICs $(x_0, p_0) = (1.83, 0.05)$ (blue star) and $(x_0, p_0) = (1.95, 0.22)$ (green diamond). (b) The evolution of the variance of the angular momentum \(p\) for the orbits shown in (a). The dashed line represents power-law behavior \(n^{1}\). Note that the green and blue points overlap in (b).}
\end{figure}

The theoretical diffusion coefficient,  \(D_1(K)\), for normal diffusion (where \( \mu = 1\)) is given as follows \citep{Izrailev1990}
\begin{equation} \label{eq:Dcl}
    D_{1}(K)=
   \begin{cases}
    \dfrac{K^2}{2}  \left\{ 1- 2B_2(K) \left[ 1-B_2(K) \right] \right\}, & \text{if } \, K \ge 4.5, \\
    0.30(K-K_{cr})^3, & \text{if } \, K_{cr} < K < 4.5,
   \end{cases}
   \end{equation}
 where \(K_{\text{cr}} \approx 0.971635\) is the critical parameter value where the invariant curve of \eqref{eq:ssm} \citep{greene1979method} breaks and \(B_2(K)\) is a Bessel function of the first kind, which depends on \(K\). On the other hand, the theoretical result for the diffusion coefficient's (\(D_1\)) as a function of \(K\) fails in regions where AMs are present, particularly for period $p=1$ AM intervals. These intervals are defined by: 
\begin{equation} \label{eq:acmdint}
    2\pi m \leq K \leq \sqrt{(2\pi m)^2 + 16}.
\end{equation}
where \(m\) is any positive integer. In these intervals, the theoretical expression for (\(D_1\)), which assumes normal diffusion (characterized by \(\mu = 1\)), no longer follows the expression given by Eq.~\eqref{eq:Dcl}. Instead, the diffusion coefficient diverges as a result of the influence of the AMs. To illustrate this, we select a parameter value \(K=6.5\), where a period \(p=1\) AM of the SM \eqref{eq:ssm} exists  (i.e., \(m=1\) in the interval given in Eq.~\ref{eq:acmdint}). In this case, the stable AM is located at $(x_0, p_0) = (1.8298, 0)$, which will be discussed further in Sect.~\ref{section:ResultsCh5}. Fig.~\ref{fig5:FigA2a} shows the phase space plot for three representative orbits: one in the chaotic sea with IC $(x_0, p_0) = (1.4, 0.1)$ (red triangle) and, two near the stable AM with ICs $(x_0, p_0) = (1.83, 0.05)$ (blue star) and $(x_0, p_0) = (1.95, 0.22)$ (green diamond). The plot is essentially a zoomed-in view of Fig.~\ref{fig5:FigA1c}. 

In Fig.~\ref{fig5:FigA2b}, we present the angular momentum \(p\) for these three orbits as a function of the number of iterations. For the orbit in the chaotic sea (red points), the variance of \(p\) increases linearly over time (i.e., proportionally to $n^{1}$ represented by the dashed line), indicating normal diffusion. In contrast, the two orbits near the stable AMs [blue and green in Fig.~\ref{fig5:FigA2a}] deviate from a linear relation, exhibiting greater variance over time than that seen in normal diffusion. We will examine these details in Sect.~\ref{section:ResultsCh5}. Note that both orbits near the stable AM exhibit the same behavior due to the influence of the AM, resulting in overlapping green and blue points in Fig.~\ref{fig5:FigA2b}. We will analyze and compare the diffusion behaviors for orbits near the stable AMs and chaotic orbits as well as regular orbits and unstable AMs in detail in Sect.~\ref{section:ResultsCh5}.   

Now, in order to quantify the diffusion rate, we calculate the effective diffusion coefficient, \(D_{\text{eff}}\), using the formula \citep{Meiss2015}
\begin{equation}\label{eq:Deff_sm}
    D_{\text{eff}} = \frac{\langle (\Delta p)^2 \rangle}{n}.
\end{equation}
where \(\langle . \rangle\) denotes the average variance of the angular momentum \(p\) for an ensemble of ICs at each iteration \(n\). 

Although the theoretical definition of the effective diffusion coefficient, \(D_\mu\) \eqref{eq:pvar} involves an infinite time limit, we calculate \(D_{\text{eff}}\) \eqref{eq:Deff_sm} numerically for large but finite values of $n$. This practical approach allows us to estimate the diffusion exponent without requiring the theoretical limit (the idea is in a way similar to our reasoning for introducing ftmLEs instead of mLE in Section~\ref{section:LEs}). Hence, we can say that  \(D_{\text{eff}}\) differs from the theoretically derived diffusion coefficient, \(D_\mu\). Earlier studies (see \citep{batistic2013intermediate, ManRob2014PRE} and references therein for details) have shown that \(D_{\text{eff}}\) exhibits peaks corresponding to parameter values where specific orbit types, known as AMs, are present. These peaks deviate significantly from the theoretical predictions based on normal diffusion.

In order to investigate the global behavior of multiple closely interacting systems, we can extend the \(2D\) SM \eqref{eq:ssm} into a model consisting of $N$ coupled SMs. In these systems, each map is described by the following $2ND$ system \citep{KG1988}
\begin{equation}\label{eq:csm}
    \begin{aligned}
x^j_{n+1} &= x^j_n + p^j_{n+1} ,\\
p^j_{n+1} &= p^j_n + K_j \sin(x^j_n) - \beta [\sin(x^{j+1}_n - x^j_n) + \sin(x^{j-1}_n - x^j_n)],
\end{aligned}
\end{equation}
where \(j = 1, 2, \ldots, N\) indexes each map, \(K_j\) represents the nonlinearity strength of the $j$-th map, and \(\beta\) quantifies the coupling strength between neighboring \(2D\) maps. The choice of each \(K_j\) value is important since it sets the initial dynamical state of each \(2D\) SM before coupling. This choice influences the initial amount of regular and chaotic motion as well as the presence of AMs in each phase plane. To preserve the symplectic nature of the coupled system, we impose periodic boundary conditions in our study (\(x^0_{n} = x^N_{n}\) and \(x^{N+1}_{n} = x^1_{n}\)).

To investigate the diffusion properties of the coupled SM systems \eqref{eq:csm}, we can generalize the diffusion coefficient expression in \eqref{eq:Deff_sm}, defining the effective diffusion coefficient of the $2ND$ map when \(n \rightarrow \infty\) as follows:

\begin{equation} \label{eq:Deff_csm}
D^N_{\text{eff}} = \frac{1}{n} \sum_{j=1}^{N} \langle (p^j_{n+1} - p^j_{0})^2 \rangle = \frac{1}{n} \sum_{j=1}^{N} \langle (\Delta p_n^j)^2 \rangle.
\end{equation}
This expression enables us to calculate the variance of the momentum for each map and then average these variances across all $N$ maps. Although each map contains the same number of ICs, the specific locations of these IC ensembles may differ between the \(2D\) SMs.

\section{Numerical results} \label{section:ResultsCh5}
\subsection{Chaos and diffusion measures} \label{section:Ch5R1}
To demonstrate the efficiency of the GALI method and compare its efficiency to the mLE in identifying and quantifying chaos in the SM, we consider various types of orbits, including regular, chaotic, as well as stable and unstable AMs, in the $2D$ SM \eqref{eq:ssm}, similar to the simplified $2D$ GC Hamiltonian analysis presented in Sect.~\ref{section:Ch3R1}. In addition, we examine small regions surrounding these orbits in the phase space to classify their diffusion properties. By doing so, we can form a relationship between the dynamics of each orbit of the SM and the corresponding diffusion properties.

To  compute GALI\(_2\) \eqref{eq:GALI} and ftmLE \eqref{eq:ftmLE}, we perform all numerical simulations by iterating the SM \eqref{eq:ssm} alongside its corresponding tangent map given as follows:  
\begin{equation}
    \begin{aligned}    \label{eq:ssm TM}         
    \delta x_{n+1} - \delta x_n &= \delta p_{n+1}   \\
    \delta p_{n+1} - \delta p_n &=  K \delta x_n \cos (x_n)
    \end{aligned}
    \end{equation}
where the deviation vectors \([\delta x_n, \delta p_n]\) are defined by the perturbations \(\delta x_n\) and \(\delta p_n\) of the corresponding canonical coordinates \(x_n\) and \(p_n\).

Furthermore, we note that we apply the modulo \(2\pi\) to the coordinates \(x\) and  \(p\) for each iteration to prevent numerical issues that could arise from unbounded orbits. As discussed in Sect.~\ref{section:GALI}, GALI\(_2\) is the only computationally feasible GALI index for the $2D$ SM, so our analysis will focus solely on GALI\(_2\).

\begin{figure}[!htbp] 
    \centering
    \subfloat[Phase space portrait for \(K = 3.1\)\label{fig5:Fig01a}]{\includegraphics[width=0.33\textwidth]{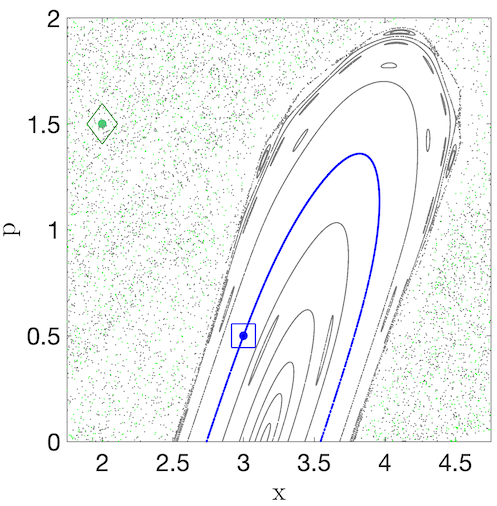}} 
    \subfloat[GALI\(_2 (n)\) for \(K = 3.1\)\label{fig5:Fig01b}]{\includegraphics[width=0.33\textwidth]{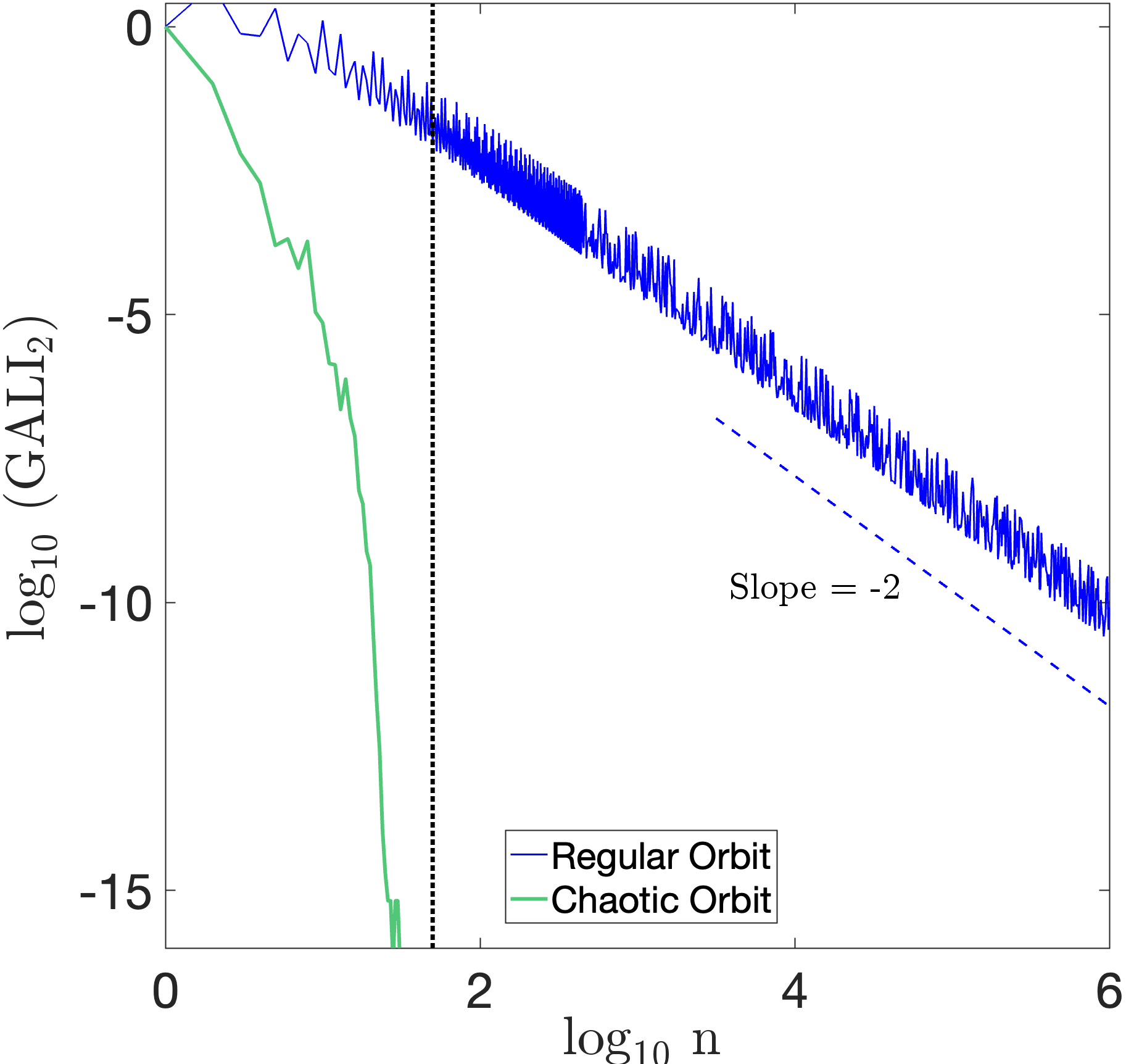}}
    \subfloat[ftmLE\((n)\) for \(K = 3.1\)\label{fig5:Fig01c}]{\includegraphics[width=0.33\textwidth]{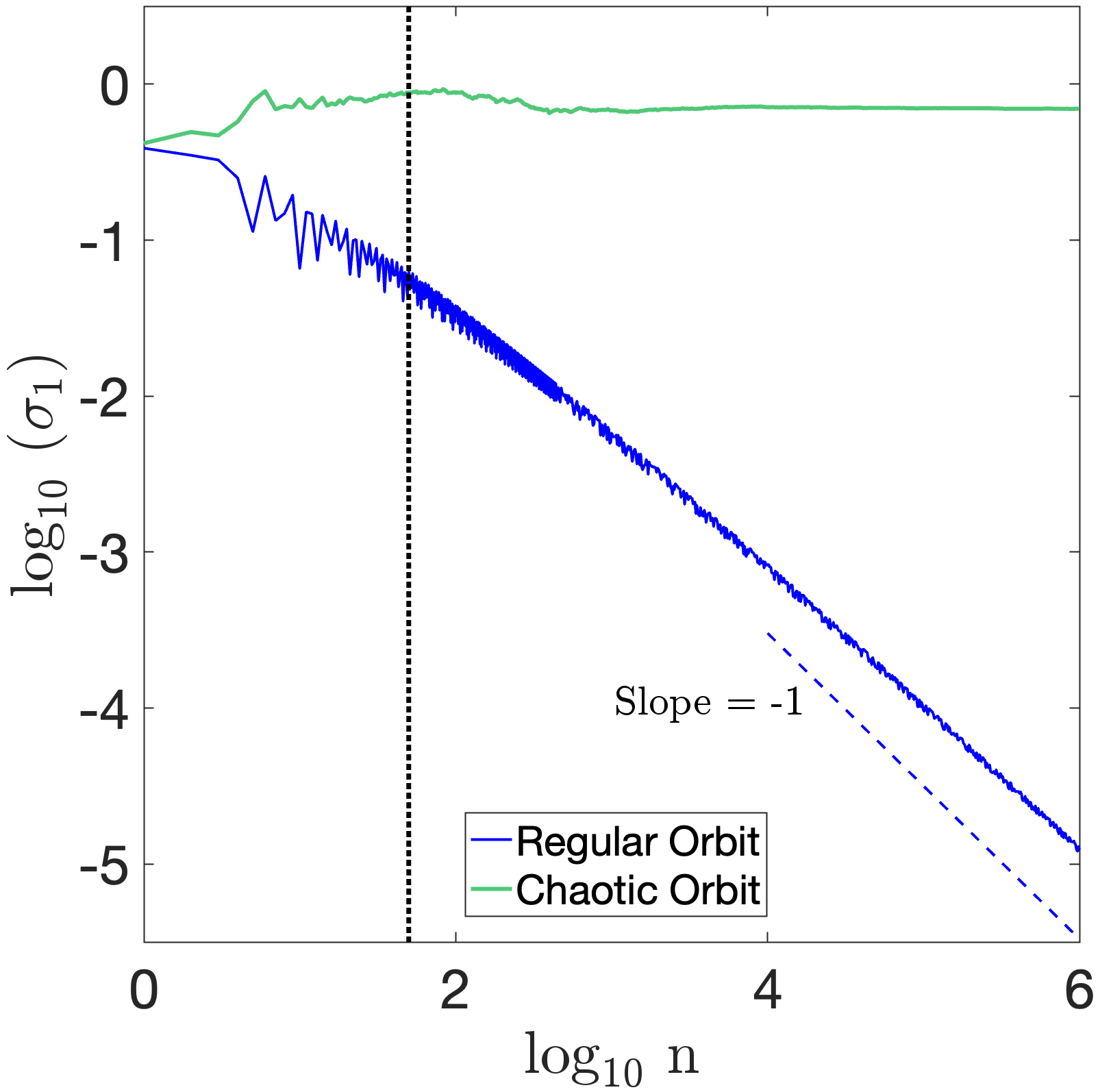}}\\
    \subfloat[Phase space portrait for \(K = 6.5\)\label{fig5:Fig01d}]{\includegraphics[width=0.33\textwidth]{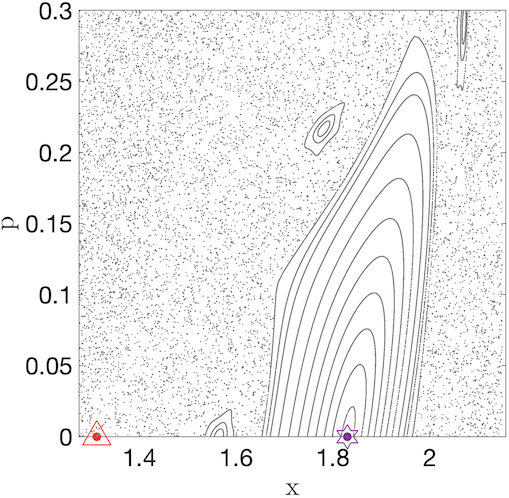}}
    \subfloat[GALI\(_2 (n)\) for \(K = 6.5\)\label{fig5:Fig01e}]{\includegraphics[width=0.33\textwidth]{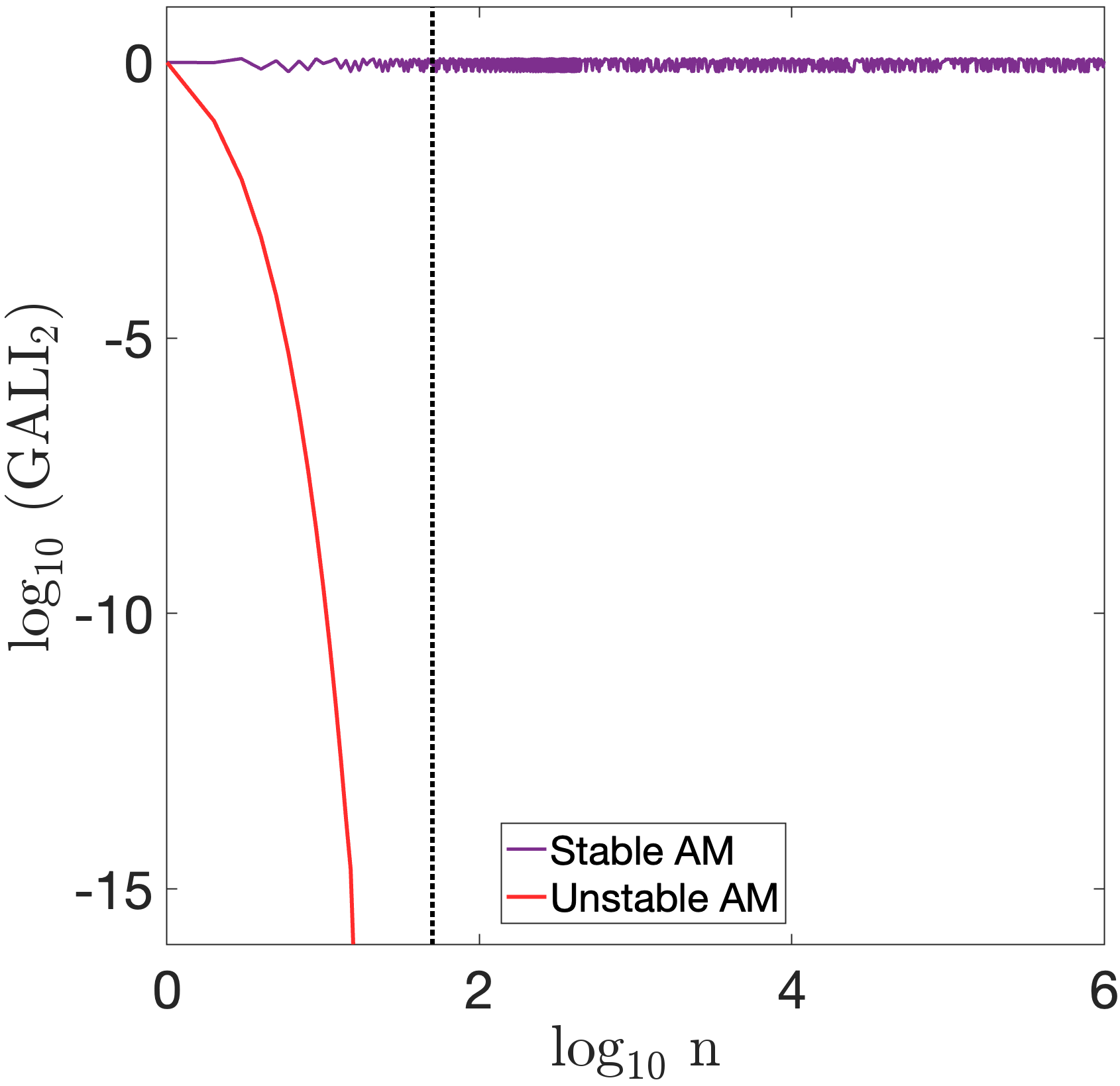}}
    \subfloat[ftmLE\((n)\) for \(K = 6.5\)\label{fig5:Fig01f}]{\includegraphics[width=0.33\textwidth]{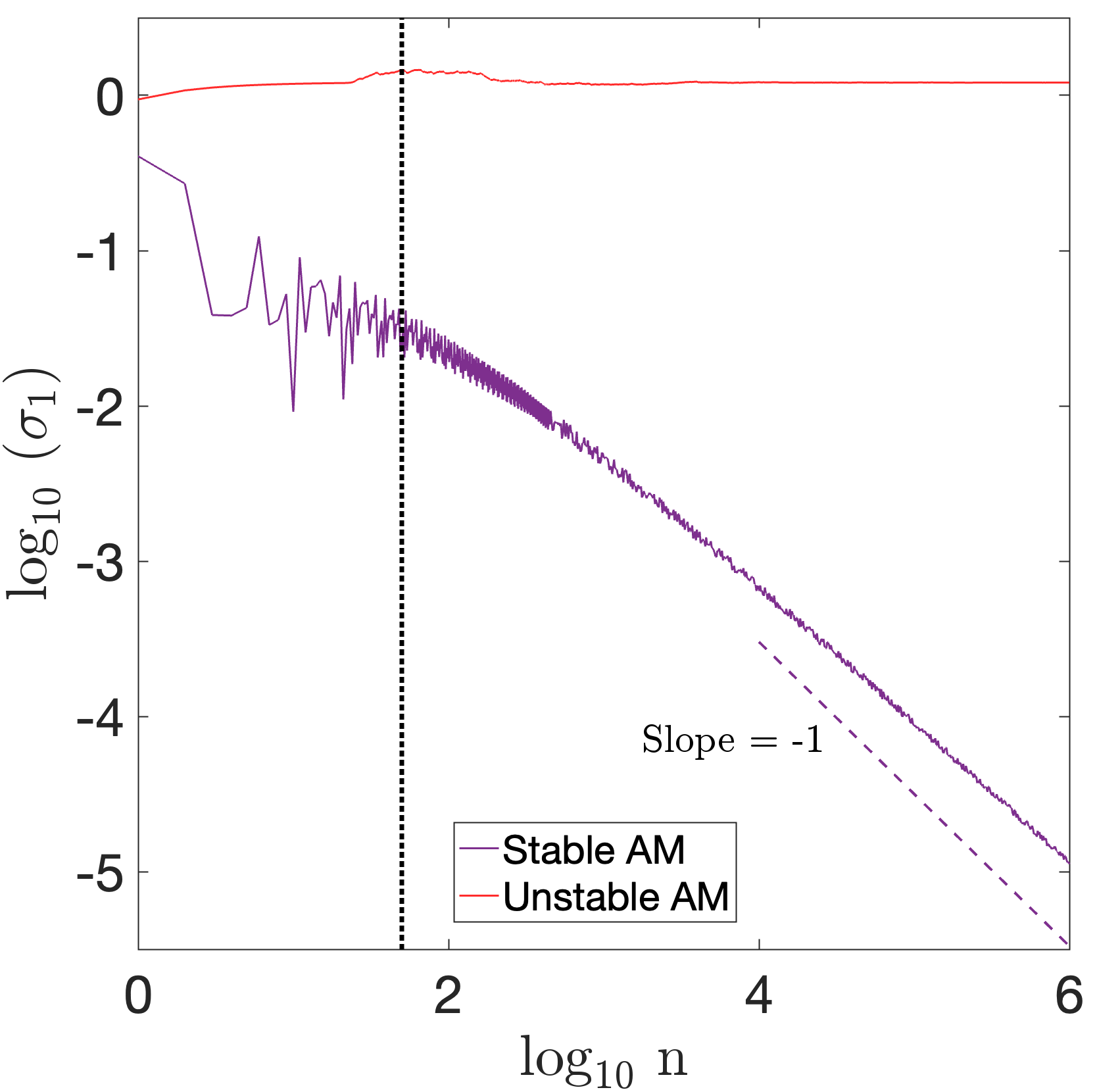}}
    \caption{Exemplary cases for the SM \eqref{eq:ssm} with [(a), (b), and (c)] for \( K = 3.1 \), and [(d), (e), and (f)] for \( K = 6.5 \). [(a) and (d)] phase space portraits generated by \( 10^6 \) iterations of various ICs (black points), with specific orbits highlighted: the chaotic orbit (green diamond) and the regular orbit (blue square) in (a), and the period $p=1$ stable AM (purple star) and the unstable AM (red triangle) in (d).  The evolution of the [(b) and (e)] GALI$_2$ \eqref{eq:GALI} and [(c) and (f)] ftmLE, \( \sigma_1 \) \eqref{eq:ftmLE}, as a function of the number of iterations, $n$, for the chaotic and regular orbits of (a) and the stable and unstable AMs of (d). The dashed line in (b) [(c) and (f)] corresponds to a function  \(\propto n^{-2} \) [\( \propto n^{-1} \)]. The black dotted vertical lines in (b), (c), (e), and (f) indicate $n=50$ (see the text for more details).}
 \label{fig5:Fig01}
\end{figure}

Figure \ref{fig5:Fig01a} displays the phase space plot of the SM with \(K=3.1\), produced by $10^6$ iterations of several ICs. The chaotic orbit with IC $(x_0, p_0) = (2, 1.5)$ (green diamond) is shown by green scattered points, while the consequents of the regular orbit with IC $(x_0, p_0) = (3, 0.5)$ (blue square) form the blue curve. The corresponding time evolution of the GALI\(_2\)  index as a function of the number of iterations $n$ is shown in Fig.~\ref{fig5:Fig01b}. The GALI\(_2\) exhibits exponential decay for the chaotic orbit (green curve) and a power law decay of the form $\text{GALI}_2 \propto n^{-2}$ (which corresponds to the dashed blue line) for the regular orbit (blue curve). This power law relation aligns with the expected behavior discussed in \eqref{Prop:GALI_2 for SM}. 

On the other hand, the time evolution of the ftmLE, \( \sigma_1 (n) \), in Fig.~\ref{fig5:Fig01c} shows a slower convergence to zero for the regular orbit (blue curve), following a power law decrease proportional to a function \(n^{-1}\) (indicated by the dashed blue line), while \( \sigma_1\) eventually saturates at a positive value for the chaotic orbit (green curve). Comparing Figs.~\ref{fig5:Fig01b} and \ref{fig5:Fig01c} highlights the efficiency of GALI\(_2\) in distinguishing between regular and chaotic orbits. In particular, the GALI\(_2\) of the chaotic orbit drops to practically zero (GALI$_2 \approx 10^{-16}$) after \(n \approx 50\) iterations, which is much smaller than the value observed for the regular orbit. However, we do not see a clear distinction from the ftmLE results shown in Fig.~\ref{fig5:Fig01c}. The ftmLE for the chaotic orbit (green curve) has not yet converged to its positive limit value by \(n \approx 50\) iterations, nor has the ftmLE for the regular orbit [blue curve in \ref{fig5:Fig01c}] shown a clear decreasing trend within this iteration interval.

It is important to note that the zero threshold GALI$_2 \approx 10^{-16}$ for the SM is significantly lower than the zero threshold we employed (GALI$_2 \le 10^{-8}$) for the GC Hamiltonian in Fig.~\ref{fig3:Fig2b}. One reason for this difference is that GC orbits often require longer simulation times to exhibit their chaotic behavior. In addition, the GALI\(_2\) remains constant for regular orbits of the GC Hamiltonian, in contrast to the case of the SM, where the GALI\(_2\) decreases to zero following a power law  \(n^{-2}\). The significant difference in the order of magnitude between the GALI\(_2\) values for regular and chaotic orbits makes it easy to distinguish between them.

Figure \ref{fig5:Fig01d} presents the phase portraits of the \(2D\) SM for stable and unstable AMs with a different parameter value, \(K=6.5\). The locations of these two period $p=1$ AMs are as follows: $(x_0, p_0) = (1.8298, 0)$ (purple star) for the stable AM and $(x_0, p_0) = (1.3118, 0)$ (red triangle) for the unstable AM. 
 The GALI\(_2\) values [Fig.~\ref{fig5:Fig01e}] show small oscillations close to a positive constant value for the stable AM and an exponential decay for the unstable AM, resembling the patterns observed for regular and chaotic orbits, respectively, in \citep{manos2012probing}. Fig.~\ref{fig5:Fig01f} demonstrates that the ftmLE for the stable (purple curve) and unstable (red curve) AMs behaves similarly to the regular and chaotic orbits shown in Fig.~\ref{fig5:Fig01c}.

As we can clearly see in Fig.~\ref{fig5:Fig01c}, the evolution of the ftmLE, denoted by the blue curve, corresponds to a typical regular orbit with \(\sigma_1 \propto n^{-1}\) (see Sect.~\ref{section:LEs}). One of the main goals of this chapter is to investigate the long-term global dynamics of SMs. Since  it is not practical to analyze individual orbits, we focus on studying the collective behavior of many orbits. For this purpose, it is important to establish an approach that will allow us to distinguish between regular and chaotic orbits based solely on the numerical values of the indices, without requiring us to analyze their time evolution. 

For $n=50$ (indicated by the black vertical dotted lines in Fig.~\ref{fig5:Fig01}), we observe that  $\log_{10}(\sigma_1) \approx -1.2042$ for the regular orbit while $\log_{10}(\sigma_1) \approx -0.0559$ for the chaotic orbit [Fig.~\ref{fig5:Fig01c}]. The difference between the orders of magnitude of the ftmLE values between the two orbits is approximately two. In comparison, the difference observed for the GALI$_2$ values at the same iteration is significantly greater, i.e. $\log_{10}\left({\rm GALI}_2\right) \approx-2.0571$ for the regular orbit and $\log_{10}\left({\rm GALI}_2\right) \approx -16.227$ for the chaotic orbit.  Furthermore, we note that the GALI$_2$ is particularly efficient in distinguishing between regular and chaotic orbits, especially in cases where the chaotic and sticky orbits exhibit a transient phase with decreasing \(\sigma_1\) values similar to those observed for regular orbits (see Fig.~\ref{fig3:Fig2a} and related discussions in Sect.~\ref{section:Ch3R1}). 

In order to determine the rate at which orbits of the system \eqref{eq:ssm} diffuse in the phase space, we analyze the behavior of many ICs around different types of orbits. In particular, we numerically evaluate the diffusion exponent $\mu$ in Eq.~\eqref{eq:pvar}. For a specific region of the SM's phase space, we use a grid of $315 \times 315$ evenly spaced ICs, i.e., in total approximately \(100,000\) ICs. We then compute the evolution of the variance $\langle (\Delta p)^2 \rangle$ as a function of the map's iterations $n$. To identify potential changes in the diffusion behavior over time, we apply a locally weighted regression smoothing algorithm \citep{CD1988} to the resulting data, which helps to reduce noise and provides a clearer view of the underlying diffusion behavior. In this work, we employ the numerical approach outlined in \citep{LBKSF2010,BLKSF2011} in order to compute \(\mu\). This approach involves applying a moving average procedure in a small window size along the total simulation time, followed by performing a linear fit on the logarithm of $\langle (\Delta p)^2 \rangle$ with respect to the logarithm of $n$. 

\begin{figure}[!htbp]   
    \centering
    \subfloat[$\langle (\Delta p)^2 (n) \rangle$\label{fig5:Fig02a}]{\includegraphics[width=0.475\textwidth]{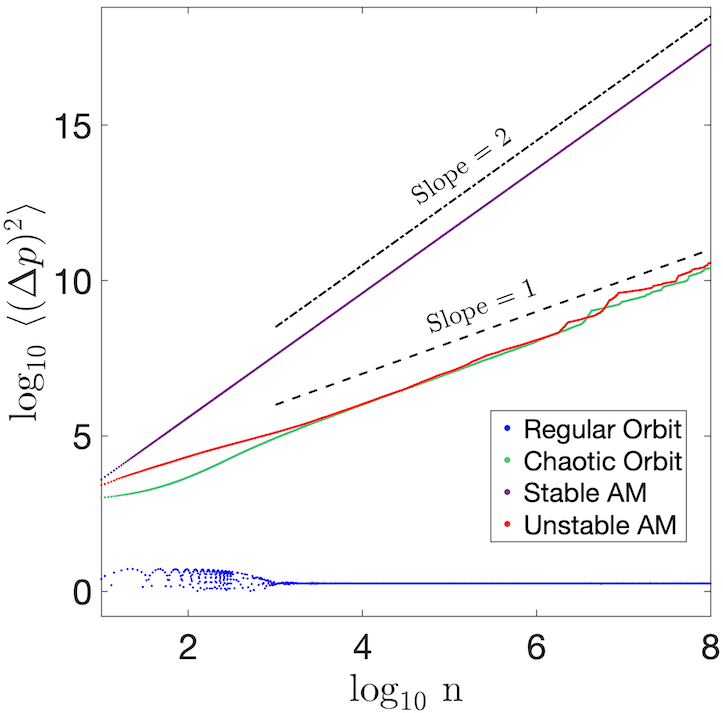}} 
    \subfloat[$\mu (n)$\label{fig5:Fig02b}]{\includegraphics[width=0.475\textwidth]{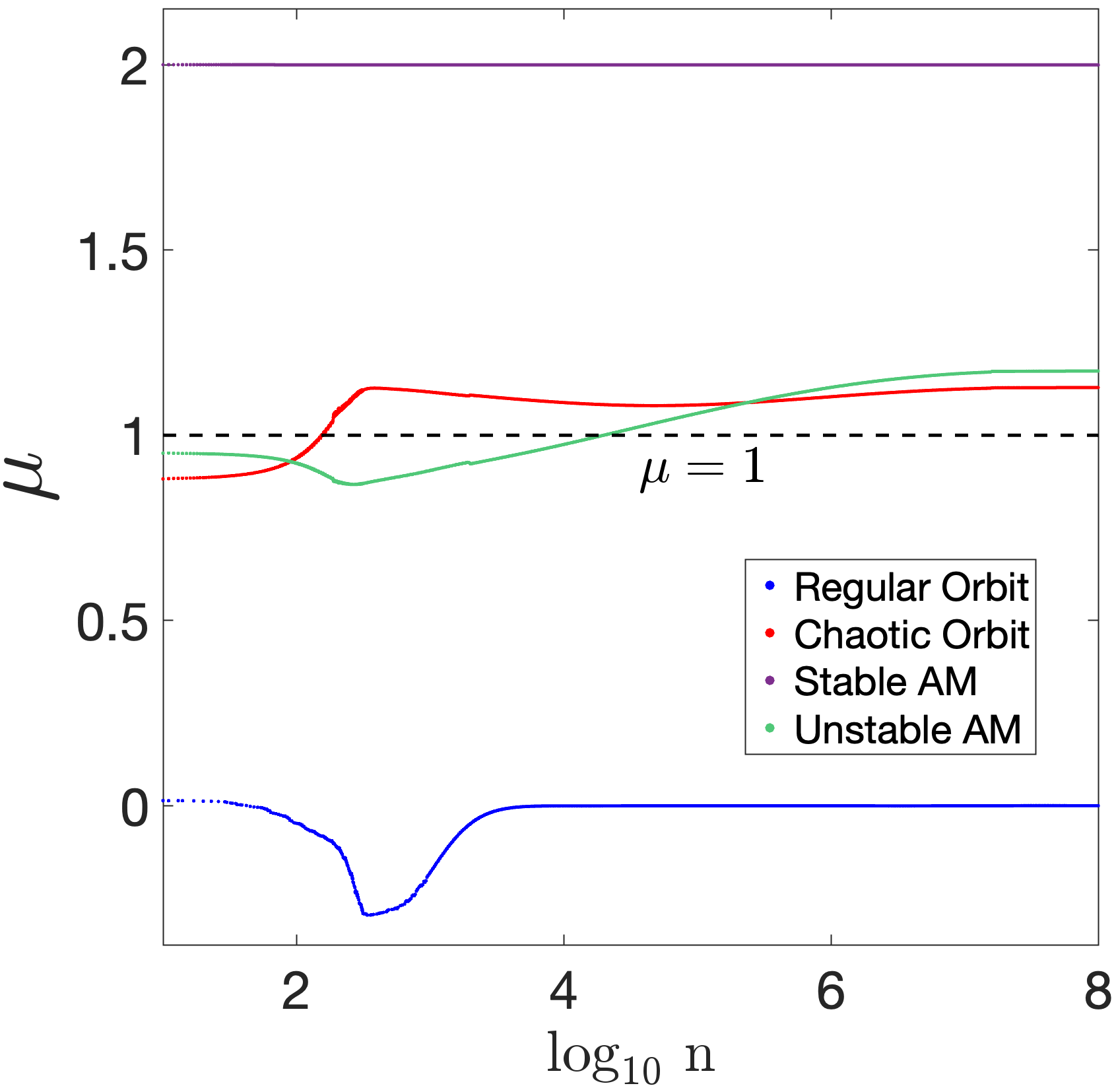}}
    \caption{(a) The variance of the momentum, $\langle (\Delta p)^2 \rangle$, as a function of the number of iterations, $n$, for small regions surrounding the four different orbits shown in Fig.~\ref{fig5:Fig01}. Each curve represents the average behavior of approximately \(100,000\) ICs arranged in a grid of the \((x, p)\) space, where the \(x\) coordinates are defined in\(\mod \, 2\pi\), while the \(y\) coordinates get positive values without restrictions, i.e. \(p \in (0, \infty)\). The blue curve corresponds to the region $[2.995, 3.005] \times [0.495, 0.505]$ around the regular orbit, the green curve to the region $[1.995, 2.005] \times [1.495, 1.505]$ surrounding the chaotic orbit, the purple curve for the region $[1.825, 1.835] \times [0.0, 0.01]$ in the vicinity of the stable AM of period $p=1$, and the red curve to the region $[1.295, 1.315] \times [0.0, 0.01]$ in the neighborhood of the unstable AM. The dashed-dotted and dashed lines indicate power-law behaviors $n^2$ and $n^1$, respectively. (b) Numerical estimation of the diffusion exponent, $\mu$, based on Eq.~\ref{eq:pvar}, computed from the curves in (a). The dashed horizontal line corresponds to normal diffusion, $\mu = 1$.}
    \label{fig5:Fig02}
   \end{figure}

In Fig.~\ref{fig5:Fig02a}, we present the results of this analysis for sets of orbits with ICs arranged in a grid of the phase space \((x, p)\) located in small regions around the four specific orbits discussed in Fig.~\ref{fig5:Fig01}. For $K = 3.1$, we consider the regions $[2.995, 3.005] \times [0.495, 0.505]$ (containing the IC of the regular orbit, shown in the blue curve) and $[1.995, 2.005] \times [1.495, 1.505]$ (containing the IC of the chaotic orbit, shown in the green curve) as depicted in Fig.~\ref{fig5:Fig01a}. For $K = 6.5$, we analyze the regions $[1.825, 1.835] \times [0.0, 0.01]$ (containing the stable AM, shown in the purple curve) and $[1.295, 1.315] \times [0.0, 0.01]$ (containing the unstable AM, shown in the red curve) as illustrated in Fig.~\ref{fig5:Fig01d}.
  
The results depicted in Fig.~\ref{fig5:Fig02a} show that the system \eqref{eq:ssm} exhibits ballistic transport in the region surrounding the stable AM of period $p=1$, where spreading occurs very fast. The increase in $\langle (\Delta p)^2 \rangle$ (purple curve) follows a function proportional to $n^2$ (dash-dotted line), which corresponds to a diffusion exponent $\mu = 2$ in Fig.~\ref{fig5:Fig02b}. On the other hand, orbits near the unstable AM (red curve) and the chaotic orbit (green curve) indicate normal diffusion ($\mu = 1$). In this case, both the red and green curves show a growth close to $\langle (\Delta p)^2 \rangle \propto n$, represented by the dashed line in Fig.~\ref{fig5:Fig02a}. As expected, orbits near the regular orbit (blue curve) display almost no diffusion, which is a characteristic of confined motion. These findings highlight the relationship between the type of orbit we considered and the corresponding diffusion behavior in the  system's phase space around these orbits. Note that the overlap of curves in Fig.~\ref{fig5:Fig02} reflects the results obtained for the neighborhoods of the chaotic (red curve) and the unstable AM (green curve). 

Our analysis in Fig.~\ref{fig5:Fig02b} reveals distinct diffusion rates associated with the different orbit types considered. Overall, our observations are as follows:

\begin{itemize}
    \item \textbf{Ballistic transport ($\mu = 2$)}: The ensemble of orbits near the stable AM with $K = 6.5$ exhibit ballistic transport, which is characterized by a rapid increase in the variance of angular momentum over time (purple curves in Fig.~\ref{fig5:Fig03}), corresponding to a diffusion exponent of $\mu = 2$.
    \item \textbf{Normal diffusion ($\mu= 1$)}: The collection of orbits around the chaotic orbit (green curves in Fig.~\ref{fig5:Fig03}) with $K = 3.1$ and the unstable AM (red curves in Fig.~\ref{fig5:Fig03}) with $K = 6.5$ display normal diffusion, where $\mu$ approaches $1$.
    \item \textbf{No diffusion ($\mu = 0$)}: The orbits associated with the regular orbit (blue curves in Fig.~\ref{fig5:Fig03}) for $K = 3.1$ show no significant diffusion, which is indicated by a $\mu$ value converging to $0$.
\end{itemize}

It is important to note that while typical regular orbits and stable AMs share similar dynamical properties, as illustrated by their ftmLEs' evolution (blue and purple curves of Figs.~\ref{fig5:Fig01c} and \ref{fig5:Fig01f}, respectively), their diffusion rates differ significantly. Regions around regular orbits exhibit negligible diffusion ($\mu \approx 0$), whereas those close to stable AMs show ballistic diffusion rates ($\mu \approx 2$), as we clearly observed in Fig.~\ref{fig5:Fig02}.

It is also worth pointing out that the diffusion exponent generally stabilizes after an initial transient period. For ballistic and normal diffusion, this stabilization typically occurs around $n \approx 10^3$ iterations. However, the application of a moving average technique allows us to detect any potential fluctuations in $\mu$ values throughout the numerical simulation. This averaging approach not only helps to identify potential changes in the diffusion process but also detects any anomalous behavior that may occur under different conditions. By applying this technique, we will gain insight into the evolution of diffusion in the system, and we can determine the required number of iterations needed for accurate diffusion characterization. This numerical computation of \(\mu\) will become more clear in subsequent figures, such as Fig.~\ref{fig5:Fig04b} and Fig.~\ref{fig5:Fig07}.

\subsection{Single standard map} \label{sec:single SMs}
In order to establish a foundation for understanding the global diffusion properties and dynamics of coupled SMs, we start with a detailed analysis of the $2D$ SM \eqref{eq:ssm}. In this section, we focus on  the long-term behavior of orbits and examine how specific orbit types of the SM, particularly AMs with different periods $p$, influence the system's dynamics and transport properties.

To get a comprehensive understanding of the diffusion process, we extend the previous study by \citep{ManRob2014PRE} by analyzing the behavior of many ICs across the entire phase space. In particular, we compute the diffusion exponent, $\mu$ [using Eq.~\eqref{eq:pvar}], as a function of the nonlinearity parameter $K$ for a $315 \times 315$ grid of evenly spaced ICs covering the entire phase space of the SM, where $x, p \in (0, 2\pi)$. For each IC, we iterate the map \eqref{eq:ssm} up to $n = 10^7$ to investigate how $\mu$ depends not only on $K$ but also on the number of iterations $n$.

\begin{figure}[!htbp]
    \centering
    \subfloat[\(\mu (K)\)\label{fig5:Fig03a}]{\includegraphics[width=0.475\textwidth]{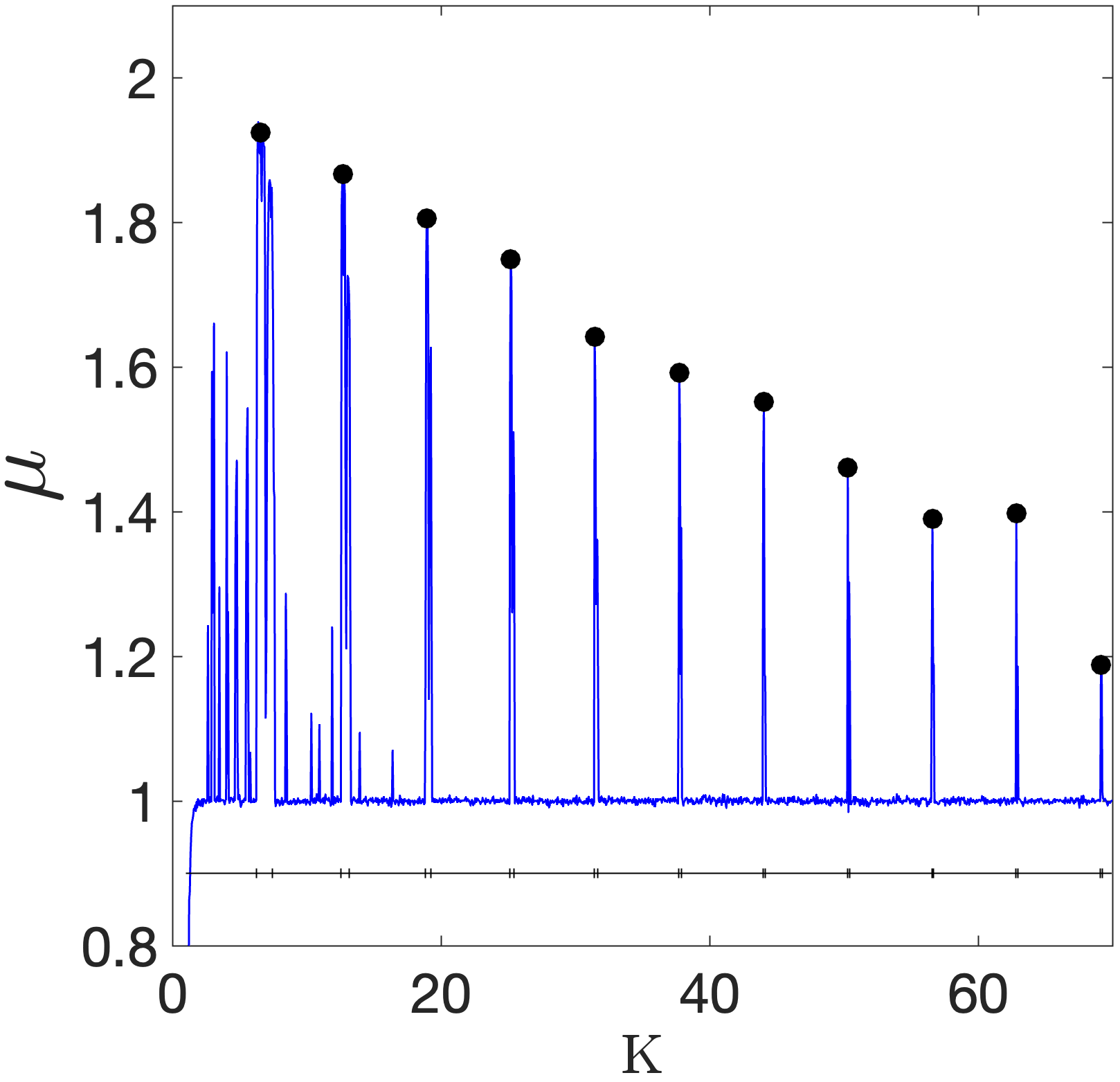}} 
    \subfloat[\( D_{\text{eff}} (K)\)\label{fig5:Fig03b}]{\includegraphics[width=0.485\textwidth]{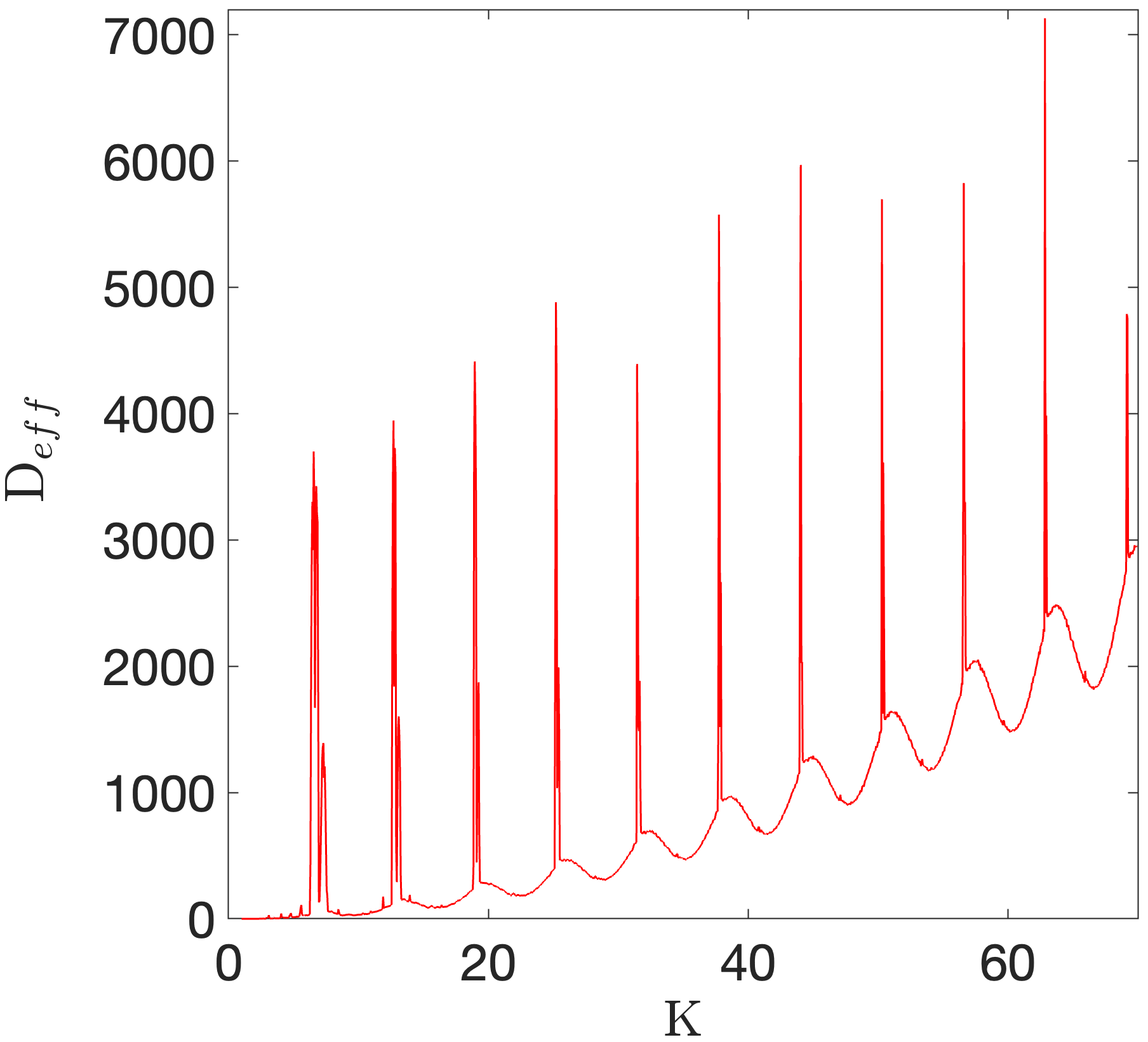}} \\
    \subfloat[\(\mu (K)\)\label{fig5:Fig03c}]{\includegraphics[width=0.485\textwidth]{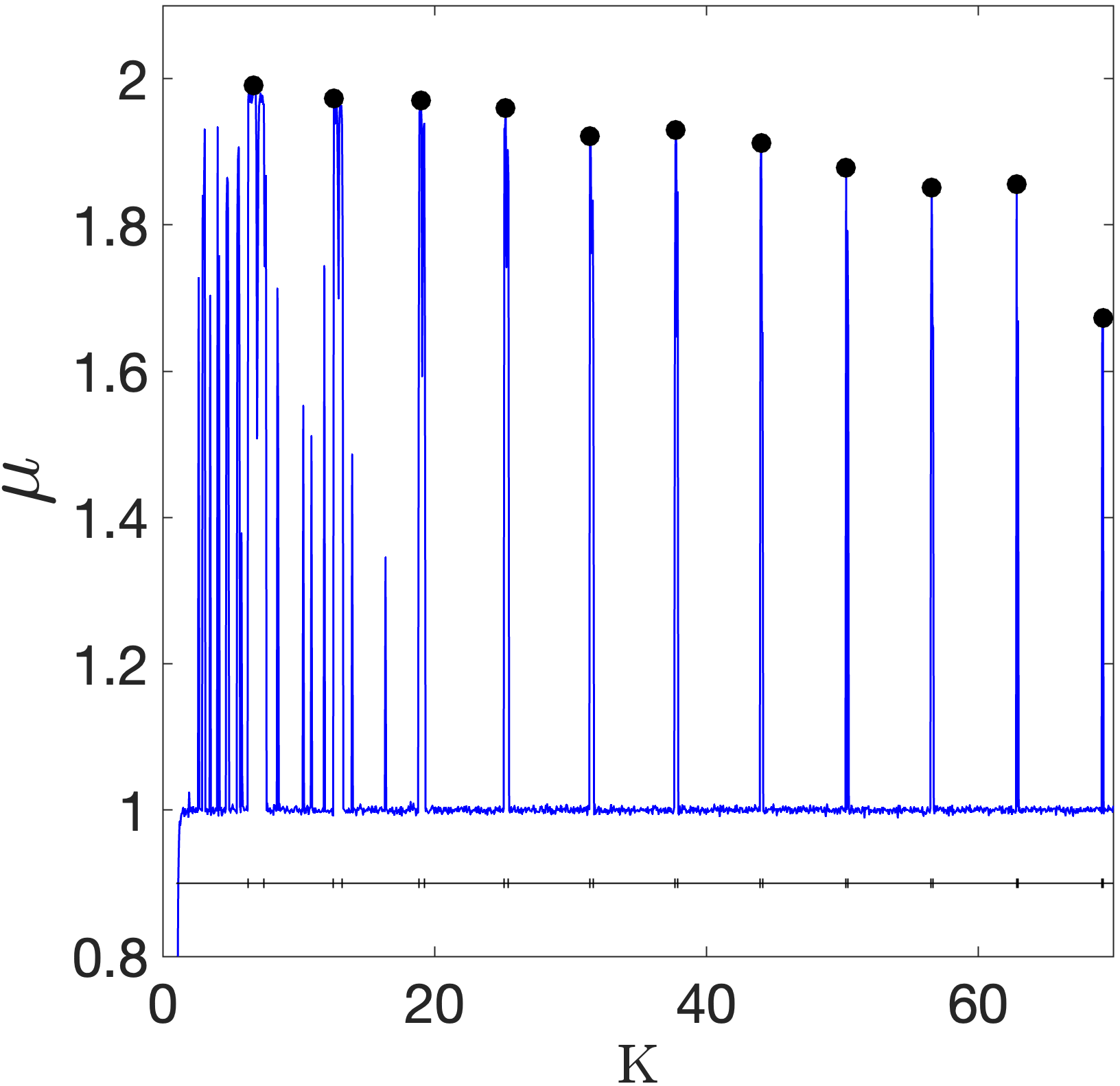}} 
    \subfloat[\( D_{\text{eff}} (K)\)\label{fig5:Fig03d}]{\includegraphics[width=0.475\textwidth]{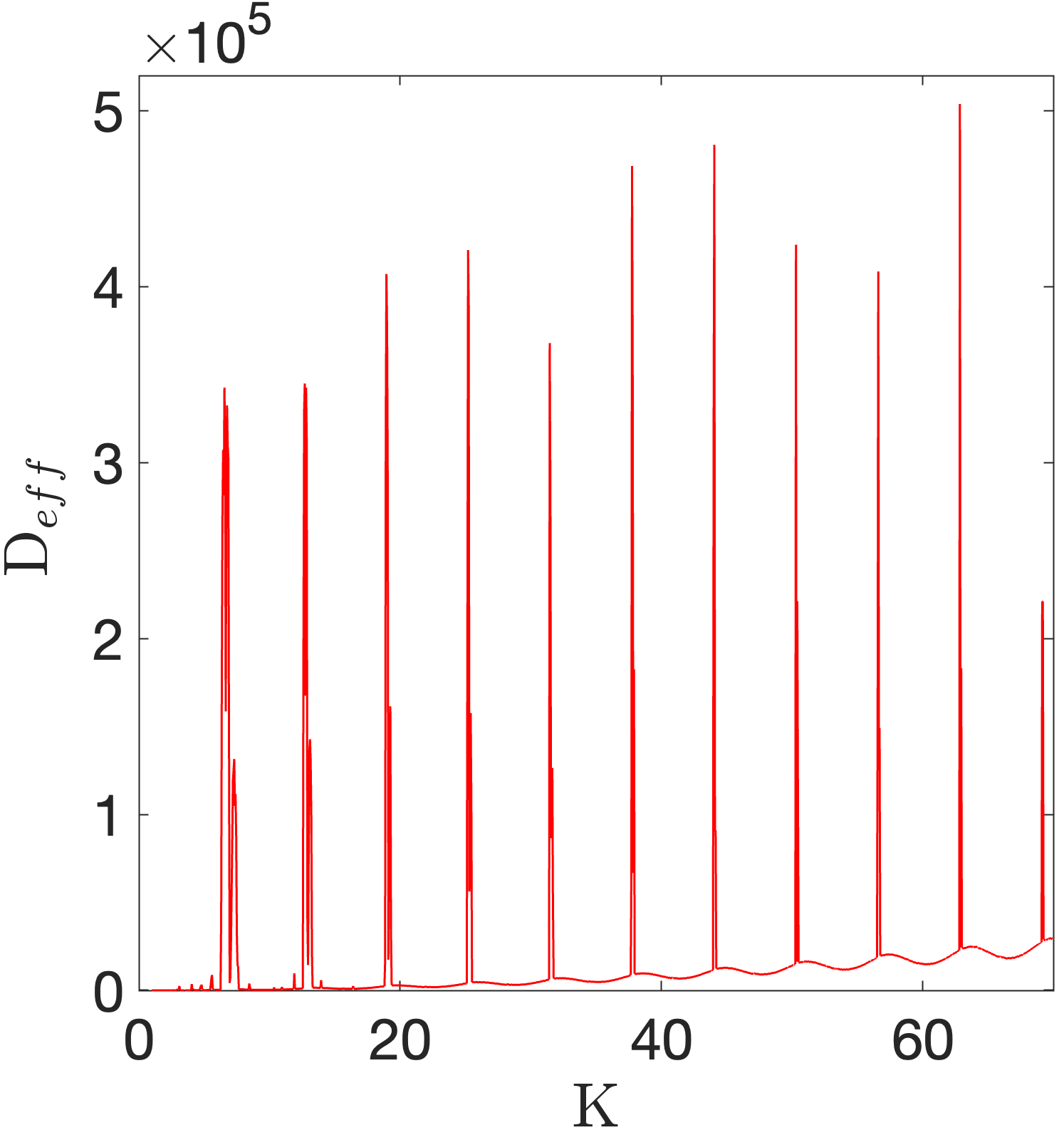}}
    \caption{[(a) and (c)] The diffusion exponent \( \mu (K) \) \eqref{eq:pvar} and [(b) and (d)] effective diffusion coefficient \( D_{\text{eff}} (K)\) \eqref{eq:Deff_sm} of the SM after [(a) and (b)] \( n = 10^4 \) and [(c) and (d)] \( n = 10^5 \) iterations. Both quantities are computed for approximately \(100,000\) ICs evenly distributed in a grid covering the map's entire phase space, with \(x \in (0, 2\pi)\) and \(p \in (0, 2\pi)\). At the bottom of (a) and (c), we indicate using small perpendicular segments the intervals where period $p=1$ AMs exist. When \( \mu \approx 1\), the system exhibits normal diffusion, whereas for the \( \mu \) values at the observed peaks marked by black circles for \( K > 2\pi \), the system displays anomalous diffusion due to the presence of period $p=1$ AMs. These same AMs are also associated with the peaks of the \(D_{\text{eff}}\) values in (b) and (d).}
    \label{fig5:Fig03}
\end{figure}

Figures \ref{fig5:Fig03a} and \ref{fig5:Fig03c} depict the diffusion exponent $\mu$ as a function of $K$ over a wide interval, $K \in [1, 70]$, for $n = 10^4$ and $n = 10^5$ iterations, respectively. To accurately compute $\mu$, we perform a linear fit of the logarithm of the variance momentum, $\langle (\Delta p)^2 \rangle$, with respect to the logarithm of the number of iterations, $n$, for each set of considered orbits, focusing on the final two decades of the simulation. For example, in the case of Fig.~\ref{fig5:Fig03c}, the fitting process is conducted on the variance momentum data over the iteration interval $10^3 < n \le 10^5$. This approach, which was previously used, for e.g., in \citep{ManRob2014PRE}, enables us to analyze how the diffusion exponent $\mu$ varies with the nonlinearity $K$ over a large range  and across increasing numbers of iterations. It also provides a more detailed picture of the diffusion processes in the SM's $2D$ phase space.

High peaks in the $\mu$ values correspond to the parameter intervals, defined by Eq.~\eqref{eq:acmdint}, where period $p=1$ AMs exist. These intervals are indicated by the horizontal black line at the bottom of Figs.~\ref{fig5:Fig03a} and \ref{fig5:Fig03c}. The peaks represent regions of anomalous diffusion, with maximum $\mu$ values approaching $\mu = 2$, which is a characteristic of ballistic transport. Within the considered range of $K \leq 70$, we observe \(11\) distinct peaks. The maximum values of $\mu$ at these peaks, denoted as $\mu^*$, are highlighted by black filled circles in Figs.~\ref{fig5:Fig03a} and \ref{fig5:Fig03c}. Comparing these plots, we observe that the associated $\mu^*$ values increase as the number of iterations \(n\) grows, implying a clear pattern of enhanced diffusion over time. 

On the other hand, regions with $\mu$ values close to \(\mu = 1\) correspond to normal diffusion. The additional peaks observed for $K \lesssim 4\pi$ are attributed to the influence of AMs of higher periods ($ p > 1$) on the SM's dynamics. It is important to note that the calculated effective diffusion coefficient, $D_{\text{eff}}$ [Eq.~\ref{eq:Deff_sm}], exhibits similar behavior [Figs.~\ref{fig5:Fig03b} and (d)], with pronounced peaks appearing at the same intervals of $K$ values. To keep the presentation simple, we omit the maximum peaks (marked by black circles) and parameter intervals (shown by the horizontal black line at the bottom) from Figs.~\ref{fig5:Fig03b} and (d), which are highlighted in Figs.~\ref{fig5:Fig03a} and (c).

The findings in Fig.~\ref{fig5:Fig03} highlight the fact that the ballistic motion, resulting from the presence of AMs, governs the global diffusion behavior of the SM. Consequently, the entire orbits in the phase space display anomalous diffusion, with the diffusion exponent ultimately converging to the maximum value of  \(\mu = 2\). On the other hand, the effective diffusion coefficient $D_{\text{eff}}$ diverges towards infinity in these regions, a similar result to that was reported in \citep{contopoulos2018}. Note that $D_{\text{eff}} \rightarrow \infty$ even for very large values of $K$ because AM islands remain present regardless of how large $K$ becomes. It is also worth noting that as $K$ increases, the size of the island surrounding stable AMs becomes progressively smaller.

\begin{figure}[!htbp]
    \centering
    \subfloat[$\langle (\Delta p)^2 (n)\rangle$\label{fig5:Fig04a}]{\includegraphics[width=0.475\textwidth]{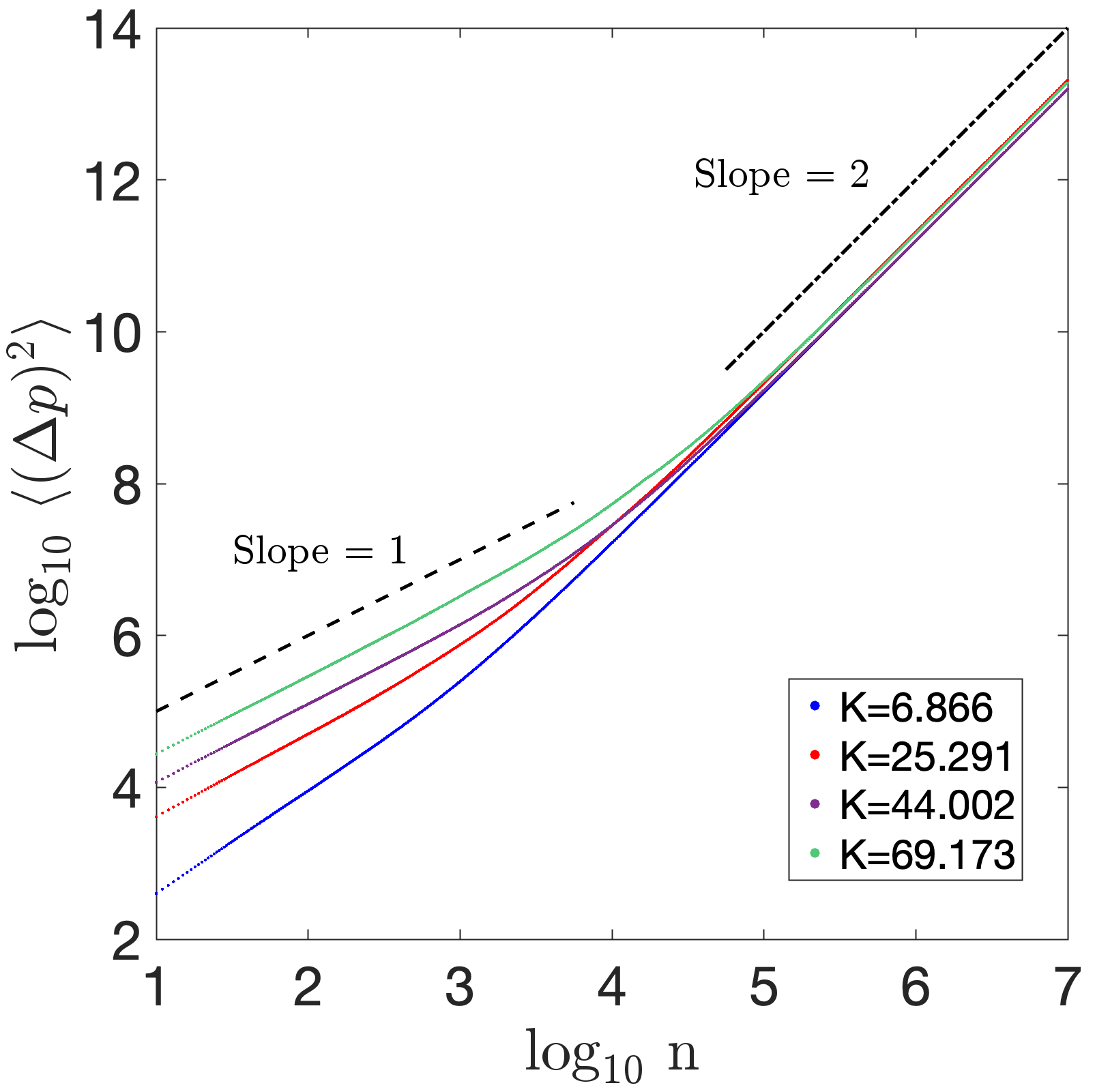}} 
    \subfloat[$\mu (n)$\label{fig5:Fig04b}]{\includegraphics[width=0.475\textwidth]{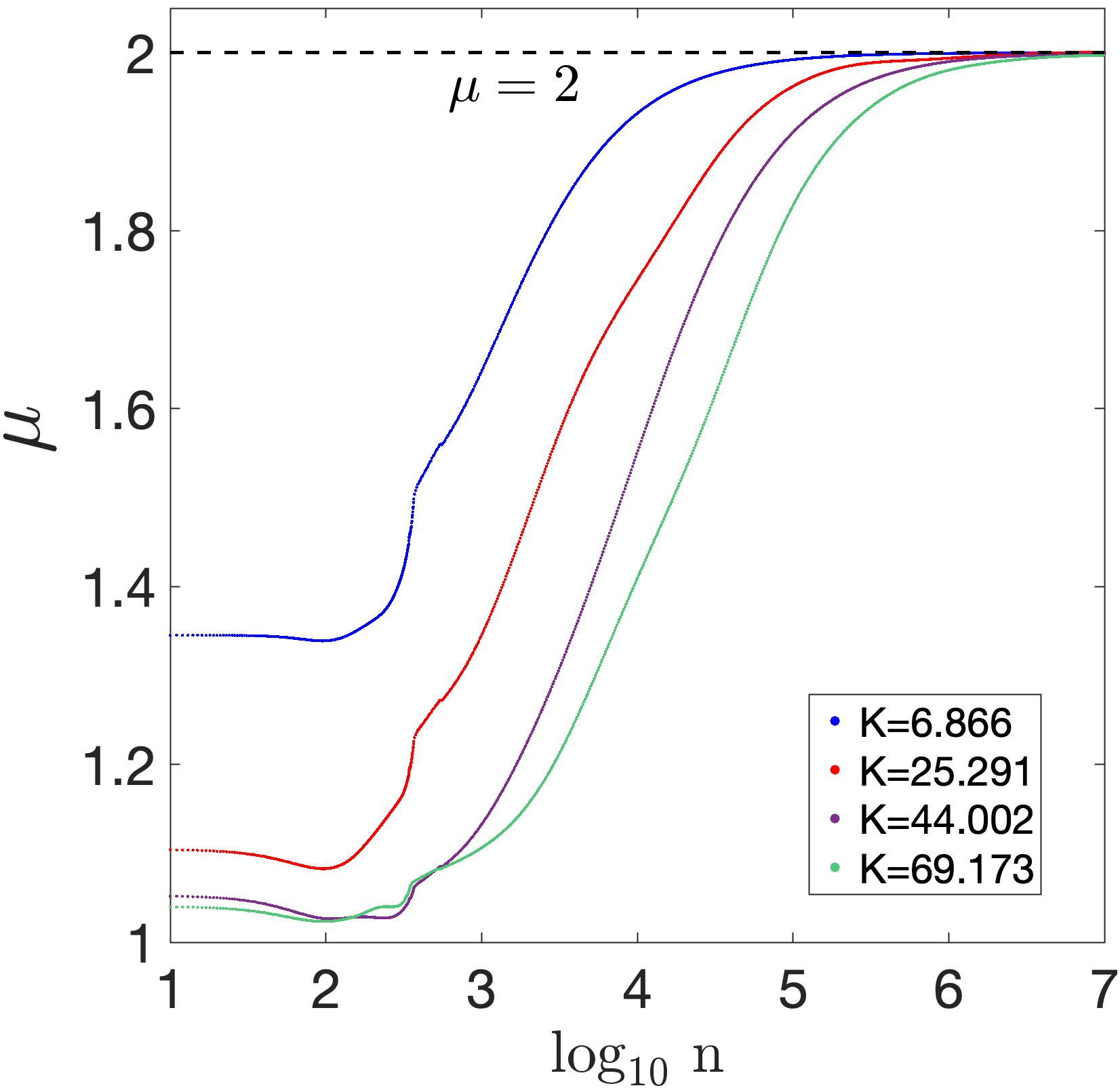}}
    \caption{(a) The average variance \( \langle (\Delta p)^2 \rangle \) \eqref{eq:pvar} and (b) the corresponding numerically estimated diffusion exponent \( \mu (n) \) \eqref{eq:pvar} of the SM for approximately \(100,000\) ICs in a grid covering the map's entire phase space \( x \in (0, 2\pi)\) and \(p \in (0, 2\pi) \). Different colors represent different values of the kicked strength parameter corresponding to specific black circle-marked peaks in Fig.~\ref{fig5:Fig03a}: \( K = 6.866 \) (blue), \( K = 25.291 \) (red), \( K = 44.002 \) (purple), and \( K = 69.173 \) (green). The dashed and dash-dotted lines in (a), respectively, indicate power law behaviors $n^1$ and $n^2$ while the horizontal dashed line in (b) corresponds to $\mu =2$.}
    \label{fig5:Fig04}
   \end{figure}

To further investigate the anomalous diffusion due to the presence of period $p=1$ AMs, in Fig.~\ref{fig5:Fig04}, we perform a detailed analysis of the SM's \eqref{eq:ssm} diffusion behavior at the \(11\) specific parameter values of \(K\) corresponding to the observed high peaks in Fig.~\ref{fig5:Fig03}. We choose only some representative cases from different peak locations, i.e. $K = 6.866$ (blue curves), $K = 25.291$ (red curves), $K = 44.002$ (purple curves), and $K = 69.173$ (green curves), to avoid overloading the plots with many curves.  Figs.~\ref{fig5:Fig04a} and (b) present, respectively, the evolution of the variance $\langle (\Delta p)^2 \rangle$ \eqref{eq:pvar} and the related diffusion exponent, $\mu$ \eqref{eq:pvar} of the system \eqref{eq:ssm} for each considered case. The results in Fig.~\ref{fig5:Fig04} demonstrate a clear trend towards ballistic diffusion ($\mu= 2$) as the number of iterations increases. In Fig.~\ref{fig5:Fig04b}, we can also see that the time required for $\mu$ to reach this asymptotic \(\mu=2\) value depends on the specific $K$ value. Larger values of the nonlinearity, $K$, require a longer iteration time for convergence to $\mu=2$, indicating that while AMs continue to impact the diffusion process, their effect becomes apparent over longer timescales as nonlinearity increases.

The system's \eqref{eq:ssm} diffusion behavior for larger $K$ values illustrates a transition from normal to anomalous diffusion. While the long-term dynamics eventually converge toward ballistic transport ($\mu=2$) due to the influence of AMs, the early stages of the diffusion process typically exhibit normal diffusion characteristics ($\mu = 1$). This is especially evident for higher $K$ values, such as those represented by the purple and green curves in both panels of Fig.~\ref{fig5:Fig04}, where $\langle (\Delta p)^2 \rangle$ initially grows proportionally to \(n\) with $\mu = 1$ [dashed line in Fig.~\ref{fig5:Fig04a}], before transitioning into the ballistic regime. This delayed onset of anomalous diffusion is also reflected on the fact that the height of the diffusion peaks at $n=10^4$ [Figs~\ref{fig5:Fig03a}], which are lower than those observed for $n=10^5$ iterations [Figs~\ref{fig5:Fig03c}].

\begin{figure}[!htbp]
    \centering
    \subfloat[\( \mu^*(K) \)\label{fig5:Fig05a}]{\includegraphics[width=0.475\textwidth]{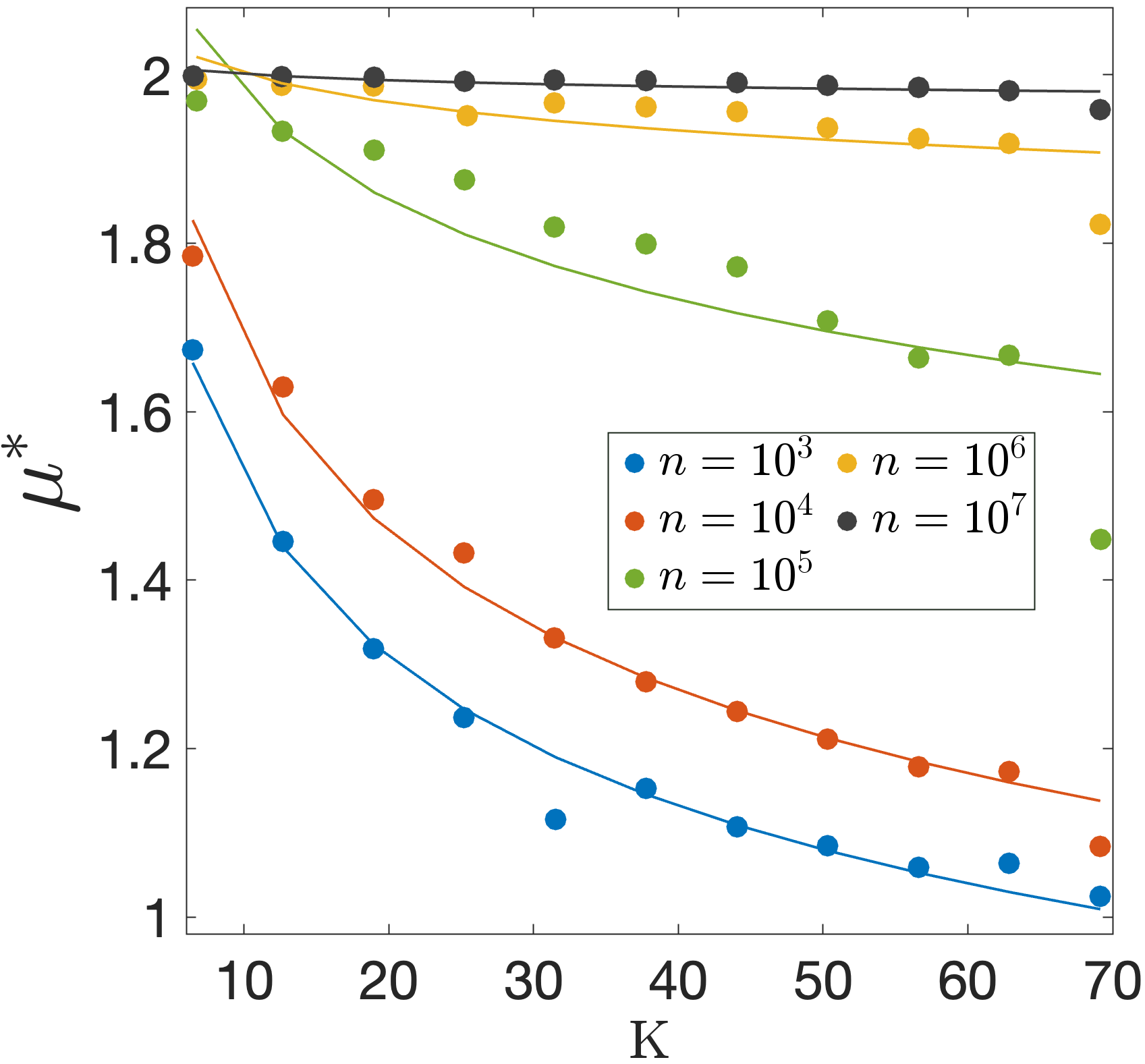}}
    \subfloat[$A(n)$ and $B(n)$\label{fig5:Fig05bc}]{\includegraphics[width=0.475\textwidth]{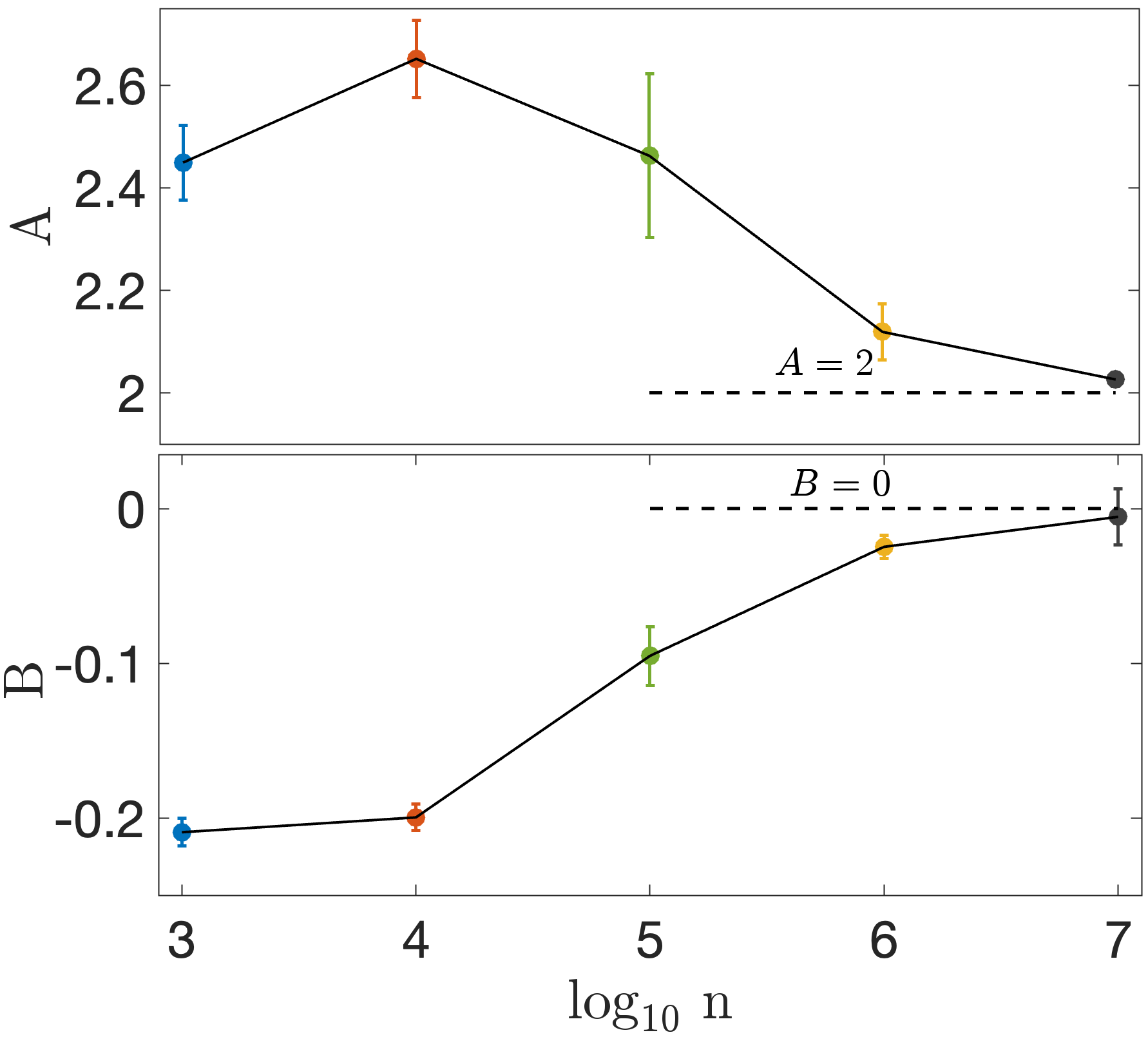}}
    \caption{(a) The maximum diffusion exponents \( \mu^*(K) \) for the first \(11\) \( K \) values where period $p=1$ AMs of the SM \eqref{eq:ssm} exist are computed at \(n = 10^3\), \(10^4\), \(10^5\), \(10^6\), and \(10^7\) iterations of the map (represented by blue, red, green, orange, and black points, respectively). \( \mu \) is computed for approximately \(100,000\) ICs evenly distributed on a grid covering the map's entire phase space \( x \in (0, 2\pi)\) and \(p \in (0, 2\pi) \). Solid curves represent fitting to the function \(\mu^*(K) = A K^B\) for each set of data points. (b) The resulting values of the fitting parameters \( A(n)\) and \( B(n)\), along with their corresponding determination errors. Horizontal dashed lines indicate \(A = 2\) and \(B = 0\).}
    \label{fig5:Fig05}
   \end{figure}

In order to further investigate the behavior of the maximum diffusion exponent values, $\mu^*$, associated with the presence of period $p=1$ AMs, we conduct a detailed analysis of the \(K\) values corresponding to the first \(11\) peaks shown in Fig.~\ref{fig5:Fig04}. In this analysis, we consider various iteration numbers: $n = 10^3$ (blue points), $n = 10^4$ (red points), $n = 10^5$ (green points), $n = 10^6$ (orange points), and $n = 10^7$ (black points) in Fig.~\ref{fig5:Fig05}. The peak values $\mu^*$ as a function of these $K$ values are plotted in Fig.~\ref{fig5:Fig05a}. Note that the obtained results for $\mu^*$ at $n = 10^4$ and $n = 10^5$ iterations  (red and green points and curves) correspond to the black circles marked in Figs.~\ref{fig5:Fig03a} and (c), respectively. To better understand the relationship between the peak values $\mu^*$ and the nonlinearity strength $K$ for each number of iterations $n$, we fit these obtained results using a power law of the form $\mu^*(K) = A K^B$. The computed fittings are shown by the colored curves in Figs.~\ref{fig5:Fig05a}, and the associated parameters values $A$ and $B$ of these fittings are presented as functions of  the map's iteration number $n$ in Fig.~\ref{fig5:Fig05bc}.

Our analysis of the peak values associated with period $p=1$ AMs provides additional insight into the diffusion process. In Fig.~\ref{fig5:Fig05a}, we have observed that the maximum diffusion exponent, $\mu^*$, for each $K$ value shows a distinct trend toward the ballistic diffusion limit ($\mu = 2$) as the number of iterations increases. This behavior is further supported by the fitting parameters $A$ and $B$ [Fig.~\ref{fig5:Fig05bc}], where we see these values approach $A = 2$ and $B = 0$ as the number of iterations increases. These asymptotic parameter values are indicated by the horizontal dashed lines in Fig.~\ref{fig5:Fig05bc}. 

The findings of Fig.~\ref{fig5:Fig05} demonstrate that the effect of AMs on the SM's diffusion process grows more pronounced with extended number of iterations, $n$. We observe that values of $\mu^*$ become nearly independent of $K$ as they converge to their maximum value $\mu^* = 2$. This convergence leads us to conclude that, while other factors may influence the initial dynamics of the SM \eqref{eq:ssm}, the long-term behavior is ultimately governed by the AM's presence, resulting in ballistic transport.
   
\begin{figure}[!htbp]
    \centering
    \subfloat[$n^\prime(K)$\label{fig5:Fig06a}]{\includegraphics[width=0.475\textwidth]{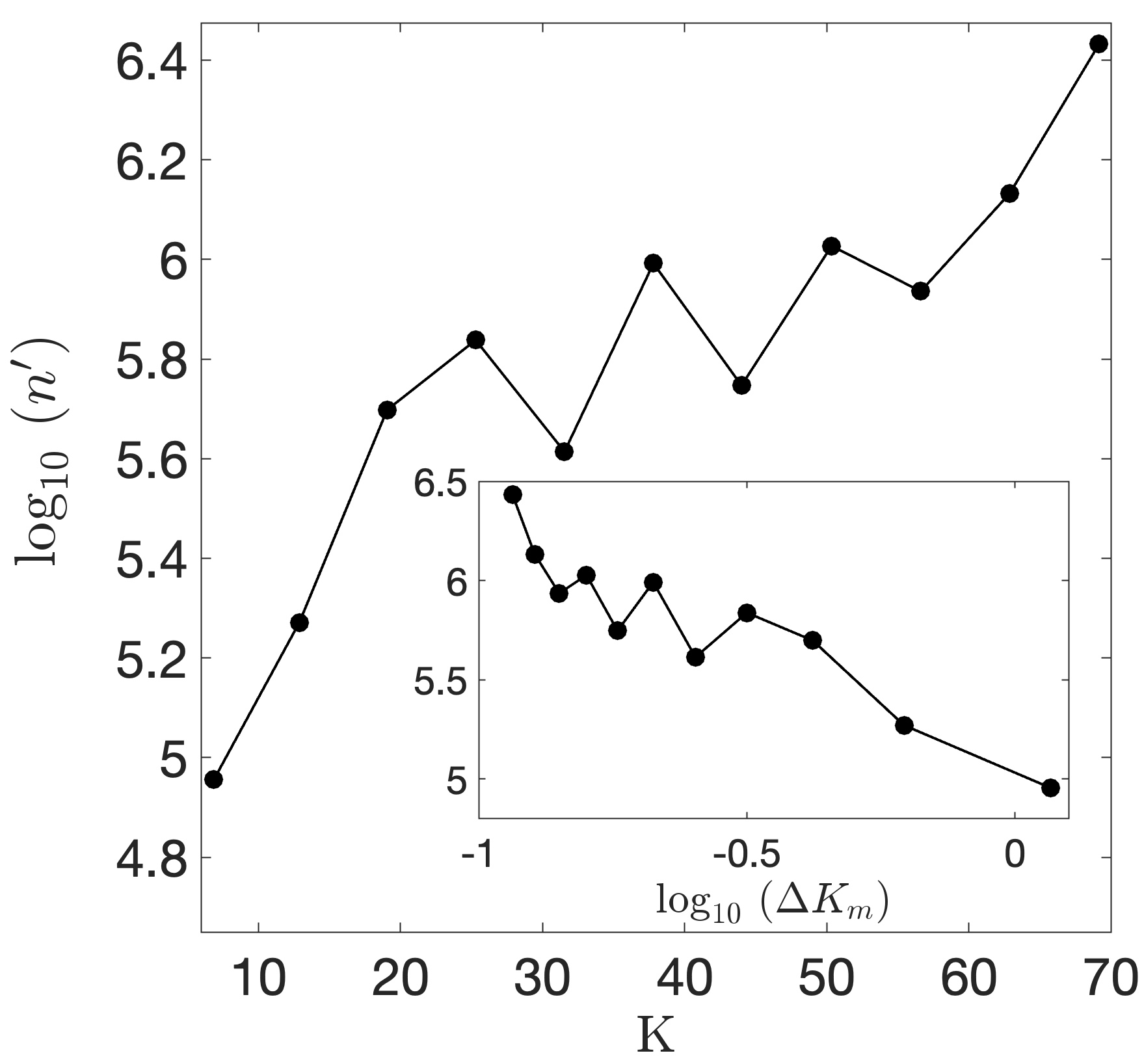}} 
    \subfloat[Average ftmLE$(n)$ \eqref{eq:ftmLE}\label{fig5:Fig06b}]{\includegraphics[width=0.475\textwidth]{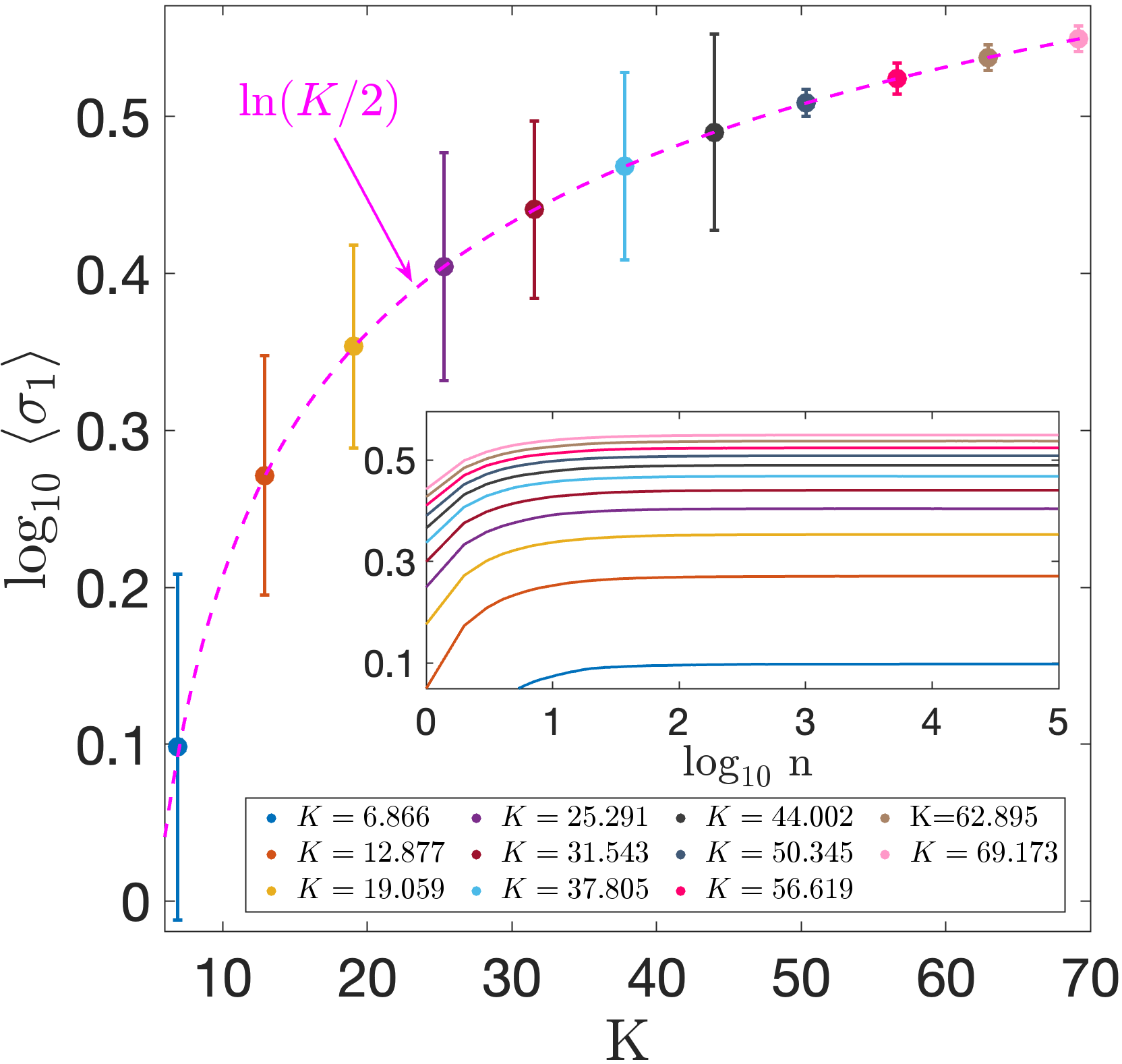}}
    \caption{(a) The number of iterations \(n^\prime(K)\) required for the diffusion exponent \(\mu\) to reach \(\mu=2\), which indicates ballistic transport, for \(K\) values $K = 6.866$, $12.877$, $19.059$, $25.291$, $31.543$, $37.805$, $44.002$, $50.345$, $56.619$, $62.895$, and $69.173$ period $p=1$, where period \(p=1\) AMs of the SM \eqref{eq:ssm} exist. Inset: \(n^\prime\) as a function of \(\Delta K_m = \sqrt{(2\pi m)^2 + 16} - 2\pi m\), which represents the width of the \(K\) intervals containing the AMs. (b) Average ftmLE, \(\langle \sigma_1 (K) \rangle\) \eqref{eq:ftmLE}, for \(10,000\) ICs on a grid covering the SM's entire phase space after \(n = 10^5\) iterations. The error bars denote one standard deviation in the average value computation, while the dashed pink curve corresponds to the power law \(\langle \sigma_1 \rangle = \ln\left( K / 2 \right)\). Inset: the time evolution of \(\langle \sigma_1 (n) \rangle\).}
    \label{fig5:Fig06}
   \end{figure}

To investigate the time required for the diffusion exponent $\mu$ to reach its asymptotic value $\mu = 2$, we examine the identified \(K\) values of the first \(11\) peaks [Fig.~\ref{fig5:Fig03}], as shown in Fig.~\ref{fig5:Fig06a}. For this analysis, we introduce a parameter, $n^\prime$, to quantify the number of iterations required for $\mu$ to reach the threshold near its asymptotic limit $\mu = 2$ (to be precise with the specific criteria being $\mu \geq 1.992$). This analysis further demonstrates the observations from Figs.~\ref{fig5:Fig04} and \ref{fig5:Fig05}, showing that the diffusion exponents associated with higher \(K\) value peaks require more iterations to converge to $\mu = 2$. As illustrated in the inset of Fig.~\ref{fig5:Fig06a}, $n^\prime$ displays a clear dependence on the width of the period $p=1$ AM intervals, defined by Eq.~\eqref{eq:acmdint} as red\(\Delta K_m = \sqrt{(2\pi m)^2 + 16} - 2\pi m\). In particular, a narrow $\Delta K_m$ interval corresponds to larger $n^\prime$ values, indicating a longer delay before reaching ballistic diffusion. 

To further understand the relationship between the delay in reaching the ballistic diffusion ($\mu = 2$) and the global dynamics of the SM in phase space, we also compute the average ftmLE \eqref{eq:ftmLE}, $\langle \sigma_1 \rangle$, for the entire phase space across a range of the nonlinearity parameter values $K$ considered in Fig.~\ref{fig5:Fig05}. Specifically, we evaluated $\langle \sigma_1 \rangle$ for $K = 6.866$, $12.877$, $19.059$, $25.291$, $31.543$, $37.805$, $44.002$, $50.345$, $56.619$, $62.895$, and $69.173$. For each $K$ value, we evolve \(10,000\) orbits with ICs on a $100 \times 100$ grid evenly distributed over the entire phase space of the map to determine the average ftmLE value, $\langle \sigma_1 \rangle$ [Fig.~\ref{fig5:Fig06b}]. 

We find that the correlation between $\langle \sigma_1 \rangle$ and the nonlinearity strength \(K\) aligns with the theoretical prediction given in \citep{Shevchenko2004}, where $\langle \sigma_1 \rangle (K) = \ln(K/2)$ for the chaotic region of the \(2D\) SM \eqref{eq:ssm} phase space. This relation demonstrates how chaoticity changes for large values of $K$ of the map. The average ftmLE values approximately follow a $\ln(K/2)$ growth pattern, as indicated by the dashed pink curve in Fig.~\ref{fig5:Fig06b}. The inset of Fig.~\ref{fig5:Fig06b} illustrates the evolution of $\langle \sigma_1 \rangle$ as a function of the number of iterations up to $n=10^5$, indicating the positive correlation between $\langle \sigma_1 \rangle$ and $K$.

In the computation of the average ftmLE, $\langle \sigma_1 \rangle$, shown in Fig.~\ref{fig5:Fig06b}, we consider a large ensemble of ICs which are evenly distributed on a grid over the entire phase space. Although the averaging process includes both regular and chaotic orbits, the significant contribution to the $\langle \sigma_1 \rangle$ value primarily comes from chaotic orbits. This is because the measure of the chaotic regions typically outweighs that of regular islands, especially at higher values of the nonlinearity parameter $K$. For large \(K\) values, chaotic motion dominates the phase space, and the addition of some regions with regular orbits (which practically have $\sigma_1 \approx 0$) does not drastically impact the computation of $\langle \sigma_1 \rangle$. By using large ensembles of around \(10,000\) orbits and long simulation times of \(10^5\) iterations, we ensure that the computed average ftmLE provides a reliable measure of the system's global chaoticity. This is indicated by the clear saturation of $\langle \sigma_1 \rangle$ to a positive number in the inset of Fig.~\ref{fig5:Fig06b}. 

Let us now examine the relation between the chaotic dynamics in the phase space of the SM \eqref{eq:ssm} and the map's diffusion properties in the presence of AMs of different periods. For this analysis, we employ the GALI$_{2}$ method due to its efficiency in accurately distinguishing between regular and chaotic orbits, as demonstrated in Fig.~\ref{fig5:Fig01}. By using the GALI$_2$, we can also produce detailed color plots that provide a global overview of the system's dynamics, at the same time quantifying the portion of the phase space covered by chaotic regions, P\(_\text{C}\). By systematically adjusting the  size of the phase space region surrounding an AM, we can control the percentage of chaotic orbits in this region, which allows for an investigation of how the proportion of chaotic to regular motion affects the SM's phase space global diffusion transport properties.  

In order to estimate the percentage of chaotic orbits for a given $K$ value using GALI$_{2}$, we first consider a \(2D\) grid with evenly spaced points covering the map's entire phase space. Each point on this grid serves as an IC for an orbit. In particular, we use a dense grid of $500 \times 500$ cells, corresponding to 250,000 ICs, covering the entire phase space \(x \in (0, 2\pi)\) and \(p \in (0, 2\pi)\). In addition, we consider subspaces of the phase space in $(0, 2\pi) \times (0, 2\pi)$ that contain AMs and surrounding chaotic areas. The size of these subspaces and the percentage of chaotic orbits, P\(_\text{C}\), may vary to analyze how different chaotic region proportions impact the SM's global diffusion properties.

For the SM \eqref{eq:ssm}, we also select specific values of $K$ for which AMs are present and use the GALI$_2$ method to distinguish between chaotic and regular regions in the system's phase space. In particular, for our analysis we chose $K = 6.5$, $K = 4.0844$, $K = 6.9115$, and $K = 3.1$, where AMs of periods \(1\) to \(4\) are observed, respectively. Then we evolve these considered orbits for $n = 50$ iterations and compute GALI$_{2}$ for each orbit. The values of the GALI$_{2}$ at $n = 50$ allow us to estimate the percentage of chaotic orbits for different values of $K$ in the SM's phase space. More specifically, orbits with GALI$_{2}$ values smaller than $10^{-8}$ are classified as chaotic, while orbits whose GALI$_{2}$ values remain greater than $10^{-8}$ after $n=50$ iterations are classified as non-chaotic.

It is also important to point out that the threshold value of zero, GALI$_2 \le 10^{-8}$, is widely accepted as a good criterion for distinguishing between chaotic and regular motion, as shown in previous studies \citep{manos2007studying,manos2009global}. Moreover, the number of $n=50$ iterations is generally sufficient to clearly differentiate between the exponential decay of the index observed in the case of chaotic motion and the power law decay indicating regular motion, as illustrated in Fig.~\ref{fig5:Fig01b} (see also Eq.~\ref{Prop:GALI_2 for SM}, where this behavior of the GALI\(_2\) was discussed). 

In order to get a more detailed understanding of the diffusion process, we perform a refined analysis of the diffusion exponent, $\mu$, for each grid point and $K$ value considered by systematically computing the variance of the angular momentum as follows: We construct an additional finer grid of $50 \times 50$ ICs to each of the original $500 \times 500$ grid cells (i.e., \(2,500\) ICs per cell) considered in order to compute the GALI$_2$. This refinement enables us to do a localized computation of $\mu$ for each of the studied \(250,000\) ICs covering the entire phase space. Then, for each one of these local grids, we numerically compute the `local' diffusion exponent $\mu$ using Eq.~\eqref{eq:pvar} and perform a linear fit of the variance $\langle (\Delta p)^{2} \rangle$ with respect to the number of iterations $n$ over the range $10^3$ to $10^4$. This technique allows us to detect small fluctuations in the diffusion rates and relate these changes with specific dynamical features in the phase space. The approach generates high-resolution color plots that capture the dynamics of chaotic and regular regions in the SM  \eqref{eq:ssm}, providing a systematic analysis of the diffusion variations in the system. 

\begin{figure}[!htbp]    
     \centering
     \subfloat[Regions around the stable AM of period $p=1$ for \(K = 6.5\)\label{fig5:Fig07a}]{\includegraphics[width=1\textwidth]{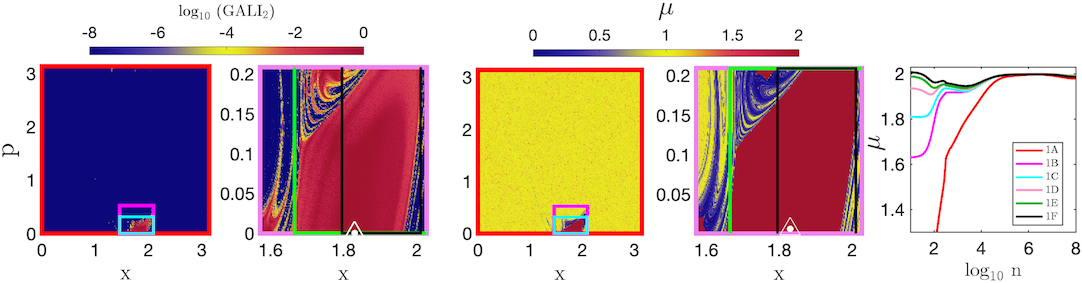}}\\  %
    \vspace{-0.15in}
    \subfloat[Regions around the stable AM of period $p=2$ fo \(K = 4.0844\)\label{fig5:Fig07b}]{\includegraphics[width=1\textwidth]{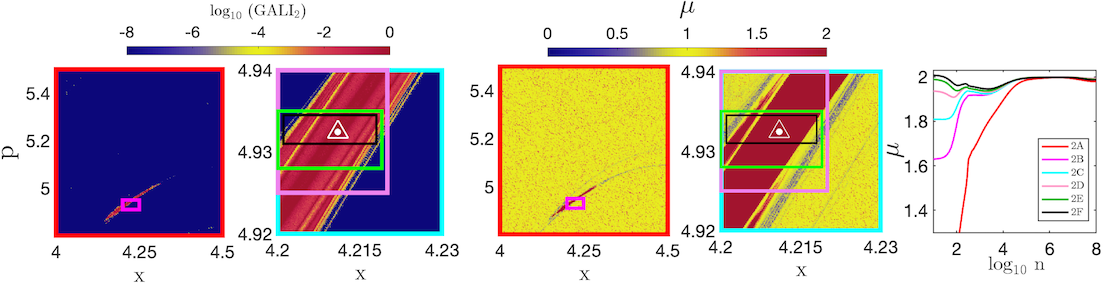}} \\ 
    \vspace{-0.15in}
    \subfloat[Regions around the stable AM of period $p=3$ for \(K = 6.9115\)\label{fig5:Fig07c}]{\includegraphics[width=1\textwidth]{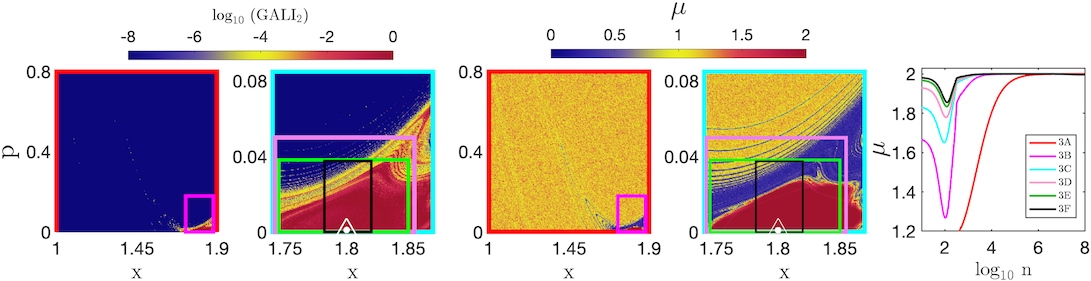}}  \\
    \vspace{-0.15in}
    \subfloat[Regions around the stable AM of period $p=4$ for \(K = 3.1\)\label{fig5:Fig07d}]{\includegraphics[width=1\textwidth]{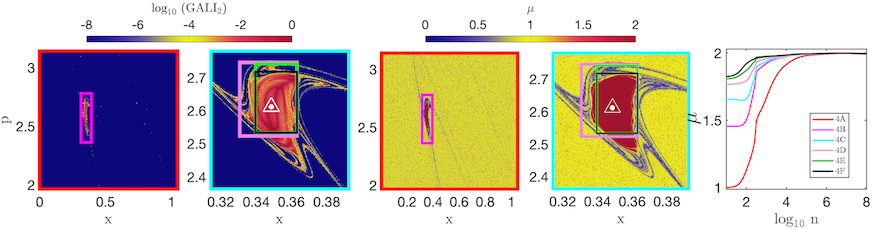}} 
     \caption{Phase space portraits of the SM \eqref{eq:ssm} where ICs on a dense grid covering specific phase space regions are colored by (the first two columns) the GALI$_2$ value after \(n = 50\) iterations and (the third and fourth columns) their diffusion exponent \(\mu\) \eqref{eq:pvar} computed over \(10^3 \leq n \leq 10^4\). In the GALI\(_2\) plots, red areas indicate regular motion, dark blue corresponds to chaotic motion, and yellowish hues represent weakly chaotic orbits. In the \(\mu\) color plots, ICs with subdiffusion rates appear in dark blue (\(0 \leq \mu < 1\)), normal diffusion in yellow/orange (\(\mu \approx 1\)), superdiffusion in lighter reds (\(1 \leq \mu < 2\)), and ballistic transport in dark red/brown (\(\mu \approx 2\)). Each row corresponds to a different \(K\) value associated with stable AMs of period $p$, with AM locations marked by small white triangles in the rightmost GALI$_2$ and \(\mu\) plots. Particularly, the AM locations are \((x, p) = (1.8298, 0)\) for \(p = 1\) in (a), \((4.211, 4.9324)\) for \(p = 2\) in (b), \((1.80, 0.0)\) for \(p = 3\) in (c), and \((0.3486, 2.6121)\) for \(p = 4\). The six phase space regions are highlighted for each \(K\) value using differently colored rectangles in the phase space, labelled from `A' to `F' according to their chaotic orbit percentage. Region `A' has the highest percentage of chaotic orbit (P\(_\text{C} \approx 99\%\)), and `F' has the lowest (P\(_\text{C} \approx 6.5\%\)). Plots in the fifth column show the evolution of the computed diffusion exponent \(\mu\) over iterations for each of the considered six regions. The colors used to identify regions `A' to `F' are indicated in the legend of the last panel in each row. Further details are provided in the text and Table~\ref{tab:ssm}.}
    \label{fig5:Fig07}
    \end{figure}

Figure \ref{fig5:Fig07} illustrates the interplay between global chaos and diffusion properties in the SM \eqref{eq:ssm} for four representative values of the nonlinearity parameter $K$ associated with different AM periods ($p = 1, 2, 3,$ and $4$), specifically $K = 6.5$, $K = 4.0844$, $K = 6.9115$, and $K = 3.1$. Each value of \(K\) corresponds to an AM of the SM with these respective periods. In each case, the first two panels present color plots of the phase space regions based on the GALI$_{2}$ values, whereas the subsequent two panels depict the same regions colored according to the computed diffusion exponent $\mu$ \eqref{eq:pvar}. In the GALI$_{2}$ color plots of Fig.~\ref{fig5:Fig07}, chaotic regions (GALI$_2  \le 10^{-8}$) appear in dark blue color, while regular regions appear in red. In the $\mu$ color plots, ballistic diffusion ($\mu \approx 2$) is shown in red color, normal diffusion ($\mu \approx 1$) appears in various shades of yellow, and subdiffusion in blue. The stable AM locations for each \(K\) are marked by a white point in a triangle in the second and fourth columns of Fig.~\ref{fig5:Fig07}. Additional examples of AMs of different periods, along with methods on how to locate them, are discussed in detail in \citep{contopoulos2018} and references therein.

To systematically analyze how varying the percentage of chaotic regions affects diffusion properties in the SM \eqref{eq:ssm}, we define six distinct phase space regions, labeled `A' to `F', each with a different percentage of chaos. These regions are marked by different colored rectangles in the GALI$_{2}$ and $\mu$ color plots of Fig.~\ref{fig5:Fig07}. The regions cover a wide range of chaotic orbit percentages, from the highest percentage of chaotic orbits (approximately $99\%$) in region `A' to the lowest (approximately $6.5\%$) in region `F'. In general, the specific percentage of chaos P$_\text{C}$ for each region is as follows: P$_\text{C} \approx 99\%$ (A), P$_\text{C} \approx 75\%$ (B), P$_\text{C} \approx 50\%$ (C), P$_\text{C} \approx 25\%$ (D), P$_\text{C} \approx 12.5\%$ (E), and P$_\text{C} \approx 6.5\%$ (F). The values, along with the locations of these regions, are given in Table \ref{tab:ssm} and visually represented by colored rectangles in Fig.~\ref{fig5:Fig07}.

For instance, in the $K = 3.1$ case shown in Fig.~\ref{fig5:Fig07d}, we define six regions of the phase space, labeled `4A' to `4F'. The corresponding rectangles appear with distinct colors in the GALI$_{2}$ and $\mu$ color plots. In particular, regions 4A and 4B are depicted by red and maroon rectangles, respectively, on the left side of both the GALI\(_2\) and \(\mu\) color plots. We highlight remaining regions, 4C, 4D, 4E, and 4F, by cyan, light pink, green, and black rectangles on the right side of these plots.
 
We compute the diffusion exponent, $\mu$, for each of the defined subspace regions as a function of the number of iterations, $n$. The time evolution of $\mu$ is shown in the rightmost column of Fig.~\ref{fig5:Fig07}, where different curves correspond to different regions, labeled `A' to `F' in the GALI$_2$ and $\mu$ color plots.

Our analysis of Fig.~\ref{fig5:Fig07} further illustrates a strong correlation between the percentage of chaotic orbits, $P_{C}$, in the SM's phase space regions and the convergence of the diffusion exponent to its asymptotic value of $\mu=2$. Regions with a higher proportion of chaotic orbits (i.e., larger $P_{C}$ values) demonstrate faster convergence toward ballistic diffusion. On the other hand, regions with a smaller ratio of chaos require more iterations to reach the asymptotic value $\mu=2$. The specific number of iterations required to reach ballistic diffusion, $n^\prime$ [see also Fig.~\ref{fig5:Fig06a}], for each rectangle in Fig.~\ref{fig5:Fig07} is also given in the last column of Table~\ref{tab:ssm}.

The GALI$_2$ and $\mu$ color plots in Fig.~\ref{fig5:Fig07} offer comprehensive overviews of the SM's dynamics at different simulation times. The GALI$_2$ plots (which are produced for $n=50$ iterations) are based on relatively short-term snapshots, which allow us to efficiently distinguish between regular and chaotic regions. On the other hand, the $\mu$ color plots require longer simulation times ($10^3$ to $10^4$ iterations) to accurately characterize the diffusion behavior across the phase space. 

    \clearpage
\begin{table}[ht]
\caption{The number of iterations of the SM \eqref{eq:ssm} required ($n^\prime$) for the diffusion exponent $\mu$ to converge to its maximum asymptotic ballistic value \(\mu = 2\), indicating ballistic transport, across various $K$ values and IC ensembles in different phase space regions, each with distinct percentage of chaotic orbits, P$_\text{C}$. The \(K\) values considered are $K = 6.5$, $K = 4.0844$, $K = 6.9115$, and $K = 3.1$, which correspond to the map's stable AMs of period $p = 1, 2, 3,$ and $4$, respectively. Each region is named based on the AM period \(p\) followed by a letter from `A' to `F'. The \(x\) and \(y\) ranges for each region are selected so that regions with the same letter have similar P$_\text{C}$ values, which decrease progressively from A to F. These regions are also illustrated in Fig.~\ref{fig5:Fig07} by differently colored rectangles, where P$_\text{C} \approx 99\%$ for region A (red), P$_\text{C} \approx 75\%$ for region B (maroon), P$_\text{C} \approx 50\%$ for region C (cyan), P$_\text{C} \approx 25\%$ for region D (pink), P$_\text{C} \approx 12.5\%$ for region E (green), and P$_\text{C} \approx 6.5\%$ for region F (black).}
\label{tab:ssm}
    \begin{tabularx}{\textwidth}{|c|c|c|X|c|c|}
    \hline
    \textbf{K} & \textbf{Period $p$} & \textbf{Ensemble Name} & \textbf{$(x, p)$ region} & \textbf{P$_\text{C}$ (\%)} & \textbf{$n^\prime$} \\
    \hline
    6.5 & 1 & 1A & [0.000, 3.142] $\times$ [0.0, 3.142] & 99.76 & $5.45 \times 10^4$ \\
        &   & 1B & [1.282, 2.094] $\times$ [0.0, 0.524] & 74.98 & $5.34 \times 10^3$ \\
        &   & 1C & [1.461, 2.094] $\times$ [0.0, 0.314] & 50.03 & $5.25 \times 10^3$ \\
        &   & 1D & [1.571, 2.027] $\times$ [0.0, 0.209] & 24.84 & $6.1 \times 10^3$ \\
        &   & 1E & [1.665, 2.027] $\times$ [0.0, 0.209] & 12.46 & $1.80 \times 10^3$ \\
        &   & 1F & [1.795, 2.010] $\times$ [0.0, 0.209] & 6.53 & $4.88 \times 10^2$ \\
    \hline
    4.0844 & 2 & 2A & [4.000, 4.500] $\times$ [4.800, 5.500] & 99.23 & $1.98 \times 10^5$ \\
           &   & 2B & [4.200, 4.250] $\times$ [4.910, 4.950] & 75.39 & $1.48 \times 10^5$ \\
           &   & 2C & [4.200, 4.230] $\times$ [4.920, 4.940] & 50.38 & $5.25 \times 10^4$ \\
           &   & 2D & [4.200, 4.220] $\times$ [4.925, 4.940] & 25.38 & $4.79 \times 10^4$ \\
           &   & 2E & [4.200, 4.210] $\times$ [4.928, 4.935] & 12.70 & $2.33 \times 10^4$ \\
           &   & 2F & [4.201, 4.218] $\times$ [4.931, 4.935] & 6.45 & $2.7 \times 10^1$ \\
    \hline
   6.9115 & 3 & 3A & [1.0, 1.9] $\times$ [0.0, 0.80] & 99.12 & $4.37 \times 10^6$ \\
        &  & 3B & [1.725, 1.88] $\times$ [0.0, 0.18] & 75.31 & $6.08 \times 10^5$ \\
        &  & 3C & [1.74, 1.87] $\times$ [0.0, 0.085] & 50.01 & $4.45 \times 10^5$ \\
        &  & 3D & [1.742, 1.855] $\times$ [0.0, 0.05] & 25.38 & $1.34 \times 10^5$ \\
        &  & 3E & [1.745, 1.85] $\times$ [0.0, 0.038] & 12.43 & $1.27 \times 10^5$ \\
        &  & 3F & [1.782, 1.82] $\times$ [0.0, 0.0375] & 6.28 & $9.12 \times 10^4$ \\
\hline
  3.1 & 4 & 4A & [0.000, 1.047] $\times$ [1.964, 3.142] & 99.28 & $6.03 \times 10^5$ \\
        &  & 4B & [0.314, 0.393] $\times$ [2.362, 2.780] & 75.20 & $1.99 \times 10^5$ \\
        &  & 4C & [0.331, 0.376] $\times$ [2.417, 2.732] & 50.40 & $1.22 \times 10^5$ \\
        &  & 4D & [0.330, 0.363] $\times$ [2.523, 2.746] & 24.93 & $4.21 \times 10^4$ \\
        &  & 4E & [0.340, 0.363] $\times$ [2.534, 2.741] & 12.31 & $2.33 \times 10^4$ \\
        &  & 4F & [0.310, 0.363] $\times$ [2.534, 2.718] & 6.46 & $1.30 \times 10^4$ \\
    \hline
    \end{tabularx}
\end{table}
It's important to note that the short-term GALI$_2$ plots presented in Fig.~\ref{fig5:Fig07} may reveal transient chaotic-like behaviors typically associated with stickiness effects (e.g.~see \citep{contopoulos2018}) along unstable POs surrounding stable islands. However, as the number of iterations increases, we expect these regions to exhibit more distinct chaotic characteristics, with the resulting GALI$_2$ value approaching zero and the diffusion exponent \(\mu\) converging to $\mu \approx 1$. More generally, for large $n$ values, both the GALI$_2$ and $\mu$ plots are expected to exhibit similar structures. The GALI$_2$ plot will mainly consist of two colors: red, representing regular motion, and dark blue, indicating chaotic motion. On the other hand, the $\mu$ plots will show two dominant colors: red, corresponding to ballistic diffusion ($\mu \approx 2$), and yellow, indicating normal diffusion ($\mu \approx 1$) in the cases where AMs are present.

\begin{figure}[!htbp]
    \centering
    \subfloat[Phase space portrait for  \(K = 6.9115\)\label{fig5:FigA_07a}]{\includegraphics[width=0.475\textwidth]{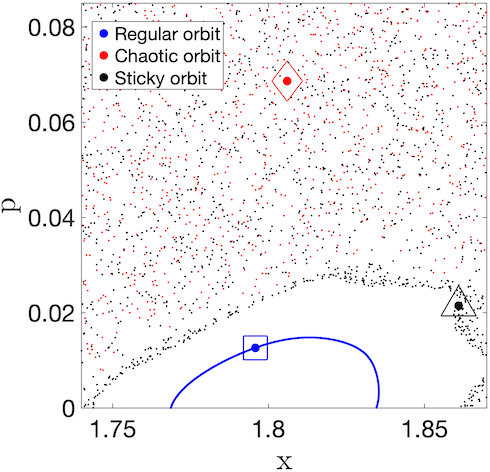}} 
    \subfloat[Fourth panel of Fig.~\ref{fig5:Fig07c}\label{fig5:FigA_07b}]{\includegraphics[width=0.475\textwidth]{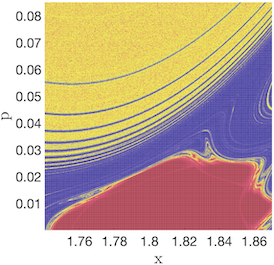}}
    \caption{(a) Phase space portrait of the SM \eqref{eq:ssm} with \(K = 6.9115\) for three ICs: $(x, p) = (1.8, 0.012)$ regular orbit (blue square point), $(1.8, 0.02)$ sticky orbit (black triangle point), and $(1.8, 0.075)$ chaotic orbit (red diamond point), with their consequents shown by the same color points. (b) The right \(\mu\) color map of Fig.~\ref{fig5:Fig07c}.}
    \label{fig5:FigA_07}
   \end{figure}

In general, Fig.~\ref{fig5:Fig07} represents a transient figure that depends on the number of iterations $n$. Yellow features within the blue region in the GALI\(_2\) color plots and blue features within the yellow region in the \(\mu\) color plots correspond to stickiness regions along unstable asymptotic curves originating from unstable POs around the islands of stability. To further differentiate the sticky orbits from regular and chaotic orbits of Fig.~\ref{fig5:Fig07c}, we consider a specific case. Fig.~\ref{fig5:FigA_07a} shows the phase space portrait of the \(2D\) SM for three ICs taken from different regions of the \(\mu\) color plot in Fig.~\ref{fig5:FigA_07b} [which is the same as the fourth panel of Fig.~\ref{fig5:Fig07c}] depicting a region around the AM with period \(p=3\) for $K=6.9115$. In particular, we consider three orbits: the regular orbit with IC \((x, p) = (1.8, 0.012)\) (blue square point), whose consequents are given by the blue curve; the sticky orbit with IC \((x, p) = (1.8, 0.02)\) (black square point), whose consequents are  formed by black scattered points; and the chaotic orbit with IC $(x, p) = (1.8, 0.075)$ (red diamond point), whose consequents are represented by red scattered points. 

After sufficiently many iterations, we expect Fig.~\ref{fig5:FigA_07b} to display only two colors in the presence of stable AMs: yellow for normal diffusion and red for ballistic motion. We also observe interesting phenomena, such as blue regions (indicating subdiffusion) within the yellow areas (indicating normal diffusion). These blue regions correspond to the black points (stickiness points) in Fig.~\ref{fig5:FigA_07a}. This phenomenon is known as ``stickiness inside chaos" \citep{contopoulos2008stickiness, contopoulos2010stickiness}. 

   \subsection{Coupled standard maps} \label{sec:coupled SMs}
Having established a comprehensive understanding of the diffusion and chaotic behaviors of the SM \eqref{eq:ssm}, we now turn our attention to the dynamics of coupled systems. The objective is to investigate how the behavior of each individual interacting map influences the general diffusion rate and respective chaoticity of the coupled system. In order to achieve this goal, we consider a system of five coupled SMs ($N=5$ in Eq.~\ref{eq:csm}), with distinct values of nonlinearity kicks, $K_j$ for $j=1, \dots, 5$. These K\(_j\) values can either be equal or different. In addition, we consider varying coupling strengths, quantified by the parameter $\beta$. The relatively small number of coupled maps ($N=5$) we consider allows for extensive computational analysis while capturing the important dynamical features of $ND$ coupled SMs.

To investigate the dynamics of the $N=5$ coupled SM \eqref{eq:csm},  we generally follow ensembles of ICs. The coordinates of these ICs in each one of the five coupled SMs are arranged on a $315 \times 315$ grid of evenly spaced points (i.e., yielding approximately \(100,000\) ICs per coupled SM). Given the nature of the interaction between neighboring maps in Eq.~\eqref{eq:csm}, we must avoid zeroing the coupling interaction. This occurs when the ICs for the angle coordinates, $x^j_{0}$, $j=1,2,\ldots,N$, of adjacent \(2D\) SMs are equal, i.e., $x^j_{0}=x^{j+1}_{0}$. In order to prevent this type of multiple zeroing coupling interactions between the neighboring maps, we employ a specific IC arrangement. In particular, we shift the initial angular position ($x^j_{0}$ in each \(2D\) map) by a fixed display, $d$, which is calculated as the phase space angular range of the chosen IC ensemble divided by the number of maps in the system. More specifically, we set $x^{j+1}_{0}=x^j_{0}+d$, where $d=\dfrac{x_{\text{max}}-x_{\text{min}}}{N}$. If the computed $x^{j+1}_0$ value exceeds the interval $[x_{\text{min}}, x_{\text{max}}]$ of the considered subspace of a specific \(2D\) map (i.e., $x^{j+1}_0>x_{\text{max}}$), we adjust it by subtracting $(x_{\text{max}}-x_{\text{min}})$ to ensure that the considered IC falls within the specific angular position range.

In order to quantify the global diffusion properties of the coupled SMs \eqref{eq:csm}, we introduce a generalized diffusion exponent, $\mu$, which we derive from what we define as the generalized diffusion coefficient of the coupled system for $n \rightarrow \infty$. This coefficient is denoted by $D^N_{\mu}$. We then calculate the exponent $\mu$ by performing a numerical power law fit of the sum of the variances of the angular momentum for all $N$ \(2D\) maps with respect to the number of iterations, $n$, according to: 
\begin{equation}\label{eq:pvarN}
    \sum_{j=1}^{N} \langle (\Delta p^j)^2 \rangle = D^N_{\mu} n^{\mu}.
\end{equation}
In order to obtain Eq.~\eqref{eq:pvarN} we simply extended Eq.~\eqref{eq:pvar}, which we previously used to compute $\mu$ for the single SM \eqref{eq:ssm}, to apply it to the case of the general $ND$ SM \eqref{eq:csm}.

We determine the diffusion exponent, $\mu$ \eqref{eq:pvarN}, and the effective diffusion coefficient, \(D_{\text{eff}}\) \eqref{eq:Deff_csm}, for the coupled system using a procedure similar to what we used for the SM \eqref{eq:ssm}, to produce Fig.~\ref{fig5:Fig03}. By comparing $D^N_{\text{eff}}$ and \(\mu\) values, we aim to identify conditions under which the coupled SMs exhibit similar diffusion properties to the (individual) SMs. In addition, we will explore how the coupling between the \(N = 5\) SMs affects the long-term diffusion properties of the system's phase space and how the presence of AMs in one or more of the individual SMs impacts the coupled system over extended time (iteration) intervals.

\subsubsection{Coupled SMs with equal nonlinearity kicks and similar proportions of chaotic orbits} \label{sec:2NDResults_K1}
\begin{figure}[!htbp]
    \centering
    \subfloat[{\( \mu (K) \) [inset: \(D^N_{\text{eff}} (K)\)] for $\beta = 10^{-4}$}\label{fig5:Fig08a}]{\includegraphics[width=0.475\textwidth]{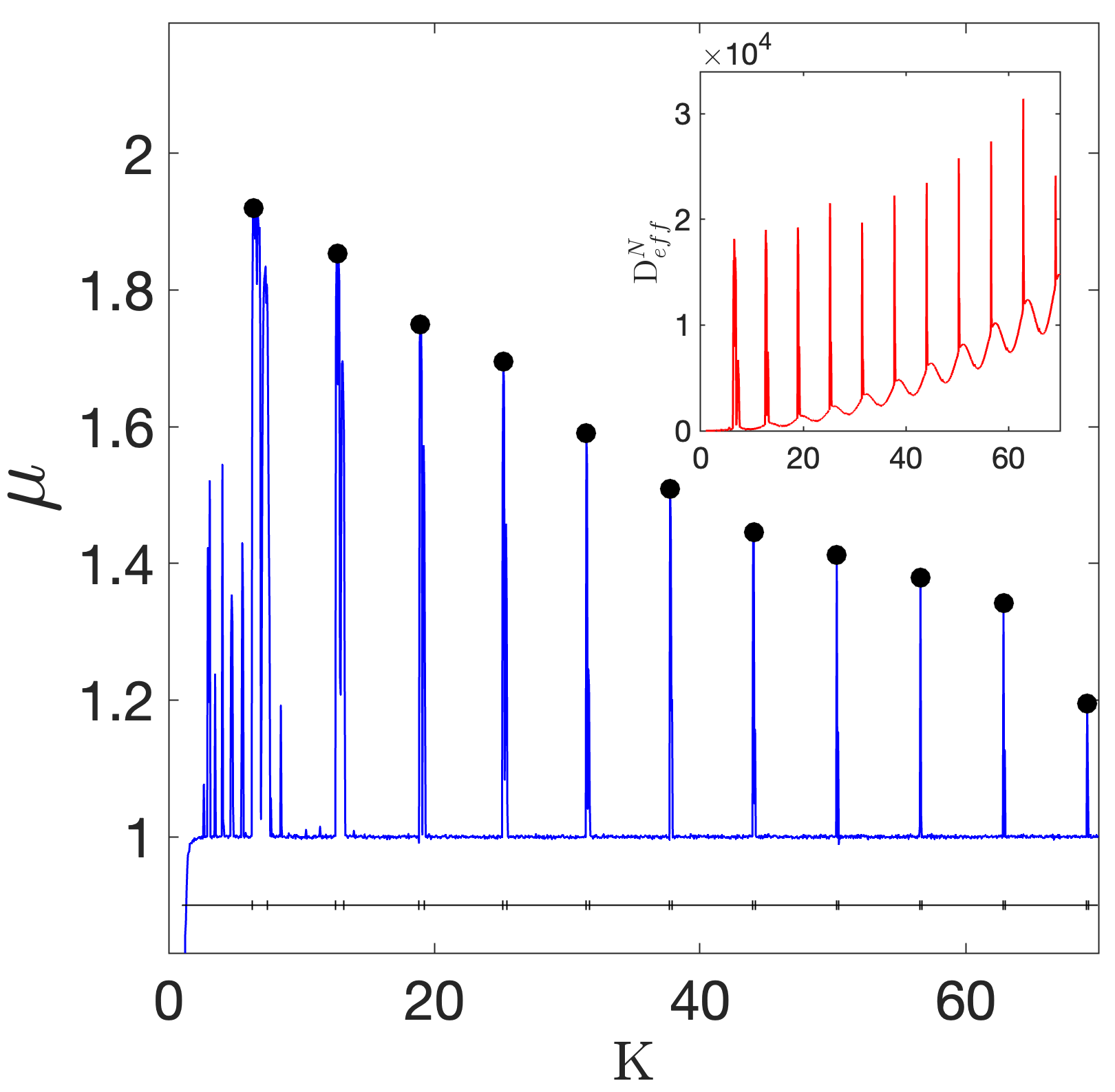}} 
    \subfloat[{\( \mu (K) \) [inset: \( D^N_{\text{eff}} (K)\)] for $\beta = 10^{-3}$}\label{fig5:Fig08b}]{\includegraphics[width=0.475\textwidth]{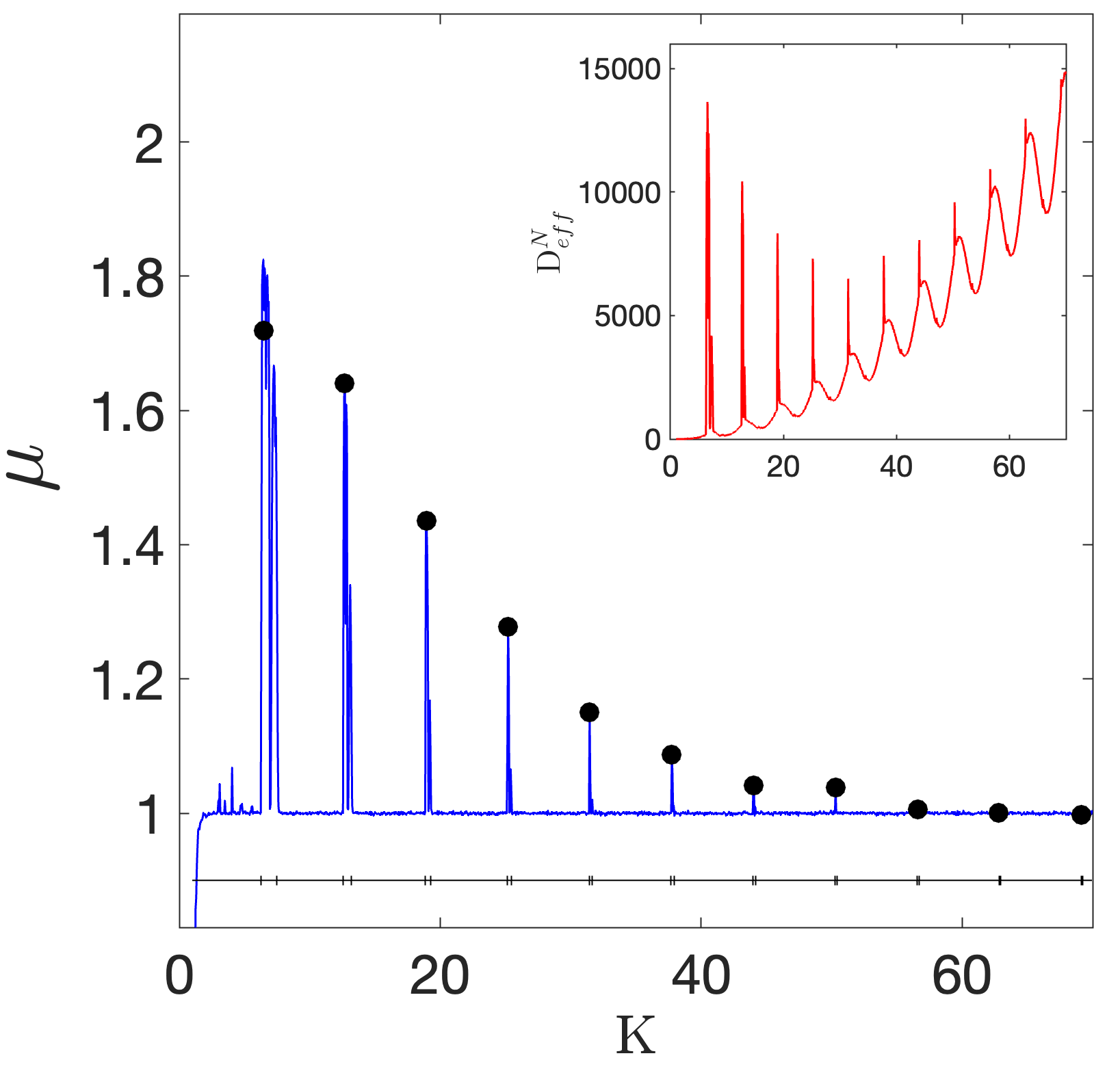}}
    \caption{The diffusion exponent \( \mu (K) \) \eqref{eq:pvarN} [insets: the effective diffusion coefficient \( D^N_{\text{eff}} (K)\) \eqref{eq:Deff_csm}] of the \(N = 5\) coupled system \eqref{eq:csm} with equal \(K_j = K = 6.5\), \(j = 1, 2, \ldots, 5\) values for (a) \(\beta = 10^{-4}\) and (b)  \(\beta = 10^{-3}\). The values of \( \mu \) are computed for approximately \(100,000\) ICs on grids covering the entire phase space of each of the five SMs. The black horizontal lines at the bottom of the panels indicate the \(K\) intervals where period $p=1$ AMs exist, similar to Fig.~\ref{fig5:Fig03}.}
    \label{fig5:Fig08}
   \end{figure}

 We initially start with a system of $N=5$ coupled SMs, where each \(2D\) map has identical setups by considering equal kick-strength parameters \(K\) across all maps, i.e., $K_j = K$, $j = 1, \dots, 5$. In Fig.~\ref{fig5:Fig08a}, we show the diffusion exponent $\mu$ \eqref{eq:pvarN} and the effective diffusion coefficient $D^N_{\text{eff}}$ \eqref{eq:Deff_csm} (inset) of the system as functions of $K$ for a specific coupling strength value, $\beta = 10^{-4}$. We chose this moderate coupling value to ensure that we observe the system's dynamical effects, such as diffusion and chaotic behavior, within reasonable computational times. Results shown in Fig.~\ref{fig5:Fig08a} are obtained using a large set of ICs (approximately \(100,000\)) on a $315 \times 315$ grid covering the entire phase space of each \(2D\) SM. 

We observed that the diffusion exponent \(\mu\) is sensitive to the coupling strength, especially for values of $K$ where period $p=1$ AMs are present. In Fig.~\ref{fig5:Fig08b}, we repeat this analysis for a larger coupling strength, $\beta = 10^{-3}$. We can clearly observe a gradual decrease in the values of the diffusion exponent toward normal diffusion $(\mu =1)$ as $\beta$ increases, particularly for \(K\) intervals containing period $p=1$ AMs. These intervals (denoted by the horizontal black line at the bottom of the panels of  Fig.~\ref{fig5:Fig08}) are in a way similar to those in Fig.~\ref{fig5:Fig03}. Furthermore, the decrease effect is more pronounced for higher values of the nonlinear parameter $K$. The results presented in Fig.~\ref{fig5:Fig08} depend on the number of iterations $n$ and, in principle, on the number of the considered ICs. This latter dependence, previously highlighted for the SM [Fig.~\ref{fig5:Fig03}], will also be discussed later. 

\begin{figure}[!htbp]
    \centering
    \subfloat[{\( \mu (K) \) [inset: \(D^N_{\text{eff}} (K)\)] for $\beta = 10^{-4}$}\label{fig5:Fig082a}]{\includegraphics[width=0.475\textwidth]{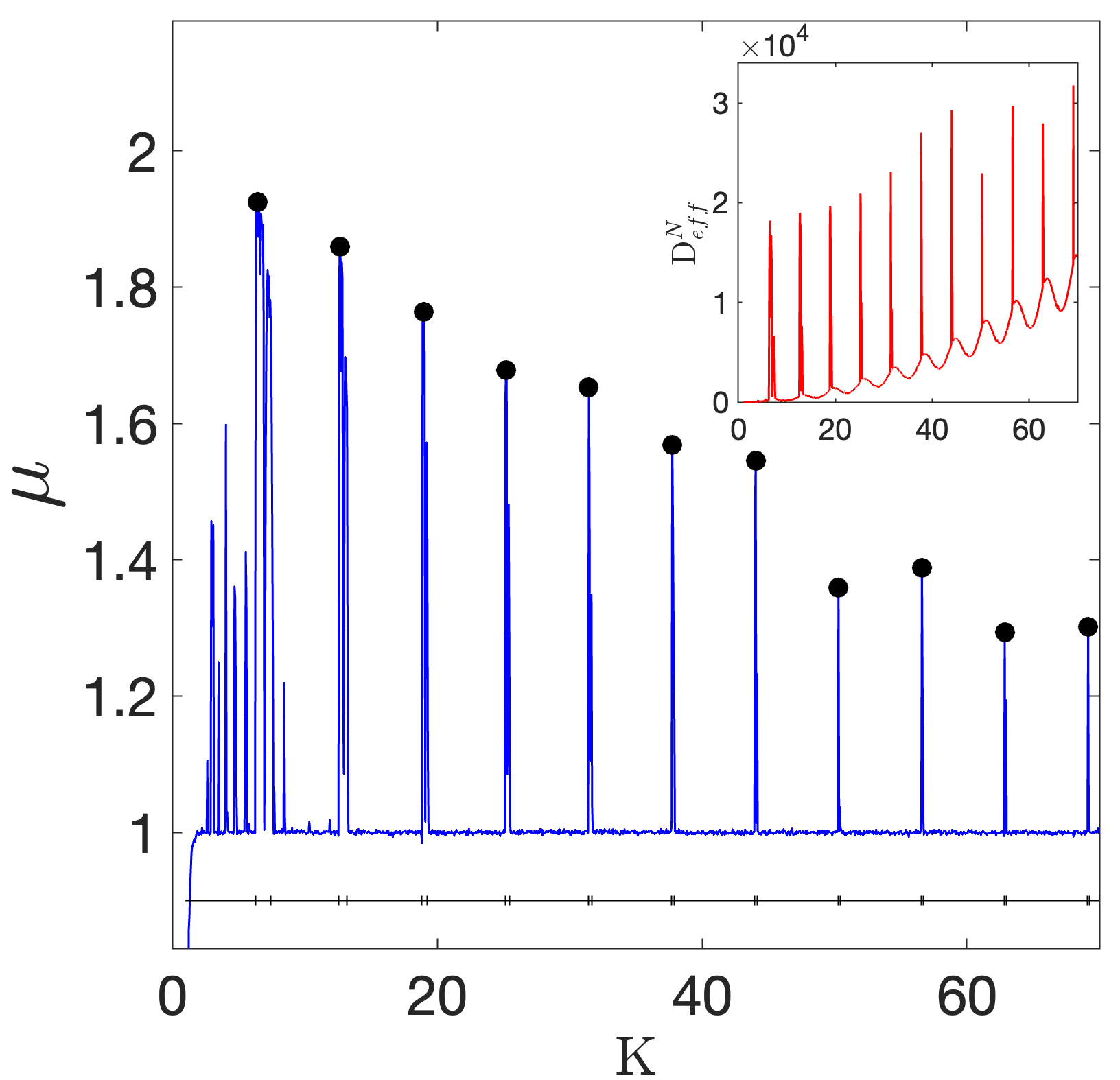}} 
    \subfloat[{\( \mu (K) \) [inset: \( D^N_{\text{eff}} (K)\)] for $\beta = 10^{-3}$}\label{fig5:Fig082b}]{\includegraphics[width=0.475\textwidth]{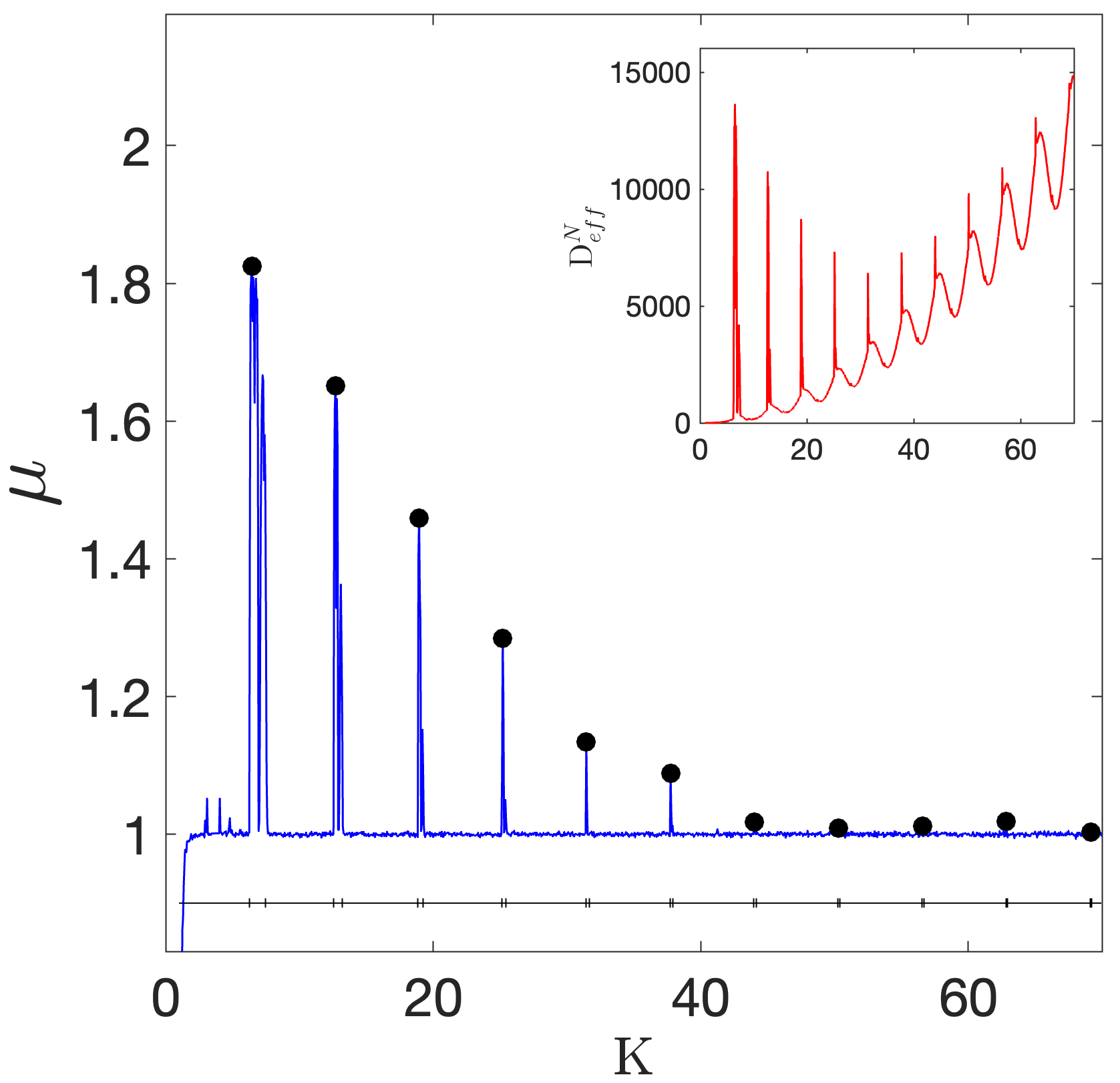}}
    \caption{Similar to Fig.~\ref{fig5:Fig08} but for approximately $50, 000$ ICs on $224 \times 224$ grids for each of the \(N=5\) SM covering the entire phase space of the system.}
    \label{fig5:Fig082}
   \end{figure}

In order to make our numerical simulations computationally feasible, we carefully chose a sufficiently large number of ICs to accurately capture the fundamental dynamical features of the coupled SM systems \eqref{eq:csm} without requiring extensive computational resources. After testing various sizes of orbit ensembles, we concluded that \(100,000\) ICs (similarly to what was used for the \(2D\) map \eqref{eq:ssm} in Sect.~\ref{sec:single SMs} can effectively capture the dynamics in a realistic computation time. As an example, Fig.~\ref{fig5:Fig082} is presented, which is similar to Fig.~\ref{fig5:Fig08}, but obtained by using half the number of ICs. In this case, we use approximately $50, 000$ ICs considered on a $224 \times 224$ grid in each \(2D\) map instead of approximately $100, 000$ ICs obtained by $315 \times 315$ grids over each SM's entire phase space. By comparing Figs.~\ref{fig5:Fig08} and \ref{fig5:Fig082}, we observe no significant difference in the presented dynamical features. To balance computational efficiency and accuracy, we opted to use  $315 \times 315$ (approximately $100,000$ ICs) grids throughout our investigations, similar to previous studies such as \citep{ManRob2014PRE}. This grid is sufficiently large to correctly capture the basic features of the dynamics in feasible computational times while avoiding excessive computational demand. 

Building upon the analysis of a coupled system of SMs with identical $K$ values, we now focus on the impact of increasing the coupling strength ($\beta$) on the system's diffusion properties. In particular, we examine the behavior of the maximum diffusion exponent $\mu^*$ \eqref{eq:pvarN} values associated with regions where AMs of period $p=1$ are present in the individual SM. These $\mu^*$ values come from the $11$ pronounced peaks in the diffusion exponent seen in both Figs.~\ref{fig5:Fig03} or \ref{fig5:Fig08}.

\begin{figure}[!htbp]
    \centering
    \subfloat[$\mu^*(K)$\label{fig5:Fig09a}]{\includegraphics[width=0.475\textwidth]{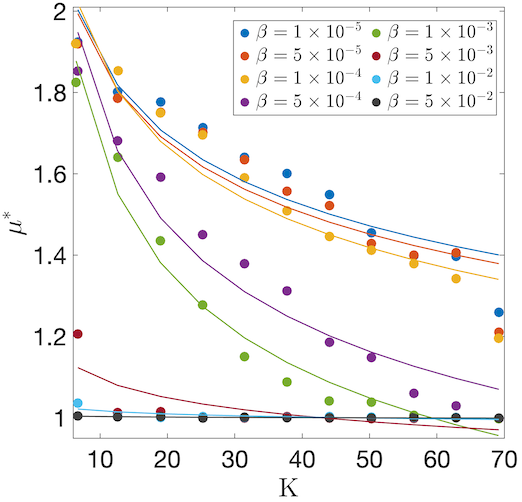}} 
    \subfloat[$A^\prime(n)$ and $B^\prime(n)$\label{fig5:Fig09bc}]{\includegraphics[width=0.475\textwidth]{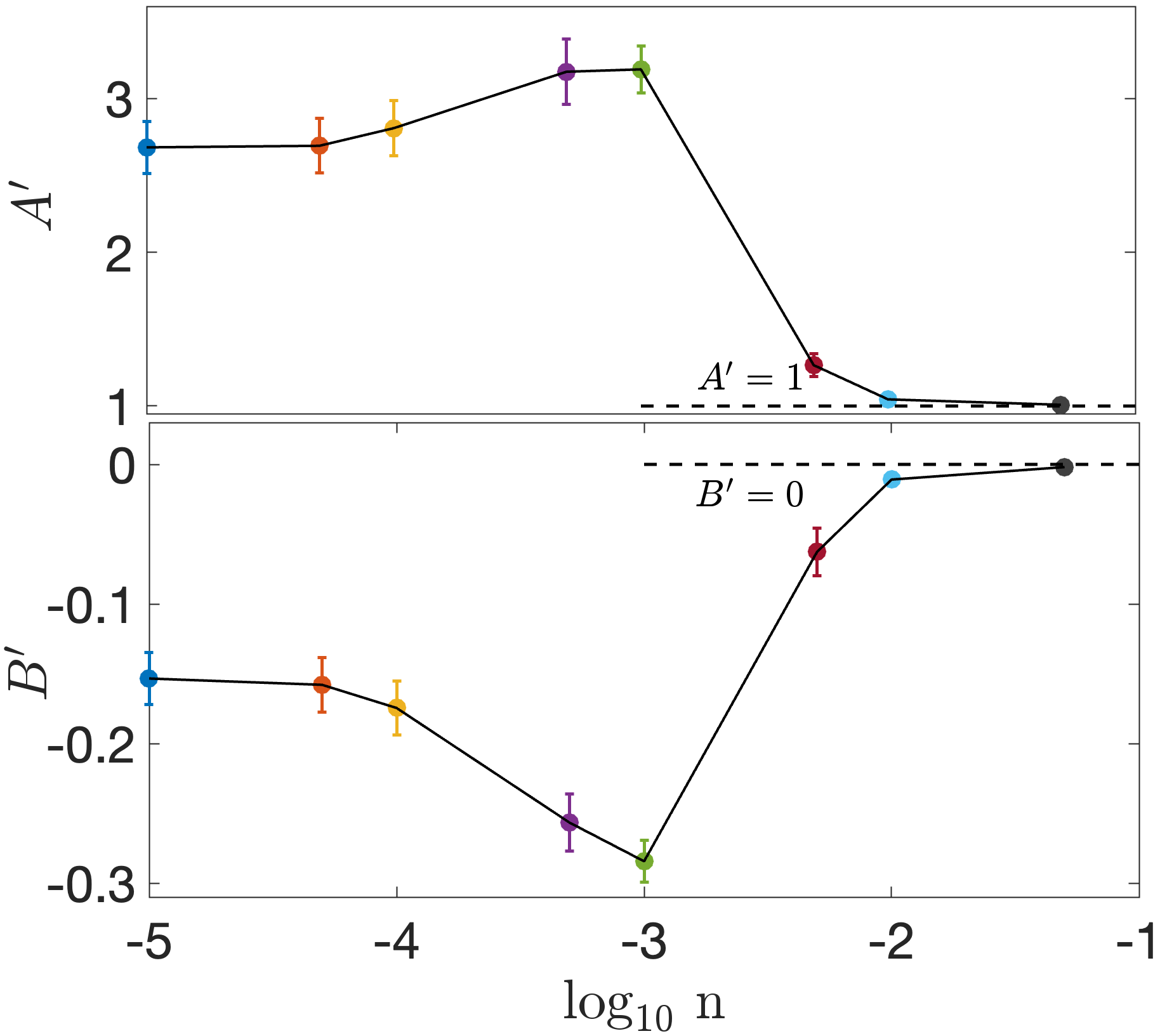}}
    \caption{(a) The maximum diffusion exponents \( \mu^*\) for  the first \(11\) \( K = K_j \), \(j=1, \dots, 5\) values where period $p=1$ AMs of the $N=5$ couples SMs system \eqref{eq:csm} exist are computed for eight  $\beta$ values on $\beta \in [10^{-5}, 5 \times 10^{-2}]$ (see the legend for the specific values) at \(n = 10^4\) iterations. \( \mu \) is computed for approximately \(100,000\) ICs on a grid covering the map's entire phase space \( x \in (0, 2\pi)\) and \(p \in (0, 2\pi) \). Solid curves represent fitting to the function  \(\mu^* = A^\prime K^{B^\prime}\) for each set of data points. (b) The resulting values of the fitting parameters \( A^\prime(n)\) and \( B^\prime(n) \), along with their corresponding determination errors. Horizontal dashed lines indicate \(A^\prime = 1\) and \(B^\prime = 0\). See text in the description for details.}
    \label{fig5:Fig09}
   \end{figure}

Figure \ref{fig5:Fig09a} shows the values of $\mu^*$ as a function of $K$ after $n = 10^4$ iterations of the coupled SM \eqref{eq:csm} for various coupling parameters $\beta$. By fitting the $\mu^*$ values to a function of the form $\mu^* = A^\prime K^{B^\prime}$ for each one of the different considered coupling strengths, we observe a clear trend towards normal diffusion in all cases. Even for moderate coupling strength values $\beta \gtrsim 5\times10^{-3}$ (brown, cyan, and black points and curves in Fig.~\ref{fig5:Fig09a}), the coupled SMs system \eqref{eq:csm} quickly reaches the global normal diffusion rates ($\mu=1$) even for small $K$ values. As the coupling strength increases, the system quickly transitions from superdiffusive behavior (\(\mu > 1\)), which is influenced by the period $p=1$ AMs, to normal diffusion (\(\mu = 1\)). This transition is shown by the decreasing fitting function, in which curves eventually reach values that indicate normal diffusion ($\mu^* = 1$) regardless of the value of $K$. The fitting parameters $A^\prime$ and $B^\prime$, along with their standard deviations, are shown in Fig.~\ref{fig5:Fig09bc} as functions of the coupling parameter $\beta$. From the results of Fig.~\ref{fig5:Fig09bc} a clear trend emerges: $\mu^*$ approaches $\mu = 1$ and becomes independent of $K$ as $A^\prime$ and $B^\prime$ approach $1$  and $0$, respectively [dashed lines in Fig.~\ref{fig5:Fig09bc}] for relatively large $\beta$ values. The results in Fig.~\ref{fig5:Fig09} highlight that strong coupling between neighboring SMs can suppress the superdiffusive behavior caused by AMs, consequently leading the global diffusion process of the coupled system to exhibit normal diffusion in SMs.

\begin{figure}[!htbp]
    \centering
    \includegraphics[width=0.5\textwidth]{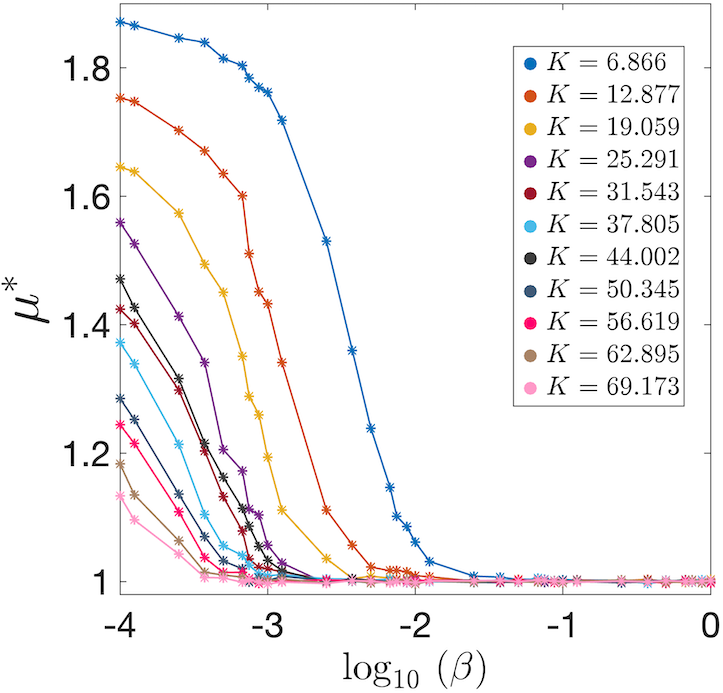}
    \caption{The maximum diffusion exponents \(\mu\) as a function of the coupling strength \(\beta\) of the $N=5$ couples SMs system \eqref{eq:csm} with values of \(K_j = K\), \(j = 1, 2, \ldots, 5\) for which pronounced superdiffusion transport takes place (values of $K$ corresponding to the black filled circles in Fig.~\ref{fig5:Fig08}).  \( \mu \) is computed for approximately \(100,000\) ICs on a grid covering the map's entire phase space \( x \in (0, 2\pi)\) and \(p \in (0, 2\pi) \) at \(n = 10^4\) iterations.}
    \label{fig5:Fig09d}
   \end{figure}

Continuing the analysis of the coupled system \eqref{eq:csm}, we now investigate how the maximum diffusion exponent values ($\mu^*$) associated with strong superdiffusion change with variations of  the coupling strength ($\beta$). By examining the behavior of $\mu^*$ for a specific set of $K$ values which produce strong superdiffusion, we found that increasing coupling strength effectively suppressed superdiffusive behavior (Fig.~\ref{fig5:Fig09d}). This trend is evident in the gradual decay of $\mu^*$ towards a diffusion rate $\mu \approx 1$ as $\beta$ increases in Fig.~\ref{fig5:Fig09d}. Such behavior typically occurs from the extended chaotic regions in the phase space of the SM due to an increase in the values of $\beta$, where normal diffusion ($\mu = 1$) is observed for ICs located further from the stable AMs regions. 

The findings in Figs.~\ref{fig5:Fig08} and \ref{fig5:Fig09} show that strong coupling between each single \(2D\) SM leads to global normal diffusion, even if there are initially local regions where superdiffusion takes place. Hence, the coupling strength $\beta$ can effectively control the transition from superdiffusion ($\mu > 1$) to almost normal diffusion ($\mu \approx 1$) in the coupled SMs model \eqref{eq:csm}. 

\begin{figure}[!htbp]
    \centering
    \subfloat[\(\sum_{j=1}^{N} \langle (\Delta p^j)^2 (n)\rangle\)\label{fig5:Fig10a}]{\includegraphics[width=0.475\textwidth]{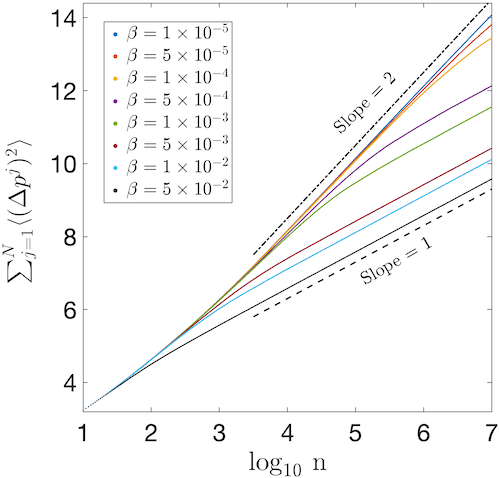}} 
    \subfloat[\(\mu (n)\)\label{fig5:Fig10b}]{\includegraphics[width=0.475\textwidth]{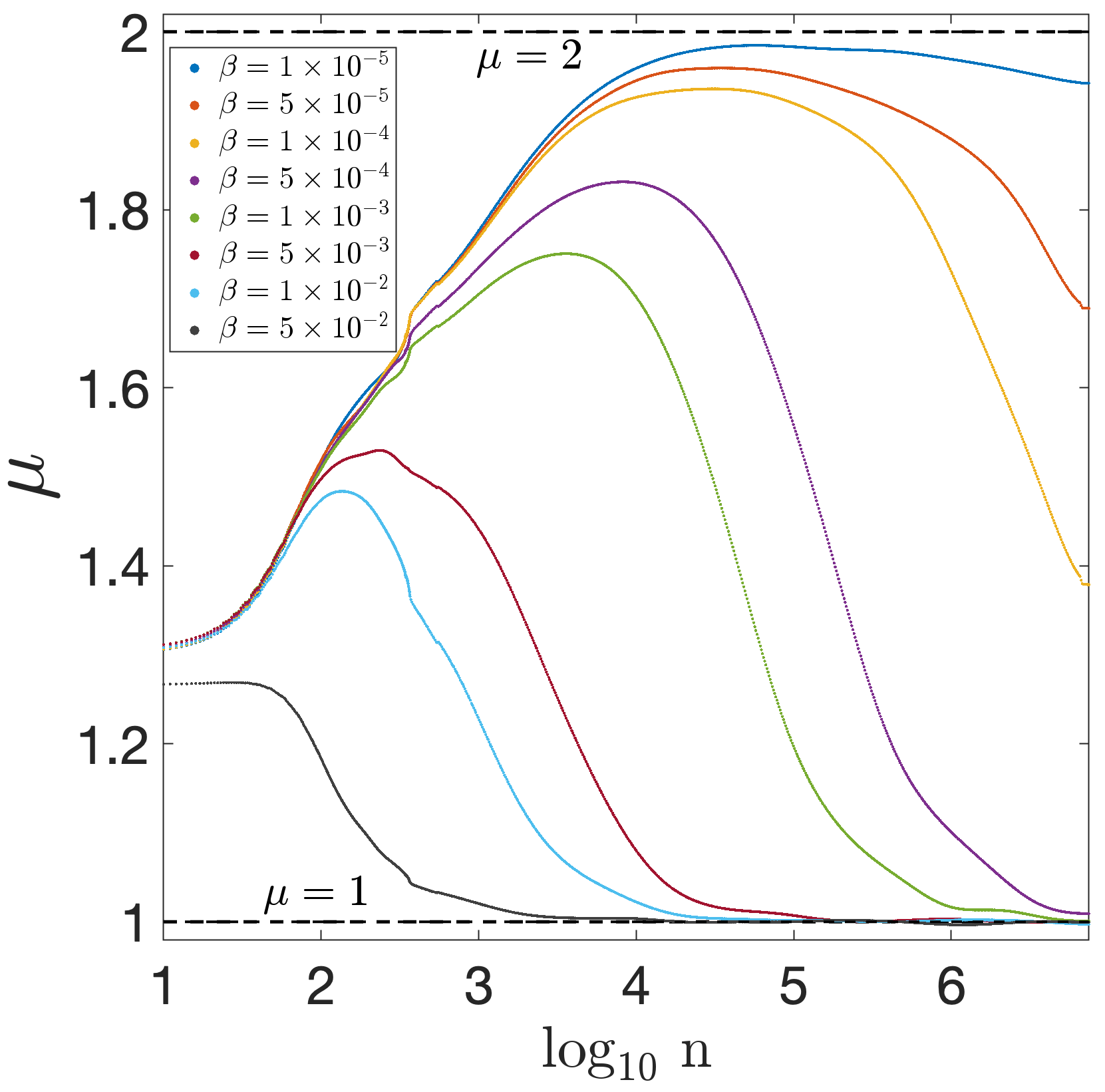}} \\
    \subfloat[Average ftmLE, \(\langle \sigma_1 (n) \rangle\)\label{fig5:Fig10c}]{\includegraphics[width=0.475\textwidth]{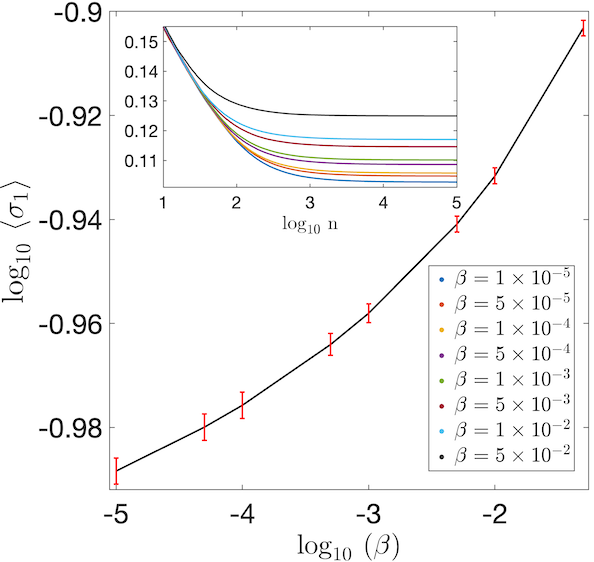}}
    \subfloat[Average GALI\(_2 (n)\)\label{fig5:Fig10d}]{\includegraphics[width=0.475\textwidth]{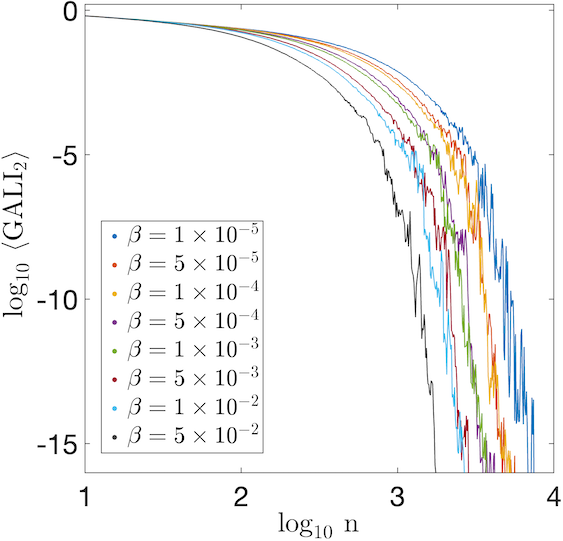}}
    \caption{(a) The variance \(\sum_{j=1}^N \langle (\Delta p^j)^2 \rangle\) \eqref{eq:pvarN} and (b) the corresponding numerically estimated diffusion exponent \(\mu (n)\) \eqref{eq:Deff_csm} of the $N=5$ coupled SMs system \eqref{eq:csm} with equal kick parameters, \(K_j = K = 6.5\), for various \(\beta\) values (see the legend for the specific values). The setup and ensemble of ICs are similar to Figs.~\ref{fig5:Fig08} and \ref{fig5:Fig09}. The black dotted and dashed lines in (a) indicate particular diffusion rates, while the horizontal dash-dotted line in (b) represents $\mu =2$. (c) The average ftmLE \eqref{eq:ftmLE}, \(\langle \sigma_1 (\beta) \rangle\), after \(n = 10^5\) iterations over approximately \(100,000\) ICs, with error bars showing one standard deviation in the computation of the average value. Inset: The evolution of \(\langle \sigma_1 (n) \rangle\). (d) The average \(\langle \text{GALI}_2 (n) \rangle\) values for the same set of ICs as in (c). The respective standard deviations (not shown here) are very small and hardly visible.}     
    \label{fig5:Fig10}
\end{figure}

Now let us explore the relationship between the coupling parameter and the time it takes for coupled SMs to reach ballistic transport ($\mu = 2$). We first plot the variance \(\sum_{j=1}^N \langle (\Delta p^j)^2 \rangle\) \eqref{eq:Deff_csm} in Fig.~\ref{fig5:Fig10a} and the diffusion exponent \eqref{eq:pvarN} in Fig.~\ref{fig5:Fig10a} for ensembles of approximately \(100, 000\) ICs of the \(N = 5\) coupled SMs system with \(K_j = K = 6.5\)), as function of the number of iterations $n$ for different coupling strengths \(\beta\). These plots allow us to investigate how $\beta$ influences the system's diffusion properties towards ballistic transport. The results show a gradual transition from superdiffusive \(\mu > 1\) to normal diffusive \(\mu = 1\) behavior for all considered $\beta$ values. It is noted that transition to normal diffusion requires fewer iterations for relatively larger coupling strengths.

Moreover, we analyze the chaotic behavior of the coupled SMs \eqref{eq:csm} by computing the average (over all considered ICs) ftmLE \eqref{eq:ftmLE}, \(\langle \sigma_1 \rangle\), [Fig.~\ref{fig5:Fig10c}] and the average (over all considered ICs) GALI\(_2\) \eqref{eq:GALI}, \(\langle \text{GALI}_2 \rangle\), [Fig.~\ref{fig5:Fig10d}] for the same ICs and parameter setup as in Figs.~\ref{fig5:Fig10a} and ~\ref{fig5:Fig10b}. The computation of the ftmLE and GALI\(_2\) indices help us understand the connection between coupling strength, chaoticity, and the time required for the global diffusion to reach ballistic transport in the coupled system. The inset of Fig.~\ref{fig5:Fig10c} shows the evolution of the \(\langle \sigma_1 \rangle\) with respect to \(n\). Larger \(\langle \sigma_1 \rangle\) values are obtained for stronger coupling strength \(\beta\) cases, which exhibit normal diffusion rates in Fig.~\ref{fig5:Fig10b}. The monotonic relation between the degree of chaos and coupling strength \(\beta\) is further supported by the exponential decay to zero of the GALI$_2$ values for all considered \(\beta\) values, which confirms global chaotic behavior of the coupled SMs system \eqref{eq:csm} [Fig.~\ref{fig5:Fig10d}]. Additionally, we can also observe a relationship between the time required for the \(\langle \text{GALI}_2 \rangle\) to reach zero and the related \(\langle \sigma_1 \rangle\) value, with larger numbers of iterations corresponding to smaller ftmLE values. This relation becomes apparent when we  compare the results obtained for the smallest coupling value we considered, \(\beta_1 = 10^{-5}\) [blue curves in Fig.~\ref{fig5:Fig10}] with the ones found for the largest value, \(\beta_8 = 5 \times 10^{-2}\) [black curves in Fig.~\ref{fig5:Fig10}]. For \(\beta_8\) \(\langle \sigma_1 \rangle\) saturates to a positive value more quickly [Fig.~\ref{fig5:Fig10c}], and at the same time \(\langle \text{GALI}_2 \rangle\) decays to zero more rapidly [Figs.~\ref{fig5:Fig10d}], indicating stronger chaotic behavior. 

A natural question arises whether all cases exhibiting the superdiffusion rates (\(\mu > 1\)) in Figs.~\ref{fig5:Fig10a} and (b) have indeed truly converged to their final diffusion exponent value of \(\mu = 1\). This question is particularly relevant given the considered number of iterations ($n=10^7$) and number of ICs ($100, 000$), especially for the smallest three $\beta$ values, namely \( \beta = 10^{-5}\) (blue curves), $\beta=5 \times 10^{-5}$ (green curves), and $\beta=10^{-4}$ (orange curves). In order to address this question, we further analyze cases with varying numbers of ICs and compute the diffusion measures for even larger $n$ values by focusing on the \(\beta=10^{-4}\) case [orange curves in Figs.~\ref{fig5:Fig10}]. 

\begin{figure}[!htbp]
    \centering
    \subfloat[\(\sum_{j=1}^{N} \langle (\Delta p^j)^2 (n)\rangle\)\label{fig5:FigA10a}]{\includegraphics[width=0.475\textwidth]{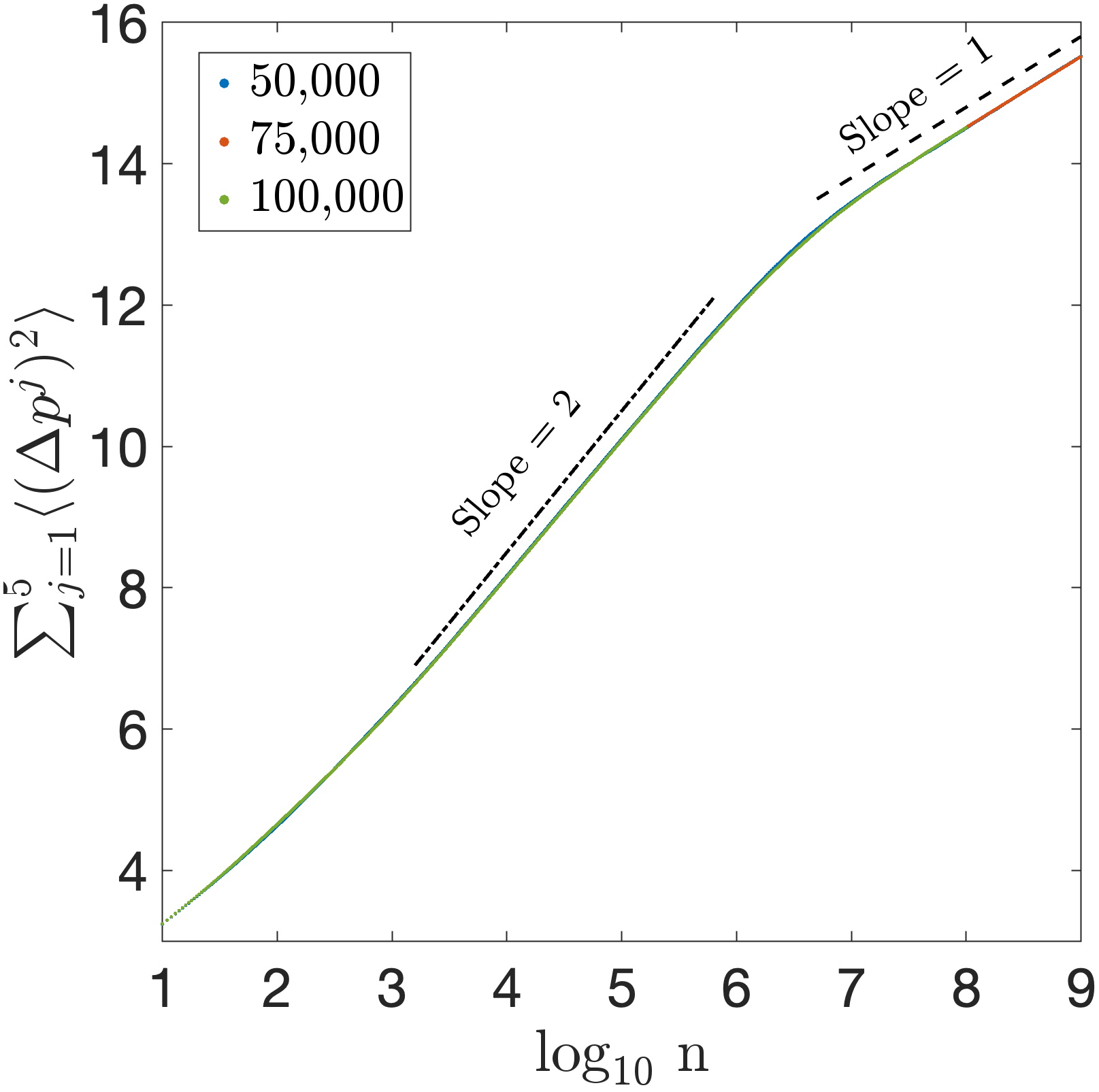}} 
    \subfloat[\(\mu (n)\)\label{fig5:FigA10b}]{\includegraphics[width=0.475\textwidth]{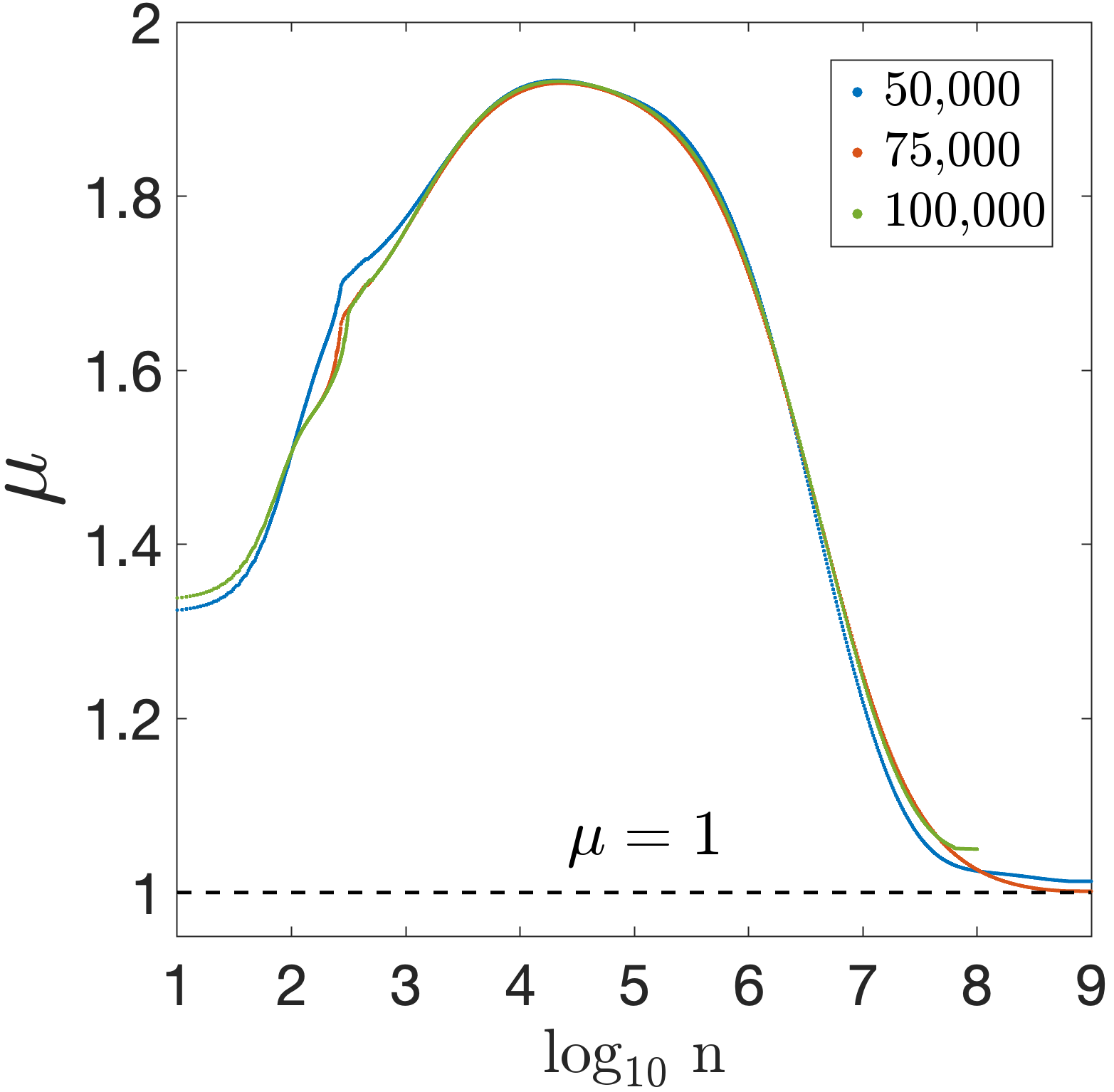}} 
    \caption{(a) and (b) similar to Figs.~\ref{fig5:Fig10a} and \ref{fig5:Fig10b}, respectively, but for $\beta=10^{-4}$ with $50,000$ (blue curves), $75,000$ (red curves), and $100,000$ (green curves) ICs on the grid. Due to computational limitations, the evolution of the ICs for the green curves was stopped earlier than the final number of iterations we considered, $n = 10^9$.}
    \label{fig5:FigA10}
\end{figure}

We present the computed variance \eqref{eq:pvarN} for ensembles of approximately \(100, 000\) ICs in Fig.~\ref{fig5:FigA10a} and the corresponding computed diffusion exponent \(\mu (n)\) \eqref{eq:pvarN} in Fig.~\ref{fig5:FigA10a} of the $N=5$ coupled SMs system \eqref{eq:csm} as functions of the iterations $n$. Here, we use equal kick parameter values \(K_j = K = 6.5\) and set  \(\beta=10^{-4}\) for $50,000$ (blue curves), $75,000$ (red curves), and $100,000$ (green curves) ICs on the grid of each SM's phase space. The results in Fig.~\ref{fig5:FigA10} clearly show that the different sets of ICs lead to similar outcomes, confirming \(\mu\) indeed eventually converges to normal diffusion ($\mu = 1$), as we deduced from the findings in Fig.~\ref{fig5:Fig10}. 

The impact of increasing the number of iterations $(n)$ in Fig.~\ref{fig5:FigA10b} is further illustrated in Fig.~\ref{fig5:Fig10b}. As $n$ increases, $\mu$ gradually approaches $\mu = 1$, with this convergence happening faster for larger $\beta$ values. It is worth noting that all curves practically overlap, especially in Fig.~\ref{fig5:Fig10a}. We observe that the green curve requires more number of iterations to reach the final time we considered. However, the trend indicates that $\mu$ converges towards $\mu = 1$.

\begin{figure}[!htbp]
    \centering
    \subfloat[\( \mu^*(K) \)\label{fig5:FigA1209a}]{\includegraphics[width=0.475\textwidth]{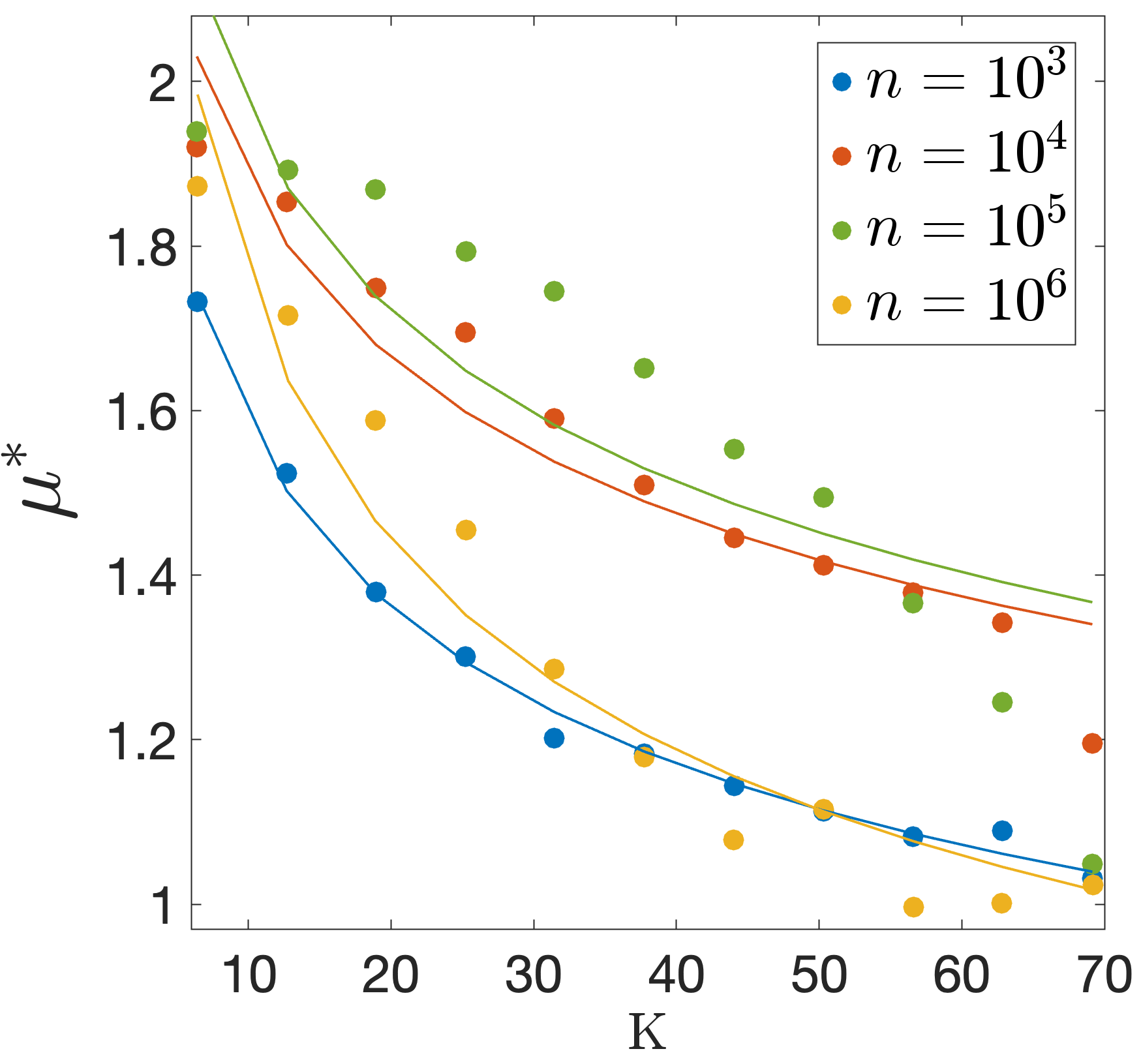}}
    \subfloat[\( \mu^*(K) \)\label{fig5:FigA1209b}]{\includegraphics[width=0.475\textwidth]{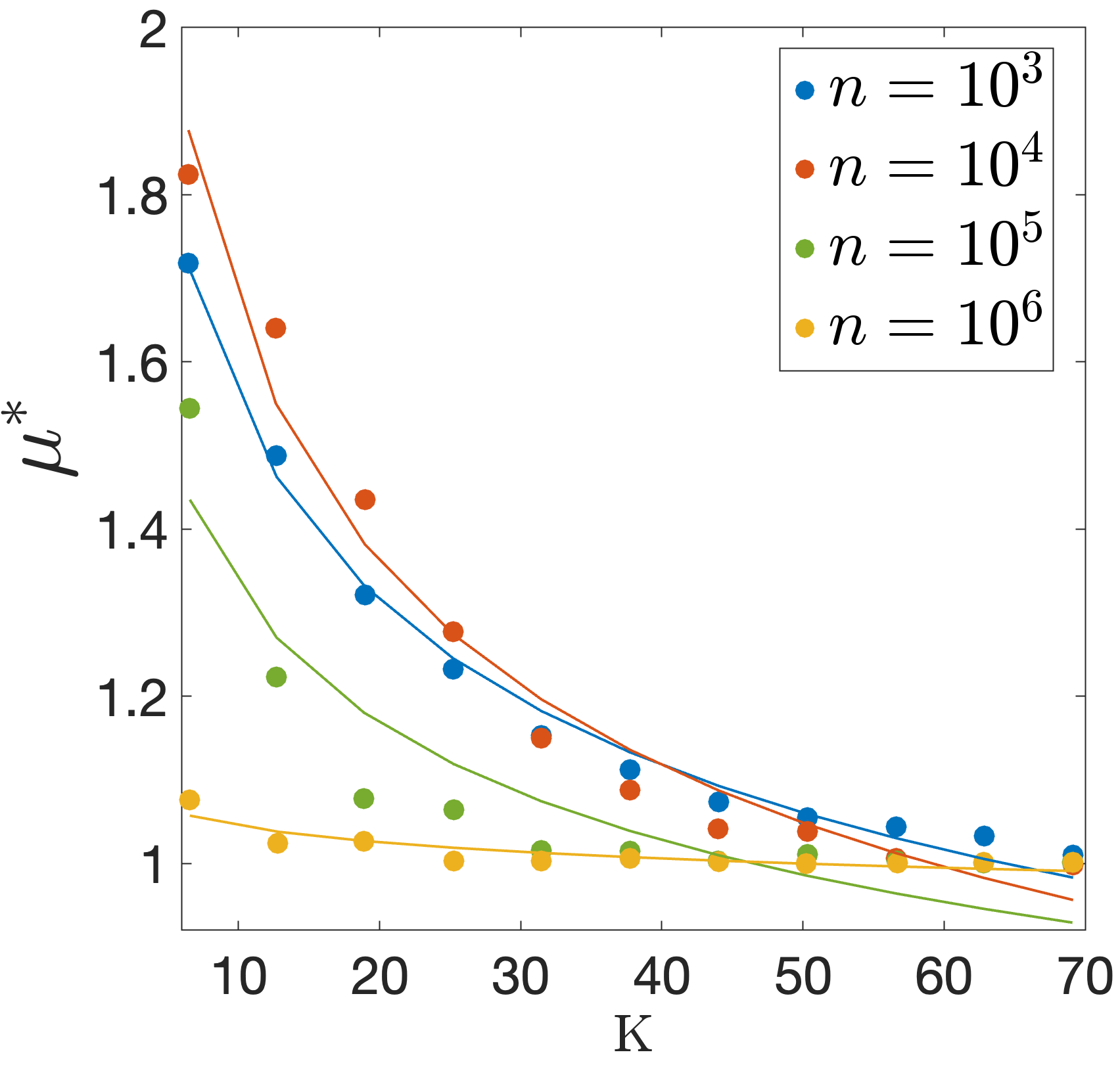}}\\
    \subfloat[$A^*(n)$ and $B^*(n)$\label{fig5:FigA1209c}]{\includegraphics[width=0.475\textwidth]{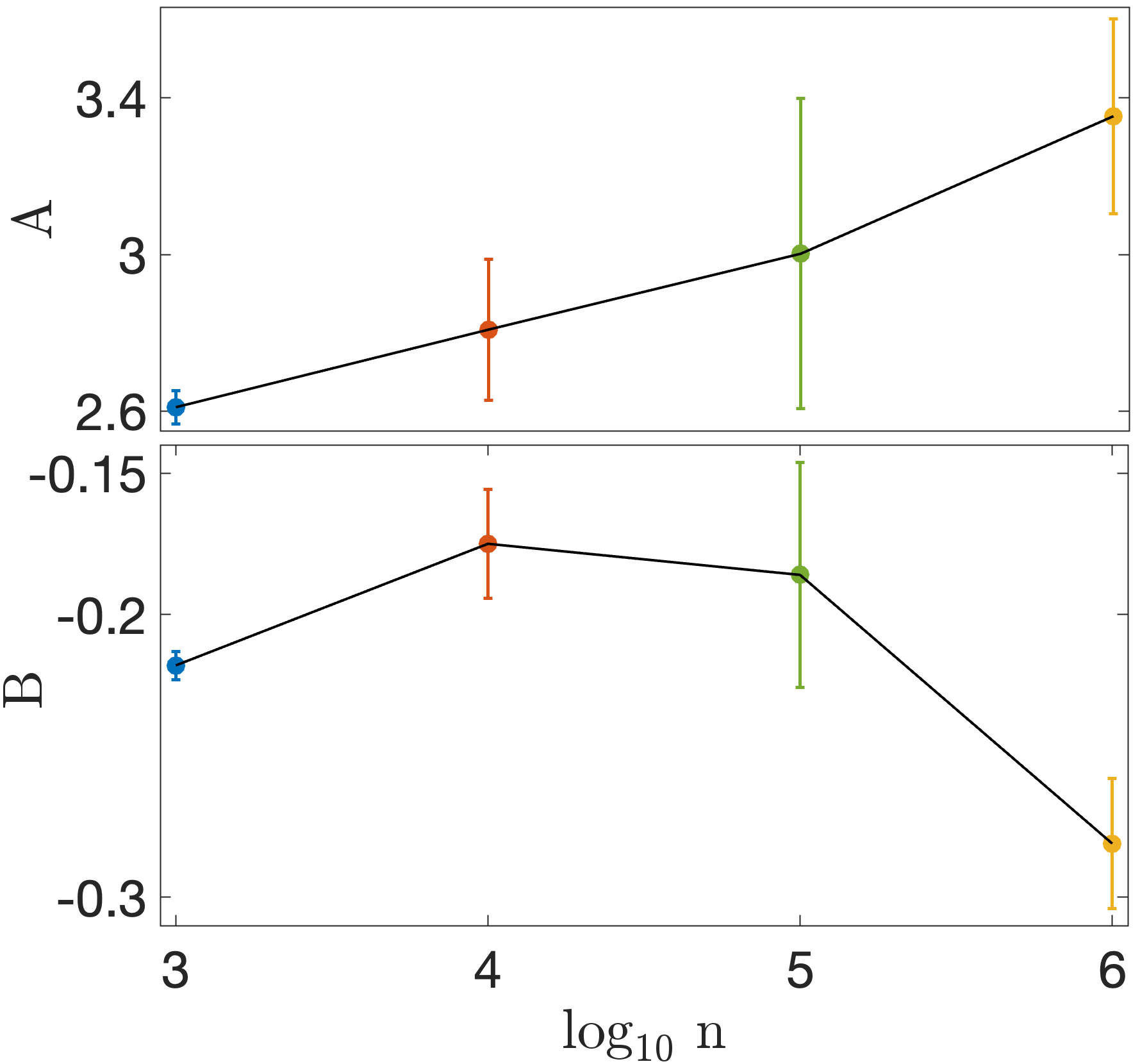}}
    \subfloat[$A^*(n)$ and $B^*(n)$\label{fig5:FigA1209d}]{\includegraphics[width=0.475\textwidth]{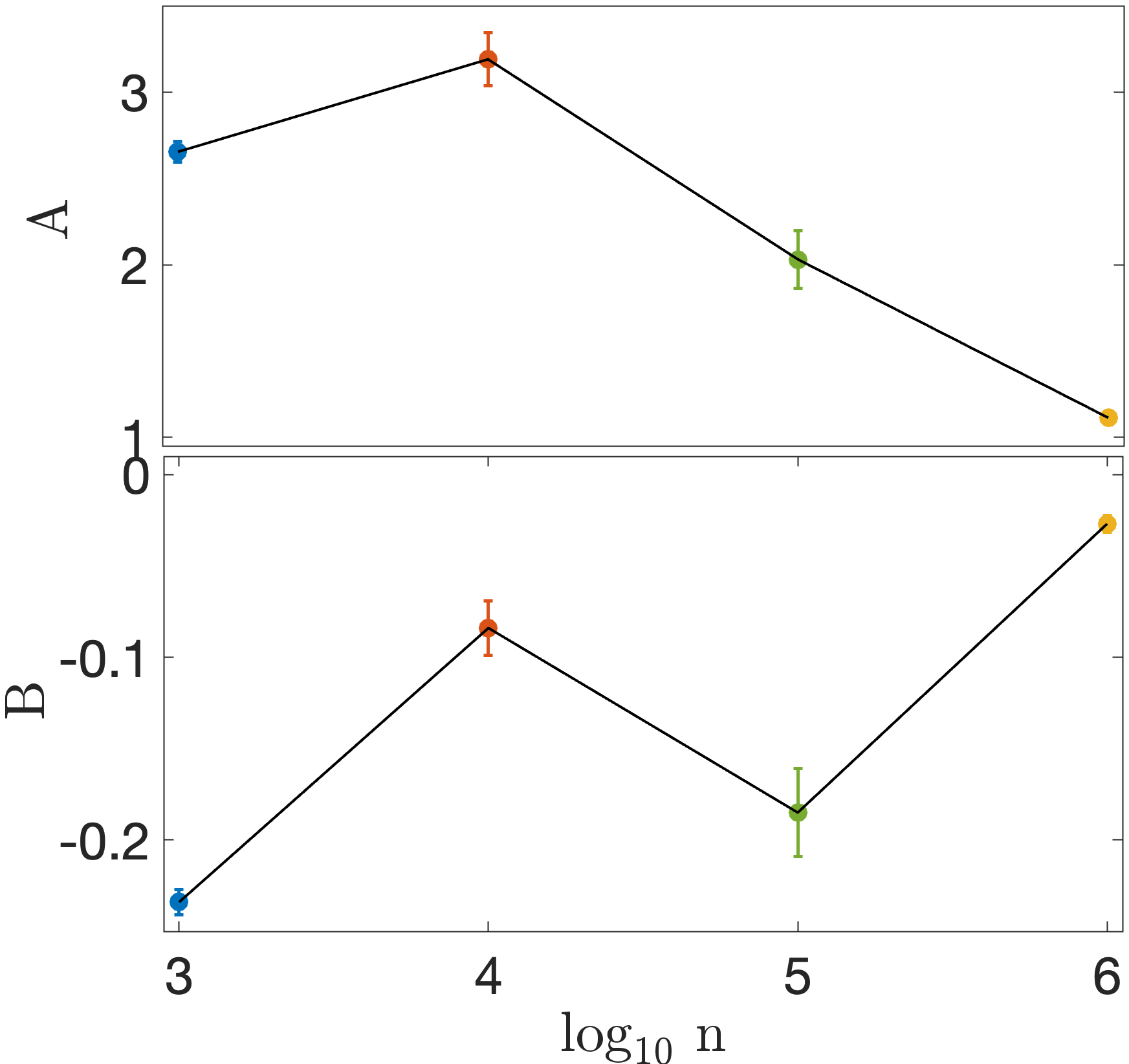}}
    \caption{The maximum diffusion exponents \( \mu^*(K) \) as a function of \(K\) values on the first \(11\) intervals where period $p=1$ AMs of the coupled SMs system \eqref{eq:ssm} exist [Fig.~\ref{fig5:Fig08}] for (a) $\beta=10^{-4}$ and (b) $\beta=10^{-3}$, computed at \(n = 10^3\), \(10^4\), \(10^5\), and \(10^6\), iterations of the maps (represented by blue, red, green, and orange points, respectively). The diffusion exponent \( \mu \) is computed using approximately \(100,000\) ICs on a grid covering the entire phase space of the map \( x \in (0, 2\pi)\) and \(p \in (0, 2\pi) \). Solid curves depict fittings of the data points to the function \(\mu^*(K) = A^* K^{B^*}\) for each iteration set. The resulting values of the fitting parameters \( A^*(n)\) and \( B^*(n)\), with their associated determination errors for (c) $\beta=10^{-4}$ and (d) $\beta=10^{-3}$.}
    \label{fig5:FigA1209}
   \end{figure}

To further explore the behavior of the maximum diffusion exponent values, $\mu^*$, associated with period $p=1$ AMs for the $N=5$ coupled SMs system \eqref{eq:csm}, we perform a similar analysis to the case of the SM \eqref{eq:ssm} shown in Fig.~\ref{fig5:Fig05}.  Fig.~\ref{fig5:FigA1209} illustrates the $\mu^*$ as a function of $K$ values in the first \(11\) intervals observed in Fig.~\ref{fig5:Fig08} for $\beta = 10^{-4}$ [Fig.~\ref{fig5:FigA1209a}] and $\beta = 10^{-3}$ [Fig.~\ref{fig5:FigA1209b}], evaluated at different iteration numbers: $n = 10^3$ (blue points and curves), $n = 10^4$ (red points and curves), $n = 10^5$ (green points and curves), and $n = 10^6$ (orange points and curves). The plots in Fig.~\ref{fig5:FigA1209} are similar to Fig.~\ref{fig5:Fig05} but now we consider the coupled system of five identical \(2D\) maps with varying \(K_j\) values. For instance, in the first set of points in Fig.~\ref{fig5:FigA1209}, $\mu^*$ can be obtained for \(K_j = K = 6.5\) from the intersection points of the $\beta = 10^{-4}$ (orange curves) and  $\beta = 10^{-3}$ (green curves) in Fig.~\ref{fig5:Fig10b} with the vertical lines at $n={10}^3,\ n={10}^4,\ n={10}^5$ and $n={10}^6$.  

Following the approach in Fig.~\ref{fig5:Fig05}, we also fit the peak $\mu^*$ values to the nonlinearity strength $K$ for each number of iterations $n$ using a power law of the form $\mu^*(K) = A^* K^{B^*}$. The resulting fittings are shown by the corresponding colored curves in Figs.~\ref{fig5:FigA1209a} and (b), along with the associated parameters $A^*$ and $B^*$ as functions of the map's iteration number $n$ presented in Figs.~\ref{fig5:FigA1209c} and (d), respectively.

In contrast to the monotonic relationship between $\mu^*$ and $n$ observed for each $K$ in Fig.~\ref{fig5:Fig05a} for the \(2D\) SM \eqref{eq:ssm}, Figs.~\ref{fig5:FigA1209a} and (b) reveal a non-monotonic behavior between  $\mu^*$ and $n$ for each $K$ in the \(N=5\) coupled SMs system \eqref{eq:csm}. The reason is that the addition of the coupling parameter \(\beta\) causes the diffusion exponent \(\mu\) to transition from superdiffusion (\(\mu > 1\)) to normal diffusion (\(\mu = 1\)) for the coupled SM systems, exhibiting this non-monotonic behavior [e.g.~see Fig.~\ref{fig5:Fig10b} for \(K_j = K = 6.5\)]. 

\begin{figure}[!htbp]
    \centering
    \includegraphics[width=0.5\textwidth]{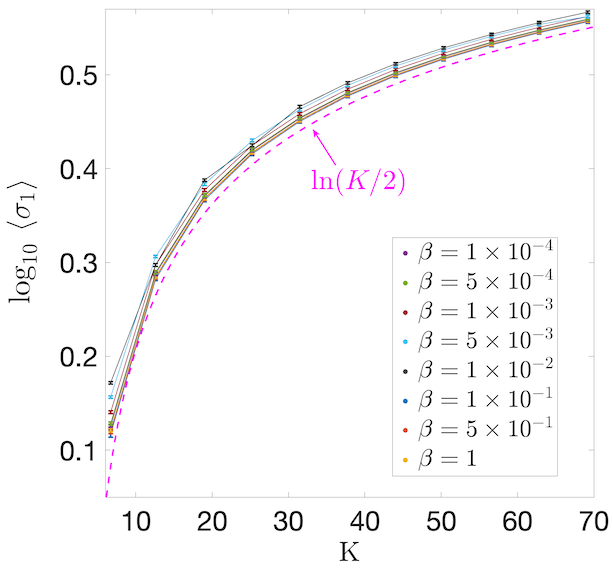}
    \caption{The average ftmLE, \(\langle \sigma_1 (K) \rangle\), computed at \(n = 10^5\) iterations of the $N=5$ coupled SMs system \eqref{eq:csm} with equal \(K_j = K=6.5\) values and varying \(\beta\) values (see the legend for specific values). The pink dashed curve represents the law \(\langle \sigma_1 (K)\rangle = \ln(K/2)\), which is also depicted in Fig.~\ref{fig5:Fig06b} for the SM system \eqref{eq:ssm}.}
    \label{fig5:Fig11}
   \end{figure}

Now let us change our focus to examine the influence of the coupling strength \(\beta\) on the coupled SMs chaoticity by analyzing the average ftmLE \(\langle \sigma_1 \rangle\) for a range of the system's kicked strength (\(K\)) and coupling strength (\(\beta\)). We consider five $\beta$ values, ranging from \( 10^{-4}\) to \( 1 \). For a fixed $\beta$,  \(\langle \sigma_1 \rangle\) is computed as the average ftmLE over \(10,000\) ICs on a \(100 \times 100\) grid covering the SM's entire phase space with \(K\) value on the interval \(6.5 \lessapprox K \lessapprox 70\) after \(n = 10^5\) iterations. Fig.~\ref{fig5:Fig11} illustrates \(\langle \sigma_1 \rangle\) increasing as \(K\) increases for a fixed coupling strength $\beta$. This monotonic behavior between the average ftmLE values and the nonlinearity kick $K$ of the coupled SMs \eqref{eq:csm} is similar to the one observed in Fig.~\ref{fig5:Fig06b} for the SM \eqref{eq:ssm}. 

For relatively large nonlinearity kick $K_j=K$ values, increasing the coupling strength \(\beta\) between \(2D\) SMs slightly increases the average ftmLE value  \(\langle \sigma_1 \rangle\). This slight monotonic relation between  \(\langle \sigma_1 \rangle\) and $\beta$ values for each \(K\) is clearly the result of a small boost in the extent of chaotic motion due to the coupling in the coupled SMs. While adding a small coupling initially affects the diffusion properties, it also has a small influence on the system's chaotic behavior, especially for large $K$ values.

   \subsubsection{Coupled SMs with equal nonlinearity kicks but different proportions of chaotic orbits} \label{sec:2NDResults_K2}
\begin{figure}[!htbp]
    \centering
    \subfloat[GALI$_2$ color map\label{fig5:Fig12a}]{\includegraphics[width=1\textwidth]{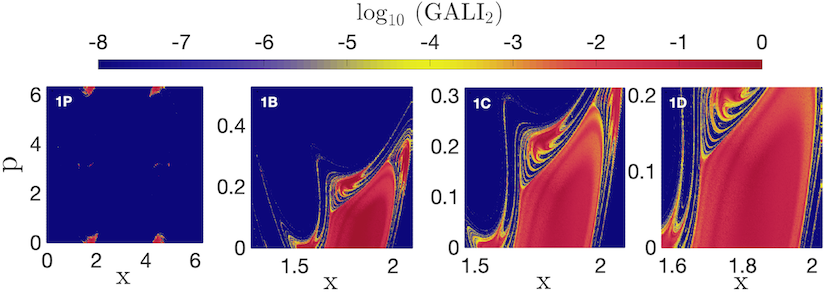}} \\
    \subfloat[Diffusion exponent $\mu$ color map\label{fig5:Fig12b}]{\includegraphics[width=1\textwidth]{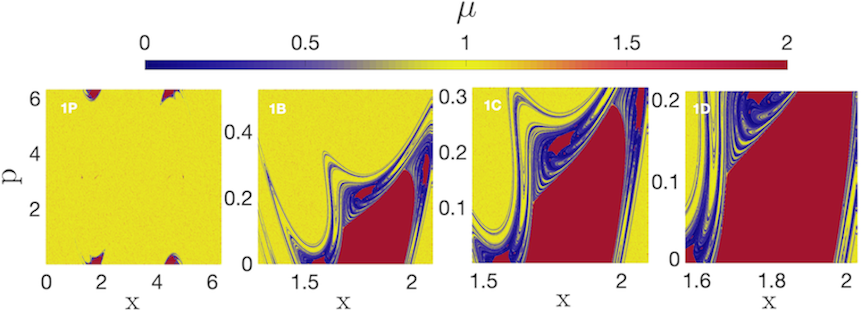}}
    \caption{The phase space portraits colored by (top row) GALI$_2$ \eqref{eq:GALI} and (bottom row) the diffusion exponent \(\mu\) \eqref{eq:Deff_csm} for the \(2D\) SM \eqref{eq:ssm} with \(K = 6.5\) for which a period \(p = 1\) AM exists. The leftmost panel in each row contains an ensemble of ICs on the entire phase space, \([0,2\pi) \times [0,2\pi)\), exhibiting extensive chaos (P\(_\text{C} \approx\) $99\%$), 1P. The following panels (1B, 1C, and 1D) show progressively smaller regions around the period \(p=1\) AM with P\(_\text{C} \approx 75\%\), P\(_\text{C} \approx 50\%\), and P\(_\text{C} \approx 25\%\), respectively.}
    \label{fig5:Fig12}
   \end{figure}

Now, let us shift our focus to investigate the influence of the percentage of chaotic ICs ($P_{C}$) on the long-term diffusion properties for the system of coupled SMs system \eqref{eq:csm} with equal nonlinearity kicks $K_j = K$. For this purpose, we consider a system of $N=5$ coupled \(2D\) maps with $K_j = K = 6.5$, where AMs of period $p=1$ are present for each individual coupled map. We begin our study by creating phase space color plots to visualize the chaoticity and the diffusion rates for the different ensembles of ICs we consider (Fig.~\ref{fig5:Fig12}). The top row of color maps displays phase space color plots based on the \( \text{GALI}_2 \) values of orbits, while the bottom row presents color plots of the diffusion exponent \(\mu\). The first panels of both the \( \text{GALI}_2 \) and the \(\mu\) color plots show an ensemble of ICs covering the entire phase space \( x \in (0, 2\pi)\) and \(p \in (0, 2\pi) \) of the \(2D\) map, which contains extensive chaotic regions (approximately P$_\text{C} \approx 99 \%$), labeled `1P'. We then focus on ensembles of progressively smaller regions around an AM of period $p=1$, gradually reducing the fraction of chaotic orbits. These maps are presented as 1B, 1C, and 1D and contain approximately P$_\text{C} \approx 75\%$, P$_\text{C} \approx 50\%$, and P$_\text{C} \approx 25\%$ chaotic orbits, respectively (see Table~\ref{tab:ssm} for more details).

In the following analysis, we construct two coupled map arrangements (`SA' and `SB') with a single (S) parameter value, $K_j=K =6.5$, using the ensembles shown in Fig.~\ref{fig5:Fig12}. For these arrangements, we compute the evolution of the variance \(\sum_{j=1}^{5} \langle (\Delta p^j)^2 \rangle\) \eqref{eq:pvarN} and the corresponding diffusion exponent $\mu$ \eqref{eq:pvarN} as well as the average ftmLE \eqref{eq:ftmLE}, \(\langle \sigma_1 \rangle\), and the average GALI\(_2\), \(\langle \text{GALI}_2 \rangle\), considering a moderate value of the coupling strength, specifically \(\beta = 10^{-3}\). These setups allow us to efficiently investigate the impact of different chaotic orbit percentages on the diffusion characteristics of the coupled SMs, particularly in the presence of period $p=1$ AMs, for reasonably long computational times.

\begin{table}[h!]
    \caption{Nomenclature of different coupled SMs arrangements in Sect.~\ref{sec:2NDResults_K2}. Fig.~\ref{fig5:Fig13} presents results for the \(N = 5\) coupled SMs system \eqref{eq:csm}, where each \(2D\) map has a single (S) nonlinear strength value \(K_j = K = 6.5\) (corresponding to a period \(p=1\) AM). In all arrangements labeled as `SA', the central map (Map 3) includes ICs that cover the entire phase space with \(P_{\text{C}} \approx 99\%\) (1P), while the other four maps contain ICs in regions with various percentages of chaotic orbits: \(P_{\text{C}} \approx 75\%\) for 1B, \(P_{\text{C}} \approx 50\%\) for 1C, and \(P_{\text{C}} \approx 25\%\) for 1D (see Fig.~\ref{fig5:Fig12} as well as Table \ref{tab:ssm}). Fig.~\ref{fig5:Fig14} shows results for an alternative arrangement type labeled as `SB', where the central map (Map 3) ICs in regions (1B, 1C, or 1D), while the other four maps include ICs covering the entire phase space (1P). In all six cases, the coupling strength parameter is set to a moderate value, \(\beta = 10^{-3}\).}
    \label{tab:Single_EqualK}
    \centering
    \renewcommand{\arraystretch}{1.2}  
    \begin{tabularx}{\textwidth}{|X|X|X|X|X|X|X|}
       \hline
       Cases & \multicolumn{3}{c|}{\textbf{Arrangement SA}} & \multicolumn{3}{c|}{\textbf{Arrangement SB}} \\ \hline
       \textbf{Map} & \textbf{S75A} & \textbf{S50A} & \textbf{S25A} & \textbf{S75B} & \textbf{S50B} & \textbf{S25B} \\ \hline
       Map 1 & 1B & 1C & 1D & \multicolumn{3}{c|}{1P} \\ \hline
       Map 2 & 1B & 1C & 1D & \multicolumn{3}{c|}{1P} \\ \hline
       Map 3 & \multicolumn{3}{c|}{1P} & 1B & 1C & 1D \\ \hline
       Map 4 & 1B & 1C & 1D & \multicolumn{3}{c|}{1P} \\ \hline
       Map 5 & 1B & 1C & 1D & \multicolumn{3}{c|}{1P} \\ \hline
       Figure & \multicolumn{3}{c|}{Fig.~\ref{fig5:Fig13}} & \multicolumn{3}{c|}{Fig.~\ref{fig5:Fig14}} \\ \hline
    \end{tabularx}
   \end{table}
It is worth noting that we can set up the \(N=5\) coupled SMs system \eqref{eq:csm} by coupling each panel (the four maps with different percentages of chaos, 1P, 1B, 1C, and 1D in Fig.~\ref{fig5:Fig12}) in many possible ways. For simplicity, we establish two distinct arrangements to systematically explore how the percentage of chaos ensembles affects the long-term diffusion properties and chaotic behavior of the coupled SMs. Both arrangements involve coupled \(2D\) SMs with equal nonlinearity parameters,  $K_j = K = 6.5$, and they are defined as Arrangement SA and Arrangement SB. 

\subsubsection*{Dynamics of coupled SMs in Arrangement SA}
In this arrangement, we initialize the ICs of the central map ($j = 3$) to cover the entire phase space, resulting in approximately P$_\text{C} \approx 99\%$ chaotic orbits. This arrangement corresponds to the leftmost panels of each row in Fig.~\ref{fig5:Fig12}. For the remaining four maps, we consider ICs drawn from regions with a fixed proportion of chaotic orbits, specifically P\(_\text{C} \approx 75\%\), P\(_\text{C} \approx 50\%\), or P\(_\text{C} \approx 25\%\). The values of ranges of these regions are given in Table~\ref{tab:ssm}.

The S75A case, for instance, consists of five coupled \(2D\) SMs systems with equal nonlinearity parameters \( K_j = K = 6.5 \) for which an AM of period $p=1$ exists. The ICs in the central map cover the entire phase space with P$\_\text{C} \approx 99\%$,  while the other four maps contain ICs drawn from predominantly chaotic regions with P\(_\text{C} \approx 75\%\).

\begin{figure}[!htbp] 
    \centering
    \subfloat[{Evolution of the variance [inset: $\mu (n)$]}\label{fig5:Fig13a}]{\includegraphics[width=0.475\textwidth]{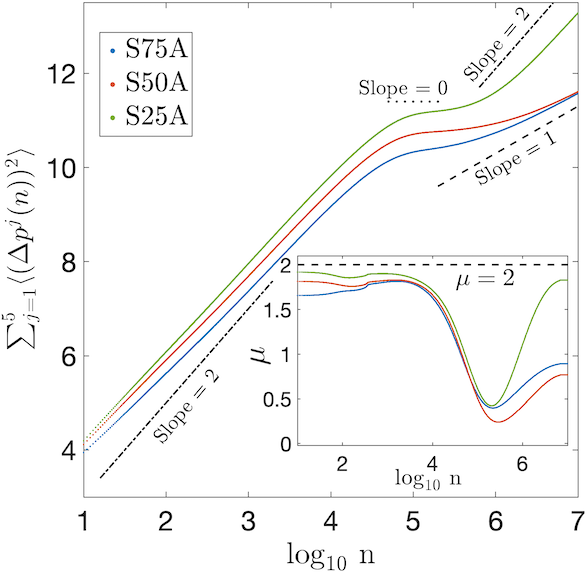}} 
    \subfloat[{Evolution of the average ftmLE [inset: average GALI$_2 (n)$]}\label{fig5:Fig13b}]{\includegraphics[width=0.475\textwidth]{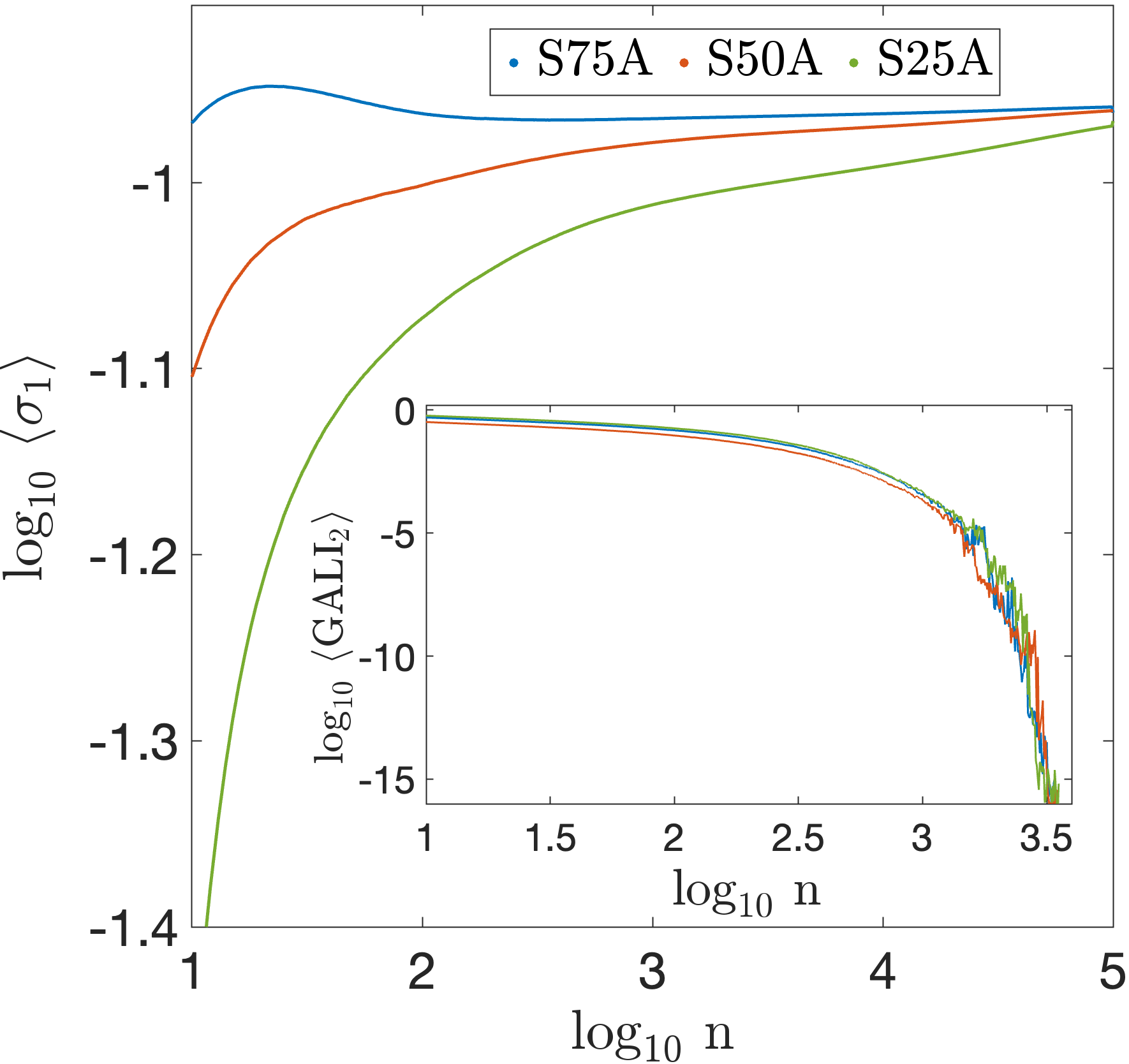}}
    \caption{(a) The evolution of the variance \(\sum_{j=1}^{5} \langle (\Delta p^j)^2 \rangle\) with respect to the number of iterations \(n\) for the $N=5$ coupled SMs system \eqref{eq:csm} evaluated over ensembles of approximately \(100,000\) ICs. Inset: The evolution of the diffusion exponent \(\mu\) \eqref{eq:Deff_csm} for the same set of ICs. The dash-dotted and dashed lines indicate power-law behaviors $n^2$ and $n^1$, respectively, while the horizontal dashed line in the inset corresponds to $\mu =2$. (b) The evolution of the average ftmLE \eqref{eq:ftmLE}, \(\langle \sigma_1 \rangle\), as a function of the number of iterations \(n\) [inset: The corresponding \(\langle \text{GALI}_2 \rangle\)] for the same cases as in (a). The blue, red, and green curves represent cases S75A, S50A, and S25A of Arrangement SA, respectively (see Table \ref{tab:Single_EqualK} for more details).}
    \label{fig5:Fig13}
   \end{figure}

In Fig.~\ref{fig5:Fig13a}, we present the evolution of the variance \(\sum_{j=1}^{5} \langle (\Delta p^j)^2 \rangle\) for the ensembles of approximately \(100,000\) ICs of the \(N=5\) coupled SMs system \eqref{eq:csm}. The S25A case exhibits convergence toward ballistic diffusion over time as indicated by a slope approximately \(2\), whereas the S75A and S50A cases converge to normal diffusion, characterized by a slope of approximately \(1\). The inset of Fig.~\ref{fig5:Fig13a} depicts the diffusion exponent \( \mu \) as a function of the number of iterations \( n \) for these three cases. On the other hand, in Fig.~\ref{fig5:Fig13b}, we illustrate the evolution of the average ftmLE \eqref{eq:ftmLE}, \( \langle \sigma_1 \rangle \), for the same three cases. All \( \langle \sigma_1 \rangle \) values converge to similar positive values, implying chaotic behavior. The rate of convergence to this asymptotic value is influenced by the proportion P\(_\text{C}\) of chaotic ICs in the off-center maps, with a larger proportion of chaotic regions resulting in faster convergence to the limiting positive values. The exponential decay of the \( \langle \text{GALI}_2 \rangle \) value in the inset of  Fig.~\ref{fig5:Fig13b} further validates the global chaotic dynamic nature of all considered cases.

\subsubsection*{Dynamics of coupled SMs in Arrangement SB}

In this case, we initialize the central map (map 3) with ICs drawn from regions exhibiting different proportions of chaotic orbits, i.e., P\(_\text{C} \approx 75\%\), P\(_\text{C} \approx 50\%\), and P\(_\text{C} \approx 25\%\). The remaining four maps, however, contain ICs that cover the entire phase space, resulting in P\(_\text{C} \approx 99\%\) chaotic orbits for each of the \(2D\) maps \eqref{eq:ssm}. For instance, the S75B case consists of a coupled SMs system \eqref{eq:csm} where the central map ($j = 3$)  has P\(_\text{C} \approx 50\%\) chaotic orbits, while ICs in the other four maps cover the entire phase space (see Table~\ref{tab:Single_EqualK} for more details).

\begin{figure}[!htbp]
    \centering
    \subfloat[{Evolution of the variance [inset: $\mu (n)$]}\label{fig5:Fig14a}]{\includegraphics[width=0.475\textwidth]{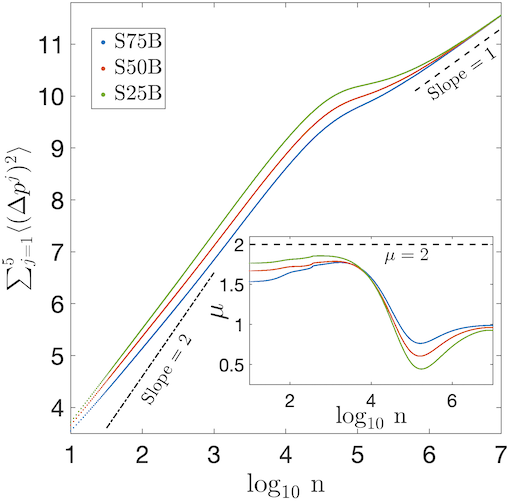}} 
    \subfloat[{Evolution of the average ftmLE [inset: average GALI$_2 (n)$]}\label{fig5:Fig14b}]{\includegraphics[width=0.475\textwidth]{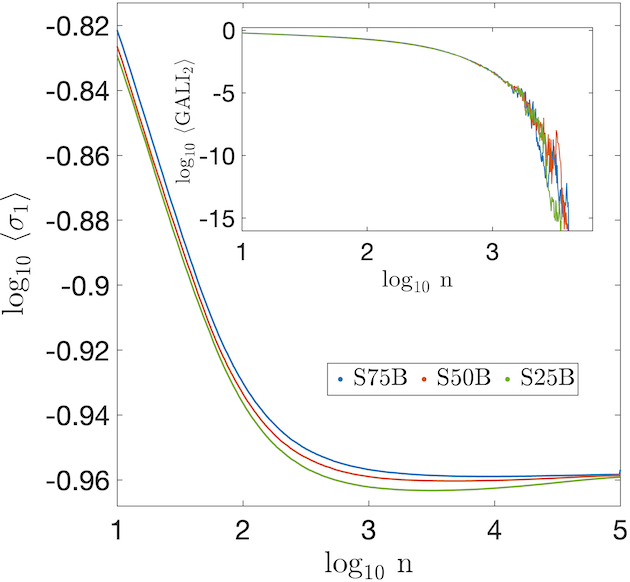}}
    \caption{Similar to Fig.~\ref{fig5:Fig13} but for cases of the Arrangement SB of the maps. The blue, red, and green curves correspond to cases S75B, S50B, and S25B, respectively (see Table~\ref{tab:Single_EqualK}) for more details for the specific cases.}
    \label{fig5:Fig14}
   \end{figure}

Figure \ref{fig5:Fig14a} illustrates results for arrangement SB, where all the considered cases exhibit a tendency of the dynamics towards normal diffusion rates (\( \mu \approx 1 \)). Unlike what was observed for Arrangement SA, here the diffusion exponent does not show signs of approaching the ballistic transport rate \( \mu \approx 2 \) within the observed simulation time.

Regarding the global chaoticity of the system, we observe a more uniform evolution for both the average ftmLE \eqref{eq:ftmLE}, \( \langle \sigma_1 \rangle \), and the \( \langle \text{GALI}_2 \rangle \) \eqref{eq:GALI} values in Fig.~\ref{fig5:Fig14b}. The \( \langle \sigma_1 \rangle \) values converge to approximately the same positive value for all considered cases. Furthermore, the \( \langle \text{GALI}_2 \rangle \) values demonstrate similar exponential decay rates, indicating consistent chaotic behavior throughout the different IC setups.

   \subsubsection{Coupled SMs with different nonlinearity kicks and periods of AMs} \label{sec:2NDResults_K3}
In order to extend our investigations, we introduce additional complexity to the coupled SMs system \eqref{eq:csm} by considering various [mixed (M)] nonlinearity $K$ values and AMs periods for individual \(2D\) maps. Following a similar approach to Sect.~\ref{sec:2NDResults_K2}, we examine diffusion rate properties and the global chaoticity for specific arrangements, what we refer as Arrangement MA and Arrangement MB. These setups included five \(2D\) maps with distinct portions of chaotic orbits and mixed $K_j$ values. The details for these configurations are given in Table~\ref{tab:A_MixedK}.
\begin{table}[h!]
    \caption{Nomenclature of different coupled SMs arrangements in Sect.~\ref{sec:2NDResults_K3}. Fig.~\ref{fig5:Fig16} presents results for the \(N = 5\) coupled SMs system \eqref{eq:csm}, where each \(2D\) map has mixed (M) nonlinear strength values. In all the arrangements labeled as `MA', the central map (Map 3) has \(K_3 = K = 3.1\), corresponding to a period \(p=4\) AM, and ICs covering the entire phase space (region 4D in Fig.~\ref{fig5:Fig15} and Table \ref{tab:ssm}). The other four maps (\(j=1, 2, 4, 5\)) have \(K_j = K = 6.5\), corresponding to a period \(p=1\) AM, and ICs in regions containing various percentages of chaotic orbits: \(P_{\text{C}} \approx 75\%\) for 1B, \(P_{\text{C}} \approx 50\%\) for 1C, and \(P_{\text{C}} \approx 25\%\) for 1D (see Fig.~\ref{fig5:Fig12} and Table \ref{tab:ssm}). Fig.~\ref{fig5:Fig17} shows results for an alternative arrangement type, labeled as `MB', where the central map (Map 3) has \(K_3 = 6.5\) and ICs in region 1P [Fig.~\ref{fig5:Fig12}], while the other four maps (\(j=1, 2, 4, 5\)) have \(K_j = K = 3.1\) and ICs in regions 4B, 4C, and 4D [Fig.~\ref{fig5:Fig15}]. In all six cases, the coupling strength parameter is set to a moderate value, \(\beta = 10^{-3}\).}
    \label{tab:A_MixedK}
    \centering
    \renewcommand{\arraystretch}{1.2}  
    \begin{tabularx}{\textwidth}{|X|X|X|X|X|X|X|}
       \hline
       Cases & \multicolumn{3}{c|}{\textbf{Arrangement MA}} & \multicolumn{3}{c|}{\textbf{Arrangement MB}} \\ \hline
       \textbf{Map} & \textbf{M75A} & \textbf{M50A} & \textbf{M25A} & \textbf{M75B} & \textbf{M50B} & \textbf{M25B} \\ \hline
       Map 1 & 1B & 1C & 1D & 4B & 4C & 4D \\ \hline
       Map 2 & 1B & 1C & 1D & 4B & 4C & 4D \\ \hline
       Map 3 & \multicolumn{3}{c|}{4A (period $p=4$, $K=3.1$)} & \multicolumn{3}{c|}{1P (period $p=1$, $K=6.5$)} \\ \hline
       Map 4 & 1B & 1C & 1D & 4B & 4C & 4D \\ \hline
       Map 5 & 1B & 1C & 1D & 4B & 4C & 4D \\ \hline
       Figure & \multicolumn{3}{c|}{Fig.~\ref{fig5:Fig16}} & \multicolumn{3}{c|}{Fig.~\ref{fig5:Fig17}} \\ \hline
    \end{tabularx}
   \end{table}

\begin{figure}[!htbp]     
    \centering
    \subfloat[GALI$_2$ color map\label{fig5:Fig15a}]{\includegraphics[width=1\textwidth]{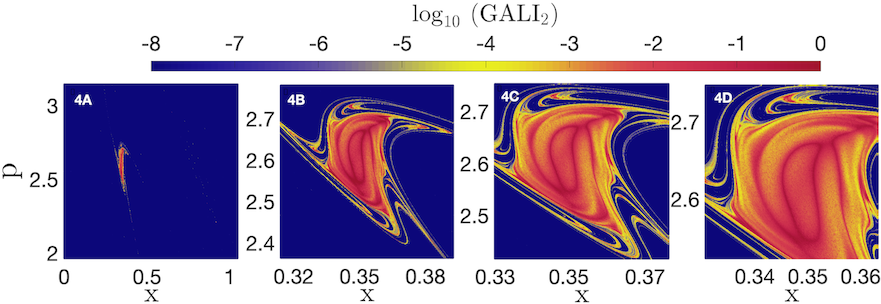}}\\ 
    \subfloat[$\mu$ color map\label{fig5:Fig15b}]{\includegraphics[width=1\textwidth]{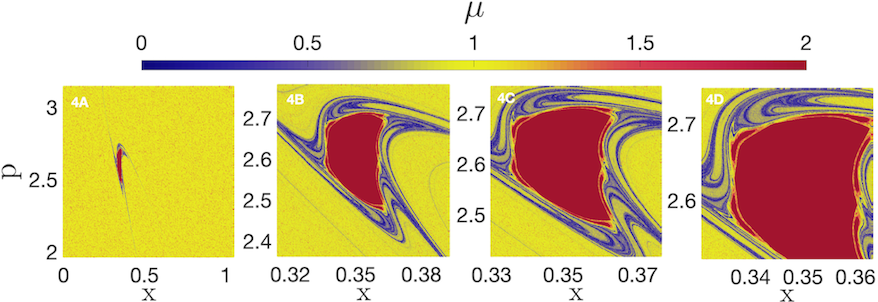}}
    \caption{The phase space portraits colored by (top row) GALI$_2$ \eqref{eq:GALI} and (bottom row) the diffusion exponent \(\mu\) \eqref{eq:Deff_csm} for the \(2D\) SM \eqref{eq:ssm} with \(K = 3.1\) for which a period \(p = 4\) AM exists. The panels 4A, 4B, 4C, and 4D correspond to P\(_\text{C} \approx 99\%\), P\(_\text{C} \approx 75\%\), P\(_\text{C} \approx 50\%\), and P\(_\text{C} \approx 25\%\), respectively.}
    \label{fig5:Fig15}
   \end{figure}

In arrangement MA, the off-center maps (\(j = 1, 2, 4, 5\)) feature ensembles from regions 1B, 1C, and 1D (as defined in Table~\ref{tab:ssm} and illustrated in Fig.~\ref{fig5:Fig12}), while maps in Arrangement MB correspond to cases 4B, 4C, and 4D in Fig.~\ref{fig5:Fig15} (see also Table~\ref{tab:ssm} for more specific ensembles of ICs). Fig.~\ref{fig5:Fig15} presents the phase space color plots based on \( \text{GALI}_2 \) (top row) and \( \mu \) (bottom row) values for the considered \(2D\) map regions, which involves stable AM of period $p=4$ of the map. The reader can refer to Fig.~\ref{fig5:Fig12} for similar color maps in the case of period $p=1$ AMs. 

For instance, Map  4A includes ensembles of ICs in the region \( x \in [0.0, 1.047]\) and \( p \in [1.964, 3.142] \), where a period $p=4$ AM is present in the \(2D\) SM system \eqref{eq:ssm} and exhibits extensive chaos, P\(_\text{C} \approx 99\% \) (see Table \ref{tab:ssm}). The regions labeled 4B, 4C, and 4D represent subsets of Map 4, with gradually decreasing percentages of chaotic orbits: P\(_\text{C} \approx 75\%\), P\(_\text{C} \approx50\%\), and P\(_\text{C} \approx 25\%\), respectively.

\subsubsection*{Dynamics of coupled SMs in Arrangement MA}
\begin{figure}[!htbp]
    \centering
    \subfloat[{Evolution of the variance [inset: $\mu (n)$]}\label{fig5:Fig16a}]{\includegraphics[width=0.475\textwidth]{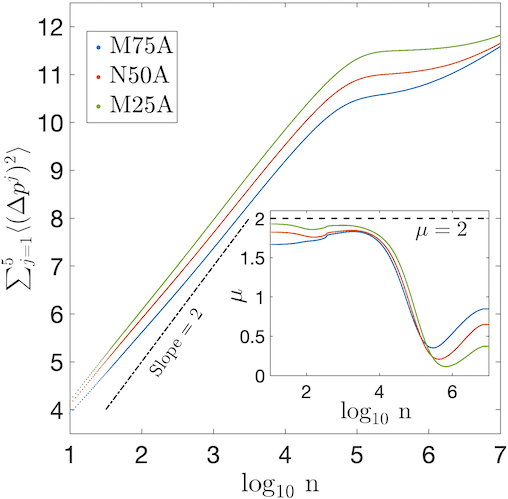}} 
    \subfloat[{Evolution of the average ftmLE [inset: average GALI$_2 (n)$]}\label{fig5:Fig16b}]{\includegraphics[width=0.475\textwidth]{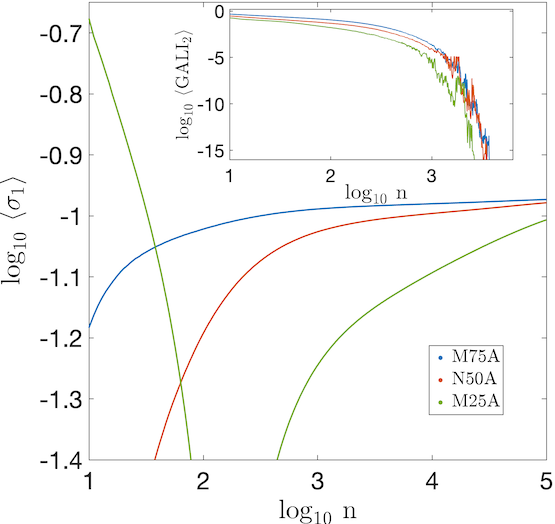}}
    \caption{Similar to Fig.~\ref{fig5:Fig13} but for the coupled SMs system \eqref{eq:csm} described in Sect.~\ref{sec:2NDResults_K3}. The blue, red, and green curves correspond to the M75A, M50A, and M25A cases, respectively, as presented in Table~\ref{tab:A_MixedK}.}
    \label{fig5:Fig16}
   \end{figure}

In Arrangement MA, we study a setup where the central map (\(j=3\)) has \( K_3 = 3.1 \) with a period $p=4$ AMs, while the remaining maps (\(j = 1, 2, 4, 5\)) have \( K_j = K =6.5 \) where period $p=1$ AMs of the \(2D\) SM are present. We follow the same analysis performed in Fig.~\ref{fig5:Fig13} and examine the evolution of the variance  \(\sum_{j=1}^{5} \langle (\Delta p^j)^2 \rangle\) \eqref{eq:pvarN} with respect to the number of iterations \(n\) for the $N=5$ coupled SMs system \eqref{eq:csm} [Fig.~\ref{fig5:Fig16a}]. On the other hand, Fig.~\ref{fig5:Fig16b} shows the evolution of the average ftmLE \eqref{eq:ftmLE}, \( \langle \sigma_1 \rangle \),  along with the \( \langle \text{GALI}_2 \rangle \) \eqref{eq:GALI} values (inset). We conduct these analyses over ensembles consisting of approximately \(100,000\) ICs. Initially, all cases exhibit superdiffusive spreading (\( \mu > 1 \)) [Fig.~\ref{fig5:Fig16a}]. However, after approximately \( n = 10^5 \) iterations, the diffusion rate drops significantly, transitioning to low subdiffusion rates. Surprisingly, toward the end of the considered simulation period, the diffusion rates begin to increase once again. These dynamic changes in diffusion behavior are further highlighted in the inset of Fig.~\ref{fig5:Fig16a}, where we plot the diffusion exponent \( \mu \) as a function of the number of iterations \( n \).

In terms of the system's global chaotic behavior, we observe that the average ftmLE, \( \langle \sigma_1 \rangle \), [Fig.~\ref{fig5:Fig16b}] shows the tendency to converge to similar positive values for all cases, comparable to the trends observed for Arrangements SA and SB in Sect.~\ref{sec:2NDResults_K2}. This convergence rate toward the asymptotic \( \langle \sigma_1 \rangle \) value depends on the proportion of chaotic ICs in the off-center maps, with a higher percentage of chaotic orbits leading to faster convergence to a positive constant value for the ftmLE. Additionally, the \( \langle \text{GALI}_2 \rangle \) values exhibit an exponential decay to zero [inset of Fig.~\ref{fig5:Fig16b}], clearly indicating a global chaotic behavior for all cases.

\subsubsection*{Dynamics of coupled SMs in Arrangement MB}
\begin{figure}[!htbp]
    \centering
    \subfloat[{Evolution of the variance [inset: $\mu (n)$]}\label{fig5:Fig17a}]{\includegraphics[width=0.475\textwidth]{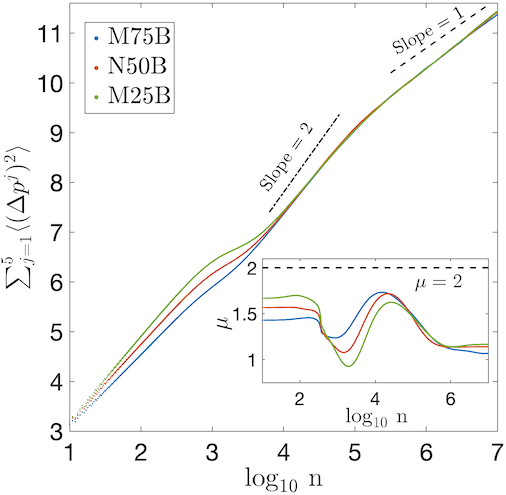}} 
    \subfloat[{Evolution of the average ftmLE [inset: average GALI$_2 (n)$]}\label{fig5:Fig17b}]{\includegraphics[width=0.475\textwidth]{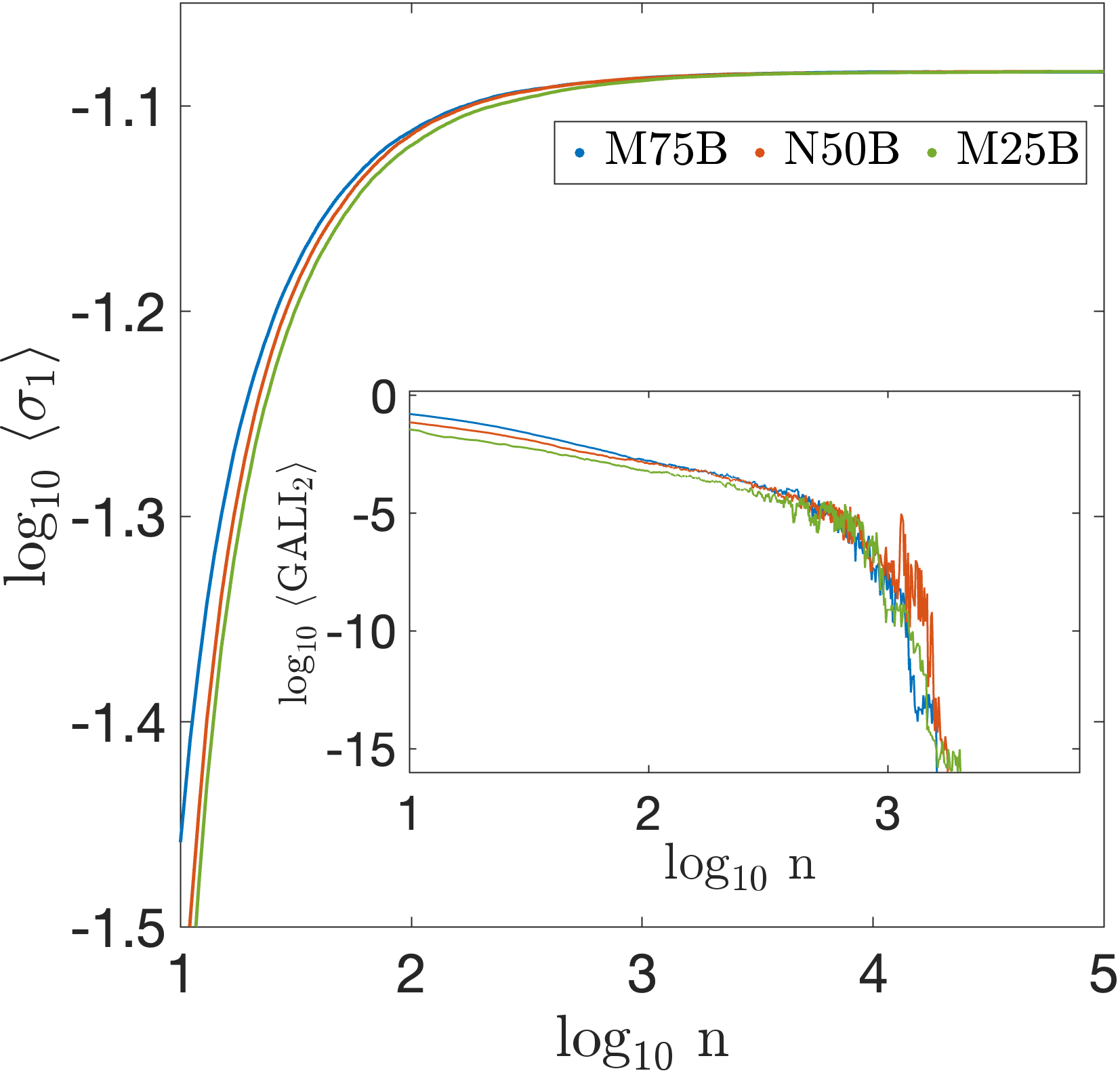}}
    \caption{Similar to Fig.~\ref{fig5:Fig12} but for the coupled SMs system \eqref{eq:csm} described in Sect.~\ref{sec:2NDResults_K3}. The blue, red, and green curves correspond to the M75B, M50B, and M25B cases, respectively, as presented in Table~\ref{tab:A_MixedK}.}
    \label{fig5:Fig17}
\end{figure}

In Arrangement MB, we investigate a system setup with mixed kick-strength values, where the central map (\(j=3\) has \( K_3 = 6.5 \) with period \(p=4\) AM of the \(2D\) map, while the off-center maps \(j=1, 2, 4, 5\)) have \( K_j  = K = 3.1 \) associated with the period \(p=1\) AMs.  We study three cases, M75B, M50B, and M25B, corresponding to ICs in the off-center maps taken from the regions 4B, 4C, and 4D, respectively [Fig.~\ref{fig5:Fig15}]. For the central map (Map 3) with \( K = 3.1 \), we consider ICs that cover the entire phase space \(x \in (0, 2\pi)\) and \(p \in (0, 2\pi)\), 1P (see Table \ref{tab:A_MixedK}). Fig.~\ref{fig5:Fig17a} illustrates the evolution of the variance \(\sum_{j=1}^{5} \langle (\Delta p^j)^2 \rangle\) \eqref{eq:pvarN} and the corresponding $\mu$ \eqref{eq:pvarN} (inset) as a function of the number of iterations $n$ using ensembles containing approximately \(100, 000\) ICs. On the other hand, Fig.~\ref{fig5:Fig17b} shows the evolution of the average ftmLE \eqref{eq:ftmLE}, \( \langle \sigma_1 \rangle \), along with the \( \langle \text{GALI}_2 \rangle \) values (inset) for the same ICs. 

Initially, all cases in Arrangement MB exhibit superdiffusive behavior (\( \mu > 1 \)) [Fig.~\ref{fig5:Fig17a} inset], although this effect is clearly less pronounced than in the case for Arrangement MA. Over an extended number of iterations $n$, we observe a gradual transition toward normal diffusion (\( \mu \approx 1 \)) for the three considered cases, M75B (blue curve), M50B (red curve), and M25B (green curve).

Regarding the system's global chaotic behavior,  the average ftmLE, \( \langle \sigma_1 \rangle \), values [Fig.~\ref{fig5:Fig17a}] show a tendency to converge to a similar asymptotic positive value for all the considered cases. The convergence rate toward this asymptotic ftmLE value is moderately influenced by the portion of chaotic ICs P\(_\text{C}\) in the central map. In addition, the \(\langle \text{GALI}_2 \rangle \) values follow the same trend of exponential decay to zero [inset of Fig.~\ref{fig5:Fig17b}], indicating consistent global chaotic behavior for all three cases we consider.

\section{Summary and conclusions} \label{section:SummaryCh5}
In this chapter, we systematically investigated the long-term diffusion and chaoticity properties of \(2D\) SMs \eqref{eq:ssm} and \(N=5\) coupled SMs system \eqref{eq:csm}. We focused on parameter values where the phase space of the systems exhibited anomalous diffusion rates due to the presence of AMs of varying periods \(p\). For the SMs (Sect.~\ref{sec:single SMs}), we reviewed the typical diffusion behavior: regular (subdiffusion, \(\mu < 1\)) and chaotic (normal diffusion, \(\mu = 1\)) orbits, along with the stable and unstable AMs of different periods, where the former exhibited ballistic transport (\(\mu=2\)) (Fig.~\ref{fig5:Fig02}). Using GALI\(_2\) \eqref{eq:GALI_chaos} and ftmLE \eqref{eq:ftmLE} indices, we quantified chaoticity for selected ICs (Fig.~\ref{fig5:Fig01}). Additionally, we investigated diffusion properties by measuring the diffusion exponent \(\mu\) \eqref{eq:pvar} and the effective diffusion coefficient D\(_\text{eff}\) \eqref{eq:Deff_sm} over a range of nonlinear kick-strength values (\(K\)) and iteration numbers (Fig.~\ref{fig5:Fig03}).

We conducted a systematic study of AMs of period \(p=1\) and observed that such modes asymptotically exhibited extreme diffusion rates (ballistic transport with \(\mu \approx 2\)) despite initially behaving as normal diffusion (\(\mu \approx 1\)) (Fig.~\ref{fig5:Fig05}). Larger \(K\) values for SMs with \(p=1\) AMs correlated with strong chaotic behavior in phase space, as indicated by the average ftmLE \(\langle \sigma \rangle\) [Fig.~\ref{fig5:Fig06b}], which followed the law \(\langle \sigma \rangle = \ln(K/2)\) \citep{Shevchenko2004}. We examined cases where AMs of different periods occurred and explored how chaotic IC ensembles influenced diffusion. Our findings showed that more chaotic ensembles delayed the convergence to ballistic transport, with \(\mu\) taking longer to approach \(\mu=2\) (Fig.~\ref{fig5:Fig07}).

We extended our study to coupled SM systems characterized by the presence of AMs of varying periods (Sect.~\ref{sec:coupled SMs}). Using the same ensembles of ICs over the entire phase space, we identified intervals of \(K\) where AMs of period \(p=1\) emerged and derived scaling laws to describe their dependence on \(K\). We calculated the global diffusion exponent \(\mu^*\) for varying coupling strengths (\(\beta\)). We observed that increasing \(\beta\) suppressed superdiffusion and propelled the system towards normal diffusion (\(\mu^* \approx 1\)) more rapidly at higher \(K\) values.

For cases where \(K_j = K = 6.5\), corresponding to ballistic diffusion (\(\mu \approx 2\)) in individual maps, we found that coupling resulted in a global diffusion rate approaching normal diffusion. This convergence occurred faster with stronger couplings and correlated with an increase in global chaoticity, as measured by ftmLE and GALI\(_2\). Additionally, we investigated different arrangements of the \(N=5\) coupled SMs \eqref{eq:csm} to understand how varying arrangements influenced global diffusion rates. These simulations highlighted conditions where ballistic transport, driven by low-period AMs, could be suppressed by neighboring maps without AMs and ensembles of chaotic ICs exhibiting normal diffusion. We conducted extensive numerical simulations for two specific arrangements to support these findings.

In the Arrangement A type of coupled maps, we explored two cases: Arrangement SA with a single \(K\) value (presence of AMs of period \(p=1\) in all SMs, Fig.~\ref{fig5:Fig13}) and Arrangement MA with mixed \(K\) values (a central SM with \(p=4\) AM and the rest having \(p=1\), Fig.~\ref{fig5:Fig16}). For both setups, we used a fixed coupling strength \(\beta=10^{-3}\). For Arrangement SA, the diffusion exponent \(\mu\) exhibited a dynamic evolution: it initially started at \(\mu \approx 2\) due to the AMs' influence, plateaued at \(\mu \approx 0\), and eventually recovered to its initial rate in systems with chaotic IC fractions \(P_C\) (e.g.~see case S25A, green curve in Fig.~\ref{fig5:Fig13}). In contrast, for higher \(P_C\) fractions (e.g.~see S50A and S75A, red and blue curves in Fig.~\ref{fig5:Fig13}), \(\mu\) converged to normal diffusion values \(\mu = 1\). In Arrangement MA, no clear trends emerged within the maximum number of iterations we considered [Fig.~\ref{fig5:Fig16a}]. Regarding global chaoticity, as quantified by \(\langle \sigma \rangle\) [\(\langle \text{GALI}_2 \rangle\)], we observed faster convergence to a similar positive value [decay to zero exponentially fast] for higher P\(_\text{C}\) values in the off-center maps - see Fig.~\ref{fig5:Fig13b} and Fig.~\ref{fig5:Fig16b}.

In the Arrangement B type of coupled maps, we investigated two cases: Arrangement SB with a single \(K\) value of period \(p=1\) (the central SM with varying P\(_\text{C}\) , while the remaining SMs have ICs in the entire phase space,  Fig.~\ref{fig5:Fig14}) and Arrangement MB with mixed \(K\) values (central SMs with AMs of period \(p=1\) and ICs in the whole phase space, while the remaining maps have AMs of period \(p=4\) with varying \(P_C\), Fig.~\ref{fig5:Fig17}). For both setups we again employed \(\beta=10^{-3}\). We observed that initial strong anomalous diffusion rate of \(\mu \approx 2\), which gradually transitioned to normal diffusion rates (\(\mu \approx 1\)) across all cases (e.g.~see S25B, S50B, S75B, M25B, M50B, M75B). Interestingly, in all cases, the \(\langle \sigma \rangle\) values asymptotically converge to similar values, while the \(\langle \text{GALI}_2 \rangle\) decays to zero exponentially fast around comparable rates [Fig.~\ref{fig5:Fig14b} and Fig.~\ref{fig5:Fig17b}].  
\clearpage

\chapter{Behavior of the GALI method in non-Hamiltonian dissipative systems} \label{chapter:six}

\section{Introduction}\label{section:introductionCh6}
Following R\"{o}ssler's seminal work in 1979 \citep{rossler1979equation}, which introduced a hyperchaotic system characterized by two positive Lyapunov exponents (LEs), researchers became increasingly interested in understanding the presence and manifestation of chaos in more detail. The subsequent development of hyperchaotic discrete maps, such as those in \citep{baier1990maximum}, further expanded the scope of hyperchaotic DSs. For instance, the study by \citep{xu2019fast} presents a fast image encryption algorithm based on a hyperchaotic map, while a recent study by \citep{wang2023image} shows how the complexity of the parameter space in a hyperchaotic map enhances security by improving the privacy of the encrypted images. We note that an overview of various hyperchaotic systems and their real-world applications can be found in \citep{sprott2010elegant}.

While the smaller alignment index (SALI) and the generalized alignment index (GALI) methods (Sect.~\ref{section:GALI}) have predominantly been applied to discriminate and classify between regular and chaotic orbits in conservative systems (see, for example, \citep{skokos2008detecting,skokos2016chaos,bountis2009application,manos2012probing}), preliminary applications of these methods (mainly of the SALI) have also been reported for time dependent dissipative models \citep{huang2012analysis,huang2013circuit,tchakui2020chaotic,yan2023analysis, maaita2024comparison}. Further studies have explored the SALI method effectiveness in identifying parameter regimes associated with regular or chaotic behavior in dissipative systems, such as in the modified Lorenz \citep{huang2013circuit}  and the L\"{U} chaotic systems \citep{huang2014numerical}. In general, all these studies emphasize the efficiency of the SALI (practically the GALI\(_2\)) method as a chaos detection technique for dissipative systems. In addition, the GALI method has already been successfully applied to detect regular and chaotic motion in time-dependent Hamiltonian equations \eqref{eq:Gen TD Ham} (see \citep{manos2013interplay,machado2016chaotic,tchakui2020chaotic,manos2022orbit}). 

This chapter systematically examines the behavior of the GALI method \eqref{eq:GALI} for both continuous and discrete time non-Hamiltonian dissipative systems. We initially classify the various dynamical behaviors, or attractors (see Sect.~\ref{section:Dissipative Systems} for more detail), observed in each system using the respective LE spectra (Sect.~\ref{section:LEs}). Given the clear characterization of orbit evolutions in dissipative systems according to their LE spectrum, we obtain a classification which forms the foundation for our study of the GALI method. By analyzing the evolution of the GALI indices of order \(k\) (GALI\(_k\)) for different types of orbits observed in these systems, we try to understand the effectiveness of the index in capturing the behavior of various dynamical regimes. In order to achieve this, we focus our investigation on three well-established models: the \(3D\) Lorenz chaotic system \citep{lorenz1963deterministic}, a modified four-dimensional (\(4D\)) Lorenz system exhibiting hyperchaotic motion \citep{yujun2010new}, and the generalized H{\'e}non map \citep{awrejcewicz2018quantifying}. These models, which exhibit diverse dynamical behaviors, provide a comprehensive framework for our study.  

Previous studies of the SALI for dissipative systems have mainly been focused on differentiating between stable fixed points (which were considered to be ordered or regular orbits) and chaotic orbits. Our study extends these investigations by computing the GALI\(_k\) (in particular for \(k=2, 3, 4\)) and analyzing their behavior across a wide range of orbits, including stable fixed points, stable periodic orbits (POs), i.e. stable limit cycles as well as chaotic (strange) and hyperchaotic attractors (See Sect.~\ref{section:Dissipative Systems} for definitions and descriptions of these attractors). While our focus is on these four orbital behaviors, it is important to note that other orbits, such as unstable periodic orbits, can also be observed in the dissipative systems we consider. However, we limit our study to orbits that are commonly found and those with distinctive LE spectrum behaviors. 

The main novelty of our work is that it constitutes (to the best of our knowledge) the first application of the GALI\(_k\) indices for both continuous and discrete time hyperchaotic systems. In addition, our work performs the first comprehensive investigation into the behavior of the GALI indices for various types of attractors in continuous and discrete time dissipative systems. Furthermore, another important aspect of our study is that we analyze and compare the behavior of these indices with the LEs for non-Hamiltonian dissipative systems exhibiting hyperchaotic motion. Our study extends previous works, which have primarily focused on conservative systems (see, for e.g., \citep{skokos2007geometrical,manos2012probing}), to include dissipative systems. This extension offers a more comprehensive understanding of the behavior of the GALI indices in general DSs. We note that the content of this chapter is based on the findings presented in \cite{moges2025GALI}.

\section{Non-Hamiltonian dissipative models} \label{section:modelCh6}
In order to examine how the GALI indices behave for various attractors in both continuous and discrete time dissipative systems, we chose three widely known models from the literature, each showcasing distinct dynamical features. We begin our study with the renowned \(3D\) Lorenz system \citep{lorenz1963deterministic}, which is widely known for its strange attractors. Next, we investigate the behavior of the GALIs by including hyperchaotic motion in our analysis, considering a modified \(4D\) Lorenz hyperchaotic system \citep{yujun2010new}. Hyperchaotic motion is characterized by trajectories exhibiting two (or more) positive LEs, which indicates the presence of exponential growth of perturbations to the system's initial conditions (ICs) in multiple directions. Lastly, we consider the generalized H{\'e}non map \citep{awrejcewicz2018quantifying}, which exhibits both chaotic and hyperchaotic attractors. The three considered models host a diverse range of orbit types, including stable fixed points, stable limit cycles, chaotic attractors, and hyperchaotic motion, allowing us to perform a comprehensive analysis of the GALI's behavior for both continuous and discrete DSs. 

\subsection{The $3D$ Lorenz system}
The Lorenz system, introduced by Edward Lorenz in 1963 \citep{lorenz1963deterministic}, is a well-known example of a chaotic DS. It consists of three coupled ODEs that describe a simplified model of atmospheric convection, and it regarded as one of the pioneer systems in dynamical meteorology \citep{lorenz1991dimension}. Despite the extensive research of the system presented in the literature (e.g.~see \citep{strogatz2018nonlinear,Shen_IJBC_2023_review}), the Lorenz system remains an invaluable model for exploring and understanding chaotic behaviors in DSs. Its applications span across fields, including science \citep{martyushev2006maximum}, engineering \citep{brogliato2007dissipative}, communication security \citep{grassi1999system,cuomo1993circuit}, and economics \citep{zhang2006discrete}. 

The Lorenz equations are given as follows: 
\begin{equation} \label{eq:3DODE} 
    \begin{aligned}
        \frac{dx}{dt} &= a (y - x), \\
        \frac{dy}{dt} &= r x - y - xz, \\
        \frac{dz}{dt} &= xy - bz
    \end{aligned}
\end{equation}
where the three variables \( x \), \( y \) and \(z\) represent the temperature differences in a fluid flow, and \(a\), \(b\) and \(r\) are non-negative parameters. In particular, \eqref{eq:3DODE} can describe how a \(2D\) fluid layer that is uniformly heated from below and cooled from above evolves over time \citep{lorenz1963deterministic}. In this particular setup \( x \) represents the temperature difference between the bottom and the top layer, \( y \) corresponds to the horizontal movement, and \( z \) denote the vertical temperature distribution. Furthermore, the behavior of the DS \eqref{eq:3DODE} is governed by the three parameters: \( a \) (Prandtl number) which is related to the fluid's viscosity and thermal conductivity;  \( b \) (Rayleigh number) that depends on the temperature difference between the ground and the atmosphere; and \( r \) (aspect ratio in the convection setup) which is determined by the ratio of the atmosphere's height to the horizontal scale of the convective cells. Note that the variables \( x \), \( y \), and \( z \) are dimensionless, and they depend on time \( t \). 

Let us quickly verify that \eqref{eq:3DODE} is indeed dissipative. Consider the vector $\mathbf{f} = \left( a (y - x), \, r x - y - x z, \, x y - b z \right)$. Then the divergence of $\mathbf{f}$ is given by
\begin{equation}
    \nabla \cdot \mathbf{f} = 
    \frac{\partial}{\partial x} [a (y - x)] 
    + \frac{\partial}{\partial y} [r x - y - x z] 
    + \frac{\partial}{\partial z} [x y - b z]
    \label{eq:divergence Lorenz}
\end{equation}
Simplifying \eqref{eq:divergence Lorenz} yields
\begin{equation}
    \nabla \cdot \mathbf{f} = - (a + 1 + b).
\end{equation}
Since $a$, $b$, and $r$ are positive parameters, the divergence is negative. This indicates that the phase space volume contracts over time, which means that the Lorenz system \eqref{eq:3DODE} is dissipative (see Sect.~\ref{section:Dissipative Systems}).

All fixed points in the system \eqref{eq:3DODE} must either be sinks or saddles, and any closed orbits, if they exist, must exhibit either stability or saddle-like behavior (e.g.~see \citep[Sect. 9.2]{strogatz2018nonlinear}. The point \((x, y, z) = (0, 0, 0)\) is a fixed point for all parameter values. When the parameter \(r > 1\), two additional fixed points appear, which are symmetrically located about the z-axis at \(\left( b(r - 1), b(r - 1), r - 1 \right)\) and \(\left( -b(r - 1), -b(r - 1), r - 1 \right)\). This symmetric fixed points are stable for \(r - 1 < a < \frac{(r + 1)^2}{4}\).

More importantly, the Lorenz system \eqref{eq:3DODE} exhibits chaotic behavior across a wide range of parameter values. For example, for \(a = 10\) and \(b = \frac{8}{3}\), the system \eqref{eq:3DODE} displays chaotic attractors of different structures depending on the value of \(r\). In particular, the classical parameter set \(a = 10\), \(b = \frac{8}{3}\), and \(r = 28\) leads to the appearance of a strange attractor in the system. This attractor is referred to as ``strange" due to its fractal structure, which means that the attractors exhibits self-similarity at different scales. This means that when we zoom in on any part of the attractor, we will see patterns that resemble the overall structure (this will be discussed in more detail in Sect.~\ref{sec:3DODE chaotic}). Since their initial introduction by Edward Lorenz \citep{lorenz1963deterministic}, these parameter values have become a standard benchmark for studying chaotic dynamics in the Lorenz system. While the Lorenz system is primarily known for its chaotic behavior, it can also display stable limit cycles under specific conditions, which can be investigated through numerical simulations. In this chapter, this model will enable us to analyze and compare the behavior of the LEs and the GALI method for stable fixed points, stable limit cycles, and chaotic orbits. The numerical results and the related discussion are presented in Sect.~\ref{sec:3DODE}

\subsection{The $4D$ Lorenz hyperchaotic system}
Next, we consider a modified \(4D\) Lorenz hyperchaotic system \citep{yujun2010new}, which exhibits hyperchaotic motion characterized by two positive LEs. This system is an extension of the \(3D\) Lorenz system introduced in \citep{qi2005analysis}, which displays chaotic behavior with a single positive LEs
\begin{equation} \label{eq:3DODE-qi}
\begin{aligned}
\frac{dx}{dt} &= a (y- x) + yz, \\
\frac{dy}{dt} &= x (c - z) - y, \\
\frac{dz}{dt} &= xy - bz.
\end{aligned}
\end{equation}

In order to observe hyperchaotic attractors, the system of ODEs in \eqref{eq:3DODE-qi} must have at least four-dimensions. The addition of this extra dimension allows the presence of more complex dynamics compared to the one exhibited by the standard system \eqref{eq:3DODE}, and also results in the introduction of hyperchaotic behavior. 

The authors of \citep{qi2005analysis} extended the chaotic system \eqref{eq:3DODE-qi} by introducing a \(4D\) phase space and adding three terms to the coupled equations, one of which is the nonlinear term \(xz\).

\begin{equation} \label{eq:4DODE}
\begin{aligned}
\frac{dx}{dt} &= a(y - x) + yz, \\
\frac{dy}{dt} &= x (c - z) - y + w, \\
\frac{dz}{dt} &= xy - bz, \\
\frac{dw}{dt} &= -xz + rw,
\end{aligned}
\end{equation}
where the state variables \(x\), \(y\), \(z\), and \(w\) define the system, while parameters \(a\), \(b\), \(c\), and \(r\) influence the system's behavior. By adjusting the control parameter \(r\), different dynamic behaviors, including various chaotic and hyperchaotic attractors can be observed in the system \eqref{eq:4DODE}.

Similarly to what was done in Sect.~\ref{sec:3DODE}, in order to verify the dissipative nature of the system \eqref{eq:4DODE}, we take vector \(\mathbf{f} = \left( \frac{dx}{dt}, \frac{dy}{dt}, \frac{dz}{dt}, \frac{dw}{dt} \right)\), and consider its divergence, which is given by:
\begin{equation}
\nabla \cdot \mathbf{f}  = \frac{\partial f}{\partial x} + \frac{\partial f}{\partial y} + \frac{\partial f}{\partial z} + \frac{\partial f}{\partial w} = -a - 1 - b + r.
\end{equation}

This indicates that the system \eqref{eq:4DODE} becomes dissipative when \(-a - 1 - b + r < 0\) (see Sect.~\ref{section:Dissipative Systems}). The authors in \citep{yujun2010new} demonstrated that the system can exhibit hyperchaotic motion by varying the parameter \(r\). For example, by setting \(a = 35\), \(b = \frac{8}{3}\), and \(c = 55\), the system \eqref{eq:4DODE} shows a wide range of chaotic behavior with a positive maximum Lyapunov exponent (mLE) \eqref{eq:mLEs}, and a particular case, such as \(r = 1.3\), leads to hyperchaotic behavior with two positive LE.

\subsection{The generalized H{\'e}non map}
The third and final model we consider is a discrete time dissipative system. We introduce the generalized hyperchaotic dissipative system, which is an extension of the well-known H{\'e}non map. The  \(2D\) map that H{\'e}non \citep{henon1976two} introduced based on the idea of stretching and folding regions in the phase space of a discrete time system. A point $(x_n, y_n)$ in a plane is mapped by the H{\'e}non map to a new point $(x_{n+1}, y_{n+1})$ using the following rule:

\begin{equation}\label{eq:2DHenMap} 
\begin{aligned}
 x_{n+1} &= a - (x_n)^2 + b y_n,  \\
 y_{n+1} &= x_n,
\end{aligned}
\end{equation}
where $x_n$ and $y_n$ are the state variables at discrete time $n$, and $a$ and $b$ are the control parameters of the map. For parameter values $a = 1.4$ and $b = 0.3$, the dynamics converge to a classical strange attractor. Examples and details on this map can be found in \citep{awrejcewicz2018quantifying,moges2020investigating}. 

For completeness' sake, let us quickly determine that the H{\'e}non map \eqref{eq:2DHenMap} is indeed dissipative by finding the determinant of its Jacobian matrix:
\begin{equation}
J = 
\begin{bmatrix}
-2x_j & b \\
1 & 0
\end{bmatrix}
\end{equation}
Actually the determinant of the \(2 \times 2\) matrix, \(J\) is 
\begin{equation} \label{eq: J 2DHenMap}
\text{det }(J) = (-2x_j)(0) - (b)(1) = -b.
\end{equation}
The determinant is always negative, indicating that the volume element in the phase space of the system \eqref{eq:2DHenMap} is contracting, confirming that the H{\'e}non map \eqref{eq:2DHenMap} is dissipative (see for, e.g., \citep[Sect. 9.2]{strogatz2018nonlinear} for more details).

The generalized H{\'e}non map is an extension of the map \eqref{eq:2DHenMap}, described by the following equations \citep{baier1990maximum}:
\begin{equation} \label{eq:NDHenMap} 
    \begin{aligned}
        x_{n+1}^{(1)} &= a - \left(x_{n}^{(j-1)}\right)^2 - b x_n^{(j)}, \\
        x_{n+1}^{(j)} &= x_{n}^{(j-1)}, \quad \text{for} \quad j = 2, 3, \dots, N
    \end{aligned}
    \end{equation}    
where \(N \geq 2\) represents the dimension of the system, \(a\) and \(b\) are non-negative control parameters, and \(x \in \mathbb{R}^N\) is an \(ND\) state vector. In \eqref{eq:NDHenMap}, there are \(N - 1\) linear equations and a single nonlinear one having a quadratic term. The dimension defined by the variable \(x_{n}^{(1)}\) is the one that undergoes stretching and folding, while all other variables experience a simple linear transformation from \(x_{n}^{(j-1)}\) to \(x_{n+1}^(j)\). Thus, map \eqref{eq:NDHenMap} is one of the simplest dissipative systems capable of exhibiting higher-dimensional chaos \citep{baier1990maximum,richter2002generalized}.

To investigate the behavior of the GALI indices across different types of attractors in discrete dissipative systems, we will focus on the \(3D\) generalized H{\'e}non map \eqref{eq:NDHenMap}. This choice will allow us to explore the following dynamical structures: stable fixed points, stable limit cycles, chaotic attractors, and hyperchaotic attractors, in the single simplified form of the system \eqref{eq:NDHenMap}. The corresponding \(3D\) map can be obtained by setting \(N=3\) in \eqref{eq:NDHenMap} and is given by the following equations:

\begin{equation} \label{eq:3DHenMap} 
 \begin{aligned}
	x_{n+1} &= a - y_n^2  - b z_n, \\
	y_{n+1} &= x_n,  \\
	z_{n+1} &= y_n. 
\end{aligned}
\end{equation}

Following \citep{richter2002generalized}, we can determine the stability of the fixed points of the map \eqref{eq:3DHenMap} by computing its Jacobian matrix and eigenvalues. The Jacobian matrix of system \eqref{eq:3DHenMap} is: 
\[
J = 
\begin{bmatrix}
\frac{\partial x_{n+1}}{\partial x_n} & \frac{\partial x_{n+1}}{\partial y_n} & \frac{\partial x_{n+1}}{\partial z_n} \\
\frac{\partial y_{n+1}}{\partial x_n} & \frac{\partial y_{n+1}}{\partial y_n} & \frac{\partial y_{n+1}}{\partial z_n} \\
\frac{\partial z_{n+1}}{\partial x_n} & \frac{\partial z_{n+1}}{\partial y_n} & \frac{\partial z_{n+1}}{\partial z_n}
\end{bmatrix}
 = 
\begin{bmatrix}
0 & -2y_n & -b \\
1 & 0 & 0 \\
0 & 1 & 0
\end{bmatrix}.
\]
Now, we can find its eigenvalues by solving the characteristic equation: \(\text{det}(J - \lambda I) = 0\), where \( I \) is the identity matrix. The characteristic equation simplifies to \(\lambda^3 + 2y_n \lambda + b = 0\). Solving this cubic equation yields the eigenvalues, which provide information on the stability and dynamics of the system's fixed points. 

It has been shown in \citep{richter2002generalized} that the fixed point \(x^* = \frac{-(1+b) + \sqrt{(1+b)^2 + 4a}}{2} \begin{pmatrix} 1, & 1, & 1 \end{pmatrix}\) is locally stable for the parameter range \(0 < a < \frac{4(1-b^2)(1+b)(3-b)}{4}\). As the parameter \( a \) increases, the fixed points become POs, eventually leading to chaotic motion (see the bifurcation diagram presented in \citep[Figure 2]{richter2002generalized}). In particular, the system \eqref{eq:3DHenMap} exhibits more complex dynamics compared to its \(2D\) counterpart \eqref{eq:2DHenMap}. For instance, the system displays motion characterized by two positive LEs, indicating the presence of hyperchaotic dynamics for the parameter values \( a = 1.76 \) and $b=0.1$.

\section{Numerical results} \label{section:ResultsCh6}
In this section, we explore the behavior of the GALI method across a spectrum of DSs. To provide a comprehensive understanding, we select representative attractors from the various models discussed in Sect.~\ref{section:modelCh6}: stable fixed points, stable PO (limit cycles), chaotic (strange), and hyperchaotic attractors The following subsections present detailed case studies, supported by phase space diagrams and computation's of the LE spectrum \eqref{section:LEs} and the GALI method \eqref{eq:GALI} for each attractor. This analysis will enable us to illustrate the GALI method's performance for each one of these dynamical cases. 

The choice of model parameters is crucial for obtaining a diverse range of dynamical behaviors in the studied systems. By carefully computing the LE spectrum, we initially identify parameter values leading to the four desired dynamical features (stable fixed points, stable limit cycles, chaotic as well as hyperchaotic attractors) in the system's phase space. An important aspect of our analysis is to establish the behavior of the GALI indices for these orbits. Computing both chaotic detection techniques, i.e., the LEs and the GALIs, further allows us to explore the relationships between both methods in the context of dissipative systems. 

\subsection{Numerical investigation of the $3D$ Lorenz system} \label{sec:3DODE}
In our numerical analyses, we begin by setting the parameters \( a = 10 \) and \( b = \frac{3}{8} \) in the \(3D\) Lorenz system \eqref{eq:3DODE}. We then vary the third control parameter \( r \) to identify the existence of different dynamical regimes for this system. In order to compute the spectrum of LEs and the GALI\(_k\) for \(k = 2\) and \(3\), we integrate both the system of ODEs \eqref{eq:3DODE} and the corresponding variational equations. To derive these variational equations, we linearize the system \eqref{eq:3DODE} by substituting \( x \) with \( x + \delta x \), \( y \) with \( y + \delta y \), and \( z \) with \( z + \delta z \) where \([ \delta x,  \delta y,  \delta z]\) represent a deviation vector. Then the set of variational equations governing the time evolution of the deviation vector is: 
\begin{equation} \label{eq:3DLorenz var} 
 \begin{aligned}
\frac{d \delta x}{dt} &= a (\delta y - \delta x), \\ 
\frac{d \delta y}{dt} &= (r - z) \delta x  - \delta y - x \delta z, \\
\frac{d \delta z}{dt} &= x \delta y + y \delta x - b \delta z.
\end{aligned}
\end{equation}

In what follows, unless otherwise specified, we numerically integrate both set of equations \eqref{eq:3DODE}  and \eqref{eq:3DLorenz var} up to \( t = 10^5 \) time units using the fourth-order Runge-Kutta (RK4) integration scheme \eqref{eq:RK} (see \citep{Hairer1993} for more detail). We choose an appropriate integration step size that is sufficiently small to accurately capture the orbit evolution but large enough to avoid extensive computational demand.  In our numerical simulations, we observed that a final integration time of \( t = 10^5 \) is adequate to capture the characteristics of each considered orbit of the system.

\subsubsection{A stable fixed point case}  \label{sec:3DODE fixed point}
We begin our investigation by considering a stable fixed point in the \(3D\) Lorenz system \citep[Sect. 9.2]{strogatz2018nonlinear}. In the Lorenz system \eqref{eq:3DODE} corresponding to an IC at \(t=0\) when discussing the asymptotic behavior of an orbit, we study what happens to this orbit as \(t\) approaches infinity. The simplest type of the Lorenz orbit is a fixed (stationary) point. For instance, in fluid dynamics, steady-state flows are characterized by convection patterns. If a fluid system starts at rest, and it is heated from below, then the system will eventually reach to a steady-state configuration where the flow motion stabilizes. This represents a stable fixed point attractor or a stable equilibrium. In this case, any IC that is relatively close to the fixed point will result in an orbit that will eventually converge to the point itself.

Figure \ref{fig6:Fig1a} shows the \(3D\) phase portrait for the parameters \( a = 10 \), \( b = \frac{8}{3} \), and \( r = 2.1 \) of an orbit with IC \( (x, y, z) = (1, 3, 6) \) of system \eqref{eq:3DODE}. The black curve depicts the final stages of the evolution of this IC, while the gray segment highlights the initial time interval of the trajectory's evolution. In addition, we show all possible \(2D\) projections of the \(3D\) phase space plot in different colors: the red curve represent the projection on the \( xy \)-plane, the purple curve correspond to the \( xz \)-plane projection, and the blue curve indicates the \( yz \)-plane.

The computed time evolution of the three ftLEs, \(\sigma_1\) (red curve), \(\sigma_2\) (blue curve) and \(\sigma_3\) (green curve), with the ordering \(\sigma_1 > \sigma_2 > \sigma_3\) \eqref{eq:LEs order}, for this orbit is shown in Fig.~\ref{fig6:Fig1b}. From the results in Fig.~\ref{fig6:Fig1b}, we observe that all LEs are negative. To be more specific, we note that $\sigma_1$ and $\sigma_2$ converge to constant values \( -1.20\) and \( -1.20 \), respectively, around $t \approx 3.5 \times 10^4$, while $\sigma_3$ reaches its constant value \( \sigma_3 \approx -11.26\) earlier, at approximately $t \approx 2.2 \times 10^2$. According to our discussion in Sect.~\ref{section:LEs}, for a stable fixed point, we expect all LEs to be strictly negative, something which indicates that any nearby trajectory of the given IC \( (x, y, z) = (1, 3, 6) \) [orange point in Fig.~\ref{fig6:Fig1b}] will eventually converge towards the fixed point \( (x^*, y^*, z^*) = (4.899, 4.899, 9)\). This convergence to the fixed point is also seen in the phase space diagram of Fig.~\ref{fig6:Fig1a}, where the orbit starting from the orange point spirals around and eventually converges to \( (x^*, y^*, z^*)\). It is important to emphasize that the sum of LEs for this orbit is negative, signifying that the system \eqref{eq:3DODE} is dissipative (see Sect.~\ref{section:LEs}). 

In Fig.~\ref{fig6:Fig1c}, we present the time evolution of the GALI\(_2\) (blue solid curve) and the GALI\(_3\) (red solid curve, inset plot) for orbits of Fig.~\ref{fig6:Fig1a}. As we can see from this figure, the GALI\(_2\) oscillates around a constant positive value, a behavior typically observed for regular orbits of conservative Hamiltonian systems [see, for example, black, orange and purple curves in Fig.~\ref{fig3:Fig2b}].On the other hand, the GALI\(_3\) [red solid curve in the inset of Fig.~\ref{fig6:Fig1c}] decays to zero exponentially fast. The decay rate of the GALI\(_3\) corresponds with the theoretical prediction \eqref{eq:GALI_chaos} \(\text{GALI}_3 \propto \exp[-(2\Lambda_1 - \Lambda_2 - \Lambda_3)\) where \( \Lambda_1 = -1.20 \), \( \Lambda_2 = -1.20 \), and \( \Lambda_3 = -11.26 \) are the estimates of the three largest LEs obtained in Fig.~\ref{fig6:Fig1b}.

\subsubsection{A stable limit cycle case}  \label{sec:3DODE limt cycle}
Another simple type of attractor that appears in the Lorenz system \eqref{eq:3DODE} is a PO, or in other words a limit cycle (see \citep[Sect. 9.2]{strogatz2018nonlinear}). For example, in a fluid system exhibiting convection patterns, we can have ICs such that the fluid's temperature gradient leads to the formation of a stable convection cell, i.e., the flow will follow a periodic trajectory. This means that after a period of time \(p\), the system returns to the same configuration it had at time \(t=0\).

A PO in system \eqref{eq:3DODE}, which can be classified as stable or unstable. A PO is stable if nearby ICs lead to orbits that eventually converge to the PO itself, while it is unstable if nearby ICs diverge from the PO. This stable PO can also be referred as a periodic attractor. In particular, the PO is considered a limit cycle if it's an isolated closed trajectory\footnote{A closed trajectory is a trajectory in the phase portrait that forms a continuous loop. An isolated closed trajectory is a type of closed trajectory where nearby trajectories do not form similar closed loops.}. If all neighboring trajectories approach the limit cycle, it is classified as stable (see \citep[Sect. 7.0]{strogatz2018nonlinear} for further details and examples). 

Figure \ref{fig6:Fig1d} illustrates the \(3D\) phase space portrait of an orbit converging to a stable limit cycle in the \(3D\) Lorenz system\eqref{eq:3DODE} with parameters \( a = 10 \), \( b = \frac{8}{3} \), and \( r = 250 \). The considered orbit has the same IC as the orbit in Fig.~\ref{fig6:Fig1a}. The time evolution of the three ftLEs for this orbit is shown in Fig.~\ref{fig6:Fig1e}. The ftmLE, \( \sigma_1 \) (red curve), asymptotically approaches zero, indicating the periodic nature of the attractor. In general, a single LE being zero reflects the existence of a direction in the phase space where trajectories neither converge nor diverge, which is typically associated with a neutral direction. On the other hand, the other two LEs, \( \sigma_2 \) (blue curve) and \( \sigma_3 \) (green curve), remain constant and negative (with the ordering \(\sigma_2 > \sigma_3\) \eqref{eq:LEs order}), indicating that the trajectory converges toward an attractor. The fact that the mLE is zero and the remaining two are negative further confirms that the limit cycle is stable (see Sect.~\ref{section:LEs}). 

Figure \ref{fig6:Fig1f} presents the time evolution of the GALI\(_2\) (blue curve) and the GALI\(_3\) (red curve) of the same orbit. Both indices decay to zero exponentially fast, with rates following the theoretical expectations of \eqref{eq:GALI_chaos}: 
\[
\text{GALI}_2 \propto \exp[-(\Lambda_1 - \Lambda_2)] \quad \text{and} \quad \text{GALI}_3 \propto \exp[-(2\Lambda_1 - \Lambda_2 - \Lambda_3)],
\] where \( \Lambda_1 = 0 \), \( \Lambda_2 = -2.67 \), and \( \Lambda_3 = -11 \) are estimations of the three LEs obtained from the results of Fig.~\ref{fig6:Fig1e}. This analysis demonstrates that for stable limit cycles of the Lorenz system \eqref{eq:3DODE}, both GALI\(_2\) and GALI\(_3\) converge to zero exponentially fast. 

\subsubsection{A chaotic (strange) attractor case}  \label{sec:3DODE chaotic}

In the \(3D\) Lorenz system \eqref{eq:3DODE}, due to its dissipative nature, an orbit typically converges to one of three types of attractors over time: a fixed point, a limit cycle or a chaotic attractor. A chaotic attractor in \eqref{eq:3DODE} represents a bounded region of phase space where the system exhibits chaotic behavior without settling into a fixed point or PO. The Lorenz system that can exhibit this type of chaotic behaviors, known as chaotic or \textit{strange attractors} \citep[Sect. 9.3]{strogatz2018nonlinear}. Again, considering our fluid dynamics example to explain the physical meaning of these attractors. We refer to the case when a fluid is heated from below, and can develop irregular flow patterns that remain spatially bounded but do not exhibit stable or periodic behavior. Instead, these patterns may demonstrate chaotic dynamics, characterized by trajectories that are very sensitive to their ICs.

\begin{figure}[!htbp]
    \centering
    \subfloat[Phase portrait of an orbit tending to a stable fixed point\label{fig6:Fig1a}]{\includegraphics[width=0.33\textwidth]{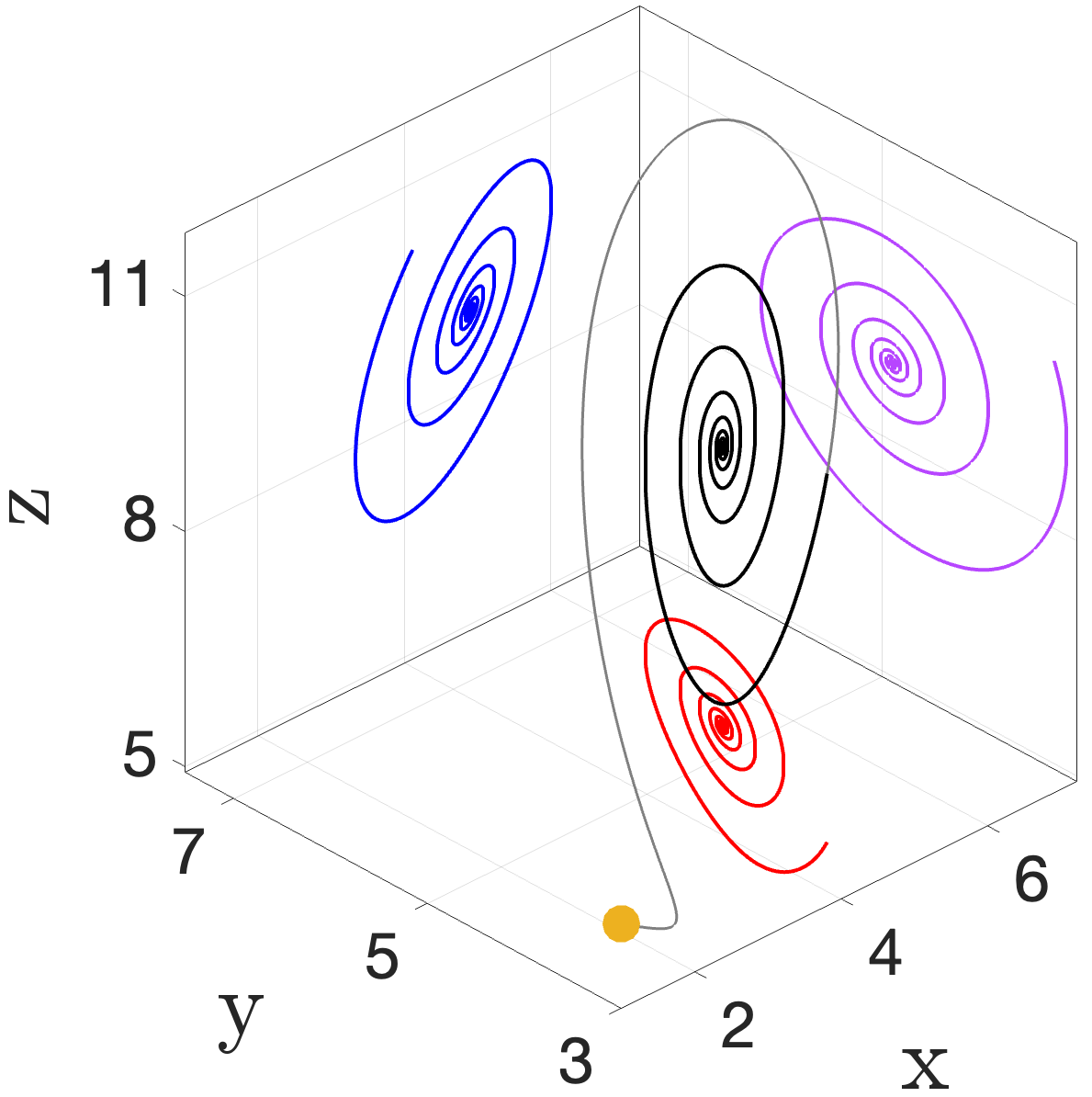}}
    \subfloat[ftLEs\((t)\) for the orbit of (a)\label{fig6:Fig1b}]{\includegraphics[width=0.33\textwidth]{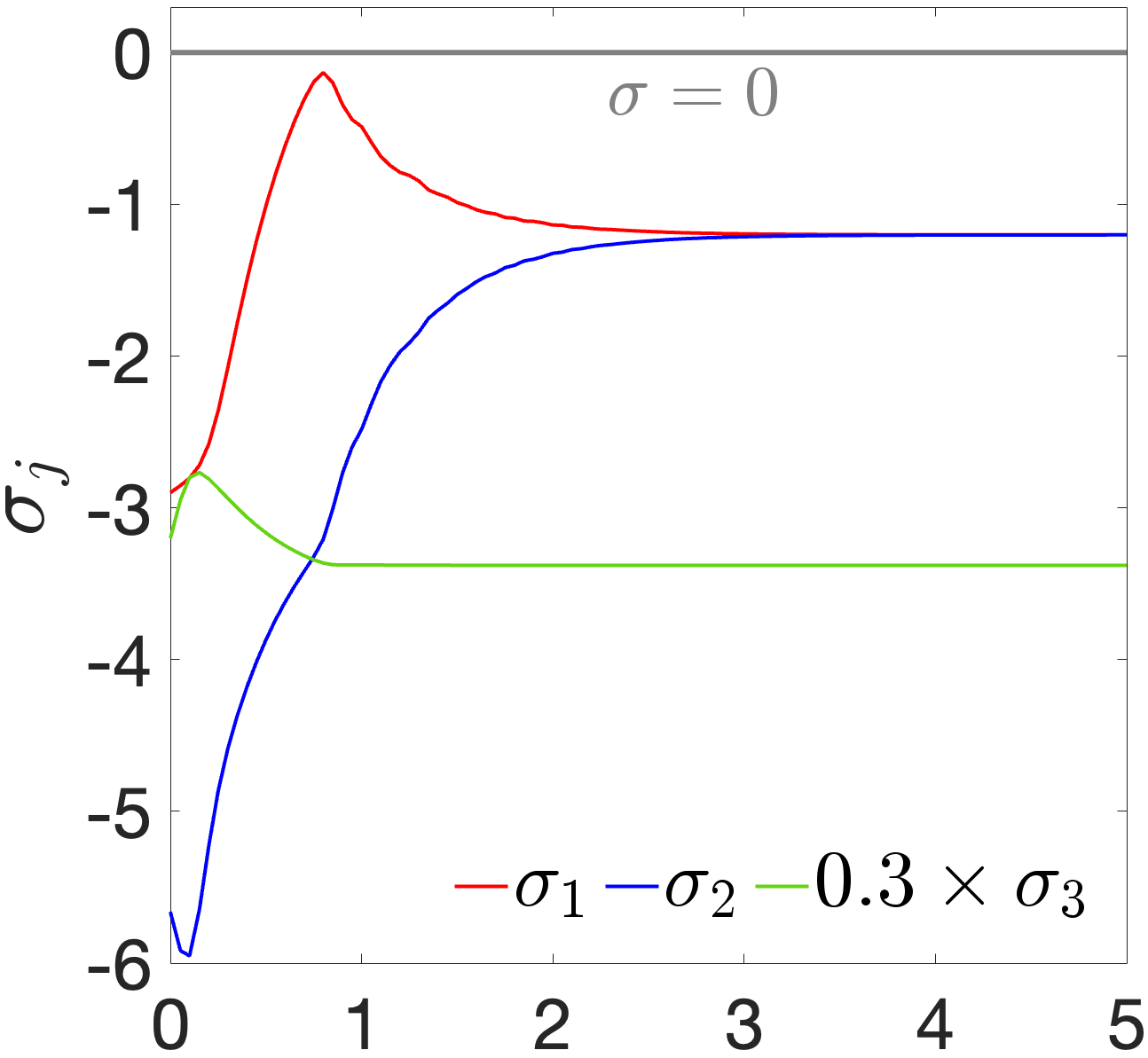}}
    \subfloat[GALI\(_{k} (t)\) for the orbit of (a)\label{fig6:Fig1c}]{\includegraphics[width=0.33\textwidth]{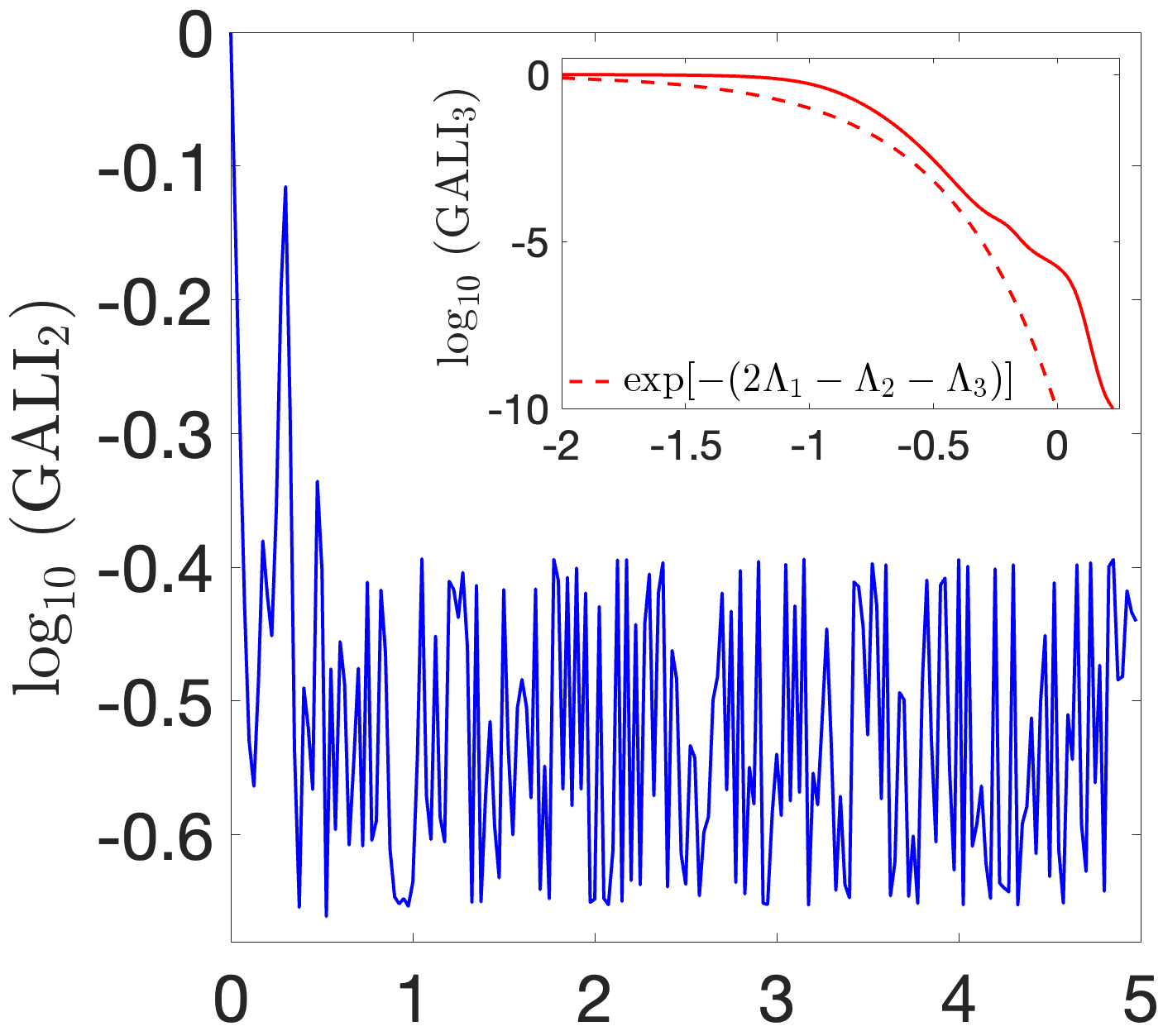}}\\
\subfloat[Phase portrait of an orbit tending to a stable limit cycle\label{fig6:Fig1d}]{\includegraphics[width=0.33\textwidth]{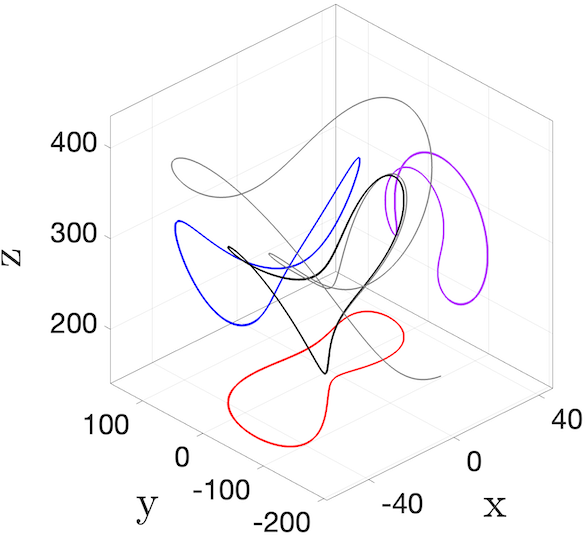}}
\subfloat[ftLEs\((t)\) for the orbit of (d)\label{fig6:Fig1e}]{\includegraphics[width=0.33\textwidth]{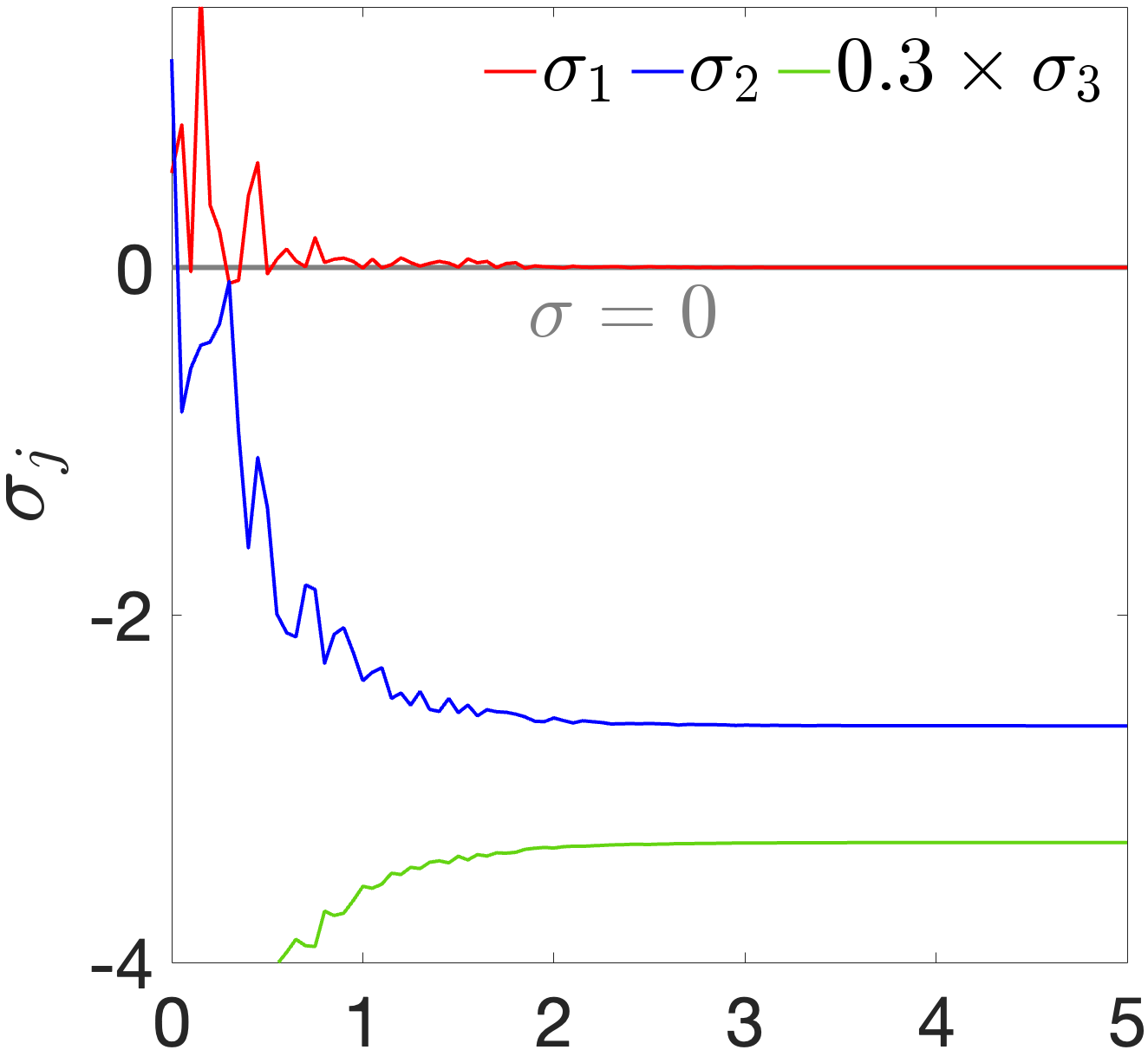}}
\subfloat[GALI\(_{k} (t)\) for the orbit of (d)\label{fig6:Fig1f}]{\includegraphics[width=0.33\textwidth]{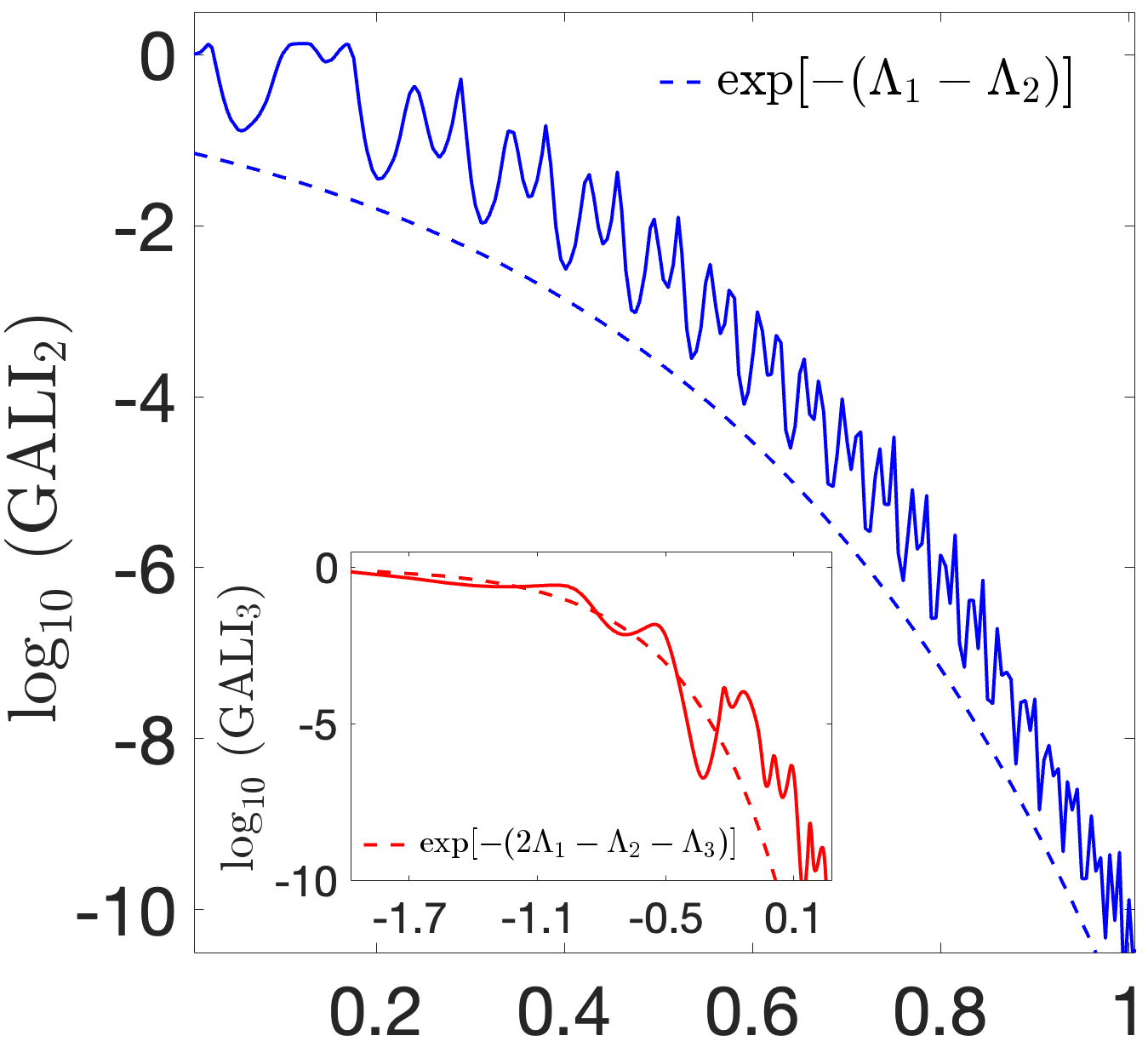}}\\

\subfloat[Phase portrait of a chaotic attractor\label{fig6:Fig1g}]{\includegraphics[width=0.33\textwidth]{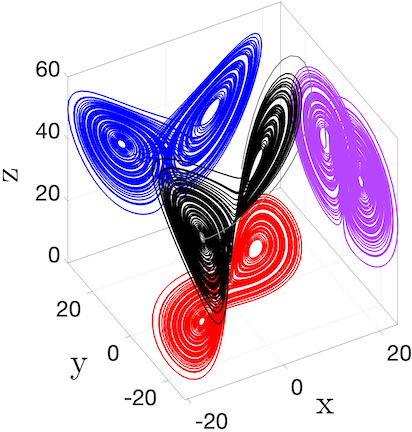}}
\subfloat[ftLEs\((t)\) of the chaotic attractor of (g)\label{fig6:Fig1h}]{\includegraphics[width=0.33\textwidth]{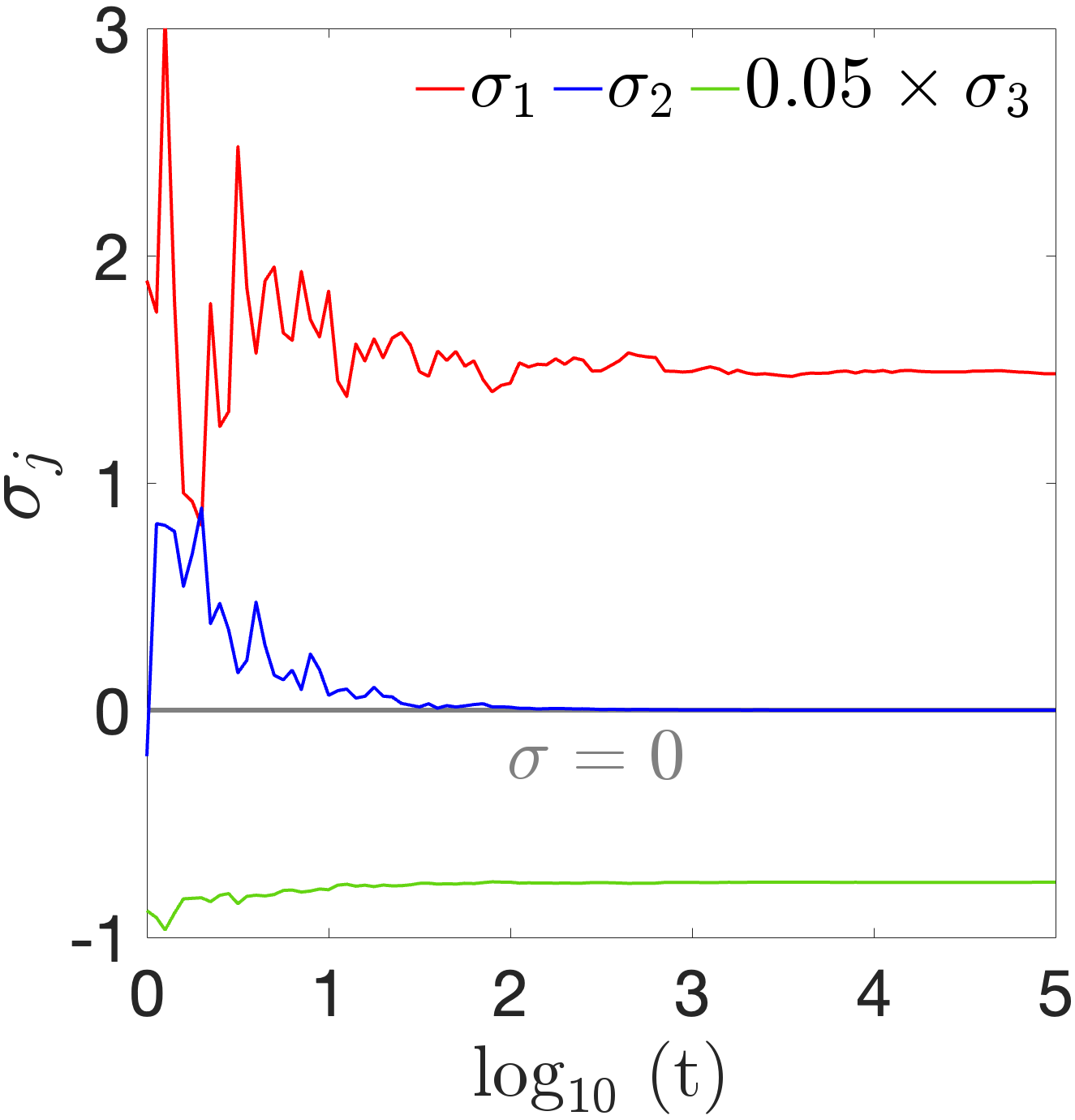}}
\subfloat[GALI\(_{k} (t)\) of the chaotic attractor of (g)\label{fig6:Fig1i}]{\includegraphics[width=0.33\textwidth]{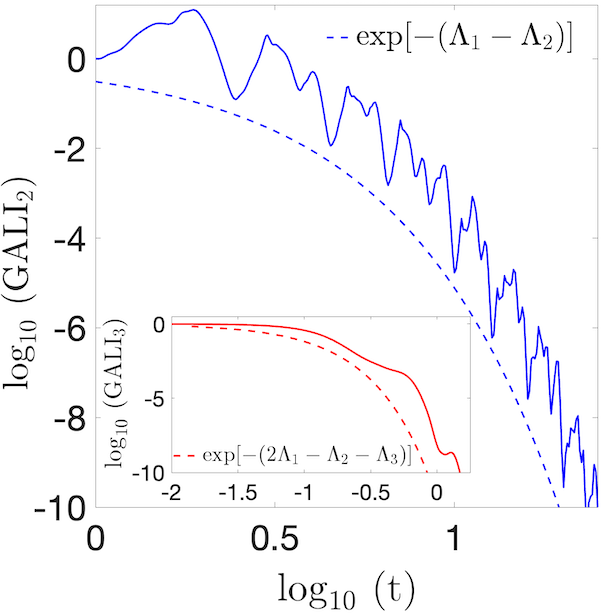}}\\
\caption{The \(3D\) phase space portraits of various orbits of the \(3D\) Lorenz system \eqref{eq:3DODE} with parameters \( a = 10 \) and \( b = \frac{8}{3} \): (a) For \( r = 2.1 \), the trajectory with the IC \( (x, y, z) = (1, 3, 6) \) (orange circle point) tends to the stable fixed point attractor \( (x^*, y^*, z^*) = (4.899, 4.899, 9)\). (d) For \( r = 250 \), we observe a stable limit cycle attractor and (g) a chaotic (strange) attractor appears for \( r = 33.3 \). In the \(3D\) phase space portraits of panels (a), (d) and (b), the black curves represent the final (asymptotic) phase, while the initial evolution of the trajectories is denoted by gray curves. The red, purple, and blue curves depict the corresponding \(2D\) projections of the orbits on the \( xy \), \( xz \), and \( yz \) planes, respectively. The time evolution of the three ftLEs \(\sigma_1 > \sigma_2 > \sigma_3\) \eqref{eq:LEs order} for the orbits in panels (a), (d) and (g) is presented in (b), (e), and (h), respectively. Note that, in these panels, the \( \sigma_3 \) values are scaled for visualization purposes, and for reference, we draw a horizontal gray line at \( \sigma = 0 \). The respective time evolution of GALI\(_2\) (blue curve) and GALI\(_3\) (red curve, inset plot) for the three orbits is shown in (c), (f), and (i). The blue and red dashed curves in these panels represent functions proportional to \( \exp\left[-(\Lambda_1 - \Lambda_2)\right] \) and \( \exp\left[-(2\Lambda_1 - \Lambda_2 - \Lambda_3)\right] \), respectively. In (c) \( \Lambda_1 = -1.20 \), \( \Lambda_2 = -1.20\), and \( \Lambda_3 = -11.27 \); in (f) \( \Lambda_1 = 0 \), \( \Lambda_2 = -2.67 \), and \( \Lambda_3 = -11 \); and in (i) \( \Lambda_1 = 1.02 \), \( \Lambda_2 = 0 \), and \( \Lambda_3 = -14.69 \). See text and legend of each panel for details.}
	\label{fig6:Fig1}
\end{figure}

To obtain a typical chaotic (strange) attractor, which is mostly known as the Lorenz strange attractor \citep{lorenz1991dimension}, let us change the parameter \( r \) to \( r = 33.3 \) by keeping the orbit's IC \( (x, y, z) = (1, 3, 6) \). Fig.~\ref{fig6:Fig1g} displays the corresponding \(3D\) phase portrait of this orbit. The strange attractor's trajectory creates intricate and non-repeating patterns in the phase space, which often resemble a butterfly. For more examples on strange attractors, we refer the interested reader to \citep[Figs. 9.3.2 and 9.3.2]{strogatz2018nonlinear}. The butterfly feature is clearly shown in the \(2D\) projections (blue, red, and purple curves) in Fig.~\ref{fig6:Fig1g}. The phase space structure shown is believed to be the origin of the famous metaphor ``butterfly effect", a notion introduced by Lorenz in his 1972 lecture, titled ``Predictability: Does the Flap of a Butterfly's Wings in Brazil Set Off a Tornado in Texas?" \citep{lorenz1972predictability}

The corresponding time evolution of the three ftLEs is shown in Fig.~\ref{fig6:Fig1h}. The ftLEs of the strange attractor exhibit the following behaviors: \( \sigma_1 > 0 \) (red curve), \( \sigma_2 = 0 \) (blue curve), and \( \sigma_3 < 0 \) (green curve), which is consistent with the properties of ftLEs for chaotic attractors discussed in Sect.~\ref{section:LEs}. In particular, the positive ftmLE (\( \sigma_1 > 0 \)) indicates the presence of chaos. The fact that the second exponent converges to zero (\( \sigma_2 = 0 \)) suggests that the Lorenz system exhibits neutral stability in the second LE direction, where the trajectory may oscillate, but neither expand nor contract on this direction. This typically results in a fractal structure, meaning the attractor has self similar patterns at different scales of the phase space. These patterns can easily be observed in the \(2D\) projections shown by the blue, red, and purple curves in Fig.~\ref{fig6:Fig1g}. Furthermore, a negative third LE\( \sigma_3\) implies that the trajectory is confined in a bounded region of the system's phase space, which is preventing the system from diverging to infinity despite its chaotic behavior. 

Both the GALI\(_2\) and the GALI\(_3\) of the orbit of the Fig.~\ref{fig6:Fig1g} decay to zero exponentially (blue and red curves in Fig.~\ref{fig6:Fig1i}, respectively). The indices follow the theoretical evolution for chaotic orbits of conservative Hamiltonian systems \eqref{eq:GALI_chaos}, respectively: \( \text{GALI}_2 \propto \exp[-(\Lambda_1 - \Lambda_2)] \) and \( \text{GALI}_3 \propto \exp[-(2\Lambda_1 - \Lambda_2 - \Lambda_3)] \), with \( \Lambda_1 = 1.018 \), \( \Lambda_2 = 0 \), and \( \Lambda_3 = -14.69 \), i.e., the values obtained from Fig.~\ref{fig6:Fig1h}. 

\subsubsection{Parametric exploration of the \(3D\) Lorenz system using the GALI method and the LEs}  \label{sec:3DODE parameter space}
To understand the influence of a parameter on the \(3D\) Lorenz system's dynamics \eqref{eq:3DODE} and to gain a deeper understanding of the GALI method's behavior for this system, we systematically vary the control parameter \( r \) over a range that will allow us to observe the three attractors discussed in Secs.~\ref{sec:3DODE fixed point} to \ref{sec:3DODE chaotic}: stable fixed points, stable limit cycles, and chaotic attractors. Since the GALI\(_3\) index decreases to zero exponentially fast in all cases, we do not expect it to provide any relevant information for identifying different types of orbits of the system. As a result, we will focus our analysis on the GALI\(_2\) which depicts different behaviors for some trajectory types.

We first analyze the effect of parameter \(r\) using the three ftLEs to identify regions with different dynamical features (see Sect.~\ref{section:LEs}). For each value of \( r \) (but for fixed \( a = 10 \) and \( b = \frac{8}{3} \) parameter), we then compute the GALI\(_2\) and compare its behavior with the corresponding ftLE spectrum (\( \sigma_{j}\), \(j=1, 2, 3\)). By systematically varying \( r \) and analyzing both the ftLEs and the GALI\(_2\), we aim to provide a comprehensive understanding of the effectiveness of the GALI\(_2\) in characterizing the dynamics of the \(3D\) dissipative system.

\begin{figure}[!htb]
    \centering
    \subfloat[ftmLEs\label{fig6:Fig2a}]{\includegraphics[width=0.51\textwidth]{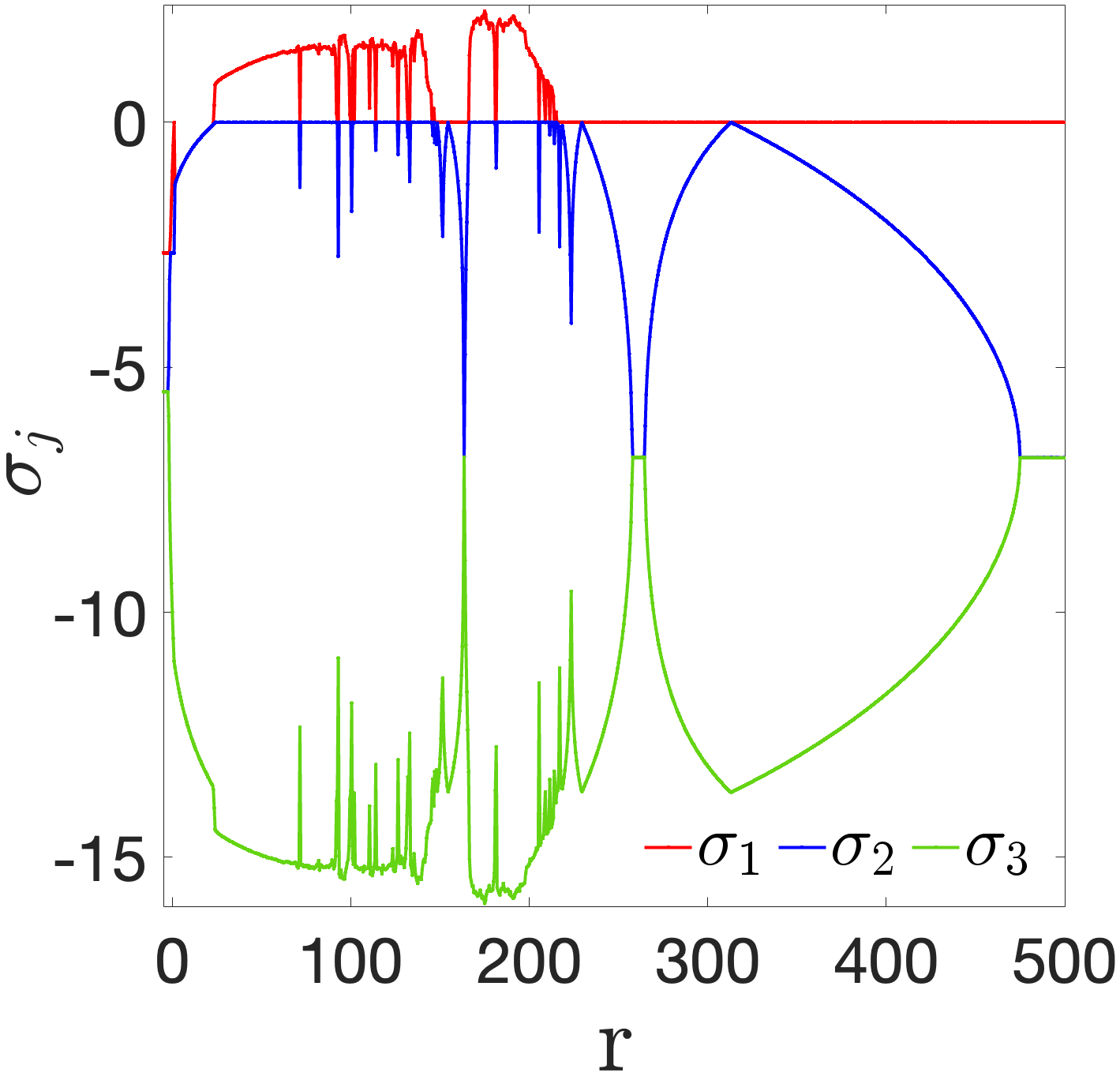}}
    \subfloat[GALI\(_{2}\)\label{fig6:Fig2b}]{\includegraphics[width=0.49\textwidth]{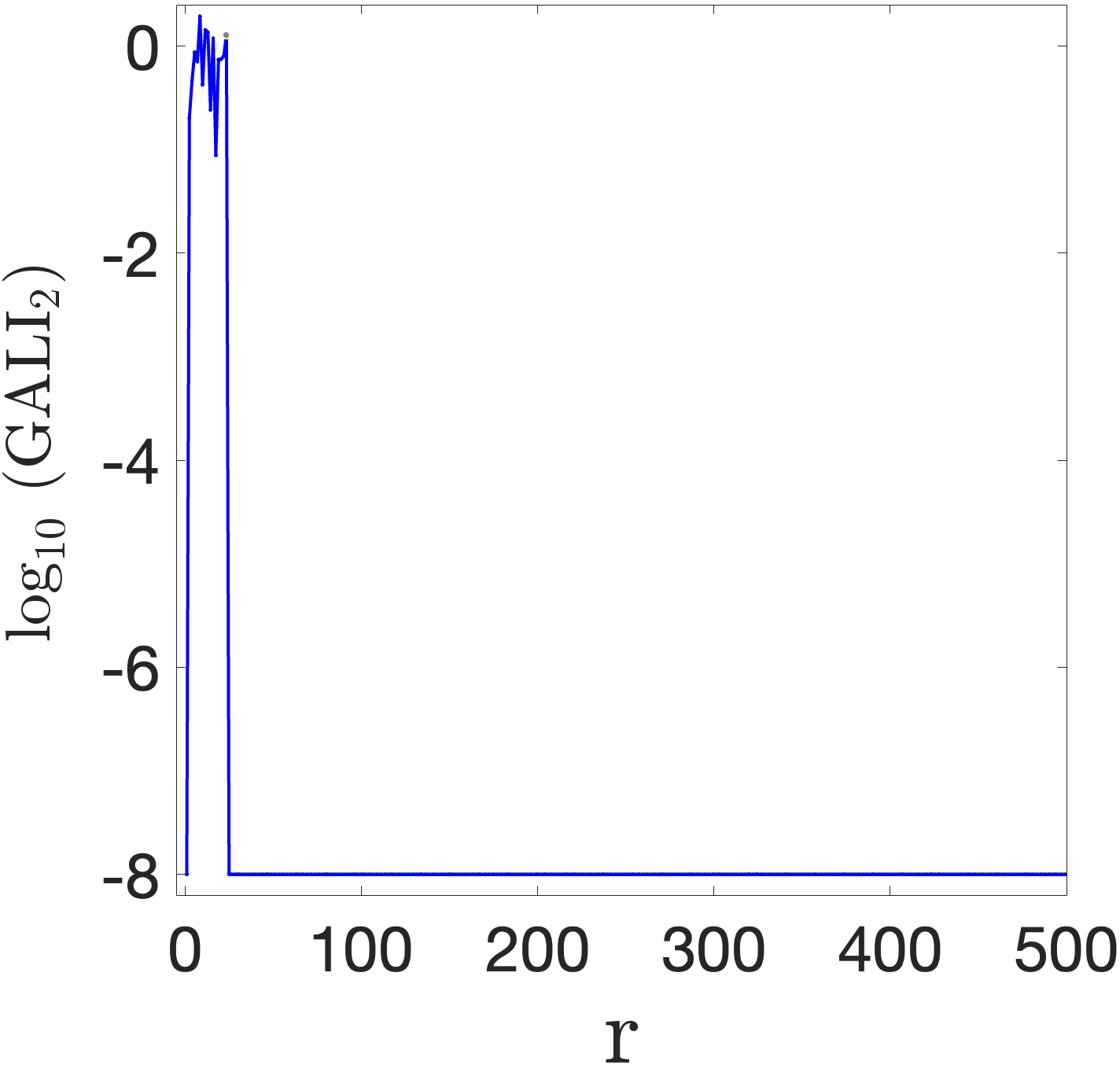}} \\
	\caption{A parametric exploration of the \(3D\) Lorenz system \eqref{eq:3DODE} for parameters \( a = 10 \) and \( b = \frac{8}{3} \) while varying \(r\). We consider a total of \(1011\) points for \( r \in [-5, 500] \). Final values of (a) the three ftLEs \(\sigma_1 > \sigma_2 > \sigma_3\) \eqref{eq:LEs order}, and (b) the GALI$_2$ \eqref{eq:GALI} at \(t=10^4\).}
\label{fig6:Fig2}
\end{figure}

In Fig.~\ref{fig6:Fig2a} and (b), we present the asymptotic values of the three ftLEs and the GALI\(_2\), respectively, computed at \(t = 10^4\) time units. Note that the final integration time is smaller compared to the representative cases shown in Fig.~\ref{fig6:Fig1}, where the integration was carried out up to \(t = 10^5\) time units. This is due to the fact that the time evolution of both the ftLEs and the GALI\(_2\) show their expected behaviors before \(t = 10^5\). In our analysis, we consider the IC \((x, y, z) = (2, 1, 5)\) and vary \(r\) over equally spaced values in the range \([-5, 500] \) by fixing \( a = 10 \) and \( b = \frac{8}{3} \). It is important noting that we use only one IC as a representative case for each value of \(r\), as other ICs yield similar results in this parameter setup. We observe that for \( r \) values in the interval \([-5, 21.3]\), the system exhibits stable fixed points as all \( \sigma_{j} < 0\). The GALI\(_2\) in Fig.~\ref{fig6:Fig2b} reaches a positive constant value for the same range of \(-5 \le r < 21.3\), for which the red and blue curves in Fig.~\ref{fig6:Fig2a}, overlap indicating \( \sigma_1 \approx \sigma_2 \). 

For the parameter values we considered in the \(3D\) Lorenz system \eqref{eq:3DODE}, the two largest ftLEs, \( \sigma_{1}\) and \( \sigma_{2}\), often converge to similar negative values, which indicates uniform contraction of the phase space along the direction associated with \( \Lambda_{1}\) and \( \Lambda_{2}\). However, the system can also contract at different rates, resulting in distinct negative ftmLE values, i.e.~\( \sigma_1 \not\approx \sigma_2 < 0\). For instance, when \( r = -2 \), we observe that \( \Lambda_1 \approx -2.67 \) and \( \Lambda_2 \approx -5.0 \). In this case, the trajectory converges to the stable point attractor more quickly in the direction of the mLE \( \Lambda_1\), which is the smaller negative exponent (\( \Lambda_1 > \Lambda_2\)). When the system \eqref{eq:3DODE} exhibits such varying contraction rates in the phase space, the GALI$_2$ does not fluctuate around a positive constant value, which has indicative of regular motion in conservative systems. Rather the index decays to zero exponentially for \(r \in [-5, 1.3)\) [\(\log_{10} \text{GALI}_2 < -8\)] in Fig.~\ref{fig6:Fig2b}. 

When increasing the parameter value in the range \(21.3 < r < 146.9\), the system exhibits chaotic attractors, which are characterized by a positive ftmLE [Fig.~\ref{fig6:Fig2a}]. This range is followed by a short window \(146.9 \le r \le  166\), where stable limit cycles emerge, which are indicated by zero ftmLE. Beyond this interval, the system returns to chaotic behavior for \(166 < r < 215.4\) before eventually leading to the creation of a limit cycle for \( r \) values in the range \([215.4, 500]\). On the other hand, the GALI$_2$ decays to zero exponentially fast for all values of \(r\) in the range \((21.3 < r \le 500)\) [Fig.~\ref{fig6:Fig2b}]. 

Overall, Fig.~\ref{fig6:Fig2} demonstrates that while GALI\(_2\) effectively distinguishes between stable fixed points and chaotic attractors in the Lorenz system \eqref{eq:3DODE}, it exhibits similar behavior for stable limit cycles and chaotic attractors. Furthermore, the GALI\(_2\) shows different behaviors even for stable fixed points, depending on whether the two largest ftLEs converge to distinct negative values (the GALI\(_2\) becomes practically zero) or converge to practically equal negative values (the GALI\(_2\) fluctuates around positive constant values). For stable fixed point attractor characterized by two distinct largest ftLEs, we find that the LE spectrum a more reliable indicator in determining the type of motion. While using the GALI$_2$ index can be valuable, for instance for quickly detecting chaotic attractors, it is essential to complement this analysis with the computation of the LE spectrum.

\begin{figure}[!htb]
    \centering
    \subfloat[ftmLEs\label{fig6:Fig3a}]{\includegraphics[width=0.36\textwidth]{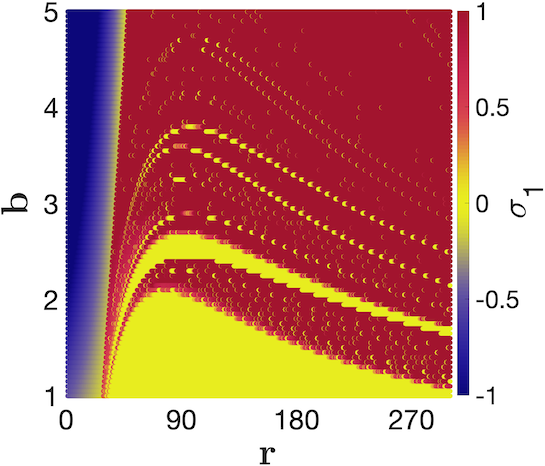}}
    \subfloat[Classification based on the three ftLEs\label{fig6:Fig3b}]{\includegraphics[width=0.325\textwidth]{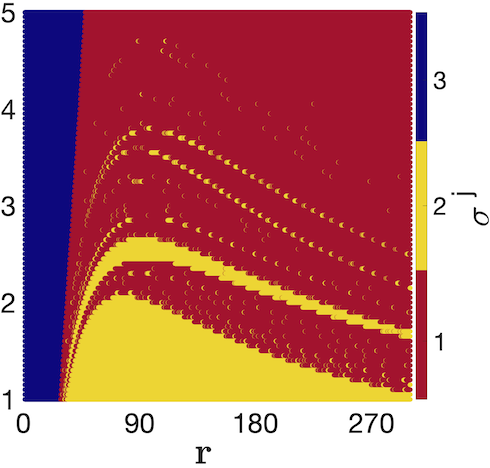}}
    \subfloat[GALI\(_{2}\)\label{fig6:Fig3c}]{\includegraphics[width=0.32\textwidth]{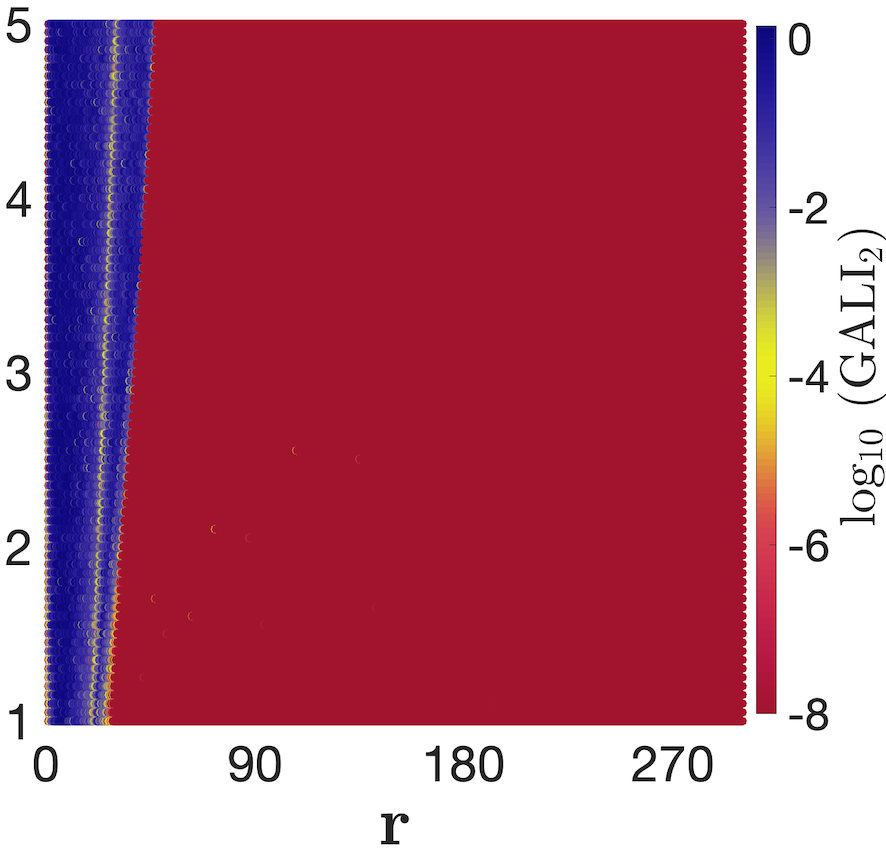}}\\
    \caption{An exploration of the \( (r, b) \) space of the Lorenz system \eqref{eq:3DODE} with \( a = 35 \). We produce color plots using a grid of \( 2991 \times 81 = 242,271 \) points in the region \( r \in [0, 300]\) and \( b \in [1, 5] \) by integrating the IC \((x, y, z) = (2, 1, 5)\)  up to \( t = 10^4 \) and recording the orbit's ftLEs \(\sigma_1 > \sigma_2 > \sigma_3\) \eqref{eq:LEs order} and the GALIs \eqref{eq:GALI}. The classification of various dynamical regimes is based on the values of (a) the ftmLE, \( \sigma_1 \), and (b) the three ftLEs, \(\sigma^j\), according to the classification in Sect.~\ref{section:LEs}. In both panels, chaotic attractors are indicated by the red region \( (\sigma_1 > 0 \text{ or } \sigma^j = 1) \), stable periodic motion (limit cycles) is represented by the yellowish/orange areas \( (\sigma_1 \approx 0 \text{ or } \sigma^j = 2) \), and stable fixed point attractors are shown in dark blue \( (\sigma_1 < 0 \text{ or } \sigma^j = 3) \). (c) A similar classification is performed using GALI$_2$ values. Note that in (a), the computed \(\sigma_1\) values are scaled to the interval \([-1, 1]\).}
\label{fig6:Fig3}
\end{figure}

In order to obtain a global understanding of how the GALI\(_2\) index behaves for the \(3D\) Lorenz system, we further conduct a bi-parametric exploration of the system's parameter space by varying both \( r \) and \( b \) while fixing the parameter \( a = 35 \). For each parameter set \(r, b\), we consider the same IC \((x, y, z) = (2, 1, 5)\) as a representative case, as other ICs lead to consistent results, similar to Fig.~\ref{fig6:Fig2}. We note that for this study, we have changed the value of the parameter of \(a\) form \(a=10\), which was used in Figs.~\ref{fig6:Fig1} to \ref{fig6:Fig3}, to \(a=35\). The reason for this change is simply to allow us to observe a more diverse set of dynamical behaviors in the system \eqref{eq:3DODE}. We also chose to vary the \(r, b\) parameters and present the relevant parameter space, rather than other possible combinations like \((r, a)\) and \((a, b)\), not for any fundamental dynamical reason but because this choice leads to a diverse visual representation of the orbits we considered. 

Figure \ref{fig6:Fig3a} displays a color plot based on the computed ftmLE, \( \sigma_1 \), value for various combinations of \( r \) and \( b \) at $t=10^4$ time units. For a clear visualization, we apply what we call a signed scaling approach to the computed \( \sigma_1 \) values, mapping them to the interval \( [-1, 1] \) to capture both their sign and proximity to zero. This scaling normalizes positive values to the range \( [0, 1] \) and negative values to \( [-1, 0] \), allowing us to focus on whether \( \sigma_1 \) is negative, zero, or positive, rather than on its actual computed values. For instance, in our studied parameter range \( r \in [0, 300]\) and \( b \in [1, 5] \), the actual value of \( \sigma_1 \) is approximately on the interval \( [-1.29, 3.08] \). Thus, we linearly scale this interval so that \(\sigma_1 = -1.29\) maps to \(-1\), and \(\sigma_1 = 3.29\) maps to \(1\). This approach ensures that we can easily classify each IC based on the sign of \( \sigma_1 \): negative values indicate stable fixed points, zero values correspond to stable limit cycles, and positive values denote chaotic orbits, which is our primary interest. The color coding in Fig.~\ref{fig6:Fig3a} differentiates between each attractor. The red regions (where \( \sigma_1 > 0 \)) represent chaotic attractors. The yellowish/orange areas (where \( \sigma_1 \approx 0 \)) indicate stable limit cycles, while dark blue regions (\( \sigma_1 < 0 \)) correspond to stable fixed points. It is important to note that there were no stable \(2D\) tori (for which two largest ftLEs asymptotically approach zero, as described in Sect.~\ref{section:LEs}) were observed under the considered parameter ranges.

To classify the different types of trajectories and attractors, we further analyzed the entire LE spectrum. The procedures we implement is the following. We first compute all three sets of ftLEs for the considered orbit in the parameter space \((r, b)\) of the \(3D\) Lorenz system \eqref{eq:3DODE}. Our objective is to classify each studied orbit in the parameter space as one converging to either a stable fixed point, a stable limit cycle, or a chaotic attractor. We further introduce a discrete quantity, denoted by \(\sigma^j\), to label all three attractors from numbers `1' to `3' according to the ftLEs (see Sect.~\ref{section:LEs}). More specifically,  \(\sigma^j =1\) represents chaotic attractors characterized by \(\sigma_1 > 0, \sigma_2 \le 0\) and \(\sigma_{3} < 0\); \(\sigma^j = 2\) indicates stable limit cycle attractors, which corresponds to \(\sigma_1 \approx 0\), while \(\sigma_2\), \(\sigma_3 < 0\); and lastly, \(\sigma^j = 3\) denotes the presence of stable fixed point attractors, which is associated with \(\sigma_j < 0\), for all \(j=1,2,3\). 

In Fig.~\ref{fig6:Fig3b}, we present the same parametric space of Fig.~\ref{fig6:Fig3a} using a three-color scale, labeled as \( \sigma^j \). Each color corresponds to a specific dynamical regime: chaotic attractors (\( \sigma^j = 1 \), red region), stable limit cycles (\( \sigma^j = 2 \), yellowish/orange region), and stable fixed points (\( \sigma^j = 3 \), blue region). Fig.~\ref{fig6:Fig3b} provides a clear visualization of how the LE spectrum classifies distinct types of dynamical behaviors in the parameter space. This analysis will be invaluable when we examine the dynamics of hyperchaotic systems \eqref{eq:4DODE} and \eqref{eq:3DHenMap}, which involve two positive LEs, in the following sections. 

We note that the color plots of the ICs based on the values of the ftmLEs [Fig.~\ref{fig6:Fig3a}] and the three ftLEs [Fig.~\ref{fig6:Fig3a}], are very similar. This suggests that the use of the ftmLE is sufficient in determining the orbits' nature for the three types of orbits (i.e., stable fixed point, stable limit cycle, and chaotic attractors) and parameter values we considered in our analysis of the \(3D\) Lorenz model \eqref{eq:3DODE}. 

Figure \ref{fig6:Fig3c} provides a similar analysis to the one done in Fig.~\ref{fig6:Fig3a} using the computed GALI$_2$ values. For small \( r \) values (in the leftmost region), the GALI$_2$ reaches non-zero positive values (blue area) associated with the presence of stable fixed point attractors. The existence and location of this area align well with the blue regions observed in Figs.~\ref{fig6:Fig3a} and (b). On the other hand, the GALI$_2$ is practically zero for all the other types of orbits: stable limit cycles and chaotic attractors [dark red regions in Fig.~\ref{fig6:Fig3c}]. This behavior points out the method's limitation in distinguishing between stable limit cycles and chaotic attractors in the system \eqref{eq:3DODE}. 

\subsection{Numerical investigation of the $4D$ hyperchaotic Lorenz system} \label{sec:4DODE}
In order to investigate the performance of the GALI method to hyperchaotic systems, we focus on the \(4D\) hyperchaotic Lorenz model \eqref{eq:4DODE}. This model can exhibit two positive LEs, a behavior which results in to more complex dynamics than the one observed in its \(3D\) counterpart \eqref{eq:3DODE}. In this section, we conduct a similar analysis to the one performed in Sect.~\ref{sec:3DODE}, focusing on representative orbit types, namely stable fixed point, stable PO, chaotic and hyperchaotic attractors for the \(4D\) system. 

\subsubsection{A stable fixed point case} \label{sec:4DODE fixed point}

\begin{figure}[!htbp]
    \centering
    \subfloat[The \((x,y,z)\) projection\label{fig6:Fig4a}]{\includegraphics[width=0.33\textwidth]{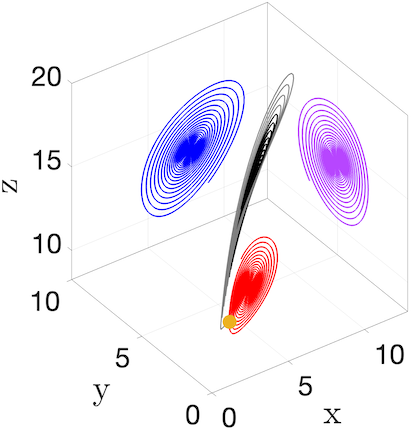}}
    \subfloat[The \((x,y,w)\) projection\label{fig6:Fig4b}]{\includegraphics[width=0.33\textwidth]{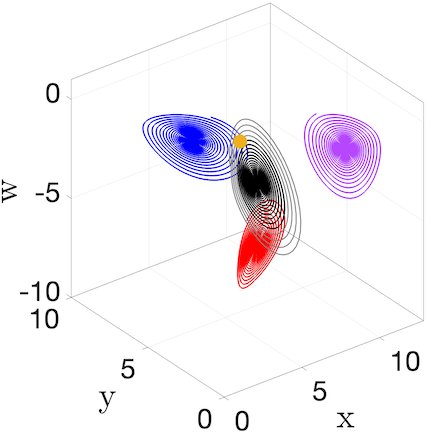}}
    \subfloat[The \((w,y,z)\) projection\label{fig6:Fig4c}]{\includegraphics[width=0.33\textwidth]{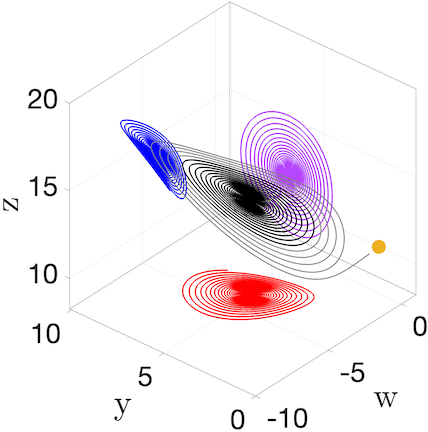}}\\
    \subfloat[ftLEs\((t)\)\label{fig6:Fig4d}]{\includegraphics[width=0.45\textwidth]{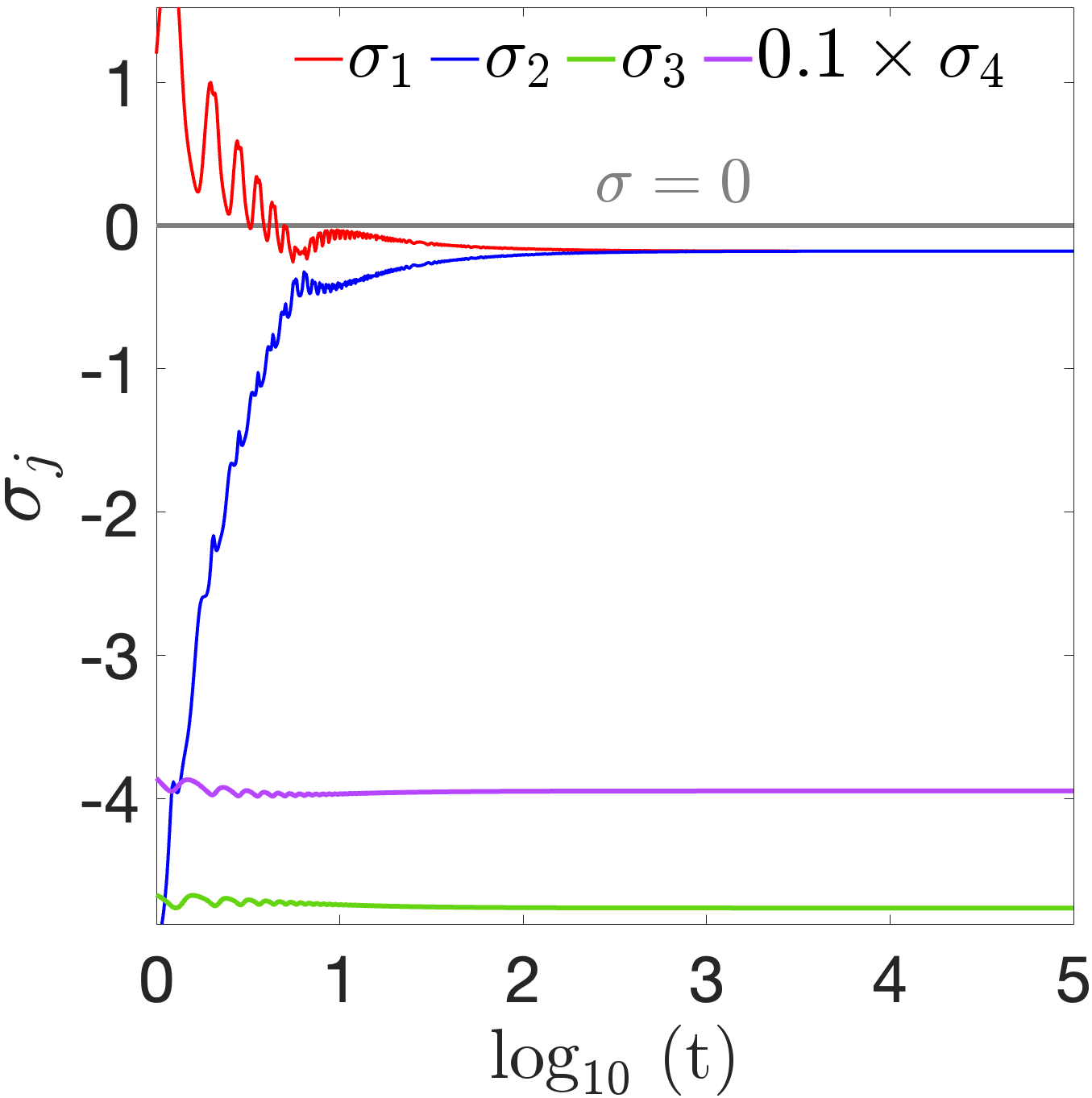}}
    \subfloat[GALI\(_{k} (t)\)\label{fig6:Fig4e}]{\includegraphics[width=0.455\textwidth]{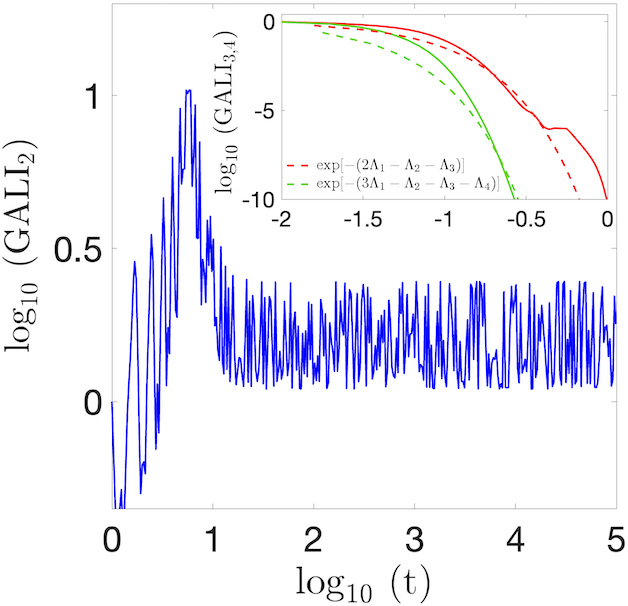}}
\caption{[(a), (b), and (c)] All possible \(3D\) projections of the \(4D\) hyperchaotic Lorenz systems \eqref{eq:4DODE} \(4D\) phase space for an orbit with IC \((x, y, z, w) = (3, 2, 10, 1)\) (orange circle points) which is approaching to a stable fixed point attractor. The black curves correspond to the final (asymptotic) phase, the gray curves show the initial evolution of the trajectory, and the colored curves depict the corresponding \(2D\) projections. The parameters of the system are fixed to $a=35$, $b=8/3$, $c=55$ and $r=-12$. (d) The time evolution of the four orbit's ftLEs, $\sigma_j$ $(j=1,2,3,4)$. (f) The time evolution of GALI$_{2}$ (blue curve), GALI$_{3}$ (red curve, inset plot), and GALI$_{4}$ (green curve, inset plot) of the same orbit. Both GALI$_{3}$ and GALI$_{4}$ decay to zero exponentially following the laws (dashed curves) given in \eqref{eq:GALI}, where the values \(\Lambda_1 = -1.23\), \(\Lambda_2 = -1.23\), \(\Lambda_3 = -11.94\), and \(\Lambda_4 = -50.66\) obtained from the results of panel (d) are used as good approximations of the LEs. For visualization purposes, in panel (d), the purple curve, $\sigma_4$, is rescaled, and we draw a horizontal gray line to represent $\sigma = 0$.}
  \label{fig6:Fig4}
\end{figure}

Let us begin by analyzing the \(4D\) Lorenz system \eqref{eq:4DODE} for the IC \((x, y, z, w) = (3, 2, 10, 1)\) and parameters \(a = 35\), \(b = 8/3\), \(c = 2\), and \(r = -12\). Since the system involves four state variables \((x, y, z, w)\), visualizing the entire \(4D\) phase space is challenging. In order to address this, we present all possible \(3D\) projections of the phase space: \((x, y, z)\), \((x, y, w)\), and \((y, z, w)\) in Figs.~\ref{fig6:Fig4a}, (b) and (c), respectively. These plots depict the evolution of the trajectory (black curves) in the \(3D\) projections of the system's \(4D\) phase space, while the orbit's IC is indicated by an orange circle point in each panel. Each \(3D\) plot also shows the corresponding possible \(2D\) projections [blue, red and purple curves in Figs.~\ref{fig6:Fig4a}, (b) and (c)] in the respective planes, similar to the arrangement of Fig.~\ref{fig6:Fig1}. In all \(3D\) subspace portraits, we observe that the orbit starting from the orange point evolves and eventually converges to the stable fixed point attractor at \((x^*, y^*, z^*, w^*) = (7.141, 5.129, 13.733, -3.923)\).

Also, similar to what was done in Sect.~\ref{sec:3DODE}, we computed the time evolution of the ftLEs, $\sigma_j$ for $j=1,2,3,4$, [Fig.~\ref{fig6:Fig4d}] as well as the evolution of the GALI$_k$ for $k=2,3$ [Fig.~\ref{fig6:Fig4e}] and $k=4$ [inset of Fig.~\ref{fig6:Fig4e}], for the same IC. We computed both the four ftLEs and the GALI\(_k\)'s up to $t=10^5$ time units. From the results of Fig.~\ref{fig6:Fig4d}, we see that all ftLEs become negative (they are all located below the horizontal gray line, $\sigma = 0$), as expected for stable fixed point attractors. All the LEs being negative denotes the contraction of the \(4D\) phase space volume. Furthermore, the two largest ftLEs, $\sigma_1$ (red curve) and $\sigma_2$ (blue curve), eventually converge to the same negative value \(\sigma_1 \approx \sigma_2 = -1.23\), suggesting a uniform contraction along the directions associated with the two largest LEs. Note that $\sigma_4$ [purple curve in Fig.~\ref{fig6:Fig4d}], attains very negative values, is scaled for visualization purposes so that its curve is shown close to the ones of the other exponents. 

In Fig.~\ref{fig6:Fig4e}, we see that the GALI$_2$ (blue curve) eventually fluctuates around a constant positive value, while the GALI$_3$ (red curve, inset plot) and GALI$_4$ (green curve, inset plot) decay to zero exponentially fast. As in the case discussed in Sect.~\ref{sec:3DODE}, the GALI$_2$ fluctuates around a positive constant value due to the fact that $\sigma_1 \approx \sigma_2$. The dashed curves in the inset of Fig.~\ref{fig6:Fig4e} represent the function of the forms corresponding to the theoretical expressions \eqref{eq:GALI_chaos} for GALI$_3$ and GALI$_4$, respectively. More specifically, we have \( \text{GALI}_3 \propto \exp[-(2\Lambda_1 - \Lambda_2 - \Lambda_3)]\) and \(\text{GALI}_4 \propto \exp[-(3\Lambda_1 - \Lambda_2 - \Lambda_3 - \Lambda_4)]\), where \(\Lambda_1 = -1.23\), \(\Lambda_2 = -1.23\), \(\Lambda_3 = -11.94\), and \(\Lambda_4 = -50.66\). These values were obtained from the results of Fig.~\ref{fig6:Fig4d} as good approximations of the LEs $\sigma_j$, $j=1,2,3,4$. 

\subsubsection{A stable limit cycle case}
\begin{figure}[!htbp]
    \centering
    \subfloat[The \((x,y,z)\) projection\label{fig6:Fig5a}]{\includegraphics[width=0.33\textwidth]{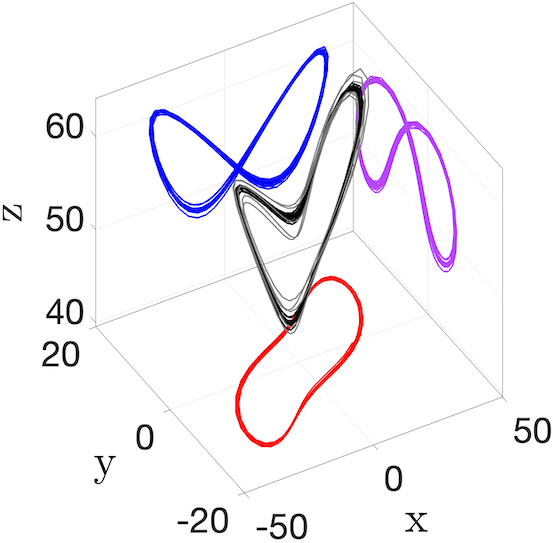}}
    \subfloat[The \((x,y,w)\) projection\label{fig6:Fig5b}]{\includegraphics[width=0.33\textwidth]{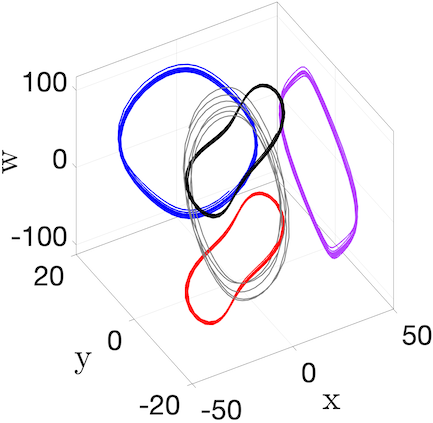}}
    \subfloat[The \((w,y,z)\) projection\label{fig6:Fig5c}]{\includegraphics[width=0.33\textwidth]{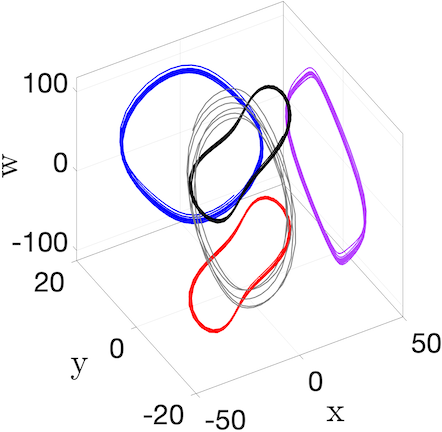}}\\
    \subfloat[ftLEs\((t)\)\label{fig6:Fig5d}]{\includegraphics[width=0.45\textwidth]{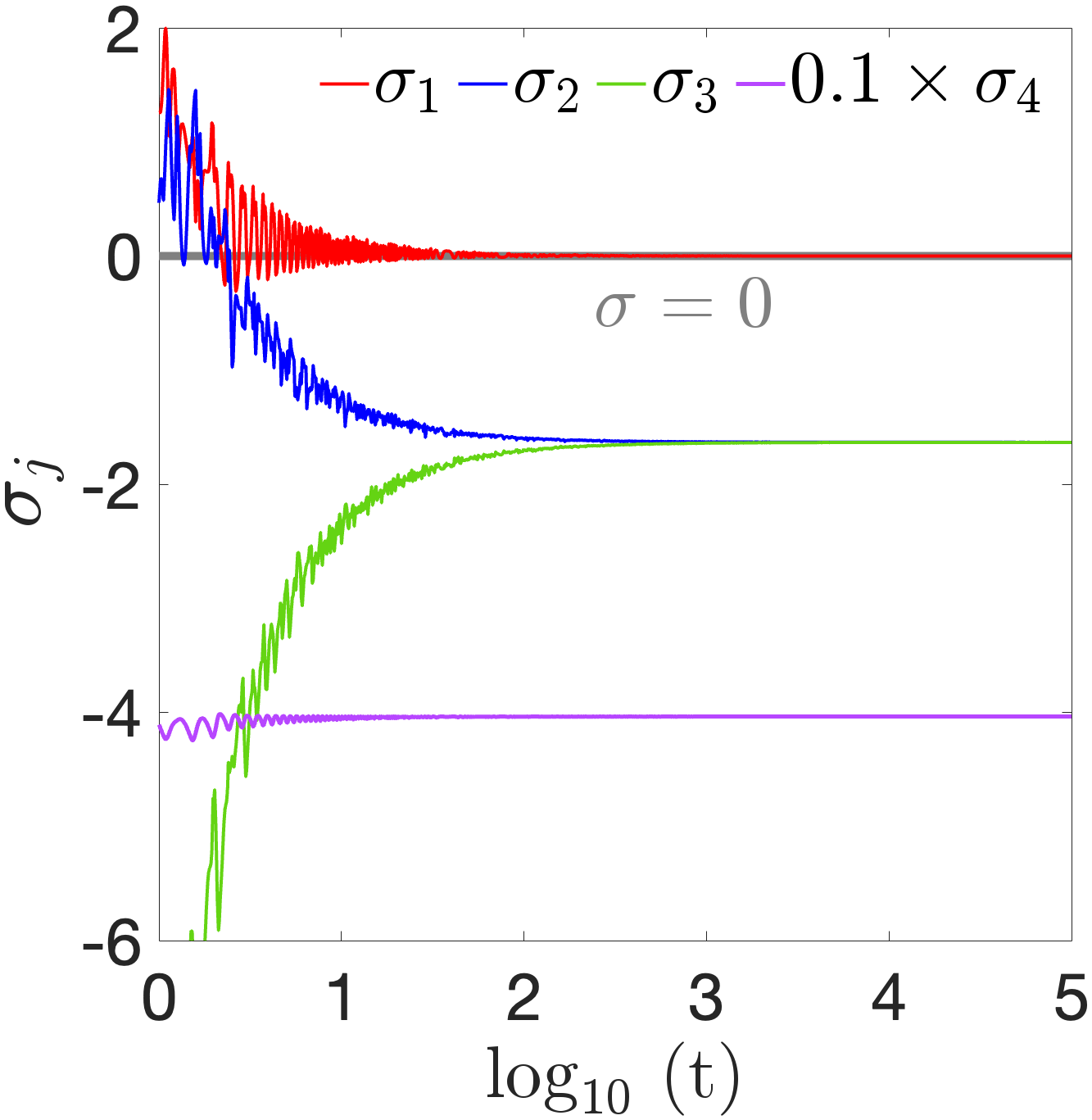}}
    \subfloat[GALI\(_{k} (t)\)\label{fig6:Fig5e}]{\includegraphics[width=0.455\textwidth]{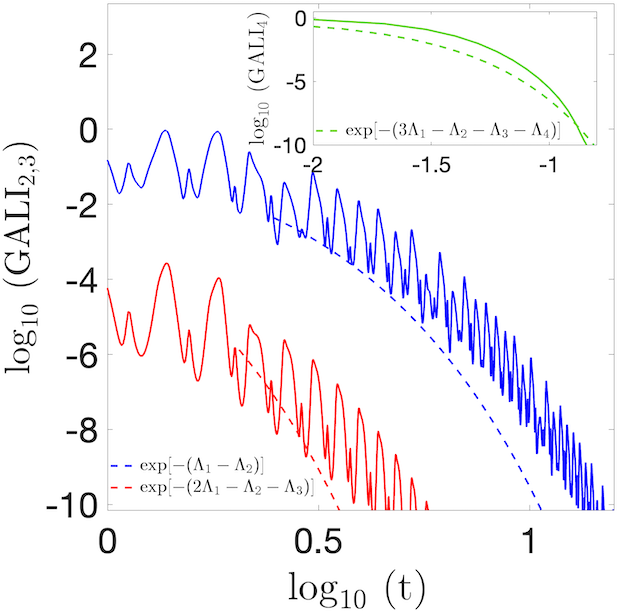}}
\caption{Similar to Fig.~\ref{fig6:Fig4} but for an orbit of the \(4D\) hyperchaotic Lorenz system which leads to a stable limit cycle with $r=-5$. The orbits IC is \((x, y, z, w) = (3, 2, 10, 1)\).  In (c),  GALI$_{2}$ (blue curve), GALI$_{3}$ (red curve) and GALI$_{4}$ (green curve, inset plot), decay to zero exponentially fast following the theoretical predicted laws \eqref{eq:GALI_chaos} (dashed curves), which are also given in the panel's legend for \(\Lambda_1 = 0\), \(\Lambda_2 = -1.63\), \(\Lambda_3 = -1.63\), and \(\Lambda_4 = -40.37\).}
  \label{fig6:Fig5}
\end{figure}

To examine a stable limit cycle attractor in the \(4D\) Lorenz system \eqref{eq:4DODE}, we set \(r = -5\) while keeping the other parameters, as well as the studied orbit's IC the same as in Sect.~\ref{sec:4DODE fixed point}. Figs.~\ref{fig6:Fig5a}, (b), and (c) present all possible \(3D\) projections of the \(4D\) phase space for this orbit, where we observe that after an initial transient phase (corresponding to gray colored part of the orbit) a convergence to closed loops (black curves) in the three \(3D\) subspaces.

The time evolution of the four ftLEs and the three GALIs is shown in Figs.~\ref{fig6:Fig5d} and (f), respectively. The largest ftLE, \(\sigma_1\) [red curve in Fig.~\ref{fig6:Fig4d}], eventually converges to zero, while the remaining exponents saturate at negative constant values. This behavior indicates that the attractor is a stable limit cycle, similar to what was seen in Fig.~\ref{fig6:Fig1e}. All GALI$_k$ indices with the GALI$_2$ and the GALI$_3$ depicted by the blue and red curves in the main Fig.~\ref{fig6:Fig1e} and the GALI$_4$ denoted by the green curve in the inset of Fig.~\ref{fig6:Fig4e} decay to zero exponentially fast. These decays are proportional to the functions described by \(\exp \left[ -\left( \Lambda_1 - \Lambda_2 \right) \right]\), \(\exp \left[ -\left( 2 \Lambda_1 - \Lambda_2 - \Lambda_3 \right) \right]\), and \(\exp \left[ -\left( 3 \Lambda_1 - \Lambda_2 - \Lambda_3 - \Lambda_4 \right) \right]\), respectively, with \(\Lambda_1 = 0\), \(\Lambda_2 = -1.63\), \(\Lambda_3 = -1.63\), and \(\Lambda_4 = -40.37\), being good approximations of the orbit's LEs. 

\subsubsection{A chaotic (strange) attractor case}
\begin{figure}[!htbp]
    \centering
    \subfloat[The \((x,y,z)\) projection\label{fig6:Fig6a}]{\includegraphics[width=0.325\textwidth]{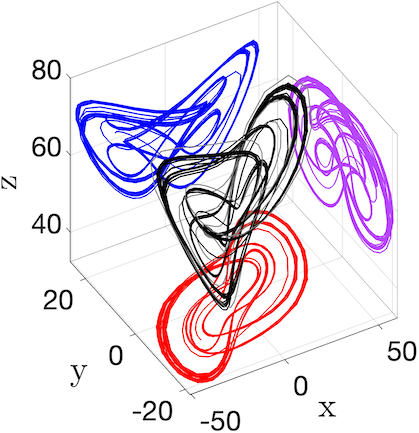}}
    \subfloat[The \((x,y,w)\) projection\label{fig6:Fig6b}]{\includegraphics[width=0.325\textwidth]{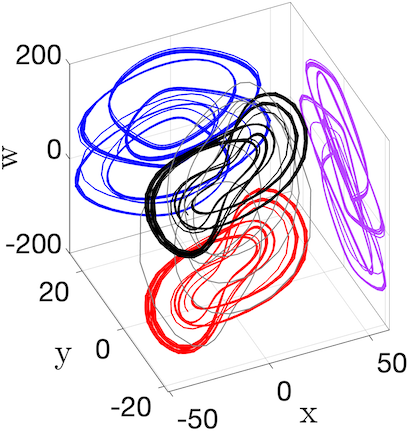}}
    \subfloat[The \((w,y,z)\) projection\label{fig6:Fig6c}]{\includegraphics[width=0.34\textwidth]{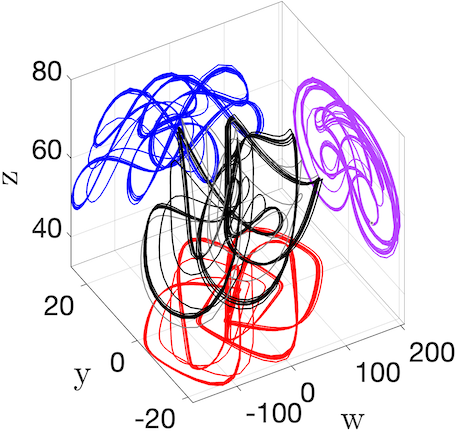}}\\
    \subfloat[ftLEs\((t)\)\label{fig6:Fig6d}]{\includegraphics[width=0.45\textwidth]{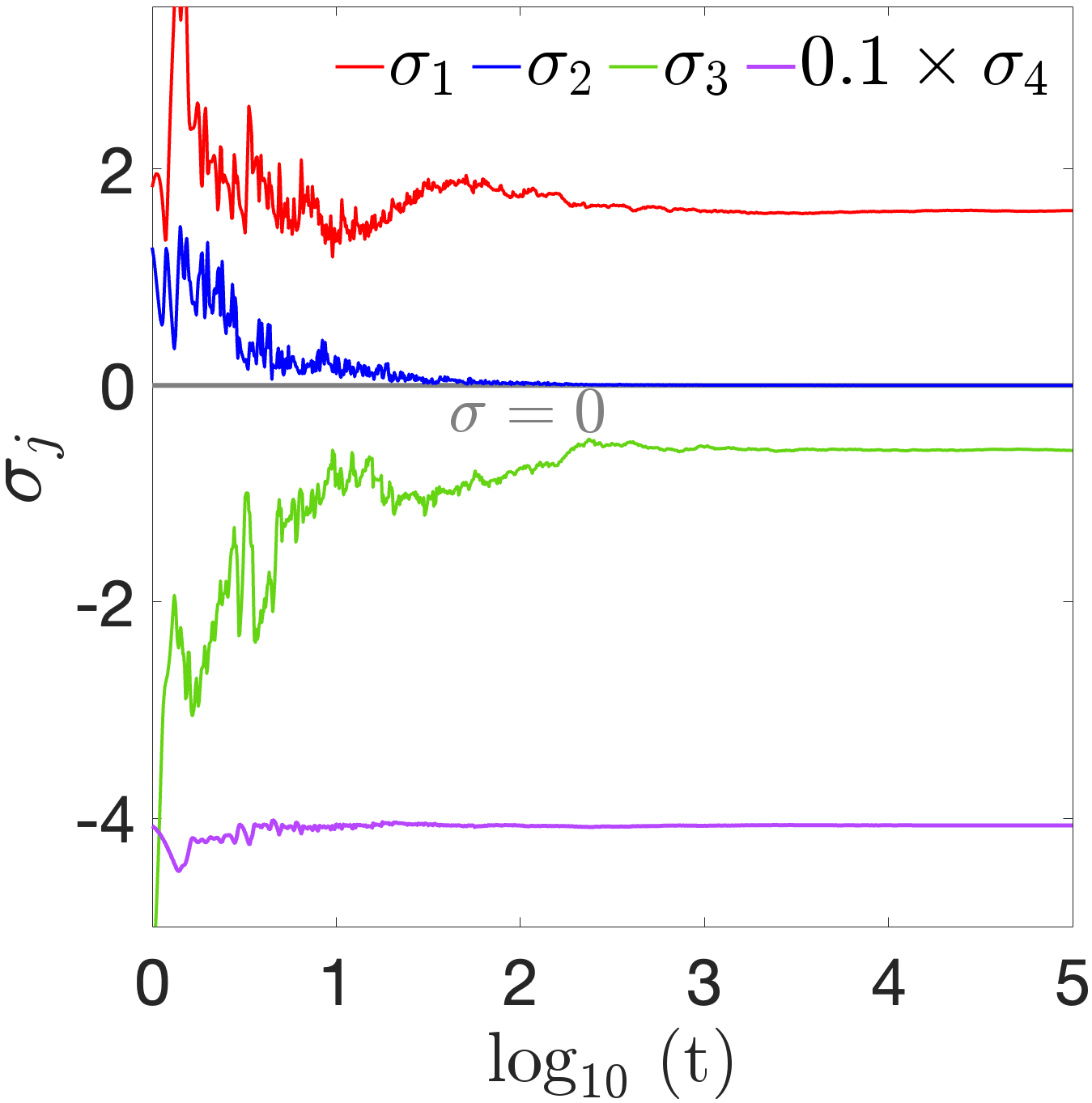}}
    \subfloat[GALI\(_{k} (t)\)\label{fig6:Fig6e}]{\includegraphics[width=0.455\textwidth]{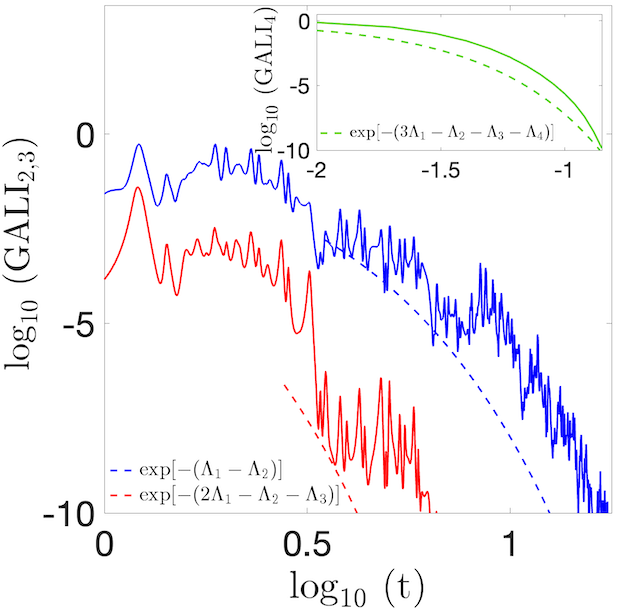}}
\caption{Similar to Fig.~\ref{fig6:Fig4} but for an orbit of the \(4D\) hyperchaotic Lorenz system which leads to a chaotic (strange) attractor with $r = -1$.  In (c),  GALI$_{2}$ (blue curve), GALI$_{3}$ (red curve) and GALI$_{4}$ (green curve, inset plot), decay to zero exponentially fast following the theoretical predicted laws \eqref{eq:GALI_chaos} (dashed curves), which are also given in the panel's legend for \(\Lambda_1 = 1.60\), \(\Lambda_2 = 0\), \(\Lambda_3 = -0.59\), and \(\Lambda_4 = -40.64\).}
  \label{fig6:Fig6}
\end{figure}

By setting \(r = -1\), while keeping all other parameters and the orbit's IC the same to the one used in Sect.~\ref{sec:4DODE fixed point}, the \(4D\) system \eqref{eq:4DODE} exhibits a chaotic attractor. Figs.~\ref{fig6:Fig6a}, (b) and (c) display the evolution of this orbit on all possible \(3D\) projections of the system's \(4D\) phase space. These plots reveal a structure which remains confined to a bounded region of the phase pace, a characteristic which has also been observed for the strange attractor of the \(3D\) model in Fig.~\ref{fig6:Fig1g}. This confined structure is more visually clear when we look at the associated \(2D\) projections (blue, red, and purple curves). 

Among all ftLEs [Fig.~\ref{fig6:Fig6d}], only \(\sigma_1\) (red curve) remain a positive asymptotically reaching constant value, \(\sigma_2\) tends to zero, while \(\sigma_3\) and \(\sigma_4\) are negative. One positive and one zero LEs indicate the motion takes place on a strange attractor (see Sects.~\ref{section:LEs} and \ref{sec:3DODE chaotic}).  Fig.~\ref{fig6:Fig6e} shows the exponential decay of GALI$_k$, \(k = 2, 3, 4\) for this orbit on the strange attractor. All GALIs decay to zero following the theoretical expectations of \eqref{eq:GALI_chaos}: \(
\text{GALI}_2 \propto \exp[-(\Lambda_1 - \Lambda_2)]\),  \(\text{GALI}_3 \propto \exp[-(2\Lambda_1 - \Lambda_2 - \Lambda_3)]\), and \(\text{GALI}_4 \propto \exp[-(3\Lambda_1 - \Lambda_2 - \Lambda_3 - \Lambda_4)]\), with \(\Lambda_1 = 1.60\), \(\Lambda_2 = 0\), \(\Lambda_3 = -0.59\), and \(\Lambda_4 = -40.64\), where these values are obtained from Fig.~\ref{fig6:Fig6d}.

\subsubsection{A hyperchaotic attractor case}
\begin{figure}[!htbp]
    \centering
    \subfloat[The \((x,y,z)\) projection\label{fig6:Fig7a}]{\includegraphics[width=0.33\textwidth]{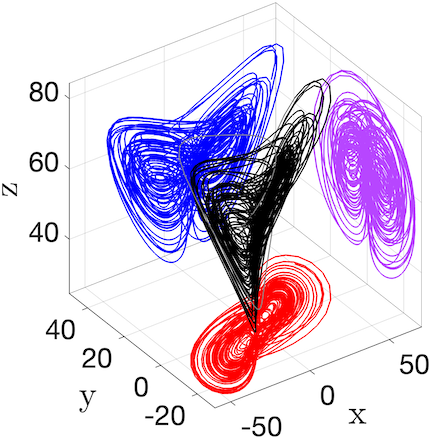}}
    \subfloat[The \((x,y,w)\) projection\label{fig6:Fig7b}]{\includegraphics[width=0.33\textwidth]{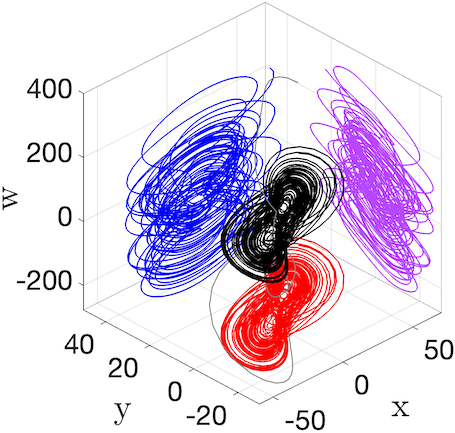}}
    \subfloat[The \((w,y,z)\) projection\label{fig6:Fig7c}]{\includegraphics[width=0.33\textwidth]{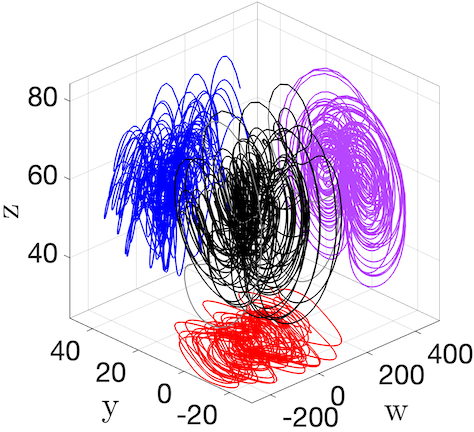}}\\
    \subfloat[ftLEs\((t)\)\label{fig6:Fig7d}]{\includegraphics[width=0.45\textwidth]{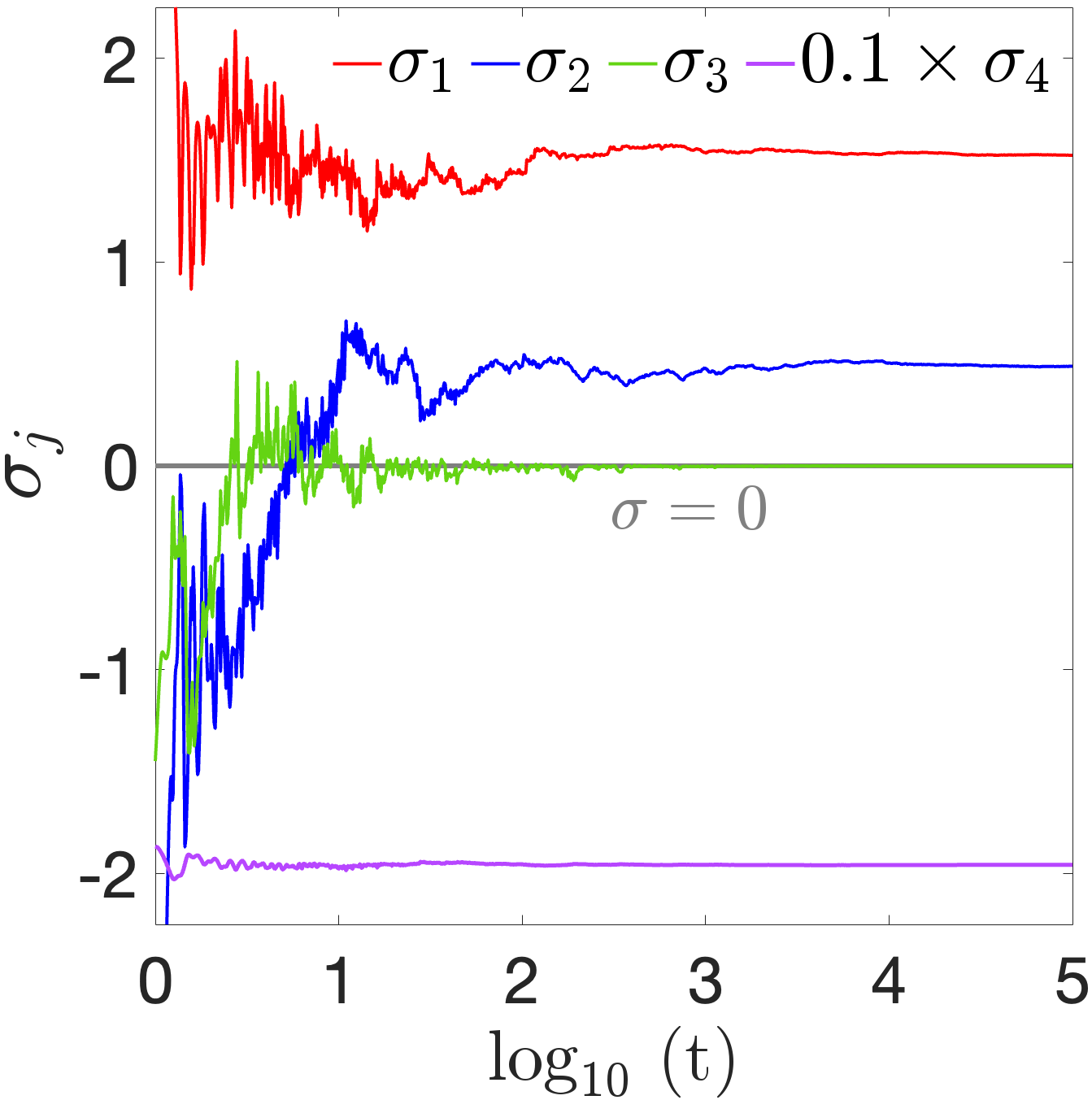}}
    \subfloat[GALI\(_{k} (t)\)\label{fig6:Fig7e}]{\includegraphics[width=0.455\textwidth]{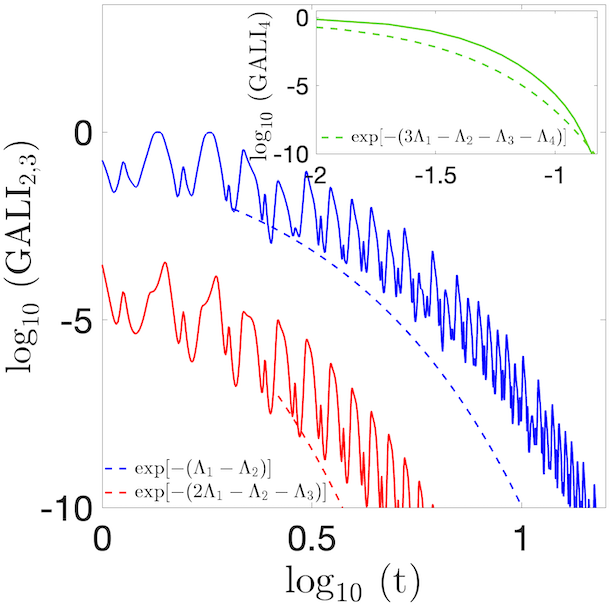}}
\caption{Similar to Fig.~\ref{fig6:Fig4} but for an orbit of the \(4D\) hyperchaotic Lorenz system \eqref{eq:4DODE} which leads to a hyperchaotic attractor with $r = 1.5$.  In (c),  GALI$_{2}$ (blue curve), GALI$_{3}$ (red curve) and GALI$_{4}$ (green curve, inset plot), decay to zero exponentially fast following the theoretical predicted laws \eqref{eq:GALI_chaos} (dashed curves), which are also given in the panel's legend for \(\Lambda_1 = 1.53\), \(\Lambda_2 = 0.51\), \(\Lambda_3 = 0\), and \(\Lambda_4 = -39.19\).}
  \label{fig6:Fig7}
\end{figure}

Lastly, let us analyze the behavior of the GALI method for a hyperchaotic attractor of the \(4D\) system \eqref{eq:4DODE}. To the best of our knowledge, this will be the first computation of the GALI method for this type of attractor in the literature. To create a hyperchaotic attractor, we set \(r = 1.5\) while keeping all the other parameter values as well as the orbit's IC the same to those in the three previous cases. Figs.~\ref{fig6:Fig7a}, (b), and (c) present the \(3D\) projections of the orbit tending to a hyperchaotic attractor. Similar to the strange chaotic attractor case in Fig.~\ref{fig6:Fig6a}, the orbit's projections are confined in specific regions of the phase space. We can also see this more clearly in the corresponding \(2D\) projections of the various \(3D\) subspace. 

Figures \ref{fig6:Fig6d} and (e) depict the time evolution of the four ftLEs, \(\sigma_j\), \(j = 1, 2, 3, 4\), and the GALI$_k$ for \(k = 2, 3, 4\), respectively. The two largest ftLEs, \(\sigma_1, \sigma_2\) [respectively, the red and blue curve in Fig.~\ref{fig6:Fig7d}], attain positive values, eventually tending to \(\sigma_1 = 1.53\) and \(\sigma_2 = 0.51\), while the remaining LEs eventually tend to \(\sigma_3 = 0\) and \(\sigma_4 = -39.19\). The presence of two positive LEs indicate that the orbit exhibits hyperchaotic motion. 

Fig.~\ref{fig6:Fig7e} shows the exponential decay of GALI$_k$, for \(k = 2, 3, 4\) for this orbit on the hyperchaotic attractor. All GALIs decay to zero, following the theoretical expectations of \eqref{eq:GALI_chaos}, i.e.,  \(\exp[-(\Lambda_1 - \Lambda_2)]\),  \(\exp[-(2\Lambda_1 - \Lambda_2 - \Lambda_3)]\), and \(\quad \exp[-(3\Lambda_1 - \Lambda_2 - \Lambda_3 - \Lambda_4)]\), respectively, where \(\Lambda_1 = 1.53\), \(\Lambda_2 = 0.51\), \(\Lambda_3 = 0\), and \(\Lambda_4 = -39.19\). It is worth mentioning that all the GALI indices decay to zero exponentially fast for the stable limit cycle case [Fig.~\ref{fig6:Fig5e}], as well as for both the chaotic (strange) [Fig.~\ref{fig6:Fig6e}] and hyperchaotic [Fig.~\ref{fig6:Fig7e}] attractors. Thus, the GALI method cannot be used to discriminate between these cases. 

\subsubsection{Parametric exploration of the \(4D\) Lorenz system using the GALI method and LEs}
\begin{figure}[!htb]
    \centering
 \includegraphics[width=0.5\textwidth]{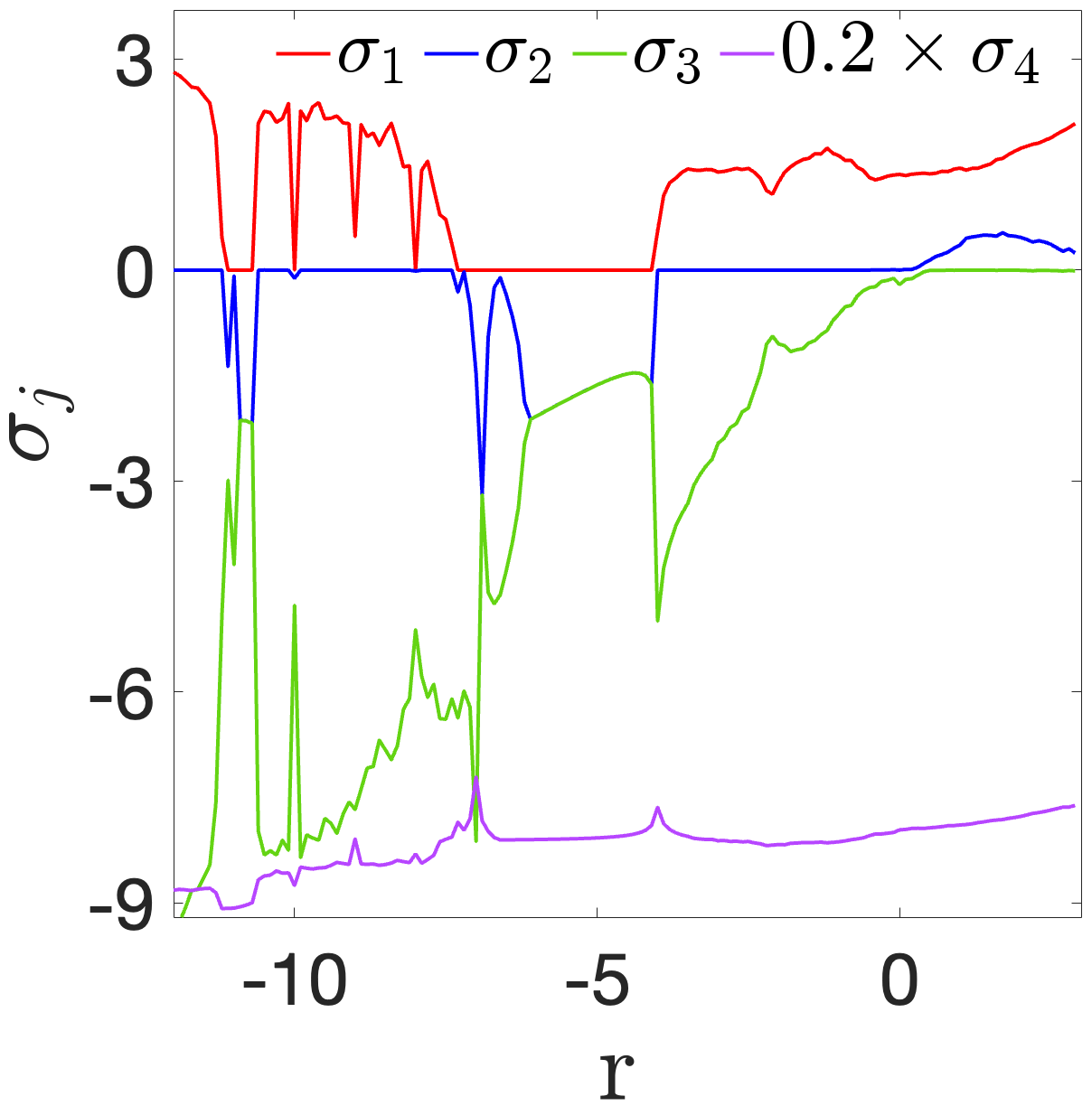}
	\caption{A parametric exploration of the four ftLEs values (\(\sigma_1 > \sigma_2 > \sigma_3 > \sigma_4\) \eqref{eq:LEs order}) of the \(4D\) hyperchaotic Lorenz system \eqref{eq:4DODE} for varying values of \(r\) when all other parameters are fixed to $a=35$, $b=\frac{8}{3}$, and $c=55$. Results are obtained for the orbit with IC \((x, y, z, w) = (2, 1, 5, 1)\). We consider a total of \(600\) equidistant values for \( r \in [-12, 3] \) and the ftLEs are computed at \(t=10^4\) time units.}
\label{fig6:Fig8}
\end{figure}

To investigate the influence of a single parameter on the dynamics of the \(4D\) hyperchaotic Lorenz system, we conduct a parametric study of the four ftLEs by varying \( r \) in the interval \([-12, 3]\), while keeping the other parameters of the system fixed to $a=35$, $b=\frac{8}{3}$, and $c=55$, and studying the behavior of  the orbit with IC \((x, y, z, w) = (2, 1, 5, 1)\). Fig.~\ref{fig6:Fig8} illustrates the computed four ftLEs values \(\sigma_1 > \sigma_2 > \sigma_3 > \sigma_4\) \eqref{eq:LEs order} for \(1011\) values of \(r\). Our main observations of results of Fig.~\ref{fig6:Fig8} are: 

\begin{enumerate}[label=\textnormal{(\Roman*)}]
    \item For values of \( r \) in the range \([-12, -11]\), the studied orbit exhibits a positive ftmLE \( \sigma_1 > 0 \) (red curve), while the second ftLE \( \sigma_2 = 0 \) (blue curve), and the other two exponents \(\sigma_3\) (green curve) and \(\sigma_4\) (purple curve) eventually converge to negative values. This indicates the existence of strange attractors in this small range of \(r\) values, which lead to chaotic motion. 
    \item Increasing \( r \) to take values in the interval \((-11, -10.65]\) results in \( \sigma_1 \) approaching zero, while the remaining ftLEs become negative (i.e., \(\sigma_j < 0\) for \(j=2,3,4\)),  implying that the system displays stable limit cycles for these values.
    \item For \( r \in  (-10.65, -7.4]\), the system again exhibits chaotic (strange) attractors, characterized by \( \sigma_1 > 0 \),  \( \sigma_2 = 0 \), and \( \sigma_3, \sigma_4 < 0 \). 
    \item For \( r \) values in the range \((-7.4, -4]\) and  \((-4, 0.65]\), the system shows stable limit cycles and chaotic (strange) attractors, respectively.
    \item Finally, for relatively larger positive values of \(r\) in the interval \((0.65, 3]\), hyperchaotic attractors occur, characterized by the two largest ftLEs being positive (\( \sigma_1, \sigma_2 > 0 \)).
\end{enumerate}

Performing the computation of the GALI$_2$ for the same orbits of the \(4D\) system \eqref{eq:4DODE} when \(r\) varies in the interval \([-12, 3]\), i.e., conducting an analysis similar to the one done for the \(3D\) system in Fig.~\ref{fig6:Fig2b}, we find that the GALI$_2$ values remain practically zero for all tested \( r \) values. This decay of the GALI\(_2\) to zero occurs because no stable fixed point attractors with two equal ftLEs are observed for the tested \(r\) values. The GALI$_3$ and the GALI$_4$ are also eventually become zero. However, all the GALI indices will go to zero at different rates, as demonstrated by the representative cases from Fig.~\ref{fig6:Fig4e} to Fig.~\ref{fig6:Fig7e}, as their decrease rate depends on different combinations of the LEs values. We chose not to present the parametric exploration for the GALI indices, as it would trivially show zero values for all \(r\) values, something which eventually does not allow the discrimination between the different types of observed attractors.

To further investigate the dynamics of the system, we performed a bi-parametric exploration of the \(4D\) hyperchaotic Lorenz system \eqref{eq:4DODE}, varying both parameters \( r \) and \( c \) in the ranges \( r \in [-12, 1] \) and \( c \in [1, 55] \), while keeping \( a = 35\), \( b = \frac{8}{3}\), and considering again the orbit with IC \((x, y, z, w) = (3, 2, 10, 1)\), performing in this way a study similar to the one presented in Fig.~\ref{fig6:Fig3} for the \(3D\) system. 

\begin{figure}[!htb]
    \centering
    \subfloat[ftmLEs\label{fig6:Fig9a}]{\includegraphics[width=0.345\textwidth]{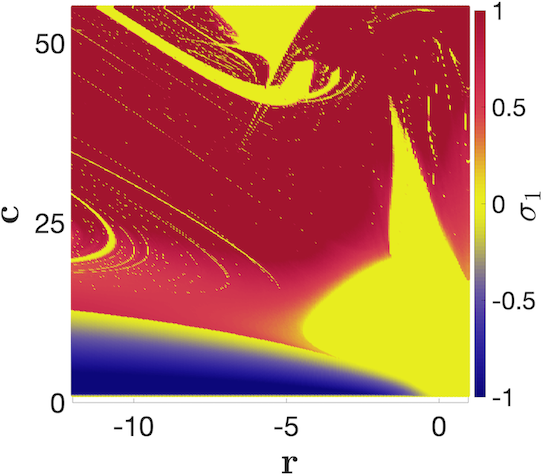}}
    \subfloat[Classification based on the three ftLEs\label{fig6:Fig9b}]{\includegraphics[width=0.325\textwidth]{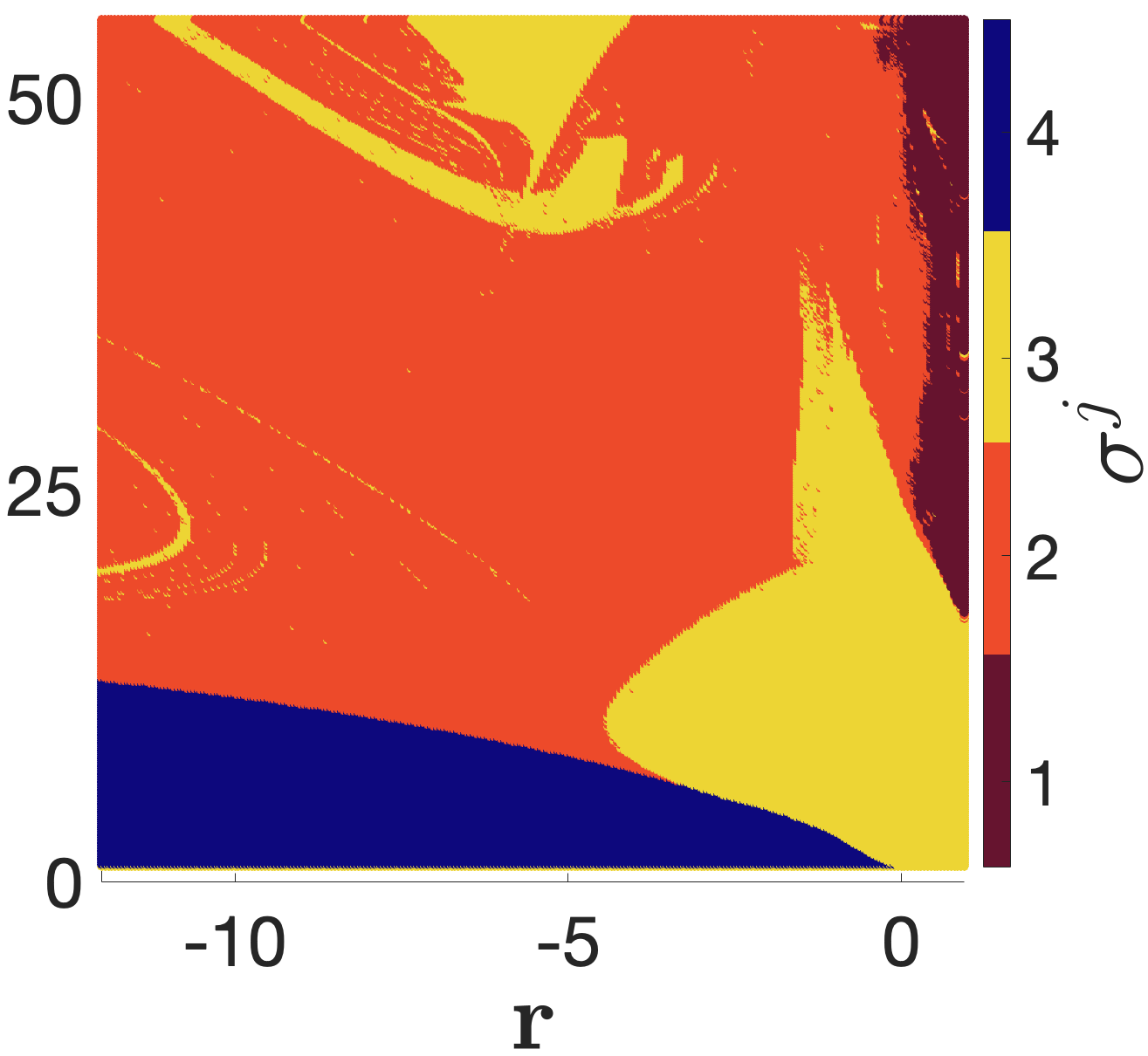}}
    \subfloat[GALI\(_{2}\)\label{fig6:Fig9c}]{\includegraphics[width=0.33\textwidth]{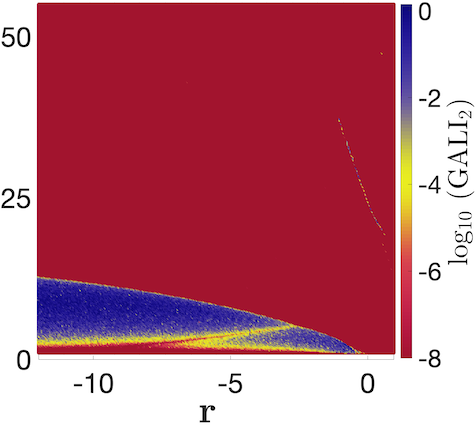}}\\
    \caption{An exploration of the \( (r, c) \) space of the hyperchaotic Lorenz system \eqref{eq:4DODE} with \( a = 35 \) and $b=\frac{8}{3}$. To produce these color plots, we used a grid of \( 590 \times 260 = 153,400 \) points in the region \( r \in [-12, 1] \) and \( c \in [1, 55] \) by integrating the IC \((x, y, z, w) = (3, 2, 10, 1)\) up to \( t = 10^4 \) and according to the orbit's ftLEs \(\sigma_1 > \sigma_2 > \sigma_3 > \sigma_4\) \eqref{eq:LEs order} and the GALIs \eqref{eq:GALI}. The classification of various dynamical regimes is based on the values of (a) the ftmLE, \( \sigma_1 \), and (b) the four ftLEs, \(\sigma^j\), according to the classification in Sect.~\ref{section:LEs}. In both panels, stable fixed point attractors are shown in dark blue \( (\sigma_1 < 0 \text{ or } \sigma^j = 4) \), and stable periodic motion (limit cycles) are represented by the yellowish/orange areas \( (\sigma_1 \approx 0 \text{ or } \sigma^j = 3) \). Contrary, in (a), red/brown regions represent either chaotic or hyperchaotic attractors (\(\sigma_1 > 1 \)), while in (b) brown and light red regions, respectively, denoted chaotic (\(\sigma^j = 2\)) and hyperchaotic (\(\sigma^j = 1\)) attractors. (c) A similar classification is performed using GALI$_2$ values. Note that in (a), the computed \(\sigma_1\) values are scaled to the interval \([-1, 1]\).}
\label{fig6:Fig9}
\end{figure}

Figure \ref{fig6:Fig9a} depicts the parametric space \((r, c)\) with each point being colored according to the value of the ftmLE, \(\sigma_1\) scaled to the range of \([-1, 1]\), similar to what was done in Fig.~\ref{fig6:Fig3a}. The color coding reveals different types of attractors: yellow/orange points indicate the existence of stable limit cycles or potentially periodic motions (\(\sigma_1 = 0\)), purple/dark blue regions denote parameter values for which stable fixed point attractors exist (\(\sigma_1 < 0\)), and brown/red regions show the presence of either chaotic or hyperchaotic attractors. Since \(\sigma_1 > 0\) for both chaotic and hyperchaotic attractors, the ftmLE value by itself is not sufficient to differentiate between them. We can address this issue by also evaluating the value of the second ftLE, \(\sigma_2\). In addition, we can also check the value of the other two ftLEs \(\sigma_3\) and \(\sigma4\). For instance, if the ftmLE \(\sigma_1\) is positive for an orbit, we can characterize it as belonging to a strange attractor if \(\sigma_{2} \approx 0 \) and \(\sigma_{3,4} < 0 \). Therefore, we find it a good practice to compute all four sets of ftLEs to correctly classify the orbits.  

The procedure we implement are the same as the one described in Fig.~\ref{fig6:Fig3b}, with the key difference being the introduction of an extra, hyperchaotic attractor. Initially, we compute the four sets of ftLEs for the considered orbit in \((r, c)\) parameter space of the \(4D\) hyperchaotic Lorenz system \eqref{eq:4DODE}. Our goal is to classify each studied orbit in the parameter space as converging to one of the four attractors: a stable fixed point, a stable limit cycle, a chaotic or a hyperchaotic attractor. Then, we label these attractors using a discrete quantity \(\sigma^j\) from numbers `1' to `4' according to the ftLEs (see Sect.~\ref{section:LEs}). In particular,  \(\sigma^j =1\) represents hyperchaotic attractors, characterized by the two largest ftLEs positive (\(\sigma_1\), \(\sigma_2 > 0\)); \(\sigma^j = 2\) indicates chaotic attractors characterized by \(\sigma_1 > 0, \sigma_2 \le 0\) and \(\sigma_3\), \(\sigma_4 < 0\); \(\sigma^j = 3\) is assigned to stable limit cycle attractors, corresponding to \(\sigma_1 \approx 0\), while \(\sigma_2\), \(\sigma_3\), \(\sigma_4 < 0\); and lastly \(\sigma^j = 4\) means the presence of stable fixed point attractors, where all ftLEs \(\sigma_j < 0\), for all \(j=1,2,3,4\)). The parametric space exploration based on this quantity is depicted in Fig.~\ref{fig6:Fig9b}. It is important to emphasize that there are somewhat related approaches reported in the literature. For instance, in \citep[Fig. 1]{barrio2015chaos}, a classification of different behaviors of the \(4D\) R\"{o}ssler model based on the values of the LEs was presented.

Brown regions in Fig.~\ref{fig6:Fig9b} represent areas of the parametric space where hyperchaotic attractors (\(\sigma^j = 1\)) exist. These attractors are characterized by \(\sigma_1 > 0\), \(\sigma_2 > 0\) and \(\sigma_{3, 4} < 0\). Light red regions correspond to the existence of chaotic attractors (\(\sigma^j = 2\)), for which \(\sigma_1 > 0\), while \(\sigma_{2,3,4} < 0\). Yellow regions denote the presence of stable limit cycles (\(\sigma^j = 3\)), with \(\sigma_1 \approx 0\) for which \(\sigma_{2,3,4} < 0\). Lastly, blue regions represent parameter values for which stable fixed point attractors (\(\sigma^j = 4\)) exist. In this case, all ftLEs have negative values. Fig.~\ref{fig6:Fig9c} presents a similar color plot of the parametric space, where points are colored according to the GALI$_2$ value of the orbits. In this representation, the purple and dark red colors indicate the existence of stable fixed attractors. However, the GALI$_2$ index does not effectively differentiate between stable limit cycle, chaotic and hyperchaotic motions.

\subsection{Numerical investigation of the generalized H{\'e}non map} \label{sec:3DHenMap}
In order to validate the generality of the behavior of the GALI method observed for continuous DSs, namely the \(3D\) and \(4D\) Lorenz models (Secs.~\ref{sec:3DODE} and \ref{sec:3DODE}, respectively), we now extend our investigation to discrete systems. In this section, we consider the generalized H{\'e}non map \eqref{eq:3DHenMap} as a representative example of such systems. This map will allow us to further explore the GALI method's behavior for hyperchaotic systems as it presents cases where hyperchaotic attractors exists. To do so, we conduct a similar analysis as in Sect.~\ref{sec:4DODE}. Note that, in all representative cases we studied, we consider the orbits with IC \(x = 0.5\), \(y = 0.4\), and \(z = 0.2\), and they are iterated up to \(n = 10^5\).

\subsubsection{A stable fixed point case}
We first present the \(3D\) phase space portrait of the map \eqref{eq:3DHenMap} with parameters \(a = 0.3\) and \(b = 0.5\), along with its \(2D\) projections for an orbit tending to a stable fixed attractor. Fig.~\ref{fig6:Fig10a} illustrates the portrait for the trajectory starting from the IC \((x, y, z) = (0.5, 0.4, 0.2)\) (orange circle point). The orbit spirals and converges towards the stable fixed point attractor \((x^*, y^*, z^*) = (0.3521, 0.3521, 0.3521)\). This stable attractor is located at the center of the spiral formed by the studied trajectories. The consequents of this trajectory are shown in Fig.~\ref{fig6:Fig10a} (sequence of black points where the gray points highlight the initial stages of the trajectory's evolution), while red, blue, and purple points denote the \((x, y)\), \((y, z)\), and \((x, z)\) projections of the orbit, respectively. 

Figures \ref{fig6:Fig10b} and (c) show the time evolution of the computed three ftLEs and the GALI$_{2}$ and GALI$_{3}$ indices, respectively, for the orbit of Fig.~\ref{fig6:Fig10a}. All ftLEs, namely  \(\sigma_1\) (red curve), \(\sigma_2\) (blue curve), and \(\sigma_3\) (green curve), converge to negative values; in particular, we get \(\sigma_1 = \sigma_2 = -0.02\) and \(\sigma_3 = -0.66\) [Fig.~\ref{fig6:Fig10b}]. The fact that the two largest ftLEs have equal negative values is reflected in the behavior of GALI$_2$, as the index fluctuates around a positive constant value [blue curve in Fig.~\ref{fig6:Fig10c}]. On the other hand, the GALI$_3$ [red curve inset plot of Fig.~\ref{fig6:Fig10c}] decays to zero exponentially fast following the theoretical rate, \(\exp \left[ -\left( 2 \Lambda_1 - \Lambda_2 - \Lambda_3 \right) \right]\) \eqref{eq:GALI_chaos} for \(\Lambda_1 = -0.02\), \(\Lambda_2 = -0.02\) and \(\sigma_3 = -0.66\). We obtain these values from the results of Fig.~\ref{fig6:Fig10b} as a good approximation of the LEs. 

\subsubsection{A stable limit cycle case}
When we set the parameter values \(a=0.3481\) and \(b=0.5\), we observe the appearance of a stable limit cycle in the H{\'e}non map \eqref{eq:3DHenMap}. The \(3D\) phase space portrait (and its \(2D\) projections) of the orbit with IC \((x, y, z) = (0.5, 0.4, 0.2)\) are shown in Fig.~\ref{fig6:Fig10d}. The circular structures observed in the plot suggest the convergence of the studied orbit towards a stable PO (limit cycle) attractor. The time evolution of the flLEs, depicted in Fig.~\ref{fig6:Fig10e}, confirms this observation; as the ftmLE, \(\sigma_1\), (red curve) converges to zero, while \(\sigma_2\) and \(\sigma_3\) (blue and green curves, respectively) remain negative. 

Both the GALI$_2$ [blue curve, main panel of Fig.~\ref{fig6:Fig10f}] and the GALI$_3$ [red curve, in the inset of Fig.~\ref{fig6:Fig10f}] decay to zero exponentially, in line with the theoretical relationships \eqref{eq:GALI_chaos} \(\text{GALI}_2 \propto \exp \left[ -\left( \Lambda_1 - \Lambda_2 \right) \right]\) and \(\text{GALI}_3 \propto \exp \left[ -\left( 2 \Lambda_1 - \Lambda_2 - \Lambda_3 \right) \right]\), with \(\Lambda_1 = 0\), \(\Lambda_2 = -0.02\), and \(\Lambda_3 = -0.67\) obtained from the results of Fig.~\ref{fig6:Fig10e}. Interestingly, despite the attractor's stable nature, all GALI\(_k\)'s decay to zero exponentially fast, which is a behavior typically associated with chaotic motion in conservative Hamiltonian systems (see Sect.~\ref{section:GALI}). We observe that this exponential decay of GALIs for stable limit cycles is consistent across the three models we considered.

\subsubsection{A chaotic (strange) attractor case}
When we set the parameters of the \(3D\) discrete map \eqref{eq:3DHenMap} to \(a = 0.75\) and \(b = 0.01\), we observe a chaotic attractor. The corresponding \(3D\) phase portrait along with the related \(2D\) projections of this attractor is shown in Fig.~\ref{fig6:Fig10g}. Fig.~\ref{fig6:Fig10h} and (i) illustrate the time evolution of the three ftLEs and the GALI$_2$ [and GALI$_3$, in the inset plot], respectively. We note that the ftmLE eventually saturates to \(\sigma_1 = 0.051 > 0\) [blue curve in Fig.~\ref{fig6:Fig10h}] indicating the chaotic nature of the attractor, while \(\sigma_2, \sigma_3 < 0\). Furthermore, both GALIs asymptotically decay to zero, following the theoretical predicted relations \(\text{GALI}_2 \propto \exp \left[ - \left( \Lambda_1 - \Lambda_2 \right) \right]\) and \(\text{GALI}_3 \propto \exp \left[ - \left( 2 \Lambda_1 - \Lambda_2 - \Lambda_3 \right) \right]\), for \(\Lambda_1 = 0.051\), \(\Lambda_2 = -0.021\), and \(\Lambda_3 = -0.723\) [Fig.~\ref{fig6:Fig10i}].  This behavior aligns with our previous observations in the continuous time dissipative systems [see Figs.~\ref{fig6:Fig1i} and Figs.~\ref{fig6:Fig10i}].

\subsubsection{A hyperchaotic attractor case}
We now examine the behavior of the GALI method for a hyperchaotic attractor of the discrete time dissipative system \eqref{eq:3DHenMap}. This analysis represents the first application of the GALI method to a hyperchaotic attractor in such systems. Furthermore, it allows us to extend our results obtained from the analysis of hyperchaotic attractors beyond the continuous time dissipative system discussed in Sect.~\ref{sec:4DODE} (see lower panels in Fig.~\ref{fig6:Fig10F2}). 

Setting parameters \(a = 1.6\) and \(b = 0.01\) in the \(3D\) H{\'e}non map \eqref{eq:3DHenMap} leads to a hyperchaotic attractor. Fig.~\ref{fig6:Fig10j} displays the \(3D\) phase space portrait (black curve points) along with the related \(2D\) projections for this attractor. The portrait reveals a distinctive paraboloid-like structure. This paraboloid is open along the second state variable, \(y\), resulting in a parabolic curve in the \((y, z)\) plane (blue curve). The \(3D\) structure formed by black points resembles the hyperchaotic attractor observed in \citep[Fig.~3]{wang2023image} for parameters \(a = 1.99\) and \(b = 0.001\) of the \(3D\) H{\'e}non map \eqref{eq:3DHenMap}.  

The time evolution of the three ftLEs is depicted in Fig.~\ref{fig6:Fig10k}. The attractor is characterized by two positive ftLEs: \(\sigma_1 = 0.19\) and \(\sigma_2 = 0.18\), while \(\sigma_3 = -4.97\) remains negative. This arrangement of the LEs confirms that the orbit tends to an attractor, with the sum of ftLEs, which is related to the contraction rate of phase space volume, being negative. Moreover, this is a hyperchaotic attractor since the two ftLEs (\(\sigma_1\) and \(\sigma_2\)) are positive. As expected, both the GALI\(_2\) and the GALI\(_3\) decay to zero exponentially fast, following the theoretically derived functions \eqref{eq:GALI_chaos}. The dashed curves in Fig.~\ref{fig6:Fig10j} represent these theoretical functions, further confirming the relationship between the time evolution of the GALIs and the values of the LEs. 


\begin{figure}[!htbp]
    \centering
    \subfloat[Phase portrait of an orbit tending to a stable fixed point\label{fig6:Fig10a}]{\includegraphics[width=0.33\textwidth]{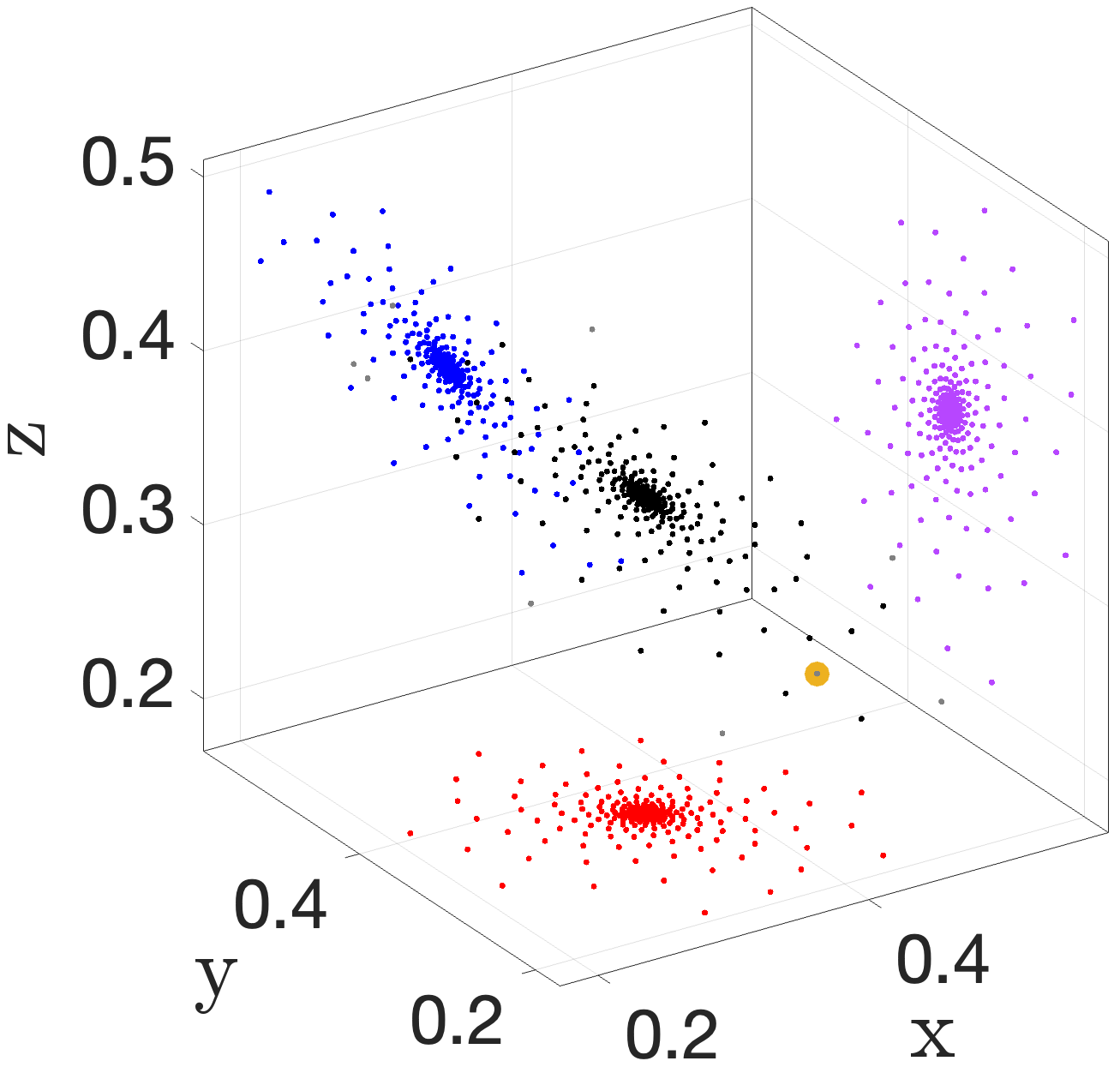}}
    \subfloat[ftLEs\((t)\) for the orbit of (a)\label{fig6:Fig10b}]{\includegraphics[width=0.33\textwidth]{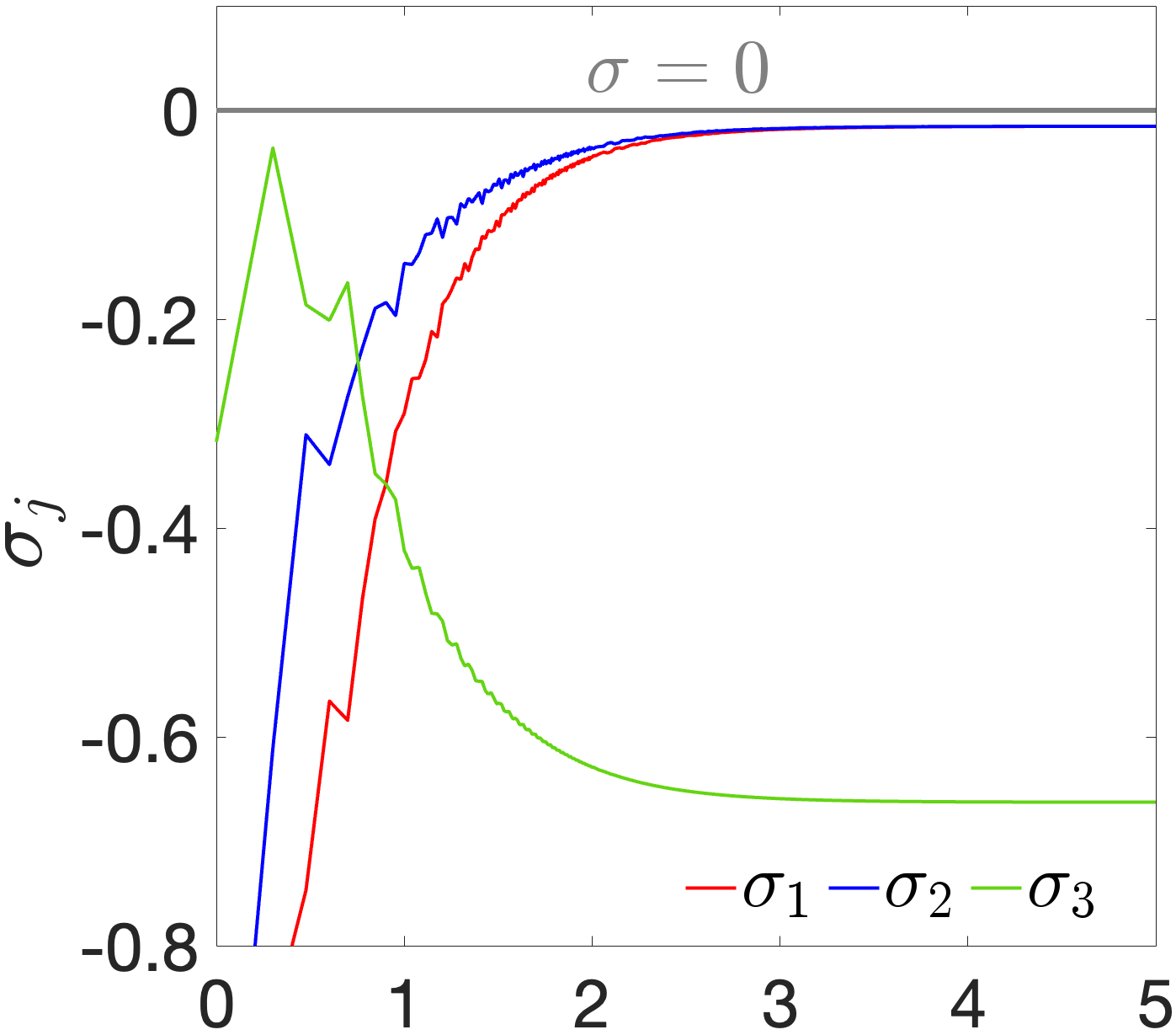}}
    \subfloat[GALI\(_{k} (t)\) for the orbit of (a)\label{fig6:Fig10c}]{\includegraphics[width=0.33\textwidth]{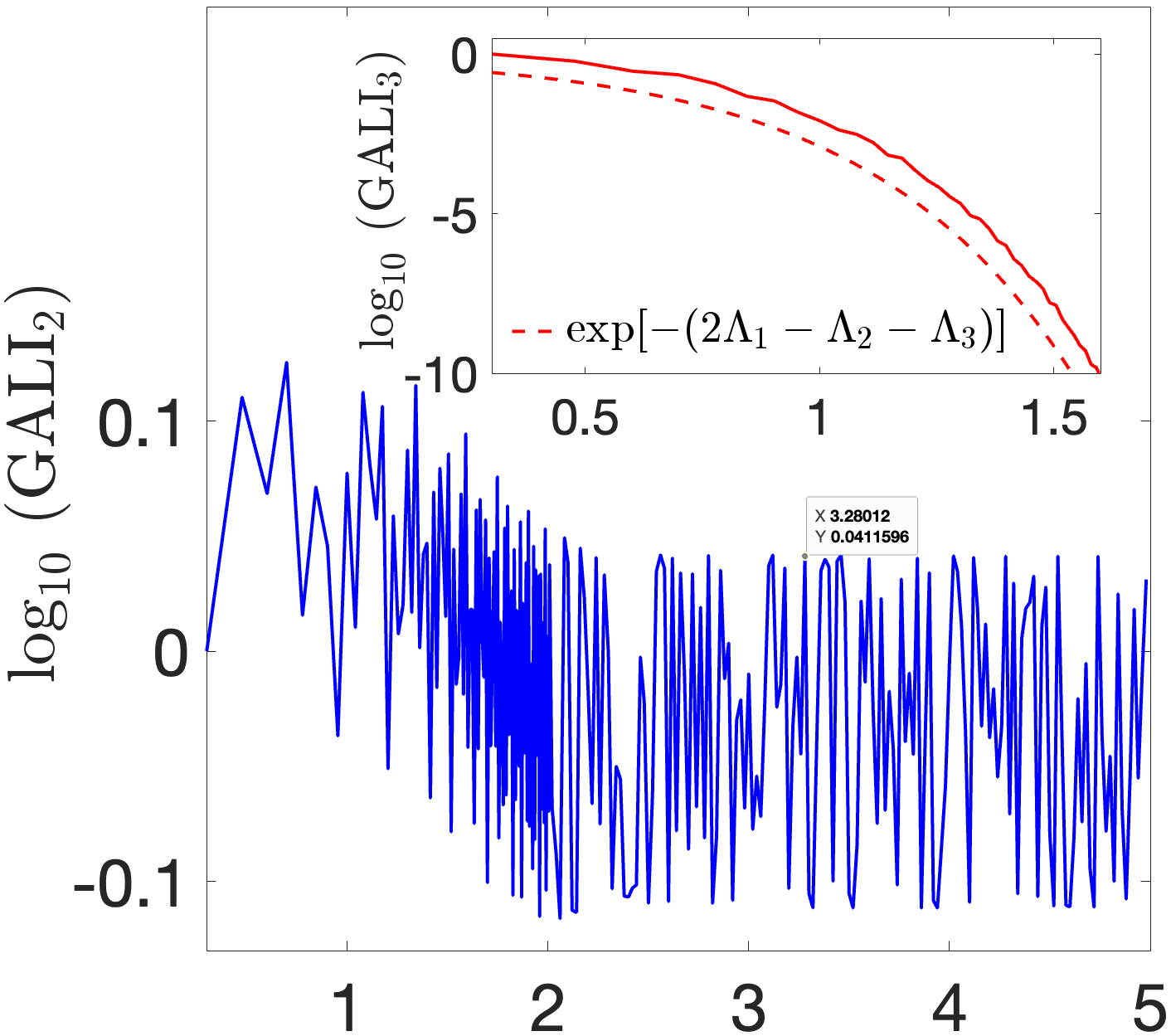}}\\
\subfloat[Phase portrait of an orbit tending to a stable limit cycle\label{fig6:Fig10d}]{\includegraphics[width=0.33\textwidth]{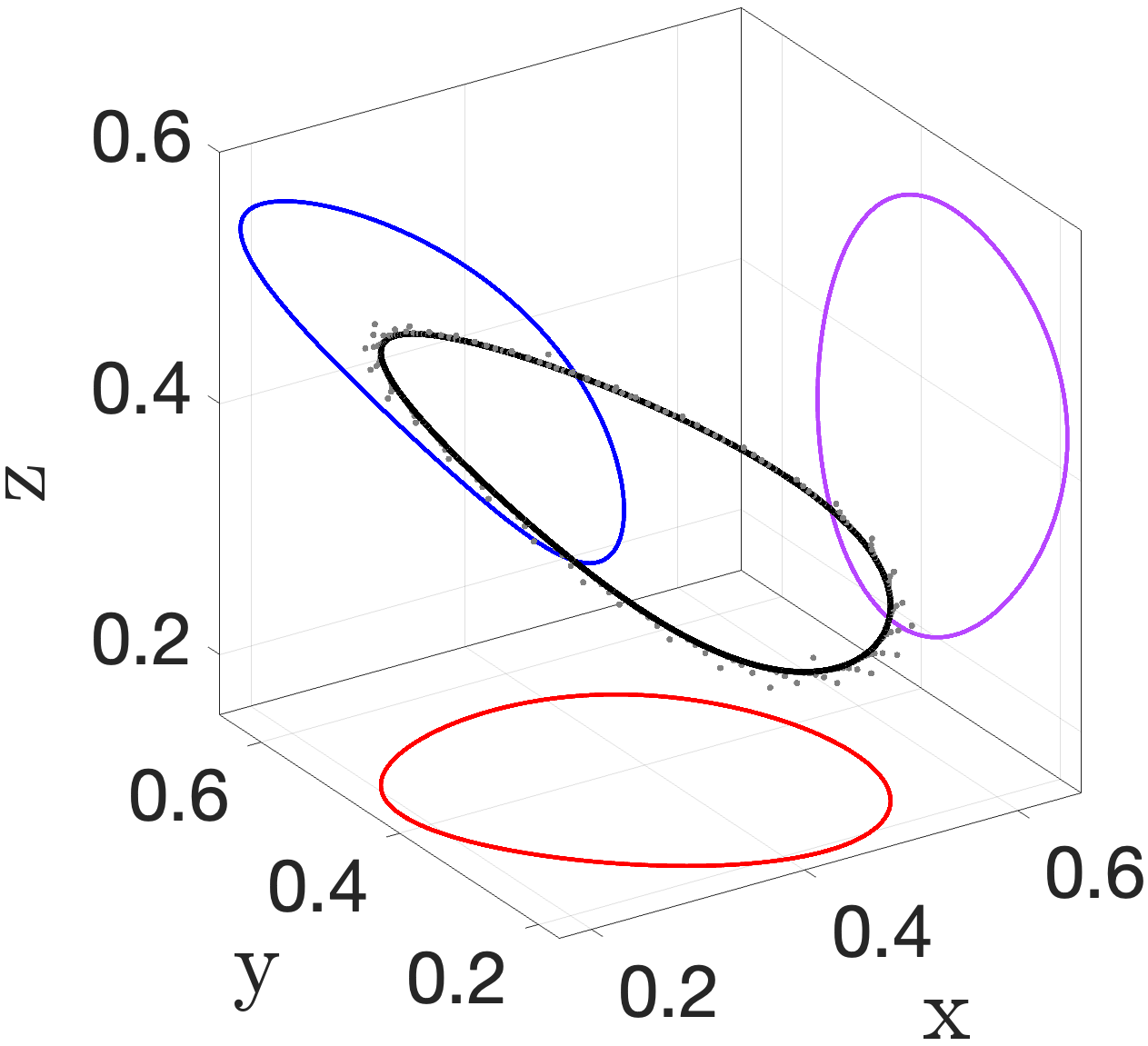}}
\subfloat[ftLEs\((t)\) for the orbit of (d)\label{fig6:Fig10e}]{\includegraphics[width=0.33\textwidth]{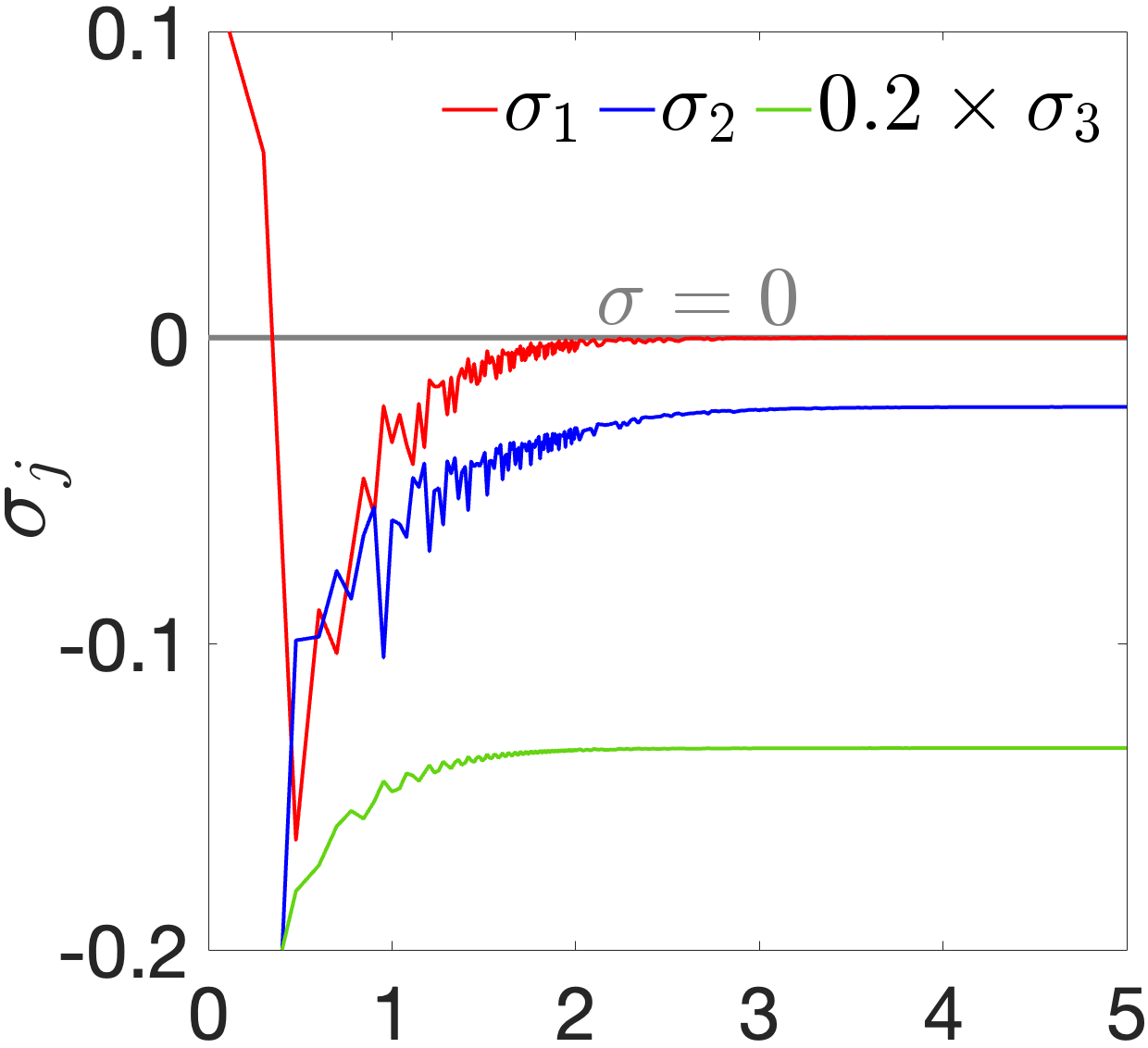}}
\subfloat[GALI\(_{k} (t)\) for the orbit of (d)\label{fig6:Fig10f}]{\includegraphics[width=0.33\textwidth]{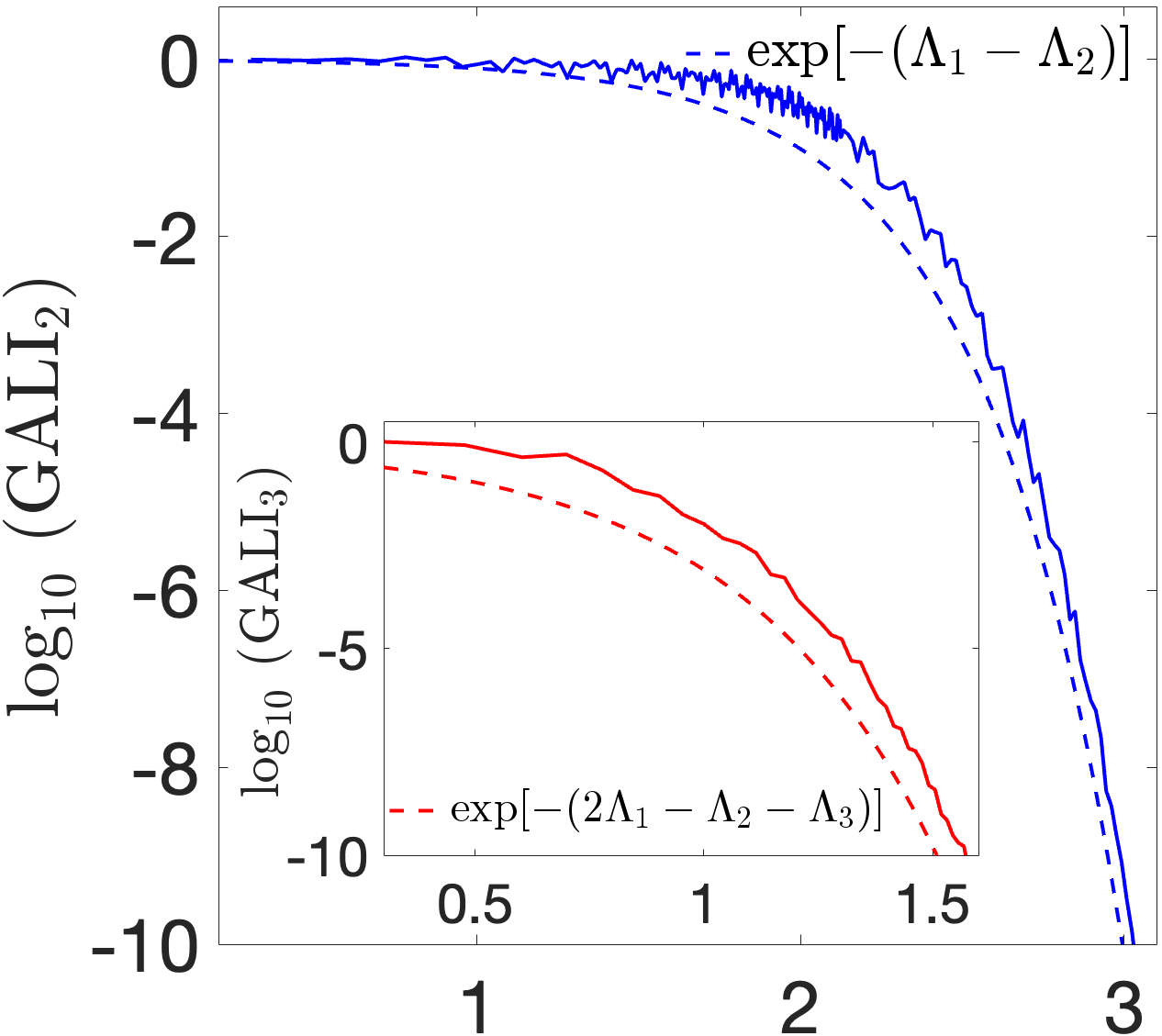}}\\
 \caption{The \(3D\) phase space portraits to different types of attractors of the \(3D\) H{\'e}non map \eqref{eq:3DHenMap} for the orbit with IC \( (x, y, z) = (0.5, 0.4, 0.2) \) which tends: (a) For \( a= 0.3\) and \( b = 0.5 \), the trajectory starting from the IC denoted by the orange circle converges to a stable fixed point attractor. (d) For \( a= 0.3481\) and \( b= 0.5 \), a stable limit cycle attractor is reached. The time evolution of the three ftLEs \(\sigma_1 > \sigma_2 > \sigma_3\) \eqref{eq:LEs order} for the orbit tending to (b) the stable fixed point, and (e) the stable limit cycle. The horizontal gray lines indicate \( \sigma = 0 \) for reference. The corresponding time evolution of the GALI\(_2\) (blue solid curve) and the GALI\(_3\) (red solid curve, inset plot) for the same attractors are shown in (c), and (f), respectively. The blue and red dashed curves represent functions proportional to \(\exp\left[-(\Lambda_1 - \Lambda_2)\right] \) and \( \propto \exp\left[-(2\Lambda_1 - \Lambda_2 - \Lambda_3)\right] \), respectively. Refer to the text for a detailed discussion and the associated $\Lambda_j$, \(j = 1, 2, 3\) values.}
 \label{fig6:Fig10F1}
\end{figure}

\begin{figure}[!htbp]
    \centering
\subfloat[Phase portrait of a chaotic attractor\label{fig6:Fig10g}]{\includegraphics[width=0.33\textwidth]{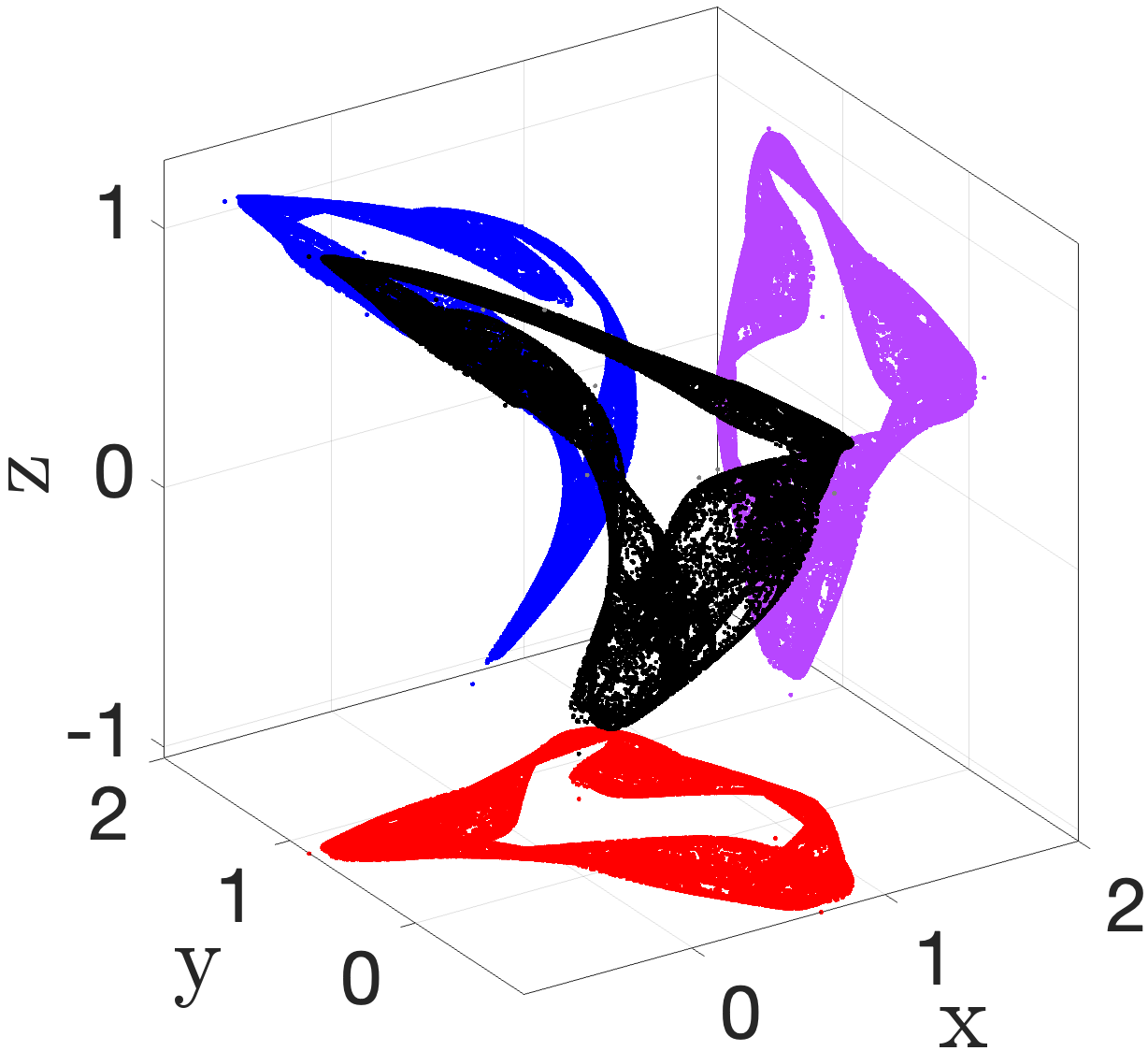}}
\subfloat[ftLEs\((t)\) for the orbit of (a)\label{fig6:Fig10h}]{\includegraphics[width=0.33\textwidth]{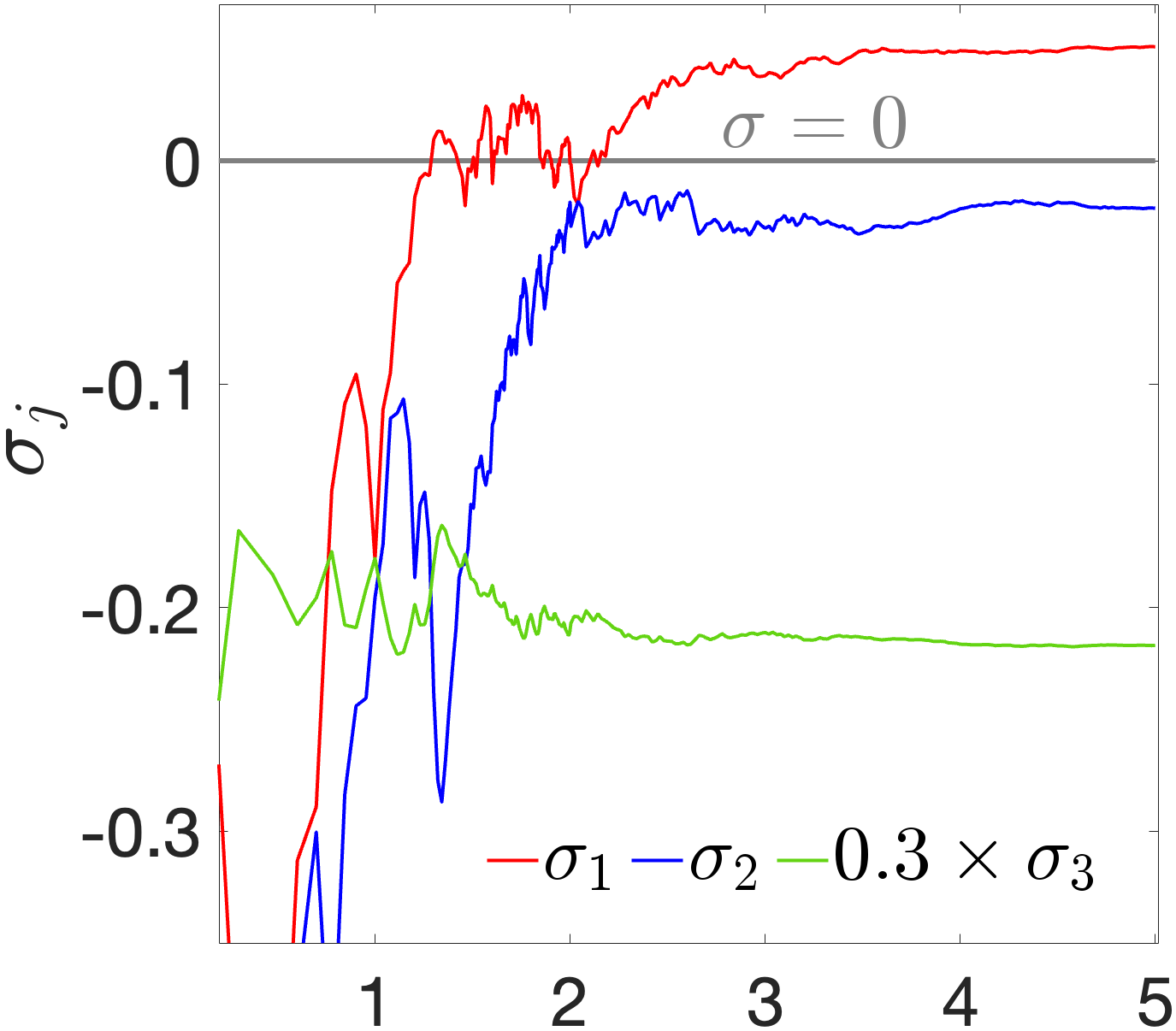}}
\subfloat[GALI\(_{k} (t)\) for the orbit of (a)\label{fig6:Fig10i}]{\includegraphics[width=0.33\textwidth]{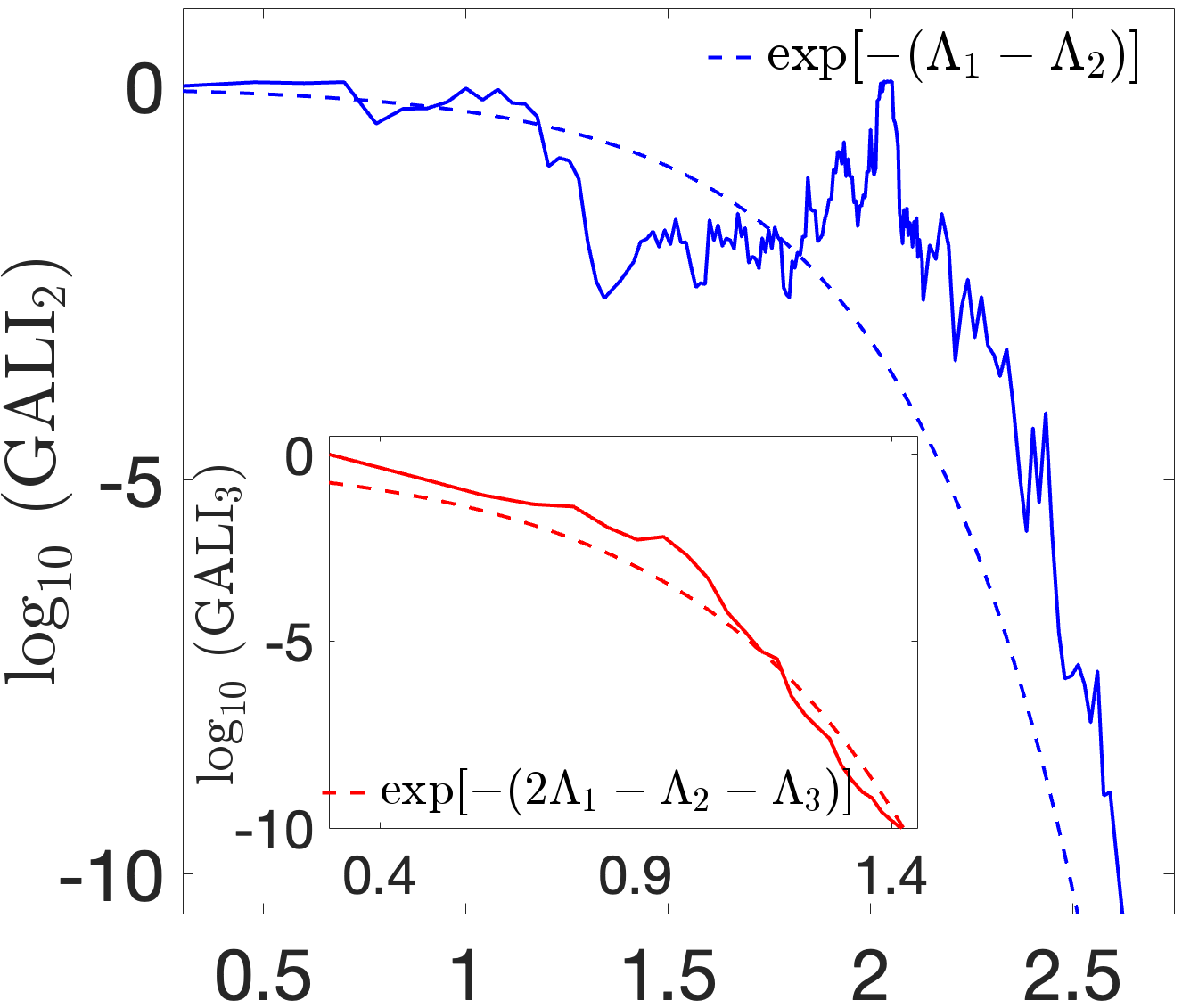}}\\
\subfloat[Phase portrait of a hyperchaotic attractor\label{fig6:Fig10j}]{\includegraphics[width=0.33\textwidth]{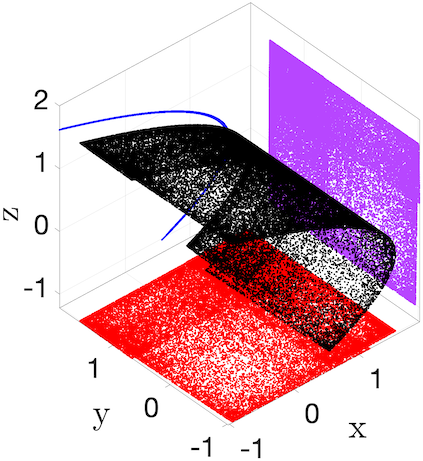}}
\subfloat[ftLEs\((t)\) for the orbit of (d)\label{fig6:Fig10k}]{\includegraphics[width=0.33\textwidth]{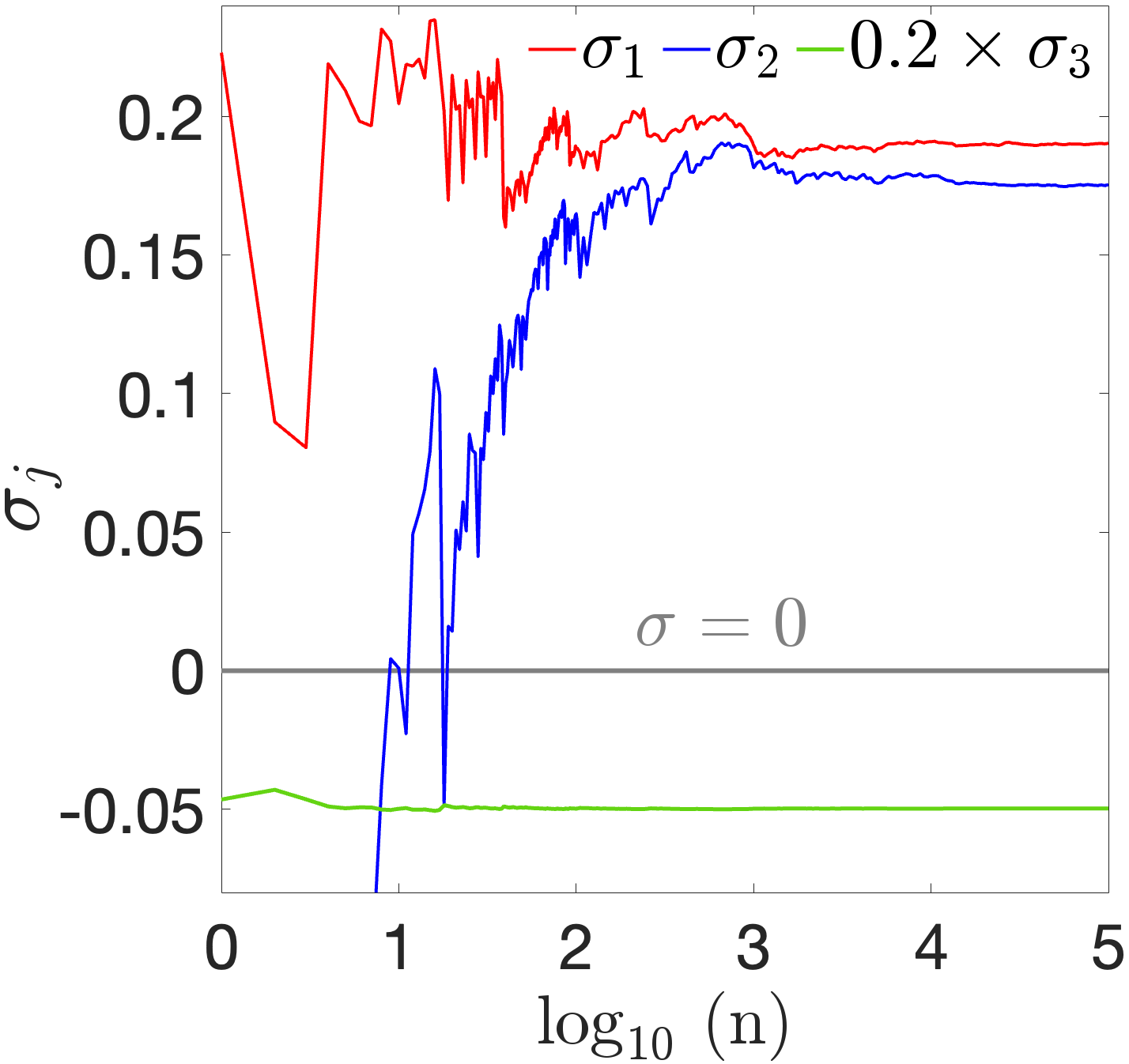}}
\subfloat[GALI\(_{k} (t)\) for the orbit of (d)\label{fig6:Figli}]{\includegraphics[width=0.33\textwidth]{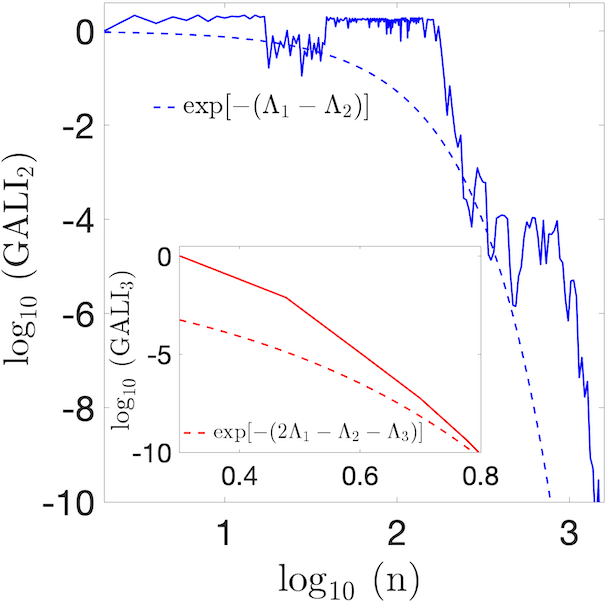}}\\
 \caption{Similar to Fig.~\ref{fig6:Fig10F1} but for the orbit lies on [a, b, and c] \( a = 0.3\) and \( b = 0.5\), a chaotic attractor, and [d, e, and f] for \( a=  1.6\)  and  \(b = 0.01\), a hyperchaotic attractor.}
\label{fig6:Fig10F2}
\end{figure}

\subsubsection{Parametric exploration of the generalized H{\'e}non map using the GALI method and LEs}
Let us now conduct a parametric study of the H{\'e}non map \eqref{eq:3DHenMap} to explore the system's dynamics by varying the parameter \(a\) from \(a = -0.0264\) to \(a = 1.5\) while fixing \(b = -0.1\). For this analysis, we employ the three ftLEs and the GALI\(_2\) index, similarly to what we did in Sect.~\ref{sec:3DODE parameter space}.

\begin{figure}[!htb]
    \centering
\subfloat[ftmLEs\label{fig6:Fig11a}]{\includegraphics[width=0.49\textwidth]{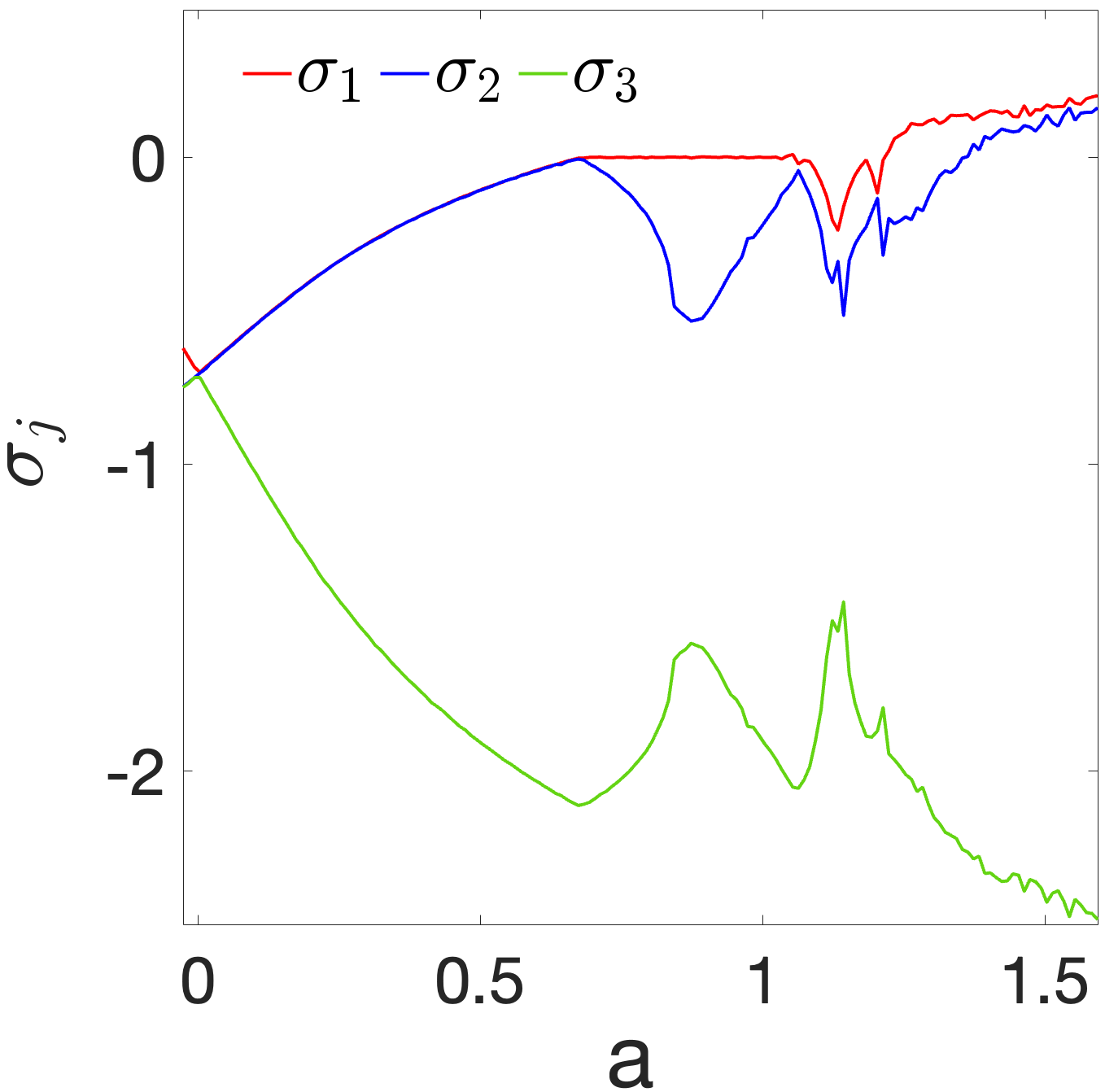}}
\subfloat[GALI\(_{2}\)\label{fig6:Fig11b}]{\includegraphics[width=0.51\textwidth]{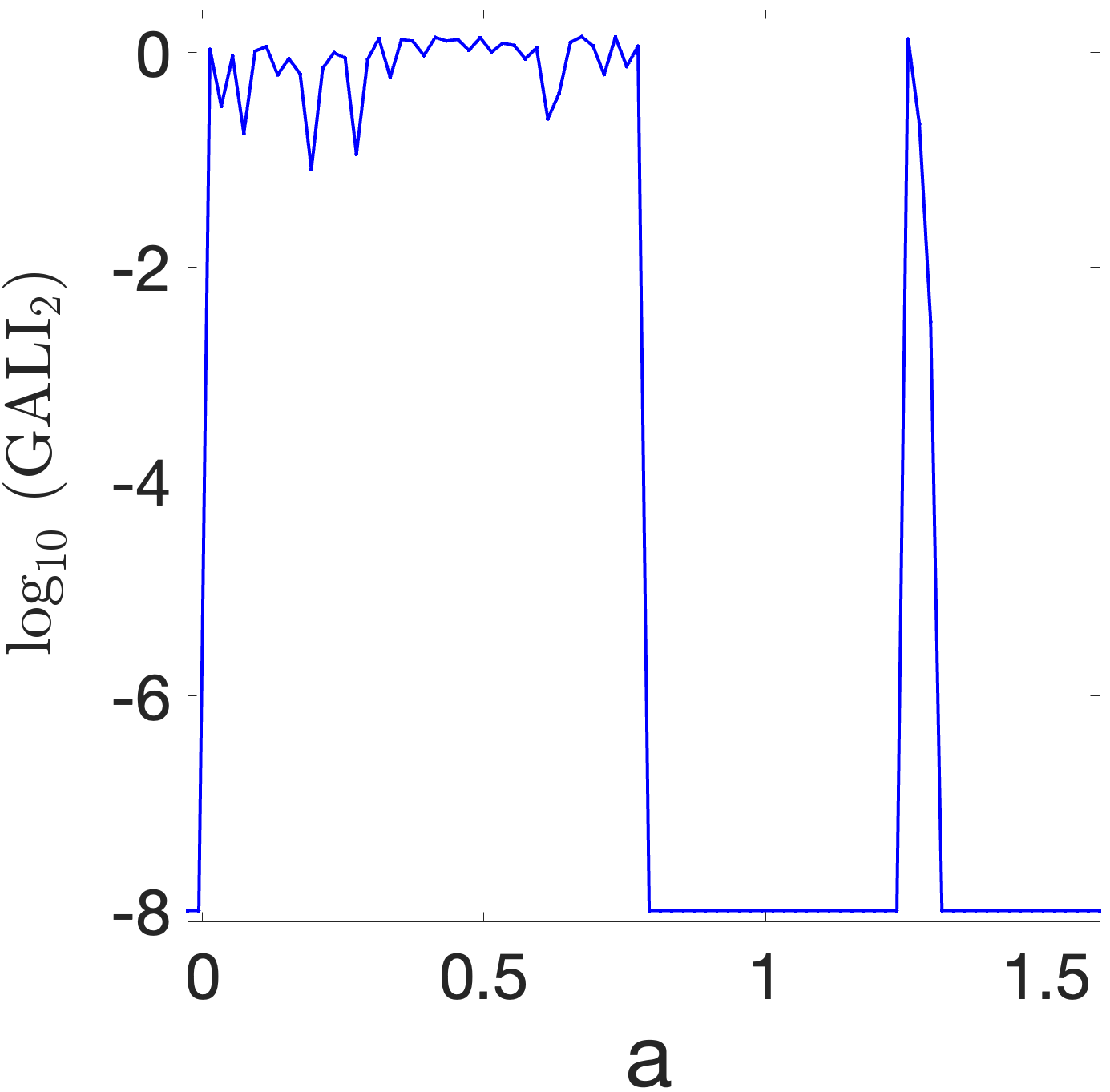}} \\
\caption{A parametric exploration of the \(3D\) H{\'e}non map \eqref{eq:3DHenMap} for parameters \( b = 0.5 \) while varying \(a\). We consider a total of \(163\) points for \( a \in [0, 1.2] \). Final values of (a) the three ftLEs \(\sigma_1 > \sigma_2 > \sigma_3\) \eqref{eq:LEs order}, and (b) the GALI$_2$ \eqref{eq:GALI} at \(t=10^4\).}
\label{fig6:Fig11}
\end{figure}

Figure \ref{fig6:Fig11a} shows the values of the three ftLEs, \(\sigma_j\), \(j=1, 2, 3\), for various tested \(a\) values of the \(3D\) system \eqref{eq:3DHenMap} after \(n = 10^4\) iterations. These exponents reveal different dynamical regimes: In the leftmost half of the \(a\) interval, where \(a \le 0.7835\), all ftLE values are negative, indicating that the system is attracted to a sink (see Fig.~\ref{fig6:Fig10a} for the phase space portrait of a representative case). As \(a\) increases, \(\sigma_1\) approaches zero, implying the emergence of stable limit cycles in the interval \(a \in (0.7835, 1.0835]\). Then, for \(a \in (1.0835, 1.3634]\) the system again displays fixed points characterized by all negative ftLEs before transitioning to chaotic behavior, indicated by the presence of one positive ftLE (\(\sigma_1 > 0\)) for \(a > 1.3634\). Finally, at \(a = 1.4835\), the system transitions to hyperchaotic behavior, with a second positive ftLE, \(\sigma_2 > 0\) appearing.

Figure \ref{fig6:Fig11b} depicts the GALI$_2$ values as a function of the parameter \(a\) for the dissipative map \eqref{eq:3DHenMap}. We see that the GALI$_2$ remains positive only in the intervals \(a \in [0.01254, 0.01284]\) and \(a \in [0.035, 0.7835]\), where stable fixed point attractors are present. In these cases, all ftLEs are negative, and the values of \(\sigma_1\) and \(\sigma_2\) are practically equal. Outside these intervals of \(a\) values, the GALI$_2$ is practically zero. It is important to note that in these cases the GALI$_2$ decays to zero exponentially in accordance with the theoretical prediction  \eqref{eq:GALI_chaos}.

We further conduct a bi-parametric exploration of the dynamics of the map \eqref{eq:3DHenMap} to classify different types of attractors based on the values of the three ftLEs and the GALI$_2$, similarly to the investigations performed for continuous systems (Figs.~\ref{fig6:Fig3} and \ref{fig6:Fig9}). We keep the IC of the considered orbit to \((x, y, z) = (0.5, 0.4, 0.2)\) the same, while varying the parameters \(a\) and \(b\) in the region \(a \in [0, 1.2]\) and  \( b \in [-0.12, 0.12]\) considering a grid of total of \(240 \times 1765 = 423,600\) parameter values.

\begin{figure}[!htb]
    \centering
    \subfloat[ftmLEs\label{fig6:Fig12a}]{\includegraphics[width=0.345\textwidth]{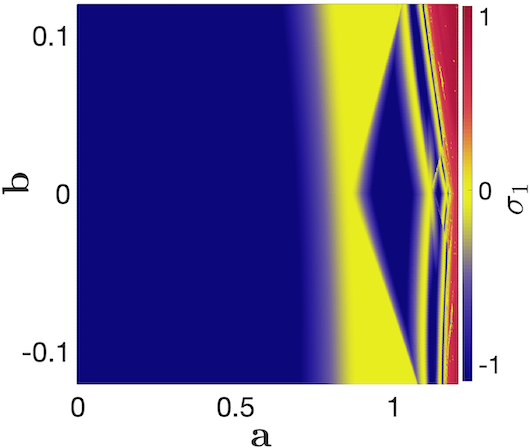}}
    \subfloat[Classification based on the three ftLEs\label{fig6:Fig12b}]{\includegraphics[width=0.325\textwidth]{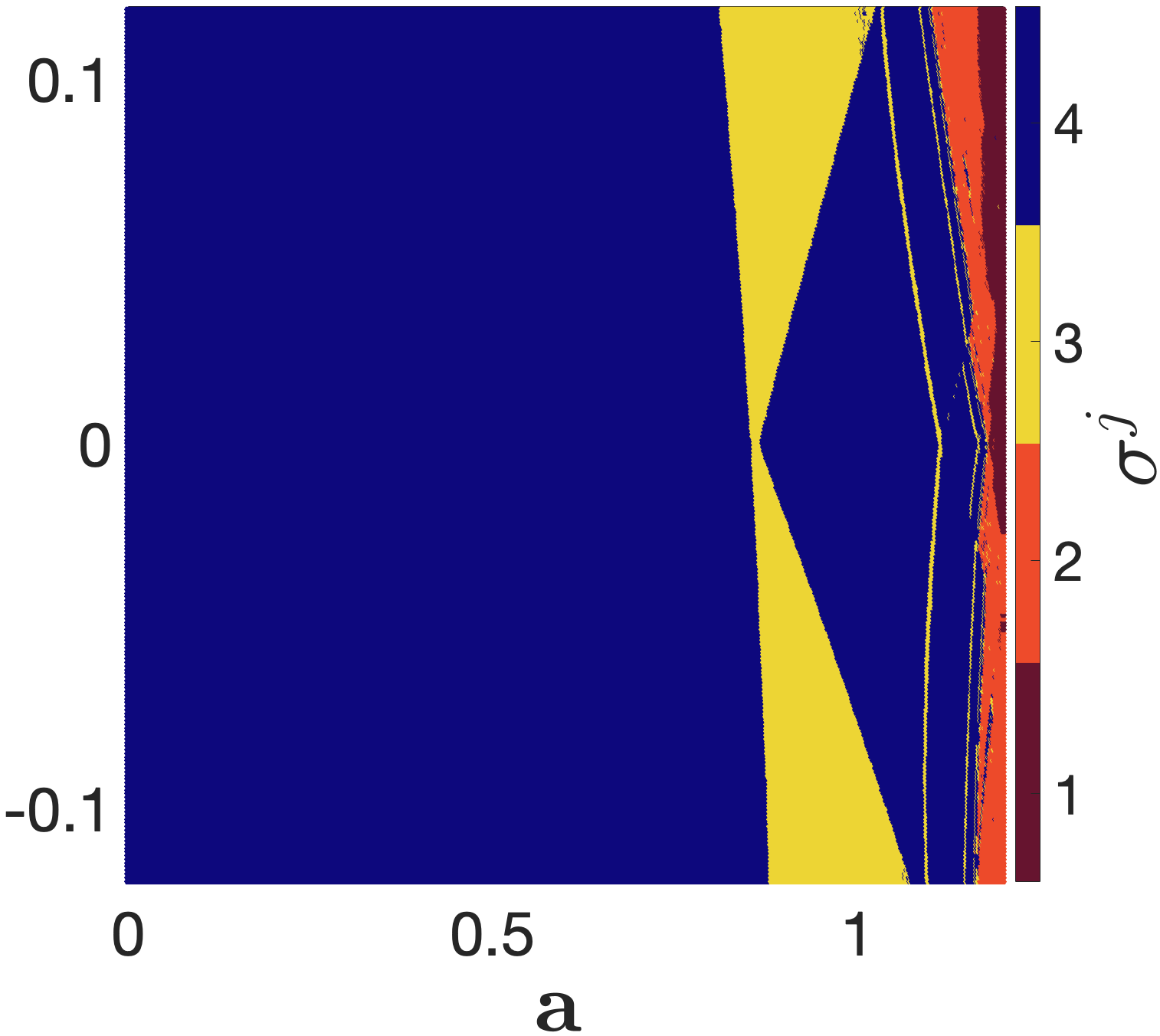}}
    \subfloat[GALI\(_{2}\)\label{fig6:Fig12c}]{\includegraphics[width=0.33\textwidth]{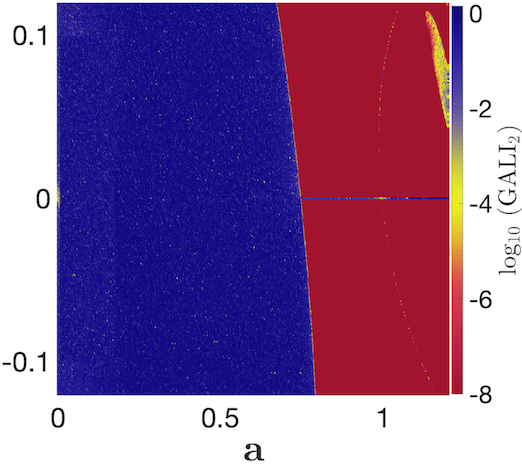}}\\
    \caption{An exploration of the \( (a, b) \) space of he H{\'e}non map \eqref{eq:3DHenMap}. We produce color plots using a grid of \( 240 \times 1765 = 423,600 \) points in the region \(a \in [0, 1.2]\) and  \( b \in [-0.12, 0.12]\) by iterating the IC \((x, y, z) = (0.5, 0.4, 0.2)\)  up to \( t = 10^4 \) and recording the orbit's ftLEs \(\sigma_1 > \sigma_2 > \sigma_3\) \eqref{eq:LEs order} and the GALIs \eqref{eq:GALI}. The classification of various dynamical regimes is based on the values of (a) the ftmLE, \( \sigma_1 \), and (b) the three ftLEs, \(\sigma^j\), according to the classification in Sect.~\ref{section:LEs}. In both panels, stable fixed point attractors are shown in dark blue \( (\sigma_1 < 0 \text{ or } \sigma^j = 4) \), and stable periodic motion (limit cycles) are represented by the yellowish/orange areas \( (\sigma_1 \approx 0 \text{ or } \sigma^j = 2) \).  In contrast, in (b) chaotic (\(\sigma^j = 2\)) and hyperchaotic (\(\sigma^j = 1 \)) attractors are indicated by brown and red regions, respectively, while in (a), brown/red regions denote either chaotic or hyperchaotic attractors (\(\sigma_1 > 0\)). (c) A similar classification is performed using GALI$_2$ values. Note that in (a), the computed \(\sigma_1\) values are scaled to the interval \([-1, 1]\).}
\label{fig6:Fig12}
\end{figure}

Figure \ref{fig6:Fig12a} illustrates the parametric space \((a, b)\) with each point colored according to the values of the ftmLE, \(\sigma_1\). This enables a `preliminary' classification between the regular and chaotic behaviors we observed. Here, the \(\sigma_1\) values are scaled within the range \([-1, 1]\) similarly to Figs.~\ref{fig6:Fig3a} and \ref{fig6:Fig9a}. Yellowish/orange regions denote the existence of a stable limit cycle or potentially periodic motions (\(\sigma_1 = 0\)), purple/dark blue areas indicate the presence of stable fixed point attractors (\(\sigma_1 < 0\)), and brown/red regions identify either chaotic or hyperchaotic motions (\(\sigma > 0\)).   

For a more accurate classification of the orbital behaviors and similarity to what was done in Fig.~\ref{fig6:Fig9a}, similarly to the approach we follow in Fig.~\ref{fig6:Fig9a}, we assign discrete values \(\sigma^j\), numbered from \(1\) to \(4\), according to the values of the three ftLEs as described in Sect.~\ref{section:LEs}. The resulting color map is shown in Fig.~\ref{fig6:Fig12b}, where each attractor type is color coded as follows: \(\sigma^j = 1\) (brown) for hyperchaotic attractors (\(\sigma_1\), \(\sigma_2 > 0\) and \(\sigma_3 < 0\)); \(\sigma^j = 2\) (red) for chaotic attractors (\(\sigma_1 > 0, \sigma_2 \le 0\) and \(\sigma_3 < 0\)); \(\sigma^j = 3\) (orange) for stable limit cycle attractors (\(\sigma_1 \approx 0\) and \(\sigma_2,\), \(\sigma_3 < 0\)); and \(\sigma^j = 4\) (blue) for stable fixed point attractors (\(\sigma_j < 0\), for all \(j=1,2,3\)).

In Fig.~\ref{fig6:Fig12c}, we preset a similar color plot of the parametric space where points are colored according to the GALI$_2$ value of the orbits. Here, the blue areas indicate the existence of stable fixed point attractors for which \(\sigma_1 \approx \sigma_2\), while the red regions (GALI$_2$ is practically zero) correspond to chaotic, hyperchaotic, or stable limit cycle attractors. Thus, it becomes apparent that the GALI$_2$ is not able to differentiate among these three types of attractors. Notably, some points in the GALI$_2$ color plot (such as those in the parameter space around \((a, b) = (1.161, 0.08)\), located in the upper right corner of Fig.~\ref{fig6:Fig12c}) are unexpectedly in colored yellow and/or blue instead of red, although the related orbital evolution takes place on chaotic attractors. This behavior indicates weakly chaotic or hyperchaotic attractors. For such cases, we may require longer iteration times for the GALI$_2$ to reliably decay to zero, thereby to fully reveal  their chaotic characteristics.

It is worth noting that none of the models or parameter values considered in our study exhibited a \(j\)-dimensional (\(j > 1\)) stable torus. For such attractors, the largest \(k\) ftLEs approach zero asymptotically, while the other exponents remain negative (see Sect.~\ref{section:LEs}). In this case as a \(j\)-dimensional stable torus, we would expect all the GALI$_k$ with \( 2 \le k \le j\), oscillate around a constant positive value.

\section{Summary and conclusions} \label{section:SummaryCh6}
In this chapter, we systematically investigated the behavior and efficiency of the GALI method in distinguishing between various types of attractors, in particular stable fixed points, stable periodic orbits (limit cycles), chaotic (strange), and hyperchaotic attractors in non-Hamiltonian dissipative systems. We focused on the \(3D\) Lorenz system \eqref{eq:3DODE}, and the \(4D\) continuous time dissipative system \eqref{eq:4DODE}, as well as the generalized H{\'e}non map \eqref{eq:3DHenMap}. Initially, we used the LE (\(\sigma_j\)) spectrum to identify the parameter values for which each type of attractor appears and compared the obtained results to the behavior of the GALI method. Our findings for both continuous and discrete time dissipative systems can be summarized as follows:

\begin{enumerate} [label=\textnormal{(\Roman*)}]
    \item Stable fixed attractors (all \(\sigma_j < 0\)): GALI$_2$ fluctuates around a positive constant value [Figs.~\ref{fig6:Fig1c}, \ref{fig6:Fig4c}, and \ref{fig6:Fig10c}], if and only if the corresponding two ftLEs have practically equal negative values, i.e., if \(\sigma_1 \approx \sigma_2\) [Figs.~\ref{fig6:Fig1b}, \ref{fig6:Fig4b}, and \ref{fig6:Fig10b}]. On the other hand, the GALI$_3$ and the GALI$_4$ decay to zero exponentially fast at rates determined by the theoretical expressions \eqref{eq:GALI_chaos} evolving the values of several LEs. 
    
    \item Stable limit cycles (\(\sigma_1 \approx 0\) with the rest ftLEs being negative): The GALI$_k$ (for $k = 2, 3, 4$) decay exponentially fast to zero at rates defined by the respective LEs [Figs.~\ref{fig6:Fig1f}, \ref{fig6:Fig5c}, and \ref{fig6:Fig10f}] in accordance with the theoretical expectations of Eq.~\eqref{eq:GALI_chaos}.

    \item Chaotic attractors (only \(\sigma_1 > 0\)) and hyperchaotic attractors (both \(\sigma_1\) and \(\sigma_2 > 0\)): The GALI$_k$ indices again decay to zero exponentially, with rates determined by the corresponding LEs. This behavior is observed in Figs.~\ref{fig6:Fig1i}, \ref{fig6:Fig6c}, and \ref{fig6:Fig10i} for chaotic attractors and in Figs.~\ref{fig6:Fig7c}, and \ref{fig6:Fig10j} for hyperchaotic attractors.

    \item No \(k\)-dimensional stable torus attractors were observed in the studied models and parameter ranges. Such attractors would exhibit \(k\) zero ftLEs, with the others being negative. On the other hand, we would expect GALI$_2$ to fluctuate around a constant positive value only when the two largest ftLEs are zero. Otherwise, the GALI$_k$ indices for $k = 2, 3, 4$, would decay to zero exponentially fast at rates defined by the LEs according to Eq.~\ref{eq:GALI_chaos}. 
\end{enumerate}

When we compared the behavior of the GALI method in conservative and dissipative DSs, we observed similar trends for chaotic trajectories but identified notable differences for other types of regular motion. In dissipative systems, the phase space volume contracts over time, which directly impacts the evolution of the hyper-volume defined by the deviation vectors used to compute the GALI indices. Since the GALI method does not involve re-orthogonalizing these deviation vectors, as was done, for example, in \citep{manos2012probing} the phase space contraction shrinks this volume, and eventually reducing it to zero. We observed this volume reduction even for regular motions, like for orbits which asymptotically converge to stable attractors (e.g., stable limit cycles). Hence, the evolution of the GALI exhibits behavior similar to that seen for chaotic trajectories, despite the underlying motion being stable. Overall, for dissipative systems, the GALI$_k$ decays to zero exponentially fast for limit cycles and chaotic/hyperchaotic attractors, following rates defined by the LEs \eqref{eq:GALI_chaos}.

For a comprehensive understanding of how control parameters affect attractor types in each dissipative system, we further conducted numerical analyzes for different parameter values that result in a variety of attractors: stable attractors (fixed point and limit cycle), chaotic, and hyperchaotic attractors. To achieve this, we produced parameter exploration plots and bi-parametric color plots for each parametric pair, using one IC per pair. The obtained plots were color-coded based on three quantities: the values of the ftmLEs \eqref{eq:mLEs}, the classification of different dynamical regimes as determined by the ftLEs  \eqref{eq:ftmLE}  according to Sect.~\ref{section:LEs}, and the GALI$_2$ \eqref{eq:GALI} [see Figs.~\ref{fig6:Fig3}, \ref{fig6:Fig9}, and \ref{fig6:Fig12}]. At the end, quantifying each parameter pair according to ftLEs proved the most effective approach for classify the four attractor types across the three dissipative systems we considered.

In summary, this chapter presented a comprehensive study of the GALI method's behavior and limitations in distinguishing different attractors in \textit{dissipative systems}. This analysis constitutes the first detailed investigation of the GALI method's effectiveness in distinguishing between regular and chaotic attractors in dissipative systems. In our study, we examined trajectories tending to attractors in the \(3D\) Lorenz system \eqref{eq:3DODE}, \(4D\) hyperchaotic Lorenz system \eqref{eq:4DODE}, and the generalized H{\'e}non map \eqref{eq:3DHenMap} to systematically explore the method's capabilities and potential limitations in analyzing dissipative system dynamics.

\clearpage


\chapter{Conclusions and outlook} \label{chapter::summary}
In this thesis, we investigated the role of chaos in various dynamical systems (DSs), including Hamiltonian systems, multidimensional maps, and non-Hamiltonian dissipative systems. A primary focus of our investigations was the application of numerical techniques, particularly the generalized alignment index (GALI) [Sect.~\ref{section:GALI}], to detect and quantify chaos. Furthermore, we compared the performance of the GALI method with that of Lyapunov  exponents (LEs) to efficiently discriminate between different dynamical behaviors [Sect.~\ref{section:LEs}].  

Our presentation began with an introduction to general DSs and numerical methods in Chap. \ref{chapter:two}, emphasizing LEs and the GALI method. In Chap.~\ref{chapter:three}, we employed the GALI method of order 2 (GALI\(_2\)) \eqref{section:GALI} as an efficient tool for detecting and quantifying kinetic and magnetic chaos in plasma physics models. Our primary objective was to compare kinetic and magnetic chaos in a toroidal plasma model. To achieve this, we analyzed the dynamics of various perturbations of the guiding center Hamiltonian \eqref{eq:GC H} representing plasma particles, and we considered representative cases of the magnetic field (MF) Hamiltonian \eqref{eq:per MFL Ham}. For all these setups, we constructed appropriate Poincar{\'e} surface of sections (PSSs), where orbits were colored according to their final GALI\(_2\) values. For low energy particles, strong similarities were observed between the kinetic and magnetic chaos, both appearing near separatrices of stable magnetic island chains. In contrast, for high energy particles, stable islands were located at different radial positions, and kinetic chaos was less pronounced due to increased particle drift across MF lines. By analyzing the three constants of motion (i.e., energy, magnetic moment, and toroidal momentum), we classified particle orbits as chaotic or regular using the GALI\(_2\) values. This approach provided valuable information for understanding energy and momentum transport, having potential applications in tokamak simulations. 

In Chapter \ref{chapter:four}, our focus shifted to the investigation of the evolution of the phase space structures in a three-dimensional (\(3D\)) bar galaxy Hamiltonian system \eqref{eq:BG H}, particularly before and after successive \(2D\) and \(3D\) pitchfork bifurcations, as well as \(3D\) period-doubling bifurcations. We studied transitions in which families of \(2D\) or \(3D\) stable periodic orbits (POs) become unstable, leading to the creation of new stable PO families. Using the color and rotation (CR) visualization technique [Sect.~\ref{section:CR Method}], we explored the system's \(4D\) phase space. We found that perturbations of stable POs lead to the formation of invariant tori, characterized by smooth color variations in any \(3D\) projections of the \(4D\) phase space. In contrast, perturbations of unstable POs often led to Figure-8 structures. The formation of these Figure-8 structures were closely related to the presence of surrounding invariant tori. As energy levels increased, these structures began to break down, leading to the formation of scattered clouds of points with mixed colors in the \(3D\) projections. This behavior indicated the introduction of chaotic motion, which we confirmed through the computation of the GALI\(_2\) index. 

In Chapter \ref{chapter:five}, we focused on the long-term diffusion and chaotic behavior in the \(2D\) standard map (SM) \eqref{eq:ssm} as well as in coupled SM systems \eqref{eq:csm}. We explored the impact of accelerator modes (AMs) of varying periods \(p\) on the diffusion behaviors of the SMs. A higher percentage of chaotic trajectories in ensembles of orbits around stable AMs delayed the convergence of the diffusion exponent \(\mu\) \eqref{eq:pvar} to its limiting value, slowing the trend toward the asymptotic ballistic transport (\(\mu = 2\)). In the case of coupled SM systems \eqref{eq:csm}, increasing the coupling strength \(\beta\) suppressed global diffusion and accelerated the convergence to normal diffusion (\(\mu = 1\)), particularly at higher kick strength values \(K_j\). This trend of faster convergence to normal diffusion was associated with an increase in global chaos, as was quantified by the mLE and the GALI\(_2\). 

Finally, in Chapter \ref{chapter:six}, we applied the GALI method \eqref{eq:GALI} to non-Hamiltonian dissipative systems, including the \(3D\) Lorenz system \eqref{eq:3DODE}, the \(4D\) hyperchaotic Lorenz system \eqref{eq:4DODE}, and the generalized H{\'e}non dissipative map \eqref{eq:NDHenMap}. We examined the GALI method's performance to distinguish among various orbit types, namely, stable fixed points, stable limit cycles, chaotic, and hyperchaotic attractors. Our results showed that the GALI$_2$ decays to zero exponentially fast at rates determined by the theoretical expression \eqref{eq:GALI_chaos} for stable limit cycles, chaotic, and hyperchaotic attractors, when the values of the largest two LEs were distinct (i.e., the GALI method was unable to discriminate between these cases). On the other hand, it fluctuates around a positive constant value for stable fixed point attractors, where the corresponding largest two LEs were nearly equal and negative. This analysis of the GALI$_2$ behavior in dissipative systems filled a gap in the existing literature.

Although our work produced valuable insights into the numerical investigation of chaos in Hamiltonian models and diffusion trajectories in multidimensional maps, several interesting research avenues remain open for future study. One potential direction involves extending the analysis of kinetic chaos we performed in Chapter \ref{chapter:three} by considering time-dependent perturbations, particularly those relevant to fusion applications. A notable future task is the analytical and numerical phase space study of the nonlinear charged particle dynamics under time-dependent, axisymmetry-breaking electromagnetic perturbations in an axisymmetric equilibrium MF in toroidal fusion plasma configurations.

The coupled SM system \eqref{eq:csm}, which was extensively analyzed in this thesis, also presents numerous opportunities for further investigations. In Chapter \ref{chapter:five}, we examined the system's chaotic and diffusion properties, including the interplay between normal and anomalous diffusion, pattern propagation, and the potential presence of localization phenomena. Our analysis explored the relationship between the diffusion coefficient \(\mu\) \eqref{eq:pvar} and the strength of chaos [quantified through the values of the mLE \eqref{eq:mLEs} and the GALI\(_2\) \eqref{eq:GALI}], as functions of key system parameters such as the kick (\(K\))  and coupling (\(\beta\)) strength. This study considered conditions with constant \(K\) values for the analyzed maps. A logical extension of this work would involve introducing randomness in the \(K\) values, possibly in cases where some maps exhibit strong chaos while others govern predominantly regular motions. Furthermore, allowing time dependence in the model's parameters could reveal additional dynamical features. For example, dynamically varying \(K\) between values for which chaotic or regular motion prevails, or introducing time-dependent behavior in the parameters of one (or more) of the coupled maps, could lead to interesting dynamical behaviors. Such studies could significantly enhance our understanding of diffusion and transport processes in coupled systems and offer potential applications in the context of complex dynamical networks.

\clearpage


\phantomsection
\addcontentsline{toc}{chapter}{\bibname}


%
%
\bibliographystyle{plainnat}
%

\bibliography{Thesis_Bibliography_DB} 

\clearpage

%
\appendix




\chapter{The components of the galactic potential studied in Chap. \ref{chapter:four}} \label{chapter:appendixA}

\section{The bulge potential}
The Bulge Potential \(V_S(x, y, z)\) is defined as: 
\begin{equation}\label{eq:Bulge Potential}
V_S(x, y, z) = -\frac{GM_S}{\sqrt{x^2 + y^2 + z^2 + \epsilon_s^2}},
\end{equation}
where \(G\) is the gravitational constant, \(M_S\) is the mass and \(\epsilon_s\) the scale length of the bulge (see Sect.~\ref{section:HamiltonianCh4} for more details). 

For simplicity, let us define \(Q\) as \(Q = x^2 + y^2 + z^2 + \epsilon_S^2\). Then, the first partial spatial derivatives of \(V_S\) \eqref{eq:Bulge Potential} are expressed follows:
\begin{equation}\label{eq:Bulge Potential FD}
    \begin{aligned}
        \frac{\partial V_S}{\partial x} &= GM_S  \dfrac{x}{ Q^{3/2}}, \\
        \frac{\partial V_S}{\partial y} &= GM_S  \dfrac{y}{ Q^{3/2}}, \\
        \frac{\partial V_S}{\partial z} &= GM_S  \dfrac{z}{ Q^{3/2}},
    \end{aligned}
\end{equation}

while the second partial spatial derivatives of  \eqref{eq:Bulge Potential} are given by:
\begin{equation}\label{eq:Bulge Potential SD}
    \begin{aligned}
        \frac{\partial^2 V_S}{\partial x^2} &= \dfrac{GM_S}{Q^{3/2}} \left( 1 - \dfrac{3x^2}{Q} \right), \\
        \frac{\partial^2 V_S}{\partial y^2} &= \dfrac{GM_S}{Q^{3/2}} \left( 1 - \dfrac{3y^2}{Q} \right), \\
        \frac{\partial^2 V_S}{\partial z^2} &= \dfrac{GM_S}{Q^{3/2}} \left( 1 - \dfrac{3z^2}{Q} \right), \\
        \frac{\partial^2 V_S}{\partial x \partial y} &= -3GM_S \dfrac{xy}{Q^{5/2}}, \\
        \frac{\partial^2 V_S}{\partial x \partial z} &= -3GM_S \dfrac{xz}{Q^{5/2}}, \\
        \frac{\partial^2 V_S}{\partial y \partial z} &= -3GM_S \dfrac{yz}{Q^{5/2}}.
    \end{aligned}
\end{equation}

\section{The disk potential}
The Miyamoto-Nagai potential \( V_D(x, y, z) \) is defined as:

\begin{equation} \label{eq:Disk Pot}
    V_D(x, y, z) = -\frac{GM_D}{\sqrt{x^2 + y^2 + \left(A + \sqrt{B^2 + z^2}\right)^2}},
\end{equation}
where \( M_D \) represents the mass associated with the disk, while \(A\) and \(B\), respectively, are the horizontal and vertical scale lengths of the system (see Sect.~\ref{section:HamiltonianCh4} for more details).

To compute the first and second partial spatial derivatives of the disk potential \( V_D\) \eqref{eq:Disk Pot}, let us denote the quantity in the square root of the denominator as \(Q = x^2 + y^2 + (A + \sqrt{B^2 + z^2})^2\). Then, the first derivatives are given by:
\begin{equation}
    \begin{aligned}
        \frac{\partial V_D}{\partial x} &= GM_D  \dfrac{x}{ Q^{3/2}}, \\
        \frac{\partial V_D}{\partial y} &= GM_D  \dfrac{y}{ Q^{3/2}}, \\
        \frac{\partial V_D}{\partial z} &= GM_D  \dfrac{z  \left( A + \sqrt{B^2 + z^2} \right)}{ Q^{3/2}  \sqrt{B^2 + z^2}},
    \end{aligned}
    \label{eq:FirstDerivativesDisk}
\end{equation}

while the second spatial derivatives are given by:
\begin{equation}\label{eq:Second Derivatives V_D}
    \begin{aligned}
        \frac{\partial^2 V_D}{\partial x^2} &= \dfrac{GM_D}{Q^{3/2}}  \left( 1 - \dfrac{3x^2}{Q^{5/2}} \right), \\
        \frac{\partial^2 V_D}{\partial y^2} &= \dfrac{GM_D}{Q^{3/2}}  \left( 1 - \dfrac{3y^2}{Q^{5/2}} \right), \\
        \frac{\partial^2 V_D}{\partial z^2} &= \dfrac{GM_D}{Q^{3/2}} \left[ \dfrac{A + \sqrt{B^2 + z^2}}{\sqrt{B^2 + z^2}} - \dfrac{3 z^2}{Q} \left( 1 + \dfrac{A}{\sqrt{B^2 + z^2}} \right)^2 - \dfrac{A z^2}{(B^2 + z^2)^{3/2}} \right], \\
        \frac{\partial^2 V_D}{\partial x \partial y} &= -3 GM_D \dfrac{xy}{Q^{5/2}}, \\
        \frac{\partial^2 V_D}{\partial x \partial z} &= -3 GM_D \dfrac{xz}{Q^{5/2}} \left( 1 + \dfrac{A}{\sqrt{B^2 + z^2}} \right), \\
        \frac{\partial^2 V_D}{\partial y \partial z} &= -3 GM_D \dfrac{yz}{Q^{5/2}} \left( 1 + \dfrac{A}{\sqrt{B^2 + z^2}} \right).
    \end{aligned}
\end{equation}

\section{The Ferrers bar potential} \label{sec:Ferrers Bar Pot}
\subsection{Introduction}
A specific form of the Ferrers bar potentials with  homogeneity degree \(n = 2\) is defined as \( V_B \) in Eq.~\eqref{eq:Vbar}, with its various components described in Sect.~\ref{section:HamiltonianCh4}. Here, we will first discuss the general form of this potential. Following the Appendix of \citep{pfenniger19843d}, the general integral formation for the bar potential: 

\begin{equation}
    \label{eq:Gen Bar Pot}
    \Phi = \frac{C}{n+1} \int_\lambda^\infty \dfrac{(1 - m^2(u))^{n+1}}{\Delta(u)} \, du,
\end{equation}
where \( M_B \) represents the total mass of the bar components, and \(n\) is the homogeneity degree, which essentially determines the bar's shape and density distribution. We have subsumed all prefactor constants into the single value \(C = -\pi G a b c \rho_c\), with the central density \(\rho_c = \dfrac{105}{32\pi} \dfrac{GM_B}{abc}\) and \(a > b > c\) being the semi-axes. In the expression of the potential in \eqref{eq:Gen Bar Pot}, we have:
\begin{equation}
    \label{eq:defm}
    m^2(u) = \frac{x^2}{a^2 + u} + \frac{y^2}{b^2 + u} + \frac{z^2}{c^2 + u},
\end{equation}
and
\begin{equation}
    \label{eq:defDelta}
    \Delta^2(u) = (a^2 + u)(b^2 + u)(c^2 + u).
\end{equation}
The parameter \( \lambda \) in \eqref{eq:Gen Bar Pot} is defined as the unique positive solution of 
\begin{equation} \label{eq:Bar lambda}
    m^2(\lambda) = 1,
\end{equation}
for \( m \geq 1 \) (outside the bar) and is set to zero for \( m < 1 \) (inside the bar). The reader is referred to Sect.~\ref{section:HamiltonianCh4} for more details. 

Note that the potential \( \Phi \) \eqref{eq:Gen Bar Pot} depends on the particle's position \((x, y, z)\) and the parameter \( \lambda \). However, differentiating \( \Phi \) with respect to \(\lambda\) gives a term \((1 - m^2(\lambda))^{n+1}\), which is zero according to \eqref{eq:Bar lambda}. Therefore, the potential satisfies a useful condition: \(\dfrac{\partial \Phi}{\partial \lambda} = 0\). 

In order to simplify the expression of the potential \eqref{eq:Gen Bar Pot}, we begin by applying a multinomial expansion of the term \((1 - m^2(u))^{n+1}\) to get
\begin{equation}\label{eq:Bar Pot multinomial_exp}
    \Phi = \dfrac{C}{n+1} \int_\lambda^\infty \dfrac{ \sum_{p=0}^{n+1} \binom{n+1}{p} (-m^2(u))^p}{\Delta(u)} \, du.
\end{equation}
The expression in \eqref{eq:Bar Pot multinomial_exp} simplifies to
\begin{equation} \label{eq:Bar Pot distributed_integral}
\Phi = \frac{C}{n+1} \sum_{p=0}^{n+1} \binom{n+1}{p} (-1)^p \int_\lambda^\infty \dfrac{(m^2(u))^p}{\Delta(u)} \, du.
\end{equation}

Substituting the expression for \(m^2(u)\) from \eqref{eq:defm} in \eqref{eq:Bar Pot distributed_integral}, we obtain: 
\begin{equation}
\Phi = \frac{C}{n+1} \sum_{p=0}^{n+1} \binom{n+1}{p} (-1)^p \int_\lambda^\infty \dfrac{\left( \frac{x^2}{a^2 + u} + \frac{y^2}{b^2 + u} + \frac{z^2}{c^2 + u} \right)^p}{\Delta(u)} \, du.
\label{eq:Bar Pot distributed_integral 2}
\end{equation}

Implementing the multinomial theorem, we get 
\begin{equation}
\left( \frac{x^2}{a^2 + u} + \frac{y^2}{b^2 + u} + \frac{z^2}{c^2 + u} \right)^p = \sum_{j+k+m=p} \frac{p!}{j! k! m!} \left( \frac{x^2}{a^2 + u} \right)^j \left( \frac{y^2}{b^2 + u} \right)^k \left( \frac{z^2}{c^2 + u} \right)^m,
\label{eq:multinomial_expansion m2}
\end{equation}
where the summation is taken over all non-negative integers \(j\), \(k\), and \(m\) such that \(j + k + m = p\), ensuring that each term in the expansion is consistent with the order \(p\).

Substituting the expanded form of \eqref{eq:multinomial_expansion m2} into the expression for \(\Phi\) in \eqref{eq:Bar Pot distributed_integral 2} yields:
\begin{equation} \label{eq:Bar Pot substituted_Phi}
\Phi = \frac{C}{n+1} \sum_{p=0}^{n+1} \binom{n+1}{p} (-1)^p \sum_{j+k+m=p} \dfrac{p!}{j! k! m!} \int_\lambda^\infty \frac{x^{2j} y^{2k} z^{2m}}{(a^2 + u)^{j} (b^2 + u)^{k} (c^2 + u)^{m} \Delta(u)} \, du.
\end{equation}
Introducing the new variable \(q = n + 1 - p\), we can combine the two sums in \eqref{eq:Bar Pot substituted_Phi} to get
\begin{equation}
\Phi = \frac{C}{n+1} \sum_{q=0}^{n+1} \binom{n+1}{q} (-1)^{n+1-q} \sum_{j+k+m+q=n+1} \dfrac{(n+1-q)!}{j! k! m!} \int_\lambda^\infty \frac{x^{2j} y^{2k} z^{2m}}{(a^2 + u)^{j} (b^2 + u)^{k} (c^2 + u)^{m} \Delta(u)} \, du.
\label{eq:combined_sums}
\end{equation}
Now, combining the sums in \eqref{eq:combined_sums} leads to a simplified form for \(\Phi\):
\begin{equation} \label{eq:combined_sum_final}
\Phi = \frac{C}{n+1} \sum_{j+k+m+q=n+1} \dfrac{(n+1)!}{j! k! m! q!} (-1)^{n+1-q} \int_\lambda^\infty 
 \frac{x^{2j} y^{2k} z^{2m}}{(a^2 + u)^{j} (b^2 + u)^{k} (c^2 + u)^{m} \Delta(u)} \, du. 
\end{equation}
Equation \eqref{eq:combined_sum_final} can be interpreted as defining the potential integral \(\Phi\) \eqref{eq:Gen Bar Pot} in terms of the functions \( W_{jkm} \),
\begin{equation}\label{eq:Wjkm}
    W_{jkm} = \int_\lambda^\infty \frac{du}{\Delta u} \frac{1}{\left(a^2+u\right)^j}\frac{1}{\left(b^2+u\right)^k}\frac{1}{\left(c^2+u\right)^m}.
\end{equation}
Thus, we can express the bar potential \(\Phi\) \eqref{eq:Gen Bar Pot} in a compact form as follows:
\begin{equation} \label{eq:Bar pot compact form}
    \displaystyle \Phi =  C \sum_{j+k+m+q=n+1} \frac{n!}{j! k! m! q!} (-1)^{n-1} \left( x^{2j} y^{2k} z^{2m} \right) W_{jkm}. 
\end{equation}
The terms $W_{jkm}$ satisfy the following:
\begin{equation}
    \label{eq: Wjkm recur. relation}
    \begin{aligned}
        W_{jkm} &= \dfrac{W_{j-1,k,m} - W_{j,k-1,m}}{a^2 - b^2}, \\
                       &= \dfrac{W_{j,k-1,m} - W_{j,k,m-1}}{b^2 - c^2}, \\
                       &= \dfrac{W_{j,k,m-1} - W_{j-1,k,m}}{c^2 - a^2}.
    \end{aligned}
\end{equation}
Using the general recurrence relations \eqref{eq: Wjkm recur. relation} and applying integration by parts in \eqref{eq:Wjkm} leads to   
\begin{equation} \label{eq: Wjkm Integ p}
\begin{aligned} 
    W_{n00} &= \frac{1}{2n-1} \left[\frac{2}{\Delta(\lambda)\left(a^2+\lambda \right)^{n-1}} - W_{n-1,1,0} - W_{n-1,0,1} \right], \\
    W_{0n0} &= \frac{1}{2n-1} \left[\frac{2}{\Delta(\lambda)\left(b^2+\lambda \right)^{n-1}} - W_{1,n-1,0} - W_{0,n-1,1} \right], \\
    W_{00n} &= \frac{1}{2n-1} \left[\frac{2}{\Delta(\lambda)\left(c^2+\lambda \right)^{n-1}} - W_{1,0,n-1} - W_{0,1,n-1} \right].
\end{aligned}    
\end{equation}

For $n=1$ (inhomogeneous bar distribution), the recurrence relation \eqref{eq: Wjkm Integ p} is reduced to 
\begin{equation} \label{eq: Wjkm Integ n=1}
\begin{aligned}
    W_{100} &= 2\left[\frac{1}{\Delta(\lambda)} - W_{0,1,0} - W_{0,0,1}\right], \\
    W_{010} &= 2\left[\frac{1}{\Delta(\lambda)} - W_{1,0,0} - W_{0,0,1}\right], \\
    W_{001} &= 2\left[\frac{1}{\Delta(\lambda)} - W_{1,0,0} - W_{0,1,0}\right].
\end{aligned}
\end{equation}
Thus, by adding the terms in \eqref{eq: Wjkm Integ n=1} we get 
\begin{equation}
    W_{100}+W_{010}+W_{001}=\frac{2}{\Delta(\lambda)}
\end{equation}
Determining \(W_{000}\), \(W_{100}\), \(W_{010}\), and \(W_{001}\) is possible using the incomplete elliptic integrals $F(\theta,q)$ and $E(\theta,q)$. The definitions of the angle \(\theta\) and the modulus \(q\) related to these integrals are give by:
\begin{align}
    \label{eq:defTheta}\sin\left(\theta\right) &= \left(\frac{a^2-c^2}{a^2+\lambda}\right)^{1/2},\\
    \label{eq:defQ}q &= \left(\frac{a^2-b^2}{a^2-c^2}\right)^{1/2}.
\end{align}
Now, the relations for \(W_{000}\), \(W_{100}\), \(W_{010}\), and \(W_{001}\) in terms of the incomplete elliptic integrals can be expressed as follows:
\begin{align}
    W_{000} &= \frac{2}{\sqrt{a^2 - c^2}} F(\theta, q), \\
    W_{100} &= \frac{2}{(a^2 - b^2) \sqrt{a^2 - c^2}} \left[ F(\theta, q) - E(\theta, q) \right], \\
    W_{010} &= \frac{2 \sqrt{a^2 - c^2}}{(a^2 - b^2)(b^2 - c^2)} E(\theta, q) - \frac{2}{(a^2 - b^2)(a^2 - c^2)} F(\theta, q) - \frac{2}{b^2 - c^2} \sqrt{\frac{c^2 + \lambda}{(a^2 + \lambda)(b^2 + \lambda)}}, \nonumber \\
    &= \frac{2}{\Delta(\lambda)} - W_{100} - W_{001}, \\
    W_{001} &= \frac{2}{b^2 - c^2} \sqrt{\frac{b^2 + \lambda}{(a^2 + \lambda)(c^2 + \lambda)}} - \frac{2}{(b^2 - c^2) \sqrt{a^2 - c^2}} E(\theta, q).
\end{align}
The remaining $W_{jkm}$'s can be successfully computed using the recurrence relations provided in \eqref{eq: Wjkm recur. relation} and \eqref{eq: Wjkm Integ p}. For the specific case of \(n=2\), the $W_{jkm}$'s up to the third order can also be directly obtained from the Appendix of \citep{pfenniger19843d}. It is worth noting that the form of the potential in Eq.~\eqref{eq:Bar pot compact form} for \(n=2\) is essentially equivalent to the bar potential \(V_B\) \eqref{eq:Vbar} considered in our study, which has been extensively studied by many researchers (e.g. see \citep{pfenniger19843d,skokos2002orbitala,skokos2002orbitalb,patsis2014phasea,manos2022orbit}). 

\clearpage
For $n=2$ (inhomogeneous bar distribution), the potential \eqref{eq:Bar pot compact form} simplifies to the final form of \(V_B\) \eqref{eq:Vbar}, which we aim to study, and reduces to:
\begin{equation}
\label{eq:final bar Potential}
\begin{split}
    V_B &= \frac{C}{6} \Bigg\{ W_{000} - 6x^2y^2z^2 W_{111} \\
    & \quad + x^2 \bigg[ x^2 \big( 3W_{200} - x^2 W_{300} \big) \\
    & \quad + 3 \big( y^2 \big( 2W_{110} - y^2 W_{120} - x^2 W_{210} \big) - W_{100} \big) \bigg] \\
    & \quad + y^2 \bigg[ y^2 \big( 3W_{020} - y^2 W_{030} \big)\\
    & \quad + 3 \big( z^2 \big( 2W_{011} - z^2 W_{012} - y^2 W_{021} \big) - W_{010} \big) \bigg] \\
    & \quad + z^2 \bigg[ z^2 \big( 3W_{002} - z^2 W_{003} \big) \\
    & \quad + 3 \big( x^2 \big( 2W_{101} - x^2 W_{201} - z^2 W_{102} \big) - W_{001} \big) \bigg] \Bigg\}.
\end{split}
\end{equation}

\subsection{Derivatives of the potential}
In order to write out the derivatives of the general Ferrers bar potential, we start by considering its compact form \eqref{eq:Bar pot compact form} [which is practically \(V_B\) \eqref{eq:Vbar}]. This potential can be expressed as the product of two components: the algebraic part \(\phi_A = x^{2j} y^{2k} z^{2m}\) (representing the scaled product of \(x\), \(y\), and \(z\) raised to certain powers) and the contribution from the \(W_{jkm}\) \eqref{eq:Wjkm}, which we will refer to as \(\phi_W\).

The partial spatial derivatives of \(\phi_A\) can be computed directly. However, for \(\phi_W\), we must consider that the \(W_{jkm}\) quantities depend on \(\lambda\). Therefore, when computing the derivatives with respect to the spatial coordinates \(x\), \(y\), and \(z\), we apply the chain rule such as 

\begin{equation} \label{eq:partial_derivative_phi_W}
\dfrac{\partial W_{jkm}}{\partial x_j} = \dfrac{\partial W_{jkm}}{\partial \lambda}  \dfrac{\partial \lambda}{\partial x_j}.
\end{equation}
The derivative \( \dfrac{\partial W_{jkm}}{\partial \lambda} \) is easy to be calculated as follows because \( W_{jkm} \) \eqref{eq:Wjkm} is defined as an integral having as lower limit \( \lambda \) 

\begin{equation} \label{eq:derivativeW}
    \dfrac{\partial W_{jkm}}{\partial \lambda} = 
    \left( \frac{1}{(a^2 + \lambda)^{j + 1/2}} \right)
    \left( \frac{1}{(b^2 + \lambda)^{k + 1/2}} \right)
    \left( \frac{1}{(c^2 + \lambda)^{m + 1/2}} \right).
\end{equation}
To find the spatial derivatives of \( \lambda \), we use implicit differentiation on the definition given by Eq.~\ref{eq:defm}, i.e., \( m{(\lambda)} = 1 \) for \( m \geq 1 \), for each coordinate. This process eventually leads us to the following results:

\begin{equation} \label{eq:lambda_derivatives}
\begin{aligned} 
    \dfrac{\partial \lambda}{\partial x} &= \dfrac{2x}{a^2 + \lambda}  \left[ \frac{x^2}{(a^2 + \lambda)^2} + \frac{y^2}{(b^2 + \lambda)^2} + \frac{z^2}{(c^2 + \lambda)^2} \right]^{-1}, \\
    \dfrac{\partial \lambda}{\partial y} &= \dfrac{2y}{b^2 + \lambda}  \left[ \frac{x^2}{(a^2 + \lambda)^2} + \frac{y^2}{(b^2 + \lambda)^2} + \frac{z^2}{(c^2 + \lambda)^2} \right]^{-1}, \\
     \dfrac{\partial \lambda}{\partial z} &= \dfrac{2z}{c^2 + \lambda}  \left[ \frac{x^2}{(a^2 + \lambda)^2} + \frac{y^2}{(b^2 + \lambda)^2} + \frac{z^2}{(c^2 + \lambda)^2} \right]^{-1}. 
\end{aligned}
\end{equation}

We can use the fact that \(\dfrac{\partial V_B}{\partial \lambda} = 0\), and by employing the shorthand notations we introduced (\(\phi_A\) and  \(\phi_W\)), we can apply the product rule to the derivative of \(V_B\) with respect to \( x \)

\begin{equation} \label{eq:partial_phi}
        \begin{aligned}
            \dfrac{\partial V_B}{\partial x} &= C \sum \dfrac{\partial \phi_A}{\partial x} \phi_W + C \sum \phi_A \dfrac{\partial \phi_W}{\partial \lambda} \dfrac{\partial \lambda}{\partial x}, \\
            &= C \sum \dfrac{\partial \phi_A}{\partial x} \phi_W + \dfrac{\partial \lambda}{\partial x} \left( \dfrac{\partial }{\partial \lambda} \left( C \sum \phi_A \phi_W \right) \right), \\
            &= C \sum \dfrac{\partial \phi_A}{\partial x} \phi_W + \dfrac{\partial \lambda}{\partial x} \left( \dfrac{\partial V_B}{\partial \lambda} \right), \\
            &= C \sum \dfrac{\partial \phi_A}{\partial x} \phi_W, \\
            &= C \sum_{j+k+m+q=n+1} \dfrac{n!}{j! k! m! q!} (-1)^{n-1} (2j) x^{2j-1} y^{2k} z^{2m} W_{jkm}.
        \end{aligned}
        \end{equation}

Naturally, the derivatives with respect to \(y\) and \(z\) are obtained following a similar approach. However, the evolution of the second derivative presents a challenge because the \(\dfrac{\partial V_B}{\partial \lambda}\) derivative does not vanish; therefore, we must include the derivative with respect to \(\lambda\) term. Taking the mixed second-ordered derivative \(\dfrac{\partial^2 V_B}{\partial x \partial y}\) as an example, we have:
\begin{equation} \label{eq:cross_derivative}
        \begin{aligned}
            \frac{\partial^2 V_B}{\partial x \partial y} &= \frac{\partial}{\partial y} \left(C \sum \frac{\partial \phi_A}{\partial x} \phi_W\right), \\
            &= C \sum \frac{\partial^2 \phi_A}{\partial x \partial y} \phi_W + C \sum \frac{\partial \phi_A}{\partial x} \frac{\partial \phi_W}{\partial \lambda} \frac{\partial \lambda}{\partial y}, \\
            &= C \sum_{j+k+m+q=n+1} \dfrac{n!}{j! k! m! q!} (-1)^{n-1} (2j)(2k) x^{2j-1} y^{2k-1} z^{2m} W_{jkm} \\ 
            & \quad + C \frac{\partial \lambda}{\partial y} \sum_{j+k+m+q=n+1} \dfrac{n!}{j! k! m! q!} (-1)^{n-1} (2j) x^{2j-1} y^{2k} z^{2m} \frac{\partial W_{jkm}}{\partial \lambda}.
        \end{aligned}
        \end{equation}
The same procedure can be implemented to find the remaining derivatives, and we end up with a complete set of all possible derivatives. Given the framework we have established, calculating the derivatives under the specific constraints is relatively simple. This process can also be carried out using computational packages such as Wolfram Mathematica. 

We would like to emphasize that the first \eqref{eq:Bar Potential FD} and second \eqref{eq:Bar Potential SD} derivatives of the Ferrers bar potential \eqref{eq:Bar pot compact form} have been previously obtained, notably in the Appendices of the PhD theses of \citep{gomez2007role} and \citep{manos2008study}. Nevertheless, for the sake of completeness, we included in our study the derivation. Thus, these first and second derivatives of the potential \(V_B\) \eqref{eq:Vbar} are: 
\begin{equation}\label{eq:Bar Potential FD}
    \begin{aligned}
\frac{\partial V_B}{\partial x} &= - C x \Big[ W_{100}(\lambda) - 2z^2 W_{101}(\lambda) + z^4 W_{102}(\lambda) 
- 2y^2 W_{110}(\lambda) + 2y^2 z^2 W_{111}(\lambda) \\
&\quad + y^4 W_{120}(\lambda) - 2x^2 W_{200}(\lambda) + 2x^2 z^2 W_{201}(\lambda) 
+ 2x^2 y^2 W_{210}(\lambda) + x^4 W_{300}(\lambda) \Big], \\
\frac{\partial V_B}{\partial y} &= - C y \Big[ W_{010}(\lambda) - 2z^2 W_{011}(\lambda) + z^4 W_{012}(\lambda) 
- 2y^2 W_{020}(\lambda) + 2y^2 z^2 W_{021}(\lambda) \\
&\quad + y^4 W_{030}(\lambda) - 2x^2 W_{110}(\lambda) + 2x^2 z^2 W_{111}(\lambda) 
+ 2x^2 y^2 W_{120}(\lambda) + x^4 W_{210}(\lambda) \Big], \\
\frac{\partial V_B}{\partial z} &= - C z \Big[ W_{001}(\lambda) - 2z^2 W_{002}(\lambda) + z^4 W_{003}(\lambda) 
- 2y^2 W_{011}(\lambda) + 2y^2 z^2 W_{012}(\lambda) \\
&\quad + y^4 W_{021}(\lambda) - 2x^2 W_{101}(\lambda) + 2x^2 z^2 W_{102}(\lambda) 
+ 2x^2 y^2 W_{111}(\lambda) + x^4 W_{201}(\lambda) \Big].
\end{aligned}
\end{equation}

\begin{equation}\label{eq:Bar Potential SD}
    \begin{aligned}
    \frac{\partial^2 V_B}{\partial x^2} &= - C \bigg\{W_{100}(\lambda) - 2z^2 W_{101}(\lambda) + z^4 W_{102}(\lambda) - 2y^2 W_{110}(\lambda) + 2y^2 z^2 W_{111}(\lambda) + y^4 W_{120}(\lambda) \\
    &\quad - 6x^2 W_{200}(\lambda) + 6x^2 z^2 W_{201}(\lambda) + 6x^2 y^2 W_{210}(\lambda) + 5x^4 W_{300}(\lambda) \\
    &\quad + \frac{\partial \lambda}{\partial x} \bigg[ x W_{100}'(\lambda) - 2xz^2 W_{101}'(\lambda) + xz^4 W_{102}'(\lambda) - 2xy^2 W_{110}'(\lambda) + 2xy^2 z^2 W_{111}'(\lambda) \\
    &\quad + xy^4 W_{120}'(\lambda) - 2x^3 W_{200}'(\lambda) + 2x^3 z^2 W_{201}'(\lambda) + 2x^3 y^2 W_{210}'(\lambda) + x^5 W_{300}'(\lambda) \bigg]\bigg\}, \\
    \frac{\partial^2 V_B}{\partial y^2} &= - C \bigg\{W_{010}(\lambda) - 2z^2 W_{011}(\lambda) + z^4 W_{012}(\lambda) - 6y^2 W_{020}(\lambda) + 6y^2 z^2 W_{021}(\lambda) + 5y^4 W_{030}(\lambda) \\
    &\quad - 2x^2 W_{110}(\lambda) + 2x^2 z^2 W_{111}(\lambda) + 6x^2 y^2 W_{120}(\lambda) + x^4 W_{210}(\lambda) \\
    &\quad + \frac{\partial \lambda}{\partial y} \bigg[ y W_{010}'(\lambda) - 2yz^2 W_{011}'(\lambda) + yz^4 W_{012}'(\lambda) - 2y^3 W_{020}'(\lambda) \\
    &\quad + 2y^3 z^2 W_{021}'(\lambda) + y^5 W_{030}'(\lambda) - 2x^2 y W_{110}'(\lambda) + 2x^2 y z^2 W_{111}'(\lambda) + 2x^2 y^3 W_{120}'(\lambda) + x^4 y W_{210}'(\lambda) \bigg]\bigg\}. \\
    \frac{\partial^2 V_B}{\partial z^2} &= - C \biggl\{ W_{001}(\lambda) - 6z^2 W_{002}(\lambda) + 5z^4 W_{003}(\lambda) - 2y^2 W_{011}(\lambda) + 6y^2 z^2 W_{012}(\lambda) \\
        &\quad + y^4 W_{021}(\lambda) - 2x^2 W_{101}(\lambda) + 6x^2 z^2 W_{102}(\lambda) + 2x^2 y^2 W_{111}(\lambda) + x^4 W_{201}(\lambda)  \\
        &\quad + \frac{\partial \lambda}{\partial z} \bigg[ z W_{001}'(\lambda) - 2z^3 W_{002}'(\lambda) + z^5 W_{003}'(\lambda) - 2y^2 z W_{011}'(\lambda) + 2y^2 z^3 W_{012}'(\lambda) \\
        &\quad + y^4 z W_{021}'(\lambda) - 2x^2 z W_{101}'(\lambda) + 2x^2 z^3 W_{102}'(\lambda) + 2x^2 y^2 z W_{111}'(\lambda) + x^4 z W_{201}'(\lambda) \bigg] \biggr\},\\
         \frac{\partial^2 V_B}{\partial x \partial y} &= C x \biggl\{ 4y W_{110}(\lambda) - 4y z^2 W_{111}(\lambda) - 4y^3 W_{120}(\lambda) - 4x^2 y W_{210}(\lambda)  \\
        &\quad - \frac{\partial \lambda}{\partial y} \bigg[ W_{100}'(\lambda) - 2z^2 W_{101}'(\lambda) + z^4 W_{102}'(\lambda) - 2y^2 W_{110}'(\lambda) + 2y^2 z^2 W_{111}'(\lambda) \\
        &\quad + y^4 W_{120}'(\lambda) - 2x^2 W_{200}'(\lambda) + 2x^2 z^2 W_{201}'(\lambda) + 2x^2 y^2 W_{210}'(\lambda) + x^4 W_{300}'(\lambda) \bigg] \biggr\},\\
        \frac{\partial^2 V_B}{\partial x \partial z} &= C x \biggl\{ 4z W_{101}(\lambda) - 4z^3 W_{102}(\lambda) - 4y^2 z W_{111}(\lambda) - 4x^2 z W_{201}(\lambda)  \\
        &\quad - \frac{\partial \lambda}{\partial z} \bigg[ W_{100}'(\lambda) - 2z^2 W_{101}'(\lambda) + z^4 W_{102}'(\lambda) - 2y^2 W_{110}'(\lambda) + 2y^2 z^2 W_{111}'(\lambda) \\
        &\quad + y^4 W_{120}'(\lambda) - 2x^2 W_{200}'(\lambda) + 2x^2 z^2 W_{201}'(\lambda) + 2x^2 y^2 W_{210}'(\lambda) + x^4 W_{300}'(\lambda) \bigg] \biggr\}, \\
        \frac{\partial^2 V_B}{\partial y \partial z} &= C y \biggl\{ 4z W_{011}(\lambda) - 4z^3 W_{012}(\lambda) - 4y^2 z W_{021}(\lambda) - 4x^2 z W_{111}(\lambda) \\
        &\quad - \frac{\partial \lambda}{\partial z} \bigg[ W_{010}'(\lambda) - 2z^2 W_{011}'(\lambda) + z^4 W_{012}'(\lambda) - 2y^2 W_{020}'(\lambda) + 2y^2 z^2 W_{021}'(\lambda) \\
        &\quad + y^4 W_{030}'(\lambda) - 2x^2 W_{110}'(\lambda) + 2x^2 z^2 W_{111}'(\lambda) + 2x^2 y^2 W_{120}'(\lambda) + x^4 W_{210}'(\lambda) \bigg] \biggr\}.
    \end{aligned}
    \end{equation}
    
\clearpage



\end{document}